\documentclass[12pt,a4paper,fleqn,portuges,english,intoc,bibliography=totoc,index=totoc,BCOR10mm,captions=tableheading,titlepage,citecolor=blue,linkcolor=teal]{scrbook}
\usepackage{lmodern}

\usepackage[T1]{fontenc}
\usepackage[latin9]{inputenc}
\usepackage{fancyhdr}
\usepackage{xcolor}
\usepackage{babel}
\usepackage{wrapfig}
\usepackage{units}
\usepackage{textcomp}
\usepackage{mathrsfs}
\usepackage{pdfpages}
\usepackage{amsmath}
\usepackage{amssymb}
\usepackage{cancel}
\usepackage{graphicx}
\usepackage{wasysym}
\usepackage{esint}
\usepackage{nomencl}
\providecommand{\printnomenclature}{\printglossary}
\providecommand{\makenomenclature}{\makeglossary}
\makenomenclature
\usepackage[unicode=true,pdfusetitle,
 bookmarks=true,bookmarksnumbered=true,bookmarksopen=true,bookmarksopenlevel=1,
 breaklinks=true,pdfborder={0 0 0},pdfborderstyle={},backref=false,colorlinks=true]
 {hyperref}
\hypersetup{
 pdfpagelayout=OneColumn, pdfnewwindow=true, pdfstartview=XYZ, plainpages=false}
\pdfoutput=1
\makeatletter

\pdfpageheight\paperheight
\pdfpagewidth\paperwidth

\@ifundefined{date}{}{\date{}}

\usepackage[figure]{hypcap}
\usepackage{cite}
\let\myTOC\tableofcontents
\renewcommand\tableofcontents{%
  \frontmatter
  \pdfbookmark[1]{\contentsname}{}
  \myTOC
  \mainmatter }

\setkomafont{captionlabel}{\bfseries}
\setcapindent{1em}

\usepackage{calc}

\let\mySection\section\renewcommand{\section}{\suppressfloats[t]\mySection}

\usepackage{tikz-feynman}
\tikzfeynmanset{compat=1.0.0}

\makeatother

\addto\captionsenglish{%
}
\addto\captionsportuges{%
}

\begin{document}


\subject{{\normalsize{}Thesis Submitted in Fulfilment of the Requirements for
the Degree of }\textit{\normalsize{}PHILOSOPHIÆ DOCTOR}{\normalsize{}
in PHYSICS}}
\publishers{\vspace{-4cm}
}
\title{Quantum Effects of Impurities and Lattice Defects in Topological Semimetals}
\publishers{\vspace{-4cm}
}
\author{João Pedro dos Santos Pires}
\publishers{\vspace{-3cm}
}
\date{\today}
\publishers{\hspace{-1.0cm}%
\begin{minipage}[t]{0.4\columnwidth}%
\includegraphics[scale=0.5]{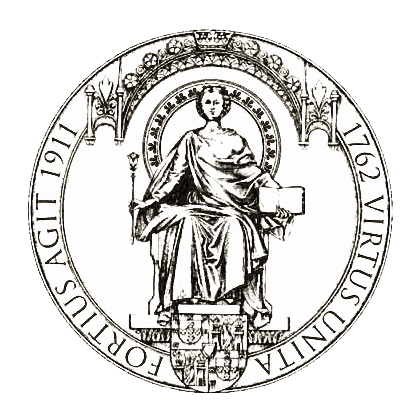}%
\end{minipage}\vspace{\baselineskip}\\
Universidade do Porto\\
{\normalsize{}Departamento de Física e Astronomia, Faculdade de Ciências}\\
{\normalsize{}Laboratory of Physics for Materials and Emergent Technologies
LaPMET}\vspace{-3cm}
}
\lowertitleback{\textbf{Chair:}\smallskip{}
\\
Prof. João M. Borregana Lopes dos Santos (Universidade do Porto, Portugal)\bigskip{}
\\
\textbf{Supervisors:}\smallskip{}
\\
Prof. João Manuel Viana Parente Lopes (Universidade do Porto, Portugal)\smallskip{}
\\
Dr. Bruno António Campos Amorim (Universidade do Minho, Portugal)\bigskip{}
\\
\textbf{Referees:}\smallskip{}
\\
Prof. Jedediah H. Pixley (Rutgers University, New Jersey, USA)\smallskip{}
\\
Prof. Miguel A. da Nova Araújo (Universidade de Évora, Portugal)\smallskip{}
\\
Prof. Alexander Altland (Universität zu Köln, Germany)\smallskip{}
\\
Prof.\,\,Maria\,\,A.\,H.\,Vozmediano\,\,(Consejo\,\,Superior\,\,de\,\,Investigaciones\,\,Científicas,\,\,Spain)\smallskip{}
\\
Prof. Ricardo G. Dias (Universidade de Aveiro, Portugal)\smallskip{}
}
\dedication{\textit{To My Late Father,}}

\maketitle
\cleardoublepage{}

\lhead{\MakeUppercase{Table of Contents}}

\rhead{}

\lfoot[\thepage]{}

\cfoot[]{}

\rfoot[]{\thepage}

\hypersetup{linkcolor=black}{\tableofcontents{}}

\hypersetup{linkcolor=teal}

\cleardoublepage{}

\lhead[]{Acknowledgments}

\cfoot[]{}

\rhead[\MakeUppercase{Acknowledgments}]{}

\chapter*{Acknowledgments}

\addcontentsline{toc}{chapter}{Acknowledgments} 

\vspace{-0.3cm}

A Ph.D. is a 4-year long marathon that can only be surpassed through
much resilience and hard-work, amidst a fair share of good luck. Mine
was not an exception but, in spite my own personal victories, there
are a number of acknowledgments that are due to people without whom
this journey would not have been as fruitful.

My biggest acknowledgement is to \textit{João Viana Lopes} for his
close supervision, constant support and the almost daily fruitful
scientific exchange of ideas. For the past four years, he has been
my closest collaborator and I have learned a great deal with him,
not only about hard-core physics, but also on how to manage expectations,
problematize and think creatively on the solution of any problem that
may come my way. He provided me with the invaluable opportunity to
take part in the growth of a new scientific group, open doors for
me to establish external collaborations and gave me the chance to
experience teaching at the University level. I am also grateful to
my co-supervisor, \textit{Bruno Amorim}, for his scientific collaboration
on the work here presented and the constructive criticism of my writtings,
presentations and scientific texts (including this thesis). The points
he implacably raised were pivotal to boost my development as a future
scientist. Besides my supervisors, there were also other collaborators
who I thank for discussing or working with me in different stages
of my research, namely: \textit{António Antunes}, \textit{Branislav
Nikoli\'{c}, Caio Lewenkopf},\textit{ Catarina Monteiro}, \textit{Dave
Perkins},\textit{ Denis Kochan, Diogo Pinheiro, Eduardo V. Castro,
Ferdinand Evers}, \textit{Hugo Terças},\textit{ \.{I}nanç Adagideli},\textit{
João M. B. Lopes dos Santos},\textit{ João Santos Silva},\textit{
Johannes Dieplinger},\textit{ Mateo González-Moreno},\textit{ João
Alendouro Pinho},\textit{ Henrique Pina Jorge},\textit{ Miguel Boultwood},\textit{
Miguel Mestre Gonçalves},\textit{ Miguel Oliveira},\textit{ Niaz Ali
Khan},\textit{ Nuno M. R. Peres},\textit{ Pedro Ribeiro},\textit{
Rafael Soares},\textit{ Ricardo Dia}s,\textit{ Roberto Raimondi, Orlando
Frazão},\textit{ Tiago Gonçalves}, and\textit{ Tatiana G. Rappoport}.
A particular thanks is directed towards \textit{Aires Ferreira} (University
of York), \textit{Eduardo R. Mucciolo} (University of Central Florida)
and \textit{Alexander Altland} (University of Cologne) for receiving
me in their respective groups as a short-term visitor. \textit{Simão
João} also holds a special place, as my PhD companion and trustworthy
collaborator.

On a more personal level, I also deeply thank \textit{Patrícia} for
her love, joyfulness and support during the hardest times of this
marathon. Without her, it would have certainly been more difficult,
if not impossible, for me to finish it successfully. I also extend
this appreciation to my mother and sister, who supported me throughout
my entire academic life, without asking anything in return. Finally,
I would like thank, hitherto unmentioned, personal friends for their
encouragement, most notably \textit{Anna Haan},\textit{ Anna Di Donato},\textit{
Bárbara Bacelar}, \textit{Carlos Fernandes},\textit{ David Lima},\textit{
Diogo Melo Pinto},\textit{ Fred Gonçalves},\textit{ Gonçalo Roboredo},\textit{
Gilberto Loureiro},\textit{ José Matos}, \textit{Maria Ramos}, \textit{Sérgio
Quelho, Sofia Maia},\textit{ Pedro Sousa Melo}, and\textit{ $\Pi$
Martins}.

At last, I acknowledge financial support from Fundação da Ciência
e Tecnologia (FCT) through the Strategic Funding No.\,\textcolor{olive}{UIDB/04650/2020},
his Personal PhD Grant No.\,\textcolor{olive}{PD/BD/142774/2018}
and Project No.\,\textcolor{olive}{POCI-01-0145-FEDER-028887}. The
numerical calculations presented in this work were performed in the
GRID FEUP High Performance Computing (HPC) infrastructure, the National
Network for Advanced Computing (RNCA) through FCT project \textcolor{olive}{CPCA/A0/7308/2020},
and also in the Viking HPC cluster of the University of York, UK.
\begin{center}
\includegraphics[scale=0.22]{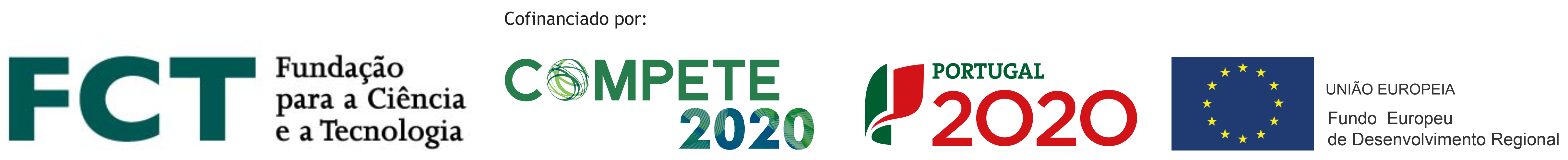}
\par\end{center}

\clearpage{}

\pagestyle{plain}

\chapter*{Abstract}

\addcontentsline{toc}{chapter}{Abstract} 

\vspace{-0.2cm}

Topological semimetals are a class of novel three-dimensional (3D)
electronic phases that feature topologically protected conical band-touchings
at the Fermi level. These band-touching points are monopoles of Berry
curvature in momentum space and effectively realize ($3\!+\!1)$-dimensional
Weyl fermions as emergent quasiparticles. Such features are robust
to perturbations but not completely insensitive to them. In this thesis,
we explore the yet fertile ground of disordered Weyl semimetals (WSMs),
most notably by analyzing the effects of on-site random fields, random
smooth potential regions, point-like scalar impurities, and lattice
point-defects in their electronic structure and electrodynamic properties.

Our starting point is the study of a Weyl (or Dirac) semimetal in
the presence of uncorrelated local random fields. At the mean-field
level, we obtain an \textit{unconventional disorder-induced critical
point}, which is characterized by a single-parameter scaling theory
for the mean density of states (DoS) near the node. Using diagrammatic
and field-theoretical formulations of the disordered single-electron
problem, we reproduce the known analytical results, and further confirm
them by real-space lattice simulations that improve over the accuracy
of published work. Despite seemingly confirmed by numerical evidence,
the reality of this disorder-induced quantum critical point have been
the subject of a long-standing debate in the literature. The controversy
lied on possible non-perturbative contributions arising from nodal
bound states that appear in statistically rare smooth regions of the
disorder landscape, leading to an \textit{avoided quantum criticality}
(AQC) scenario. To better investigate this effect, we examine a tailor-made
model of randomly placed spherical scatterers (of random strength)
that forces the existence of random smooth potential regions, enhancing
their effects in, and around, the nodal energy. Combining continuum
scattering theory with lattice simulations, we pinpoint a precise
stability criterion for the semi-metallic phase within this disorder
model, and further propose a physical mechanism that explains why
fine-tuned nodal bound states are endowed with statistical significance
for the nodal DoS. While these conclusions seem to be at odds with
the lack of evidence for AQC found in our unbiased simulations in
truly disordered WSM lattices, we are still able to reconcile them.
The key point to understand this discrepancy is to re-examine at the
AQC phenomena from a mesoscopic point-of-view, in which the nodal
states can appear as bound states of a small number of local potential
fluctuations. From this analysis, we conclude that rare-event states
in a lattice that hosts a random Anderson potential, may arise by
two main mechanisms: \textit{(i)} smooth regions of a few adjacent
sites that trap nodal electrons, and \textit{(ii)} the hybridization
of a large potential fluctuation with its disordered environment.
Both are of these suppressed for the bounded on-site disorder distributions
used in our early numerics, which explains their absence in our results\@.
Thereby, the phenomenon of AQC is to be seen as a \textit{non-universal
disorder effect}, that is actually a property of specific disorder
models.

In the last part of this work, we present the first theoretical study
of opto-electronics and charge transport in WSMs with lattice vacancies;
A common type of disorder in real semimetal samples, but which have
been so far overlooked in the literature. Unlike what we have seen
in the previous cases, this type of disorder is shown to \textit{strongly
enhance} the DoS at the Weyl node, further endowing it with a \textit{comb
of quasi-localized resonances} caused by inter-vacancy hybridization.
These resonances are shown to be insensitive to magnetic fields and,
importantly, have a strongly suppressed quantum diffusivity that leads
to a previously unseen oscillatory dependence of the bulk dc conductivity
on the charge carrier density. Moreover, the optical response of a
slightly doped semimetal is also affected by vacancy-induced states,
giving rise to a plateau-shaped dissipative response, below the inter-band
threshold, which is proportional to the vacancy concentration. The
predicted transport and optical signatures provide realistic ways
of experimentally assessing the existence of the aforementioned vacancy-induced
bound states at the nodes of real-life 3D topological semimetals.
All the original results presented in this thesis are currently published
in Refs.\,\cite{Pires2021,Pires2022a,Pires2022b}.

\clearpage{}

\pagestyle{plain}

\selectlanguage{portuges}%

\chapter*{Resumo}

Os semimetais topológicos são uma nova classe de fases electrónicas
tridimensionais (3D), que se caracterizam pela existência de pontos
cónicos de intersecção entre bandas, topologicamente protegidos ao
nível de Fermi. Esses pontos são monopólos da curvatura de Berry no
espaço dos momento, em torno dos quais emergem fermiões de Weyl a
$(3\!+\!1)$-dimensões. Muitas características destes sistemas topológicos
são robustas a perturbações, mas não completamente insensíveis a elas.
Nesta tese, exploramos o tópico fértil dos semimetais de Weyl (ou
Dirac) desordenados, analisando os efeitos de campos locais aleatórios,
regiões de potencial suave (mas aleatório), impurezas escalares pontuais
e defeitos pontuais de rede na sua estrutura electrónica e propriedades
eletrodinâmicas.

O nosso ponto de partida é o estudo de um semi-metal de Weyl (ou Dirac)
na presença de campos locais aleatórios não correlacionados no espaço.
Em campo médio, concluímos que esta desordem induz um \textit{ponto
crítico não convencional}, que é caracterizado por uma teoria crítica
em que a densidade de estados média é o parâmetro de ordem. Usando
formulações diagramáticas e de teoria de campo para abordar o problema
de electrões desordenados, reproduzimos os resultados analíticos conhecidos
sobre este ponto crítico e ainda os confirmamos através de simulações
de rede, no espaço real, que superam a precisão do trabalho anteriormente
publicado. Apesar de estar aparentemente confirmado por evidências
numéricas, a existência deste ponto crítico quântico tem sido objeto
de um grande debate na literatura recente. A controvérsia reside em
possíveis contribuições não perturbativas decorrentes de estados nodais
ligados, que aparecem em regiões suaves (mas estatisticamente raras)
da paisagem de desordem, levando a um cenário de \textit{criticalidade
quântica evitada} (CQE). Para melhor investigar este efeito, examinamos
um modelo, feito à medida, de difusores esféricos colocados aleatoriamente
(e com intensidade aleatória) que modo a forçar a existência de regiões
de potencial suaves aleatórias e, assim, amplificar os seus efeitos
em torno da energia nodal. Combinando uma teoria quântica de espalhamento
com simulações de rede, identificamos um critério preciso de estabilidade
para a fase semi-metálica no contexto deste modelo de desordem, propondo
ainda um mecanismo físico que explica a significância estatística
dos estados nodais ligados para a densidade de estados no nodo. Embora
estas conclusões pareçam estar em desacordo com a falta de evidência
de CQE nas nossas simulações em redes de Weyl verdadeiramente desordenadas,
somos ainda assim capazes de as reconciliar. O ponto fulcral para
entender esta discrepância é reexaminar o fenómeno de CQE, a partir
de um ponto de vista mesoscópico, no qual os estados nodais aparecem
como estados ligados de um pequeno número de flutuações no potencial
local. A partir desta análise, concluímos que estados nodais devido
a eventos raros num potencial aleatório de Anderson, surgem por dois
mecanismos diferentes: \textit{(i)} regiões suaves de potencial, compostas
por alguns pontos adjacentes, que prendem electrões nodais e \textit{(ii)}
a hibridização de uma grande flutuação de potencial com seu ambiente
desordenado. Ambos os efeitos são suprimidos para as distribuições
limitadas da desordem local que usamos nas nossas primeiras simulações,
o que explica sua ausência nos resultados. Deste modo, o fenómeno
de CQE deve ser visto como um \textit{efeito de desordem não universal}
que, na verdade, é uma propriedade de modelos específicos de desordem. 

Na última parte deste trabalho, apresentamos o primeiro estudo teórico
de optoeletrónica e transporte de carga em semi-metais de Weyl com
lacunas de rede. Este é tipo comum de desordem em amostras semi-metálicas
reais, mas que até agora tem sido negligenciado na literatura. Ao
contrário do que vimos nos casos anteriores, este tipo de desordem
é capaz de \textit{aumentar fortemente} a densidade de estados no
nodo de Weyl, para além de o dotar de um \textit{pente composto por
ressonâncias quase-localizadas}, como consequência da hibridização
entre lacunas. Essas ressonâncias são insensíveis a campos magnéticos
e, mais importante, têm uma difusividade quântica fortemente suprimida,
o que leva a uma inédita dependência oscilatória da condutividade
estática na densidade de portadores de carga. Além disso, a resposta
óptica linear de um semi-metal levemente dopado é também afetada pelos
estados nodais induzidos por lacunas, dando origem a uma resposta
dissipativa em forma de patamar, abaixo do limiar de transições inter-banda,
que é proporcional à concentração de defeitos. Tanto os efeitos de
transporte, como as assinaturas ópticas previstas aqui fornecem maneiras
realistas de avaliar experimentalmente a existência dos estados ligados
induzidos por defeitos pontuais em semi-metais topológicos reais.
Todos os resultados originais apresentados nesta tese estão atualmente
publicados nas Refs.\foreignlanguage{english}{\,\cite{Pires2021,Pires2022a,Pires2022b}.}\selectlanguage{english}

\clearpage{}

\lfoot[\thepage]{}

\cfoot[]{}

\rfoot[]{\thepage}

\lhead[\chaptername~\thechapter]{\rightmark}

\rhead[\leftmark]{}

\lfoot[\thepage]{}

\cfoot{}

\rfoot[]{\thepage}

\chapter*{Published Work}

\addcontentsline{toc}{chapter}{List of Publications} 

\paragraph*{Peer-Reviewed Papers (Included in this Thesis):}
\begin{labeling}{00.00.0000}
\item [{\cite{Pires2021}}] \textcolor{brown}{J. P. Santos Pires}, B. Amorim,
Aires Ferreira, \.{I}. Adagideli, E. R. Mucciolo and J. M Viana Parente
Lopes, ``Breakdown of universality in three-dimensional Dirac semimetals
with random impurities'', \textit{Physical Review Research} \textbf{3},
013183 (2021)
\item [{\cite{Pires2022a}}] \textcolor{brown}{J. P. Santos Pires}, S.
M. João, Aires Ferreira, B. Amorim and J. M Viana Parente Lopes, ``Nodal
vacancy bound states and resonances in three-dimensional Weyl semimetals'',
\textit{Physical Review B} \textbf{106} (18), 184201 (2022)
\item [{\cite{Pires2022b}}] \textcolor{brown}{J. P. Santos Pires}, S.
M. João, Aires Ferreira, B. Amorim and J. M Viana Parente Lopes, ``Anomalous
Transport Signatures in Weyl Semimetals with Point Defects'', \textit{Physical
Review Letters} \textbf{129} (19), 196601 (2022)
\end{labeling}

\paragraph*{Peer-Reviewed Papers (Not Included in this Thesis):}
\begin{labeling}{00.00.0000}
\item [{\cite{Khan19}}] N. A. Khan, J. M. Viana Parente Lopes, \textcolor{brown}{J.
P. Santos Pires} and J. M. B. Lopes dos Santos, ``Spectral functions
of one-dimensional systems with correlated disorder'', \textit{Journal
of Physics: Condensed Matter} \textbf{31} (17), 175501 (2019)
\item [{\cite{Pires19}}] \textcolor{brown}{J. P. Santos Pires}, N. A.
Khan, J. M. Viana Parente Lopes and J. M. B. Lopes dos Santos, ``Global
delocalization transition in the de Moura\textendash Lyra model'',
\textit{Physical Review B} 9\textbf{9} (20), 205148 (2019)
\item [{\cite{Pires2020}}] \textcolor{brown}{J. P. Santos Pires}, B. Amorim
and J. M Viana Parente Lopes, ``Landauer transport as a quasisteady
state on finite chains under unitary quantum dynamics'', \textit{Physical
Review B} \textbf{101} (10), 104203 (2020) 
\item [{\cite{Pires2021b}}] R. Peixoto, \textcolor{brown}{J. P. Santos
Pires}, C. S. Monteiro, M. Raposo, P. A. Ribeiro, S. O. Silva, O.
Frazão and J. M Viana Parente Lopes, ``Environmental Sensitivity
of Fabry-Perot Microcavities Induced by Layered Graphene-Dielectric
Hybrid Coatings\textit{'', Physical Review Applied} \textbf{16} (4),
044041 (2021)
\item [{\cite{Joao2022b}}] S. M. João,\textcolor{gray}{{} }\textcolor{brown}{J.
P. Santos Pires}, J. M. Viana Parente Lopes, ``A new microscopic
look into pressure fields and local buoyancy'', \textit{American
Journal of Physics} \textbf{90}, 179-186 (2022)
\item [{\cite{Suresh2022}}] A. Suresh, R. D. Soares, P. Mondal, \textcolor{brown}{J.
P. Santos Pires}, J. M. Viana Parente Lopes, Aires Ferreira, A. E.
Feiguin, P. Plechá\v{c}, and B. K. Nikoli\'{c}, ``Electron-mediated
entanglement of two distant macroscopic ferromagnets within a non-equilibrium
spintronic device'', ArXiv: 2210.06634 {[}cond-mat.str-el{]} (2022)
{[}Submitted to \textit{PRX Quantum}{]}
\item [{\cite{Pinho2022}}] J. M. Alendouro Pinho, \textcolor{brown}{J.
P. Santos Pires}, Simão M. João, B. Amorim and J. M. Viana Parente
Lopes, ``From Bloch Oscillations to a Steady-State Current in Strongly
Biased Mesoscopic\,\,Devices'',\,ArXiv:\,2212.05574 {[}cond-mat.mes-hall{]}
(2022) {[}Submitted to \textit{Physical Review B}{]}
\end{labeling}

\paragraph*{Peer-Reviewed Conference Proceedings:}
\begin{labeling}{00.00.0000}
\item [{\cite{Pires2020b}}] \textcolor{brown}{J. P. Santos Pires}, B.
Amorim and J. M Viana Parente Lopes, ``Numerical Simulation of Non-Equilibrium
Stationary Currents Through Nano-scale One-Dimensional Chains'',
\textit{EPJ Web of Conferences} \textbf{233}, 05010-1 to 05010-6 (2020)
\item [{\cite{Khan2020}}] N. A. Khan, J. M. Viana Parente Lopes, \textcolor{brown}{J.
P. Santos Pires} and J. M. B. Lopes dos Santos, ``Probing the Global
Delocalization Transition in the de Moura-Lyra Model with the Kernel
Polynomial Method'', \textit{EPJ Web of Conferences} \textbf{233},
05011-1 to 05011-5 (2020)
\item [{\vspace{0.5cm}}]~
\end{labeling}
\begin{wrapfigure}[8]{l}{0.2\columnwidth}%
\includegraphics[scale=0.09]{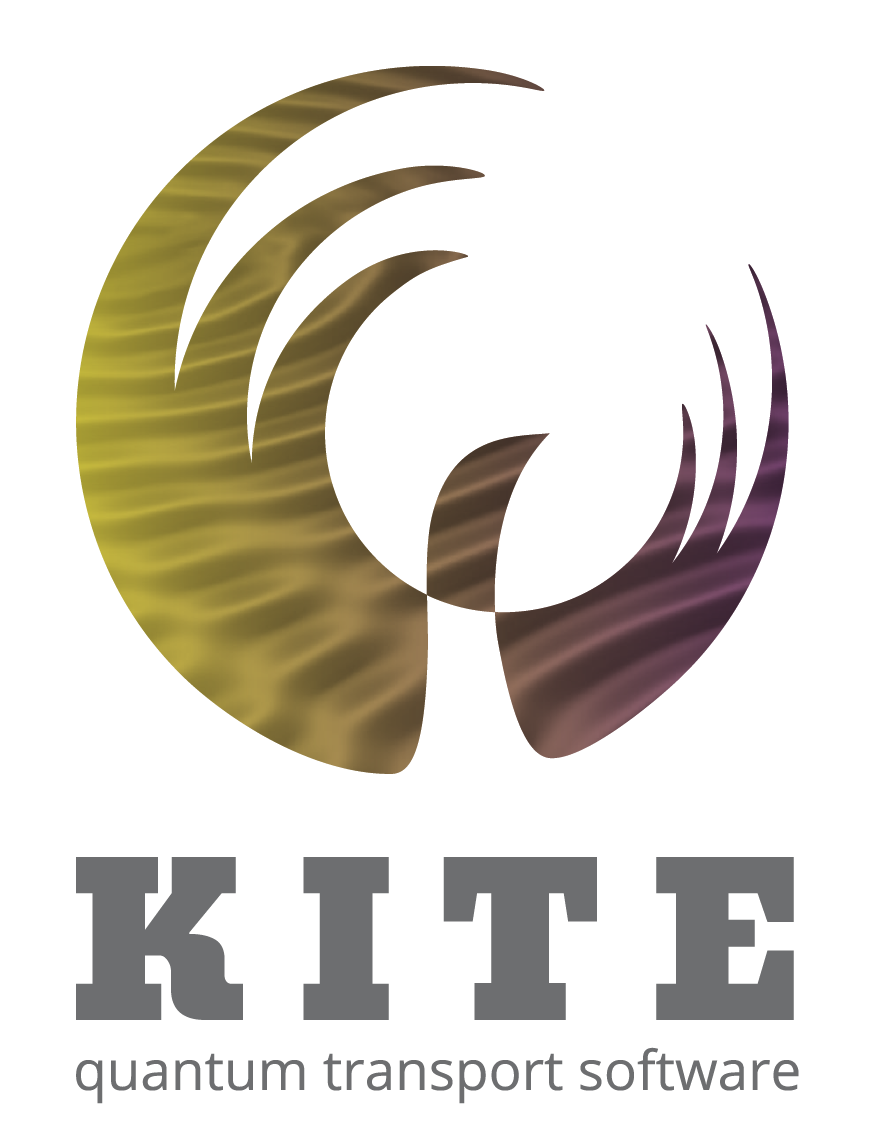}\end{wrapfigure}%

$ $

Besides the aforementioned published work, I would also like to point
out my ongoing collaboration with the \textit{\textcolor{black}{\href{https://quantum-kite.com/index.html}{KITE Project}}}.
This is an open-source software designed to perform highly efficient
and CPU-parallelized spectral calculations in real-space tight-binding
models of arbitrary complexity. The potentialities of this software
were pivotal for the work presented in this thesis and my own contribution
as a developer is also an output of my PhD work.

\clearpage{}

\pagestyle{fancy}

\lhead[\chaptername~\thechapter]{\rightmark}

\global\long\def\vect#1{\overrightarrow{\mathbf{#1}}}%

\global\long\def\abs#1{\left|#1\right|}%

\global\long\def\av#1{\left\langle #1\right\rangle }%

\global\long\def\ket#1{\left|#1\right\rangle }%

\global\long\def\bra#1{\left\langle #1\right|}%

\global\long\def\tensorproduct{\otimes}%

\global\long\def\braket#1#2{\left\langle #1\mid#2\right\rangle }%

\global\long\def\omv{\overrightarrow{\Omega}}%

\global\long\def\inf{\infty}%

\lhead[]{\MakeUppercase{Introduction}}

\rhead[\MakeUppercase{Introduction}]{}

\lfoot[\thepage]{}

\cfoot[]{}

\rfoot[]{\thepage}

\chapter*{\label{chap:Introd}Introduction}

\addcontentsline{toc}{chapter}{Introduction} 

Topological semimetals are a class of novel three-dimensional (3D)
electronic phases that feature linear band-crossings at the Fermi
level. An important type of topological semimetals are the 3D Weyl
(Dirac) systems that have point-like Fermi surfaces around which they
effectively realize exotic chiral ($3\!+\!1)$-dimensional Weyl (Dirac)
fermions\,\cite{Weyl29} as low-energy quasiparticles. The band-crossing
points, or Weyl nodes, materialize monopoles (or anti-monopoles) of
Berry curvature in momentum space\,\cite{Volovik2009} which are
analogous to the \textit{``diabolical points''} described by Berry\,\cite{Berry84}
in a generic two-level quantum system. Therefore, isolated Weyl nodes
are topologically-protected band degeneracies which show a great degree
of robustness to Hamiltonian deformations. The topological character
of Weyl semimetals yields some important physical consequences, ranging
from the existence of \textit{surface Fermi arcs}\,\cite{Wan2011,Hosur2012,Witten2015,Haldane2014,Hashimoto2017}
that connect Weyl nodes in the surface-projected first Brillouin zone
(fBz), to the \textit{negative longitudinal magnetoresistance}\,\cite{Son2013,Zhang2016}
which is driven by a condensed matter realization of the celebrated\textit{
chiral anomaly} of QED\,\cite{Adler1969,Bell1969}. Nonetheless,
perhaps the most remarkable of all is their resilience to the effects
of unavoidable perturbations, such as disorder or crystal defects.
While the nodal points may endure disorder, that does not mean that
all their physics is insensitive to it. As a matter of fact, most
recent theoretical research clearly indicates that disordered topological
semimetals are a fertile ground for novel physical phenomena to emerge,
not in spite of disorder, but rather because of it. In this thesis,
we follow on this path and explore the effects of different types
of disorder in the electronic structure and electrodynamic properties
of point-node  semimetals, most notably by focusing on scalar random
potentials and lattice point defects. 

Since the seminal work of Fradkin\,\cite{Fradkin86a,Fradkin86b},
three-dimensional gapless fermions have been known to support unconventional
critical points induced by random fields, but which precede Anderson
localization. We revisit this issue for dirty \textit{3D Weyl/Dirac
semimetals }(DWSMs), analyzing the effects of a random on-site potential
in their mean density of states (DoS). We start by reproducing the
mean-field results which predict a critical behavior of the DoS\,\cite{Goswami11}
that is characterized by a single-parameter scaling theory\,\cite{Kobayashi14}
with universal exponents\,\cite{Syzranov16}. This result is further
confirmed by real-space simulations of a disordered tight-binding
model that improve over the accuracy achieved in previously published
results\,\cite{Pixley16b,Bera16}. Physically, the aforementioned
critical point marks a disorder-induced phase transition from an incompressible
semi-metallic state, with a vanishing nodal DoS, to a conventional
diffusive metallic phase. Despite seemingly confirmed by both numerical
and analytical means, the survival of this semimetal-to-metal transition
was recently questioned\,\cite{Nandkishore14} by the inclusion of
instantonic saddle-point solutions of the effective \textit{Statistical
Field Theory}\,\cite{Coleman77} for a dirty DWSM. Those solutions
supposedly describe nodal bound states of smooth rare-regions in a
disordered landscape, and are predicted to yield a finite (albeit
exponentially small) nodal DoS within the semimetal phase. The claimed
scenario of \textit{Avoided Quantum Criticality} (AQC) soon received
numerical support\,\cite{Pixley16a} but remained a contested paradigm
until very recently\,\cite{Gurarie17,Buchhold18a,Buchhold18b,Ziegler18,Wilson20}.

To assess the rare-region-induced AQC in isolation, we investigate
the nodal DoS of a semimetal hosting randomly placed scalar spherical
scatterers of random strength. This tailor-made model forces random
smooth regions to appear and naturally enhances their effect at the
nodal energy. By combining an analytical scattering theory with accurate
lattice simulations, we establish a criterion for the nodal DoS to
be lifted in the presence of random regions\,\cite{Pires2021}. These
conclusions are complemented by a physical mechanism that endows fine-tuned
nodal bound states with a statistical significance for mean nodal
DoS (which partly resolves the dispute of Refs.\,\cite{Buchhold18b,Ziegler18}).
Remarkably, while instantons are shown to contribute to the nodal
DoS of a DWSM, we find no trace of rare-region effects in our unbiased
numerical results for a disordered lattice. To reconcile these results,
we set about describing the AQC phenomenon in a disordered lattice
from a \textit{mesoscopic point-of-view}. We employ a projected Green's
function (pGF) formalism to study the changes caused in the DoS by
a few atomic-sized impurities, including all coherent multiple-scattering
effects. While isolated impurities in the lattice do not create bound
states, we demonstrate that quantum coherent scattering inside fine-tuned
configurations of two (or more) impurities gives rise to nodal bound
states (confirming earlier results by Buchhold \textit{et al}.\,\cite{Buchhold18b}).
From this characterization, we conclude that rare-event states in
a disordered lattice can arise from two different mechanisms: \textit{(i)
}from smooth regions of a few adjacent sites, and \textit{(ii)} the
hybridization of a large atypical fluctuation of the potential with
its (typical) disordered environment. Both mechanisms are suppressed
(or may even be absent) when the on-site potential is drawn from a
bounded distributions. This fact explains why no signs of AQC were
found in our numerical simulations, while Pixley \textit{et al}.\,\cite{Pixley16a}
were able to pinpoint them using local gaussian distributions\@.

Disorder is ubiquitous in real samples but generally much more complex
than a random on-site potential. In Chapter \ref{chap:Vacancies},
we finally turn our attention to the more realistic case of vacancies,
that is, resonant point defects that are created by the removal of
random lattice sites. Using both the pGF formalism and full spectrum
lattice simulations we demonstrate that isolated vacancies can easily
trap Weyl fermions around them into localized wavefunctions having
$r^{{\scriptscriptstyle -2}}$ tails\,\cite{Pires2022b,Pires2022a}.
Unlike random potentials, lattice vacancies strongly enhance the mean
DoS, leading to a nodal peak that is broadened by coherent inter-vacancy
scattering at finite defect concentrations. In addition to the broadening,
these interference effects also produce a \textit{comb of subsidiary
resonances} that are made of quasi-localized states. The consequences
of these vacancy-induced states are also explored. Namely, we show
that the electrical conductivity displays an oscillatory behavior
as a function of the Fermi energy, which traces itself back to the
comb of quasi-localized resonances that emerges around the Weyl node.
This behavior contrasts with the monotonic dependence observed upon
varying the electronic density of analogous 2D Dirac systems, not
easily replicated by non-resonant disorder\,\cite{Katsnelson2006,Pereira08,Ferreira2015}.
An anomalous behavior is also seen in the linear optical conductivity
of a slightly doped semimetal, which now features an emergent plateau-shaped
real response below the inter-band threshold, proportional to the
vacancy concentration.

\subsection*{Brief Overview of the Thesis}

To guide the reader, we hereby include a non-extensive overview of
the main components of this work. In short, the dissertation can be
divided into two main parts: \textit{(i)} a context part provided
in Chapters\,\,\ref{chap:Introduction} and \ref{chap:Mean-Field-Quantum-Criticality},
and \textit{(ii)} an original contribution that is presented in Chapters\,\,\ref{chap:Instability_Smooth_Regions}
to \ref{chap:Vacancies}. In Chapter\,\ref{chap:Introduction}, we
take a bird's eye view of the topological properties of condensed
matter systems, starting from general considerations, but aiming at
an overview of known properties and the most outstanding phenomenology
of \textit{clean three-dimensional topological semimetals}. Once convinced
of the intrinsic interest of studying the physics of three-dimensional
Dirac-Weyl semimetals, the reader ought to take Chapter\,\,\ref{chap:Mean-Field-Quantum-Criticality}
as a state-of-the-art review that shifts the focus of discussion towards
the main issue to be addressed here: \textit{the effects of disorder
in Dirac and Weyl semimetals}. Finally, the bulk of the thesis follows
along Chapters\,\,\ref{chap:Instability_Smooth_Regions} to \ref{chap:Vacancies},
in which the main results from the authors' original research are
presented in detail. Most (but not all) novel results presented here
are published in Refs.\,\cite{Pires2021,Pires2022b,Pires2022a}.
Our presentation is summed up in Chapter\,\,\ref{chap:Concluding-Remarks},
where the main conclusions are listed and new interesting research
paths are proposed to be built on top of this work.

\global\long\def\vect#1{\overrightarrow{\mathbf{#1}}}%

\global\long\def\abs#1{\left|#1\right|}%

\global\long\def\av#1{\left\langle #1\right\rangle }%

\global\long\def\ket#1{\left|#1\right\rangle }%

\global\long\def\bra#1{\left\langle #1\right|}%

\global\long\def\tensorproduct{\otimes}%

\global\long\def\braket#1#2{\left\langle #1\mid#2\right\rangle }%

\global\long\def\omv{\overrightarrow{\Omega}}%

\global\long\def\inf{\infty}%

\lhead[\MakeUppercase{\chaptername}~\MakeUppercase{\thechapter}]{\MakeUppercase{\rightmark}}

\rhead[\MakeUppercase{\leftmark}]{}

\lfoot[\thepage]{}

\cfoot[]{}

\rfoot[]{\thepage}

\chapter{\label{chap:Introduction}Overview of Topological Matter}

\vspace{-0.2cm}

This thesis studies effects of lattice disorder in three-dimensional
topological semimetals, which can give rise to physical phenomena
that strikingly contrast with what would be expected in perfect crystalline
samples. Despite enjoying considerable theoretical and experimental
interest in recent literature, this is still a niche theme in physics
which can be better appreciated upon a general contextualization.
This chapter is designed to provide the reader with such an overview
of the phenomenology and basic theoretical notions that underlie the
physics of 3D topological semimetals. We start by introducing the
concept of topological insulators and semimetals in a pedagogical
way that is deeply rooted on standard band theory of electrons and
its topological properties. As the chapter comes to its close, the
presentation is narrowed down a more specific discussion of the known
and more interesting properties of Dirac-Weyl semimetals, the class
of solid-state systems that will be the subject of all upcoming results
and discussions. In addition to provide a common ground on the main
theme, some of the notions introduced in this chapter will also be
useful to explain some of the arguments presented in upcoming chapters.

\vspace{-0.55cm}

\section{Electrons in the Solid State: Basic Notions}

\vspace{-0.25cm}

A solid state system is composed of interacting atomic nuclei and
electrons whose collective dynamics is described by standard quantum
mechanics. In spite of the apparent simplicity of the problem, the
sheer number of components in a macroscopic sample turns its exact
microscopic description into a formidable task. A far more advantageous
strategy is to assume a top-down approach\,\cite{Anderson77} that
treats simplified models which are striped down of all degrees of
freedom not important for the physical phenomenon to be described.
Early in the history of modern physics, Felix Bloch realized\,\cite{bloch29}
that the essential electrodynamic properties of solids could be understood
from the study of a basic quantum mechanical problem: a gas of independent
spin-$\nicefrac{1}{2}$ fermions moving across a potential landscape
with the periodicity of a Bravais lattice $\mathcal{L}$. Most often\,\footnote{Here, we exclude the presence of external fields or any dynamics involving
the electronic spin.}, this problem amounts to solving the non-relativistic single-particle
Hamiltonian, 

\vspace{-0.8cm}

\begin{equation}
\mathcal{H}_{e}\!=\!-\frac{\hbar^{2}}{2m_{e}}\nabla_{\!\mathbf{r}}^{2}+V_{c}(\mathbf{r}),\label{eq:He}
\end{equation}

\vspace{-0.2cm}

where $m_{e}$ is the electron's mass and $\hbar$ is Planck's constant.
Note that the underlying static Bravais lattice\,\footnote{Mathematically, one can think of a $d$-dimensional Bravais lattice
as a discrete subgroup of the full continuous translation group in
$d$-dimensions.} of atomic nuclei is collectively described as an overall periodic
crystal-field potential $V_{c}(\mathbf{r})$, where $V_{c}(\mathbf{r}+\mathbf{R})\!=\!V_{c}(\mathbf{r})$
for any lattice translation $\mathbf{R}\!\in\!\mathcal{L}$. This
crystal-field is system-dependent and encapsulates all electrostatic
forces caused by the atomic nuclei (as well as inner electron shells)
in the propagating electron, as well as some effects of electron-electron
repulsion at the mean-field level\,\cite{Slater28,Fock30,Hartree35}.
Notwithstanding, the single-electron states and energy levels are
determined by diagonalizing $\mathcal{H}_{e}$ and the ground-state
of the many-electron system is then built as a Slater determinant
that fills the spectrum up to its Fermi level. Thereby, the most fundamental
problem of solid-state quantum electronics is to diagonalize Eq.\,\eqref{eq:He},
something that would be a cumbersome task, were it not simplified
by exploiting the lattice-translation symmetry that system still has.
This culminates in the celebrated Bloch's Theorem (BT\nomenclature{BT}{Bloch's Theorem})\,\cite{bloch29},
which states that any eigenstate of $\mathcal{H}_{e}$ takes the form

\vspace{-0.7cm}
\begin{equation}
\Psi_{s\mathbf{k}}(\mathbf{r})\!=\!\braket{\mathbf{r},s}{\Psi_{n\mathbf{k}}}\!=\!\frac{1}{\sqrt{N_{c}}}e^{i\mathbf{k}\cdot\mathbf{r}}\Phi_{s\mathbf{k}}(\mathbf{r}),\label{eq:BlochTheorem}
\end{equation}
where $N_{c}$ is the total number of unit cells\,\footnote{For book-keeping, we consider the lattice to be finite and supplemented
by periodic boundary conditions. The limit of infinite lattice can
be easily taken whenever if becomes necessary.}, $\mathbf{k}$ is the crystal momentum defined inside the first Brillouin
zone (fBz) of $\mathcal{L}$, $s$ is a set of integers labelling
the energy levels for each $\mathbf{k}$, and $\Phi_{n\mathbf{k}}(\mathbf{r})$
is a lattice periodic function that obeys $\Phi_{s\mathbf{k}}(\mathbf{r}\!+\!\mathbf{R})=\Phi_{s\mathbf{k}}(\mathbf{r})\text{ for any }\mathbf{R}\!\in\!\mathcal{L}.$
There are two important consequences of BT; Firstly, it shows that
a free electron travels across the periodic potential, essentially,
as a propagating plane-wave with a spectrum that is composed of continuous
energy-bands labeled by integers. Secondly, it also shows that the
eigenstates feature a phase-twisted periodicity in the crystal's unit
cell, that is 

\vspace{-0.7cm}

\begin{equation}
\Psi_{s\mathbf{k}}(\mathbf{r}+\mathbf{R})=\Psi_{s\mathbf{k}}(\mathbf{r})e^{i\mathbf{k}\cdot\mathbf{R}},\label{eq:Twistee}
\end{equation}
where $\mathbf{R}$ stands for a lattice translation vector. The latter
implies that one is allowed to compactify the (infinite-space) problem,
$\mathcal{H}_{e}\Psi_{E}(\mathbf{r})\!=\!E\Psi_{E}(\mathbf{r})$,
into the partial differential equation (PDE),

\vspace{-0.7cm}

\begin{equation}
\left[-\frac{\hbar^{2}}{2m_{e}}\nabla_{\!\mathbf{r}}^{2}+V_{c}(\mathbf{r})\right]\Psi_{s\mathbf{k}}(\mathbf{r})=E_{s}(\mathbf{k})\Psi_{s\mathbf{k}}(\mathbf{r}),\label{eq:TwistedBloch}
\end{equation}
which lives within the unit cell of $\mathcal{L}$, complemented by
\textit{twisted boundary conditions} that are fixed for each $\mathbf{k}\in\text{fBz}$
through Eq.\,\eqref{eq:Twistee}. Alternatively, one may also use
the form of Eq.\,\eqref{eq:BlochTheorem} to re-write the eigenvalue
problem as a PDE for the lattice-periodic function $\Phi_{n\mathbf{k}}(\mathbf{r})$,
\textit{i.e.},

\vspace{-0.7cm}
\begin{align}
\left[-\frac{\hbar^{2}}{2m_{e}}\nabla_{\!\mathbf{r}}^{2}+V_{c}(\mathbf{r})\right]\Psi_{n\mathbf{k}}(\mathbf{r})=E_{n}(\mathbf{k})\Psi_{n\mathbf{k}}(\mathbf{r})\label{eq:He-1}\\
\Longleftrightarrow\left[-\frac{\hbar^{2}}{2m_{e}}\nabla_{\!\mathbf{r}}^{2}+\frac{\hbar^{2}\abs{\mathbf{k}}^{2}}{2m_{e}}+V_{c}(\mathbf{r})\right] & \Phi_{n\mathbf{k}}(\mathbf{r})=E_{n}(\mathbf{k})\Phi_{n\mathbf{k}}(\mathbf{r}),\nonumber 
\end{align}
which is again restricted to the unit cell of $\mathcal{L}$, but
now complemented by \textit{periodic boundary conditions}. Note that
Eq.\,\eqref{eq:He-1} defines a new $\mathbf{k}$-dependent Hamiltonian 

\vspace{-0.7cm}
\begin{equation}
\mathcal{H}(\mathbf{k})\!=\!-\frac{\hbar^{2}}{2m_{e}}\nabla_{\!\mathbf{r}}^{2}+\frac{\hbar^{2}\abs{\mathbf{k}}^{2}}{2m_{e}}+V_{c}(\mathbf{r}),\label{eq:BlochHam}
\end{equation}

\vspace{-0.3cm}which is rightfully called the \textit{Bloch Hamiltonian}. 

Whether we reduce the full single-electron problem in a periodic potential
to a PDE in the unit cell with $\mathbf{k}$-dependent twisted boundaries
or a periodic cell with a $\mathbf{k}$-dependent Bloch Hamiltonian
is a matter of choice, and does not change the qualitative picture
of the eigenstates provided by BT. In both cases, BT allows us to
break up the solution of the time-independent Schrödinger equation
over all space into a set of independent eigenvalue problems (one
for each $\mathbf{k}\!\in\!\text{fBz}$) confined to the unit-cell
of $\mathcal{L}$. Hence, it can be generally argued that the spectrum
must be discrete for any $\mathbf{k}$, which demonstrates why electrons
form energy bands in a crystal.

\vspace{-0.5cm}

\section{\label{sec:Band-Truncation-and}Band Truncation and Tight-Binding
Models}

\vspace{-0.15cm}

\begin{wrapfigure}{o}{0.53\columnwidth}%
\vspace{-0.55cm}
\begin{centering}
\includegraphics[scale=0.19]{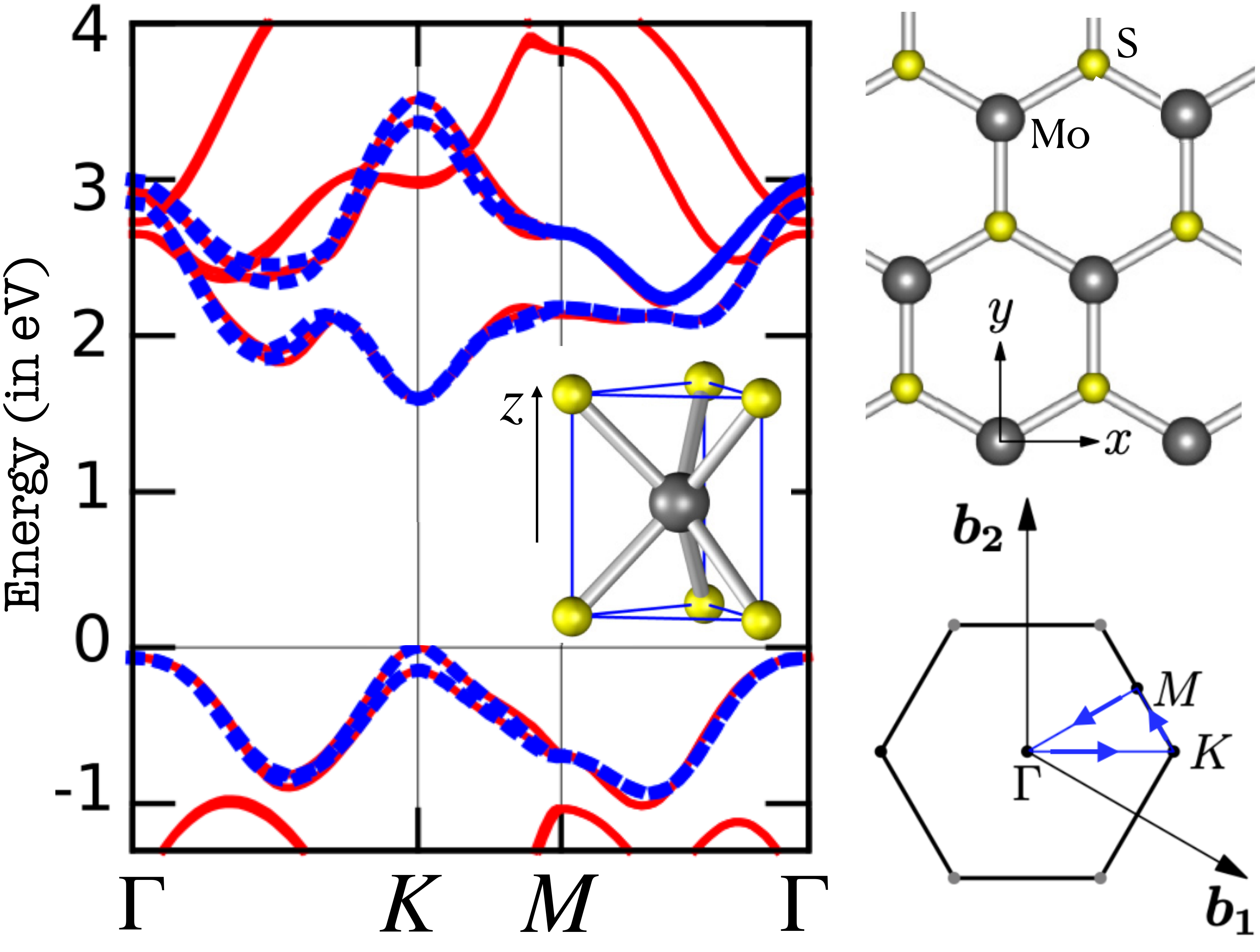}
\par\end{centering}
\vspace{-0.3cm}

\caption{\label{fig:Electronic-band-structure}Electronic band structure of
an $\text{MoS}_{2}$ monolayer, whose lattice and fBz are shown on
the right. The red curves are low-energy bands from \textit{ab initio
}calculations, while the blue correspond to an effective five-orbital
tight-binding model. Figure adapted from Liu \textit{et al.}\,\cite{Liu2013}.}

\vspace{-0.45cm}\end{wrapfigure}%
The use of Bloch's theorem allows us to greatly reduce the complexity
of the eigenvalue problem for an electron in a lattice. However, one
still does not escape the need to diagonalize an infinite-dimensional
matrix for each $\mathbf{k}$. Equivalently, a given crystal-field
potential $V_{c}(\mathbf{r})$ will always generate an infinite (albeit
discrete) number of energy bands. In turn, most electronic processes
only involve transitions between states in the vicinity of the Fermi
level ($E_{F}$), because higher-energy bands require too much energy
to be excited, while lower-energy bands are made inert by Pauli's
exclusion principle. Therefore, in most theoretical studies, one works
with effective models that truncate the space of bands to a minimal
set of $n_{b}$ isolated bands that are closest to the Fermi level.
This truncated Hilbert space is then generated by the Bloch states
$\ket{\Psi_{n\mathbf{k}}}$ with $\mathbf{k}\!\in\!\text{fBz}$ and
$s=1,2,\cdots n_{b}$, which can be promptly turned into a set orbitals
in real-space via the unitary transformation,

\vspace{-0.7cm}
\begin{equation}
\ket{\phi_{\alpha\mathbf{R}}}\!=\!\frac{1}{\sqrt{N_{c}}}\sum_{\mathbf{k}}\sum_{s=1}^{n_{b}}a_{\alpha s}e^{-i\mathbf{k}\cdot\mathbf{R}}\!\ket{\Psi_{s\mathbf{k}}},\label{eq:WannierDefinition}
\end{equation}

\vspace{-0.3cm}which defines the so-called Wannier orbitals, $\ket{\phi_{\alpha\mathbf{R}}}$\,\cite{Wannier37}.
In Eq.\,\eqref{eq:WannierDefinition}, the complex coefficients $a_{\alpha n}$
implement a general linear transformation in the truncated band-space
which is only constrained by $\sum_{s}a_{\alpha s}^{*}a_{\beta s}=\delta_{\alpha\beta}$,
so as to assure the new basis remains orthonormal, \textit{i.e.},
$\braket{\phi_{\alpha\mathbf{R}}}{\phi_{\beta\mathbf{R}^{\prime}}}=\delta_{\alpha\beta}\delta_{\mathbf{R},\mathbf{R}^{\prime}}$.
In addition, the Wannier orbitals may also be written as a linear
combination of the lattice-periodic functions $\Phi_{n\mathbf{k}}(\mathbf{r})$
{[}defined in Eq.\,\eqref{eq:BlochTheorem}{]} as follows:

\vspace{-0.7cm}
\begin{align}
\phi_{\alpha\mathbf{R}}(\mathbf{r})\! & =\!\frac{1}{N_{c}}\sum_{\mathbf{k}}e^{i\mathbf{k}\cdot\left(\mathbf{r}-\mathbf{R}\right)}\sum_{s=1}^{n_{b}}a_{\alpha s}\Phi_{s\mathbf{k}}(\mathbf{r})\Longleftrightarrow\label{eq:Wannier_RealSpace}\\
 & \phi_{\alpha\mathbf{R}}(\mathbf{R}+\mathbf{r})\!=\frac{1}{N_{c}}\sum_{\mathbf{k}}e^{i\mathbf{k}\cdot\mathbf{r}}\sum_{s=1}^{n_{b}}a_{\alpha s}\Phi_{s\mathbf{k}}(\mathbf{r}).\nonumber 
\end{align}
Equation\,\eqref{eq:Wannier_RealSpace} is important because it states
that the $\alpha^{\text{th}}$ Wannier function will always look the
same when centered on any lattice position, which is reminiscent of
the localized atomic orbitals employed in the tight-binding approach
to band-structure calculations\,\cite{Slater54,Wallace57}. However,
there is nothing here that tells us $\phi_{\alpha\mathbf{R}}(\mathbf{r})$
are actually localized wavefunctions in real-space. In many systems,
it is possible to construct Wannier orbitals that are exponentially
localized\,\cite{Wannier37} in real-space (akin isolated atomic
orbitals) but there are situations in which this is prevented by topological
properties of the isolated bands\,\cite{Thouless84,Marzari2012}.
Further details on this process lie outside the scope of this thesis
and, from now on, we will assume that our starting point is an effective
real-space tight-binding model with $n_{b}$ orbitals per unit cell,
that correctly reproduces the electronic structure and dynamics of
electrons close to the Fermi surface. A concrete example of such an
effective tight-binding model is illustrated in Fig.\,\ref{fig:Electronic-band-structure}
for the case of $\text{MoS}_{2}$ monolayer, an important quasi-2D
transition-metal dichalcogenide.

\vspace{-0.3cm}

\subsection{Basics of Tight-Binding Models}

Our upcoming theoretical analysis will often be based on effective
tight-binding model Hamiltonians that describe the dynamics of electrons
hopping on a lattice. The single-electron Hilbert space will then
be generated by all the Wannier states of the model and any physical
quantity will be represented by an operator that acts on this restricted
space. This is the moment to provide the reader with some general
properties and tools that will help clarify the future handling of
such quantum lattice models. 

We start with the general form of a periodic tight-binding Hamiltonian
supported on an underlying Bravais lattice $\mathcal{L}$, \textit{i.e.},

\vspace{-0.6cm}

\begin{align}
\mathcal{H}_{\text{tb}}\! & =\!\sum_{\mathbf{R}\in\mathcal{L}}\sum_{\Delta\mathbf{R}}\boldsymbol{\Upsilon}{}_{\mathbf{R}}^{\dagger}\!\cdot\!\mathbf{T}_{\mathbf{\Delta\mathbf{R}}}\!\cdot\!\boldsymbol{\Upsilon}{}_{\mathbf{R}+\Delta\mathbf{R}},\label{eq:Per_Ham-1}
\end{align}

\vspace{-0.3cm}

where $\boldsymbol{\Upsilon}{}_{\mathbf{R}}^{\dagger}\!=\!\left[c_{1\mathbf{R}}^{\dagger},c_{2\mathbf{R}}^{\dagger},\cdots,c_{n_{b}\mathbf{R}}^{\dagger}\right]$\,\footnote{Note that we have switched to a second-quantization language, where
$c_{\alpha\mathbf{R}}^{\dagger}$ ($c_{\alpha\mathbf{R}}$) is a fermionic
creation (annihilation) operator relative to the Wannier state $\ket{\phi_{\alpha\mathbf{R}}}$.} and $\mathbf{T}_{\mathbf{\Delta\mathbf{R}}}$ is the complex-valued
$n_{b}\!\times\!n_{b}$ hopping matrix. Since $\mathcal{H}_{\text{tb}}$
is an hermitian operator, the hopping matrix is due to obey the condition
$\mathbf{T}_{\mathbf{\Delta\mathbf{R}}}\!=\!\mathbf{T}_{-\mathbf{\Delta\mathbf{R}}}^{\dagger}$.
In addition, since $\mathbf{T}_{\mathbf{\Delta\mathbf{R}}}$ is only
a function of position differences in $\mathcal{L}$, the Hamiltonian
of Eq.\,\eqref{eq:Per_Ham-1} is still lattice-periodic and can be
reduced to a block-diagonal form in the fBz\,\footnote{In passing, it is worth commenting that usually the effective tight-binding
model for a system is constructed so as to respect all symmetries
of the original system, \textit{i.e}, it must have the same underlying
Bravais lattice and also the same point-symmetry group.}. This is done via the transformation,

\vspace{-0.7cm}

\begin{subequations}
\begin{align}
\boldsymbol{\Upsilon}_{\mathbf{R}}\! & =\!\frac{1}{\sqrt{N_{c}}}\sum_{\mathbf{k}}\!e^{i\mathbf{k}\cdot\mathbf{R}}\,\boldsymbol{\Upsilon}_{\mathbf{k}}\label{eq:BlochRep}\\
\boldsymbol{\Upsilon}_{\mathbf{R}}^{\dagger}\! & =\!\frac{1}{\sqrt{N_{c}}}\sum_{\mathbf{k}}\!e^{-i\mathbf{k}\cdot\mathbf{R}}\,\boldsymbol{\Upsilon}_{\mathbf{k}}^{\dagger},
\end{align}
\end{subequations}

which yields,

\vspace{-0.7cm}

\begin{equation}
\mathcal{H}_{\text{tb}}\!=\!\!\sum_{\mathbf{k}}\!\boldsymbol{\Upsilon}{}_{\mathbf{k}}^{\dagger}\!\cdot\!\mathcal{H}\left(\mathbf{k}\right)\!\cdot\!\boldsymbol{\Upsilon}{}_{\mathbf{k}}\longrightarrow\mathcal{H}\left(\mathbf{k}\right)\!=\!\!\sum_{\Delta\mathbf{R}}\left(\mathbf{T}_{\mathbf{\Delta\mathbf{R}}}e^{i\mathbf{k}\cdot\Delta\mathbf{R}}\right).\label{eq:BlochHam-1}
\end{equation}
Equation\,\eqref{eq:BlochHam-1} defines a finite-dimensional Bloch
Hamiltonian, $\mathcal{H}(\mathbf{k})$, which is analogous to the
one defined in Eq.\,\eqref{eq:BlochHam} for the continuum Schrödinger
equation with a periodic potential. Provided the tight-binding model
only has short-ranged hoppings, the summation over $\Delta\mathbf{R}$
has a small number of elements and can be often done by hand. Thereby,
the construction of a Bloch Hamiltonian from the periodic real-space
tight-binding model is a straightforward task and its diagonalization
can usually be done analytically for a generic $\mathbf{k}$. From
now on, we assume that the eigenvalue problem,

\vspace{-0.7cm}
\begin{equation}
\sum_{\beta=1}^{n_{\text{b}}}\mathcal{H}_{\alpha\beta}(\mathbf{k})\phi_{\mathbf{k}s}^{\beta}=\varepsilon_{\mathbf{k}}^{s}\phi_{\mathbf{k},s}^{\alpha},
\end{equation}
can always be solved, defining $\varepsilon_{\mathbf{k}}^{s}$ as
the energy dispersion of band $s$, and the eigenstates of $\mathcal{H}_{\text{tb}}$
as being

\vspace{-0.7cm}
\begin{equation}
\ket{\Psi_{\mathbf{k}}^{s}}\!=\!\!\sum_{\mathbf{R}\in\mathcal{L}}\sum_{\alpha=1}^{n_{\text{b}}}\phi_{\mathbf{k}s}^{\alpha}e^{i\mathbf{k}\cdot\mathbf{R}}\ket{\phi_{\mathbf{R},\alpha}}.\label{eq:Eigenstates_Bloch}
\end{equation}

Note that the previous diagonalization procedure relies on the lattice
translation invariance of $\mathcal{H}_{\text{tb}}$ that is built
into Eq.\,\eqref{eq:Per_Ham-1}. Real samples, even mono-crystalline
ones, always have this symmetry slightly broken by the presence of
substitutional impurities, vacant sites or other lattice defects.
Such symmetry-breaking terms could be included in the continuum single-electron
Hamiltonian of Eq.\,\eqref{eq:He}, immediately preventing the use
of BT. A far more advantageous approach to study these effects is
to include them in the effective tight-binding Hamiltonian in which
they take the form of non-periodic real-space perturbations. In a
generic case, this leads to the disordered real-space Hamiltonian,

\vspace{-0.7cm}

\begin{align}
\mathcal{H}_{\text{tb}}^{\text{d}}\! & =\!\sum_{\mathbf{R}\in\mathcal{L}}\sum_{\mathbf{R}^{\prime}\in\mathcal{L}}\boldsymbol{\Upsilon}{}_{\mathbf{R}^{\prime}}^{\dagger}\!\cdot\!\mathbf{T}_{\mathbf{\mathbf{R}^{\prime}\mathbf{R}}}\!\cdot\!\boldsymbol{\Upsilon}{}_{\mathbf{R}},\label{eq:Per_Ham-1-1}
\end{align}
which now features a hopping matrix that is no longer translation
invariant. Clearly, the eigenstates of $\mathcal{H}_{\text{tb}}^{\text{d}}$
will not take the form established by BT {[}Eq.\,\eqref{eq:Eigenstates_Bloch}{]}
but we can still study it numerically by considering $\mathcal{L}$
to be a finite lattice. Assuming the finite lattice has $N_{c}$ unit
cells, the Hamiltonian of Eq.\,\eqref{eq:Per_Ham-1-1} can be represented
as a sparse hermitian $(n_{\text{b}}N_{c}\times n_{\text{b}}N_{c})$-dimensional
matrix that can be diagonalized to obtain the single-electron eigenstates
or, more generally, be used to evaluate any (observable) function
of the disordered Hamiltonian. This will be one of the major lines
we will follow to tackle the properties of disordered topological
semimetals.

\paragraph*{Tight-Binding Observables:}

In Appendix\,\ref{chap:Crash-Course-KPM}, it is shown that almost
all observables related to static or transport properties of free-electrons
can be written in terms of the position operator, velocity operator,
and the Hamiltonian itself. Since they will be eventually useful,
here we introduce the tight-binding expression for both the position
and velocity operators. The expression of the lattice position operator
depends on the precise boundary conditions imposed on the finite lattice.
Nevertheless, we can write it unambiguously for the infinite lattice
as

\vspace{-0.7cm}

\begin{equation}
\boldsymbol{\mathcal{R}}\!=\!\!\!\sum_{\mathbf{R}\in\mathcal{L}}\!\,\boldsymbol{\Upsilon}{}_{\mathbf{R}}^{\dagger}\!\cdot\left(\mathbf{R}\mathcal{I}_{n_{\text{b}}\!\times n_{\text{b}}}\!+\!\text{\ensuremath{\boldsymbol{\Delta}}}\right)\cdot\!\boldsymbol{\Upsilon}{}_{\mathbf{R}},
\end{equation}
where $\mathcal{I}_{n\times n}$ is the $n\!\times\!n$ identity matrix,
and $\boldsymbol{\Delta}=\text{diag}(\boldsymbol{\delta}_{1},\cdots,\boldsymbol{\delta}_{n_{\text{b}}})$
is a diagonal matrix containing the relative positions of the different
Wannier orbitals within the unit cell. As for the velocity operator,
one can define it from the commutator of the Hamiltonian with the
position operator. That leads to

\vspace{-0.7cm}
\begin{equation}
\boldsymbol{\mathbf{\mathcal{V}}}\!=\!\frac{1}{i\hbar}\left[\boldsymbol{\mathcal{R}},\mathcal{H}\right]=\frac{1}{i\hbar}\sum_{\alpha,\beta,\gamma=1}^{n_{\text{b}}}\sum_{\mathbf{R}\in\mathcal{L}}\sum_{\mathbf{R}^{\prime}\in\mathcal{L}}\sum_{\mathbf{L}\in\mathcal{L}}\left(\mathbf{L+\boldsymbol{\delta}_{\!\alpha}}\right)T_{\mathbf{R}^{\prime}\mathbf{R}}^{\beta,\gamma}\left[c_{\alpha\mathbf{L}}^{\dagger}c_{\alpha\mathbf{L}},c_{\beta\mathbf{R}^{\prime}}^{\dagger}c_{\gamma\mathbf{R}}\right],\label{eq:Velocity}
\end{equation}
such that one can employ the identity,

\vspace{-0.7cm}
\begin{equation}
\left[c_{\alpha\mathbf{L}}^{\dagger}c_{\alpha\mathbf{L}}\,,\,c_{\beta\mathbf{R}^{\prime}}^{\dagger}c_{\gamma\mathbf{R}}\right]=\delta_{\alpha\beta}\delta_{\mathbf{L}\mathbf{R}^{\prime}}c_{\alpha\mathbf{L}}^{\dagger}c_{\gamma\mathbf{R}}-\delta_{\alpha\gamma}\delta_{\mathbf{L}\mathbf{R}}c_{\beta\mathbf{R}^{\prime}}^{\dagger}c_{\alpha\mathbf{L}},\label{eq:Commutatort}
\end{equation}
to arrive at

\vspace{-0.7cm}
\begin{align}
\boldsymbol{\mathbf{\mathcal{V}}}\! & =\!\frac{1}{i\hbar}\sum_{\alpha,\beta=1}^{n_{\text{b}}}\sum_{\mathbf{R}\in\mathcal{L}}\sum_{\mathbf{R}^{\prime}\in\mathcal{L}}c_{\alpha\mathbf{R}^{\prime}}^{\dagger}\left[\mathbf{R}^{\prime}\!-\mathbf{R}\!+\!\boldsymbol{\delta}_{\!\alpha}\!-\!\boldsymbol{\delta}_{\!\beta}\right]T_{\mathbf{R}^{\prime}\mathbf{R}}^{\alpha\beta}c_{\beta\mathbf{R}}.\label{eq:Velocity-1}
\end{align}
In fact, Eq.\,\eqref{eq:Velocity-1} defines an operator that is
very similar to the tight-binding Hamiltonian. The only change is
that the hopping matrix gets multiplied by the hopping displacement
vector times $1/i\hbar$. The construction of lattice observables
as well as calculation methods will be left for later discussion.

\section{\label{sec:Generic-Symmetries}Generic Symmetries of Bloch Hamiltonians}

Until now, the only symmetry we have explored was the lattice translation
symmetry that any ideal crystal has. However, typical solid-state
systems have additional symmetries that can yield further important
physical consequences. Some of those form what is called the \textit{crystalline
point-symmetry group}, which may include operations such as discrete
rotations, planar-reflections and space-inversion\,\footnote{In some cases, the system may have additional spacial symmetries (called
non-symmorphic) that are combinations of point-group operations and
non-primitive space translations.}. In addition to spatial symmetries one may also have non-spacial
ones, such as time-reversal symmetry, sublattice symmetries (related
to exchange of orbitals within a unit cell) and also spectral symmetries,
such as particle-hole. We will see that some these symmetries are
essential for the realization of stable gapless electronic phases
in three-dimensions. Furthermore, it is known that some effects of
disorder can be assessed quite generally by looking at the presence
(or absence) of certain non-spacial symmetries, leading to the Altland-Zirnbauer
scheme of classifying random Hamiltonians\,\cite{Altland98,Altland2002,Heinzner2005,Chiu2016}.
This section is devoted to a characterization of four important and
common types of symmetry \textemdash{} \textit{spacial inversion},
\textit{time-reversal}, \textit{particle-hole} and \textit{chiral
symmetry} \textemdash{} which will be shown to translate into very
recognizable features of the tight-binding Bloch Hamiltonian and its
band-structure.

\vspace{-0.4cm}

\subsection{Inversion Symmetry}

Spacial inversion (or \textit{parity}) is a very common symmetry of
real crystals and refers to an invariance of the lattice Hamiltonian
upon the reversal of all spacial directions. Such a system is called
\textit{centrosymmetric} because there is a point in space relative
to which the operation, $\mathcal{P}:\mathbf{r}\to-\mathbf{r}$, leaves
the system invariant. Such an operation can be implemented in the
Hilbert space by means of an unitary operator $\mathcal{U}_{i}$,
such that

\vspace{-0.7cm}
\begin{equation}
\mathcal{U}_{i}\mathcal{H}_{\text{tb}}\mathcal{U}_{i}^{-1}\!=\!\mathcal{H}_{\text{tb}}\Rightarrow\mathcal{U}_{i}\mathcal{H}\left(\mathbf{k}\right)\mathcal{U}_{i}^{-1}\!=\!U_{i}\mathcal{H}\left(-\mathbf{k}\right)U_{i}^{-1}\!=\!\mathcal{H}\left(\mathbf{k}\right),\label{eq:InversionSymmetry}
\end{equation}
where $U_{i}$ is a unitary matrix in the space of Wannier orbitals
within a unit cell\,\footnote{This transformation depends on the atomic basis of the lattice model.
For example, in monolayer graphene inversion symmetry exchanges sublattices,
\textit{i.e.}, $U_{i}\!=\!\sigma_{x}$.}. Since a unitary transformation that commutes with the Hamiltonian
does not change its spectrum, Eq.\,\eqref{eq:InversionSymmetry}
leads to the conclusion that there will always be two bands, $s$
and $s^{\prime}$, such that $\varepsilon_{\mathbf{k}}^{s}=\varepsilon_{-\mathbf{k}}^{s^{\prime}}$.
In other words, the existence of an inversion center in the system
guarantees that for each $\mathbf{k}$, there is a Bloch eigenstate
of the same energy at $-\mathbf{k}$ that may, or may not, belong
to the same energy band.

\vspace{-0.4cm}

\subsection{\label{subsec:Time-Reversal-Symmetry}Time-Reversal Symmetry}

Time-reversal symmetry (TRS), sometimes called \textit{reciprocity},
is also a quite general property of quantum and classical systems
alike. Typically, its breaking is related to the presence of magnetism
of some kind (or external magnetic fields) in the system. In quantum
mechanics, this symmetry cannot be implemented as a unitary operator,
but rather must be an anti-unitary\,\cite{Wigner32} that is written
as $\mathcal{T}=\mathcal{U}_{t}\mathcal{C},$ where $\mathcal{U}_{t}$
is an unitary operator and $\mathcal{C}$ is the complex-conjugation.
Unlike inversion symmetry, time-reversal leaves $\mathbf{r}$ unchanged
and acts trivially in the space of in-cell orbitals. The same is not
true in regard to the spin-$\nicefrac{1}{2}$ of the electron, which
gets rotated by a $\sigma_{y}$ Pauli operator. Therefore, if we explicitly
consider each eigenstate $\ket{\Psi_{\mathbf{k}}^{s}}$ as being a
two-component spinor, acting with the time-reversal operator yields
the following transformation on the Bloch state:

\vspace{-0.7cm}
\begin{equation}
\mathcal{T}\ket{\vphantom{\Psi_{-\mathbf{k}}^{s}}\Psi_{\mathbf{k}}^{s}}=\sigma_{y}\,\mathcal{C}\ket{\vphantom{\Psi_{-\mathbf{k}}^{s}}\Psi_{\mathbf{k}}^{s}}=\sigma_{y}\ket{\Psi_{-\mathbf{k}}^{s}},\label{eq:Timereversal}
\end{equation}
where we have used the fact that the conjugation operation, $\mathcal{C}$,
reverses the $\mathbf{k}$ vector. Note that Eq.\,\eqref{eq:Timereversal}
implies that $\mathcal{T}^{2}=-1$, which would not be possible if
$\mathcal{T}$ were a unitary operator and, in this case, it is a
direct consequence of the spinful nature of electrons\,\footnote{While this is certainly the case, the property of $\mathcal{T}^{2}=-1$
is not exclusive to systems with spin-$\nicefrac{1}{2}$ degrees of
freedom.}. In addition, it is also easily seen that $\varepsilon_{\mathbf{k}}^{s}\!=\!\varepsilon_{-\mathbf{k}}^{s},$
which shows that the implication of TRS in the band structure is similar
to what is produced by parity, but now one is guaranteed to stay in
the same band.

\vspace{-0.4cm}

\paragraph*{Kramer's Degeneracy:}

Time-reversal and inversion symmetry may be deceitfully similar symmetry
operations but can have dramatically different consequences. The most
remarkable difference is perhaps \textit{Kramer's theorem}\,\cite{Kramers30,Klein52}.
In a general point of the fBz, the condition $\varepsilon_{\mathbf{k}}^{s}=\varepsilon_{-\mathbf{k}}^{s}$
only guarantees an equality between energies of orthogonal states
at different momenta. However, there are special $\mathbf{k}$-points
which are equivalent (by a dual lattice translation) to $\mathbf{-k}$,
such that TRS guarantees $\ket{\Psi_{\mathbf{k}}^{s}}$ and $\ket{\Phi_{\mathbf{k}}^{s}}\!=\!\mathcal{T}\ket{\Psi_{\mathbf{k}}^{s}}$
are states with the exactly same energy. Therefore, we may find one
of two situations: \textit{(i)} $\ket{\Phi_{\mathbf{k}}^{s}}$ and
$\ket{\Psi_{\mathbf{k}}^{s}}$ are actually the same (TRS invariant)
quantum state, or \textit{(ii)} $\ket{\Phi_{\mathbf{k}}^{s}}$ and
$\ket{\Psi_{\mathbf{k}}^{s}}$ are orthogonal and degenerate states.
If $\mathcal{T}^{2}=1$ both situations can happen and the two states
may only differ by a global phase factor, \textit{i}.e., $\ket{\Phi_{\mathbf{k}}^{s}}=e^{i\theta}\ket{\Psi_{\mathbf{k}}^{s}}$.
In contrast, if $\mathcal{T}^{2}=-1$ the state $\ket{\Phi_{\mathbf{k}}^{s}}$
must be orthogonal to $\ket{\Psi_{\mathbf{k}}^{s}}$ or otherwise,

\vspace{-0.7cm}
\begin{equation}
\mathcal{T}^{2}\ket{\Psi_{\mathbf{k}}^{s}}=e^{-i\theta}\mathcal{T}\ket{\Psi_{\mathbf{k}}^{s}}=e^{-i\theta}\ket{\Phi_{\mathbf{k}}^{s}}=\ket{\Psi_{\mathbf{k}}^{s}},
\end{equation}
which implies $\mathcal{T}^{2}\!=\!1$ and contradicts the original
statement. Therefore, if the system is time-reversal symmetric and
$\mathcal{T}^{2}=-1$, then the band-structure must be degenerate
at any time-reversal invariant momentum (TRIM). In this context, $\ket{\Phi_{\mathbf{k}}^{s}}$
and $\ket{\Psi_{\mathbf{k}}^{s}}$ are usually called a \textit{Kramer's
Pair}.

\vspace{-0.4cm}

\paragraph*{$\mathcal{PT}$\textendash \,Symmetry:}

At this point, it is interesting to explore the case in which both
TRS and parity are present in the system (then called \textit{$\mathcal{PT}$-symmetric}).
In that situation, one can use both symmetries to conclude that the
electronic band structure must obey $\varepsilon_{\mathbf{k}}^{s}=\varepsilon_{\mathbf{k}}^{s^{\prime}}$.
Once again, this condition may be a trivial statement if $\mathcal{P}\mathcal{T}\ket{\Psi_{\mathbf{k}}^{s}}$
is the same quantum state as $\ket{\Psi_{\mathbf{k}}^{s}}$. However,
like $\mathcal{T}$ one may also have $\left(\mathcal{P}\mathcal{T}\right)^{2}=-1$
which guarantees, by Kramer's theorem, that $\ket{\Psi_{\mathbf{k}}^{s}}$
is a degenerate eigenstate for any $\mathbf{k}\in\text{fBz}$. Finally,
since any single-electron system always features a spin-$\nicefrac{1}{2}$
degree of freedom, the presence of $\mathcal{PT}$-symmetry guarantees
that every band is two-fold degenerate, even when accounting for spin-orbit
coupling. In case there is spin-rotation symmetry, this two-fold degeneracy
is interpreted as being due to two independent spin-sectors within
each band, such that spin will enter as a simple degeneracy factor
in all calculations.

\vspace{-0.4cm}

\subsection{Generalized Spectral Symmetries}

Besides the standard symmetries presented before, there are also two
others that are of great importance in condensed matter physics\,\cite{Chiu2016}.
These are not quantum-mechanical symmetries of the single-particle
problem, in the strict sense, as they are not represented by operators
that commute with the corresponding Hamiltonian. However, they are
commonly found in solid-state systems and have important implications
on the general form of the single-particle spectrum.

\vspace{-0.4cm}

\paragraph{Particle-Hole Symmetry:}

One such symmetry is \textit{particle-hole symmetry} (PHS), which
can be defined as an operation that converts creation into annihilation
operators of opposite momentum, \textit{i.e.},

\vspace{-0.7cm}
\begin{align}
c_{\alpha\mathbf{k}}^{\dagger}\! & \to\!\mathcal{A}\,c_{\alpha\mathbf{k}}^{\dagger}\mathcal{A}^{-1}\!=\!U_{\text{ph}}^{\alpha\beta}c_{\beta-\mathbf{k}}\label{eq:ParticleHole}\\
c_{\alpha\mathbf{k}}\! & \to\!\mathcal{A}\,c_{\alpha\mathbf{k}}\mathcal{A}^{-1}\!=\!\left(U_{\text{ph}}^{\alpha\beta}\right)^{*}c_{\beta-\mathbf{k}}^{\dagger}.
\end{align}
For this symmetry operation to be well-defined, the anti-commutation
relations between fermionic operators must be preserved, which means
that

\vspace{-0.7cm}

\begin{align}
\mathcal{A}\left\{ c_{\alpha\mathbf{k}},c_{\beta\mathbf{k}^{\prime}}^{\dagger}\right\} \mathcal{A}^{-1} & =\left[\mathbf{U}_{\text{ph}}\!\cdot\mathbf{U}_{\text{ph}}^{\dagger}\right]^{\beta\alpha}=\delta_{\alpha\beta}\delta_{\mathbf{k}\mathbf{k}^{\prime}},
\end{align}
thus imposing $\mathbf{U}_{\text{ph}}$ to be a unitary matrix. While
preserving the fermionic statistics of single-particle excitations,
the PHS must also leave the full Hamiltonian invariant, which boils
down to the following condition on the Bloch Hamiltonian:

\vspace{-0.7cm}
\begin{align}
\!\!\!\!\!\!\mathcal{H}_{\text{tb}}\!=\!\mathcal{A}\,\mathcal{H}_{\text{tb}}\mathcal{A}^{-1} & =\sum_{\mathbf{k}}U_{\text{ph}}^{\alpha\gamma}c_{\gamma-\mathbf{k}}\left[\mathcal{H}\!\left(\mathbf{k}\right)\right]^{\alpha\beta}\left(U_{\text{ph}}^{\beta\delta}\right)^{*}c_{\delta-\mathbf{k}}^{\dagger}\!\!\!\!\!\!\nonumber \\
 & =\sum_{\mathbf{k}}U_{\text{ph}}^{\alpha\gamma}\left[\mathcal{H}\!\left(\mathbf{k}\right)\right]^{\alpha\beta}\left(U_{\text{ph}}^{\beta\gamma}\right)^{*}\!\!-\!\sum_{\mathbf{k}}U_{\text{ph}}^{\alpha\gamma}c_{\delta-\mathbf{k}}^{\dagger}\left[\mathcal{H}\!\left(\mathbf{k}\right)\right]^{\alpha\beta}\left(U_{\text{ph}}^{\beta\delta}\right)^{*}c_{\gamma-\mathbf{k}}\nonumber \\
 & =\sum_{\mathbf{k}}\text{Tr}\left[\mathcal{H}\!\left(\mathbf{k}\right)\right]-\sum_{\mathbf{k}}U_{\text{ph}}^{\alpha\gamma}c_{\gamma\mathbf{k}}^{\dagger}\left[\mathcal{H}\!\left(-\mathbf{k}\right)\right]^{\alpha\beta}\left(U_{\text{ph}}^{\beta\delta}\right)^{*}c_{\delta\mathbf{k}}\nonumber \\
 & =\sum_{\mathbf{k}}\text{Tr}\left[\mathcal{H}\!\left(\mathbf{k}\right)\right]-\sum_{\mathbf{k}}c_{\gamma\mathbf{k}}^{\dagger}\left(U_{\text{ph}}^{\beta\delta}\right)^{*}\left(\left[\mathcal{H}\!\left(-\mathbf{k}\right)\right]^{\beta\alpha}\right)^{*}U_{\text{ph}}^{\alpha\gamma}c_{\delta\mathbf{k}}\nonumber \\
 & =\sum_{\mathbf{k}}\text{Tr}\left[\mathcal{H}\!\left(\mathbf{k}\right)\right]-\sum_{\mathbf{k}}\boldsymbol{\Upsilon}{}_{\mathbf{k}}^{\dagger}\!\cdot\mathbf{U}_{\text{ph}}^{\dagger}\cdot\mathcal{H}^{*}\!\left(-\mathbf{k}\right)\cdot\mathbf{U}_{\text{ph}}\cdot\boldsymbol{\Upsilon}{}_{\mathbf{k}}.\label{eq:ConditionPHS}
\end{align}
Equation\,\eqref{eq:ConditionPHS} can be verified if the Bloch Hamiltonian
is constrained to transform as $\mathcal{H}(\mathbf{k})=-\mathbf{U}_{\text{ph}}^{\dagger}\cdot\mathcal{H}^{*}\left(-\mathbf{k}\right)\cdot\mathbf{U}_{\text{ph}},$
which trivially guarantees that $\text{Tr}\left[\mathcal{H}\left(\mathbf{k}\right)\right]\!=\!0$
because $\mathcal{H}\!\left(\mathbf{k}\right)\!=\!\left[\mathcal{H}\left(\mathbf{k}\right)\right]^{\dagger}$.
Hence, when acting in space of bands at a given $\mathbf{k}$, the
particle-hole operation is anti-unitary, $\mathcal{A}=\mathcal{U}_{\text{ph}}\mathcal{C}$,
and has the following action:

\vspace{-0.7cm}
\begin{equation}
\mathcal{A}\mathcal{H}(\mathbf{k})\mathcal{A}^{-1}=\mathcal{U}_{\text{ph}}\mathcal{C}\mathcal{H}\left(\mathbf{k}\right)\mathcal{C}^{-1}\mathcal{U}_{\text{ph}}^{-1}=\mathbf{U}_{\text{ph}}^{\dagger}\!\cdot\!\mathcal{H}^{*}\left(\mathbf{k}\right)\!\cdot\!\mathbf{U}_{\text{ph}}=-\mathcal{H}\left(-\mathbf{k}\right).
\end{equation}
Therefore, this anti-unitary operation is not a proper quantum-mechanical
symmetry of the Bloch Hamiltonian, but still it characterizes its
spectrum: given an eigenstate $\ket{\Psi_{s\mathbf{k}}}$ of $\mathcal{H}(\mathbf{k})$
with energy $\varepsilon_{\mathbf{k}}^{s}$, then $\ket{\Theta_{s\mathbf{k}}}=\mathcal{A}\ket{\Psi_{s\mathbf{k}}}$
is also an eigenstate that has crystal momentum $-\mathbf{k}$ and
energy $-\varepsilon_{\mathbf{k}}^{s}$. If the state's energy is
non-zero, then $\varepsilon_{\mathbf{k}}^{s},\neq-\varepsilon_{-\mathbf{k}}^{s}$
and PHS guarantees that every positive energy eigenstate has a corresponding
partner at a symmetric energy and with symmetric momentum; eigenstates
come in particle-hole pairs. In contrast, zero-energy states are special
in systems with PHS; If one appears at a generic point $\mathbf{k}$
in the fBz, then there will be another placed at $-\mathbf{k}$. The
notable exception happens when $\mathbf{k}$ is a TRIM for which a
doublet of eigenstates is only guaranteed provided $\mathcal{A}^{2}=-1$.

\vspace{-0.4cm}

\paragraph{Chiral Symmetry:}

PHS is a generalized symmetry that is implemented as an anti-unitary
operator, while TRS is a proper symmetry that is also anti-unitary.
By combining these two operations, one builds a generalized symmetry
that appears as an unitary operator that anti-commutes with the single-particle
Hamiltonian: the \textit{chiral symmetry} (CS). This is implemented
by the operator $\mathcal{S}\!=\!\mathcal{T}\mathcal{A}$ which, when
applied to the Bloch Hamiltonian, yields

\vspace{-0.7cm}

\begin{align}
\mathcal{S}\mathcal{H}(\mathbf{k})\mathcal{S}^{-1}=\mathcal{T}\mathcal{A}\mathcal{H}(\mathbf{k})\mathcal{A}^{-1}\mathcal{T}^{-1} & =\mathcal{T}\mathcal{U}_{\text{ph}}\mathcal{H}^{*}(\mathbf{k})\mathcal{U}_{\text{ph}}^{-1}\mathcal{T}^{-1}\\
 & =\mathcal{U}_{\text{t}}\left(\mathcal{U}_{\text{ph}}\mathcal{H}^{*}(-\mathbf{k})\mathcal{U}_{\text{ph}}^{\dagger}\right)^{*}\mathcal{U}_{\text{t}}^{\dagger}.\nonumber 
\end{align}
The system is said to have chiral symmetry if $\mathcal{S}\mathcal{H}(\mathbf{k})\mathcal{S}^{-1}=-\mathcal{H}\left(\mathbf{k}\right)$
and a major consequence is that, given a Bloch eigenstate $\ket{\Psi_{\mathbf{k}}^{s}}$,
then $\ket{\Xi_{\mathbf{k}}^{s}}\!=\!\mathcal{S}\ket{\Psi_{\mathbf{k}}^{s}}$
is also an eigenstate of the tight-binding Hamiltonian with energy
$-\varepsilon_{\mathbf{k}}^{s}$. Note that, unlike what happened
in PHS, here there are no zero-energy Kramer pairs, because the CS
is unitary and always squares to the identity. A second important
by-product of CS on a generic Bloch Hamiltonian is that it can always
be (unitarily) transformed into the off-diagonal form,

\vspace{-0.7cm}
\begin{equation}
\mathcal{\mathcal{H}}(\mathbf{k})=\left[\begin{array}{cc}
0 & h(\mathbf{k})\\
h^{^{\!\dagger}}\!(\mathbf{k}) & 0
\end{array}\right],\label{eq:OffDiagona=0000E7}
\end{equation}
where $h(\mathbf{k})$ is a $\mathbf{k}$-dependent $n\!\times\!m$
matrix. In order to see how this relation comes about, one can start
by verifying that the off-diagonal Bloch Hamiltonian of Eq.\,\eqref{eq:OffDiagona=0000E7}
is chiral symmetric, with the following unitary operator carrying
out the transformation:

\vspace{-0.7cm}

\begin{equation}
\mathcal{S}\to\left[\!\begin{array}{cc}
\mathcal{I}_{{\scriptscriptstyle n\times n}} & 0\\
0 & -\mathcal{I}_{{\scriptscriptstyle m\times m}}
\end{array}\!\right]\Longrightarrow\mathcal{S}\cdot\mathcal{S}=\mathcal{I}_{{\scriptscriptstyle (n+m)\times(n+m)}}\text{ and }\mathcal{S}\cdot\mathcal{\mathcal{H}}(\mathbf{k})\cdot\mathcal{S}=-\mathcal{\mathcal{H}}(\mathbf{k}).\label{eq:ChiralSymmetryOperator}
\end{equation}
The reciprocal implication can be obtain by assuming that $\mathcal{S}$
takes the block-diagonal form of Eq.\,\eqref{eq:ChiralSymmetryOperator},
but now the Bloch Hamiltonian reads

\vspace{-0.7cm}

\begin{equation}
\mathcal{\mathcal{H}}(\mathbf{k})=\left[\begin{array}{cc}
f(\mathbf{k}) & g(\mathbf{k})\\
g^{\!^{\dagger}}\!(\mathbf{k}) & w(\mathbf{k})
\end{array}\right],\label{eq:OffDiagona=0000E7-1}
\end{equation}
with $f(\mathbf{k})\text{ and }w(\mathbf{k})$ being $n\!\times\!n$
and $m\!\times\!m$ hermitian blocks, and $g(\mathbf{k})$ a $\mathbf{k}$-dependent
$n\!\times\!m$ matrix. In this case, we impose CS in the system through
the condition

\vspace{-0.7cm}

\begin{equation}
\mathcal{S}\!\cdot\!\mathcal{\mathcal{H}}(\mathbf{k})\!\cdot\!\mathcal{S}=-\mathcal{\mathcal{H}}(\mathbf{k})\Longrightarrow\left[\begin{array}{cc}
f(\mathbf{k}) & -g(\mathbf{k})\\
-g^{^{\!\dagger}}\!(\mathbf{k}) & w(\mathbf{k})
\end{array}\right]=\left[\begin{array}{cc}
-f(\mathbf{k}) & -g(\mathbf{k})\\
-g^{\!^{\dagger}}\!(\mathbf{k}) & -w(\mathbf{k})
\end{array}\right],\label{eq:Chiral Symmetry}
\end{equation}
which can only happen if $f(\mathbf{k})=w(\mathbf{k})=0$. Note that,
while Eq.\,\eqref{eq:Chiral Symmetry} proves that a chiral symmetric
$\mathcal{H}(\mathbf{k})$ must be off-diagonal in case $\mathcal{S}$
takes the block-diagonal form of Eq.\,\eqref{eq:ChiralSymmetryOperator},
the result is completely general. Since $\mathcal{S}$ is a $\mathbf{k}$-independent
unitary matrix, it can be brought into a diagonal form upon a unitary
transformation\,\footnote{In this context, such is interpreted as a unitary transformation in
the space of orbitals within a unit-cell.} and, needing to obey $\mathcal{S}^{\dagger}\cdot\mathcal{S}\!=\!1$,
it must always have the form,

\vspace{-0.7cm}
\begin{equation}
\mathcal{S}\!=\!U^{\dagger}\cdot\left[\!\begin{array}{cc}
\mathcal{I}_{{\scriptscriptstyle n\times n}} & 0\\
0 & -\mathcal{I}_{{\scriptscriptstyle m\times m}}
\end{array}\!\right]\cdot U,
\end{equation}
with a unitary matrix $U$. The above proof can then be applied to
the transformed Bloch Hamiltonian, $\tilde{\mathcal{H}}(\mathbf{k})\!=\!U^{\dagger}\mathcal{H}(\mathbf{k})U$,
without any adaptation.

Chiral symmetry is ubiquitous in Hamiltonians from a wide variety
of contexts. However, for our purposes, we are only interested on
its realization in single-electron tight-binding problems, for which
an off-diagonal form of the Bloch Hamiltonian indicates there is a
choice of the local orbital basis that evidences a sublattice symmetry\,\cite{Gade91,Gade93}.
Perhaps the simplest example of that is the celebrated one-dimensional
\textit{Su-Schrieffer-Heeger model}\,\cite{Su79}, even though many
others of higher dimensionality are also known (\textit{e.g.}, 2D
graphene and 3D Dirac semimetals).

\vspace{-0.4cm}

\subsection{Classification based on Altland-Zirnbauer Symmetries}

The last three aforementioned \textit{Altland-Zirnbauer symmetries\,}\cite{Altland98}
\textemdash{} time-reversal, particle-hole and chiral \textemdash{}
are of particular importance because they are quite generic in condensed
matter Hamiltonians. In particular, a tenfold classification of such
Hamiltonians can be made based on the presence or absence of these
symmetries, together with the fact that an anti-unitary representation
of TRS and PHS can square to $\pm1$. This symmetry-based classification
can be used to obtain very general information from the Hamiltonian,
most notably to classify all possible $d$-dimensional bulk topological
gapped phases\,\cite{Schnyder2008,Kitaev2009} which, in turn, can
also be used to infer the existence of gapless modes bound to boundaries
or around topological defects of any dimension\,\cite{Teo2010,Essin2011}.
Without going into much detail, at a given dimension $d$, a gapped
topological phase can be either impossible ($0$) or else be labeled
by a $\mathbb{Z}$-, $2\mathbb{Z}$- or $\mathbb{Z}_{2}$-topological
index (this classification is shown in Table\,\ref{tab:(a)-Classification-of}\,a
for one-, two- and three-dimensional systems). These topological indices
can be thought as a classification of all topologically distinct ground
states that a fermionic Hamiltonian with those symmetries can have.
Transitions between distinct ground-states can be made to happen by
deforming the Hamiltonian without changing its symmetry class, but
they must always proceed by closing the spectral gap. In addition,
the topological indices also indicate the number of protected gapless
modes that appear at the boundary between different bulk topological
insulating phases. This bulk-boundary correspondence may actually
be extended to the analysis of modes appearing around any topological
defect characterized by a co-dimension $\delta$\,\footnote{The co-dimension of a defect is defined as $\delta\!=\!d\!-\!d_{\text{d}}$,
where $d$ is the bulk dimension and $d_{\text{d}}$ is the defect's
dimension. For example, a line defect in three-dimensions has $\delta\!=\!2$,
while a planar boundary has $\delta\!=\!1$. For a comprehensive review
on the general theory of lattice defects see Mermin's review\,\cite{Mermin79}.} (see Table\,\ref{tab:(a)-Classification-of}\,b). This gives rise
to a series of so-called \textit{Index Theorems} that constrain the
number of such gapless states and may even protect them against scattering
effects\,\cite{Atiyah1963,Jackiw76,Weinberg81,Fukui2010,Fukui2010b,Teo2010,Essin2011}.

\begin{table}[t]
\vspace{-0.4cm}
\begin{centering}
\includegraphics[scale=0.22]{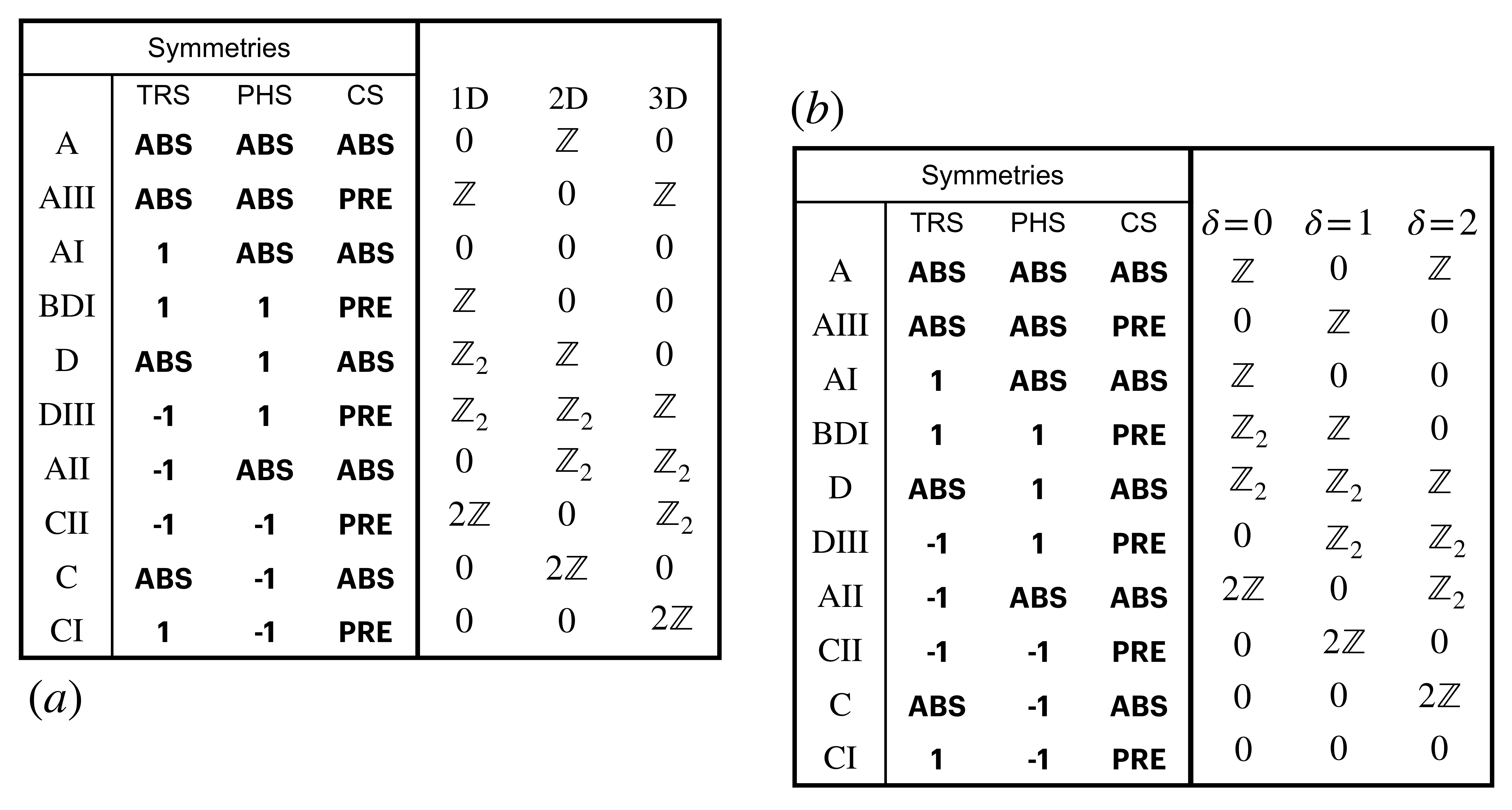}
\par\end{centering}
\caption{\label{tab:(a)-Classification-of}(a) Classification of all bulk gapped
phases from topological indices.\,\,{[}\textbf{Notation:} ABS\,=\textit{\,\textquotedblright absent\textquotedblright},\,\,PRE\,=\textit{\,\textquotedblright present}\textquotedblright ,\,\,$\pm1$\,=\textit{\,\textquotedblright Present\,+\,Squares
to $\pm1$}\textquotedblright{]} (b) Altland-Zirnbauer classification
of real-space topological defects of low co-dimension.}

\vspace{-0.8cm}
\end{table}

\vspace{-0.4cm}

\section{\label{sec:Topological-Properties}Topological Properties of Bloch
Hamiltonians}

Topology has been a cornerstone of quantum solid-state matter since
the early work of Thouless \textit{et al}.\,\cite{Thouless82} and
Haldane\,\cite{Haldane88} on the theory of the integer quantized
Hall effect\,\cite{Klitzing80}. Then, it was recognized that mathematical
properties of the whole Brillouin zone in a 2D electron system can
guarantee a precise quantization of transverse currents as integer
multiples of $e^{2}/\hbar$, for a wide range of deformations of the
system's Hamiltonian. Soon, building on the ideas of Berry\,\cite{Berry84},
those properties were connected to the fundamental geometry of the
band-structure and the quantized Hall conductivity was associated
to topological invariants that can only change via a new kind of quantum
phase transition: a \textit{topological phase transition}. In this
section, we give a brief overview on the topological properties of
general quantum mechanics, highlighting their consequences to the
physics of independent electrons in a crystal. This discussion will
allow us to naturally introduce the central concepts of a topological
insulator and semimetal.

\vspace{-0.4cm}

\subsection{Geometric Phases in Quantum Mechanics}

The importance of topology in quantum mechanics arises whenever a
quantum system is described by a continuous set of parameters, $\boldsymbol{\lambda}=\left(\lambda_{1},\cdots\lambda_{d}\right)$,
that determine its Hamiltonian, $H(\boldsymbol{\lambda})$. For each
$\boldsymbol{\lambda}$, the Hamiltonian has a set of eigenstates
$\ket{\psi_{\boldsymbol{\lambda}}^{n}}$, labelled by an index $n$
which encapsulates a complete set of good quantum numbers. For now,
we will consider that $H(\boldsymbol{\lambda})$ is finite-dimensional
and all eigenstates,

\vspace{-0.7cm}
\begin{equation}
H(\boldsymbol{\lambda})\ket{\psi_{\boldsymbol{\lambda}}^{n}}=E_{\boldsymbol{\lambda}}^{n}\ket{\psi_{\boldsymbol{\lambda}}^{n}},\label{eq:ParametricHam}
\end{equation}
are non-degenerate ($n$ then specifies the energy). Even though we
are keeping the discussion general, it is important to note that a
Bloch Hamiltonian, $\mathcal{H}(\mathbf{k})$, falls in this scheme
perfectly, where $\mathbf{k}\in\text{fBz}$ plays the role of $\boldsymbol{\lambda}$
and the eigenstates are the periodic parts of the Bloch wavefunctions,
$\ket{\chi_{\mathbf{k}}^{s}}$, which are labelled by a discrete band
index $s$.

The topology of $H(\boldsymbol{\lambda})$ comes from the properties
of its spectrum as it gets deformed within its parameter space ($\boldsymbol{\lambda}$-space).
In order to see this, we assume that $\boldsymbol{\lambda}\to\boldsymbol{\lambda}(\tau)$
acquires a time-dependence that moves it around a curve $\mathcal{C}$
in $\boldsymbol{\lambda}$-space. If the system starts at the eigenstate
$\ket{\psi_{\boldsymbol{\lambda}(0)}^{m}}$ then a sufficiently slow
variation of $\boldsymbol{\lambda}$ will not lead to a mixing with
other eigenstates\,\footnote{Which is guaranteed by the \textit{Adiabatic Theorem}\,\cite{Born28}.}
and therefore,

\vspace{-0.7cm}
\begin{equation}
\ket{\psi_{\boldsymbol{\lambda}(\tau)}^{m}}=\mathcal{A}_{m}\left(\tau\right)\ket{\psi_{\boldsymbol{\lambda}(0)}^{m}},
\end{equation}
where $\mathcal{A}_{m}\left(\tau\right)$ is a time-dependent complex
phase. The general expression of this phase factor can be obtained
from the Schrödinger equation,

\vspace{-0.7cm}
\begin{equation}
i\hbar\frac{d}{d\tau}\ket{\psi_{\boldsymbol{\lambda}(\tau)}^{m}}=H\left(\boldsymbol{\lambda}(\tau)\right)\ket{\psi_{\boldsymbol{\lambda}(\tau)}^{m}}\Rightarrow\mathcal{A}_{m}\left(\tau\right)=e^{i\gamma_{m}\left(\tau\right)}\exp\left[\frac{1}{i\hbar}\int_{0}^{\tau}\!\!\!\!du\,E_{\boldsymbol{\lambda}(u)}^{m}\right],\label{eq:GeometricPhase}
\end{equation}
where the second term is conventionally known as the \textit{dynamical
phase-factor} and the first is called the \textit{geometrical phase-factor}.
The function $\gamma_{m}\left(\tau\right)$ is not entirely arbitrary
and must obey the following ordinary differential equation (ODE):

\vspace{-0.7cm}
\begin{equation}
\frac{d}{d\tau}\gamma_{m}\left(\tau\right)=i\braket{\psi_{\boldsymbol{\lambda}(\tau)}^{m}}{\partial_{\tau}\psi_{\boldsymbol{\lambda}(\tau)}^{m}}.\label{eq:GeometricPhaseEq}
\end{equation}
In a sense, Eq.\,\eqref{eq:GeometricPhaseEq} already justifies the
geometrical nature of $\gamma_{m}$, as it abstractly measures how
much the eigenstate $\ket{\psi_{\boldsymbol{\lambda}(\tau)}^{m}}$
misaligns as it gets transported along $\mathcal{C}$ without changing
its eigenvalue index. This is the quantum-mechanical notion of parallel
transport along the manifold of parameters that determines the Hamiltonian.
Most often, quantum-mechanical properties are insensitive to global
phase-factors and indeed in this case, we can see that a different
gauge choice for the basis $\ket{\psi_{\boldsymbol{\lambda}}^{n}}$,
namely

\vspace{-0.7cm}
\begin{equation}
\ket{\psi_{\boldsymbol{\lambda}}^{n}}\to e^{i\chi^{n}\!(\boldsymbol{\lambda})}\ket{\psi_{\boldsymbol{\lambda}}^{n}}\label{eq:GaugeTransformation}
\end{equation}
leads to different geometric phases for the same curve in parameter
space, \textit{i.e.},

\vspace{-0.7cm}
\begin{equation}
\gamma_{m}\left(\tau\right)\to\tilde{\gamma}_{m}\left(\tau\right)=\gamma_{m}\left(\tau\right)+\chi^{m}\left(\boldsymbol{\lambda}(0)\right)-\chi^{m}\left(\boldsymbol{\lambda}(\tau)\right).
\end{equation}
This means that one can usually eliminate all $\exp(i\gamma_{m}(\tau))$
factors simply by a \textit{global gauge transformation}. However,
there is a caveat to this statement in case $\boldsymbol{\lambda}(0)\!=\!\boldsymbol{\lambda}(\tau)$.
Then, we cannot choose $\chi^{m}\left(\boldsymbol{\lambda}(0)\right)-\chi^{m}\left(\boldsymbol{\lambda}(\tau)\right)$
to take on any value, because the same gauge must be used for both
$\ket{\psi_{\boldsymbol{\lambda}(0)}^{n}}$ and $\ket{\psi_{\boldsymbol{\lambda}(\tau)}^{n}}=\ket{\psi_{\boldsymbol{\lambda}(0)}^{n}}$.
Hence, as first observed by Berry\,\cite{Berry84}, the geometrical
phase factor of Eq.\,\eqref{eq:GeometricPhase} can yield physical
effects if it is found to be nonzero on a closed curve in $\boldsymbol{\lambda}$-space. 

Having demonstrated that $\gamma_{m}(\tau)$ is gauge-independent
for closed curves, we now show that it may be written as closed contour
integral in $\boldsymbol{\lambda}$-space. For that, we pick-up its
integral definition, 

\vspace{-0.7cm}
\begin{equation}
\gamma_{m}(\tau)\!=\!i\!\int_{0}^{\tau}\!\!\!du\braket{\psi_{\boldsymbol{\lambda}(u)}^{m}}{\partial_{u}\psi_{\boldsymbol{\lambda}(u)}^{m}}\!=\!i\!\int_{0}^{\tau}\!\!\!du\left[\braket{\psi_{\boldsymbol{\lambda}(u)}^{m}}{\boldsymbol{\nabla}_{\!\!\boldsymbol{\lambda}}\psi_{\boldsymbol{\lambda}}^{m}}\right]\partial_{u}\boldsymbol{\lambda}(u),
\end{equation}
and upon defining the \textit{Berry Connection} as $\boldsymbol{\mathcal{A}}_{\boldsymbol{\lambda}}^{m}\!=\!i\braket{\psi_{\boldsymbol{\lambda}}^{m}}{\boldsymbol{\nabla}_{\!\!\boldsymbol{\lambda}}\psi_{\boldsymbol{\lambda}}^{m}}$,
we can write

\vspace{-0.7cm}
\begin{equation}
\gamma_{m}^{\mathcal{C}}\!=\!\ointctrclockwiseop_{\mathcal{C}}\boldsymbol{\mathcal{A}}_{\boldsymbol{\lambda}}^{m}\cdot\boldsymbol{d\lambda}\label{eq:BerryPhaseContour}
\end{equation}
which identifies the geometric phase as a contour integral of the
Berry connection along a closed curve in $\boldsymbol{\lambda}$-space.
Note that we could drop the dependence on $\tau$, as the quantity
only depends on the curve itself, not on its precise parametrization.

\subsection{Berry Curvatures and Invariants in the Brillouin Zone}

Bearing in mind the definition of a Berry Connection, and Eq.\,\eqref{eq:BerryPhaseContour},
this is the point where we will specialize the discussion to systems
of Bloch electrons. For that, we replace $\boldsymbol{\lambda}\to\mathbf{k}$
and $H(\boldsymbol{\lambda})\to\mathcal{H}(\mathbf{k})$ and restrict
the discussion to two- or three-dimensional $\mathbf{k}$-spaces.
Then, we define the Berry connection for a band $s$ as

\vspace{-0.7cm}
\begin{equation}
\boldsymbol{\mathcal{A}}^{\!s}(\mathbf{k})\!=\!i\braket{\xi_{\mathbf{k}}^{s}}{\boldsymbol{\nabla}_{\!\!\mathbf{k}}\xi_{\mathbf{k}}^{s}},
\end{equation}
which is a vector field in $\mathbf{k}$-space with the same dimensionality.
Therefore, by considering our parameter space as two-dimensional (three-dimensional)
we can use \textit{Green's Theorem} (\textit{Stokes' Theorem}) to
re-write Eq.\,\eqref{eq:BerryPhaseContour} as a surface integral
that defines a new gauge-invariant scalar (vector) field, the \textit{Berry
Curvature}. Respectively, we have 

\vspace{-0.7cm}
\begin{equation}
\gamma_{s}^{\mathcal{C}}\!=\!\iint_{\text{int(\!\ensuremath{\mathcal{C}}\!)}}\!\!\!\!\!d^{{\scriptscriptstyle (2)}}\!\mathbf{k}\ \Omega_{s\mathbf{k}}\text{ with }\Omega_{s\mathbf{k}}\!=\!\partial_{k_{x}}\mathcal{A}_{y}^{\!s}(\mathbf{k})-\partial_{k_{y}}\mathcal{A}_{x}^{\!s}(\mathbf{k})
\end{equation}
for a two-dimensional system, and

\vspace{-0.7cm}

\begin{equation}
\gamma_{s}^{\mathcal{C}}\!=\!\iint_{S_{\mathcal{C}}}\!\!\!\boldsymbol{\Omega}_{s\mathbf{k}}\cdot\boldsymbol{dS}\text{ with }\boldsymbol{\Omega}_{s\mathbf{k}}\!=\!\boldsymbol{\nabla}_{\!\!\mathbf{k}}\times\boldsymbol{\mathcal{A}}^{\!s}(\mathbf{k})
\end{equation}
in the three-dimensional case, where $S_{\mathcal{C}}$ is any surface
with a boundary supported in $\mathcal{C}$. In both cases, once the
Berry curvature field is defined in the entirety of the fBz, we can
compute the Berry phase over any closed curve in $\mathbf{k}$-space
just by evaluating the flux of curvature through its inside. In the
remainder of this section, we will restrict the discussion to the
two-dimensional case, leaving the other as the basis for the next
section.

Now that we have defined the Berry curvature as a scalar field in
a 2D Brillouin zone, we set out to prove that its integral over the
entire fBz is a topological invariant. For that, we begin by remarking
that a Bloch state {[}Eq.\,\eqref{eq:BlochTheorem}{]} can be written
as,

\vspace{-0.7cm}
\begin{equation}
\ket{\psi_{\mathbf{k}}^{s}}\!=\!\sum_{\mathbf{R}\in\mathcal{L}}e^{i\mathbf{k}\cdot\mathbf{R}}\ket{\xi_{\mathbf{k}}^{s}}\tensorproduct\ket{\mathbf{R}},
\end{equation}
where $\mathcal{L}$ is the real-space lattice and $\mathbf{k}\in\text{fBz}$.
Since the full tight-binding Hamiltonian is invariant under lattice
translations, one requires that $\ket{\vphantom{\psi_{\mathbf{k}+\mathbf{K}}^{s}}\psi_{\mathbf{k}}^{s}}$
and $\ket{\psi_{\mathbf{k}+\mathbf{K}}^{s}}$ are the same quantum
state, for any vector $\mathbf{K}$ of the dual lattice $\mathcal{L}^{*}$.
As usual, this equivalence must be seen as modulo a global phase-factor
and, in particular, we can write that 

\vspace{-0.7cm}
\begin{equation}
\ket{\psi_{\mathbf{k}+\mathbf{K}}^{s}}=e^{i\vartheta_{s}(\mathbf{k})}\ket{\vphantom{\psi_{\mathbf{k}+\mathbf{K}}^{s}}\psi_{\mathbf{k}}^{s}}\Leftrightarrow\ket{\xi_{\mathbf{k}+\mathbf{K}}^{s}}=e^{i\vartheta_{s}(\mathbf{k})}\ket{\vphantom{\psi_{\mathbf{k}+\mathbf{K}}^{s}}\xi_{\mathbf{k}}^{s}}.
\end{equation}
At first sight, it may seem that the phase-factor can always be \textit{``gauged-away''}
upon a suitable choice of global phases for each $\ket{\xi_{\mathbf{k}}}$.
However, this is not always feasible by a gauge transformation that
is regular across the entire fBz\,\footnote{This is what one refers to as being a topological obstruction.}.
So, bearing this in mind, we can imagine that, instead of integrating
$\Omega^{s}\!(\mathbf{k})$ over the fBz, we use the fact that

\vspace{-0.7cm}
\begin{equation}
\iint_{\text{fBz}}\!\!\!\!d^{{\scriptscriptstyle (2)}}\!\mathbf{k}\,\Omega_{s\mathbf{k}}=\ointctrclockwise_{\mathcal{X}}\!\boldsymbol{\mathcal{A}}^{\!s}(\mathbf{k})\cdot\boldsymbol{dk}\label{eq:IntegralBerry}
\end{equation}
and integrate the Berry connection around the path $\mathcal{X}$
which follows the border of the fBz in a clockwise sense. The fBz
of a 2D Bravais lattice is always a polyhedron such that for each
side there is an opposite side whose $\mathbf{k}$s are obtained from
the first by a dual lattice translation. Hence, we can write the integral
in Eq.\,\eqref{eq:IntegralBerry} as a sum over all non-equivalent
sides of the fBz, \textit{i.e.},

\vspace{-0.7cm}
\begin{align}
\ointctrclockwise_{\mathcal{X}}\!\boldsymbol{\mathcal{A}}^{\!s}(\mathbf{k})\cdot\boldsymbol{dl} & =\int_{\text{0}}^{1}\!\!\!\boldsymbol{\mathcal{A}}^{\!s}(x\mathbf{b}_{1})\,dx+\int_{\text{0}}^{1}\!\!\!\boldsymbol{\mathcal{A}}^{\!s}(\mathbf{b}_{1}+x\mathbf{b}_{2})\,dx\\
 & \qquad-\int_{\text{0}}^{1}\!\!\!\boldsymbol{\mathcal{A}}^{\!s}(x\mathbf{b}_{1}+\mathbf{b}_{2})\,dx-\int_{\text{0}}^{1}\!\!\!\boldsymbol{\mathcal{A}}^{\!s}(x\mathbf{b}_{2})\,dx,\nonumber 
\end{align}
where we have already included the sense of motion in the signs of
the integrals. Now, we can use the definition of the Berry connection
to relate the two terms, namely

\vspace{-0.7cm}
\[
\boldsymbol{\mathcal{A}}^{\!s}(\mathbf{k}+\mathbf{K}_{\text{side}})=i\braket{\xi_{\mathbf{k}+\mathbf{K}_{\text{side}}}^{s}}{\boldsymbol{\nabla}_{\!\!\mathbf{k}}\xi_{\mathbf{k}+\mathbf{K}_{\text{side}}}^{s}}=\boldsymbol{\mathcal{A}}^{\!s}(\mathbf{k})-\boldsymbol{\nabla}_{\!\!\mathbf{k}}\vartheta_{s}(\mathbf{k})
\]
which yields,

\vspace{-0.7cm}

\begin{equation}
\iint_{\text{fBz}}\!\!\!\!d^{{\scriptscriptstyle (2)}}\!\mathbf{k}\,\Omega_{s\mathbf{k}}=\ointctrclockwise_{\mathcal{X}}\!\boldsymbol{\mathcal{A}}^{\!s}(\mathbf{k})\cdot\boldsymbol{dl}=\ointctrclockwise_{\mathcal{X}}\!\left[\boldsymbol{\nabla}_{\!\!\mathbf{k}}\vartheta_{s}(\mathbf{k})\right]\cdot\boldsymbol{dl}\label{eq:ContourGrad}
\end{equation}
or, in direct terms, the integral of the Berry curvature in a 2D fBz
is equivalent to the contour integral of a gradient field around its
border\,\cite{Thouless84}. This has sticking implications because,
if $\vartheta_{s}(\mathbf{k})$ were an arbitrary single-valued function
defined in $\mathcal{X},$ the integral of Eq.\,\eqref{eq:ContourGrad}
would have to be zero. However, since $\vartheta_{s}(\mathbf{k})$
is actually a phase-angle\,\footnote{At this point, we recap our earlier comment regarding the possibility
of gauging-away the $\mathbf{k}$-dependent phase factor.}, as one goes around the closed loop there is a possibility of returning
to same state but with a dephasing of $2\pi n_{s}$, where $n_{s}\!\in\!\mathbb{Z}$.
Therefore, we conclude that

\vspace{-0.7cm}
\begin{equation}
\iint_{\text{fBz}}\!\!\!d^{{\scriptscriptstyle (2)}}\mathbf{k}\,\Omega_{s\mathbf{k}}=2\pi n_{s},\label{eq:ChernNumber}
\end{equation}
meaning that the integral of the Berry curvature is a quantized number
for each band. This number cannot change its value upon deformations
of the Bloch Hamiltonian unless our adiabatic theory breaks down,
namely, if two bands cross at some point of the fBz (\textit{i.e.},
a \textit{``diabolical point''}\,\cite{Berry84}) entailing a topological
phase transition. Otherwise, the integer $n_{s}$ in Eq.\,\eqref{eq:ChernNumber}
defines a topological invariant called the (first) \textit{Chern Number}.
Before proceeding further, we remark two important properties of the
Berry curvature which can guarantee the topological triviality in
the presence of certain symmetries. First of all, it can be easily
shown that if $\mathcal{P}\mathcal{T}$ is a symmetry of the Hamiltonian,
we have 

\vspace{-0.7cm}
\begin{equation}
\Omega_{s\mathbf{k}}=-\Omega_{s\mathbf{k}}\text{ (in 2D) or }\boldsymbol{\Omega}_{s\mathbf{k}}=-\boldsymbol{\Omega}_{s\mathbf{k}}\text{ (in 3D),}
\end{equation}
which immediately renders the Berry curvature, and the Chern number
of any band, identically zero. This hints that solid-state systems
with non-trivial topology will generally require to have a broken
time-reversal symmetry or be noncentrosymmetric. Secondly, it is also
important to state that the Berry curvature scalar\,\footnote{An analogous expression exists for the vector case.}
may be written in terms of the eigenstates of the Bloch Hamiltonian,
thus taking the form

\vspace{-0.7cm}{\footnotesize{}
\begin{equation}
\!\!\!\!\!\Omega_{s\mathbf{k}}\!=\!i\sum_{s^{\prime}\neq s}\frac{\bra{\xi_{\mathbf{k}}^{s^{\phantom{\prime}}}}\!\partial_{k_{x}}\!\mathcal{H}(\mathbf{k})\!\ket{\xi_{\mathbf{k}}^{s^{\prime}}}\!\bra{\xi_{\mathbf{k}}^{s^{\prime}}}\!\partial_{k_{y}}\!\mathcal{H}(\mathbf{k})\!\ket{\xi_{\mathbf{k}}^{s^{\phantom{\prime}}}}\!-\!\bra{\xi_{\mathbf{k}}^{s^{\phantom{\prime}}}}\!\partial_{k_{y}}\!\mathcal{H}(\mathbf{k})\!\ket{\xi_{\mathbf{k}}^{s^{\prime}}}\!\bra{\xi_{\mathbf{k}}^{s^{\prime}}}\!\partial_{k_{x}}\!\mathcal{H}(\mathbf{k})\!\ket{\xi_{\mathbf{k}}^{s^{\phantom{\prime}}}}}{\left(\varepsilon_{s\mathbf{k}}-\varepsilon_{s^{\prime}\mathbf{k}}\right)^{2}}.\label{eq:Omega_Eigenstates}
\end{equation}
}As pointed out by Thouless \textit{et al.}\,\cite{Thouless82},
the fBz-integral of Eq.\,\eqref{eq:Omega_Eigenstates} is proportional
to the Kubo formula\,\cite{Kubo57} for the antisymmetric (Hall)
component of the conductivity tensor in a clean 2D crystal,

\vspace{-0.7cm}

\begin{equation}
\sigma_{\text{H}}\!=\!\frac{e^{2}}{4\pi h}\iint_{\text{fBz}}\!\!\!\!d^{{\scriptscriptstyle (2)}}\!\mathbf{k}\,\sum_{s}\Omega_{s\mathbf{k}}=\frac{e^{2}}{4\pi\hbar}\sum_{s}n_{s}
\end{equation}
Thus, a nonzero Chern number is what guarantees the existence of a
quantized Hall effect. In the theory of the integer Quantum Hall Effect
with external magnetic fields, this topological integer is historically
called the \textit{TKNN Invariant}\,\cite{Thouless82} (the Chern
number associated to magnetic Bloch bands).

\subsection{\label{subsec:Dynamics-of-Bloch}Dynamics of Bloch Electrons in 3D
Topological Bands}

The existence of Berry curvature in a two-dimensional band structure
can lead to quantized topological invariants that find their most
important consequences in the quantized Hall conductivity of 2D electron
gases and the existence of robust chiral edge states. In 3D, one can
perform analogous reasonings which historically took us far onto the
physics of 3D topological insulators (TIs), both in their classification
by $\mathbb{Z}_{2}$ topological invariants\,\cite{Fu2006,Fu2007,Fukui2007,Moore2007,Fukui2008,Roy2009},
as well as in generalizing the bulk-boundary correspondence to these
three-dimensions\,\cite{Zhang2009}. Likewise, even in gapless 3D
systems one can define a topological invariant (dubbed topological
charge) which is related to the flux of the Berry curvature field
across a closed surface in the fBz. Excellent reviews on the subject
are available from Hasan and Kane\,\cite{Hasan2010}, and Qi and
Zhang\,\cite{Qi2010}, but the content lies mostly outside the scope
of this work. However, long before the advent of topological insulators,
it was known that the existence of a non-zero Berry curvature field
in three-dimensions\,\footnote{Which generically happens whenever $\mathcal{PT}$-symmetry is absent. }
would have observable consequences on the semiclassical dynamics of
Bloch states under external electromagnetic fields. Even though the
so-called \textit{anomalous velocity effects} in the dynamics of solid-state
electrons were known before long (\textit{e.g.}, see Refs.\,\cite{Karplus54,Kohn57,Blount62}),
the modern interpretation in terms of Berry curvature of the band
structure was only described in a sequence of papers from the early
1990's until the early 2000's\,\cite{Resta92,KingSmith93,Chang95,Chang96,Sundaram99,Xiao2005,Chang2008}.
Since this theory is pivotal to understand the outstanding transport
phenomenology present in topological semimetals, we will present here
a brief overview of these concepts. For a comprehensive review, we
refer the reader to Xiao \textit{et al.}\,\cite{Xiao2010}.

For external electromagnetic fields ($\mathbf{E}(\mathbf{r},t)$ and
$\mathbf{B}(\mathbf{r},t)$) that vary slowly compared to all microscopic
space and time scales, the evolution of wide wave-packets of Bloch
states in a given band can be studied via the equations of motion
for its central momentum, $\mathbf{k}(t)$, and position, $\mathbf{r}(t)$.
This yields a semi-classical description of the quantum dynamics of
a wave-packet in a six-dimensional phase-space $(\mathbf{k},\mathbf{r})$
whose Hamilton's equations read,\,\footnote{For shortness, we will suppress the time- and space-dependence of
the electromagnetic field in our equations.}

\vspace{-0.7cm}

\begin{subequations}
\begin{align}
\dot{\mathbf{r}}_{s} & =\frac{1}{\hbar}\boldsymbol{\nabla}_{\!\mathbf{k}}\varepsilon_{s\mathbf{k}}-\frac{1}{\hbar}\left(\mathbf{B}\cdot\!\boldsymbol{\nabla}_{\mathbf{k}}\right)\mathbf{m}_{s\mathbf{k}}+\boldsymbol{\Omega}_{s\mathbf{k}}\!\times\!\dot{\mathbf{k}}_{s}\label{eq:drdt-2}\\
\dot{\mathbf{k}}_{s} & =-\frac{e}{\hbar}\mathbf{E}+\frac{e}{\hbar}\mathbf{B}\times\dot{\mathbf{r}}_{s},\label{eq:dkdt-2}
\end{align}
\end{subequations}

where $\dot{\phantom{r}}$ stands for a time-derivative, $\varepsilon_{s\mathbf{k}}$
($\boldsymbol{\Omega}_{s\mathbf{k}}$) is the dispersion relation
(Berry curvature) of the band $s$, and $\mathbf{m}_{s\mathbf{k}}$
is the \textit{intrinsic magnetic moment} of a Bloch state in the
band $s$. It is usual to define the group velocity of such a wave-packet
as 

\vspace{-0.7cm}
\begin{equation}
\mathbf{v}_{s\mathbf{k}}=\frac{1}{\hbar}\boldsymbol{\nabla}_{\!\mathbf{k}}\varepsilon_{s\mathbf{k}}-\frac{1}{\hbar}\left(\mathbf{B}\cdot\boldsymbol{\nabla}_{\mathbf{k}}\right)\mathbf{m}_{s\mathbf{k}},\label{eq:VelocityMag}
\end{equation}
which includes a magnetic correction to the band-structure dispersion.
Rather than taking Eqs.\,\eqref{eq:drdt-2}-\eqref{eq:dkdt-2} in
their present form, it is useful to cast them in the alternative form\,\footnote{This form can be derived straightforwardly by iterating Eqs.\,\eqref{eq:drdt-2}-\eqref{eq:dkdt-2}
and using the triple-product identity, $(\mathbf{a}\!\times\!\mathbf{b})\!\times\!\mathbf{c}=(\mathbf{a}\!\cdot\!\mathbf{c})\mathbf{b}-(\mathbf{b}\!\cdot\!\mathbf{c})\mathbf{a}$.},

\vspace{-0.7cm}

\begin{subequations}
\begin{align}
\dot{\mathbf{r}}_{s} & =\frac{1}{\Delta_{s\mathbf{k}}}\left[\mathbf{v}_{s\mathbf{k}}-\frac{e}{\hbar}\mathbf{E}\times\boldsymbol{\Omega}_{s\mathbf{k}}+\frac{e}{\hbar}\left(\boldsymbol{\Omega}_{s\mathbf{k}}\cdot\mathbf{v}_{s\mathbf{k}}\right)\mathbf{B}\right]\label{eq:drdt-1-2}\\
\dot{\mathbf{k}}_{s} & =\frac{1}{\Delta_{s\mathbf{k}}}\left[-\frac{e}{\hbar}\mathbf{E}+\frac{e}{\hbar}\mathbf{B}\times\!\mathbf{v}_{s\mathbf{k}}+\frac{e^{2}}{\hbar^{2}}\left(\mathbf{E}\cdot\mathbf{B}\right)\boldsymbol{\Omega}_{s\mathbf{k}}\right],\label{eq:dkdt-1-1}
\end{align}
\end{subequations}
where $\Delta_{s\mathbf{k}}\!=\!1+\frac{e}{\hbar}\left(\boldsymbol{\Omega}_{s\mathbf{k}}\cdot\mathbf{B}\right)$.
Note that in Eqs.\,\eqref{eq:drdt-1-2}-\eqref{eq:dkdt-1-1} there
are several terms which would vanish for topologically trivial bands.
Namely, the $\mathbf{E}\cdot\mathbf{B}$ and $\mathbf{E}\times\boldsymbol{\Omega}$
terms, as well as the magnetic contribution to the velocity {[}Eq.\,\eqref{eq:VelocityMag}{]}
would all be absent without Berry curvature. In addition, non-trivial
band topology also leads to the denominators $\Delta_{s\mathbf{k}}$,
in the semiclassical equations. As was shown by Xiao \textit{et al.}\,\cite{Xiao2005},
these represent a change in the semi-classical phase-space invariant
measure which also implies a breakdown of \textit{Liouville's Theorem}\,\cite{Xiao2005,Duval2006,Xiao2006}.
In particular, this means that for a phase-space density $f_{\mathbf{r},\mathbf{k}}^{s}(t)$
that evolves according to a Boltzmann Equation, the charge carrier
and current densities are calculated as,

\vspace{-0.7cm}

\begin{subequations}
\begin{align}
\rho(\mathbf{r},t)\! & =\!-e\int\frac{d^{(3)}\mathbf{k}}{8\pi^{3}}\sum_{s}\mathbf{r}_{s}\Delta_{s\mathbf{k}}f_{\mathbf{r},\mathbf{k}}^{s}(t)\\
\mathbf{J}(\mathbf{r},t)\! & =\!-e\int\frac{d^{(3)}\mathbf{k}}{8\pi^{3}}\sum_{s}\dot{\mathbf{r}}_{s}\Delta_{s\mathbf{k}}f_{\mathbf{r},\mathbf{k}}^{s}(t),\label{eq:CurrentDensity}
\end{align}
\end{subequations}

which identifies $p_{s}(\mathbf{r},\mathbf{k},t)\!=\!\Delta_{s\mathbf{k}}f_{\mathbf{r},\mathbf{k}}^{s}(t)/8\pi^{3}$
as the proper probability density in $\mathbf{k}$-space. Equations\,\eqref{eq:drdt-2}-\eqref{eq:CurrentDensity}
form the foundation of the semi-classical transport theory that will
be used to derive some of the most relevant phenomenology associated
to topological semimetals in Sec.\,\ref{sec:Observable-Signatures-of}.
At the moment, we suspend this discussion and switch gears to discuss
the more pressing subject of\textit{ topological phases of matter}.

\vspace{-0.5cm}

\subsection{Two-Dimensional Topological Phases}

\begin{figure}[t]
\vspace{-0.5cm}
\begin{centering}
\includegraphics[scale=0.21]{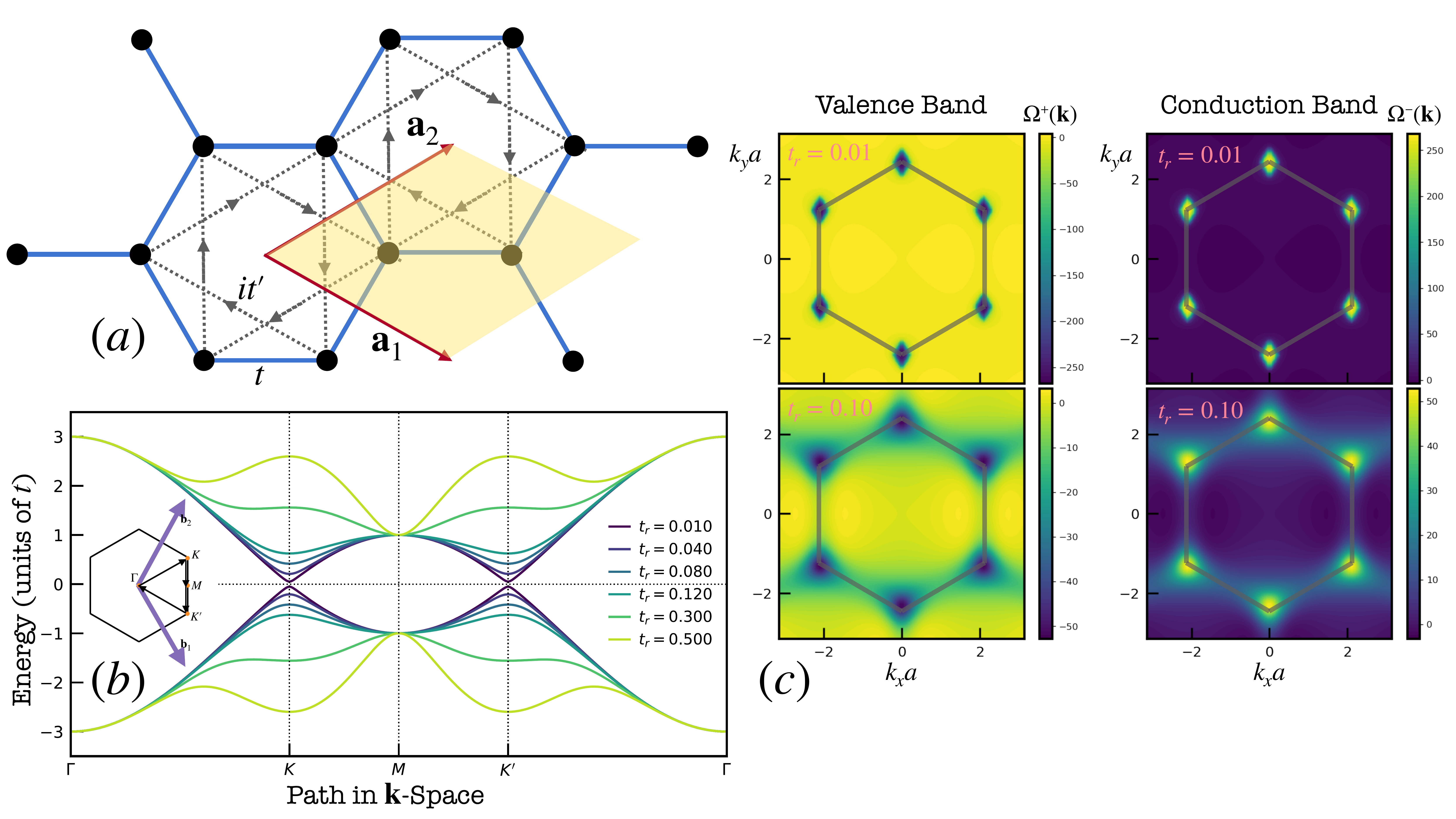}
\par\end{centering}
\vspace{-0.1cm}

\caption{\label{fig:Haldane1}(a) Haldane model in the real-space honeycomb
lattice. (b) Band structure of the model, represented along the path
indicated as an inset. The different curves correspond to different
values of $t_{r}\!=\!t^{\prime}/t$. (c) Contour-plots of the Berry
curvature scalar for the valence band (left panels) and conduction
band (right panels), in the first Brillouin zone for two values of
$t_{r}$.}

\vspace{-0.5cm}
\end{figure}

The simplest single-electron model having a non-trivial topology is
the so-called \textit{Haldane model}\,\cite{Haldane88}, which intrinsically
breaks time-reversal symmetry. This is a two-band tight-binding model
in the honeycomb lattice that has nearest-neighbor (NN\nomenclature{NN}{Nearest-Neighbor})
hoppings of strength $t\!>\!0$, and a purely imaginary next-nearest-neighbor
(NNN\nomenclature{NNN}{Next-Nearest-Neighbor}) hopping of the form
$\pm it^{\prime}$\,\footnote{Note that the original Haldane model, the NNN hopping is of the form
$t'e^{\pm i\varphi}$. For the sake of simplicity, we consider only
the case in which $\varphi\!=\!\nicefrac{\pi}{2}$.}. This lattice model is depicted in Fig.\,\ref{fig:Haldane1}\,a,
together with the unit cell and the primitive vectors of the triangular
Bravais lattice,

\vspace{-0.7cm}
\begin{align}
\mathbf{a}_{1}\! & =\!a\left(\frac{3}{2}\mathbf{x}+\frac{\sqrt{3}}{2}\mathbf{y}\right)\text{ and }\mathbf{a}_{2}\!=\!a\left(\frac{3}{2}\mathbf{x}-\frac{\sqrt{3}}{2}\mathbf{y}\right),
\end{align}
with $a$ being the NN distance. The corresponding Bloch Hamiltonian
can be written directly as 

\vspace{-0.7cm}
\begin{equation}
\mathcal{H}_{\text{H}}(\mathbf{k})=t\left(\begin{array}{cc}
t_{r}f_{\mathbf{k}} & -g_{\mathbf{k}}\\
-g_{\mathbf{k}}^{*} & -t_{r}f_{\mathbf{k}}
\end{array}\right),\label{eq:Haldane}
\end{equation}
where $t_{r}=t^{\prime}/t$ and the functions $f_{\mathbf{k}}$ and
$g_{\mathbf{k}}$ are defined as 

\vspace{-0.7cm}

\begin{subequations}
\begin{align}
g_{\mathbf{k}}= & i\left(1+e^{-i\mathbf{k}\cdot\mathbf{a}_{2}}+e^{-i\mathbf{k}\cdot\mathbf{a}_{1}}\right)\\
f_{\mathbf{k}}= & \sin\left(\mathbf{k}\cdot\mathbf{a}_{1}\right)\!-\!\sin\left(\mathbf{k}\cdot\mathbf{a}_{2}\right)\!-\!\sin\left(\mathbf{k}\cdot(\mathbf{a}_{1}-\mathbf{a}_{2})\right).
\end{align}
\end{subequations}

\vspace{-0.4cm}

The Hamiltonian of Eq.\,\eqref{eq:Haldane} yields the dispersion
relation, $\varepsilon_{\pm\mathbf{k}}\!\!=\!\pm t\sqrt{t_{r}^{2}f_{\mathbf{k}}^{2}+\abs{g_{\mathbf{k}}}^{2}}$
which correspond to the following Bloch eigenstates,

\vspace{-0.7cm}

\begin{equation}
\ket{\chi_{\mathbf{k}}^{\pm}}=\frac{1}{\sqrt{2t_{r}^{2}f_{\mathbf{k}}^{2}+2\abs{g_{\mathbf{k}}}^{2}\pm2t_{r}f_{\mathbf{k}}\sqrt{t_{r}^{2}f_{\mathbf{k}}^{2}+\abs{g_{\mathbf{k}}}^{2}}}}\left[\!\!\begin{array}{c}
t_{r}f_{\mathbf{k}}\pm\sqrt{t_{r}^{2}f_{\mathbf{k}}^{2}+\abs{g_{\mathbf{k}}}^{2}}\\
-g_{\mathbf{k}}^{*}
\end{array}\!\!\right].\label{eq:BlochEigenstates}
\end{equation}
The band structure of the model is represented in Fig.\,\ref{fig:Haldane1}\,b,
along a path in $\mathbf{k}$-space that goes through all the high-symmetry
points of the fBz. From the Bloch eigenstates, we can compute the
Berry curvature which is shown in Fig.\,\ref{fig:Haldane1}\,c across
the entire fBz. These results demonstrate two important points: \textit{(i)}
both bands are topologically non-trivial with a Chern number $n_{{\scriptscriptstyle \pm}}\!\!=\!\pm\text{sign}\,t_{r}$,
and \textit{(ii)} for small enough $t_{r}$, most of the curvature
is concentrated around the $K$ and $K^{\prime}$ points, where the
spectral gap is the narrowest. In fact, if we derived a low-energy
continuum theory for this system (by expanding around the point $K$
or $K^{\prime}$), the result would be a 2D Dirac Hamiltonian with
mass-terms that have opposite signs in each valley\,\cite{Girvin2019}.
The two bands are then said to be \textit{inverted} in the Haldane
model.

This model is important because it exemplifies the simplest non-trivial
topological insulating phase \textemdash{} a \textit{2D\,}\nomenclature{2D}{Two-dimensional}\textit{
Quantum Hall Insulator} (QHI)\,\cite{Nagaosa10} \textemdash{} which
was first realized experimentally by Chang \textit{et al.} in $(\text{Bi},\text{Sb})_{2}\!\text{Te}_{3}$
thin-films doped with chromium\,\cite{Chang13} or vanadium\,\cite{Chang15}.
If one assumes the model has the lower (upper) band occupied (empty)
then this nontrivial topology gives rise to an anomalous quantized
Hall conductivity, 

\vspace{-0.7cm}

\begin{equation}
\sigma_{H}=\frac{e^{2}}{h}n_{{\scriptscriptstyle -}},
\end{equation}
which appears without an applied magnetic field\,\footnote{The topological interpretation of the normal QHE is very similar to
this, but the band structure may be considered as generated by magnetic
lattice translations.}. Associated to this bulk transport property is the fact that any
exposed surface in this system will support localized in-gap edge
states that propagate in a well-defined direction along that edge.
Even though we do not intend a further pursuit of this discussion,
it is worth mentioning that these edge-modes are a paradigmatic example
of a \textit{bulk-edge correspondence}, which is a characteristic
feature of topological systems that is also present in 3D topological
insulators and semimetals.

\paragraph*{Topological Phase Transitions:}

The Haldane model of Eq.\,\eqref{eq:Haldane} already features a
topological phase transition; When $t_{r}$ changes sign the spectral
gap is closed so that the two bands are allowed to exchange their
Chern numbers. In fact, with minor changes to the model, it is possible
to obtain a richer phase-diagram that includes both trivial and non-trivial
topological phases. Looking at $\mathcal{H}_{\text{H}}(\mathbf{k})$,
it is obvious that the Hamiltonian is akin to that of graphene but
including a $\sigma_{z}$-mass term that depends on $\mathbf{k}$.
Because it changes sign when $\mathbf{k}\to-\mathbf{k}$, this mass
is not a trivial mass and, therefore, we are free to include a further
\textit{normal Dirac mass}, $m\sigma_{z}$, turning the Bloch Hamiltonian
into 
\begin{figure}[t]
\vspace{-0.5cm}
\begin{centering}
\includegraphics[scale=0.21]{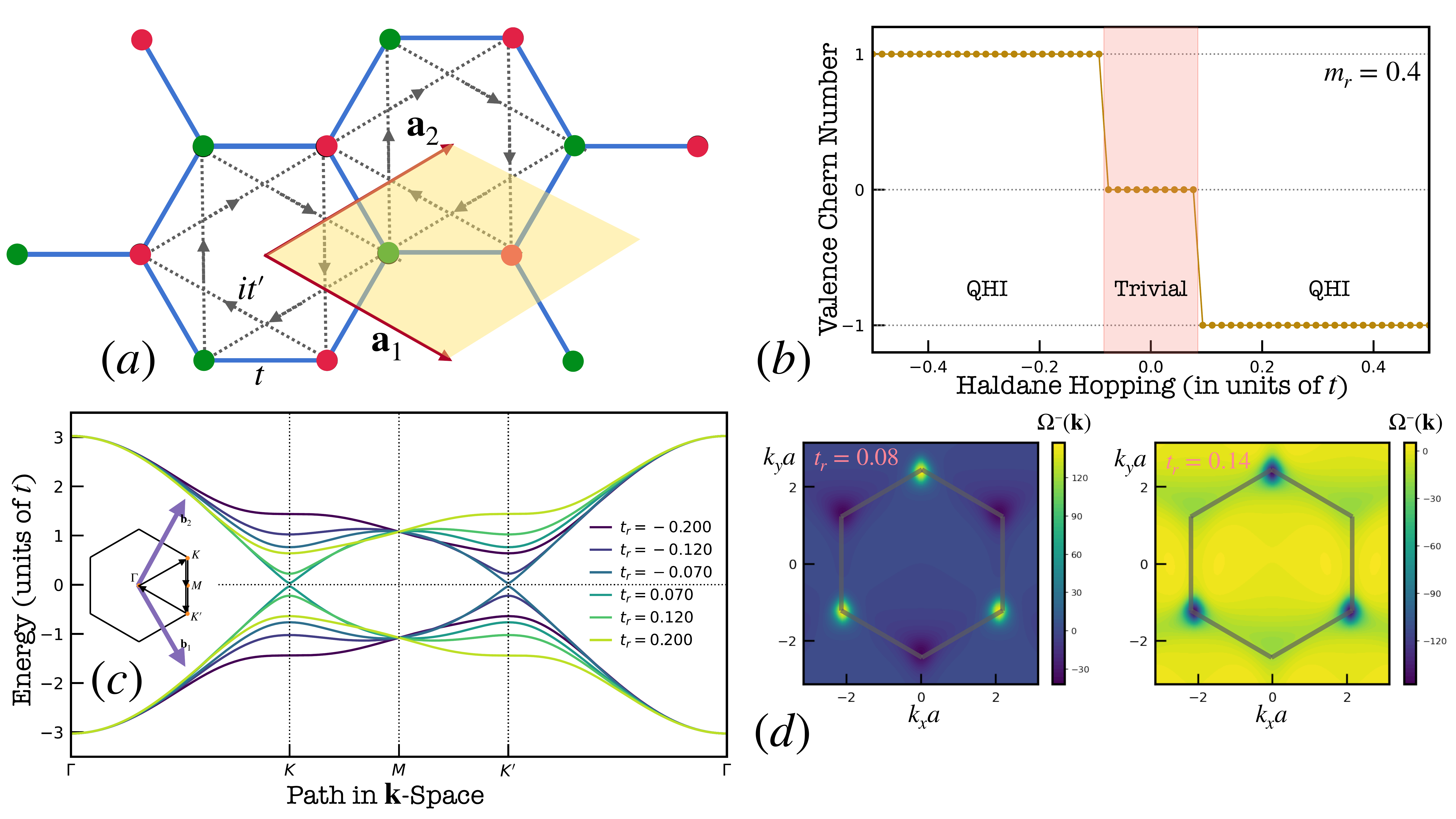}
\par\end{centering}
\vspace{-0.1cm}

\caption{\label{fig:Haldane2}(a) Haldane model with broken inversion symmetry.
(b) Chern number of the valence band as a function of $t_{r}$ for
$m_{r}=0.4$. Three topologically distinct phases can be observed.
(c) Band structure of the model, represented along the path indicated
as an inset. The different curves correspond to different values of
$t_{r}\!=\!t^{\prime}/t$, for $m_{r}=0.4$. (d) Contour-plots of
the Berry curvature scalar in the first Brillouin zone for the valence
band, before and after the transition from the trivial insulating
phase to the QHI with negative Chern number.}
\end{figure}

\vspace{-0.7cm}

\begin{equation}
\mathcal{H}_{\text{H2}}(\mathbf{k})=t\left(\begin{array}{cc}
t_{r}f_{\mathbf{k}}+m_{r} & -g_{\mathbf{k}}\\
-g_{\mathbf{k}}^{*} & -t_{r}f_{\mathbf{k}}-m_{r}
\end{array}\right),\label{eq:Haldane-1}
\end{equation}
where $m_{r}=m/t$. In the lattice model, this change can be realized
by adding symmetric on-site energies ($\pm m$) to each sublattice,
as shown in Fig.\,\ref{fig:Haldane2}\,a, which immediately breaks
the system's inversion center. In Fig.\,\ref{fig:Haldane2}\,b,
we show the lower band's Chern number calculated, for $m_{r}=0.4$,
as a function of the relative hopping $t_{r}$. From there, we see
that the model can now realize three distinct insulating phases: a
trivial insulator for $t_{r}$ close to $0$, flanked by two QHI phases
of opposite Chern numbers. By looking at the band structure {[}Fig.\,\ref{fig:Haldane2}\,c{]},
one realizes that each transition is accompanied by a gap closing
at either $K$ or $K^{\prime}$, thus allowing the Chern number in
each band to jump by $\pm1$. In contrast, the presence of inversion
symmetry, requires both gaps to close simultaneously and, therefore,
only variations of $\pm2$ in the Chern number are allowed. The changes
in the Berry curvature field are also shown in Fig.\,\ref{fig:Haldane2}\,d,
before and after a transition point, confirming the previous interpretation.

\vspace{-0.7cm}

\paragraph{Quantum Spin Hall Insulating Phases:}

Interestingly, note that there are no topological phases of the Haldane
model if $t_{r}\!=\!0$. This is surprising, as we have previously
stated that a non-zero Berry curvature can arise when either \textit{time-reversal
or inversion symmetry} are absent. It turns out that solely breaking
inversion symmetry in the model of Eq.\,\eqref{eq:Haldane-1} is
not enough to do this. The dilemma was finally solved by introducing
another class of 2D topological systems, the \textit{Quantum Spin
Hall Insulator }(QSHI), first predicted by Kane and Mele\,\cite{Kane2005a,Kane2005b,Fu2006a,Moore2007}
and which only admit a two-fold $\mathbb{Z}_{{\scriptscriptstyle 2}}\!-$classification
(odd or even parity). Note that, unlike the Haldane models, these
time-reversal invariant 2D topological phases require one to devise
a four-band minimal model that can be seen as two time-reversed copies
of the Haldane model (with spin-$\nicefrac{1}{2}$) which are coupled
by a NN spin-orbit term.

\vspace{-0.5cm}

\section{Topological Insulators and Semimetals}

In Sect.\,\ref{sec:Topological-Properties}, we have reviewed the
nontrivial topology that can emerge in the band structure of Bloch
electrons provided $\mathcal{PT}$-symmetry is not present. For simplicity,
we limited the discussion to 2D systems where we found two classes
of topologically non-trivial phases: the QHI, which admits a $\mathbb{Z}$-fold
classification, and the QSHI, with a $\mathbb{Z}_{2}$-classification.
Every time there is a transition between these topological phases,
a gap must be closed somewhere in the band structure, thus giving
rise to a gapless electronic phase\,\footnote{Graphene can be seen as the transitional state between the different
QHI phases of the centrosymmetric Haldane model.}. In this section, we move on our discussion to the subject of \textit{Three-Dimensional
Topological Insulators} (TIs\nomenclature{TI}{Topological Insulator}),
a different class of gapped topological phases that can be realized
in (even $\mathcal{PT}$-symmetric) 3D systems and are characterized
by a set of $4$ $\mathbb{Z}_{{\scriptscriptstyle 2}}$ topological
indices only changeable by closing spectral gaps. The bulk of this
thesis will be concerned with the physics of these \textit{transitional
3D gapless phases}, so that a full and comprehensive discussion on
TIs left to one of many excellent published reviews, \textit{e.g.},
Hasan and Kane\,\cite{Hasan2010}, Hasan and Moore\,\cite{Hasan2011},
Qi and Zhang\,\cite{Qi2011} and Ando\,\cite{Ando2013}. Instead,
here we will provide some context through a paradigmatic example \textemdash{}
the \textit{Fu-Kane-Mele model}\,\cite{Fu2007} \textemdash{} which
realizes four different time-reversal invariant 3D topological gapped
phases. A slight deformation of this model\,\cite{Murakami07,Murakami2008}
will serve as the cornerstone for our discussion of topological semimetals.

\vspace{-0.5cm}

\subsection{The Fu-Kane-Mele Model and 3D Topological Phases}

\begin{figure}[t]
\vspace{-0.55cm}
\begin{centering}
\includegraphics[scale=0.225]{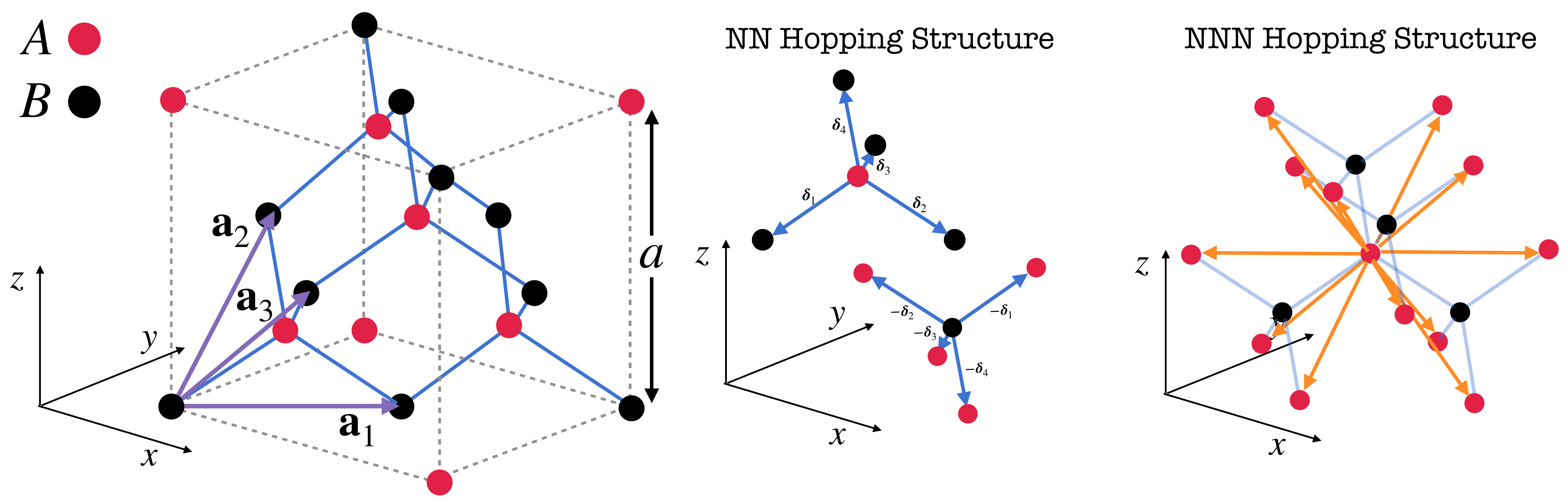}
\par\end{centering}
\vspace{-0.15cm}

\caption{\label{fig:FKM_Struct}Structure of the Fu-Kane-Mele tight-binding
model. On the right, we depict the structure of NN and NNN hoppings.}

\vspace{-0.55cm}
\end{figure}

The Fu-Kane-Mele model is a $\mathcal{P}\mathcal{T}$-symmetric tight-binding
model of independent spinful electrons in a 3D diamond lattice. The
structure is represented in Fig.\,\ref{fig:FKM_Struct}, which consists
of two interpenetrating face-centered cubic (fcc) lattices ($A$ and
$B$) that have a relative displacement by the vector $\boldsymbol{\delta}_{1}$,
defined in Eq.\,\eqref{eq:delta1}. The primitive vectors that generate
underlying Bravais lattice are

\vspace{-0.7cm}

\begin{align}
\mathbf{a}_{1} & =\left(\frac{a}{2},\frac{a}{2},0\right)\text{, }\mathbf{a}_{2}=\left(0,\frac{a}{2},\frac{a}{2}\right)\text{ and }\mathbf{a}_{3}=\left(\frac{a}{2},0,\frac{a}{2}\right)
\end{align}
with the hopping vectors to (from) each site in sublattice A (B) reading,

\vspace{-0.7cm}

\begin{subequations}
\begin{align}
\boldsymbol{\delta}_{1} & =\left(-\frac{a}{4},-\frac{a}{4},-\frac{a}{4}\right),\label{eq:delta1}\\
\boldsymbol{\delta}_{2} & =\boldsymbol{\delta}_{1}+\mathbf{a}_{1}=\left(\frac{a}{4},\frac{a}{4},-\frac{a}{4}\right),\\
\boldsymbol{\delta}_{3} & =\boldsymbol{\delta}_{1}+\mathbf{a}_{2}=\left(-\frac{a}{4},\frac{a}{4},\frac{a}{4}\right),\\
\boldsymbol{\delta}_{4} & =\boldsymbol{\delta}_{1}+\mathbf{a}_{3}=\left(\frac{a}{4},-\frac{a}{4},\frac{a}{4}\right).
\end{align}
\end{subequations}

The Hamiltonian of this model, $\mathcal{H}_{\!{\scriptscriptstyle \text{FKM}}}$,
is a four-band model with spin-degenerate bands, which has two parts:
\begin{enumerate}
\item A deformed NN hopping part, which reads

\vspace{-0.7cm}
\begin{equation}
\mathcal{H}_{\!{\scriptscriptstyle \text{FKM}}}^{1}=t\sum_{\mathbf{R}}\left(\sum_{i=1}^{4}\boldsymbol{\Psi}_{\mathbf{R}+\boldsymbol{\delta}_{i}}^{\dagger}\!\cdot\!\boldsymbol{\Psi}_{\!\mathbf{R}}+\frac{\delta t}{t}\boldsymbol{\Psi}_{\mathbf{R}+\boldsymbol{\delta}_{1}}^{\dagger}\!\cdot\!\boldsymbol{\Psi}_{\!\mathbf{R}}\right),
\end{equation}
where $\mathbf{R}$ is summed over the A sublattice positions, $t,\delta t\in\mathbb{R}$
are hopping parameters, and $\boldsymbol{\Psi}_{\!\mathbf{R}}^{\dagger}=\left(c_{\uparrow\mathbf{R}}^{\dagger},c_{\downarrow\mathbf{R}}^{\dagger}\right)$
is a two-component fermion creation operator (that accounts for electron
spin).
\item A complex spin-orbit NNN hopping part, whose Hamiltonian is written
as 

\vspace{-0.7cm}
\begin{equation}
\mathcal{H}_{\!{\scriptscriptstyle \text{FKM}}}^{2}\!=\!ia^{2}\lambda_{{\scriptscriptstyle \text{SO}}}\sum_{\mathbf{R}}\sum_{i=1}^{6}g_{i}^{j}\left(\boldsymbol{\Psi}_{\mathbf{R}+\boldsymbol{\Delta}_{i}}^{\dagger}\!\cdot\sigma_{j}\cdot\!\boldsymbol{\Psi}_{\!\mathbf{R}}\!-\!\boldsymbol{\Psi}_{\mathbf{R}-\boldsymbol{\delta}_{1}+\boldsymbol{\Delta}_{i}}^{\dagger}\!\cdot\sigma_{j}\cdot\!\boldsymbol{\Psi}_{\!\mathbf{R}-\boldsymbol{\delta}_{1}}\right)\label{eq:FKM_SpinOrbit}
\end{equation}
where $\lambda_{{\scriptscriptstyle \text{SO}}}$ is the scale associated
to the spin-orbit coupling energy, $\boldsymbol{\sigma}=\left(\sigma_{x},\sigma_{y},\sigma_{z}\right)$
is a vector of Pauli matrices acting on the spin components, and the
coefficients $\mathbf{g}_{i}=\left(g_{i}^{x},g_{i}^{y},g_{i}^{z}\right)$\,\footnote{Summation over $j=x,y,z$ is also implicit in Eq.\,\eqref{eq:FKM_SpinOrbit}.}
will be defined later. For now, it important to refer that $\boldsymbol{\Delta}_{i}$
are the NNN hopping vectors\,\footnote{We are only considering 6 of them, as the rest are obtained by inversion.}
defined as

\vspace{-0.7cm}
\begin{align}
\boldsymbol{\Delta}_{1} & =\mathbf{a}_{1}=\boldsymbol{\delta}_{2}-\boldsymbol{\delta}_{1}\text{ , }\quad\boldsymbol{\Delta}_{4}=\mathbf{a}_{2}-\mathbf{a}_{1}=\boldsymbol{\delta}_{3}-\boldsymbol{\delta}_{2},\nonumber \\
\boldsymbol{\Delta}_{2} & =\mathbf{a}_{2}=\boldsymbol{\delta}_{3}-\boldsymbol{\delta}_{1}\text{ , }\quad\boldsymbol{\Delta}_{5}=\mathbf{a}_{3}-\mathbf{a}_{2}=\boldsymbol{\delta}_{4}-\boldsymbol{\delta}_{3},\\
\boldsymbol{\Delta}_{3} & =\mathbf{a}_{3}=\boldsymbol{\delta}_{4}-\boldsymbol{\delta}_{1}\text{ , }\quad\boldsymbol{\Delta}_{6}=\mathbf{a}_{1}-\mathbf{a}_{3}=\boldsymbol{\delta}_{2}-\boldsymbol{\delta}_{4},\nonumber 
\end{align}
which only connect points of the same sublattice. Finally, by using
the more convenient double-index notation, $\boldsymbol{\Delta}_{i}\to\boldsymbol{\Delta}_{kl}=\boldsymbol{\delta}_{k}\!-\!\boldsymbol{\delta}_{l}$,
the strength of each NNN hopping can be analytically calculated through
$g_{i}^{j}\to g_{kl}^{j}=\nicefrac{8}{a^{2}}\left(\boldsymbol{\delta}_{l}\times\boldsymbol{\delta}_{k}\right)_{j}$,
which yields

\vspace{-0.9cm}
\begin{align}
g_{1}^{j} & \!=\!\left(-1,1,0\right)\text{ , }\quad\:g_{4}^{j}\!=\!\left(-1,0,-1\right)\nonumber \\
g_{2}^{j} & \!=\!\left(0,-1,1\right)\text{ , }\quad\:g_{5}^{j}\!=\!\left(-1,-1,0\right)\\
g_{3}^{j} & \!=\!\left(1,0,-1\right)\text{ , }\quad\:g_{6}^{j}\!=\!\left(0,-1,-1\right).\nonumber 
\end{align}

\vspace{-0.4cm}
\begin{figure}[t]
\vspace{-0.5cm}
\begin{centering}
\includegraphics[scale=0.22]{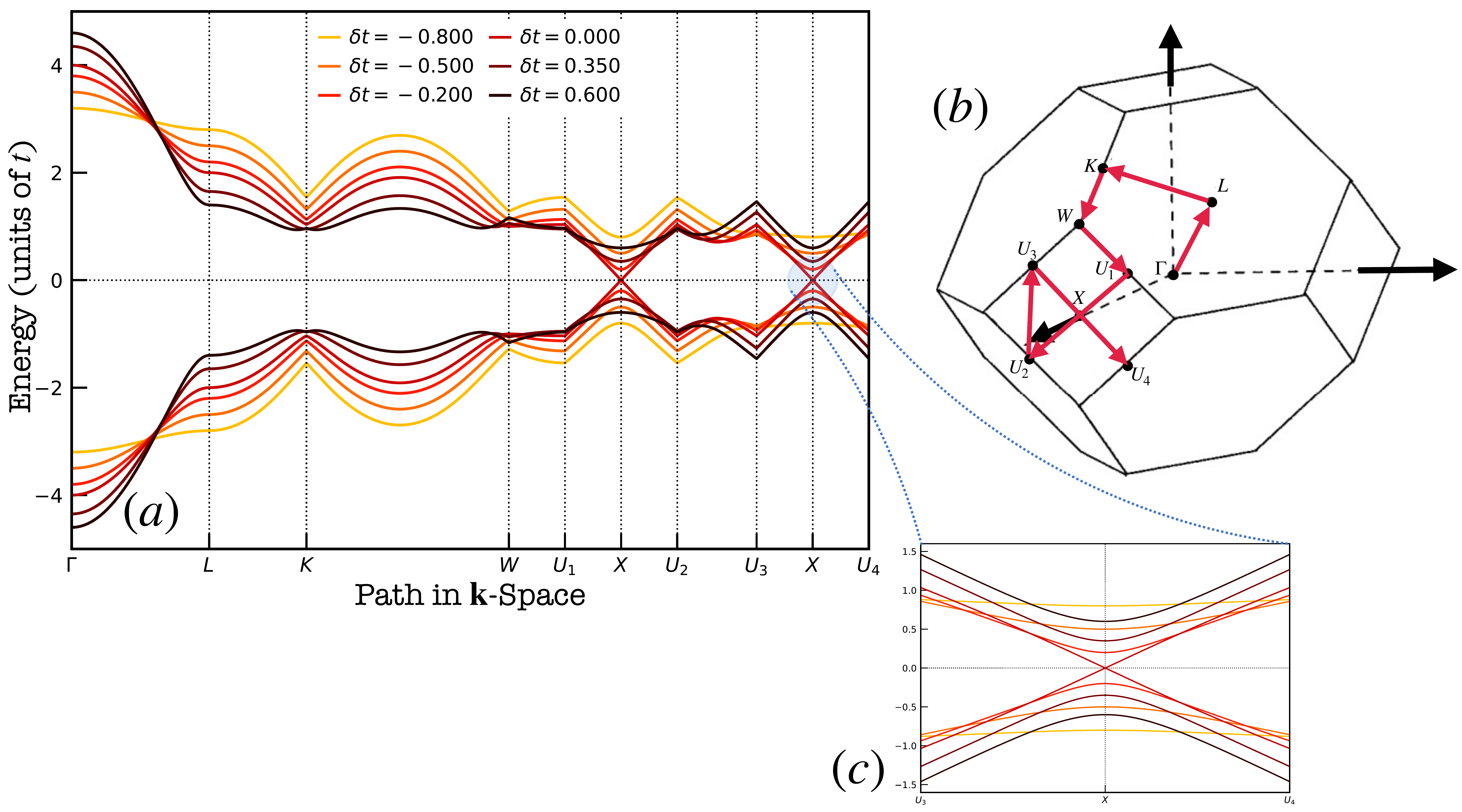}
\par\end{centering}
\vspace{-0.1cm}

\caption{\label{fig:FKM_Band}(a) Band Structure of the Fu-Kane-Mele tight-binding
model at the transition point ($\delta t\!=\!0$). (b) Path in the
fBz in which the bands are represented. (c) Close-up of the $X$ point,
where the gap-closing occurs.}

\vspace{-0.3cm}
\end{figure}

\end{enumerate}
In Fig.\,\eqref{fig:FKM_Band}, we show the band structure of the
Fu-Kane-Mele Model along the indicated path in the fBz, for $\lambda_{\text{SOC}}\!=\!t$
and different values of the deformation parameter $\delta t$. First
of all, we observe with no surprise that the FKM model has spin-degenerate
bands, which is a consequence of $\mathcal{PT}$-symmetry {[}as discussed
in Sect.\,\ref{sec:Generic-Symmetries}{]}. Interestingly, we also
observe that there is a gap-closing transition when $\delta t$ changes
sign which, according to the topological classification of Fu \textit{et
al.}\,\cite{Fu2007}, corresponds to a phase transition from a weak
TI phase ($\delta t\!>\!0$) to a strong TI phase ($\delta t\!<\!0$).
Precisely at $\delta t\!=\!0$, the system features three isotropic
linear band-crossings at the high-symmetry points $X$ of the fBz.
Like the transition points found in the 2D Haldane model, these band-crossings
at the Fermi level realize a fine-tuned gapless electronic phase with
Dirac points that, unlike the 2D case, involve doubly-degenerate bands.
Strictly speaking, the Fu-Kane-Mele model with $\delta t\!=\!0$ realizes
a \textit{Three-dimensional Dirac Semimetal} (DSM) with three inequivalent
valleys, around which, the low-energy quasiparticles behave as ultra-relativistic
Dirac fermions in $(3\!+\!1)$\textemdash dimensions. In the following
analysis, we shall see that such a 3D DSM is not a topologically protected
phase (unlike the gapped phases of the same model) but are stabilized
by point-group symmetries, as shown in Refs.\,\cite{Wang12,Young12,Wang13,Steinberg14}.

\subsection{The Murakami-Kuga Model for a 3D Weyl Semimetal}

\begin{figure}[t]
\vspace{-0.5cm}
\begin{centering}
\includegraphics[scale=0.22]{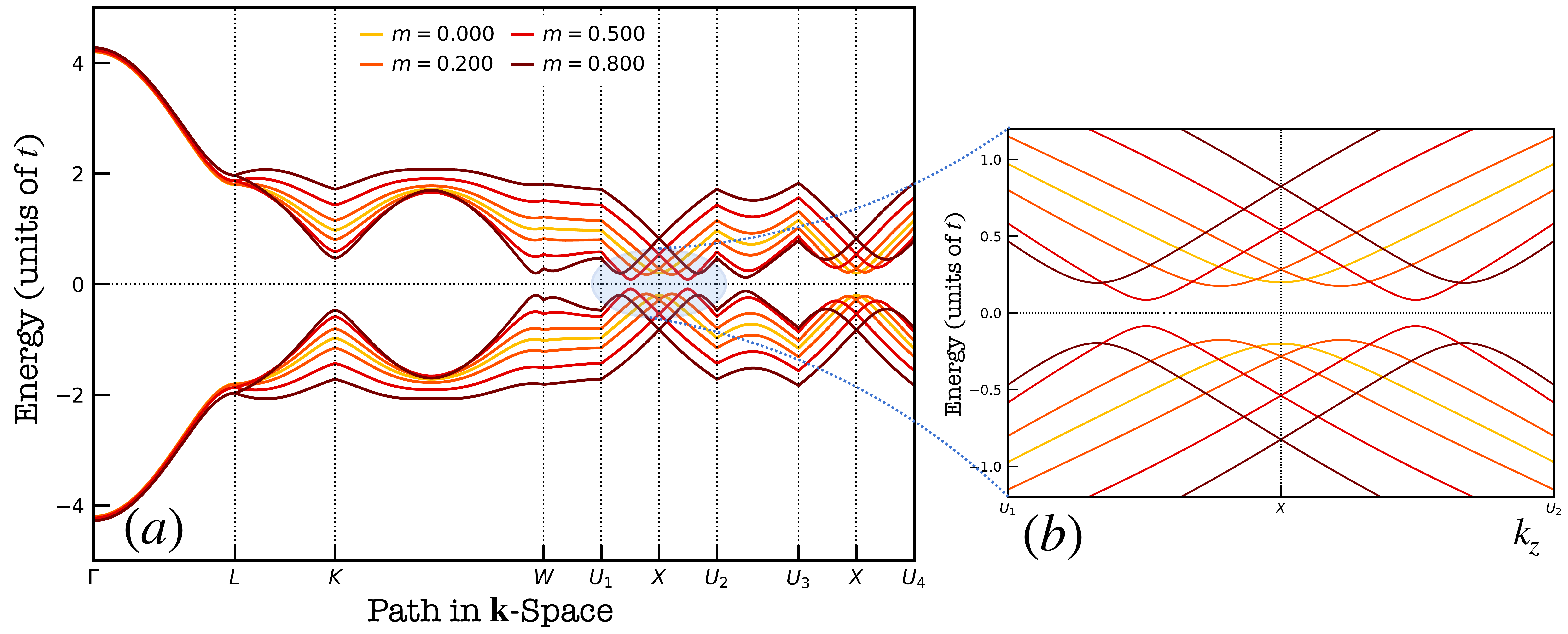}
\par\end{centering}
\vspace{-0.1cm}

\caption{\label{fig:MK_Insulator}\textit{(a)} Band structure of the Murakami-Kuga
model in the strong TI phase, i.e $\delta t=0.2t$, as a function
of the inversion symmetry breaking parameter $m$. The path in the
fBz is the same represented in Fig.\,\ref{fig:FKM_Band}\,b.\textit{
(b)} Close-up of the topological gap, which seems to be displaced
along the $k_{z}$-direction with an increasing $m$.}

\vspace{-0.5cm}
\end{figure}

The Fu-Kane-Mele model is only able to generate fine-tuned gapless
phases, because it is $\mathcal{P}\mathcal{T}$-symmetric. Inspired
by our earlier study of the Haldane model, we now present a slightly
modified system where the center of symmetry is broken by a staggered
potential that has values $\pm m$ in each sublattice of the diamond
structure\,\footnote{In truth, now the lattice structure became zincblende.}.
This tight-binding model, first proposed by Murakami and Kuga\,\cite{Murakami2008},
was based upon earlier ideas by Murakami\,\cite{Murakami07} and
provided the first example of a topologically stable gapless phases
in a three-dimensional lattice model. The full Hamiltonian of this
system reads,

\vspace{-0.7cm}

\begin{align}
\mathcal{H}_{\!{\scriptscriptstyle \text{MK}}}\!\! & =t\sum_{\mathbf{R}}\left(\sum_{i=1}^{4}\boldsymbol{\Psi}_{\mathbf{R}+\boldsymbol{\delta}_{i}}^{\dagger}\!\cdot\!\boldsymbol{\Psi}_{\!\mathbf{R}}+\frac{\delta t}{t}\boldsymbol{\Psi}_{\mathbf{R}+\boldsymbol{\delta}_{1}}^{\dagger}\!\cdot\!\boldsymbol{\Psi}_{\!\mathbf{R}}+\frac{m}{t}\boldsymbol{\Psi}_{\mathbf{R}}^{\dagger}\!\cdot\sigma_{z}\cdot\!\boldsymbol{\Psi}_{\!\mathbf{R}}\right)+\\
 & \qquad\qquad+i\lambda_{{\scriptscriptstyle \text{SO}}}\sum_{\mathbf{R}}\sum_{i=1}^{6}g_{i}^{j}\left(\boldsymbol{\Psi}_{\mathbf{R}+\boldsymbol{\Delta}_{i}}^{\dagger}\!\cdot\sigma_{j}\cdot\!\boldsymbol{\Psi}_{\!\mathbf{R}}\!-\!\boldsymbol{\Psi}_{\mathbf{R}-\boldsymbol{\delta}_{1}+\boldsymbol{\Delta}_{i}}^{\dagger}\!\cdot\sigma_{j}\cdot\!\boldsymbol{\Psi}_{\!\mathbf{R}-\boldsymbol{\delta}_{1}}\right),\nonumber 
\end{align}
which simply introduces one further control parameter with respect
to the original Fu-Kane-Mele model, that is the inversion-breaking
$m$. In Fig.\,\ref{fig:MK_Insulator}, we show the band structure
obtained for the \textit{Murakami-Kuga model}, upon the deformation
of a strong Fu-Kane-Mele TI phase by ever stronger values of $m$.
From the plots, two things are clear:\textit{ (i)} the valence and
conduction bands are no longer two-fold degenerate, and\textit{ (ii)}
the topological gap gets displaced away from the the $X$ points in
the fBz. In spite of this more complex band-structure, the Murakami-Kuga
term does not seem to change any qualitative feature of the model,
\textit{i.e.}, one still has a robust TI phase in this case. 

\begin{figure}[t]
\vspace{-0.5cm}
\begin{centering}
\includegraphics[scale=0.22]{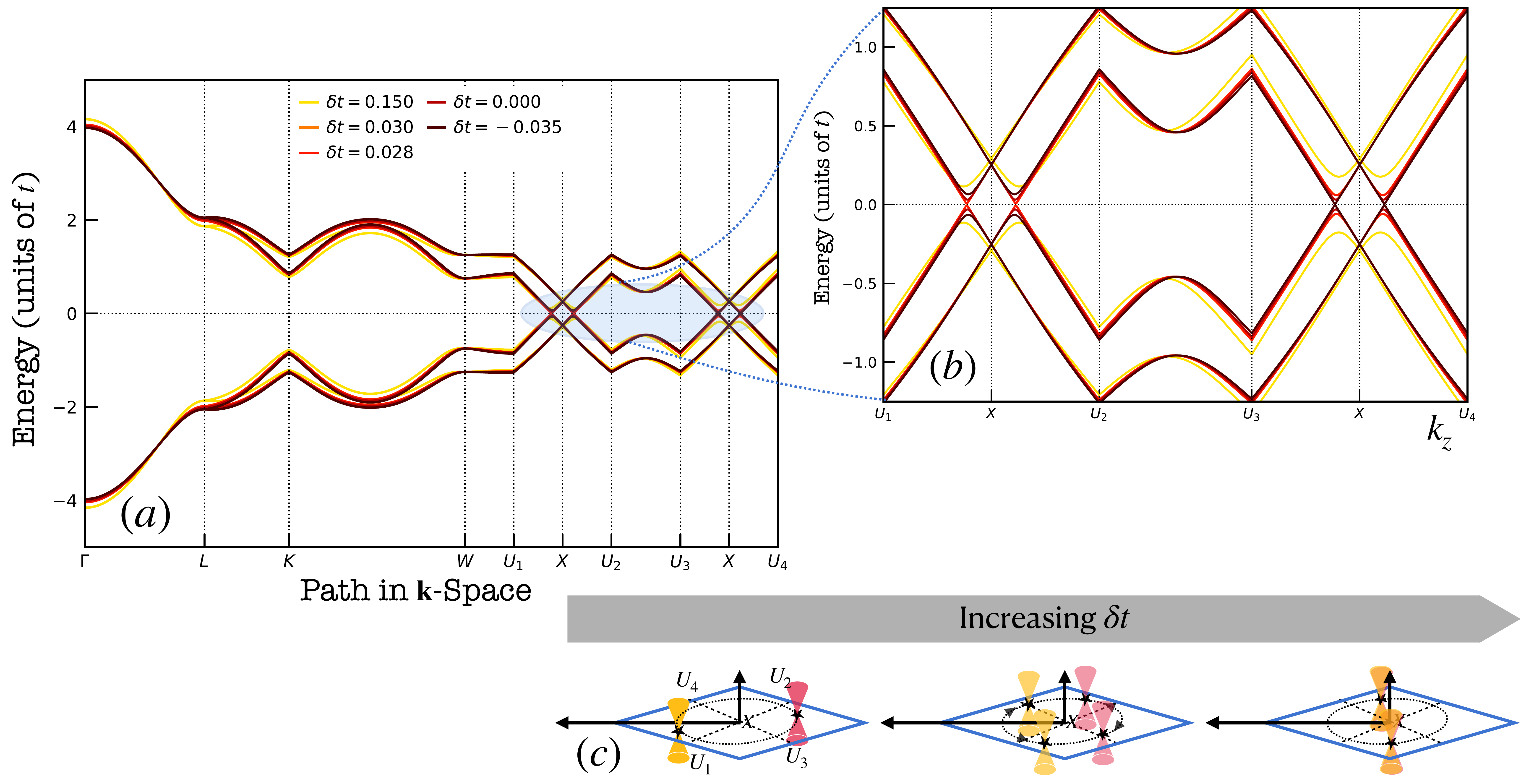}
\par\end{centering}
\vspace{-0.1cm}

\caption{\label{fig:MK_Insulator-1}\textit{(a)} Band structure of the Murakami-Kuga
model in the transition stage between a strong and a weak TI phase,
for $m=0.3$ as a function of the parameter $\delta t$. The path
in the fBz is the same represented in Fig.\,\ref{fig:FKM_Band}\,b.
\textit{(b)} Close-up of the gap-closing points. \textit{(c)} Scheme
of the two pairs of Weyl nodes wandering around a high-symmetry $X$
point.}

\vspace{-0.3cm}
\end{figure}

Having analyzed the model deep in a TI phase, we now set out to analyze
the effects of the staggered potential close to a transition point
between different gapped phases. Just like in the Fu-Kane-Mele model,
we find that a gap-closing transition also occurs in this system but
the gaps now close at six points of the fBz that are located slightly
offset from the $X$ points. However, unlike the centrossymetric Fu-Kane-Mele
model, the spectral gap can be shown to remain closed for a sizable
range of parameters $\delta t$\,\cite{Murakami2008}. As $\delta t$
is increased, the topological phase transition in this model then
proceeds through the following steps: \textit{(i)} all offset gaps
are closed forming a set of six four-fold degenerate Dirac points;
\textit{(ii)} each Dirac point is then broken into two non-degenerate
Weyl points that move away from each other, encircling the corresponding
$X$ point\,\footnote{According to Ref.\,\cite{Murakami2008} the trajectory traces an
almost perfect circle within the fBz's side, around the $X$ point.}; \textit{(iii)} two Weyl points of opposite chirality (originated
from different gap closing points) meet and mutually annihilate in
order to re-open the spectral gap. This intermediate stage of wandering
Weyl nodes, schematically shown in Fig.\,\ref{fig:MK_Insulator-1}\,c,
places the system in a stable gapless electronic phase which features
12 Weyl nodes in the band structure. Contrary to the four-fold band-crossings
of the Fu-Kane-Mele model, next we will see that these simple band-crossings
are protected by a topological charge and are an example of a (topological)
Weyl semimetal. Finally, we remark that the band structure presented
in Figs.\,\ref{fig:MK_Insulator-1}\,a and b confirm the wandering
Weyl nodes' scenario for this transitional phase.

\section{\label{subsec:Topological-Properties}The Theory Three-Dimensional
Band Crossings}

The transitional gapless state of the Murakami-Kuga model provided
us with a concrete example of a three-dimensional Weyl semimetal (WSM).
Even though this may seem like a very exotic situation, it has been
known from the early days of solid-state physics that electronic band
structures often cross in crystals\,\cite{Herring37} and, provided
the Fermi level can be tuned to such a point, this opens up the possibility
of realizing gapless semi-metallic phases. In 3D crystals, these band-crossings
may happen at isolated points, lines or surfaces in the fBz (see Lv
\textit{et al.}\,\cite{Lv21} for a comprehensive review). Here,
we will only be concerned with systems having point-like band-crossings,
that we shall refer to as \textit{three-dimensional topological semimetals}.
Even this restricted class of band-crossings can host a wide variety
of qualitatively different emergent quasiparticles. However, one must
be aware that not all band crossing are realizable in practice. From
general quantum mechanics, we know that level crossings are usually
avoided by strong hybridization, if there are no selection rules preventing
it. In crystals, a prime source of hybridization arises from \textit{spin-orbit
coupling} (SOC) which is, strictly speaking, always present for moving
electrons. A prime example is 2D graphene, which has two Dirac cones
that are known to be gapped by SOC while having a very tiny induced
spectral gap ($\sim1\text{\textmu eV}$)\,\cite{Dresselhaus65,Wang2007}.

One of the most outstanding properties of three-dimensional topological
semimetals is that they can actually be robust to SOC even in the
absence of any spatial symmetry. Two-fold degenerate linear band-crossings
can be robust (and quite generic) in the fBz of a system that has
a broken time-reversal or inversion symmetry. The transitional phase
of the Murakami-Kuga model\,\cite{Murakami2008} and transition-metal
monoarsenides\,\cite{Huang2015a} are two examples of Weyl semimetals
in this class. The physical properties of these systems are mostly
determined by their low-energy emergent quasiparticles which are $N_{v}$
uncoupled \textit{flavors} of independent fermions that obey a general
Hamiltonian of the form

\vspace{-0.7cm}

\begin{equation}
\mathcal{H}(\mathbf{q})=\hbar f_{0}\left(\mathbf{q}\right)+\hbar\!\!\sum_{i=x,y,z}\!\!\!\sigma_{i}f_{i}\left(\mathbf{q}\right)\label{eq:GenericBandCrossing}
\end{equation}
where $\mathbf{q}$ is the momentum-space shift from the point where
band-crossing is located. The coefficients $f_{0xyz}$ are real-valued
functions of $\mathbf{q}$ with the dimensions of an inverse-time.
As usual, we are only interested in the limit $\abs{\mathbf{q}}\!\approx\!0$
which, assuming that the band-crossing point is placed at zero energy,
we can express generically as

\vspace{-0.7cm}
\begin{equation}
\mathcal{H}(\mathbf{q})\approx\hbar\mathbf{q}\cdot\!\left.\boldsymbol{\nabla}f_{0}\left(\mathbf{q}\right)\right|_{\mathbf{q=0}}+\!\hbar\!\sum_{i=x,y,z}\!\!\sigma_{i}\mathbf{q}\cdot\left.\boldsymbol{\nabla}f_{i}\left(\mathbf{q}\right)\right|_{\mathbf{q=0}}.
\end{equation}
For future convenience, we shall call $\mathbf{v}_{i}\!=\!\left.\boldsymbol{\nabla}f_{i}\left(\mathbf{q}\right)\right|_{\mathbf{q=0}}$
the \textit{generalized velocities} and $\mathbf{c}\!=\!\left.\boldsymbol{\nabla}f_{0}\left(\mathbf{q}\right)\right|_{\mathbf{q=0}}$
the \textit{tilt vector} of the node. The justification for these
names will shortly become evident, but for now, it is important to
observe that this notation turns the linear band-crossing Hamiltonian
into the simplified form,

\vspace{-0.7cm}

\begin{equation}
\mathcal{H}(\mathbf{q})\approx\hbar\mathbf{c}\!\cdot\!\mathbf{q}+\!\hbar\!\!\sum_{i=x,y,z}\!\!\sigma_{i}\mathbf{v}_{i}\cdot\mathbf{q}.\label{eq:Weylhamiltonian}
\end{equation}
If $\mathbf{c}=\mathbf{0}$, Equation\,\eqref{eq:Weylhamiltonian}
describes emergent Weyl fermions that are entirely analogous to their
namesake high-energy counterparts. In order to see this, we must perform
an invertible linear transformation, $\mathbb{T}$, such that 
\begin{equation}
\mathbb{T}\!\cdot\!\mathbf{v}_{i}=\pm v_{\text{F}}\,\mathbf{x}_{i}\to\mathbb{T}\cdot\left[\begin{array}{ccc}
\partial_{x}f_{x} & \partial_{y}f_{x} & \partial_{z}f_{x}\\
\partial_{x}f_{y} & \partial_{y}f_{y} & \partial_{z}f_{y}\\
\partial_{x}f_{z} & \partial_{y}f_{z} & \partial_{z}f_{z}
\end{array}\right]=\pm\left[\begin{array}{ccc}
v_{\text{F}} & 0 & 0\\
0 & v_{\text{F}} & 0\\
0 & 0 & v_{\text{F}}
\end{array}\right],\label{eq:Jacobian}
\end{equation}
where $\mathbf{x}_{i}$ are the cartesian unit vectors, and $v_{\text{F}}$
is a positive constant. Note that Eq.\,\eqref{eq:Jacobian} defines
$\mathbb{T}$ as the inverse Jacobian matrix ($\mathcal{J}$) of $\left[f_{x}\left(\mathbf{q}\right),f_{y}\left(\mathbf{q}\right),f_{z}\left(\mathbf{q}\right)\right]$,
up to a factor of $\pm v_{\text{F}}$. The sign must be chosen in
order to maintain orientation of the basis upon transformation,\textit{
i.e.},

\vspace{-0.7cm}
\begin{equation}
\mathbb{T}=v_{\text{F}}\,\text{sign}\left(\det\left[\mathcal{J}\right]\right)\mathcal{J}^{-1}
\end{equation}
and therefore, we get to the transformed Hamiltonian,

\vspace{-0.7cm}

\begin{equation}
\mathcal{H}(\mathbf{q})\approx\hbar v_{\text{F}}\,\text{sign}\left(\det\left[\mathcal{J}\right]\right)\!\sum_{i=x,y,z}\sigma_{i}\underset{(k_{x},k_{y},k_{z})}{\underbrace{\mathbf{q}\cdot\mathcal{J}^{-1}}}\!\!\cdot\mathbf{x}_{i}\equiv\chi\hbar v_{\text{F}}\boldsymbol{\sigma}\cdot\mathbf{k},\label{eq:Weylhamiltonian-1}
\end{equation}
where the chirality of the Weyl fermions ($\chi$) is defined as $\text{sign}\left(\det\left[\mathcal{J}\right]\right).$
It is important to emphasize that the chirality is an intrinsic property
of the original Hamiltonian {[}Eq.\,\eqref{eq:Weylhamiltonian}{]},
which can also be determined from the generalized velocities as follows:

\vspace{-0.7cm}
\begin{equation}
\det\left[\mathcal{J}\right]=\left|\begin{array}{ccc}
v_{x}^{1} & v_{x}^{2} & v_{x}^{3}\\
v_{y}^{1} & v_{y}^{2} & v_{y}^{3}\\
v_{z}^{1} & v_{z}^{2} & v_{z}^{3}
\end{array}\right|=\mathbf{v}_{x}\cdot\left(\mathbf{v}_{y}\times\mathbf{v}_{z}\right).
\end{equation}
\begin{figure}[t]
\vspace{-0.5cm}
\begin{centering}
\includegraphics[scale=0.23]{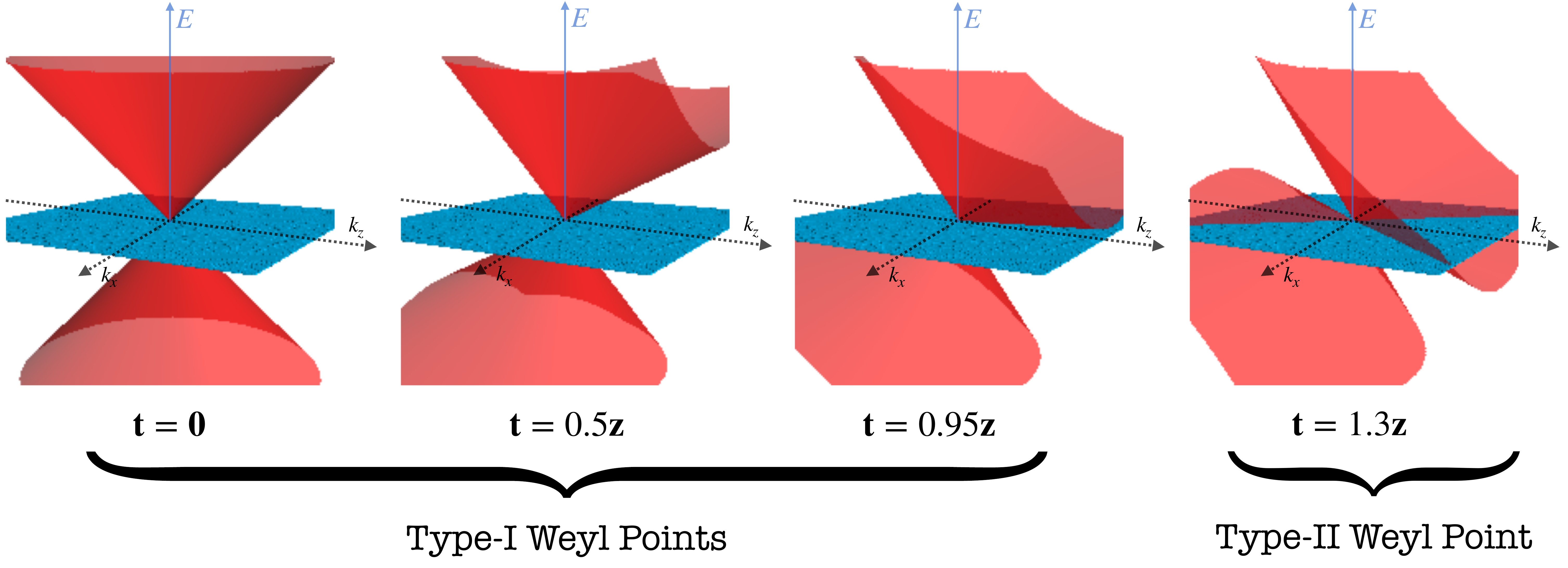}
\par\end{centering}
\vspace{-0.1cm}

\caption{\label{fig:Type_I_II_WP} Depiction of the low-energy dispersion relation
of a tilted Weyl point in the plane $k_{x}-k_{z}$. From left to right,
we present the non tilted case ($\mathbf{t}=\mathbf{0}$), two tilted
type-I Weyl points ($\mathbf{t}=0.5\mathbf{z}$ and $0.95\mathbf{z}$),
and a type-II Weyl point ($\mathbf{t}=1.3\mathbf{z}$).}

\vspace{-0.3cm}
\end{figure}

Up until this point, we have disregarded the $\mathbf{c}\cdot\mathbf{q}$
term in the low-energy Hamiltonian. Without this term, the low energy
dispersion relation gives rise to a distorted Weyl cone where each
band has a monotonic dispersion along all $\mathbf{q}$ directions.
In other words, if $E_{F}\!=\!0$, the Fermi surface is composed of
a single point at $\mathbf{q}\!=\!\mathbf{0}$ (for each valley).
As first pointed out by Soluyanov \textit{et al.}\,\cite{Soluyanov2015},
when there a finite tilt vector is present, $\abs{\mathbf{c}}\!>\!0$,
the Hamiltonian may not be mappable to a simple isotropic Weyl cone,
in which case it leads to a new kind of (Lorentz-invariance breaking)
quasiparticles. In order to see this, we reconsider the Hamiltonian
of Eq.\,\eqref{eq:Weylhamiltonian} including the tilt term,

\vspace{-0.7cm}
\begin{equation}
\mathcal{H}(\mathbf{k})=\chi\hbar v_{\text{F}}\left[\begin{array}{cc}
k_{z}+\mathbf{t}\!\cdot\!\mathbf{k} & k_{x}-ik_{y}\\
k_{x}+ik_{y} & -k_{z}+\mathbf{t}\!\cdot\!\mathbf{k}
\end{array}\right],
\end{equation}
where $\mathbf{t}=\mathbf{c}/v_{\text{F}}$ is the tilt-vector written
in the transformed coordinates. The dispersion relation of this Hamiltonian
can be obtained simply as $\varepsilon_{\pm\mathbf{k}}=\hbar v_{\text{F}}\left(\mathbf{t}\!\cdot\!\mathbf{k}\pm\abs{\mathbf{k}}\right).$
As shown in Fig.\,\ref{fig:Type_I_II_WP}, the existence of a non-zero
$\mathbf{t}$ has the effect of \textit{``tilting the Weyl cone in
energy}'' along the direction defined by the latter in $\mathbf{k}$-space.
To see what is the effect of this tilt on the emergent Weyl fermions,
it is useful to calculate the corresponding band-velocity,

\vspace{-0.7cm}

\begin{equation}
\mathbf{v}_{s\mathbf{k}}\!=\!\frac{1}{\hbar}\boldsymbol{\nabla}_{\mathbf{k}}\varepsilon_{n\mathbf{k}}=v_{\text{F}}\left(\mathbf{t}+s\frac{\mathbf{k}}{\abs{\mathbf{k}}}\right),
\end{equation}
where $s=\pm1$ labels the band. For an non tilted cone, the velocities
might be anisotropic in $\mathbf{k}$-space, but will always have
the symmetry $\mathbf{v}_{s\mathbf{k}}\!=\!-\mathbf{v}_{s-\mathbf{k}}$.
In the presence of a tilt, this is no longer true and, in particular,
states that are along the axis defined by $\mathbf{t}$ will have
a greater velocity for $\mathbf{k}\cdot\mathbf{t}\!>\!0$. If $\abs{\mathbf{t}}\!<\!1$,
the sign of the corresponding velocities is still preserved within
each band $s$ and the model is equivalent to Eq.\,\eqref{eq:Weylhamiltonian-1}.
However, if $\abs{\mathbf{t}}$ exceeds unity the situation gets hugely
modified with $s=+1$ ($s=-1$) having only right (left) moving Bloch
states along the $\mathbf{t}$-axis. The later situation defines a
\textit{type-II Weyl point}\,\footnote{Note that we do not use the term semimetal here. This is because,
unlike in the case of a type-I Weyl point, in this case the density
of states is not zero at the Weyl point. Hence, it is not strictly
a semi-metallic phase.}, as opposed to the hitherto discussed type-I, which happens for $\abs{\mathbf{t}}\!<\!1$.
Both cases are shown in Fig.\,\ref{fig:Type_I_II_WP}. \begin{wrapfigure}{o}{0.4\columnwidth}%
\vspace{-0.2cm}

\includegraphics[scale=0.15]{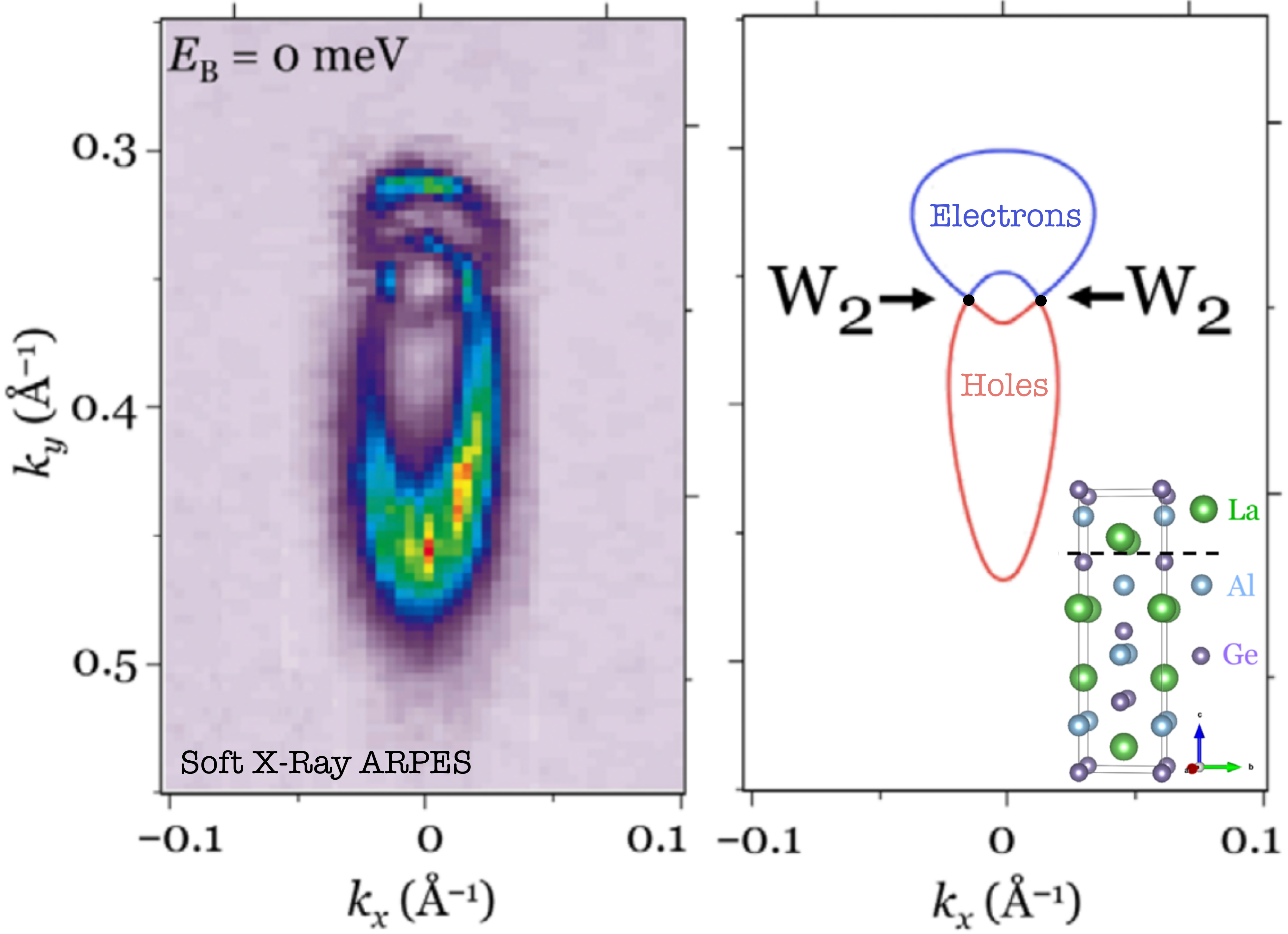}

\vspace{-0.2cm}

\caption{\label{fig:ObservationTypeIISM}Observation of the Fermi surface associated
to a pair of type-II Weyl points in $\text{La}\text{Al}\text{Ge}$.
Pictures adapted from Xu \textit{et al}.\,\cite{Xu2017}.}

\vspace{-0.5cm}\end{wrapfigure}%
The transitional case, where $\mathbf{t}$ is a unit vector, was dubbed
a \textit{type-III Weyl point}\,\cite{Huang2018}.

Finally, it is also enlightening to look at these two situations in
terms of their Fermi surfaces at low doping. Around a type-I Weyl
point, if the Fermi level lies slightly above (below) the band-crossing
a spherical Fermi surface of electron-like (hole-like) quasiparticles
will appear around it. The effect of a small tilt or Fermi-velocity
anisotropy will be to off-center and deform this sphere. In stark
contrast, even if a type-II Weyl point lies at the Fermi level, the
Fermi surface is not point-like but instead consists of two adjacent
lobes made up of electron- and hole-like excitations. The Fermi-surface
of $\text{La}\text{Al}\text{Ge}$, an experimentally realized type-II
Weyl material, is shown in Fig.\,\ref{fig:ObservationTypeIISM} as
measured by Xu \textit{et al}.\,\cite{Xu2017} using soft X-ray ARPES.

\subsection{Topology of Weyl Points}

We have referred that two-fold degenerate band-crossings, or Weyl
points, can be present in the band-structure of a solid-state system
even without being protected by a space-group symmetry. This is due
to the remarkable fact that these points act as monopoles of the Berry
curvature in $\mathbf{k}$-space, to which a \textit{topological charge}
can be associated, thus making them removable only through a mutual
annihilation process\,\footnote{One such example is provided by the phase-transition of the Murakami-Kuga
model observed in Fig.\,\ref{fig:MK_Insulator-1}.}. For completeness, we will characterize the topology of a Weyl node
by calculating its Berry curvature field and intrinsic orbital magnetic
moment. Without loss of generality, we consider an isotropic Weyl
point with a tilt-vector $\mathbf{t}=t\mathbf{z}$, whose $2\!\times\!2$
Bloch Hamiltonian reads,

\vspace{-0.7cm}
\begin{equation}
\mathcal{H}(\mathbf{k})=\chi\hbar v_{\text{F}}\left[\begin{array}{cc}
\left(t+1\right)k_{z} & k_{x}-ik_{y}\\
k_{x}+ik_{y} & \left(t-1\right)k_{z}
\end{array}\right],
\end{equation}
$\chi=\pm1$ being the chirality, and whose eigenstates, $\ket{\xi_{{\scriptscriptstyle \chi s}\mathbf{k}}}$,
can be written as\,\footnote{Note that the eigenstates do not depend in the Weyl cone's tilt, at
all.}

\vspace{-0.7cm}

\begin{subequations}
\begin{align}
\ket{\xi_{{\scriptscriptstyle ++}\mathbf{k}}} & =\ket{\xi_{{\scriptscriptstyle --}\mathbf{k}}}=\left[\:\:\,\begin{array}{cc}
\cos\frac{\theta_{\mathbf{k}}}{2},e^{i\varphi_{\mathbf{k}}}\!\! & \sin\frac{\theta_{\mathbf{k}}}{2}\end{array}\right]^{\text{T}}\label{eq:States11}\\
\ket{\xi_{{\scriptscriptstyle +-}\mathbf{k}}} & =\ket{\xi_{{\scriptscriptstyle -+}\mathbf{k}}}=\left[\begin{array}{cc}
-\sin\frac{\theta_{\mathbf{k}}}{2},e^{i\varphi_{\mathbf{k}}}\!\! & \cos\frac{\theta_{\mathbf{k}}}{2}\end{array}\right]^{\text{T}},\label{eq:States22}
\end{align}
\end{subequations}

with the spherical angles $\theta_{\mathbf{k}}$ and $\varphi_{\mathbf{k}}$
defined as

\vspace{-0.7cm}

\begin{equation}
\theta_{\mathbf{k}}\!=\!\arccos\left[\!\frac{k_{z}}{\sqrt{k_{x}^{2}+k_{y}^{2}+k_{z}^{2}}}\!\right]\text{ and }\tan\varphi_{\mathbf{k}}=\frac{k_{y}}{k_{x}}.
\end{equation}
Using these exact Bloch states, we can easily obtain the 3D Berry
connection and curvature fields, as well as the intrinsic orbital
magnetic moment near such a band-crossing point. More precisely, we
have

\vspace{-0.7cm}

\begin{subequations}
\begin{align}
\boldsymbol{\mathcal{A}}_{\chi s\mathbf{k}} & =\frac{\left(k_{y},-k_{x},0\right)}{2\abs{\mathbf{k}}\left(\abs{\mathbf{k}}+\chi sk_{z}\right)}\\
\boldsymbol{\Omega}_{\chi s\mathbf{k}} & =\boldsymbol{\nabla}_{\!\mathbf{k}}\!\times\!\boldsymbol{\mathcal{A}}_{\chi s\mathbf{k}}=\frac{\chi s}{2\abs{\mathbf{k}}^{3}}\mathbf{k}\label{eq:BerryCurv}
\end{align}
\end{subequations}

which clearly shows that the Weyl node is a monopole of the field
$\boldsymbol{\Omega}_{\chi s}(\mathbf{k})$ with a topological charge
$\chi\eta$\,\footnote{To be precise, usually one speaks about the topological charge by
looking at the conduction band, i.e., $\eta=+1$, thus defining the
positive chirality as a positive charge.}. At the same time, since Weyl systems are never $\mathcal{PT}$-symmetric,
we can also calculate the intrinsic magnetic moment associated to
the emergent Weyl fermions\,\cite{Chang96,Sundaram99}. This quantity
is evaluated from the $\ket{\xi_{{\scriptscriptstyle \chi}s\mathbf{k}}}$
states as well, through the following formula,

\vspace{-0.7cm}

\begin{equation}
\boldsymbol{\mathcal{M}}_{\chi s}(\mathbf{k})=-\chi\,e\,\hbar\,v_{\text{F}}\,\Im\left[\bra{\boldsymbol{\nabla}_{\!\mathbf{k}}\xi_{{\scriptscriptstyle \chi s}\mathbf{k}}}\times\left(\left[\boldsymbol{\sigma}\!\cdot\!\mathbf{k}-\chi s\abs{\mathbf{k}}\mathbb{I}_{2\times2}\right]\ket{\boldsymbol{\nabla}_{\!\mathbf{k}}\xi_{{\scriptscriptstyle \chi s}\mathbf{k}}}\right)\right],
\end{equation}
which ends up yielding a very simple result,

\vspace{-0.7cm}

\begin{equation}
\boldsymbol{\mathcal{M}}_{\chi s}(\mathbf{k})=\chi\frac{ev_{\text{F}}\mathbf{k}}{\abs{\mathbf{k}}^{2}}.\label{eq:magneticMoment}
\end{equation}
Equation\,\eqref{eq:magneticMoment} indicates that a wave-packet
built from Weyl quasiparticles is self-rotating and, thus features
an intrinsic magnetic moment (other than the electronic spin!) which
is locked to its crystal momentum. Its direction may be along or opposite
to the propagation direction, depending if the chirality of the node
is positive or negative, respectively. In the end, the results of
Eqs.\,\eqref{eq:BerryCurv} and \eqref{eq:magneticMoment} allow
us to conclude that Weyl points in the band-structure of a crystal
are topologically protected, as one can think of them as being a sources
(or sinks) of the Berry curvature field or, equivalently, as hedgehog
defects in the fBz. In addition, we can also write down the semiclassical
equations that govern the dynamics of a wave-packet built from a superposition
of emergent lattice Weyl fermions:

\vspace{-0.7cm}

{\small{}
\begin{subequations}
{\small{}
\begin{align}
\!\!\!\!\!\!\!\!\!\!\!\!\!\!\dot{\mathbf{r}}_{\chi s} & =\frac{1}{\Delta_{s\mathbf{k}}}\left[v_{\text{F}}s\frac{\mathbf{k}}{\abs{\mathbf{k}}}-v_{\text{F}}\chi e\left(\mathbf{B}\cdot\boldsymbol{\nabla}_{\mathbf{k}}\right)\frac{\mathbf{k}}{\abs{\mathbf{k}}^{2}}-\frac{e\chi s}{\hbar\abs{\mathbf{k}}^{3}}\mathbf{E}\!\times\mathbf{k}+\frac{e\chi s}{\hbar\abs{\mathbf{k}}^{3}}\left(\mathbf{k}\cdot\mathbf{v}_{s\mathbf{k}}\right)\mathbf{B}\right]\label{eq:drdt-1-1-2}\\
\!\!\!\!\!\!\!\!\!\!\!\!\!\!\dot{\mathbf{k}}_{\chi s} & =\frac{1}{\Delta_{s\mathbf{k}}}\left[-\frac{e}{\hbar}\mathbf{E}+s\frac{v_{\text{F}}e}{\abs{\mathbf{k}}\hbar}\mathbf{B}\times\!\mathbf{k}+\frac{\chi se^{2}}{\abs{\mathbf{k}}^{3}\hbar^{2}}\left(\mathbf{E}\cdot\mathbf{B}\right)\mathbf{k}+2v_{\text{F}}\chi e\frac{\left(\mathbf{B}\times\mathbf{k}\right)\left(\mathbf{B}\cdot\mathbf{k}\right)}{\abs{\mathbf{k}}^{4}}\right],\label{eq:dkdt-1-1-2}
\end{align}
}
\end{subequations}
}{\small\par}

where $\mathbf{E}$ ($\mathbf{B}$) is the applied electric (magnetic)
field, and $\Delta_{\chi s\mathbf{k}}\!=\!1\!+\!\hbar^{-1}\!\abs{\mathbf{k}}^{-3}\!\!e\chi s\left(\mathbf{k}\cdot\mathbf{B}\right)$
is the topological correction to the phase-space volume\,\footnote{To simplify the Lorentz-term in Eq.\,\eqref{eq:dkdt-1-1-2}, we have
used the fact that 
\[
\left(\mathbf{B}\cdot\boldsymbol{\nabla}_{\mathbf{k}}\right)\frac{\mathbf{k}}{\abs{\mathbf{k}}^{2}}=\mathbf{B}/\abs{\mathbf{k}}^{4}-2\mathbf{k}\left(\mathbf{k}\cdot\mathbf{B}\right)/\abs{\mathbf{k}}^{4}.
\]
}. The effect of interaction between the magnetic field and the intrinsic
magnetic moment of the Bloch wave-packet is already included (\textit{e.g.},
see Knoll \textit{et al}.\,\cite{Knoll2020}). Equations\,\eqref{eq:drdt-1-1-2}
and \eqref{eq:dkdt-1-1-2} will be crucial to derive some important
transport effects that arise from these topological features of a
single Weyl node. This discussion will be undertaken in Sect.\,\ref{sec:Observable-Signatures-of}.

\vspace{-0.3cm}

\subsection{Alternative Stable Band-Crossings}

Up to this point, our discussion have focused on point-like band crossings
that: \textit{(i)} are linear and \textit{(ii)} only involve only
non-degenerate bands. These restrictions clearly do not cover all
possibilities so we leave here some further comments on alternative
cases of band-crossing points. First of all, even if only non-degenerate
bands are involved, one may ask why are linear band-crossings particularly
important. The reason is because they are generic, in the sense that
any higher-order contact points will require at least one of the first
derivatives of the $f$-functions in Eq.\,\eqref{eq:GenericBandCrossing}
to be exactly zero. This a fine-tuned situation that will not, in
general, be independent of microscopic details. However, as shown
by Fang \textit{et al.}\,\cite{Fang2012}, there are systems in which
higher-order band-crossing points (dubbed \textit{multi-Weyl points})
are enforced by the existence of discrete rotation axis as point-group
symmetries; more precisely, a four-fold (six-fold) rotation axis can
stabilize a double-Weyl (triple-Weyl) point that features a quadratic
(cubic) dispersion in the plane perpendicular to that axis, whilst
maintaining a linear dispersion along it. In that case, the most general
(untilted) low-energy Hamiltonian reads\,\cite{Lv21}

\vspace{-0.7cm}

\begin{equation}
\mathcal{H}_{N}(\mathbf{k})=\chi\,\hbar\left[\begin{array}{cc}
v_{\parallel}k_{z} & v_{\perp}\left(k_{x}-ik_{y}\right)^{N}\\
v_{\perp}\left(k_{x}+ik_{y}\right)^{N} & -v_{\parallel}k_{z}
\end{array}\right],\label{eq:GeneralHam}
\end{equation}
where $N\!=\!1,2,3$ labels the order of the Weyl point, $\chi\!=\!\pm1$
is the chirality, $v_{\perp}$/$v_{\parallel}$ are positive constants,
and the rotation axis is assumed to lie along $k_{z}$. For $N\!=\!2,3$,
this Hamiltonian can be shown to yield a monopole of the Berry curvature
field with a topological charge $2\chi$ or $3\chi$, respectively.

Besides the existence of higher order band-touchings, one must also
allow for the possibility of degenerate bands that touch somewhere
in the fBz. To the untrained eye, this situation may seem even more
unlikely than the aforementioned multi-Weyl points but it can actually
be quite generic. As shown in Sect.\,\ref{sec:Generic-Symmetries},
the mere presence of both time-reversal and inversion symmetries is
enough to impose an exact two-fold degeneracy on spin-$\nicefrac{1}{2}$
electronic bands over the entire fBz. This implies that a $\mathcal{PT}$-symmetric
band-touching point always amounts to a four-fold degeneracy that
realizes massless Dirac fermions, as opposed to Weyl fermions as emergent
quasiparticles. In this case, the Hamiltonian at low energies is the
$4\!\times\!4$ matrix,

\vspace{-0.7cm}
\begin{equation}
\mathcal{H}_{\text{D}}(\mathbf{k})=\hbar v_{\text{F}}\left[\begin{array}{cc}
\boldsymbol{\sigma}\cdot\mathbf{k} & \mathbf{\mathbb{O}}_{2\times2}\\
\mathbf{\mathbb{O}}_{2\times2} & -\boldsymbol{\sigma}\cdot\mathbf{k}
\end{array}\right].\label{eq:DiracPoint}
\end{equation}
Equation\,\eqref{eq:DiracPoint} defines a band-crossing that may
be seen as two superimposed and uncoupled Weyl points of opposite
chirality. Therefore, a Dirac point carries no topological charge
and is amenable to gap-opening homogeneous perturbations. Unlike the
single Weyl point, this is a chiral symmetric Hamiltonian that can
be put into an off-diagonal form upon the unitary transformation to
the so-called \textit{Dirac representation} of the $\alpha$-matrices
(see Appendix\,\,\ref{chap:Dirac_Spherical} for further details).
Despite not enjoying the same topological protection as isolated Weyl
points, Dirac points can be symmetry-enforced in systems that have
nonsymmorphic space groups. More precisely, one must ensure that the
symmetry group (doubled by electron spin) has a four-dimensional irreducible
representation, such that a stable Dirac point happens at high-symmetry
points of the fBz. This scenario was first proposed by Young \textit{et
al}.\,\cite{Young12} and, since then, several stable Dirac semimetal
phases have been confirmed in materials such as $\text{Na}_{3}\text{Bi}$\,\cite{Liu2014a,Xu2015},
$\text{Cd}_{3}\text{As}_{2}$\,\cite{Liu2014b,Neupane2014,Borisenko2014},
and black phosphorous\,\cite{Kim2015a} (multi-layered phospherene).

Here, we will focus solely on the physics of type-I Weyl semimetals
(WSMs) and Dirac semimetals (DSMs). These systems provide condensed
matter realizations of massless Weyl and Dirac fermions which are
known to be possible in the current framework of relativistic Quantum
Field Theory. Nevertheless, as solid-state is not bound to be symmetric
with respect to the Poincaré group, there are many alternative possibilities
which have been classified by Bradlyn \textit{et al}.\,\cite{Bradlyn2016}.
For example, stable three-fold, four-fold, six-fold and eight-fold
degenerate band-crossing points can be stabilized by axial point-group
symmetries of the crystal. When these happen close to the Fermi level,
they lead to unconventional emergent massless quasiparticles that
behave as either spin-$\nicefrac{1}{2}$, spin-$1$ or spin-$\nicefrac{3}{2}$
fermions\,\footnote{Interestingly, the latter (known as \textit{massless Rarita-Schwinger
fermions}) are related to a very profound result on relativistic Quantum
Field Theory: if there are massless Rarita-Schwinger fermions, then
that particle must be a gravitino\,\cite{Das76}. Of course, in our
context, Poincaré group symmetry is absent altogether and, the result
does not apply.}. As a matter of fact, it was recently shown by Lv \textit{et al}.\,\cite{Lv2019},
that these different flavors of quasiparticles can even be found within
the band structure of the same material: $\text{Pd}\text{Bi}\text{Se}$.

\vspace{-0.4cm}

\section{\label{sec:Observable-Signatures-of}Observable Signatures of Topological
Semimetals}

We have shown that 3D Weyl semimetals are gapless phases that have
a non-trivial band topology which is made up of Berry curvature monopoles.
In this section, we will show that these topological properties actually
lead to remarkable measurable signatures that greatly enhance the
experimental interest on these systems. First of all, the nonzero
Berry curvature interplays nontrivially with any external electromagnetic
fields, which materializes into \textit{(i) }an unexpected Landau
spectrum caused by strong uniform magnetic fields\,\cite{Nielsen83}
and \textit{(ii) }a set of unconventional magneto-transport effects
such as the \textit{Chiral Magnetic Effect}\,\cite{Nielsen83}, a
\textit{Negative Longitudinal Magnetoresistance}\,\cite{Son2013,Zhang2016},
and the \textit{Planar Hall Effect}. Most of these effects can be
traced back to the fact that WSMs effectively realize the celebrated
chiral anomaly of QED\,\cite{Adler1969,Bell1969}. Besides bulk effects,
the topology of Weyl nodes is likewise responsible by boundary effects,
which amount to the existence of surface Fermi arc states\,\cite{Wan2011,Hosur2012,Witten2015,Haldane2014,Hashimoto2017}
that connect pairs of Weyl points through the surface-projected fBz.
The rest of this chapter will be devoted to providing the reader with
a helpful overview of this important phenomenology.

\vspace{-0.35cm}

\subsection{\label{subsec:Landau-Quantization}Landau Quantization and the Chiral
Anomaly}

The nontrivial topology of the Fermi surface in a WSM finds its first
characteristic signature in the structure of energy levels that is
generated by imposing a strong uniform magnetic field $\mathbf{B}$.
In the presence of a perpendicular magnetic field, a 2D quantum gas
of charged particles has its continuous spectrum collapsed into a
discrete set of macroscopically degenerate \textit{Landau levels}.
Landau quantization in 3D is slightly different because the momentum
parallel to $\mathbf{B}$ remains a good quantum number (for any gauge
choice), which generates an infinite set of macroscopically degenerate
\textit{Landau bands} that disperse along that axis. Following Nielsen
and Ninomiya\,\cite{Nielsen83}, we derive here the Landau bands
of a single Weyl node of chirality $\chi,$ which is described by
the minimally-coupled Hamiltonian,

\vspace{-0.7cm}

\begin{equation}
\mathcal{H}_{{\scriptscriptstyle \text{W}}}^{\prime}\!=\!-i\chi\hbar v_{\text{F}}\boldsymbol{\sigma}\cdot\left(\boldsymbol{\nabla}_{\mathbf{r}}+i\frac{e}{\hbar}\mathbf{A}(\mathbf{r})\right),\label{eq:WeylHam-1-1}
\end{equation}

where $e$ is the elementary charge, $c$ is the speed of light, and
$\mathbf{A}(\mathbf{r})$ is a vector potential such that $\boldsymbol{\nabla}\!\times\!\mathbf{A}\!=\!\mathbf{B}$.
For concreteness, we assume that $\mathbf{B}=B\mathbf{z}$ is oriented
along the $z$ direction\,\footnote{This implies no loss of generality, as the free Weyl Hamiltonian is
spherically symmetric.} and express the vector potential in the Landau gauge, $\mathbf{A}(x,y)=-By\mathbf{x}$.
This choice preserves the full translation invariance along $x$ and
$z$, maintaining $k_{x}$ and $k_{z}$ as good quantum numbers of
the problem. With this in mind, we express an arbitrary eigenwavefunction
of energy $E$ as

\vspace{-0.7cm}

\begin{equation}
\Psi_{E,k_{x},k_{z}}(\mathbf{r})=\left(\!\!\begin{array}{c}
f_{1}(y)\\
f_{2}(y)
\end{array}\!\!\right)e^{ik_{x}x+ik_{z}z},
\end{equation}
where $f_{1}(y)/f_{2}(y)$ are undetermined functions of $y$, which
are required to obey the ODE

\vspace{-0.7cm}

\begin{equation}
-i\chi\hbar v_{\text{F}}\left[i\sigma_{x}\left(k_{x}-\frac{eB}{\hbar c}y\right)+\sigma_{y}\partial_{y}+i\sigma_{z}k_{z}\right]\left(\!\!\begin{array}{c}
f_{1}(y)\\
f_{2}(y)
\end{array}\!\!\right)=E\left(\!\!\begin{array}{c}
f_{1}(y)\\
f_{2}(y)
\end{array}\!\!\right).\label{eq:Schroedingeryy-2}
\end{equation}
This equation may be written in dimensionless form by measuring all
distances in units of the magnetic length, $l_{m}\!=\!\sqrt{\hbar/eB}$
and energies in units of $\hbar v_{\text{F}}/l_{m}$. This yields
the dimensionless system

\vspace{-0.7cm}

\begin{equation}
\left(\!\!\begin{array}{cc}
q_{z} & -\partial_{u}+\left(q_{x}-u\right)\\
\partial_{u}+\left(q_{x}-u\right) & -q_{z}
\end{array}\!\!\right)\cdot\left(\!\!\begin{array}{c}
f_{1}(u)\\
f_{2}(u)
\end{array}\!\!\right)=\chi\varepsilon\left(\!\!\begin{array}{c}
f_{1}(u)\\
f_{2}(u)
\end{array}\!\!\right),\label{eq:Schroedingeryy-1-4}
\end{equation}
where $\varepsilon\!=\!El_{m}/\hbar v_{\text{F}}$, $q_{x}\!=\!l_{m}k_{x}$,
$q_{z}\!=\!l_{m}k_{z}$ and $u\!=\!y/l_{m}$. At this point, it is
important to remark that 

\vspace{-0.7cm}
\begin{equation}
\mathcal{O}^{\dagger}/\mathcal{O}\!=\!\pm\frac{1}{\sqrt{2}}\partial_{u}+\frac{1}{\sqrt{2}}\left(q_{x}-u\right)
\end{equation}
are creation/annihilation operators for the eigenstates of a one-dimensional
quantum harmonic oscillator (1DQHO) centered in $u\!=\!q_{x}$. More
precisely, by considering the orthonormal set of wavefunctions (as
functions of $u\in\mathbb{R}$),

\vspace{-0.7cm}
\begin{equation}
\phi_{n}^{q_{x}}(u)=\frac{\pi^{-\frac{1}{4}}}{\sqrt{2^{n}n!}}H_{n}\left(u-q_{x}\right)\exp\left[-\frac{\left(u-q_{x}\right)^{2}}{2}\right],
\end{equation}
where $H_{n}(x)$ is the $n^{\text{th}}$ Hermite polynomial, we can
prove that 

\vspace{-0.7cm}

\begin{subequations}
\begin{align}
\mathcal{O}^{\dagger}\phi_{n}^{q_{x}}(u) & =\sqrt{n+1}\phi_{n+1}^{q_{x}}(u)\\
\mathcal{O}\,\,\phi_{n}^{q_{x}}(u) & =\sqrt{n}\phi_{n-1}^{q_{x}}(u)
\end{align}
\end{subequations}

and also that $\left[\mathcal{O}^{\dagger},\mathcal{O}\right]=\mathcal{O}^{\dagger}\mathcal{O}-\mathcal{O}\mathcal{O}^{\dagger}=\left[\partial_{u},\left(q_{x}-u\right)\right]=1$.
Using this definition, we can re-write the eigenvalue problem of Eq.\,\eqref{eq:Schroedingeryy-1-4}
as

\vspace{-0.7cm}

\begin{equation}
\left(\!\!\begin{array}{cc}
q_{z} & \sqrt{2}\mathcal{O}\\
\sqrt{2}\mathcal{O}^{\dagger} & -q_{z}
\end{array}\!\!\right)\cdot\left(\!\!\begin{array}{c}
f_{1}(u)\\
f_{2}(u)
\end{array}\!\!\right)=\chi\varepsilon\left(\!\!\begin{array}{c}
f_{1}(u)\\
f_{2}(u)
\end{array}\!\!\right).\label{eq:Schroedingeryy-1-4-1}
\end{equation}
Finally, because the system has particle-hole symmetry, one might
as well study the eigenstates of $\mathcal{H}_{{\scriptscriptstyle \text{W}}}^{\prime2}$
instead. Any eigenstate of $\mathcal{H}_{{\scriptscriptstyle \text{W}}}^{\prime2}$
is also an eigenstate of $\mathcal{H}_{{\scriptscriptstyle \text{W}}}^{\prime}$,
but the eigenvalue problem {[}Eq.\,\eqref{eq:Schroedingeryy-1-4-1}{]}
is recast as a simpler one:

\vspace{-0.7cm}

\begin{equation}
\left(\!\!\begin{array}{cc}
q_{z}^{2}+2\mathcal{O}\mathcal{O}^{\dagger} & 0\\
0 & q_{z}^{2}+2\mathcal{O}^{\dagger}\mathcal{O}
\end{array}\!\!\right)\cdot\left(\!\!\begin{array}{c}
f_{1}(u)\\
f_{2}(u)
\end{array}\!\!\right)=\varepsilon^{2}\left(\!\!\begin{array}{c}
f_{1}(u)\\
f_{2}(u)
\end{array}\!\!\right).
\end{equation}
Since $\mathcal{O}^{\dagger}\mathcal{O}$ is a number operator and
$\mathcal{O}\mathcal{O}^{\dagger}=1+\mathcal{O}^{\dagger}\mathcal{O}$,
we conclude that 

\vspace{-0.7cm}

\begin{subequations}
\begin{align}
\varepsilon\! & =\!\pm\sqrt{2n_{1}+q_{z}^{2}}\qquad\to f_{1}(u)=\phi_{n_{1}}^{q_{x}}(u)\\
\varepsilon\! & =\!\pm\sqrt{2n_{2}+2+q_{z}^{2}}\thinspace\to f_{2}(u)=\phi_{n_{2}}^{q_{x}}(u),
\end{align}
\end{subequations}

which requires that $n=n_{2}=n_{1}+1>0$, for the sake of consistency.
For each value of $n\!=\!1,2,\cdots$, we have 

\vspace{-0.7cm}
\begin{subequations}
\begin{align}
f_{1}(u) & =A_{n}\phi_{n-1}^{q_{x}}(u)\,\,\,\,\text{ and }\,f_{2}(u)=B_{n}\phi_{n}^{q_{x}}(u)\text{ \,\,\,\,}\text{for }\varepsilon>0\\
f_{1}(u) & =A_{-n}\phi_{n-1}^{q_{x}}(u)\,\text{ and }\,f_{2}(u)=B_{-n}\phi_{n}^{q_{x}}(u)\,\text{ for }\varepsilon<0
\end{align}
\end{subequations}

where the coefficients $A_{n}/B_{n}$ are required to verify the following
relation

\vspace{-0.7cm}
\begin{equation}
\frac{B_{n}}{A_{n}}=\text{sign}(n)\left(\chi\sqrt{1\!+\left(\frac{q_{z}}{2n}\right)^{2}}-\frac{q_{z}}{2n}\right)\equiv\Xi_{n}^{q_{z}},
\end{equation}
in order to guarantee that $\left(f_{1}(u),f_{2}(u)\right)^{T}$ is
a solution of Eq.\,\eqref{eq:Schroedingeryy-1-4-1}. In conclusion,
the eigenvalues of the Weyl node in the presence of a magnetic field
$\mathbf{B}=B\mathbf{z}$ are 

\vspace{-0.7cm}

\begin{equation}
E_{n}(k_{z})=\text{sign}(n)v_{\text{F}}\sqrt{2\,\hbar\,e\,B\,n\!+\hbar^{2}k_{z}^{2}},\text{ with \ensuremath{n=\cdots,-2,-1,1,2,\cdots}},\label{eq:Spectrum}
\end{equation}
with the corresponding eigenstates 
\begin{figure}[t]
\vspace{-0.5cm}
\begin{centering}
\includegraphics[scale=0.22]{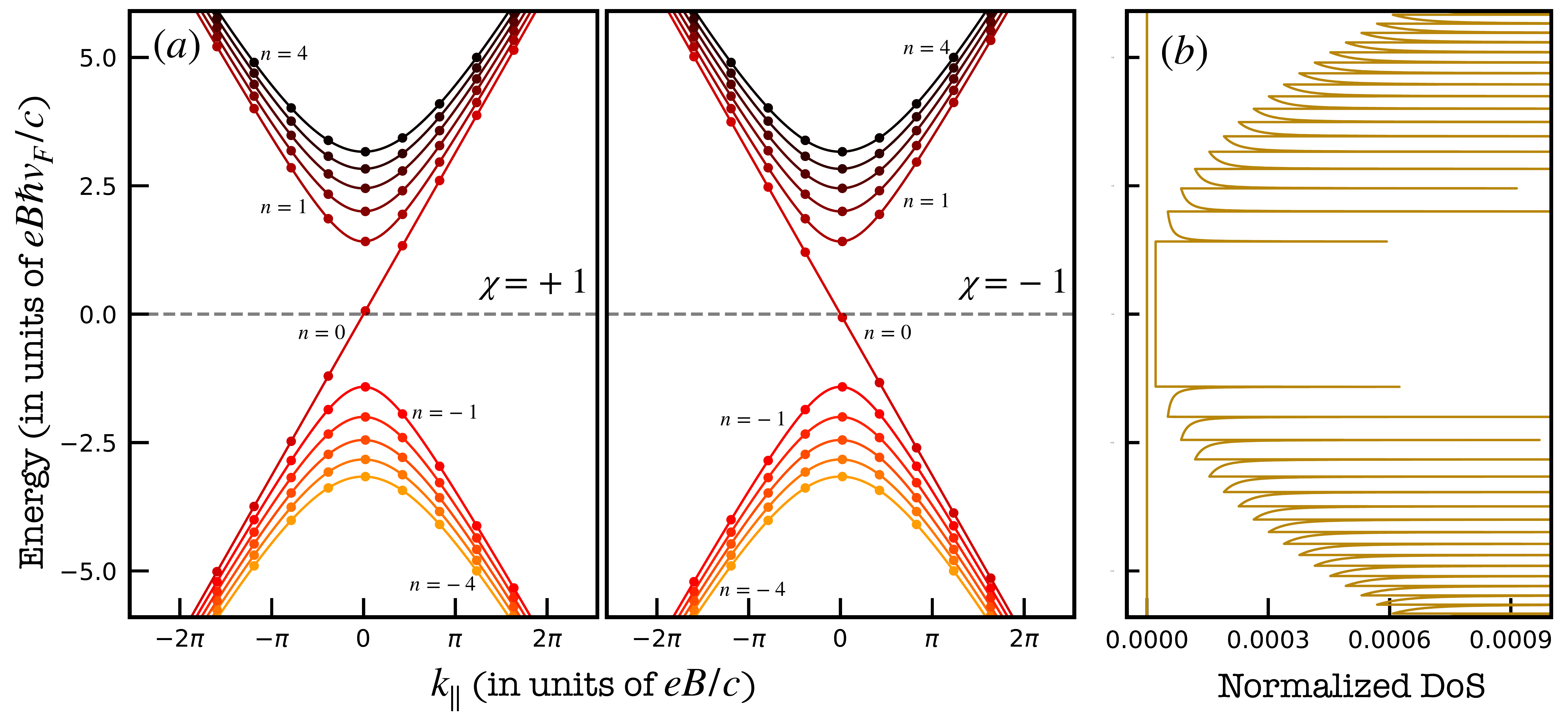}
\par\end{centering}
\vspace{-0.1cm}

\caption{\label{fig:MagneticLL}(a) Dispersion of the lowest Landau bands for
a single Weyl node of positive (left) and negative (right) chirality.
(b) Representation of the normalized density of states in the presence
of a magnetic field.}

\vspace{-0.4cm}
\end{figure}

\vspace{-0.7cm}
\begin{equation}
\Psi_{n,k_{x},k_{z}}(x,y,z)=\mathcal{N}\left(\!\!\begin{array}{c}
\phi_{n-1}^{l_{m}k_{x}}\left(\frac{y}{l_{m}}\right)\\
\Xi_{n}^{l_{m}k_{z}}\phi_{n}^{l_{m}k_{x}}\left(\frac{y}{l_{m}}\right)
\end{array}\!\!\right)e^{ik_{x}x+ik_{z}z},
\end{equation}
where $\mathcal{N}$ is an irrelevant normalization constant. The
spectrum of Eq.\,\eqref{eq:Spectrum} does not depend on chirality
at all, meaning that WSMs containing two opposite chirality nodes
will simply double the degeneracy of these dispersing Landau levels.
In addition, we see that the dispersion along $k_{z}$ is Dirac-like,
with a field-dependent mass-gap that is given by $2v_{\text{F}}\sqrt{2e\hbar Bn}$.

\vspace{-0.4cm}

\paragraph{Chiral Zero-Energy Landau Modes:}

As first shown by Nielsen and Ninomiya\,\cite{Nielsen83}, the previous
procedure fails to capture some eigenstates of the problem which depend
on chirality. These states are 3D analogues to the well-known nodal
Landau levels of graphene\,\cite{McClure56,Checkelsky2008,CastroNeto2009}.
More precisely, we can return back to Eq.\,\eqref{eq:Schroedingeryy-1-4-1},
and cast it into the form

\vspace{-0.7cm}

\begin{equation}
\begin{cases}
\sqrt{2}\mathcal{O}f_{2}(u)-\left(q_{z}-\chi\varepsilon\right)f_{1}(u)=0\\
\sqrt{2}\mathcal{O}^{\dagger}f_{1}(u)-\left(q_{z}+\chi\varepsilon\right)f_{2}(u)=0
\end{cases}.
\end{equation}
For a generic $\varepsilon$, this is a coupled system that must be
solved as before, \textit{i.e.}, by squaring it and then use the algebraic
solution of the 1DQHO. However, there is a special case in which $\varepsilon\!=\!\pm q_{z}$,
for which the equations may decouple. In fact, if we assume a positive
chirality ($\chi=+1$) for the Weyl node, we can have $\varepsilon=q_{z}$
which yields

\vspace{-0.7cm}
\begin{equation}
\mathcal{O}f_{2}(u)=0\to f_{2}(u)=\phi_{0}^{q_{x}}(u),\text{ with \ensuremath{f_{1}(u)=0}},
\end{equation}
which is a square normalizable solution. Instead, if $\varepsilon=-q_{z}$
is chosen, this yields

\vspace{-0.7cm}

\begin{equation}
\mathcal{O}^{\dagger}f_{1}(u)=0,
\end{equation}
which has no physically acceptable solution. Clearly, the situation
would appear reversed for $\chi=-1$ and thus, we conclude that inside
the mass-gap of normal dispersing Landau bands, there is additional
($n\!=\!0$) magnetic state whose dispersion with $k_{z}$ depends
explicitly on the topological charge of the Weyl node, \textit{i.e.,}

\vspace{-0.7cm}

\begin{equation}
E_{0}\left(k_{z}\right)=\chi\hbar v_{\text{F}}k_{z}.
\end{equation}
Note that this state is very special for it has a definite handedness!
More clearly, if $\chi=1$ ($\chi=-1$) all positive energy states
move in the positive (negative) direction of $k_{z}$. The sense of
propagation along the $z$ direction is then determined by the topological
charge of the Weyl node. In addition, if one looks at the wavefunctions,
it becomes clear that all states in this Landau band are polarized,
with only one of the spinor components being non-zero.

Picking up on all previous results, we conclude that a Weyl node of
chirality $\chi$, in the presence of an uniform magnetic field $\mathbf{B}$,
acquires a spectrum with an infinite number of Landau bands that disperse
along the field axis as follows,

\vspace{-0.7cm}

\begin{equation}
E_{n}(\mathbf{k})=\begin{cases}
\text{sign}(n)v_{\text{F}}\sqrt{2\,\hbar\,e\,B\,\abs n\!+\frac{\hbar^{2}}{B^{2}}\left(\mathbf{B}\cdot\mathbf{k}\right)^{2}} & n\neq0\\
\chi\frac{\hbar v_{\text{F}}}{B}\left(\mathbf{B}\cdot\mathbf{k}\right) & n=0
\end{cases}.
\end{equation}
These bands are represented in Fig.\,\ref{fig:MagneticLL}\,a. The
density of states (DoS) associated to these Landau bands can also
be calculated and is obviously $E\to-E$ symmetric. Therefore, we
may focus solely on the positive energy branch, which can be obtained
as follows

\vspace{-0.7cm}
\begin{align}
\rho_{\text{LL}}(E>0) & =\!\int\!\!\frac{dk}{2\pi}\,\delta\!\left(E\!-\!\chi\hbar v_{\text{F}}k\right)+\!\sum_{n=1}^{\infty}\int\!\frac{dk}{2\pi}\,\delta\!\left(E\!-\!\hbar v_{\text{F}}\sqrt{\frac{2\,e\,B\,n}{\hbar\,c}\!+\!k^{2}}\right),
\end{align}
where the integral is over the momentum parallel to the magnetic field.
These integrals can be done analytically, yielding

\vspace{-0.7cm}
\begin{equation}
\rho_{\text{LL}}\left(E\right)=\frac{1}{hv_{\text{F}}}\left(1+\sum_{n=1}^{\infty}\frac{\Theta\left(\varepsilon-\sqrt{2n}\right)+\Theta\left(\varepsilon+\sqrt{2n}\right)}{\sqrt{1-\frac{2n}{\varepsilon^{2}}}}\right),\label{eq:DoSLandau}
\end{equation}
where $\varepsilon=l_{m}E/\hbar v_{\text{F}}$ and $\Theta(x)$ is
the Heaviside function. The DoS of Eq.\,\eqref{eq:DoSLandau} is
represented in Fig.\,\ref{fig:MagneticLL}\,b, where we see two
distinctive features that are characteristic of a WSM (or a DSM) in
the regime of well-defined Landau levels: \textit{(i)} There is a
symmetric plateau in the DoS that arises from the chiral $n=0$ Landau
level, and \textit{(ii)} there is a set of sharp peaks placed at energies
$E\!=\!\pm v_{\text{F}}\sqrt{\frac{2\,\hbar\,e\,B\,n}{c}}$, with
$n\!=\!1,2,\cdots$, that are one-dimensional van Hove singularities
at edges of the non-chiral Landau bands. These results will be useful
to contextualize some results in Chapter\,\ref{chap:Vacancies}.

\vspace{-0.4cm}

\paragraph{Chiral Transport and the Lowest Landau Levels:}

A second topological consequence of the presence of a Weyl point near
the Fermi energy is the \textit{Chiral Anomaly}. This effect is well-known
in high-energy theory and corresponds to a spontaneously broken chiral
symmetry in \textit{Quantum Electrodynamics }(QED)\,\cite{Adler1969,Bell1969}.
In practice, what this anomaly does in a condensed matter context
is to allow a pumping of emergent Weyl fermions from one Weyl node
to a partner of opposite chirality, by the action of collinear magnetic
and electric fields. Depending on whether the two fields are parallel
or anti-parallel, the pumping of charge carrier can be done in one
or the opposite direction. To be more precise, we can assign a charge
density $\rho_{i}$ to each (slightly doped) Weyl node, which can
be positive or negative depending on weather the node is dominated
by electron or hole excitations. Then, in the presence of simultaneous
electric and magnetic fields, one finds that,

\vspace{-0.7cm}

\begin{equation}
\frac{d\rho_{i}}{dt}=\chi_{i}\frac{e^{2}}{4\pi^{2}\hbar^{2}}\mathbf{E}\cdot\mathbf{B},
\end{equation}
where $\chi_{i}$ defines the topological charge of the $i^{\text{th}}$
Weyl node in momentum space, while $\mathbf{E}$/$\mathbf{B}$ are
external electric/magnetic fields. The global charge, $\sum_{i}\rho_{i}(t)$,
is obviously conserved because the topological charges are precisely
compensated ($\sum_{i}\chi_{i}\!=\!0$) in any lattice realization
of a Weyl semimetal.

\vspace{-0.4cm}

\subsection{Transport Signatures of Weyl Physics}

The peculiar Landau quantization that happens near a Weyl node in
the presence of external magnetic field effectively realizes the celebrated
chiral anomaly found in quantum electrodynamics by Adler\,\cite{Adler1969},
Bell and Jackiw\,\cite{Bell1969}. In principle, the peculiar magnetic
quantized spectrum of a WSM will yield experimental consequences,
even though these may only become evident at very extreme conditions.
As usual, temperature must be very low so that the system has a sharp
Fermi surface. In addition to this, the number of occupied (unoccupied)
Landau bands above (below) the nodal energy must also be kept small,
which guarantees that the spacing between the last occupied and first
unoccupied Landau bands $\delta E_{B}$ obeys $\delta E_{B}\apprge\abs{E_{F}}$.
Since the $n\neq0$ Landau bands of an isolated Weyl node have the
dispersion relation

\vspace{-0.7cm}

\begin{equation}
E_{n}(\mathbf{k},\mathbf{B})=\text{sign}(n)\,v_{\text{F}}\,\sqrt{2\,\hbar\,e\,B\,\abs n\!+\frac{\hbar^{2}}{B^{2}}\left(\mathbf{B}\cdot\mathbf{k}\right)^{2}},
\end{equation}
we conclude that $\delta E_{B}\!\approx\!v_{\text{F}}\sqrt{\hbar\,e\,B/2\abs n}$
if $n\gg1$, which implies an external magnetic,

\vspace{-0.7cm}
\begin{equation}
B\apprge\frac{\hbar}{2e\abs n}\left(\!\frac{E_{F}}{\hbar v_{\text{F}}}\!\right)^{2}
\end{equation}
in order for Landau quantization to be relevant. The Fermi velocity
of known Dirac-Weyl semimetals is of the order $10^{5-6}\text{m/s}$\,\cite{Dolui2014},
while $E_{F}$ can realistically be placed within the $10\text{meV}$
range (\textit{e.g.}, NbAs has a Fermi level placed $25\text{meV}$
below the band-touching points\,\cite{Naumann2021}). This leads
to a rough estimate of

\vspace{-0.7cm}
\begin{equation}
B\gtrsim\frac{\hbar}{2e}\left(\!\frac{E_{F}}{\hbar v_{\text{F}}}\!\right)^{2}\!\!\approx\!30T.
\end{equation}
Therefore, one can assume that most magnetic response experiments
are done in the semiclassical regime, in which Landau quantization
is irrelevant by definition. In this regime, magneto-transport can
be studied by linear-response theory of the Bloch electrons affected
by applied electric and magnetic fields, through the formalism introduced
in Sect.\,\ref{subsec:Dynamics-of-Bloch}. From the works of Stephanov
and Yin\,\cite{Stephanov2012}, Kim \textit{et al}.\,\cite{Kim2013,Kim2014},
Son and Yamamoto\,\cite{Son2012aa}, and Son and Spivak\,\cite{Son2013},
it has been established that signatures of the chiral anomaly will
remain in the bulk magneto-transport\,\footnote{It is worth pointing out that quantum anomalies realized in Weyl semimetals
yield a rich array of consequences in other cross-transport effects,
in the presence of simultaneous magnetic fields, thermal gradients\,\cite{Chernodub2014,Chernodub2018,Chernodub2022}
and/or elastic strain\,\cite{Cortijo2015,Cortijo2016,Grushin2016,Arjona2018},
as well as in the linear optical response\,\cite{Ashby2014,deJuan2017,Levy2020}.
For conciseness, we will not discuss those effects in this work.} independently of a precise Landau quantization. In fact, these chiral
anomaly effects can be seen from semiclassical transport calculations
based on the Boltzmann Equation (BE) with dynamics that properly include
all Berry curvature effects. In order to see how this comes about,
we start from the semiclassical dynamics equations {[}Eqs.\,\eqref{eq:drdt-1-2}
and \eqref{eq:dkdt-1-1}{]} in the presence of external electromagnetic
fields and a Berry curvature field:

\vspace{-0.7cm}

\begin{subequations}
\begin{align}
\dot{\mathbf{r}}_{s} & =\frac{1}{\Delta_{s\mathbf{k}}}\left[\mathbf{v}_{s\mathbf{k}}\!-\frac{e}{\hbar}\mathbf{E}\!\times\mathbf{\Omega}_{s\mathbf{k}}-\frac{e}{\hbar}\left(\mathbf{\Omega}_{s\mathbf{k}}\cdot\mathbf{v}_{s\mathbf{k}}\right)\mathbf{B}\right]\label{eq:drdt-1-1-1}\\
\dot{\mathbf{k}}_{s} & =\frac{1}{\Delta_{s\mathbf{k}}}\left[-\frac{e}{\hbar}\mathbf{E}+\frac{e}{\hbar}\mathbf{B}\times\!\mathbf{v}_{s\mathbf{k}}+\frac{e^{2}}{\hbar^{2}}\left(\mathbf{E}\cdot\mathbf{B}\right)\mathbf{\Omega}_{s\mathbf{k}}\right],\label{eq:dkdt-1-1-1}
\end{align}
\end{subequations}

where $s=\pm1$ labels the band, $\mathbf{v}_{s\mathbf{k}}$ is the
wave-packet velocity (including the orbital magnetic moment contribution),
and $\Delta_{s\mathbf{k}}\!=\!1+\frac{e}{\hbar}\left(\mathbf{\Omega}_{s\mathbf{k}}\cdot\mathbf{B}\right)$
is the invariant measure of phase-space. Even though $\mathbf{E}$
and $\mathbf{B}$ are generic time- and position-dependent fields,
for shortness, we suppress this dependence in all upcoming equations.
At the same time, the mean density of particles in phase-space is
given as $n_{s\mathbf{k}\mathbf{r}}(t)=\Delta_{s\mathbf{k}}f_{s\mathbf{k}\mathbf{r}}(t)/8\pi^{3}$,
in terms of the invariant measure and the probability density in phase-space
$f_{s\mathbf{k}\mathbf{r}}(t)$. Following Stephanov and Yin\,\cite{Stephanov2012},
we can easily show that

\vspace{-0.7cm}

\begin{subequations}
\begin{align}
\,\boldsymbol{\nabla}_{\mathbf{r}}\cdot\left(\Delta_{s\mathbf{k}\mathbf{r}}\dot{\mathbf{r}}_{s}\right) & =-\frac{e}{\hbar}\partial_{t}\mathbf{B}\cdot\mathbf{\Omega}_{s\mathbf{k}}\\
\!\boldsymbol{\nabla}_{\mathbf{k}}\cdot\left(\Delta_{s\mathbf{k}}\dot{\mathbf{k}}_{s}\right) & =\frac{e^{2}}{\hbar^{2}}\left(\mathbf{E}\cdot\mathbf{B}\right)\left(\boldsymbol{\nabla}_{\mathbf{k}}\cdot\mathbf{\Omega}_{s\mathbf{k}}\right),
\end{align}
\end{subequations}

and, consequently,

\vspace{-0.7cm}
\begin{equation}
\partial_{t}\Delta_{s\mathbf{k}}+\boldsymbol{\nabla}_{\mathbf{r}}\cdot\left(\Delta_{s\mathbf{k}\mathbf{r}}\dot{\mathbf{r}}_{s}\right)+\boldsymbol{\nabla}_{\mathbf{k}}\cdot\left(\Delta_{s\mathbf{k}}\dot{\mathbf{k}}_{s}\right)=\frac{e^{2}}{\hbar^{2}}\left(\mathbf{E}\cdot\mathbf{B}\right)\left(\boldsymbol{\nabla}_{\mathbf{k}}\cdot\mathbf{\Omega}_{s\mathbf{k}}\right).\label{eq:Semiclassical}
\end{equation}
Equation\,\eqref{eq:Semiclassical} may be easily adapted to describe
the dynamics of the mean density of particles in phase-space through
the replacement, $\Delta_{s\mathbf{k}}\!\to\!\Delta_{s\mathbf{k}}f_{s\mathbf{k}\mathbf{r}}\left(t\right)$.
This way, we arrive at

\vspace{-0.7cm}
\begin{align}
\partial_{t}\left(\Delta_{s\mathbf{k}}f_{s\mathbf{k}\mathbf{r}}(t)\right)+ & \boldsymbol{\nabla}_{\mathbf{r}}\!\cdot\!\left(\Delta_{s\mathbf{k}\mathbf{r}}\dot{\mathbf{r}}_{s}f_{s\mathbf{k}\mathbf{r}}(t)\right)+\boldsymbol{\nabla}_{\mathbf{k}}\!\cdot\!\left(\Delta_{s\mathbf{k}}\dot{\mathbf{k}}_{s}f_{s\mathbf{k}\mathbf{r}}(t)\right)=\label{eq:BE}\\
 & f_{\eta\mathbf{k}\mathbf{r}}(t)\left[\partial_{t}\Delta_{s\mathbf{k}}+\boldsymbol{\nabla}_{\mathbf{r}}\cdot\left(\Delta_{s\mathbf{k}\mathbf{r}}\dot{\mathbf{r}}_{s}\right)+\boldsymbol{\nabla}_{\mathbf{k}}\cdot\left(\Delta_{s\mathbf{k}}\dot{\mathbf{k}}_{s}\right)\right]+\nonumber \\
 & \qquad\Delta_{s\mathbf{k}\mathbf{r}}\left[\partial_{t}f_{s\mathbf{k}\mathbf{r}}(t)+\dot{\mathbf{r}}_{s}\cdot\boldsymbol{\nabla}_{\mathbf{r}}f_{s\mathbf{k}\mathbf{r}}(t)+\dot{\mathbf{k}}_{s}\cdot\boldsymbol{\nabla}_{\mathbf{k}}f_{s\mathbf{k}\mathbf{r}}(t)\right]\nonumber 
\end{align}
In the absence of scattering, the second term in Eq.\,\eqref{eq:BE}
is the usual BE (in the absence of collision integrals) which yields
zero and, hence,

\vspace{-0.7cm}

\begin{align}
\partial_{t}n_{s\mathbf{k}\mathbf{r}}(t)+\boldsymbol{\nabla}_{\mathbf{r}}\cdot\left(\dot{\mathbf{r}}_{s}n_{s\mathbf{k}\mathbf{r}}(t)\right)+\boldsymbol{\nabla}_{\mathbf{k}}\cdot\left(\dot{\mathbf{k}}_{s}n_{s\mathbf{k}\mathbf{r}}(t)\right)=\qquad\qquad\label{eq:Continuity1}\\
\frac{e^{2}}{8\pi^{3}\hbar^{2}}\left(\mathbf{E}\cdot\mathbf{B}\right)\left(\boldsymbol{\nabla}_{\mathbf{k}}\cdot\mathbf{\Omega}_{s\mathbf{k}}\right)f_{s\mathbf{k}\mathbf{r}}(t).\nonumber 
\end{align}
For this results, we have already used that $\boldsymbol{\nabla}_{\mathbf{k}}\!\cdot\!\mathbf{\Omega}_{s\mathbf{k}}\!=\!2\pi\chi s\delta^{{\scriptscriptstyle (3)}}\left(\mathbf{k}\right)$
for a single Weyl node. At last, we can integrate Eq.\,\eqref{eq:Continuity1}
in $\mathbf{k}$ and sum over the bands, which leads to

\vspace{-0.7cm}

\begin{equation}
\partial_{t}\rho\left(\mathbf{r},t\right)\!+\!\boldsymbol{\nabla}_{\mathbf{r}}\!\cdot\!\mathbf{J}\left(\mathbf{r},t\right)\!=\!-\frac{e^{3}}{8\pi^{3}\hbar^{2}}\left(\mathbf{E}\cdot\mathbf{B}\right)\int\!d^{{\scriptscriptstyle (3)}}\mathbf{k}\sum_{s}\left(\boldsymbol{\nabla}_{\mathbf{k}}\!\cdot\!\mathbf{\Omega}_{\mathbf{k}s}\right)f_{s\mathbf{k}\mathbf{r}}(t).\label{eq:Continuity1-1}
\end{equation}
Usually, even in topological insulators, $\boldsymbol{\nabla}_{\mathbf{k}}\cdot\mathbf{\Omega}_{s\mathbf{k}}\!=\!0$
and therefore the total charge would be locally conserved. However,
we have already seen that Weyl nodes work as monopoles of the Berry
curvature field in $\mathbf{k}$-space. This means that $\boldsymbol{\nabla}_{\mathbf{k}}\cdot\mathbf{\Omega}_{s\mathbf{k}}\!=\!-2\pi\chi s\delta^{{\scriptscriptstyle (3)}}\!\left(\mathbf{k}\right)$
and, consequently,

\vspace{-0.7cm}

\begin{equation}
\partial_{t}\rho\left(\mathbf{r},t\right)+\boldsymbol{\nabla}_{\mathbf{r}}\cdot\mathbf{J}\left(\mathbf{r},t\right)=\frac{\chi e^{3}}{4\pi^{2}\hbar^{2}}\left(\mathbf{E}\cdot\mathbf{B}\right)\left[f_{+\boldsymbol{0}\mathbf{r}}\left(t\right)-f_{-\boldsymbol{0}\mathbf{r}}\left(t\right)\right].
\end{equation}
This result is equivalent to the following modified continuity equation
for the electric charge around the Weyl node:

\vspace{-0.7cm}
\begin{equation}
\partial_{t}\rho(\mathbf{r},t)+\boldsymbol{\nabla}_{\mathbf{r}}\cdot\mathbf{J}(\mathbf{r},t)=\frac{E_{F}}{\abs{E_{F}}}\frac{\chi e^{3}}{4\pi^{2}\hbar^{2}}\left(\mathbf{E}\cdot\mathbf{B}\right),\label{eq:Continuity}
\end{equation}
where $E_{F}$ is the chemical potential. Unsurprisingly, the total
charge for excitations around an isolated Weyl node is not a conserved
quantity if $\mathbf{E}\cdot\mathbf{B}\neq0$ since the continuity
equation features an anomalous source-term. In the following, we will
make use Eqs.\,\ref{eq:Continuity1}-\ref{eq:Continuity} to derive
the existence of a \textit{chiral magnetic effect} (CME), a \textit{negative
longitudinal magnetoresistance} (NLMR), and a \textit{planar quantum
Hall effect} (PQHE) in Weyl systems.

\vspace{-0.5cm}

\paragraph{Chiral Magnetic Effect:}

The most outstanding effect of the chiral anomaly in transport is,
perhaps, the \textit{chiral magnetic effect} (CME). As first shown
by Vilenkin\,\cite{Vilenkin80}, the chiral anomaly of QED allows
Weyl fermions to have a dissipassionless flow and without any driving
electric field. More precisely, this current is driven by a magnetic
field $\mathbf{B}$ and takes the form,

\vspace{-0.7cm}
\begin{equation}
\mathbf{J}_{{\scriptscriptstyle \text{CME}}}\!\!=\!-\frac{e^{2}}{4\pi^{2}\hbar^{2}}\chi E_{F}\,\mathbf{B},\label{eq:CME_Current}
\end{equation}

where $\chi=\pm1$ is the chirality of the Weyl fermions, and $E_{F}$
the Fermi energy measured with respect to the Weyl node. In a lattice
WSM, there will always be more than one Weyl node and, provided inter-node
scattering is negligible, the total CME current is simply,

\vspace{-0.7cm}
\begin{equation}
\mathbf{J}_{{\scriptscriptstyle \text{CME}}}\!\!=\!-\frac{e^{2}}{4\pi^{2}\hbar^{2}}\sum_{i=1}^{N_{v}}\chi_{i}\mu_{i}\mathbf{B},\label{eq:CME_Current-1}
\end{equation}
which sums up to zero, if $\mu_{i}=E_{F}$ for all Weyl nodes. This
implied the absence of a CME in any equilibrium situation with a common
chemical potential on all Weyl nodes. Thereby, the inevitable presence
of inter-node scattering would seem to render this effect unobservable
in any real system, as chiral unbalances would quickly relax to equilibrium.
Notwithstanding, there are two ways to go around this issue: \textit{(i)}
to create a temporary non-equilibrium state by optical excitation
in the THz-range\,\cite{Levy2020}, controlled non-local transport
in multi-terminal nano-devices\,\cite{Parameswaran2014}, or non-linear
transport effects\,\cite{Morimoto2016,Nandy2020}, or \textit{(ii)}
to use the chiral anomaly itself as \textit{``charge carrier pump''}
between nodes of opposite chirality\,\cite{Nielsen83,Son2013,Li2016b}.
As we will see shortly, the latter leads to a negative correction
on the steady-state magnetoresistance, in the presence of a longitudinal
magnetic field. 

For completeness, we present here a semiclassical derivation of the
CME, where we closely follow the treatment of Stephanov and Yin\,\cite{Stephanov2012},
that employs the same BE formalism used to obtain the charge continuity
equation {[}Eq.\,\eqref{eq:Continuity}{]} including the anomalous
$\mathbf{E}\cdot\mathbf{B}$ source term. For this purpose, it is
enough to use Eq.\,\eqref{eq:dkdt-1-1} to explicitly write the electrical
current density (per unit volume),

\vspace{-0.7cm}
\begin{align}
\mathbf{J}(\mathbf{r},t)\! & =\!-\frac{e}{8\pi^{3}}\int\!d^{{\scriptscriptstyle (3)}}\mathbf{k}\,\sum_{s}\Delta_{s\mathbf{k}}\dot{\mathbf{r}}_{s}f_{s\mathbf{k},\mathbf{r}}(t)=-\frac{e}{8\pi^{3}}\int\!d^{{\scriptscriptstyle (3)}}\mathbf{k}\,\sum_{s}\mathbf{v}_{s\mathbf{k}}f_{s\mathbf{k},\mathbf{r}}(t)\label{eq:CurrentCME}\\
 & \,\,+\frac{e^{2}}{8\pi^{3}\hbar}\mathbf{E}\!\times\!\int\!d^{{\scriptscriptstyle (3)}}\mathbf{k}\,\sum_{s}\mathbf{\Omega}_{\mathbf{k}s}f_{s\mathbf{k},\mathbf{r}}(t)+\frac{e^{2}}{8\pi^{3}\hbar}\left[\int\!d^{{\scriptscriptstyle (3)}}\mathbf{k}\sum_{s}\left(\mathbf{\Omega}_{\mathbf{k}s}\!\cdot\!\mathbf{v}_{s\mathbf{k}}\right)f_{s\mathbf{k},\mathbf{r}}(t)\right]\mathbf{B}.\nonumber 
\end{align}
If all fields are homogeneous and the single-Weyl node is isotropic,
with chirality $\chi$, and is in equilibrium with a chemical potential
$\mu$, then

\vspace{-0.7cm}
\begin{equation}
f_{\mathbf{k},\mathbf{r}}(t)\equiv f_{{\scriptscriptstyle \text{FD}}}\!\left(\hbar v_{\text{F}}\abs{\mathbf{k}}\!-\!\mu\right)=\left[1+\exp\left(\left[\hbar v_{\text{F}}\abs{\mathbf{k}}\!-\!\mu\right]/k_{\text{B}}T\right)\right]^{-1}
\end{equation}
which automatically renders the first two integrals in Eq.\,\eqref{eq:CurrentCME}
zero by $\mathbf{k}\to-\mathbf{k}$ anti-symmetry. In contrast, the
last integral is symmetric with $\mathbf{\Omega}_{\mathbf{k}s}=-\nicefrac{1}{2}\chi s\mathbf{k}/k^{3}$
and $\mathbf{v}_{s\mathbf{k}}=sv_{\text{F}}\mathbf{k}/k+\mathcal{O}(B)$\,\footnote{Here, we are ignoring the (linear in $\mathbf{B}$) contribution to
the velocity due to the intrinsic orbital angular momentum of the
self-rotating Bloch wave-packets.}. Therefore, we arrive at

\vspace{-0.7cm}
\begin{equation}
\mathbf{J}_{\text{eq}}\!=\!-\chi\frac{e^{2}}{4\pi^{2}\hbar^{2}}\left[\int_{0}^{\infty}\!\!dx\,f_{{\scriptscriptstyle \text{FD}}}\!\left(x\!-\!\mu\right)\right]\mathbf{B}\approx\underset{\mathbf{J}_{\text{CME}}}{\underbrace{-\chi\frac{e^{2}}{4\pi^{2}\hbar^{2}}\mu\mathbf{B}}}+\mathcal{O}\left[k_{\text{B}}T\right]\label{eq:CME}
\end{equation}

\vspace{-0.4cm}

where we have used that $s^{2}\!=\!1$. This result describes a steady-state
dissipationless current ($\mathbf{J}\cdot\mathbf{E}\!=\!0$) that
appears along the applied magnetic field.

\vspace{-0.5cm}

\paragraph{Anomalous Magneto-Transport Effects:}

Emergent fermions in WSMs can definitely drift along an applied magnetic
field but, in practice, this can only happen in non-equilibrium situations
with unbalanced chiralities. Such cases are often short-lived transients
which only allow a true CME to be observed at very short time-scales.
However, we have also referred that the chiral anomaly (caused by
non-orthogonal $\mathbf{E}$ and $\mathbf{B}$ fields) may be used
to \textit{``pump''} charge density between different Weyl nodes,
which effectively leads to a \textit{dynamically-induced chiral unbalance}
that activates the CME. In any real system, this pumping process is
compensated by inter-node scattering that attempts to restore equilibrium,
thus allowing for a dynamical steady-state to be established. This
is the physical mechanism behind the two most well-known transport
effects expected to arise due to the chiral anomaly in Weyl semimetals:
the \textit{negative longitudinal magnetoresistance} (NLMR) and the
\textit{planar Hall effect} (PHE).

Here, we will give a straightforward derivation\,\cite{Nielsen83}
of the aforementioned effects using a simple version of the semiclassical
transport formalism introduced in the beginning of this section {[}Eq.\,\eqref{eq:Continuity}{]}.
Despite its simplicity, this derivation is able to yield qualitatively
correct results, as confirmed by the more rigorous derivations presented,
at different levels of approximation\,\footnote{Using semiclassical as well as full quantum transport calculations
based on the Kubo formula.}, in the works of Son and and Yamamoto\,\cite{Son2012aa}, Son and
Spivak\,\cite{Son2013}, Spivak and Andreev\,\cite{Spivak2016},
Aji\,\cite{Aji2012}, Kim \textit{et al}.\,\cite{Kim2014}, Burkov\,\cite{Burkov2014,Burkov2015,Burkov2017},
and Nandy \textit{et al}.\,\cite{Nandy2017}, among others. For simplicity,
we assume the system to have a single pair of compensated bulk Weyl
nodes, which are connected by inter-node scattering processes which
are treated within the relaxation time approximation ($\tau$)\,\footnote{We assume that the intra-node scattering time is much shorter and,
therefore, each node is in an instantaneous equilibrium state with
its own chemical potential.}. The electronic system is then characterized by two time-dependent
charge densities, $\rho_{+}(t)$ and $\rho_{-}(t)$, which are associated
to chiralities $\chi\!=\!\pm1$ and, which we assume homogeneous in
space. Both charge densities evolve according to\,\footnote{There is a small subtlety here. The term $\mu/\abs{\mu}$ disappeared
for two reasons: i) If $\mu>0$ then it is simply $+1$ and ii) if
$\mu<0$ it is $-1$ but the quasiparticles emerging around each node
must be seen as holes, instead of electrons. }

\vspace{-0.7cm}
\begin{align}
\frac{d}{dt}\rho_{\pm}(t)= & \pm\frac{e^{3}}{4\pi^{2}\hbar^{2}}\left(\mathbf{E}\cdot\mathbf{B}\right)-\frac{1}{\tau}\left(\rho_{\pm}(t)-\nicefrac{1}{2}\rho_{\text{eq}}\right),\label{eq:NielsenChiral}
\end{align}
where the second term accounts for inter-node scattering that interconverts
chiralities, and $\rho_{\text{eq}}$ is the \textit{global equilibrium
density} which features a common chemical potential ($\mu$) in both
valleys. In a steady-state, the chiral anomalous source-term precisely
balances the scattering term, and the pair of coupled equations has
the non-trivial and chiral-asymmetric solution,

\vspace{-0.7cm}
\begin{equation}
\rho_{\pm}^{0}=\frac{\rho_{\text{eq}}}{2}\left(1\pm\frac{\tau e^{3}}{2\pi^{2}\hbar^{2}\rho_{\text{eq}}}\left(\mathbf{E}\cdot\mathbf{B}\right)\right).
\end{equation}
More conveniently, we can follow Fukushima \textit{et al}.\,\cite{Fukushima2008,Yang2020}
and define a chiral charge density, 

\vspace{-0.8cm}

\begin{equation}
\overline{\rho}=\rho_{+}^{0}-\rho_{-}^{0}=\frac{\tau e^{3}}{2\pi^{2}\hbar^{2}}\left(\mathbf{E}\cdot\mathbf{B}\right),
\end{equation}
which naturally gives rise to a non-zero chiral chemical potential,
$\overline{\mu}=\mu_{+}\!-\!\mu_{-}$, which measures the chemical
potential difference between the two Weyl nodes. Following the derivation
of Fukushima \textit{et al}.\,\cite{Fukushima2008,Yang2020}, this
chiral chemical potential is related to the chiral charge density
as,

\vspace{-0.7cm}
\begin{equation}
\overline{\rho}=-\frac{e\mu^{2}\overline{\mu}}{3\pi^{2}\hbar^{3}v_{\text{F}}^{3}}\left[1+\left(\frac{\overline{\mu}}{\mu}\right)^{2}\right],
\end{equation}
in the limit of zero temperature (see also Ref.\,\cite{Yang2020}).
For weak applied fields but significant natural doping ($\overline{\mu}\ll\mu$),
one can approximate $\overline{\mu}\approx-3\pi^{2}\overline{\rho}\hbar^{3}v_{\text{F}}^{3}/e\mu^{2}$,
which yields the following CME current:

\vspace{-0.7cm}
\begin{equation}
\mathbf{J}_{{\scriptscriptstyle \text{CME}}}\!\!=-\frac{e^{2}}{h^{2}}\,\overline{\mu}\,\mathbf{B}=\!\frac{3e\hbar v_{\text{F}}^{3}}{16\pi^{2}\mu^{2}}\,\overline{\rho}\,\mathbf{B}=\varsigma_{\mu}\mathbf{B}\left(\mathbf{E}\cdot\mathbf{B}\right),\label{eq:CME_Current-1-1}
\end{equation}
where $\varsigma_{\mu}\!=\!3\tau e^{4}v_{\text{F}}^{3}/\left(32\pi^{4}\hbar\mu^{2}\right)$
is a parameter inversely proportional to the squared chemical potential.
Note that Eq.\,\eqref{eq:CME_Current-1-1} yields a linear response
equation in $\mathbf{E}$, which defining a bulk dc-conductivity conductivity
tensor of the following form\,\footnote{With no loss of generality, we take $\mathbf{E}$ and $\mathbf{B}$
to lie in the plane $x-y$.}:

\vspace{-0.7cm}
\begin{equation}
\mathbf{J}\approx\boldsymbol{\Delta\sigma}_{B}\mathbf{E}\Rightarrow\boldsymbol{\Delta}\!\boldsymbol{\sigma}_{B}=\varsigma_{\mu}\left(\begin{array}{cc}
B_{x}^{2} & B_{x}B_{y}\\
B_{x}B_{y} & B_{y}^{2}
\end{array}\right).\label{eq:Magnetoconductivity}
\end{equation}

Obviously, $\sigma_{B}$ does not describe the whole conductivity
of the system as it completely neglects intra-valley processes. The
latter give rise to a usual isotropic Drude conductivity, $\sigma_{0}$,
which we assume additive to the previous effect. Hence the total linear
dc-conductivity tensor, in the presence of a magnetic field $\mathbf{B}$
takes the form, 

\vspace{-0.7cm}
\begin{equation}
\boldsymbol{\sigma}_{B}\!=\!\boldsymbol{\sigma}_{0}+\boldsymbol{\Delta\sigma}_{B}=\left(\begin{array}{cc}
\sigma_{0}+\varsigma_{\mu}B_{x}^{2} & \varsigma_{\mu}B_{x}B_{y}\\
\varsigma_{\mu}B_{x}B_{y} & \sigma_{0}+\varsigma_{\mu}B_{y}^{2}
\end{array}\right).\label{eq:MagnetoCond}
\end{equation}
Equation\,\eqref{eq:MagnetoCond} already describes two crossed effects
that arise from the simultaneous application of electric and magnetic
fields to a Weyl system. If $\mathbf{B}\parallel\mathbf{E}$, the
anomalous conductivity tensor is diagonal and isotropic, with the
chiral anomaly-driven particle pumping between different nodes leading
to a \textit{positive correction to the magneto-conductivity} that
goes as $B^{2}$. This is not the expected behavior for ordinary 3D
non-magnetic metals, in which a magnetic field generally leads to
a positive correction to the longitudinal resistance {[}\textit{e.g.},
see Sondheimer and Wilson\,\cite{Sondheimer47} or chapter 5 of Abrikosov's
book\,\cite{Abrikosov88}{]}. Instead, if the electric and magnetic
fields have an angle $\gamma$ between them, the positive magneto-conductivity
is reduced by a factor of $\cos^{2}\gamma$ and a new off-diagonal
contribution appears, which is proportional to $B^{2}\sin2\gamma$.
This is the so-called \textit{planar Hall effect} (PHE)\,\cite{Burkov2017,Nandy2017}
which (despite its name) is not a true Hall effect, as it does not
entail an antissymetric component of the conductivity tensor. Rather,
it describes an anisotropic magneto-resistance effect which is driven
by an imperfect (non-collinear) chiral anomaly effect. For completeness,
it is relevant to re-write the conductivity tensor of Eq.\,\eqref{eq:MagnetoCond}
as a resistivity tensor, which reads,

\vspace{-0.7cm}
\begin{equation}
\rho(B,\gamma)=\rho_{\perp}\left[\begin{array}{cc}
1 & 0\\
0 & 1
\end{array}\right]\!-\!\left(\rho_{\parallel}\!-\!\rho_{\perp}\right)\left[\!\begin{array}{cc}
\cos^{2}\gamma & 2\sin2\gamma\\
2\sin2\gamma & \cos^{2}\gamma
\end{array}\!\right],\label{eq:ChiralTransport-1}
\end{equation}
with $\rho_{\perp}$ ($\rho_{\parallel}$) being the longitudinal
resistivity for $\mathbf{B}\perp\mathbf{E}$ ($\mathbf{B}\!\parallel\!\mathbf{E}$).
Note that Eq.\,\eqref{eq:ChiralTransport-1} is the usual way\,\cite{Lv21}
of presenting the effect of NLMR and PHE in experimental studies of
transport in Weyl semimetals. 

The presence of a negative magneto-resistance was initially thought
of as an unambiguous experimental sign of emergent Weyl physics in
a three-dimensional crystal. However, this statement was soon questioned
by more detailed studies, mainly due to two main reasons. Firstly,
it was quickly recognized that extrinsic effects (\textit{jetting
current effects}\,\cite{Yoshida75}) can lead to \textit{``false
positives''} in the measurement of the magneto-resistivity\,\cite{Reis2016,Parameswaran2014},
which can appear to be positive even in normal metals. Secondly, a
more detailed account of the semiclassical magnetic dynamics of Bloch
states unveiled a complex interplay between \textit{(i)} the intrinsic
magnetic moment of the Bloch states and the external magnetic field\,\cite{Knoll2020},
and \textit{(ii)} of intra- and inter-node scattering processes\,\cite{Sharma2020}.
In some circumstances, both effects are able to turn $\sigma_{B}$
negative even with a chiral anomaly in the system. It is worth remarking
that similar problems are known to plague the assessment of Weyl physics
through the PHE, as well.

In spite of its somewhat ambiguous theoretical foundation, a NLMR
has been experimentally observed in different Weyl and Dirac semimetals,
including $\text{Ta}\text{As}$\,\,\cite{Huang2015}, $\text{Nb}\text{As}$,
$\text{Nb}\text{P}$\,\cite{Li2017}, $\text{Cd}_{3}\text{As}_{2}$\,\cite{Li2016a},
$\text{Na}_{3}\text{Bi}$\,\cite{Xiong2015,Liang2018}, $\text{Gd}\,\text{Pt}\,\text{Bi}$\,\,\cite{Hirschberger2016,Liang2018},
$\text{Zr}\text{Te}_{5}$\,\cite{Li2016b}, at a gap-closing phase
transition in the 3D topological insulator, $\text{Bi}_{1-x}\text{Sb}_{x}$\,\cite{Kim2013,Shin2017}.
Even though not as well studied as the NLMR effect, the recent observation
of a (possibly) chiral-anomaly-driven PHE was also reported in Weyl
semimetals\,\cite{Yang2019,Shama2020,Zhang2020}.

\vspace{-0.3cm}

\subsection{Fermi-Arc States: Dirac Strings in $\mathbf{k}$\textendash \,Space}

The topological nature of a gapped electronic phase is best demonstrated
at an exposed boundary, where in-gap surface-localized states\,\cite{Isaev2011}
(or edge-states\,\cite{Hatsugai93} in two-dimensions) can propagate
in a very robust way\,\cite{Hasan2010}. Even though these states
are confined to the surface, their existence depends crucially on
the topological properties of the bulk, which are characterized by
robust and precisely quantized invariants. This \textit{bulk-to-boundary
correspondence} of topological insulators is, in fact, a more general
result that translates into gapless topological phases (such as Weyl
semimetals) as well. In these systems, the surface states became known
as \textit{Fermi Arcs}\,\cite{Wan2011,Balents2011,Xu2011,Okugawa2014,Haldane2014,Vafek2014,Witten2015}
and are analogous to the celebrated \textit{Dirac strings}\,\cite{Dirac31,Dirac48},
an inevitable consequence of modifying Maxwell's equations to allow
the existence of magnetic monopoles.

\begin{figure}[t]
\vspace{-0.5cm}
\begin{centering}
\includegraphics[scale=0.23]{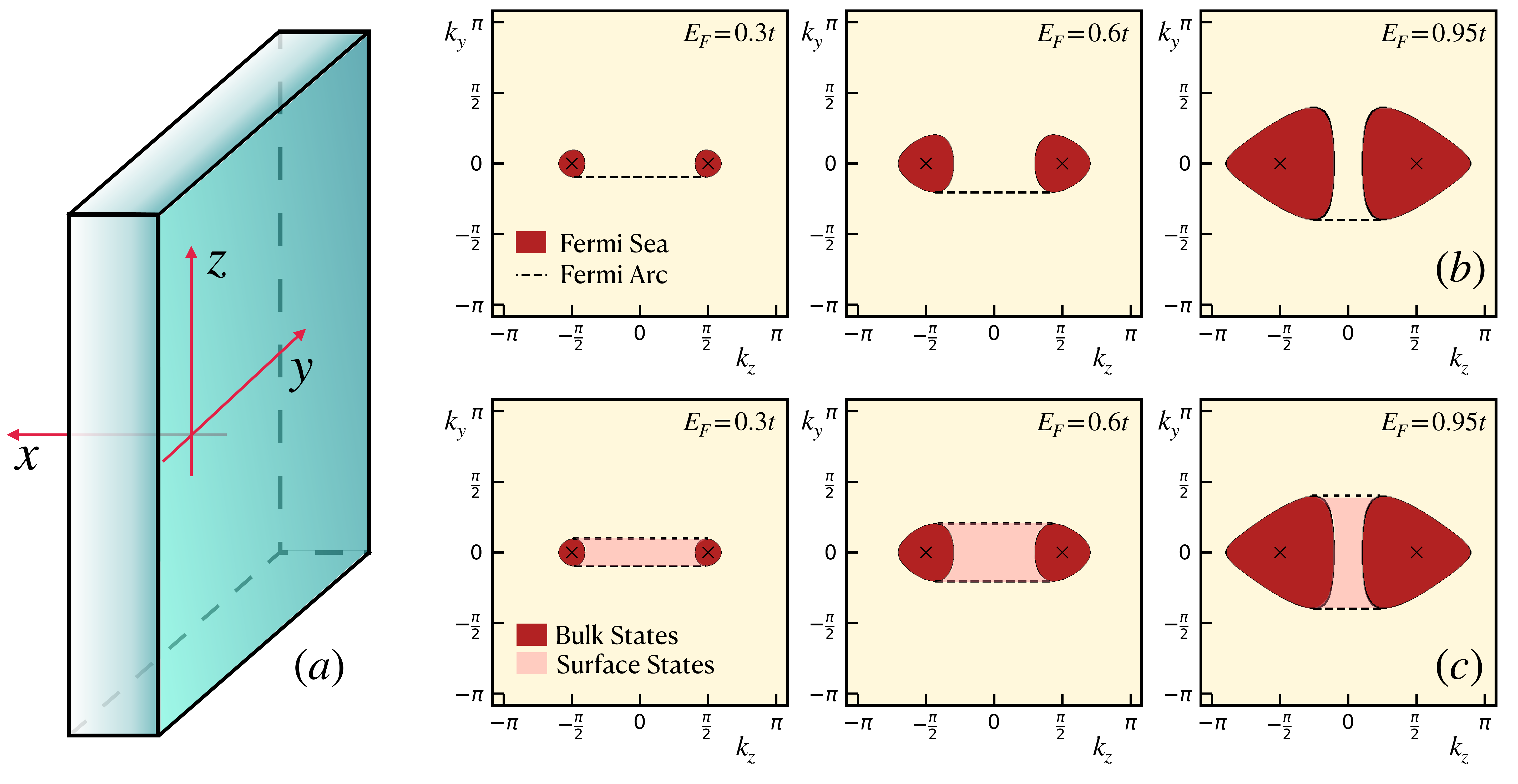}
\par\end{centering}
\vspace{-0.1cm}

\caption{\label{fig:FermiArcs}(a) Cartoon representing a Fermi Arc state in
propagating in the exposed surface of a WSM slab. (b) Surface fBz
of the semi-infinite WSM model with a single boundary. The surface-projected
Fermi sea is shown for three Fermi energies, together with the points
where Fermi arc surface states exists. (c) Projected Fermi sea for
a thick WSM slab, having two uncoupled boundaries.}

\vspace{-0.4cm}
\end{figure}

Instead of moving right into general statements, we begin by analyzing
the appearance of surface states in the specific model of a simple
cubic Weyl semimetal that features only two Weyl points separated
along the $k_{z}$-axis. Thereby, our starting point will be the Bloch
Hamiltonian\,\cite{Armitage18},

\vspace{-0.7cm}

\begin{align}
\mathcal{H}_{m}(\mathbf{k}) & \!=\!t\left(2-\cos k_{x}-\cos k_{y}-\cos k_{z}\right)\sigma_{z}+t\sin k_{x}\sigma_{x}+t\sin k_{y}\sigma_{y},\label{eq:Model_2Weyl}
\end{align}
which has zero-energy Weyl points located at $\mathbf{k}^{*}\!=\!\left(0,0,\pm\nicefrac{\pi}{2}\right)$.
In order to analyze the states at a sharp boundary, we reconsider
this model in a lattice that is \textit{semi-infinite }in the positive
$x$-direction, whilst retaining full translation invariance in both
$y$- and $z$-\,directions (see Fig.\,\ref{fig:FermiArcs}a). The
Hamiltonian can then be written as

\vspace{-0.7cm}

\begin{align}
\mathcal{H}_{m}^{\prime}\left(k_{y},k_{z}\right)\!=\! & \sum_{x=0}^{\infty}\boldsymbol{\Psi}_{x,k_{y}k_{z}}^{\dagger}\cdot\left(f_{k_{y}}\sigma_{y}\!+\!g_{k_{y}k_{z}}\sigma_{z}\right)\cdot\boldsymbol{\Psi}_{x,k_{y}k_{z}}\label{eq:Hamiltonian-1}\\
 & \qquad\qquad-\frac{t}{2}\sum_{x=0}^{\infty}\boldsymbol{\Psi}_{x+1,k_{y}k_{z}}^{\dagger}\cdot\left(\sigma_{z}\!-\!i\sigma_{x}\right)\cdot\boldsymbol{\Psi}_{x,k_{y}k_{z}}\nonumber \\
 & \qquad\qquad\qquad\qquad-\frac{t}{2}\sum_{x=0}^{\infty}\boldsymbol{\Psi}_{x,k_{y}k_{z}}^{\dagger}\cdot\left(\sigma_{z}\!+\!i\sigma_{x}\right)\cdot\boldsymbol{\Psi}_{x+1,k_{y}k_{z}},\nonumber 
\end{align}
in a mixed $\mathbf{k}$- and real-space representation, with $f_{k_{y}}\!=\!t\sin k_{y}$,
$g_{k_{y}k_{z}}\!=\!t\left(2\!-\!\cos k_{y}\right.$ $\left.-\!\cos k_{z}\right)$,
and $\boldsymbol{\Psi}_{x,k_{y}k_{z}}^{\dagger}\!\!=\!\left[a_{xk_{y}k_{z}}^{\dagger},b_{xk_{y}k_{z}}^{\dagger}\right]$
being a two-orbital fermionic creation operator. For point $\mathbf{q}\!=\!\left(k_{y},k_{z}\right)$
in the surface first Brillouin zone (s-fBz\nomenclature{s-fBz}{Surface First Brillouin Zone}),
the Hamiltonian of Eq.\,\eqref{eq:Hamiltonian-1} effectively describes
a semi-infinite 1D tight-binding model with two bands and open boundary
conditions at $x\!=\!0$. Since we are looking for localized eigenstates
at the boundary, we solve the eigenvalue problem,

\vspace{-0.7cm}
\begin{equation}
\mathcal{H}_{m}^{\prime}\left(\mathbf{q}\right)\ket{\Psi_{\mathbf{q}}^{E}}=E\ket{\Psi_{\mathbf{q}}^{E}},\label{eq:LocalizedEvalue}
\end{equation}
$E$ being the energy in units of $t$, through the \textit{ansatz},

\vspace{-0.7cm}

\begin{equation}
\ket{\Psi_{\mathbf{q}}^{E}}=\!\sum_{x=0}^{\infty}\boldsymbol{\Psi}_{0}e^{-\kappa x}\ket{\mathbf{q},x},
\end{equation}
where $\boldsymbol{\Psi}_{0}\!=\!\left[\psi_{0}^{a},\psi_{0}^{b}\right]^{T}$
is a complex bispinorial amplitude, and $\kappa$ is an inverse localization
length. With this \textit{ansatz}, Eq.\,\eqref{eq:LocalizedEvalue}
breaks down into two coupled linear equations, 

\vspace{-0.7cm}

\begin{equation}
\begin{cases}
\left[f_{k_{y}}\sigma_{y}\!+\!g_{k_{y}k_{z}}\sigma_{z}-\frac{t}{2}\left(\sigma_{z}\!+\!i\sigma_{x}\right)e^{-\kappa}\right]\cdot\!\boldsymbol{\Psi}_{0}=E\,\boldsymbol{\Psi}_{0} & \text{for \ensuremath{x\!=\!0}}\\
\left[f_{k_{y}}\sigma_{y}\!+\!g_{k_{y}k_{z}}\sigma_{z}-\frac{t}{2}\left(\sigma_{z}\!-\!i\sigma_{x}\right)e^{\kappa}-\frac{t}{2}\left(\sigma_{z}\!+\!i\sigma_{x}\right)e^{-\kappa}\right]\cdot\!\boldsymbol{\Psi}_{0}=E\,\boldsymbol{\Psi}_{0} & \text{for \ensuremath{x\!>\!0}}
\end{cases},\label{eq:System-1}
\end{equation}
which, together, imply that

\vspace{-0.7cm}
\begin{equation}
\left(\sigma_{z}\!-\!i\sigma_{x}\right)\cdot\boldsymbol{\Psi}_{0}=0\to\boldsymbol{\Psi}_{0}=\left(\begin{array}{c}
\psi_{0}\\
-i\psi_{0}
\end{array}\right),
\end{equation}
where $\psi_{0}$ is a complex number. Thereby, we can recast the
first Eq.\,\eqref{eq:System-1} simply as

\vspace{-0.7cm}

\begin{equation}
\left[f_{k_{y}}\sigma_{y}\!+\!g_{k_{y}k_{z}}\sigma_{z}-\frac{t}{2}\left(\sigma_{z}\!+\!i\sigma_{x}\right)e^{-\kappa}\right]\cdot\left(\begin{array}{c}
1\\
-i
\end{array}\right)=E\left(\begin{array}{c}
1\\
-i
\end{array}\right),
\end{equation}
which boils down to two simple expressions:

\vspace{-0.7cm}

\begin{equation}
k_{y}(E)\!=\!\arcsin\left(-E\right)\text{ and }\kappa(E,k_{z})\!=\!\ln\left[\frac{1}{2-\cos k_{y}-\cos k_{z}}\right].\label{eq:ConditionsFermiArcs}
\end{equation}
The first condition in Eq.\,\eqref{eq:ConditionsFermiArcs} indicates
that the surface state's energy uniquely determines the value of $k_{y}$\,\footnote{This condition also tells us that Fermi arc states have a well-defined
sense of propagation at the surface, because the sign of $k_{y}$
is entirely determined by its energy.} in the s-fBz. Nevertheless, for a given $k_{y}$ to support a localized
surface state, the inverse length-scale $\kappa$ must also be\textit{
finite} and \textit{positive}. From Eq.\,\eqref{eq:ConditionsFermiArcs},
we see this happens if $\cos k_{y}+\cos k_{z}\!>\!2$, which restricts
the surface-localized Fermi arcs to have,

\vspace{-0.7cm}

\begin{equation}
k_{z}\in\left[-\arcsin\left(1\!-\!\sqrt{1\!-\!E^{2}}\right),\arcsin\left(1\!-\!\sqrt{1\!-\!E^{2}}\right)\right],\label{eq:kz_Limits}
\end{equation}
which defines a finite straight line segment in the s-fBz\,\footnote{Note that Eq.\,\eqref{eq:kz_Limits} also guarantees there are no
surface states in this model, for energies $E>1$.}. In Fig.\,\ref{fig:FermiArcs}b, we represent the line of Fermi
arc states within the s-fBz of model, at three different energies,
together with the \textit{projected bulk Fermi surface}. Even though
this $\mathbf{k}$-space geometry is specific to this toy-model, there
are some features which are actually universal, namely: \textit{(i)}
they form finite open curves that connect regions of the projected
bulk Fermi surface around Weyl nodes of opposite chirality, and \textit{(ii)}
they attach to the bulk Fermi surface in a tangent direction\,\cite{Haldane2014}. 

In the previous calculation, we have determined the surface states
assuming that the bulk WSM exists for $x\!>\!0$. If we had chosen
otherwise (\textit{i.e.}, a semi-infinite bulk for $x\!<\!0$), we
would have obtained the following conditions:

\vspace{-0.7cm}
\begin{equation}
k_{y}(E)\!=\!\arcsin\left(E\right)\text{ and }k_{z}\in\left[-\arcsin\left(1\!-\!\sqrt{1\!-\!E^{2}}\right),\arcsin\left(1\!-\!\sqrt{1\!-\!E^{2}}\right)\right].\label{eq:ConditionsFermiArcs-1}
\end{equation}
These conditions are similar to the ones obtained in Eq.\,\eqref{eq:kz_Limits},
but with an important difference: $k_{y}(E)\to-k_{y}(E)$. In a finite
slab geometry\,\footnote{Assuming the width is sufficiently large to ignore inter-surface coupling.},
this difference is a crucial result for it guarantees that the projected
Fermi surface is actually closed, as shown in Fig.\,\ref{fig:FermiArcs}c.
The Fermi arc states (localized in both boundaries) then serve to
fill the void in between the two Fermi surface components which are
otherwise disjoint in the bulk fBz. In the semi-metallic limit ($E_{\text{F}}\!=\!0$)
the volume of the surface-projected Fermi sea shrinks to a straight
line that connects the projections of the Weyl nodes.

\begin{figure}[t]
\vspace{-0.5cm}
\begin{centering}
\includegraphics[scale=0.21]{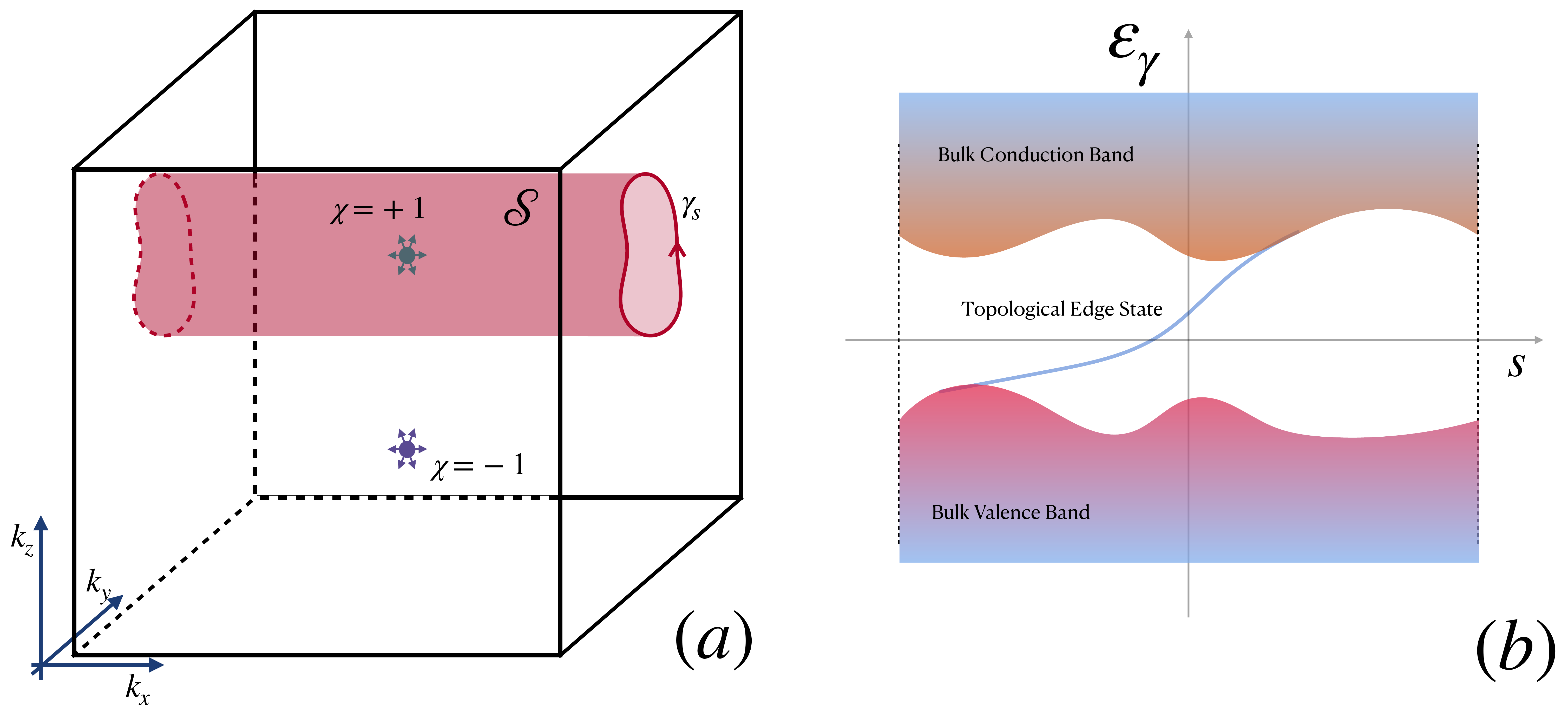}
\par\end{centering}
\vspace{-0.1cm}

\caption{\label{fig:FermiArcs2}(a) Scheme of a three-dimensional fBz containing
two Weyl nodes of opposite chirality. In magenta, an embedded two-dimensional
fBz (\textit{i.e.}, a $2$-torus) is represented which encloses a
single Weyl node. (b) Edge band-structure associated to the projection
of the gapped bands of $\mathcal{S}$ into the curve $\gamma_{s}$.
Since the Chern number of the two bands in $\mathcal{S}$ are $\pm1$,
the bulk-edge correspondence implies the existence of an edge state
which crosses the gap.}

\vspace{-0.4cm}
\end{figure}

The analysis of the model in Eq.\,\eqref{eq:Model_2Weyl} clearly
unveilled that localized states appear in exposed surfaces of a WSM,
connecting disjoint parts of the bulk's Fermi surface. Akin to the
surface/edge states of topological gapped phases, we now show that
these surface Fermi arcs are caused by the nontrivial topology of
the band-structure and, thereby, must appear in generic WSMs of different
geometry and with a larger number of nodes. Following Wan \textit{et
al}.\,\cite{Wan2011}, we look at the cartoon shown in Fig.\,\ref{fig:FermiArcs2}a,
where a three-dimensional fBz is shown to host a pair of Weyl points
with opposite chirality. Within this generic fBz, one can consider
a smooth curve $\gamma_{s}\!=\!(0,k_{y}(s),k_{z}(s))$, parametrized
by $s\!=\![-\pi,\pi]$, which is the base of a cylindrical surface,
$\mathcal{S}$, along the $k_{x}$-axis (shown in magenta). Since
the fBz has periodic boundaries, $\mathcal{S}$ is actually an embedded
$2$-torus which effectively defines the fBz zone of a two-dimensional
gapped system\,\footnote{Provided the surface \textit{does not} intersects any Weyl point.}.
From this analogy, we see that the Fermi arc states in the $k_{y}-k_{z}$
plane can be seen as \textit{topological edge states} of this effective
2D system which, therefore, are subordinated to a non-trivial topology
(nonzero Chern number) of the band-structure restricted to $\mathcal{S}$.
This 2D Chern number of the valence band is simply the flux of Berry
curvature piercing $\mathcal{S}$, \textit{i.e.},

\vspace{-0.7cm}
\begin{equation}
n_{\mathcal{S}}\!=\!\iint_{\mathcal{S}}\!\!\boldsymbol{\Omega}_{-\mathbf{k}}\!\cdot\!\boldsymbol{dS}=\iiint_{\text{int}\mathcal{S}}\!\!\!d^{{\scriptscriptstyle (3)}}\mathbf{k}\boldsymbol{\nabla}_{\mathbf{k}}\!\times\!\boldsymbol{\Omega}_{-\mathbf{k}},
\end{equation}
where Stokes' theorem was employed. Since the Berry curvature field
is divergenceless everywhere, except at Weyl points, we are faced
with two distinct situations: \textit{(i)} if \textit{none} or \textit{both
Weyl points} lie inside $\mathcal{S}$, then $n_{\mathcal{S}}\!=\!0$
and the 2D band-structure is trivial, or \textit{(ii)} if a Weyl point
of chirality $\chi$ is inside $\mathcal{S}$ then $n_{\text{S}}\!=\!\chi$
and the 2D band-structure describes a QHI. In the latter case, an
exposed edge perpendicular to the $x$-axis will support a localized
state for any energy inside the 2D spectral gap. In other words, as
represented in Fig.\,\ref{fig:FermiArcs2}b, there will be a value
of $s$ {[}or, equivalently, a point $(0,k_{y}(s),k_{z}(s))${]} for
which a surface state exists at any energy $E$ inside the 2D gap.
By considering any surface containing just one of the Weyl points,
one reconstructs the whole Fermi arc in s-fBz. 

The existence of Fermi arc states manifests the non-trivial topology
of a Weyl semimetal, being the consequence of a bulk-boundary correspondence
that is analogue to the one found in topological gapped phases. The
number and precise shape of these arcs in the s-fBz is not a universal
feature, but the fact they connect disjoint chiral components of the
bulk Fermi surface, merging with them tangentially are believed to
be general for any WSM. In contrast, the precise shape of the arcs
is known to be determined by details of the boundary condition imposed
at the boundary surface\,\cite{Okugawa2014,Li2015} (see Hashimoto\,\cite{Hashimoto2017}
for a classification of all open boundary conditions) and can even
be changed upon the application of in-plane external magnetic fields\,\cite{Tchoumakov2016}.
This caveat is particularly important in the semi-metallic limit where,
unlike what happened in Fig.\,\ref{fig:FermiArcs}c, the Fermi surface
associated to the surface states may retain a finite area. In that
case, the WSM's boundary will retain a metallic character, with a
finite surface density of charge carriers in spite of the semi-metallic
bulk. From the start, these Fermi arc states were predicted to yield
unmistakable signatures in the form of Friedel oscillations of surface
local density of states due to surface defects\,\cite{Hosur2012,Potter2014}.

\begin{wrapfigure}{o}{0.4\columnwidth}%
\vspace{-0.55cm}
\begin{centering}
\includegraphics[scale=0.2]{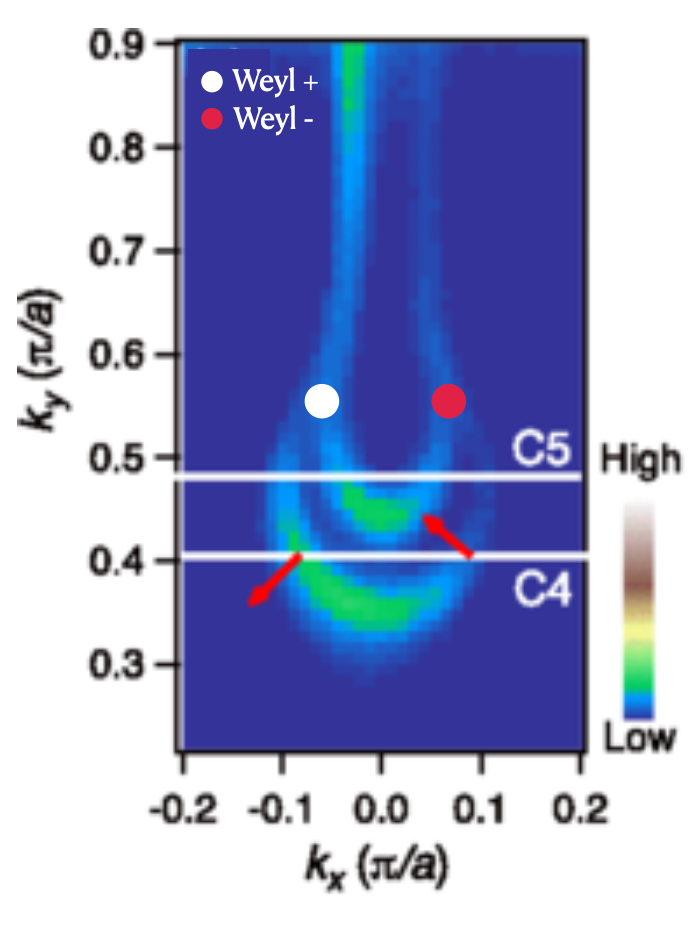}
\par\end{centering}
\vspace{-0.3cm}

\caption{\label{fig:FermiArcsObservation}Observation of Fermi arcs in the
surface of a $\text{Ta}\text{As}$ sample, using ARPES. This picture
was adapted from Lv \textit{et al}.\,\cite{Lv2015}.}

\vspace{-0.45cm}\end{wrapfigure}%
With the experimental realization of WSM phases in crystalline samples,
came along a great interest in studying Fermi arc states both from
the theoretical and experimental side. Theoretically, there have been
plenty of recent work\,\cite{Zhang2016b,Slager2017,Wilson2018,Brillaux2021}
mostly with the aim of assessing the robustness of these topological
states to the presence of disorder. Unlike what happens with the surface
states of three-dimensional TIs, the Fermi arc states seem to be much
more sensitive to disorder and are believed to quickly \textit{dissolve
into the bulk} as the disorder strength gets increased (\textit{e.g.},
see Slager \textit{et al}.\,\,\cite{Slager2017} or Wilson \textit{et
al}.\,\cite{Wilson2018}). On the experimental side, the Fermi arc
states have been originally detected by Xu \textit{et al}.\,\cite{Xu2015}
using surface ARPES measurements in $\text{Na}_{3}\text{Bi}$, a crystalline
Dirac semimetal. Since then, similar observations in different systems
have been reported within the works of Lv \textit{et al}.\,\cite{Lv2015},
Xu \textit{et al}.\,\cite{Xu2016}, Deng \textit{et al}.\,\cite{Deng2016},
Wu \textit{et al}.\,\cite{Wu2016}, Zheng \textit{et al}.\,\cite{Zheng2021},
Sakhya \textit{et al}.\,\cite{Sakhya2022}, and many others (see
Fig.\,\eqref{fig:FermiArcsObservation}).

\global\long\def\vect#1{\overrightarrow{\mathbf{#1}}}%

\global\long\def\abs#1{\left|#1\right|}%

\global\long\def\av#1{\left\langle #1\right\rangle }%

\global\long\def\ket#1{\left|#1\right\rangle }%

\global\long\def\bra#1{\left\langle #1\right|}%

\global\long\def\tensorproduct{\otimes}%

\global\long\def\braket#1#2{\left\langle #1\mid#2\right\rangle }%

\global\long\def\omv{\overrightarrow{\Omega}}%

\global\long\def\inf{\infty}%

\lhead[\MakeUppercase{\chaptername}~\MakeUppercase{\thechapter}]{\MakeUppercase{\rightmark}}

\rhead[\MakeUppercase{Mean-Field Criticality}]{}

\lfoot[\thepage]{}

\cfoot[]{}

\rfoot[]{\thepage}

\chapter{\label{chap:Mean-Field-Quantum-Criticality}Mean-Field Quantum Criticality
in Disordered Semimetals}

From the concepts introduced in Chap.\,\ref{chap:Introduction}\,{[}Sect.\,\ref{subsec:Topological-Properties}{]},
it becomes clear that three-dimensional (3D\nomenclature{3D}{three-dimensional})
electronic systems that feature linear band-crossings near the Fermi
level are very robust to additional perturbations which may be included
in the Bloch Hamiltonian. Such is guaranteed by topological properties
of the bulk bands (for WSMs\nomenclature{WSMs}{Weyl Semimetals})
or by crystal symmetries (in stable DSMs\nomenclature{DSMs}{Dirac Semimetals}).
In either case, this stability is derived with respect to uniform
(or smooth) perturbations that somehow preserve lattice translation
symmetry. This is not a realistic scenario in real systems, where
electrons are subject to non-uniform (or random) perturbations that
can be caused by \textit{imperfect stoichiometry}, \textit{structural
lattice defects} or the \textit{activation of phonon modes} at finite
temperatures. All these mechanisms effective break the translation-invariance
of the crystal and serve as \textit{disorder sources} with the potential
of qualitatively altering the way electrons propagate through the
lattice. Analyzing some of these changes is the main theme of this
thesis, and the issue we turn to now.

In this chapter, we formally introduce the essential models, basic
results and central concepts which will be employed to obtain and
interpret the original results of this thesis, which we present in
Chaps.\,\ref{chap:Instability_Smooth_Regions}-\ref{chap:Vacancies}.
Furthermore, we also re-derive some known mean-field results concerning
the effects of random on-site disorder in the density of states (DoS)
of a Weyl (or Dirac) semimetal with uncoupled nodes\,\footnote{Here, we assume that large-$\Delta\mathbf{k}$ scattering processes
are suppressed and the spinor structure of the disorder will not couple
the two Weyl sectors of a DSM node. Under these assumptions, we will
collectively call our systems Dirac-Weyl semimetals (DWSMs\nomenclature{DWSMs}{Dirac-Weyl Semimetals})
and the number of Weyl cones ($N_{v}$\nomenclature{$N_{v}$}{Number of Weyl Cones})
will simply act as a degeneracy factor in all physical observable.}. At this level of approximation, we demonstrate that these systems
host an unconventional disorder-induced critical point, which precedes
Anderson localization, and separates a \textit{semi-metallic phase}
(with a vanishing nodal DoS) from a \textit{diffusive metal phase}
(with a finite nodal DoS). These results are presented in a three-fold
way: First, we give an heuristic scaling argument which hints that
the semimetal phase must remain stable in the weak disorder limit.
Second, we showcase a\textit{ mean-field theory} that determines the
disorder-averaged DoS in the presence of a white-noise potential,
clearly showing its anticipated critical behavior at a finite disorder
strength. This mean-field treatment is presented in a diagrammatic
(\textit{Self-Consistent Born Approximation}), as well as in a statistical
field-theory language, and their results are used to build a \textit{critical
scaling theory} for this phase transition. Finally, we present some
numerical results for the mean DoS of a WSM lattice model hosting
scalar Anderson disorder which show, to numerical accuracy, that a
\textit{Semimetal-to-Metal Transition} (SMT) indeed exists in this
system. The three complementary points-of-view agree among themselves.

\section{\label{sec:Continuum-Model}Continuum Model of 3D Weyl Electrons}

Provided one is interested in physical phenomena that involves only
Fermi level excitations, it is enough to consider the continuum approximation
for the single-electron Hamiltonian. For a 3D single-node Weyl semimetal,
that Hamiltonian reads

\vspace{-0.7cm}

\begin{equation}
\mathcal{H}_{c}^{0}\!=\!-i\chi\hbar v_{\text{F}}\int\!d\mathbf{r}\Psi_{a\mathbf{r}}^{\dagger}\left(\boldsymbol{\sigma}^{ab}\!\cdot\!\boldsymbol{\nabla}_{\mathbf{r}}\right)\Psi_{b\mathbf{r}},\label{eq:Contin_SingleNodeAHam}
\end{equation}

where $\hbar$ is Planck's constant, $\chi$ is the chirality, $v_{\text{F}}$
is the Fermi velocity, $\boldsymbol{\sigma}$ is a vector of Pauli
matrices and $\Psi_{\mathbf{r}}^{\dagger}=\left[c_{1\mathbf{r}}^{\dagger},c_{2\mathbf{r}}^{\dagger}\right]$
is a two-component fermionic creation operator. Since Eq.\,\eqref{eq:Contin_SingleNodeAHam}
describes a translation-invariant system, we can obtain the energy
eigenstates through BT, by transforming the fermionic operators to
$\mathbf{k}$-space, 

\vspace{-0.7cm}

\begin{subequations}
\begin{align}
\Psi_{a\mathbf{r}}^{\dagger} & =\int\!\frac{d\mathbf{k}}{\sqrt{8\pi^{3}}}e^{i\mathbf{k}\cdot\mathbf{r}}\Psi_{a\mathbf{k}}^{\dagger}\label{eq:kSpacePsiDagger}\\
\Psi_{a\mathbf{r}} & =\int\!\frac{d\mathbf{k}}{\sqrt{8\pi^{3}}}e^{-i\mathbf{k}\cdot\mathbf{r}}\Psi_{a\mathbf{k}}\label{eq:kSpacePsi}
\end{align}
\end{subequations}

which yields

\vspace{-0.7cm}
\begin{equation}
\mathcal{H}_{c}^{0}\!=\!\chi\hbar v_{\text{F}}\int\!d\mathbf{k}\Psi_{a\mathbf{k}}^{\dagger}\left(\boldsymbol{\sigma}^{ab}\!\cdot\!\mathbf{k}\right)\Psi_{b\mathbf{k}}.\label{eq:SingleValleyHam}
\end{equation}
Equation\,\eqref{eq:SingleValleyHam} defines a $2\!\times\!2$ Bloch
Hamiltonian, $\mathcal{H}_{c}(\mathbf{k})\!=\!\chi\hbar v_{\text{F}}\boldsymbol{\sigma}\!\cdot\!\mathbf{k}$,
which describes a gapless two-band model with a single Weyl node located
in $\mathbf{k\!=\!0}$. The properties of independent particles described
by $\mathcal{H}_{c}^{0}$ are all contained in its \textit{single-particle
Green's function} (SPGF\nomenclature{SPGF}{Single-Particle Green's Function})
which we can write explicitly as,

\vspace{-0.8cm}

\begin{equation}
G_{ab}(E;\mathbf{k},\mathbf{q})\!=\!\bra{\mathbf{k},a}\left[E\!-\!\mathcal{H}_{c}^{0}\right]^{-1}\ket{\mathbf{q},b}\!=\!\delta\left(\mathbf{q}\!-\!\mathbf{k}\right)\frac{\tilde{E}\delta_{ab}-\chi\hbar v_{\text{F}}\boldsymbol{\sigma}^{ab}\!\cdot\!\mathbf{k}}{\tilde{E}^{2}-\hbar^{2}v_{\text{F}}^{2}\abs{\mathbf{k}}^{2}},\label{eq:GreenFunction}
\end{equation}
where $\tilde{E}\!=\!E\pm i0^{+}$ is a complex energy parameter whose
imaginary part depends on whether the SPGF is retarded or advanced.
Simply put, the quantity $G_{ab}(E;\mathbf{k},\mathbf{q})$ is the
time-domain Fourier transform of the transition probability amplitude
between states $\ket{\mathbf{q},b}$ and $\ket{\mathbf{k},a}$. Unsurprisingly,
it is also diagonal in $\mathbf{k}$ which reflects the momentum conservation
law imposed by the translation symmetry\,\footnote{Here-forth, we suppress the $\mathbf{q}$ argument in the $\mathbf{k}$-space
SPGF of any translation invariant model.}. Importantly, it is possible to extract two physical observables
of great interest from Eq.\,\eqref{eq:GreenFunction}: \textit{(i)}
the DoS\nomenclature{DoS}{Density of States (per unit volume and unit energy)}
of a clean Weyl semimetal, and \textit{(ii)} the real-space propagator.
While the former describes the spectral structure of the system's
eigenstates, the latter completely characterizes the way a single
electron (of a given energy) propagates across the system. Both these
quantities will be important for our future discussions.

\paragraph*{Clean Density of States:}

The DoS is defined as the number of eigenstates per unit energy and
unit volume. It can be easily related to the imaginary part of the
(traced) SPGF, \textit{i.e.},

\vspace{-0.7cm}
\begin{equation}
\rho_{0}(E)\!=\!-\frac{1}{\pi\mathscr{V}}\Im\left[\int\!\!d\mathbf{k}\left(G_{11}^{\text{r}}(E;\mathbf{k})+G_{22}^{\text{r}}(E;\mathbf{k})\right)\right]=\frac{1}{8\pi^{3}}\int\!\!d\mathbf{k}\,\delta(E\!-\!\mathcal{H}_{c}(\mathbf{k})),\label{eq:DoSContinuum}
\end{equation}
where $\mathscr{V}$ is the volume of the whole system. In our case,
the integral of Eq.\,\eqref{eq:DoSContinuum} can be calculated analytically
as,

\vspace{-0.7cm}
\begin{equation}
\rho^{0}(E>0)=\frac{1}{8\pi^{3}}\int\!\!d\mathbf{k}\,\delta(E\!-\!\hbar v_{\text{F}}\abs{\mathbf{k}})=\frac{E^{2}}{2\pi^{2}\hbar^{3}v_{\text{F}}^{3}}=\rho^{0}(-E),
\end{equation}
where we used the knowledge that the exact energy levels of $\mathcal{H}_{c}(\mathbf{k})$
are $\varepsilon_{\pm}(\mathbf{k})\!=\!\pm\hbar v_{\text{F}}\abs{\mathbf{k}}$,
for both chiralities. Note also that we made use of the particle-hole
symmetry in the system, which requires that $\rho_{0}(E)=\rho_{0}(-E)$.
Meanwhile, the generalization of the previous result for $N_{v}$
uncoupled Weyl nodes is also trivial:

\vspace{-0.7cm}
\begin{equation}
\rho^{0}(E)=\frac{N_{v}E^{2}}{2\pi^{2}\hbar^{3}v_{\text{F}}^{3}}.\label{eq:DoS_Clean_ManyValleys}
\end{equation}
The important fact to highlight from Eq.\,\eqref{eq:DoS_Clean_ManyValleys}
is the fact that the DoS vanishes quadratically as $E\!\to\!0$. This
defines the system as a semimetal and the exponent with which the
DoS vanishes at the\textit{ nodal energy} ($E\!=\!0$) is characteristic
of the space dimensionality.

\vspace{-0.5cm}

\paragraph*{Real-Space Propagator:}

Another important quantity which we can derive from the (retarded)
SPGF is the single-particle propagator (SPP) in real-space, defined
as follows:

\vspace{-0.7cm}
\begin{equation}
G_{ab}^{\text{0r}}(E;\boldsymbol{\Delta r})\!=\!\bra{\boldsymbol{\Delta r},a}\left[E\!-\!\mathcal{H}_{c}^{0}\right]^{-1}\!\ket{\mathbf{0},b}\!=\!\frac{1}{8\pi^{3}}\int\!\!d\mathbf{k}\,G_{ab}^{\text{r}}(E;\mathbf{k})e^{i\mathbf{k}\cdot\boldsymbol{\Delta r}}.\label{eq:RealSpacePropagator}
\end{equation}
This quantity describes the probability amplitude for an isolated
particle of energy $E$ to propagate through a displacement $\boldsymbol{\Delta r}$
in real-space. Since the system is translation-invariant, the SPP
is a function of differences in position only. Furthermore, from the
expression of Eq.\,\eqref{eq:RealSpacePropagator}, we can see that
$G_{ab}^{\text{0r}}(E;\boldsymbol{\Delta r})$ has a simple mathematical
structure, 

\vspace{-0.8cm}

\begin{equation}
G_{ab}^{\text{0r}}(E;\boldsymbol{\Delta r})\!=\!\left(\delta_{ab}+\chi\frac{i}{\kappa}\boldsymbol{\sigma}_{\!ab}\!\cdot\!\boldsymbol{\nabla}_{\!\boldsymbol{\Delta r}}\right)J_{1}(E;\boldsymbol{\Delta r}),\label{eq:RealSpacePropagator-1}
\end{equation}
which is completely determined by the single (spherically symmetric)
$\mathbf{k}$-space integral,

\vspace{-0.8cm}
\begin{equation}
J_{1}(E;\boldsymbol{\Delta r})\!=\!\frac{\kappa}{8\hbar v_{\text{F}}\pi^{3}}\int\!\!d\mathbf{k}\frac{e^{i\mathbf{k}\cdot\boldsymbol{\Delta r}}}{\kappa^{2}-\abs{\mathbf{k}}^{2}}=\frac{\kappa}{2\hbar v_{\text{F}}\pi^{2}\Delta r}\int_{0}^{\infty}\!\!\!dk\frac{k\sin\left(k\,\Delta r\right)}{\kappa^{2}\!-\!k^{2}},\label{eq:J1Deff}
\end{equation}
where $\kappa\!=\!E/\hbar v_{\text{F}}$ is the natural (inverse)
length scale of this continuum model. Note that the last integral
in Eq.\,\eqref{eq:J1Deff} is well-defined for $\Delta r\!>\!0$\,\cite{Buchhold18b}
and yields simply,

\vspace{-0.7cm}
\begin{equation}
J_{1}(E;\Delta r\!>\!0)\!=\!-\frac{\kappa e^{i\kappa\Delta r}}{4\pi\hbar v_{\text{F}}\Delta r}.
\end{equation}
Thereby, the full SPP at a finite distance, $\Delta r$, simply reads,

\vspace{-0.5cm}
\begin{align}
G_{ab}^{\text{0r}}(E;\boldsymbol{\Delta r})\: & =\:\left(\delta_{ab}+\frac{\chi i}{\kappa\Delta r}\boldsymbol{\sigma}_{\!ab}\!\cdot\!\boldsymbol{\Delta r}\frac{\partial}{\partial\Delta r}\right)J_{1}(E;\Delta r)\label{eq:RealSpacePropagator-1-1}\\
 & =\:\frac{e^{i\kappa\Delta r}}{4\pi\hbar v_{\text{F}}\Delta r}\left[-\delta_{ab}\frac{E}{\hbar v_{\text{F}}}+\chi i\boldsymbol{\sigma}_{\!ab}\!\cdot\!\boldsymbol{\Delta r}\left(\frac{\hbar v_{\text{F}}-iE\Delta r}{\hbar v_{\text{F}}\Delta r^{2}}\right)\right].\nonumber 
\end{align}
The clear SPP of Eq.\,\eqref{eq:RealSpacePropagator-1-1} has some
basic (but generic) features which are worth mentioning, most notably
in regard to its asymptotic behavior for $\Delta r\to0,+\infty$.
For any finite energy, $E\neq0$, $G_{ab}^{\text{0r}}(E;\boldsymbol{\Delta r})$
decays asymptotically as $1/\Delta r$, indicating that, over time,
a finite-energy Weyl electron will propagate as an outgoing free spherical
wave. In contrast, if $E\!=\!0$ the SPP decays as $1/\Delta r^{2}$
which will prove essential for our discussion of vacancies in Chap.\,\ref{chap:Vacancies}.
For now, we look at the on-site SPP, $G_{ab}^{\text{0r}}(E;\boldsymbol{0})$,
which we clearly see is ill-defined in this model. To be concrete, 

\vspace{-0.7cm}
\begin{equation}
G_{ab}^{\text{0r}}(E;\boldsymbol{0})\:=\:\delta_{ab}\frac{\kappa}{2\hbar v_{\text{F}}\pi^{2}}\int_{0}^{\infty}\!\!\!dk\frac{k^{2}}{\kappa^{2}\!-\!k^{2}},\label{eq:G_Reg}
\end{equation}
is a \textit{UV-divergent quantity}. This divergence is automatically
regularized in any lattice realization of Weyl electrons, where a
natural limitation on single-particle wavelength provided by the border
of the fBz. Nevertheless, since we will be working within the continuum
model we artificially regularize Eq.\,\eqref{eq:G_Reg} in the hope
that physical results may be independent of the regularization scheme.
In practice, the most direct way to do this is by introducing a \textit{hard
cut-off} for large momenta, $\Lambda$, such that 

\vspace{-0.7cm}
\begin{equation}
\int_{0}^{\infty}\!\!\!dk\frac{k^{2}}{\kappa^{2}\!-\!k^{2}}\to\int_{0}^{\Lambda}\!\!\!dk\frac{k^{2}}{\kappa^{2}\!-\!k^{2}}=\frac{\kappa}{2}\log\left[\frac{\kappa+\Lambda}{\kappa-\Lambda}\right]-\Lambda
\end{equation}
already gives a finite expression. However, we shall also adopt an
alternative \textit{``smooth cut-off''} regularization scheme (based
on Buchhold \textit{et al}.\,\cite{Buchhold18b}) that consists of
modulating the integrand by a Lorentzian envelope,

\vspace{-0.7cm}

\begin{equation}
\int_{0}^{\infty}\!\!\!dk\frac{k^{2}}{\left(\kappa+i0^{+}\right)^{2}\!-\!k^{2}}\to\int_{0}^{\infty}\!\!\!dk\frac{k^{2}}{\left(\kappa+i0^{+}\right)^{2}\!-\!k^{2}}\left(\frac{M^{2}}{M^{2}+k^{2}}\right)=-\frac{i\pi M^{2}}{2\kappa+2iM},
\end{equation}
controlled by a large inverse length scale $M$. Within this last
scheme, the on-site SPP takes on the following expression

\vspace{-0.7cm}

\begin{equation}
G_{ab}^{\text{0r}}(E;\boldsymbol{0},M)\:=\:-\frac{\delta_{ab}}{4\pi\hbar v_{\text{F}}}\left(\hbar v_{\text{F}}M+iE\right)\frac{M^{2}E}{E^{2}+\hbar^{2}v_{\text{F}}^{2}M^{2}},
\end{equation}
which can be separated into complex parts,

\vspace{-0.7cm}

\begin{subequations}
\begin{align}
\Re\left[G_{ab}^{\text{0r}}(E;\boldsymbol{0},M)\right] & =-\frac{\delta_{ab}\kappa M}{4\pi\hbar^{2}v_{\text{F}}^{2}}\frac{1}{1+\kappa^{2}/M^{2}}\approx-\frac{\delta_{ab}E\,M}{4\pi\hbar^{3}v_{\text{F}}^{3}}+\mathcal{O}\left[1\right]\\
\Im\left[G_{ab}^{\text{0r}}(E;\boldsymbol{0},M)\right] & =\frac{\delta_{ab}E^{2}}{4\pi\hbar^{3}v_{\text{F}}^{3}}\frac{1}{1+\kappa^{2}/M^{2}}\approx-\frac{\delta_{ab}E^{2}}{4\pi\hbar^{3}v_{\text{F}}^{3}}\!+\!\mathcal{O}\left[\frac{\kappa}{M}\right]^{2}\label{eq:DoS}
\end{align}
\end{subequations}

evidencing that, while the real part diverges linearly with $M$,
the imaginary part remains finite as $M\!\to\!\infty$. This must
be the case, since the trace of Eq.\,\eqref{eq:DoS} corresponds
to the $-1/\pi$ times the clean DoS we have calculated previously. 

\vspace{-0.4cm}

\section{\label{sec:WeaklyDisorderedNode}The Weakly Disordered Weyl Node}

The translation-invariance of the continuum model in Eq.\,\eqref{eq:Contin_SingleNodeAHam}
allowed us to completely describe the behavior of non-interacting
emergent Weyl fermions. However, real materials always host disorder
sources and, therefore, have an imperfect crystalline symmetry that
amounts to non-homogenous (often random) perturbations to the Hamiltonian
of Eq.\,\eqref{eq:Contin_SingleNodeAHam}. On general grounds, if
the disorder is strong enough, all single-particle eigenstates will
become \textit{exponentially localized}\,\cite{Anderson57,Kramer93},
and the system turns into a so-called \textit{Anderson Insulator}.
However, for 3D systems, the onset of Anderson localization across
the entire spectrum only happens for a very strong disorder, which
means that any weak disorder perturbation will only cause two major
effects: \textit{(i)} generate finite scattering times for the propagating
electrons, and \textit{(ii)} cause deformations in the global density
of states (well represented by an ensemble-average). 

Within this weak disorder regime, we focus on describing the changes
caused in DoS by a random perturbation to the continuum Hamiltonian
of a single Weyl node. To be precise, we will consider a simple disordered,
containing an uncorrelated Anderson scalar potential, which reads
as,

\vspace{-0.7cm}

\begin{equation}
\mathcal{H}\!=\!\mathcal{H}_{c}^{0}\!+\!\mathcal{V}_{d}=\hbar v_{\text{F}}\int\!d\mathbf{k}\Psi_{a\mathbf{k}}^{\dagger}\left(\boldsymbol{\sigma}^{ab}\!\cdot\!\mathbf{k}\right)\Psi_{b\mathbf{k}}\!+\!\int\!d\mathbf{r}\Psi_{a\mathbf{r}}^{\dagger}V(\mathbf{r})\Psi_{a\mathbf{r}},\label{eq:Dirty_Model}
\end{equation}
where the repeated indices are summed, and $V(\mathbf{r})$ is a random
scalar field in the continuum. This random field is assumed to have
a \textit{gaussian-like statistics} characterized by 

\vspace{-0.7cm}
\begin{equation}
\overline{V(\mathbf{r})}\!=\!0\text{ and }\overline{V(\mathbf{r}_{1})V(\mathbf{r}_{2})}=W^{2}f\left(\frac{\abs{\mathbf{r}_{2}\!-\!\mathbf{r}_{1}}}{\xi}\right),\label{eq:RandomFieldStat}
\end{equation}
where $\overline{\cdots}$ stands for an ensemble-average over disorder
realizations, $f(0)\!=\!1$ and $f(x)\underset{x\to\infty}{\longrightarrow}\exp(-x)$.
In this model, $\xi$ is the spacial correlation length of the field
values and $W$ provides a suitable measure of its local strength. 

By definition, if $W\!=\!0$ the system is not disordered and has
a vanishing DoS in the node. Therefore, the most natural procedure
to deal with the presence of the random field in Eq.\,\eqref{eq:Dirty_Model}
is to consider the effects of $\mathcal{V}_{d}$ in perturbation theory
(using $W$ as a small parameter). However, before doing that, we
will give a simple heuristic (but instructive) argument, due to Nandkishore
\textit{et al}.\,\cite{Nandkishore14}, in favor of a stable Weyl
semimetal phase in the weak disorder limit; Consider a propagating
Weyl electron of energy $E$, whose \textit{de Broglie wavelength}
is simply,

\vspace{-0.7cm}
\begin{equation}
\lambda_{E}\!=\!\frac{2\pi}{\abs{\mathbf{k}}}\!=\!\frac{hv_{\text{F}}}{\abs E}.\label{eq:DeBroglie}
\end{equation}
Then, the effect of $V(\mathbf{r})$ on this state can be roughly
estimated using the statistical variance of its values over a three-dimensional
box of volume $\lambda_{E}^{3}$, \textit{i.e.},

\vspace{-0.7cm}
\begin{equation}
\smash{W_{\text{eff}}^{2}\left(E\right)\!=\!\overline{\left[\frac{1}{\lambda_{E}^{3}}\int_{\lambda_{E}^{3}}\!\!\!d\mathbf{r}\,V(\mathbf{r})\right]^{2}}-\left[\overline{\frac{1}{\lambda_{E}^{3}}\int_{\lambda_{E}^{3}}\!\!\!d\mathbf{r}\,V(\mathbf{r})}\right]^{2},}
\end{equation}
This quantity measures the effective strength of the \textit{coarse-grained
random field} which is seen from the perspective of a propagating
Weyl electron at that energy. Naturally, as $E\!\to\!0$ the corresponding
wavelengths becomes larger and larger, up to the point when $\lambda_{E}^{3}$
contains several correlation volumes, $\xi^{3}$. In that case, the
central limit theorem can be applied to compute $W_{\text{eff}}$,
yielding

\vspace{-0.7cm}

\begin{equation}
W_{\text{eff}}\left(E\right)\!\sim\!\sqrt{\overline{\left[\frac{1}{\lambda_{E}^{3}}\int_{\lambda_{E}^{3}}\!\!\!d\mathbf{r}\,V(\mathbf{r})\right]^{2}}}\underset{\underset{\text{Theorem}}{{\scriptscriptstyle \text{Central Limit}}}}{\longrightarrow}W\left(\frac{\xi}{\lambda_{E}}\right)^{\frac{3}{2}}\sim\frac{W\abs E^{\frac{3}{2}}\xi^{\frac{3}{2}}}{h^{\frac{3}{2}}v_{\text{F}}^{\frac{3}{2}}}.
\end{equation}
To quantify how important the random field in $\mathcal{H}$ really
is, the effective strength must be compared to the \textit{kinetic
energy-scale}, $\abs E$. The relative strength of the two terms in
Eq.\,\eqref{eq:Dirty_Model} is simply

\vspace{-0.7cm}
\begin{equation}
\frac{W_{\text{eff}}\left(E\right)}{\abs E}\sim\frac{W\abs E^{\frac{1}{2}}\xi^{\frac{3}{2}}}{h^{\frac{3}{2}}v_{\text{F}}^{\frac{3}{2}}}\underset{E\to0}{\longrightarrow}0.\label{eq:Comparison_Scale4e}
\end{equation}
This heuristic comparison of scales hints that the effects of disorder
will likely be minor as one gets closer to a Weyl (or Dirac) node.
In other words, the disordered Weyl node \textit{``cleans-up'' }in
the vicinity of the nodal energy and, therefore, the semi-metallic
character of uncoupled Weyl nodes is expected to be robust to weak
disorder.

\vspace{-0.4cm}

\subsection{\label{subsec:DisorderAveragingDiagrammatics}Disorder-Averaging
Diagramatics}

The previous argument is physically sensible but still heuristic in
nature. Now, we aim at a more quantitative description of weak disorder
effects by employing a standard perturbation theory in the disorder
strength, $W$. We begin by considering $\mathcal{G}^{0}(E)=\left[\tilde{E}\!-\!\mathcal{H}_{c}^{0}\right]^{-1}$
as the being unperturbed SPGF operator, while $\mathcal{G}(E)=\left[\tilde{E}\!-\!\mathcal{H}_{c}^{0}\!-\!\mathcal{V}_{d}\right]^{-1}$
is the perturbed one. Moreover, we also write the random potential
term in $\mathbf{k}$-space,

\vspace{-0.7cm}

\begin{equation}
\mathcal{V}_{d}=\!\int\!d\mathbf{r}\,\,\Psi_{a\mathbf{r}}^{\dagger}V(\mathbf{r})\Psi_{a\mathbf{r}}\!=\!\!\int\!d\mathbf{k}d\mathbf{q}\,\,\Psi_{a\mathbf{k}}^{\dagger}V_{\mathbf{k}-\mathbf{q}}\Psi_{a\mathbf{q}}\text{ with }V_{\mathbf{p}}\!=\!\int\!\frac{d\mathbf{r}}{8\pi^{3}}\,V(\mathbf{r})e^{i\mathbf{p}\cdot\mathbf{r}},
\end{equation}
which will shortly prove to be a convenient transformation. At this
point, it is clear that the effect of $\mathcal{V}_{d}$ on the plane-wave
states is to scatter between different $\mathbf{k}$, with a transferred
momentum determined by the amplitude of the Fourier transform of $V(\mathbf{r})$.
Besides the sample-specific form of the random potential in $\mathbf{k}$-space,
it will be also important to characterize its ensemble statistics.
More precisely, assuming the gaussian-like statistics of Eq.\,\eqref{eq:RandomFieldStat},
we can calculate the pairwise contractions of $V_{\mathbf{p}}$,

\vspace{-0.7cm}
\begin{align}
\overline{V_{\mathbf{p}}V_{\mathbf{q}}} & =\int\!\frac{d\mathbf{r}}{8\pi^{3}}\int\!\frac{d\mathbf{r}'}{8\pi^{3}}\overline{V(\mathbf{r})V(\mathbf{r}')}e^{i\mathbf{p}\cdot\mathbf{r}}e^{i\mathbf{q}\cdot\mathbf{r}'}=W^{2}\int\!\frac{d\mathbf{l}}{8\pi^{3}}f\left(\frac{\abs{\mathbf{l}}}{\xi}\right)e^{i(\mathbf{p}-\mathbf{q})\cdot\mathbf{l}/2}\times\nonumber \\
 & \qquad\qquad\qquad\qquad\int\!\frac{d\mathbf{R}}{8\pi^{3}}e^{i(\mathbf{p}+\mathbf{q})\cdot\mathbf{R}}=W^{2}\delta\left(\mathbf{p}\!+\!\mathbf{q}\right)\int\!\frac{d\mathbf{l}}{8\pi^{3}}f\left(\!\frac{\abs{\mathbf{l}}}{\xi}\!\right)e^{i\mathbf{p}\cdot\mathbf{l}}\label{eq:Correlatork}
\end{align}
which mean that only $\overline{V_{\mathbf{p}}V_{-\mathbf{p}}}=V_{\mathbf{p}}^{2}\neq0$.
In addition, if we consider the \textit{white-noise limit} for the
random potential, $\overline{V(\mathbf{r}_{1})V(\mathbf{r}_{2})}\!=\!W^{2}\delta(\mathbf{r}_{2}\!-\!\mathbf{r}_{1})$,
then Eq.\,\eqref{eq:Correlatork} reduces to 

\vspace{-0.7cm}
\begin{equation}
V_{\mathbf{p}}^{2}=W^{2}\int\!\frac{d\mathbf{l}}{8\pi^{3}}\delta(\mathbf{l})e^{i\mathbf{p}\cdot\mathbf{l}}=W^{2},\label{eq:Uncorrelated}
\end{equation}
whereas for a general two-point correlator, we would have

\vspace{-0.7cm}
\begin{equation}
V_{\mathbf{p}}^{2}\!=\!\frac{W^{2}}{2\pi^{2}\abs{\mathbf{p}}}\int_{0}^{\infty}\!dl\,lf\left(\frac{l}{\xi}\right)\sin\left(\abs{\mathbf{p}}l\right)=\frac{W^{2}\xi^{2}}{2\pi^{2}\abs{\mathbf{p}}}\int_{0}^{\infty}\!\!\!\!\!dx\,xf\left(x\right)\sin\left(\xi\abs{\mathbf{p}}x\right),
\end{equation}
that depends only on the magnitude of the transferred momentum, $\abs{\mathbf{p}}$.
Nevertheless, even in this general case, by taking the limit $\abs{\mathbf{p}}\ll\xi^{-1}$,
this expression reduces to

\vspace{-0.7cm}

\begin{equation}
V_{\mathbf{p}}^{2}\!\approx\!\frac{W^{2}\xi^{3}}{2\pi^{2}}\int_{0}^{\infty}\!\!\!\!\!dx\,x^{2}f\left(x\right)=CW^{2}\xi^{3},\label{eq:V^2_Corr}
\end{equation}
where $C$ is a constant. Note that Eq.\,\eqref{eq:V^2_Corr} essentially
recovers the result for an uncorrelated disorder {[}Eq.\,\eqref{eq:Uncorrelated}{]},
albeit with a slightly altered disorder strength. In all upcoming
calculations, we shall assume that the statistics of the disordered
potential is determined by the contraction rule,

\vspace{-0.7cm}
\begin{equation}
\overline{V_{\mathbf{p}}V_{\mathbf{q}}}=\delta\left(\mathbf{p}\!+\!\mathbf{q}\right)CW^{2}\xi^{3},\label{eq:CorrelatorinP}
\end{equation}
such that all \textit{multiple-point correlators} will either be zero
(for an odd number of $V$s) or Wick combinations of two-point correlators.
Using this simplified gaussian disorder model, we can build a standard
perturbation expansion\,\cite{Rammer04} for the ensemble-averaged
SPGF. In terms of Feynman diagrams, this is written as follows:

\vspace{-0.7cm}

\begin{align}
\begin{tikzpicture}  
\begin{feynman}     
\node [dot] (i);     
\node [dot,right= 1cm of i] (f);
\diagram* {
(i) -- [fermion,line width=2pt, edge label=\(\mathbf{k}\)] (f),
};
\end{feynman} 
\end{tikzpicture}
\,=\,
\begin{tikzpicture}
\begin{feynman}     
\node [dot] (i);     
\node [dot,right= 1cm of i] (f);
\diagram* {
(i) -- [fermion, edge label=\(\mathbf{k}\)] (f),
};
\end{feynman} 
\end{tikzpicture}
\,+\,
\begin{tikzpicture}   
\begin{feynman}     
\node [dot] (i);     
\node [dot,right= 1cm of i] (c1); 
\node [dot,right= 1cm of c1] (c2); 
\node [dot,right= 1cm of c2] (f);
\node [crossed dot,above right=0.8cm and 0.5cm of c1,label=right:\(V_{\mathbf{k}_1-\mathbf{k}}^2\)] (f1);
\diagram* {
(i) -- [fermion, edge label=\(\mathbf{k}\)] (c1)
    -- [fermion, edge label=\(\mathbf{k}_1\)] (c2)
    -- [fermion, edge label=\(\mathbf{k}\)] (f),
(c1) -- [scalar] (f1) -- [scalar] (c2),
};
\end{feynman} 
\end{tikzpicture}
\,\,+\,\,\qquad\qquad\label{eq1}\\
\begin{tikzpicture}   
\begin{feynman}     
\node [dot] (i);     
\node [dot,right= 1cm of i] (c1); 
\node [dot,right= 1cm of c1] (c2); 
\node [dot,right= 1cm of c2] (c3); 
\node [dot,right= 1cm of c3] (c4); 
\node [dot,right= 1cm of c4] (f);
\node [crossed dot,above=0.8cm of c2,label=left:\(V_{\mathbf{k}_1-\mathbf{k}}^2\)] (f1);
\node [crossed dot,above=0.8cm of c3,label=right:\(V_{\mathbf{k}-\mathbf{k}_1}^2\)] (f2);
\diagram* {
(i) -- [fermion, edge label=\(\mathbf{k}\)] (c1)
    -- [fermion, edge label=\(\mathbf{k}_1\)] (c2)
    -- [fermion, edge label=\(\mathbf{k}_2\)] (c3)
    -- [fermion, edge label=\(\mathbf{k}_1\)] (c4)
    -- [fermion, edge label=\(\mathbf{k}\)] (f),
(c1) -- [scalar] (f1) -- [scalar] (c3),
(c2) -- [scalar] (f2) -- [scalar] (c4),
};
\end{feynman} 
\end{tikzpicture}
\,\,+\,\,\qquad\nonumber\\
\begin{tikzpicture}   
\begin{feynman}     
\node [dot] (i);     
\node [dot,right= 1cm of i] (c1); 
\node [dot,right= 1cm of c1] (c2); 
\node [dot,right= 1cm of c2] (c3); 
\node [dot,right= 1cm of c3] (c4); 
\node [dot,right= 1cm of c4] (f);
\node [crossed dot,above right=1.0cm and 0.5cm of c2,label=right:\(V_{\mathbf{k}-\mathbf{k}}^2\)] (f1);
\node [crossed dot,above right=0.6cm and 0.5cm of c2,label=right:\(V_{\mathbf{k}_3-\mathbf{k}_1}^2\)] (f2);
\diagram* {
(i) -- [fermion, edge label=\(\mathbf{k}\)] (c1)
    -- [fermion, edge label=\(\mathbf{k}_1\)] (c2)
    -- [fermion, edge label=\(\mathbf{k}_2\)] (c3)
    -- [fermion, edge label=\(\mathbf{k}_3\)] (c4)
    -- [fermion, edge label=\(\mathbf{k}\)] (f),
(c1) -- [scalar] (f1) -- [scalar] (c4),
(c2) -- [scalar] (f2) -- [scalar] (c3),
};
\end{feynman} 
\end{tikzpicture}
\,\,+\,\,\qquad\!\nonumber\\
\begin{tikzpicture}   
\begin{feynman}     
\node [dot] (i);     
\node [dot,right= 1cm of i] (c1); 
\node [dot,right= 1cm of c1] (c2); 
\node [dot,right= 1cm of c2] (c3); 
\node [dot,right= 1cm of c3] (c4); 
\node [dot,right= 1cm of c4] (f);
\node [crossed dot,above right=1.0cm and 0.5cm of c1,label=right:\(V_{\mathbf{k}_1-\mathbf{k}}^2\)] (f1);
\node [crossed dot,above right=1.0cm and 0.5cm of c3,label=right:\(V_{\mathbf{k}_1-\mathbf{k}}^2\)] (f2);
\diagram* {
(i) -- [fermion, edge label=\(\mathbf{k}\)] (c1)
    -- [fermion, edge label=\(\mathbf{k}_1\)] (c2)
    -- [fermion, edge label=\(\mathbf{k}\)] (c3)
    -- [fermion, edge label=\(\mathbf{k}_1\)] (c4)
    -- [fermion, edge label=\(\mathbf{k}\)] (f),
(c1) -- [scalar] (f1) -- [scalar] (c2),
(c3) -- [scalar] (f2) -- [scalar] (c4),
};
\end{feynman} 
\end{tikzpicture}
\,\,+\cdots\nonumber,
\end{align}.

In these diagrams, the directed lines represent single-electron propagators
(SPP), with the bare case being shown as a thin line and the disorder-averaged
one as a thick line. The dashed lines are \textit{impurity insertions}
that turn into effective interaction lines upon averaging. As usual,
the perturbative series in Eq.\,\eqref{eq1} can be suitably reorganized
by defining the \textit{disorder-induced self-energy} as,

\vspace{-0.7cm}

\begin{equation}
\begin{tikzpicture}   
\begin{feynman} 
\node [empty dot, scale=6.0, label=center:\(\Sigma_{E, \mathbf{k}}\)] (b);
\diagram*{
};
\end{feynman} 
\end{tikzpicture}
\,=\,
\begin{tikzpicture}   
\begin{feynman}     
\node [dot] (c1); 
\node [dot,right= 1cm of c1] (c2); 
\node [crossed dot,above right=0.8cm and 0.5cm of c1,label=right:\(V_{\mathbf{k}_1-\mathbf{k}}^2\)] (f1);
\diagram* {
(c1) -- [fermion, edge label=\(\mathbf{k}_1\)] (c2),
(c1) -- [scalar] (f1) -- [scalar] (c2),
};
\end{feynman} 
\end{tikzpicture}
\!\!\!\!\!\!\!\!\!+\,\,
\begin{tikzpicture}   
\begin{feynman}     
\node [dot] (c1); 
\node [dot,right= 1cm of c1] (c2); 
\node [dot,right= 1cm of c2] (c3); 
\node [dot,right= 1cm of c3] (c4);
\node [crossed dot,above=0.8cm of c2,label=left:\(V_{\mathbf{k}_1-\mathbf{k}}^2\)] (f1);
\node [crossed dot,above=0.8cm of c3,label=right:\(V_{\mathbf{k}-\mathbf{k}_1}^2\)] (f2);
\diagram* {
(c1)-- [fermion, edge label=\(\mathbf{k}_1\)] (c2)
    -- [fermion, edge label=\(\mathbf{k}_2\)] (c3)
    -- [fermion, edge label=\(\mathbf{k}_1\)] (c4),
(c1) -- [scalar] (f1) -- [scalar] (c3),
(c2) -- [scalar] (f2) -- [scalar] (c4),
};
\end{feynman} 
\end{tikzpicture}
\,\,+\,\,
\begin{tikzpicture}   
\begin{feynman}     
\node [dot] (c1); 
\node [dot,right= 1cm of c1] (c2); 
\node [dot,right= 1cm of c2] (c3); 
\node [dot,right= 1cm of c3] (c4);
\node [crossed dot,above right=1.0cm and 0.5cm of c2,label=right:\(V_{\mathbf{k}-\mathbf{k}}^2\)] (f1);
\node [crossed dot,above right=0.6cm and 0.5cm of c2,label=right:\(V_{\mathbf{k}_3-\mathbf{k}_1}^2\)] (f2);
\diagram* {
(c1)-- [fermion, edge label=\(\mathbf{k}_1\)] (c2)
    -- [fermion, edge label=\(\mathbf{k}_2\)] (c3)
    -- [fermion, edge label=\(\mathbf{k}_3\)] (c4),
(c1) -- [scalar] (f1) -- [scalar] (c4),
(c2) -- [scalar] (f2) -- [scalar] (c3),
};
\end{feynman} 
\end{tikzpicture}
\,\,+\cdots \label{eq:eq11}
\end{equation}

which is related to the averaged propagator by the following equation

\vspace{-0.7cm}
\begin{equation}
\mathcal{\overline{G}}(E)=\mathcal{G}^{0}(E)+\mathcal{G}^{0}(E)\Sigma_{E}\mathcal{\overline{G}}(E)=\left[\tilde{E}-\mathcal{H}_{0}-\Sigma_{E}\right]^{-1}.
\end{equation}
Since both $\mathcal{H}_{0}$ and $\Sigma_{E}$ are translation-invariant
operators, they are always diagonal in $\mathbf{k}$-space and, therefore,

\vspace{-0.7cm}
\begin{equation}
\overline{G(E;\mathbf{k})}=\frac{1}{\tilde{E}-\mathcal{H}(\mathbf{k})-\Sigma_{E,\mathbf{k}}}.\label{eq:G_Average}
\end{equation}
represents the \textit{disorder-averaged single-particle Green's function}
in $\mathbf{k}$-space. This is the \textit{disorder-dressed} SPGF
{[}Eq.\,\eqref{eq:GreenFunction}{]}, which we conclude that is completely
determined by knowing the disorder self-energy as a function of $(E,\mathbf{k})$\,\footnote{It is important to remark that the averaged propagator does not contain
all information about the disordered system. In fact, transport properties
and localization phenomena require the averaging of products of propagators
which are statistically correlated. These so-called vertex corrections
are now important for analyzing the behavior of the density of states
and will therefore be ignored.}. In fact, the disorder self-energy encapsulates all information about
the single-particle excitations of the disordered system, including
changes in the spectrum and the broadening of the clean energy levels
due to the finite life-time of plane-wave states moving through the
disordered landscape. If one develops the perturbative expansion for
the self-energy itself, the result is just Eq.\,\eqref{eq:eq11}.
However, if we are a bit more sophisticated, we can realize that among
these diagrams there are all the terms in which the bare propagators
of Eq.\,\eqref{eq:eq11} are decorated by an arbitrary number of
self-energy insertions placed in series. In simpler words, we can
just take the original series and replace,

\vspace{-0.7cm}

\hspace{-1.0cm}\begin{equation}
\begin{tikzpicture}   
\begin{feynman} 
\node [dot] (a);
\node [dot, right=1.5cm of a] (b);
\vertex [below=0.45cm of b] (c);
\diagram*{
(a) -- [fermion, edge label=\(\mathbf{k}\)] (b),
(b) -- [opacity=0.0] (c)
};
\end{feynman} 
\end{tikzpicture}
\to
\begin{tikzpicture}   
\begin{feynman} 
\node [dot] (a);
\node [empty dot, right=1cm of a, scale=6.0, label=center:\(\Sigma_{E, \mathbf{k}}\)] (b);
\node [dot, right=1cm of b] (c);
\diagram*{
(a) -- [fermion, edge label=\(\mathbf{k}\)] (b) 
    -- [fermion, edge label=\(\mathbf{k}\)] (c),
};
\end{feynman} 
\end{tikzpicture}
\mathrel{\raisebox{2.5ex}{+}}
\begin{tikzpicture}   
\begin{feynman} 
\node [dot] (a);
\node [empty dot, right=1cm of a, scale=6.0, label=center:\(\Sigma_{E, \mathbf{k}}\)] (b);
\node [empty dot, right=1.5cm of b, scale=6.0, label=center:\(\Sigma_{E, \mathbf{k}}\)] (c);
\node [dot, right=1cm of c] (d);
\diagram*{
(a) -- [fermion, edge label=\(\mathbf{k}\)] (b) 
    -- [fermion, edge label=\(\mathbf{k}\)] (c)
    -- [fermion, edge label=\(\mathbf{k}\)] (d),
};
\end{feynman} 
\end{tikzpicture}
\mathrel{\raisebox{2.5ex}{+}}
\begin{tikzpicture}   
\begin{feynman} 
\node [dot] (a);
\node [empty dot, right=1cm of a, scale=6.0, label=center:\(\Sigma_{E, \mathbf{k}}\)] (b);
\node [empty dot, right=1.5cm of b, scale=6.0, label=center:\(\Sigma_{E, \mathbf{k}}\)] (c);
\node [empty dot, right=1.5cm of c, scale=6.0, label=center:\(\Sigma_{E, \mathbf{k}}\)] (d);
\node [dot, right=1cm of d] (e);
\diagram*{
(a) -- [fermion, edge label=\(\mathbf{k}\)] (b) 
    -- [fermion, edge label=\(\mathbf{k}\)] (c)
    -- [fermion, edge label=\(\mathbf{k}\)] (d)
    -- [fermion, edge label=\(\mathbf{k}\)] (e),
};
\end{feynman} 
\end{tikzpicture}
\mathrel{\raisebox{2.5ex}{+...}}\hrulefill\label{eq:eq2}
\end{equation}

without changing the result. This is exactly the same as replacing
all internal bare propagators by disorder-averaged ones, which gives
rise to the following version of the perturbative series for the disorder
self-energy:

\vspace{-0.7cm}

\begin{equation}
\begin{tikzpicture}   
\begin{feynman} 
\node [empty dot, scale=6.0, label=center:\(\Sigma_{E, \mathbf{k}}\)] (b);
\diagram*{
};
\end{feynman} 
\end{tikzpicture}
\mathrel{\raisebox{2.5ex}{=}}
\begin{tikzpicture}   
\begin{feynman}     
\node [dot] (c1); 
\node [dot,right= 1cm of c1] (c2); 
\node [crossed dot,above right=0.8cm and 0.5cm of c1,label=right:\(V_{\mathbf{k}_1-\mathbf{k}}^2\)] (f1);
\diagram* {
(c1) -- [fermion,line width=2pt, edge label=\(\mathbf{k}_1\)] (c2),
(c1) -- [scalar] (f1) -- [scalar] (c2),
};
\end{feynman} 
\end{tikzpicture}
\mathrel{\raisebox{2.5ex}{\!\!\!\!\!\!+}}
\begin{tikzpicture}   
\begin{feynman}     
\node [dot] (c1); 
\node [dot,right= 1cm of c1] (c2); 
\node [dot,right= 1cm of c2] (c3); 
\node [dot,right= 1cm of c3] (c4);
\node [crossed dot,above=0.8cm of c2,label=left:\(V_{\mathbf{k}_1-\mathbf{k}}^2\)] (f1);
\node [crossed dot,above=0.8cm of c3,label=right:\(V_{\mathbf{k}-\mathbf{k}_1}^2\)] (f2);
\diagram* {
(c1)-- [fermion,line width=2pt, edge label=\(\mathbf{k}_1\)] (c2)
    -- [fermion,line width=2pt, edge label=\(\mathbf{k}_2\)] (c3)
    -- [fermion,line width=2pt, edge label=\(\mathbf{k}_1\)] (c4),
(c1) -- [scalar] (f1) -- [scalar] (c3),
(c2) -- [scalar] (f2) -- [scalar] (c4),
};
\end{feynman} 
\end{tikzpicture}
\mathrel{\raisebox{2.5ex}{\!\!\!+}}
\begin{tikzpicture}   
\begin{feynman}     
\node [dot] (c1); 
\node [dot,right= 1cm of c1] (c2); 
\node [dot,right= 1cm of c2] (c3); 
\node [dot,right= 1cm of c3] (c4);
\node [crossed dot,above right=1.5cm and 0.5cm of c2,label=right:\(V_{\mathbf{k}-\mathbf{k}}^2\)] (f1);
\node [crossed dot,above right=0.8cm and 0.5cm of c2,label=right:\(V_{\mathbf{k}_3-\mathbf{k}_1}^2\)] (f2);
\diagram* {
(c1)-- [fermion,line width=2pt, edge label=\(\mathbf{k}_1\)] (c2)
    -- [fermion,line width=2pt, edge label=\(\mathbf{k}_2\)] (c3)
    -- [fermion,line width=2pt, edge label=\(\mathbf{k}_3\)] (c4),
(c1) -- [scalar] (f1) -- [scalar] (c4),
(c2) -- [scalar] (f2) -- [scalar] (c3),
};
\end{feynman} 
\end{tikzpicture}
\mathrel{\raisebox{2.5ex}{\!\!+...}} \label{eq:eq12}.
\end{equation}.

Formally, Eq.\,\eqref{eq:eq12} defines the disorder self-energy
as a \textit{Self-Consistent Functional} of the disorder-averaged
propagator in $\mathbf{k}$-space. The lowest-order contribution to
this series is the so-called \textit{Self-Consistent Born Approximation}
(SCBA\nomenclature{SCBA}{Self-Consistent Born Approximation}), which
takes the following diagrammatic form:

\vspace{-0.7cm}

\begin{equation}
\begin{tikzpicture}   
\begin{feynman} 
\node [empty dot, scale=6.0, label=center:\(\Sigma_{E, \mathbf{k}}\)] (b);
\diagram*{
};
\end{feynman} 
\end{tikzpicture}
\mathrel{\raisebox{2.5ex}{=}}
\begin{tikzpicture}   
\begin{feynman}     
\node [dot] (c1); 
\node [dot,right= 1cm of c1] (c2); 
\node [crossed dot,above right=0.8cm and 0.5cm of c1,label=right:\(V_{\mathbf{k}_1-\mathbf{k}}^2\)] (f1);
\diagram* {
(c1) -- [fermion,line width=2pt, edge label=\(\mathbf{k}_1\)] (c2),
(c1) -- [scalar] (f1) -- [scalar] (c2),
};
\end{feynman} 
\end{tikzpicture}
\label{eq:eq13}.
\end{equation}

or, equivalently, 

\vspace{-0.7cm}

\begin{equation}
\Sigma_{E,\mathbf{k}}=\int d\mathbf{q}\frac{V_{\mathbf{q}-\mathbf{k}}^{2}}{\tilde{E}-\chi\hbar v_{\text{F}}\boldsymbol{\sigma}\cdot\mathbf{q}-\Sigma_{E,\mathbf{q}}}.\label{eq:SCBA}
\end{equation}
Notice that, in spite of being diagrammatic, the validity SCBA is
not limited to a perturbative regime in which the disorder strength
is a very small parameter. In principle, by solving Eq.\,\ref{eq:SCBA}
self-consistently, the solution will effectively take into account
the re-summation of an infinite sub-series of diagrams and, as will
be shown in Sect.\,\ref{sec:FieldTheory}, it is equivalent to taking
a \textit{mean-field approach} to the disordered problem.

\subsection{\label{subsec:SCBA}The Self-Consistent Born Approximation}

In order to obtain an approximate expression for the disorder-averaged
SPGF, the self-consistent expression for the disorder self-energy
{[}Eq.\,\eqref{eq:SCBA}{]} may be solved iteratively. For our purposes,
we focus on studying the disorder-averaged DoS around the Weyl node,
which is intimately related to the imaginary part of the disorder-averaged
SPGF. For this calculation, we will assume the limit of Eq.\,\eqref{eq:V^2_Corr}
for the disordered potential (which implies a $\mathbf{k}$-independent
self-energy) and, since the self-energy is a generic $2\times2$ matrix,
the SCBA equation for the (retarded) self-energy takes the explicit
form,

\vspace{-0.7cm}
\begin{equation}
\left[\!\!\begin{array}{cc}
\Sigma_{E}^{11} & \Sigma_{E}^{12}\\
\Sigma_{E}^{21} & \Sigma_{E}^{22}
\end{array}\!\!\right]\!=\!CW^{2}\xi^{3}\!\!\int\!\!d\mathbf{q}\!\left[\!\!\begin{array}{cc}
\tilde{E}\!-\!\chi\hbar v_{\text{F}}q_{z}\!-\!\Sigma_{E}^{11} & \!\!\!-\!\chi\hbar v_{\text{F}}q_{x}-i\!\chi\hbar v_{\text{F}}q_{y}-\Sigma_{E}^{12}\\
-\!\chi\hbar v_{\text{F}}q_{x}+i\!\chi\hbar v_{\text{F}}q_{y}-\Sigma_{E}^{21} & \!\!\!\tilde{E}\!+\!\chi\hbar v_{\text{F}}q_{z}\!-\Sigma_{E}^{22}
\end{array}\!\!\right]^{-1}\!\!\!\!\!,
\end{equation}
in which the matrix inversion can be performed analytically. Before
doing that, we make the \textit{ansatz} that the retarded self-energy
is proportional to the identity matrix \textemdash{} $\Sigma_{E}^{ab}=\Sigma_{E}\delta_{ab}$
\textemdash{} and, therefore,

\vspace{-0.7cm}
\begin{align}
\Sigma_{E}\! & =\!\frac{CW^{2}\xi^{3}}{8\pi^{3}}\!\int\!\!d\mathbf{q}\frac{E\,-\!\Sigma_{E}}{\left(E\,-\!\Sigma_{E}\right)^{2}\!-\!\hbar^{2}v_{\text{F}}^{2}\abs{\mathbf{q}}^{2}}
\end{align}
which reduces to the much simpler equation,

\vspace{-0.7cm}

\begin{align}
\Sigma_{E}\! & =\!\frac{CW^{2}\xi^{3}}{2\pi\hbar^{2}v_{\text{F}}^{2}}\left(E\,-\!\Sigma_{E}\right)\!\int_{0}^{\infty}\!\!dq\frac{q^{2}}{\left(E\,-\!\Sigma_{E}\right)^{2}/\hbar^{2}v_{\text{F}}^{2}\!-\!q^{2}}.\label{eq:DimensionlessSCBA}
\end{align}
Note that the integral in Eq.\,\eqref{eq:DimensionlessSCBA} is UV-problematic,
similarly to the on-site SPP of the clean Weyl model {[}Eq.\,\eqref{eq:G_Reg}{]}.
Therefore, we can fix this divergence by using the same smooth cut-off
technique, which yields

\vspace{-0.7cm}
\begin{align}
\Sigma_{E}\! & =\!\frac{CW^{2}\xi^{3}M}{2\hbar^{2}v_{\text{F}}^{2}}\:\frac{\left(E\,-\!\Sigma_{E}\right)}{i\left(E\,-\!\Sigma_{E}\right)/M\hbar v_{\text{F}}-1},\label{eq:DimensionlessSCBA-1}
\end{align}
and that can be put into the dimensionless form,

\vspace{-0.7cm}

\begin{align}
\varsigma_{\epsilon}\! & =\!g_{W}\frac{\epsilon\,-\!\varsigma_{\epsilon}}{i\left(\epsilon\,-\!\varsigma_{\epsilon}\right)-1},\label{eq:DimensionlessSCBA-1-1}
\end{align}
by defining $\epsilon\!=\!E/M\hbar v_{\text{F}}$, $\varsigma_{\epsilon}\!=\!\Sigma_{E}/M\hbar v_{\text{F}}$,
and a dimensionless coupling parameter,

\vspace{-0.7cm}

\begin{equation}
g_{W}=\frac{CW^{2}\xi^{3}M}{2\hbar^{2}v_{\text{F}}^{2}},\label{eq:DimensionlessCoupling}
\end{equation}
that is an effective disorder strength parameter. Equation\,\eqref{eq:DimensionlessSCBA-1-1}
can be solved analytically

\vspace{-0.7cm}

\begin{align}
\varsigma_{\epsilon}\! & =\!\frac{1}{2}\left(\epsilon+i\left(1\!-\!g_{W}\right)\pm\sqrt{\epsilon^{2}\!+\!2i(1\!+\!g_{W})\epsilon\!-\!(1\!-\!g_{W})^{2}}\right),\label{eq:Sol_SCBA}
\end{align}
which are generically two energy-dependent complex solutions. As long
as we deal with retarded propagators or self-energies, the proper
choice of solution must guarantee its imaginary part to be positive,
which yields

\vspace{-0.7cm}

\begin{align}
\varsigma_{\epsilon}\! & =\!\frac{1}{2}\left(\epsilon+i\left(1\!-\!g_{W}\right)-\text{Sign}(\epsilon)\sqrt{\epsilon^{2}\!+\!2i(1\!+\!g_{W})\epsilon\!-\!(1\!-\!g_{W})^{2}}\right).\label{eq:Sol_SCBA-1}
\end{align}

The (retarded) solutions of the SCBA\,\footnote{As discussed with Jed Pixley during the public presentation of this
thesis, by choosing only the retarded solutions of the SCBA equations
one is effectively breaking a \textit{retarded-advanced symmetry of
the system's Green's functions} which is a typical step of of any
mean-field theory calculation.} could be represented as function of the dimensionless parameter,
$g_{W}$, or energy. However, our interest is not on the self-energy
itself, but rather on the mean DoS in the presence of disorder. Therefore,
we take Eq.\,\eqref{eq:Sol_SCBA-1} and build the disorder-averaged
(retarded) SPGF, \textit{i.e.},

\vspace{-0.7cm}

\begin{equation}
\overline{G(E;\mathbf{k})}=\frac{1}{\hbar v_{\text{F}}}\left[\frac{M\left(\epsilon-\varsigma_{\epsilon}\right)\delta_{ab}-\chi\boldsymbol{\sigma}^{ab}\!\cdot\!\mathbf{k}}{M^{2}\left(\epsilon-\varsigma_{\epsilon}\right)^{2}-\abs{\mathbf{k}}^{2}}\right]
\end{equation}
and, with it, we can obtain the mean DoS as follows 
\begin{figure}[t]
\vspace{-0.3cm}
\begin{centering}
\includegraphics[scale=0.21]{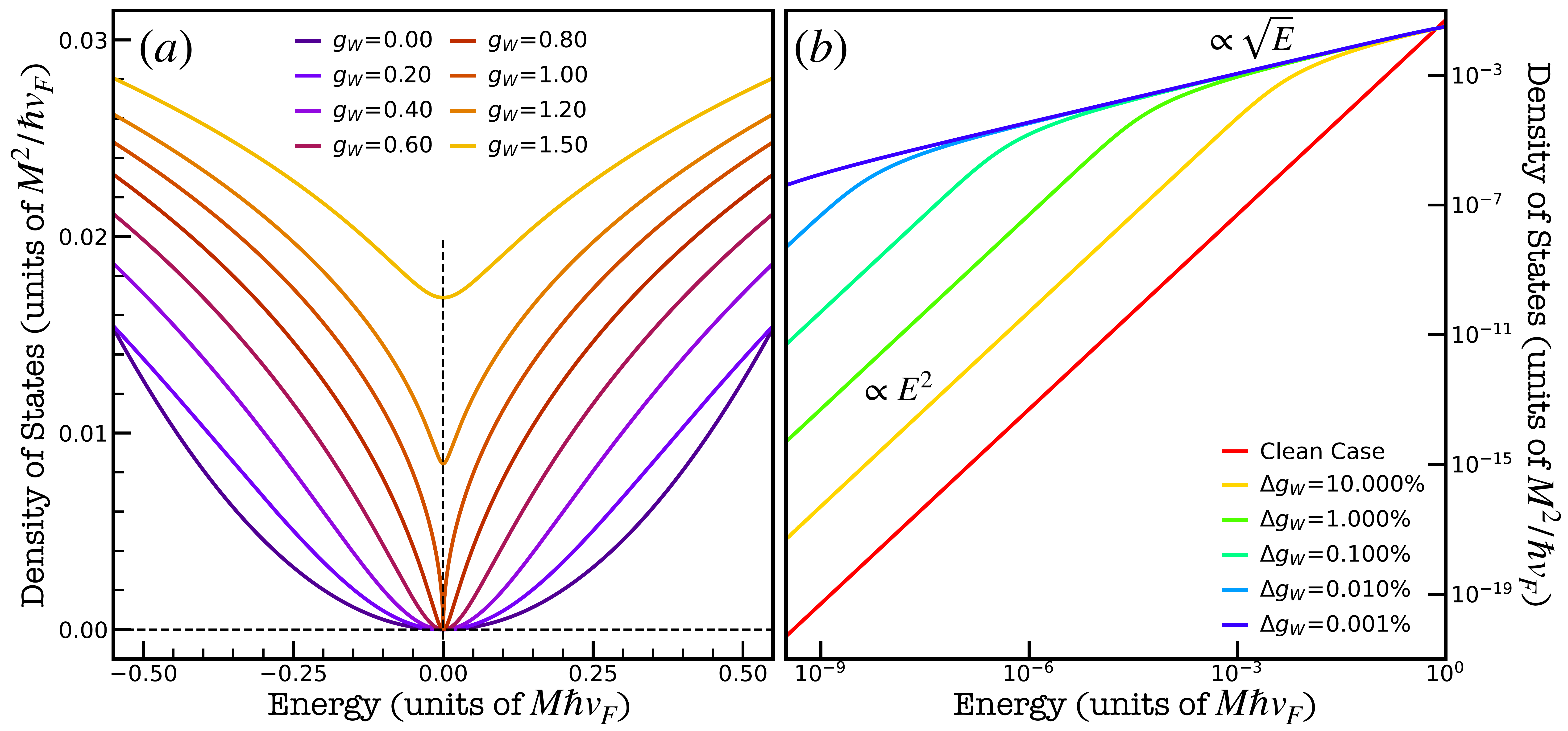}
\par\end{centering}
\vspace{-0.1cm}

\caption{\label{fig:BehaviorDoS1}Behavior of the Mean Density of States within
the Self-Consistent Born Approximation. (a) Plots of the DoS as a
function of energy for different values of the dimensionless parameter
$g_{W}$, below and above the critical value ($g_{W}\!=\!1.0$) for
the SMMT\nomenclature{SMMT}{Semimetal-to-Metal Transition}. (b) Representation
of the DoS for positive energies in a $\log\!-\!\log$ scale. A clear
a change is visible from a quadratic DoS near the nodal energy to
a critical energy behavior, $\overline{\rho(E)}\propto\sqrt{E}$.}

\vspace{-0.3cm}
\end{figure}

\vspace{-0.7cm}

\begin{align}
\overline{\rho(\varepsilon)} & =\frac{M}{4\pi^{4}\hbar v_{\text{F}}}\int\!\!d\mathbf{q}\Im\left[\frac{\epsilon-\varsigma_{\epsilon}}{M^{2}\left(\epsilon-\varsigma_{\epsilon}\right)^{2}-\abs{\mathbf{q}}^{2}}\right]\label{eq:MeanDoSSCBA}\\
 & \qquad\qquad\qquad\qquad\qquad=\frac{M^{2}}{2\pi^{2}\hbar v_{\text{F}}}\left[\frac{\left(\epsilon-\varsigma_{\epsilon}^{\prime}\right)^{2}+\varsigma_{\epsilon}^{\prime\prime}\left(1+\varsigma_{\epsilon}^{\prime\prime}\right)}{\left(\epsilon-\varsigma_{\epsilon}^{\prime}\right)^{2}+\left(1-\varsigma_{\epsilon}^{\prime\prime}\right)^{2}}\right]\nonumber 
\end{align}

where $\varsigma_{\epsilon}\!=\!\varsigma_{\epsilon}^{\prime}+i\varsigma_{\epsilon}^{\prime\prime}$.
Note that the integral over $k$ was also regularized with a smooth
cut-off. In Fig.\,\ref{fig:BehaviorDoS1}, we show the mean DoS calculated
from the SCBA, as a function of energy, for different values of the
dimensionless coupling. All energy (length) scales were also made
dimensionless by using $M\hbar v_{\text{F}}$ ($1/M$) as a natural
energy (length) unit.

\subsection{\label{subsec:SCBACriticality}SCBA: Criticality in the Mean Density
of States}

\begin{figure}[t]
\vspace{-0.3cm}
\begin{centering}
\includegraphics[scale=0.21]{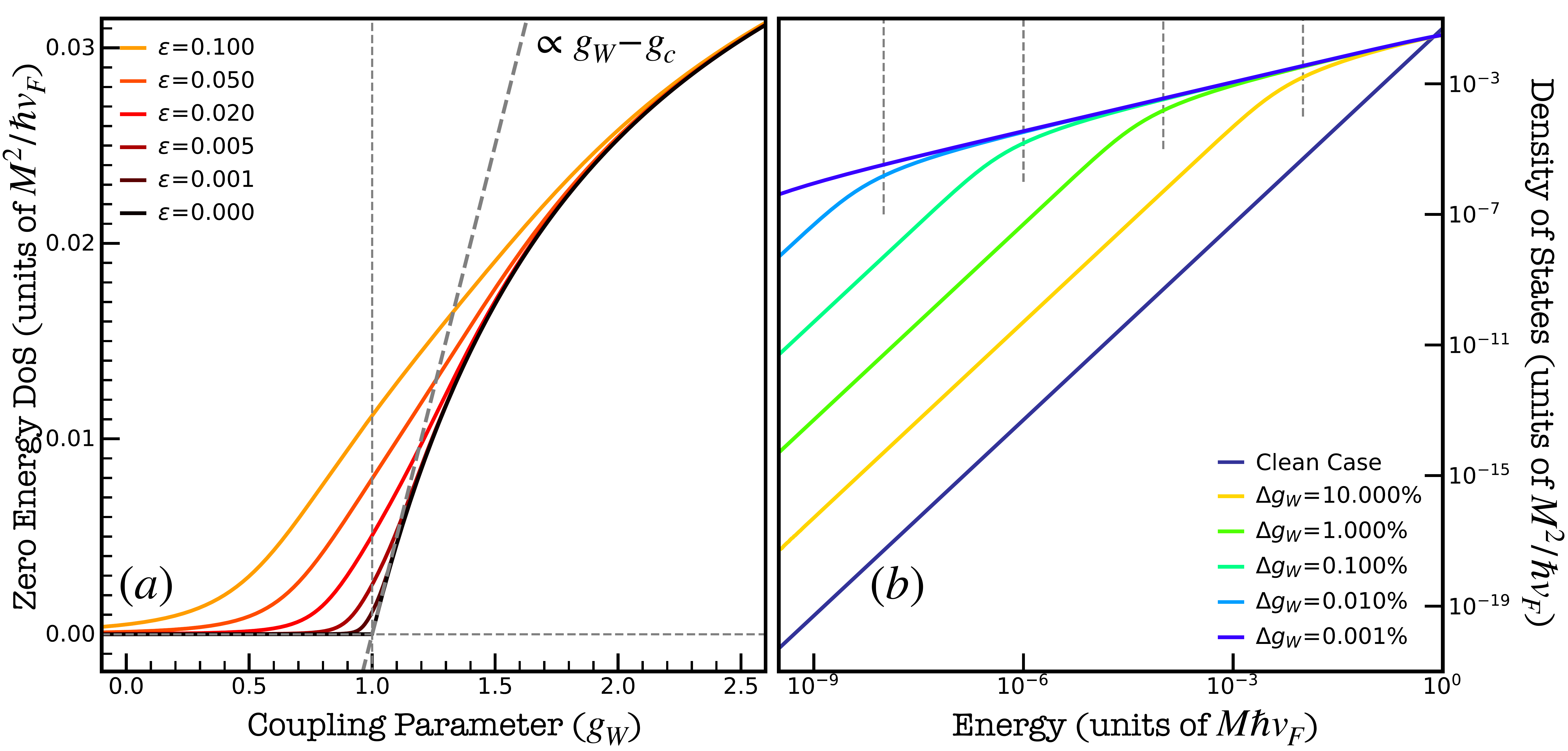}
\par\end{centering}
\vspace{-0.1cm}

\caption{\label{fig:BehaviorDoS2}Critical Behavior of the nodal DoS within
the Self-Consistent Born Approximation. (a) Dependence of the mean
DoS, $\overline{\rho(\varepsilon)}$ as a function of the dimensionless
coupling $g_{W}$ for values of energy approving the Weyl node. The
mean nodal DoS has a critical behavior at $g_{W}\!=\!g_{c}\!=\!1.0$
above which it starts growing linearly with $g_{W}$. (b) As the critical
value for the SMMT is approached from below, the energy at which there
is a transition between the clean Weyl DoS behavior ($\overline{\rho(\varepsilon)}\approx\varepsilon^{2}$
) and the critical behavior, $\overline{\rho(\varepsilon)}\propto\sqrt{\varepsilon}$
approaches zero linearly as well.}

\vspace{-0.3cm}
\end{figure}
The results of Eqs.\,\eqref{eq:Sol_SCBA-1} and \eqref{eq:MeanDoSSCBA},
depicted in Fig.\,\ref{fig:BehaviorDoS1}, unveil a rather interesting
feature of 3D semimetals in the presence of disorder, which was first
discovered by Fradkin\,\cite{Fradkin86b,Fradkin86a} in the 1980's;
There is an\textit{ unconventional disorder-induced critical point}
that precedes Anderson localization in these systems. The plots of
Fig.\,\ref{fig:BehaviorDoS2} sum up the essential features of this
phase transition: \textit{(i)} below the critical point ($g_{W}\!=\!1$)
the DoS gets progressively deformed around the Weyl node as the disorder
strength is increased, and \textit{(ii)} above the critical point
the mean nodal DoS, $\overline{\rho(\varepsilon\!=\!0)}$, acquires
a finite value (the system becomes a diffusive metal).

The DoS deformation that precedes the unconventional transition can
be observed in Fig.\,\ref{fig:BehaviorDoS2}b and amounts to a strong
renormalization of the quadratic curvature for $\epsilon\!\approx\!0$,
that is accompanied by an emergent $\sqrt{\epsilon}$\,-\,dependence
after some finite energy scale, $\epsilon_{0}$. As $g_{W}\!\to\!1^{-}$,
the curvature of the DoS around the node steadily increases and the
transitional energy scale steadily decreases, \textit{i.e.}, $\epsilon_{0}\!\to\!0$\,\nomenclature{$\epsilon_{0}$}{Transitional Energy Scale between $\rho(\epsilon)\propto \epsilon^{2}$ and $\rho(\epsilon)\propto \sqrt{\epsilon}$}.
Prior to $g_{W}\!=\!1$, the system remains a semimetal with $\overline{\rho(\varepsilon\!=\!0)}\!=\!0$,
thus confirming the validity of our earlier heuristic argument regarding
its robustness to weak disorder. Notwithstanding, when $g_{W}\!=\!1$
(defining the critical value, $g_{c}$), the mean DoS becomes a \textit{non-analytic
function of energy} at the nodal energy and, from this point on, the
system starts having $\overline{\rho(\varepsilon\!=\!0)}\!>\!0$ and
becomes a\textit{ ``conventional''} diffusive metal. This is clearly
shown in Fig.\,\ref{fig:BehaviorDoS2}a. As a matter of fact, the
mean DoS at the nodal energy can be seen as a proper order parameter
which describes a disorder-induced continuous quantum phase-transition
that precedes the more common \textit{Anderson Metal-to-Insulator
Transition} (MIT\,\nomenclature{MIT}{Metal-to-Insultator Transition (Anderson Transition)})\,\cite{Syzranov18}.
Before proceeding, it is important to highlight that a critical behavior
seen in the mean DoS is in stark contrast to what happens in a MIT.
It is widely known that the mean DoS is insensitive to the localization
of eigenstates and, therefore, does not show a critical behavior at
an MIT. The latter can only be observed in the probability distribution
of local quantities\,\cite{Altshuler89,Lerner88,VanRossum94,Mirlin96}
(such as the local DoS or the conductivity) or, alternatively, is
signaled by the geometric average of the DoS (the \textit{typical
density of states}\,\cite{Janssen2001,Pixley16b}) that effectively
serves a proper order parameter\,\cite{Schubert10}.

\vspace{-0.5cm}

\section{\label{sec:CriticalTheory}Critical Theory of the Semimetal-to-Metal
Transition}

\vspace{-0.2cm}

The mean DoS shows a critical behavior at the Semimetal-to-Metal transition
that falls within the usual scaling formalism of continuous phase
transitions. Here, we follow Kobayashi \textit{et al}.\,\cite{Kobayashi14}
and derive a scaling theory that describes the universal features
of this critical point. To accomplish this, we start by identifying
the relevant scales and couplings. The coupling parameter which controls
the distance to the critical point is undoubtedly $g_{W}$, whose
critical value is $g_{c}\!=\!1$. As shown in Fig.\,\ref{fig:BehaviorDoS2}b,
the DoS continues to be quadratically vanishing for $g_{W}<g_{c}$,
but a different square-root behavior, $\overline{\rho(E)}\!\propto\!\abs E^{\nicefrac{1}{2}}$,
emerges away from the nodal energy. The change of the DoS from a $E^{2}$
to a $E^{\nicefrac{1}{2}}$ function defines a disorder-induced energy
scale, $E_{0}$, that decreases as $g_{W}\to g_{c}$. This energy
scale can be trivially transformed into an effective \textit{``correlation
length''}, $\xi=\hbar v_{\text{F}}/E_{0}$, that diverges as $g_{W}\!\to\!g_{c}$
from below. At the same time, the value of $\overline{\rho(0)}$ serves
as a continuous order parameter that is finite only for $g_{W}>g_{c}$.

Assuming that $\xi$ is the only relevant length scale (\textit{single-parameter
scaling} \textit{hypothesis}), and $g_{W}$ the only relevant coupling
for this phase transition, we can build up a critical theory for the
mean DoS on dimensional grounds alone. For a start, the (dimensionless)
mean number of states up to an energy $E$ can be written as,

\vspace{-0.7cm}
\begin{equation}
\mathcal{N}_{E}\!=\!\mathscr{V}\xi^{-3}f\left(\frac{E}{E_{0}}\right)
\end{equation}
which is a function of the volume $\mathscr{V}$, the dimensionless
energy $E/E_{0}$, and the correlation length $\xi$. If we assume
that this transition is described by a single-parameter scaling imposes
that 

\vspace{-0.7cm}
\begin{equation}
E_{0}\!=\!\xi^{-z},
\end{equation}
which defines the \textit{dynamical critical exponent,} $z$. Moreover,
we also know that $\xi$ diverges as a power-law when approaching
the transition point from bellow. This allows us to define

\vspace{-0.7cm}
\begin{equation}
\xi\!\propto\!\delta^{-\nu},
\end{equation}
where $\delta\!=\!\abs{\left(g_{0}\!-\!g_{c}\right)/g_{c}}$. Putting
all this together, we have $E_{0}\propto\delta^{-z\nu}$ and also

\vspace{-0.7cm}

\begin{equation}
\mathcal{N}_{E}\!\propto\!\mathscr{V}\delta^{3\nu}g(E\delta^{-z\nu}),
\end{equation}
which, in turn, implies the following scaling law for the mean DoS:

\vspace{-0.7cm}

\begin{equation}
\rho(E,\delta)=\frac{1}{V}\frac{d}{dE}\mathcal{N}_{E}\propto\delta^{\left(3-z\right)\nu}g'(E\delta^{-z\nu}).
\end{equation}
Finally, the arbitrariness in the function $g$ can be removed from
the knowledge of the precise critical behavior of the DoS obtained
in Subsect.\,\ref{subsec:SCBA}. Since the mean DoS is still quadratically
vanishing for $\abs E\ll E_{0}\sim\delta^{z\nu}$,, we have

\vspace{-0.7cm}
\begin{equation}
\rho(E,\delta\!<\!0)\sim\delta^{3\left(1-z\right)\nu}E^{2}
\end{equation}
to leading order in $E/E_{0}$. However, this expression is only valid
if $E_{0}$ is finite, that is for $g_{W}<g_{c}$. Within the diffusive
metal phase, one has a finite $\rho(E\!=\!0)$ instead, which gives
rise to

\vspace{-0.7cm}

\begin{equation}
\rho(E,\delta\!>\!0)\propto\delta^{\nu(3-z)}.
\end{equation}
Precisely at the transition point ($\delta\!=\!0$), the system has
no intrinsic length scales and, therefore,

\vspace{-0.7cm}
\begin{equation}
g'(E\xi^{z})\xi^{-(3-z)}\!=\!h(E)\Longrightarrow g'(x)\sim x^{\frac{3-z}{z}},
\end{equation}
where $h$ is an arbitrary function of the energy. In conclusion,
the scaling function $g$ is a power-law determined only by the dynamical
critical exponent. Hence, near $E\!=\!0$ and close to the transition
point, the mean DoS is expected to have the scaling form,

\vspace{-0.7cm}
\begin{equation}
\!\!\!\!\!\!\!\!\!\!\!\rho(E,\delta)\propto\begin{cases}
\delta^{3\nu(1-z)}E^{2} & \!\!\!\!\delta\!<\!0\\
\abs E^{\frac{3-z}{z}} & \!\!\!\!\delta\!=\!0\\
\delta^{\nu(3-z)} & \!\!\!\!\delta\!>\!0
\end{cases},\!\!\!\label{eq:DOSBranches}
\end{equation}
which is controlled by the pair of universal exponents, $z$ and $\nu$.
From our earlier SCBA results, we can actually conclude that,

\vspace{-0.7cm}
\begin{equation}
\!\!\!\!\!\!\!\!\!\!\!z=2\text{ and }\nu\!=\!1,
\end{equation}
which are the same values obtained in the original work by Fradkin\,\cite{Fradkin86b}.
These critical exponents have been probed in several field-theoretical
studies\,\cite{Goswami11,Syzranov15a,Syzranov15b,Syzranov16,Syzranov18,Klier19}
as well as numerical studies\,\cite{Sbierski14,Bera16,Pixley16b,Wilson20,Pixley21}.
In all the cases, their values seem to be over-estimated by the mean-field
(or SCBA) approach, which demonstrates the importance of considering
\textit{loop-corrections}. For the record, Syzranov \textit{et al}.\,\cite{Syzranov15b}
have shown that $z\!=\!11/8$ and $\nu=2/3$, if a perturbative renormalization
group procedure is used up to a double-loop order.

\subsection{\label{subsec:ObservableSignatures}Observable Signatures of Quantum
Criticality}

It is important to remark that the scaling properties obtained for
the density of states near the node, far from being theoretical benchmarks,
have concrete implications in measurable electronic quantities. Most
notably, in spite of the quantum critical point being a zero temperature
($T\!=\!0$) property of the system, the electronic contribution to
the specific heat (dominant over phonon contributions at low temperatures)
shows a distinctively different behavior with $T$, depending on which
phase the system is. This can be easily seen by considering that the
specific heat at constant volume can be expressed as,

\vspace{-0.7cm}
\begin{equation}
C_{v}\!=\!\frac{\partial}{\partial T}\av E_{T}=-k_{B}\beta^{2}\frac{\partial}{\partial\beta}\int_{-\Lambda}^{\Lambda}\!\!\!\!d\varepsilon\frac{\rho(\varepsilon)\varepsilon}{1+\exp\beta\varepsilon},
\end{equation}
where $k_{B}$ is Boltzmann's constant, and where we have taken the
semimetal as undoped ($E_{\text{F}}\!=\!0$). Since we are only interested
in the scaling of $C_{v}$ with $T$, near $T\!=\!0$, we can avoid
doing the precise energy-integration and simply extract the dependence
in $\beta$. Assuming that $\rho(\varepsilon)\sim\varepsilon^{\alpha}$,
where the exponent $\alpha$ depends on the specific scaling of $\rho(\varepsilon)$
with energy, we conclude that

\vspace{-0.7cm}

\begin{equation}
C_{v}\!=\!-k_{B}\beta^{2}\frac{\partial}{\partial\beta}\frac{1}{\beta^{2+\alpha}}\int_{-\beta\Lambda}^{\beta\Lambda}\!\!\!\!dx\frac{\rho(x)x}{1+\exp x}=k_{B}\left(\frac{2+\alpha}{\beta^{1+\alpha}}\right)C(\beta)-k_{B}\frac{C^{\prime}(\beta)}{\beta^{2+\alpha}},\label{eq:CalorEsp}
\end{equation}
where 

\vspace{-0.7cm}
\begin{equation}
C(\beta,\Lambda)=\int_{-\beta\Lambda}^{\beta\Lambda}\!\!\!\!dx\frac{\rho(x)x}{1+\exp x}=\int_{0}^{\beta\Lambda}\!\!\!\!dx\left[\frac{\rho(x)x}{1+\exp x}-\frac{\rho(x)x}{1+\exp-x}\right].
\end{equation}

Clearly, the dependence of $C$ in $\beta$ is exponentially small
as $\beta\to\infty$ which allows the conclusion that $C_{v}(T)\propto T^{\alpha+1}$,
at low temperatures. Therefore, if we take the three branches of the
scaling function obtained for the DoS {[}Eq.\,\eqref{eq:DOSBranches}{]},
we arrive at the conclusion that

\vspace{-0.7cm}
\[
C_{v}(T)\propto\begin{cases}
T^{2} & \text{Semimetal}\\
T^{\frac{3}{2}} & \text{Critical Phase}\\
T & \text{Metal}
\end{cases}.
\]

\begin{wrapfigure}[18]{o}{0.54\columnwidth}%
\vspace{-0.2cm}

\includegraphics[scale=0.19]{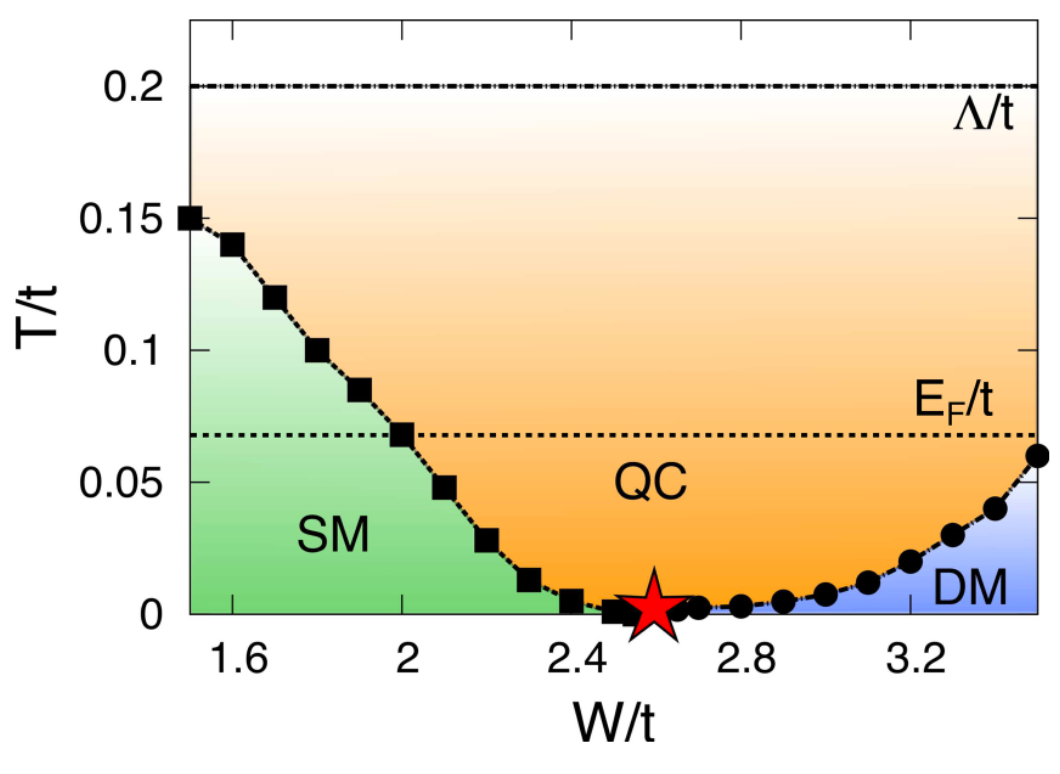}

\vspace{-0.5cm}

\caption{\label{fig:FiniteT_PhaseDiagram}Finite-temperature phase diagram
of a simple-cubic lattice DWSM with an Anderson random scalar potential
of strength $W$. The diagram was numerically obtained by Pixley \textit{et
al}.\,\cite{Pixley16b} using the $T$-dependence of the electronic
specific heat. All quantities are rescaled by the nearest-neighbor
hopping, $t\!=\!\hbar v_{\text{F}}/a$.}
\end{wrapfigure}%
Therefore, in going through the SMMT, the system undergoes a \textit{change
in the temperature-scaling of the specific heat}, which goes from
a $C_{v}\propto T^{2}$ dependence (typical of a 3D DWSM) to the normal
$C_{v}\propto T$ found in free electron gases. However, this transition
proceeds through an intermediate phase, in which, the specific heat
has an anomalous $C_{v}\propto T^{2}$ scaling. More precisely, when
observed at a finite temperature, the system undergoes a \textit{rounded
crossover regime}\,\cite{Pixley16b,Bera16} (dubbed a \textit{``critical
fan''}), as a function of disorder, in which some characteristics
of the zero-temperature critical point are still preserved. In conclusion,
an observable consequence of this unconventional phase transition
would be an \textit{anomalous temperature dependence} of the specific
heat, that would result in the phase-diagram shown in Fig.\,\ref{fig:FiniteT_PhaseDiagram}.

\vspace{-0.6cm}

\section{\label{sec:FieldTheory}Field-Theory of a Disordered Weyl Semimetal}

The previous analysis has shown, by means of a self-consistent diagrammatic
approach, that a DWSM supports an unconventional disorder-induced
quantum phase transition that is marked by a critical behavior in
the mean density of states around the node. These results were obtained
from a standard mean-field many-body formalism\,\cite{Rammer04}
which has some intrinsic limitations, most notably, because \textit{(i)}
it does not set clear limits of applicability, and \textit{(ii)} it
is not amenable to improvement, \textit{e.g.}, by using Renormalization
Group (RG\nomenclature{RG}{Renormalization Group}) methods. These
are the reasons why most literature on the subject, including Fradkin's
seminal work\,\cite{Fradkin86a,Fradkin86b}, opt to cast the ensemble-average
of the SPGF (or other $n$-point correlators) into the form of an
effective \textit{Statistical Field Theory} (SFT\nomenclature{SFT}{Statistical Field Theory (Euclidean QFT)})\,\cite{Popov83}
which can also be treated in the mean-field approximation, but further
improved by using perturbative RG. Historically, this formalism was
introduced by Wegner\,\cite{Wegner79}, as a way to study the Anderson
MIT\,\cite{Lee85} within the so-called \textit{replica trick}\,\footnote{Later on, alternative and more advantageous formulations were also
developed, most notably on the supersymmetric formalism of Efetov\,\cite{Efetov83},
which does not imply replicas.}. In this section, we will review this method with the objective of
using it to formulate the problem of the disorder-induced criticality
in the DoS. In practice, the field-theoretic formalism can become
quite involved if the aim is to study objects that are more complex
that the ensemble-averaged SPGF (\textit{e.g.}, $4$-point correlators
that determine linear response functions). However, for our purposes,
we can (and will) limit ourselves to a simplified replica formulation\,\cite{Cardy78,Wegner79,Fradkin86a,Fradkin86b,Lerner03}
that serves the purpose of calculating the DoS in the presence of
disorder. One example of a simplified SFT formulation is the \textit{Disorder
Saddle-Point Method}, developed by Yaida\,\cite{Yaida16} to analyze
the appearance of\textit{ Lifshitz tails} in semiconducting gaps.

\vspace{-0.5cm}

\subsection{\label{subsec:PathIntegralFormulation}Path-Integral Formulation
for the Single-Particle Propagator}

The starting point for an SFT formulation of the disordered electron
problem is to establish a \textit{Path-Integral Representation}\,\cite{Popov83}
of the SPGF for a given disorder realization. We assume that the clean
system has a generic single-particle Hamiltonian, $\mathcal{H}_{0}$,
to which a random scalar field $V(\mathbf{r})$ is added. Like in
the beginning of this chapter, we can establish the real-space basis
as composed of the states $\ket{\mathbf{r},a}$, where $a$ indicates
any additional quantum numbers. The retarded real-space SPP is then
defined as, 

\vspace{-0.7cm}
\begin{equation}
G_{ab}^{\text{r}}(E;\mathbf{r},\mathbf{r}^{\text{\ensuremath{\prime}}})=\bra{\mathbf{r}^{\text{\ensuremath{\prime}}},b}\left[E\!+\!i\eta\!-\!\mathcal{H}\right]^{-1}\ket{\mathbf{r},a}
\end{equation}
for the full Hamiltonian $\mathcal{H}\!=\!\mathcal{H}_{0}+V$, and
where $\eta$ is a positive infinitesimal. Whilst keeping a generic
argument, one may always use the Källén\textendash Lehmann representation,

\vspace{-0.7cm}

\begin{equation}
G_{a,b}^{\text{r}}(E;\mathbf{r},\mathbf{r}^{\text{\ensuremath{\prime}}})=\sum_{\alpha}\frac{\psi_{\alpha,b}(\mathbf{r}^{\prime})\psi_{\alpha,a}^{*}(\mathbf{r})}{E+i\eta-E_{\alpha}},\label{eq:Lehmann}
\end{equation}
where $\psi_{\alpha,a}(\mathbf{r})=\braket{\mathbf{r},a}{\psi_{\alpha}}$
and $\{E_{\alpha},\ket{\psi_{\alpha}}\}$ are the eigenpairs of a
particular realization of the disordered Hamiltonian. Taking $G^{r}$
in this eigenstate basis, we realize that the scalar function, $1/\left(E+i\eta-E_{\alpha}\right)$,
can actually be represented as a double gaussian integral over complex
variables (see Lerner\,\cite{Lerner03}), \textit{i.e.},

\vspace{-0.7cm}
\begin{align}
\frac{1}{E+i\eta-E_{\alpha}} & =i\iint_{-\infty}^{\infty}\frac{dxdy}{\pi}e^{i\left(E+i\eta-E_{\alpha}\right)\left(x^{2}+y^{2}\right)}\label{eq:LernerRep}\\
 & \qquad\qquad=\iint\frac{du_{\alpha}^{*}du_{\alpha}}{2\pi}e^{iu_{\alpha}^{*}\left(E+i\eta-E_{\alpha}\right)u_{\alpha}}.\nonumber 
\end{align}
In the last equality, we have changed variables to $u\!=\!x\!+\!iy$
and introduced a new index $\alpha$ in the integration variables,
for future convenience. From Eq.\,\eqref{eq:LernerRep}, it becomes
clear that the product $\prod_{\alpha}\left(E+i\eta-E_{\alpha}\right)^{-1}$
can be represented as

\vspace{-0.7cm}
\begin{align}
\mathcal{Z}(E)=\prod_{\alpha}\frac{1}{E+i\eta-E_{\alpha}} & =\prod_{\alpha}\iint\frac{du_{\alpha}^{*}du_{\alpha}}{2\pi}e^{iu_{\alpha}^{*}\left(E+i\eta-E_{\alpha}\right)u_{\alpha}}\label{eq:PartitionFunction}\\
 & =\left(\iint\prod_{\alpha}\frac{du_{\alpha}^{*}du_{\alpha}}{2\pi}\right)e^{i\sum_{\alpha}u_{\alpha}^{*}\left(E+i\eta-E_{\alpha}\right)u_{\alpha}},\nonumber 
\end{align}
which we name, $\mathcal{Z}(E)$, in close analogy to a partition
function. The last line in Eq.\,\eqref{eq:PartitionFunction} can
be formulated as a path-integral in the complex variables $u$, from
using the identity,

\vspace{-0.7cm}
\begin{align}
\sum_{\alpha}u_{\alpha}^{*}\left(E+i\eta-E_{\alpha}\right)u_{\alpha} & =\sum_{\alpha,\beta}u_{\alpha}^{*}\left(E+i\eta-E_{\alpha}\right)u_{\beta}\delta_{\alpha\beta}\label{eq:262}\\
 & =\sum_{\alpha,\beta}u_{\alpha}^{*}\left(E+i\eta-E_{\alpha}\right)u_{\beta}\sum_{a}\int d\mathbf{r}\psi_{\alpha,a}^{*}(\mathbf{r})\psi_{\beta,a}(\mathbf{r}),\nonumber 
\end{align}
which rests on the orthonormality of the single-particle eigenstates
$\ket{\psi_{\alpha}}$. Moreover, since $\mathcal{H}\ket{\psi_{\alpha}}\!=\!E_{\alpha}\ket{\psi_{\alpha}}$
we can re-write Eq.\,\eqref{eq:262} as

\vspace{-0.7cm}

\begin{align}
\!\!\!\!\!\!\!\!\!\!\sum_{\alpha}u_{\alpha}^{*}\left(E\!+\!i\eta\!-\!E_{\alpha}\right)u_{\alpha} & \!\!=\!\!\sum_{a,b}\!\int\!d\mathbf{r}\left[\!\sum_{\alpha}u_{\alpha}^{*}\psi_{\alpha,a}^{*}(\mathbf{r})\left(E\!+\!i\eta\!-\!\mathcal{H}_{ab}(\mathbf{r})\right)\sum_{\beta}u_{\beta}\psi_{\beta,b}(\mathbf{r})\!\right]\!\!\!\!\\
 & \qquad\qquad\qquad=\sum_{a,b}\int d\mathbf{r}\left[\Psi_{a}^{*}(\mathbf{r})\left(E\!+\!i\eta\!-\!\mathcal{H}_{ab}(\mathbf{r})\right)\Psi_{b}(\mathbf{r})\right],\nonumber 
\end{align}
where we define $\Psi_{a}(\mathbf{r})\!=\!\sum_{\alpha}u_{\alpha}\psi_{\alpha,a}(\mathbf{r})$.
At last, the $\infty$\,-\,dimensional integration over $u_{\alpha}^{*}/u_{\alpha}$
can be recast in the form of a path-integral over the complex-valued
local fields, $\Psi_{a}^{*}(\mathbf{r})/\Psi_{a}(\mathbf{r})$, such
that

\vspace{-0.7cm}
\begin{equation}
\mathcal{Z}(E)\!=\!\iint\!\!\mathcal{D}\boldsymbol{\Psi}^{\dagger}\!(\mathbf{r})\mathcal{D}\boldsymbol{\Psi}\!(\mathbf{r})e^{i\mathcal{S}\left[\boldsymbol{\Psi}^{\dagger},\boldsymbol{\Psi}\right]},\label{eq:Det_G}
\end{equation}
with the effective action,

\vspace{-0.7cm}
\begin{equation}
\mathcal{S}\left[\boldsymbol{\Psi}^{\dagger},\boldsymbol{\Psi}\right]\!=\!\!\int\!\!d\mathbf{r}\:\boldsymbol{\Psi}^{\dagger}\!(\mathbf{r})\cdot\left[\left(E\!+\!i\eta\right)\mathcal{I}\!-\!\mathcal{H}(\mathbf{r})\right]\cdot\boldsymbol{\Psi}\!(\mathbf{r}).
\end{equation}
By this point, we have done nothing more than representing the functional
determinant, $\det\left[\left(E\!+\!i\eta\right)\mathcal{I}\!-\!\mathcal{H}(\mathbf{r})\right]^{-1}$,
as the formal path-integral over complex-valued fields shown in Eq.\,\eqref{eq:Det_G}.
In order to obtain a proper generalization to the real-space SPP,
we have to write down the following expressions:

\vspace{-0.7cm}

\begin{align}
\mathcal{J}_{\beta\gamma}(E)\! & =\!\left(\iint\prod_{\alpha}\frac{du_{\alpha}^{*}du_{\alpha}}{2\pi}\right)u_{\beta}^{*}u_{\gamma}e^{i\sum_{\alpha}u_{\alpha}^{*}\left(E+i\eta-E_{\alpha}\right)u_{\alpha}}\label{eq:J_int}\\
 & \qquad=\delta_{\beta\gamma}\left[\iint\frac{du_{\beta}^{*}du_{\beta}}{2\pi}u_{\beta}^{*}u_{\beta}e^{iu_{\beta}^{*}\left(E+i\eta-E_{\beta}\right)u_{\beta}}\right]\times\nonumber \\
 & \qquad\qquad\qquad\times\!\!\left[\prod_{\alpha\neq\beta}\!\iint\!\!\frac{du_{\alpha}^{*}du_{\alpha}}{2\pi}e^{iu_{\alpha}^{*}\left(E+i\eta-E_{\alpha}\right)u_{\alpha}}\right]\nonumber 
\end{align}
where each double integral over $u_{\beta}^{*}$ and $u_{\beta}$
can be expressed as,

\vspace{-0.7cm}
\begin{equation}
\iint\frac{du_{\beta}^{*}du_{\beta}}{2\pi}u_{\beta}^{*}u_{\beta}e^{iu_{\beta}^{*}\left(E+i\eta-E_{\beta}\right)u_{\beta}}=-i\frac{1}{\left(E+i\eta-E_{\beta}\right)^{2}}.\label{eq:EnergyDenom}
\end{equation}
Jointly, Eqs.\,\ref{eq:J_int} and \ref{eq:EnergyDenom} can be simplified
into

\vspace{-0.7cm}
\begin{equation}
\mathcal{J}_{\beta\gamma}(E)\!=\!-\frac{i\delta_{\beta\gamma}}{E+i\eta-E_{\beta}}\left[\underset{\mathcal{Z}(E)}{\underbrace{\prod_{\alpha}\!\iint\!\!\frac{du_{\alpha}^{*}du_{\alpha}}{2\pi}e^{iu_{\alpha}^{*}\left(E+i\eta-E_{\alpha}\right)u_{\alpha}}}}\right]
\end{equation}
or, equivalently, 

\vspace{-0.7cm}
\begin{equation}
\frac{\delta_{\beta\gamma}}{E+i\eta-E_{\beta}}=\frac{i\mathcal{J}_{\beta\gamma}(E)}{\mathcal{Z}(E)}.\label{eq:Representation_SingleTerm}
\end{equation}
Therefore, by using the previous Källén\textendash Lehmann expression
for the real-space SPP {[}Eq.\,\eqref{eq:Lehmann}{]}, together with
Eq.\,\eqref{eq:Representation_SingleTerm}, we can write the propagator
as

\vspace{-0.7cm}

\begin{align}
G_{ab}^{\text{r}}(E;\mathbf{r},\mathbf{r}^{\text{\ensuremath{\prime}}}) & =\sum_{\alpha\beta}\psi_{\alpha,b}(\mathbf{r}^{\prime})\frac{\delta_{\alpha\beta}}{E+i\eta-E_{\alpha}}\psi_{\alpha,a}^{*}(\mathbf{r})\label{eq:Lehmann-1}\\
 & =\frac{i}{\mathcal{Z}(E)}\sum_{\alpha\beta}\psi_{\alpha,b}(\mathbf{r}^{\prime})\mathcal{J}_{\alpha\beta}(E)\psi_{\alpha,a}^{*}(\mathbf{r}),\nonumber 
\end{align}
which means that

\vspace{-0.7cm}

\begin{equation}
G_{ab}^{\text{r}}(E;\mathbf{r}^{\prime},\mathbf{r}^{\prime\prime})\!=\!\frac{i}{\mathcal{Z}_{1}}\!\!\iint\!\!\mathcal{D}\overline{\boldsymbol{\Psi}}\!(\mathbf{r})\mathcal{D}\boldsymbol{\Psi}\!(\mathbf{r})\overline{\Psi}_{a}(\mathbf{r}^{\prime})\Psi_{b}(\mathbf{r}^{\prime\prime})e^{i\mathcal{S}\left[\boldsymbol{\Psi}^{\dagger},\boldsymbol{\Psi}\right]}.\label{eq:Lehmann-1-1}
\end{equation}
Equation\,\eqref{eq:Lehmann-1-1} is exactly the path-integral representation
we have been seeking for a particular realization of the disordered
Hamiltonian. Since we are working exclusively with non-interacting
fields, the precise statistics of $\boldsymbol{\Psi}(\mathbf{r})$
(bosonic or fermionic) do not play a role. Nevertheless, we will still
need to specify the statistics because, upon ensemble-averaging, we
will generate an \textit{effectively interacting action} on these
fields. Hence, from now on, we shall consider that $\overline{\boldsymbol{\Psi}}/\boldsymbol{\Psi}$
are \textit{Grassmann Fields} that correctly reproduce the fermionic
statistics of the particles. 

Before moving to the corresponding representation of the ensemble-averaged
propagator, it is important to rephrase Eq.\,\eqref{eq:Lehmann-1-1}
as a functional derivative\,\cite{Popov83}. For that, we re-define
the generating functional as

\vspace{-0.7cm}
\begin{equation}
\mathcal{Z}_{1}\left[\boldsymbol{J}^{\dagger}(\mathbf{r}),\boldsymbol{J}(\mathbf{r})\right]\!=\!\iint\!\!\mathcal{D}\overline{\boldsymbol{\Psi}}\!(\mathbf{r})\mathcal{D}\boldsymbol{\Psi}\!(\mathbf{r})e^{i\tilde{\mathcal{S}}\left[\overline{\boldsymbol{\Psi}},\boldsymbol{\Psi},\boldsymbol{J}^{\dagger},\boldsymbol{J}\right]},
\end{equation}
with a modified action,

\vspace{-0.7cm}

\begin{equation}
\tilde{\mathcal{S}}\left[\overline{\boldsymbol{\Psi}},\boldsymbol{\Psi},\boldsymbol{J}^{\dagger},\boldsymbol{J}\right]\!=\!\!\int\!\!d\mathbf{r}\:\overline{\boldsymbol{\Psi}}\!(\mathbf{r})\cdot\left[\left(E\!+\!i\eta\right)\mathcal{I}\!-\!\mathcal{H}(\mathbf{r})\right]\cdot\boldsymbol{\Psi}\!(\mathbf{r})\!+\!\overline{\boldsymbol{J}}\!(\mathbf{r})\!\cdot\!\boldsymbol{\Psi}\!(\mathbf{r})\!+\!\overline{\boldsymbol{\Psi}}\!(\mathbf{r})\!\cdot\!\boldsymbol{J}(\mathbf{r}),
\end{equation}
that now includes a set of complex \textit{source-term fields}, $\overline{\boldsymbol{J}}(\mathbf{r})/\boldsymbol{J}(\mathbf{r})$,
which couple linearly to the dynamical fields. In the presence of
these auxiliary terms, we are allowed to write down the real-space
SPP as,

\vspace{-0.7cm}
\begin{equation}
G_{ab}^{\text{r}}(E;\mathbf{r}^{\prime},\mathbf{r}^{\prime\prime})\!=\!-i\frac{\delta^{2}}{\delta\overline{J}_{a}(\mathbf{r}^{\prime})\,\delta J_{b}(\mathbf{r}^{\prime\prime})}\ln\mathcal{Z}_{1}\left[\overline{\boldsymbol{J}}(\mathbf{r}),\boldsymbol{J}(\mathbf{r})\right],\label{eq:G_Av}
\end{equation}
which is a typical result of quantum field theory. With Eq.\,\eqref{eq:G_Av},
we have achieved the goal of rephrasing the retarded real-space SPP
in a SFT language. To close the section, we remark that an equivalent
(but redundant) representation could have been achieved for the advanced
propagator, which simply involves setting $\eta\to-\eta$ in all previous
equations. As referred before, the simultaneous use of both representations
\textit{is crucial} for calculating the relevant quantities for the
Anderson transition\,\cite{Wegner79,Efetov83,Lee85,Lerner03}, \textit{e.g.},
conductivities or local densities of states, in which vertex-corrections
have to be included. For our purposes, we can get away with only one
of them.

\subsection{\label{subsec:DisorderAveraging}Disorder-Averaging and the Replica
Method}

The great usefulness of a path-integral representation of the SPGF
is that it becomes much easier to average it over random field configurations.
For that, we assume that $\mathcal{V}V(\mathbf{r})$ is a continuum
random field, with a spinor structure given by the operator $\mathcal{V}$,
and whose local values follow a \textit{gaussian statistics}, \textit{i.e.},

\vspace{-0.7cm}
\begin{equation}
\mathcal{P}[V(\mathbf{r})]=\mathcal{N}\exp\left[-\frac{1}{2w^{2}}\!\iint\!\!d\mathbf{r}d\mathbf{r}^{\prime}V(\mathbf{r}^{\prime})C(\mathbf{r}^{\prime}\!-\!\mathbf{r})V(\mathbf{r})\right],\label{eq:RandomContinuumPot}
\end{equation}
where $\mathcal{N}$ is a normalization constant, the integral kernel
$C(\mathbf{r}^{\prime}\!-\!\mathbf{r})$ is the \textit{normalized
space-correlator} of the disorder potential (\textit{e.g.}, see Refs.\,\cite{Khan19,Pires19}
for details on correlated disordered landscapes), and $w$ measures
the local disorder strength. If we aim to evaluate the average of
$G^{r}$ with respect to the potential landscape $V(\mathbf{r})$,
we can formally write

\vspace{-0.7cm}
\begin{align}
\overline{G_{a,b}^{\text{r}}(E;\mathbf{r}^{\prime},\mathbf{r}^{\prime\prime})}\! & =\!-i\mathcal{N}\frac{\delta^{2}}{\delta\overline{J}_{a}(\mathbf{r}^{\prime})\,\delta J_{b}(\mathbf{r}^{\prime\prime})}\int\!\!\mathcal{D}V(\mathbf{r})\ln\mathcal{Z}_{1}\!\left[V(\mathbf{r}),\overline{\boldsymbol{J}}(\mathbf{r}),\boldsymbol{J}(\mathbf{r})\right]\label{eq:AvProp}\\
 & \qquad\qquad\qquad\qquad\qquad\exp\left[\!-\frac{1}{2w^{2}}\!\iint\!\!d\mathbf{r}d\mathbf{r}^{\prime}V(\mathbf{r}^{\prime})C(\mathbf{r}^{\prime}\!-\!\mathbf{r})V(\mathbf{r})\right]\!,\nonumber 
\end{align}
with a generating functional that reads,

\vspace{-0.7cm}
\begin{equation}
\mathcal{Z}_{1}\!\left[V(\mathbf{r}),\overline{\boldsymbol{J}}(\mathbf{r}),\boldsymbol{J}(\mathbf{r})\right]\!=\!\iint\!\!\mathcal{D}\overline{\boldsymbol{\Psi}}\!(\mathbf{r})\mathcal{D}\boldsymbol{\Psi}\!(\mathbf{r})e^{i\tilde{\mathcal{S}}\left[V,\overline{\boldsymbol{\Psi}},\boldsymbol{\Psi},\overline{\boldsymbol{J}},\boldsymbol{J}\right]}
\end{equation}
in terms of the following action (with sources):

\vspace{-0.7cm}

\begin{align}
\tilde{\mathcal{S}}\left[V,\overline{\boldsymbol{\Psi}},\boldsymbol{\Psi},\overline{\boldsymbol{J}},\boldsymbol{J}\right]\! & =\!\!\int\!\!d\mathbf{r}\:\overline{\boldsymbol{\Psi}}\!(\mathbf{r})\cdot\left[\left(E\!+\!i\eta\right)\mathcal{I}\!-\!\mathcal{H}_{0}(\mathbf{r})\!-\!\mathcal{V}V(\mathbf{r})\right]\cdot\boldsymbol{\Psi}\!(\mathbf{r})\\
 & \qquad\qquad\qquad\qquad\qquad+\!\overline{\boldsymbol{J}}\!(\mathbf{r})\!\cdot\!\boldsymbol{\Psi}\!(\mathbf{r})\!+\!\overline{\boldsymbol{\Psi}}\!(\mathbf{r})\!\cdot\!\boldsymbol{J}(\mathbf{r}).\nonumber 
\end{align}
Looking at Eq.\,\eqref{eq:AvProp}, an important observation comes
to mind; if we had $\mathcal{Z}_{1}$, instead of $\ln\mathcal{Z}_{1}$,
we could hope to perform an integration over $V(\mathbf{r})$, because
the path-integral would be gaussian. However, this is not the case
and a clever trick is required to go around this problem. One such
method is the celebrated \textit{Replica Trick}\,\cite{Edwards75,Parisi79,Lerner03},
where one formally writes

\vspace{-0.7cm}
\begin{equation}
\ln\mathcal{Z}_{1}\!\!\left[V(\mathbf{r}),\overline{\boldsymbol{J}}(\mathbf{r}),\boldsymbol{J}(\mathbf{r})\right]=\lim_{n\to0}\left[\frac{\mathcal{Z}_{1}^{n}\!\!\left[V(\mathbf{r}),\overline{\boldsymbol{J}}(\mathbf{r}),\boldsymbol{J}(\mathbf{r})\right]-1}{n}\right]\label{eq:ReplicaTrick}
\end{equation}
and identifies $\mathcal{Z}_{1}^{n}$ with the product of functionals
for $n$ uncoupled replicas of the disordered system, each replica
featuring the same disorder realization. To make this identification
clearer, we \textit{``expand''} the internal quantum-number space
of our local fields by means of an additional \textit{``replica label''}
in $\Psi_{a}^{*}(\mathbf{r})/\Psi_{a}(\mathbf{r})$. Thereby, we define
the replicated field as follows:

\vspace{-0.7cm}

\begin{equation}
\boldsymbol{\Phi}(\mathbf{r})=\left[\Psi_{1,1}(\mathbf{r}),\Psi_{2,1}(\mathbf{r}),\cdots,\Psi_{n_{s},1}(\mathbf{r}),\Psi_{1,2}(\mathbf{r}),\Psi_{2,2}(\mathbf{r}),\cdots,\Psi_{n_{s},n}(\mathbf{r})\right],
\end{equation}
where the first index labels the internal quantum numbers of a single
replica ($a=1,\cdots,n_{s}$) and the second index labels the replica
itself. Since all replicas are uncoupled, the total action for this
extended field is just the sum of individual actions,

\vspace{-0.7cm}

\begin{align}
\tilde{\mathcal{S}}\left[V,\overline{\boldsymbol{\Phi}},\boldsymbol{\Phi},\overline{\boldsymbol{J}},\boldsymbol{J}\right]\! & =\!\!\int\!\!d\mathbf{r}\:\overline{\boldsymbol{\Phi}}(\mathbf{r})\cdot\left[\left(E\!+\!i\eta\right)\mathbf{I}\!-\!\boldsymbol{\Gamma}V(\mathbf{r})\!-\!\mathbf{H}_{0}(\mathbf{r})\right]\cdot\boldsymbol{\Phi}(\mathbf{r})\\
 & \qquad\qquad\qquad\qquad\qquad+\!\overline{\boldsymbol{J}}(\mathbf{r})\!\cdot\!\boldsymbol{\Phi}(\mathbf{r})\!+\!\overline{\boldsymbol{\Phi}}(\mathbf{r})\!\cdot\!\boldsymbol{J}(\mathbf{r}),\nonumber 
\end{align}
where $\mathbf{I}$ is an identity matrix in the expanded space, $\boldsymbol{\Gamma}\!=\!\mathcal{V}\tensorproduct\mathcal{I}_{n\times n}$
describes the \textit{local (spinor) structure of the random potential}
in the expanded Hilbert space, and $\mathbf{H}_{0}$ is the clean
expanded Hamiltonian. Note that $\mathbf{H}_{0}$ ($\boldsymbol{\Gamma}$)
is an $nn_{s}\times nn_{s}$ block-diagonal matrix, composed of $n_{s}\times n_{s}$
blocks which are the Hamiltonians of a clean replica (the matrix structure
of the random potential). The source fields were also expanded into
the space of $n$ replicas. Finally, we have the following expression
for the averaged $n$\,-\,replica functional:

\vspace{-0.7cm}
\begin{align}
\!\!\!\!\!\!\!\!\!\!\!\!\!\!\!\!\!\!\!\!\!\!\!\!\!\!\!\!\!\!\!\!\!\!\!\!\!\!\!\!\!\!\!\!\!\!\!\!\!\!\!\!\!\!\!\!\!\!\!\!\!\!\!\!\overline{\mathcal{Z}_{n}}\! & =\!\mathcal{N}\!\iint\!\!\mathcal{D}\overline{\boldsymbol{\Phi}}(\mathbf{r})\mathcal{D}\boldsymbol{\Phi}(\mathbf{r})\!\!\int\!\!\mathcal{D}V(\mathbf{r})\exp\left[i\int\!\!d\mathbf{r}\:\overline{\boldsymbol{\Phi}}(\mathbf{r})\!\cdot\!\left[\left(E\!+\!i\eta\right)\mathbf{I}\!-\!\boldsymbol{\Gamma}V(\mathbf{r})\!-\!\mathbf{H}_{0}(\mathbf{r})\right]\!\cdot\boldsymbol{\Phi}(\mathbf{r})\right.\!\!\!\!\!\nonumber \\
 & \qquad\qquad\qquad\left.\!+\!\overline{\boldsymbol{J}}(\mathbf{r})\!\cdot\!\boldsymbol{\Phi}(\mathbf{r})\!+\!\overline{\boldsymbol{\Phi}}(\mathbf{r})\!\cdot\!\boldsymbol{J}(\mathbf{r})\right]\exp\left[-\frac{1}{2w^{2}}\!\iint\!\!d\mathbf{r}d\mathbf{r}^{\prime}\,V(\mathbf{r}^{\prime})C(\mathbf{r}^{\prime}\!-\!\mathbf{r})V(\mathbf{r})\right]
\end{align}
which can be neatly split into two parts,
\begin{align}
\overline{\mathcal{Z}_{n}}\! & =\!\mathcal{N}\!\iint\!\!\mathcal{D}\overline{\boldsymbol{\Phi}}(\mathbf{r})\mathcal{D}\boldsymbol{\Phi}\!(\mathbf{r})e^{i\tilde{\mathcal{S}}\left[0,\overline{\boldsymbol{\Phi}},\boldsymbol{\Phi},\overline{\boldsymbol{J}},\boldsymbol{J}\right]}\!\\
 & \times\,\int\!\!\mathcal{D}V(\mathbf{r})\exp\left[\!-\!\int\!\!d\mathbf{r}\left(i\overline{\boldsymbol{\Phi}}(\mathbf{r})\!\cdot\!\boldsymbol{\Gamma}\!\cdot\boldsymbol{\Phi}(\mathbf{r})\!+\!\frac{1}{2w^{2}}\!\int\!d\mathbf{r}^{\prime}V(\mathbf{r}^{\prime})C(\mathbf{r}^{\prime}\!-\!\mathbf{r})\right)V(\mathbf{r})\right]\nonumber 
\end{align}
such that the dependence on $V(\mathbf{r})$ is all contained within
the second term. Furthermore, the path-integral over $V(\mathbf{r})$
is of the gaussian type, \textit{i.e.}, the only path integral type
we can solve analytically. To explicitly compute it, we begin by defining
$C^{-1}(\mathbf{r}^{\prime}\!-\!\mathbf{r})$ as the inverse correlator
of the random potential, that is,

\vspace{-0.7cm}
\begin{equation}
\int\!\!d\mathbf{r}C^{-1}\!(\mathbf{r}^{\prime}\!-\!\mathbf{r})C(\mathbf{r}\!-\!\mathbf{r}^{\prime\prime})=\!\!\int\!\!d\mathbf{r}C(\mathbf{r}^{\prime}\!-\!\mathbf{r})C^{-1}\!(\mathbf{r}\!-\!\mathbf{r}^{\prime\prime})=\delta(\mathbf{r}^{\prime}\!-\!\mathbf{r}^{\prime\prime})
\end{equation}
and, at the same time, we can also recast the exponent into the following
form

\vspace{-0.7cm}
\begin{align}
\!\!\!\!\!\!\!\iint\!\!d\mathbf{r}d\mathbf{r}^{\prime}V(\mathbf{r}^{\prime})\left(i\overline{\boldsymbol{\Phi}}(\mathbf{r})\!\cdot\!\boldsymbol{\Gamma}\!\cdot\!\boldsymbol{\Phi}(\mathbf{r})\,\delta(\mathbf{r}^{\prime}\!-\!\mathbf{r})\!+\!\frac{1}{2w^{2}}C(\mathbf{r}^{\prime}\!-\!\mathbf{r})\right)V(\mathbf{r})=\qquad\qquad\qquad\qquad\\
=\!\frac{1}{2w^{2}}\!\iint\!\!d\mathbf{r}d\mathbf{r}^{\prime}\!\left(\!V(\mathbf{r}^{\prime})\!+\!iw\!\!\int\!\!d\mathbf{r}^{\prime\prime}C^{-1}\!(\mathbf{r}^{\prime}\!-\!\mathbf{r}^{\prime\prime})\overline{\boldsymbol{\Phi}}(\mathbf{r}^{\prime\prime})\!\cdot\!\boldsymbol{\Gamma}\!\cdot\!\boldsymbol{\Phi}(\mathbf{\mathbf{r}^{\prime\prime}})\!\right)\qquad\qquad\nonumber \\
\times C\!(\mathbf{r}^{\prime}\!\!-\!\mathbf{r})\!\left(\!V(\mathbf{r})\!+\!iw\!\!\int\!\!d\mathbf{r}^{\prime\prime}C^{-1}\!(\mathbf{r}\!-\!\mathbf{r}^{\prime\prime})\overline{\boldsymbol{\Phi}}(\mathbf{r}^{\prime\prime})\!\cdot\!\boldsymbol{\Gamma}\!\cdot\!\boldsymbol{\Phi}(\mathbf{\mathbf{r}^{\prime\prime}})\!\right)\qquad\nonumber \\
+\frac{w^{2}}{2}\!\iint\!\!d\mathbf{r}d\mathbf{r}^{\prime}\,\overline{\boldsymbol{\Phi}}(\mathbf{r}^{\prime})\!\cdot\!\boldsymbol{\Gamma}\!\cdot\!\boldsymbol{\Phi}(\mathbf{\mathbf{r}^{\prime}})C^{-1}\!(\mathbf{r}^{\prime}\!-\!\mathbf{r})\overline{\boldsymbol{\Phi}}(\mathbf{r})\!\cdot\!\boldsymbol{\Gamma}\!\cdot\!\boldsymbol{\Phi}(\mathbf{\mathbf{r}})\nonumber 
\end{align}
which, under the change of variables $V(\mathbf{r})\!\to\!V(\mathbf{r})\!+\!iw\!\int\!\!d\mathbf{r}^{\prime\prime}\,C^{-1}\!(\mathbf{r}\!-\!\mathbf{r}^{\prime\prime})\boldsymbol{\Phi}^{\dagger}(\mathbf{r}^{\prime\prime})\!\cdot\boldsymbol{\Phi}(\mathbf{\mathbf{r}^{\prime\prime}})$,
leads to the following result:

\vspace{-0.7cm}
\begin{align}
\int\!\!\mathcal{D}V(\mathbf{r})\exp\left[\!-\!\int\!\!d\mathbf{r}\left(i\overline{\boldsymbol{\Phi}}\!(\mathbf{r})\!\cdot\!\boldsymbol{\Gamma}\!\cdot\!\boldsymbol{\Phi}(\mathbf{r})\!+\!\frac{1}{2w^{2}}\!\int\!d\mathbf{r}^{\prime}V(\mathbf{r}^{\prime})C(\mathbf{r}^{\prime}\!-\!\mathbf{r})\right)V(\mathbf{r})\right]\qquad\qquad & \!\!\!\!\!\!\!\!\\
=\frac{1}{\mathcal{N}}\exp\left[-\frac{w^{2}}{2}\!\iint\!\!d\mathbf{r}d\mathbf{r}^{\prime}\overline{\boldsymbol{\Phi}}\!(\mathbf{r}^{\prime})\!\cdot\!\boldsymbol{\Gamma}\!\cdot\!\boldsymbol{\Phi}\!(\mathbf{\mathbf{r}^{\prime}})C^{-1}\!(\mathbf{r}^{\prime}\!-\!\mathbf{r})\overline{\boldsymbol{\Phi}}\!(\mathbf{r})\!\cdot\!\boldsymbol{\Gamma}\!\cdot\!\boldsymbol{\Phi}\!(\mathbf{\mathbf{r}})\right] & .\nonumber 
\end{align}
Finally, we arrived at our central result: the ensemble-averaged generating
functional for $n$ replicas of the system can be written as the path-integral,

\vspace{-0.7cm}
\begin{equation}
\overline{\mathcal{Z}_{n}}\left[\boldsymbol{J}^{\dagger},\boldsymbol{J}\right]\!=\!\!\iint\!\!\mathcal{D}\overline{\boldsymbol{\Phi}}\!(\mathbf{r})\mathcal{D}\boldsymbol{\Phi}\!(\mathbf{r})e^{-\overline{\mathcal{S}}\left[\overline{\boldsymbol{\Phi}},\boldsymbol{\Phi},\overline{\boldsymbol{J}},\boldsymbol{J}\right]}\!,\label{eq:n_ReplicaFunctional}
\end{equation}
with the effective action,

\vspace{-0.7cm}
\begin{align}
\!\!\!\!\!\!\!\!\!\!\!\!\!\overline{\mathcal{S}}\left[\overline{\boldsymbol{\Phi}},\boldsymbol{\Phi},\overline{\boldsymbol{J}},\boldsymbol{J}\right]= & -i\int\!\!d\mathbf{r}\:\overline{\boldsymbol{\Phi}}\!(\mathbf{r})\cdot\left[\left(E\!+\!i\eta\right)\mathbf{I}\!-\!\mathbf{H}_{0}(\mathbf{r})\right]\cdot\boldsymbol{\Phi}\!(\mathbf{r})-i\overline{\boldsymbol{J}}\!(\mathbf{r})\!\cdot\!\boldsymbol{\Phi}\!(\mathbf{r})\label{eq:n_ReplicaAction}\\
\qquad-i\overline{\boldsymbol{\Phi}}\!(\mathbf{r})\!\cdot\!\boldsymbol{J}(\mathbf{r}) & +\frac{w^{2}}{2}\!\iint\!\!d\mathbf{r}d\mathbf{r}^{\prime}\overline{\boldsymbol{\Phi}}(\mathbf{r}^{\prime})\!\cdot\!\boldsymbol{\Gamma}\!\cdot\!\boldsymbol{\Phi}(\mathbf{\mathbf{r}^{\prime}})C^{-1}\!(\mathbf{r}^{\prime}\!-\!\mathbf{r})\overline{\boldsymbol{\Phi}}(\mathbf{r})\!\cdot\!\boldsymbol{\Gamma}\!\cdot\!\boldsymbol{\Phi}(\mathbf{\mathbf{r}}).\nonumber 
\end{align}
Equations\,\eqref{eq:n_ReplicaFunctional}-\eqref{eq:n_ReplicaAction}
now describe the dynamics of a Grassmann field, with $n_{s}n$ local
components, that correspond to the $n$ replicas of the original fermionic
fields. Interestingly, we have managed to trade-off the inconvenient
\textit{sample-specific random field}, $V(\mathbf{r})$, by a \textit{non-random
statistically-induced quartic interaction term} among the components
of the replicated fields.

\vspace{-0.5cm}

\subsection{\label{subsec:Statistical-Field-Theory}Statistical Field-Theory
of Disordered Weyl Electrons}

By joining together our early path-integral representation of the
real-space propagator {[}Eq.\,\eqref{eq:AvProp}{]}, the replica
trick stated in Eq.\,\eqref{eq:ReplicaTrick}, and the previous representation
of $\overline{\mathcal{Z}_{n}}$, we are now in position to consider
a proper SFT description for the disorder-averaged SPGF with an uncorrelated
(white-noise) disordered field. This simpler case corresponds to taking
the space-correlator of the random potential as $C^{-1}\!(\mathbf{r}^{\prime}\!-\!\mathbf{r})=\delta(\mathbf{r}^{\prime}\!-\!\mathbf{r})$,
which allows us to write down

\vspace{-0.7cm}

\begin{align}
\overline{G_{a,b}^{\text{r}}(E;\mathbf{r}^{\prime},\mathbf{r}^{\prime\prime})}\! & =\lim_{n\to0}\left(\frac{i}{n\overline{\mathcal{Z}_{n}}}\!\iint\!\!\mathcal{D}\overline{\boldsymbol{\Phi}}\!(\mathbf{r})\mathcal{D}\boldsymbol{\Phi}\!(\mathbf{r})\overline{\boldsymbol{\Phi}}(\mathbf{r}^{\prime\prime})\!\cdot\!\boldsymbol{\Phi}\!(\mathbf{r}^{\prime})e^{-\int\!d\mathbf{r}\mathcal{L}\left[\overline{\boldsymbol{\Phi}},\boldsymbol{\Phi}\right]}\right)\label{eq:AVERAGEDPorp}\\
 & =i\!\lim_{n\to0}\left(\!\frac{1}{\overline{\mathcal{Z}_{n}}}\iint\!\!\mathcal{D}\overline{\boldsymbol{\Phi}}\!(\mathbf{r})\mathcal{D}\boldsymbol{\Phi}\!(\mathbf{r})\overline{\psi}_{b,1}(\mathbf{r}^{\prime\prime})\psi_{a,1}(\mathbf{r}^{\prime})e^{-\int\!d\mathbf{r}\mathcal{L}\left[\overline{\boldsymbol{\Phi}},\boldsymbol{\Phi}\right]}\right),\nonumber 
\end{align}
with an effective Lagrangian density given as 

\vspace{-0.7cm}
\begin{equation}
\mathcal{L}\left[\overline{\boldsymbol{\Phi}},\boldsymbol{\Phi}\right]=-i\overline{\boldsymbol{\Phi}}(\mathbf{r})\cdot\left[\left(E\!+\!i\eta\right)\mathbf{I}\!-\!\mathbf{H}_{0}(\mathbf{r})\right]\cdot\boldsymbol{\Phi}\!(\mathbf{r})+\frac{w^{2}}{2}\!\left(\overline{\boldsymbol{\Phi}}(\mathbf{r})\cdot\!\boldsymbol{\Gamma}\!\cdot\boldsymbol{\Phi}(\mathbf{\mathbf{r}})\right)^{2}
\end{equation}
in terms of the hybrid $n\times n_{s}$-component $\boldsymbol{\Phi}^{\dagger}/\boldsymbol{\Phi}$
fields. Note that, in Eq.\,\eqref{eq:AVERAGEDPorp}, the factor $1/n$
(from the replica trick) was eliminated, as there is nothing distinguishing
different replicas of the system, \textit{i.e.}, for any finite $n$
one must have $\av{\boldsymbol{\Phi}^{\dagger}(\mathbf{r}^{\prime\prime})\cdot\!\boldsymbol{\Gamma}\!\cdot\boldsymbol{\Phi}(\mathbf{r}^{\prime})}\!=\!n\av{\sum_{a}\psi_{a,1}^{*}(\mathbf{r}^{\prime\prime})\Gamma_{ab}\psi_{b,1}(\mathbf{r}^{\prime})}$.\,To
specialize the general SFT framework to our case of interest, we replace
the clean Hamiltonian, $\mathbf{H}_{0}$, by a replicated Weyl Hamiltonian
with $N_{v}$ valleys. This yields the Lagrangian

\vspace{-0.7cm}
\begin{align}
\!\!\!\!\!\!\!\!\mathcal{L}\left[\overline{\boldsymbol{\Phi}},\boldsymbol{\Phi}\right] & =\sum_{r=1}^{n}\sum_{a,b}\left[\left(\eta\!-\!iE\right)\overline{\psi}_{a,r}(\mathbf{r})\psi_{a,r}(\mathbf{r})\!+\!\hbar v_{\text{F}}\overline{\psi}_{a,r}(\mathbf{r})\boldsymbol{\sigma}^{ab}\!\!\cdot\!\boldsymbol{\nabla}_{\mathbf{r}}\psi_{b,r}(\mathbf{r})\right]+\label{eq:EffectiveL}\\
 & \qquad\qquad\qquad\qquad\qquad\qquad\qquad\qquad+\frac{w^{2}}{2}\left(\sum_{r=1}^{n}\sum_{a,b}\overline{\psi}_{a,r}(\mathbf{r})\Gamma_{ab}\psi_{b,r}(\mathbf{r})\right)^{2}\nonumber 
\end{align}
In this context, the local quantum numbers have two distinct physical
origins: \textit{(i)} the conduction/valence band for each Weyl node,
and \textit{(ii)} the labeling of the Weyl node itself. To make this
concrete, we now unwrap the local indices into an index $\sigma\!=\!\pm1$
(labelling the band), and $i=1,\cdots,N_{v}$ (labelling the valley).
This turns Eq.\,\eqref{eq:EffectiveL} into the more explicit form,

\vspace{-0.7cm}
\begin{align}
\!\!\!\!\!\!\!\!\mathcal{L}\!\left[\overline{\boldsymbol{\Phi}},\boldsymbol{\Phi}\right]\!\! & =\!\!\!\sum_{r=1}^{n}\!\sum_{\sigma=\pm}\!\sum_{j=1}^{N_{v}}\!\left[\left(\eta\!-\!iE\right)\overline{\psi}_{aj,r}(\mathbf{r})\psi_{aj,r}(\mathbf{r})\!+\!\hbar v_{\text{F}}\overline{\psi}_{\sigma j,r}(\mathbf{r})\boldsymbol{\sigma}^{\sigma\sigma'}\!\!\!\!\cdot\!\boldsymbol{\nabla}_{\mathbf{r}}\psi_{\sigma'\!j,r}(\mathbf{r})\right]\!+\label{eq:EffectiveL-2}\\
 & \qquad\qquad\qquad\qquad\qquad\qquad\qquad+\frac{w^{2}}{2}\left(\sum_{r=1}^{n}\sum_{\sigma\sigma'}\sum_{i,j=1}^{N_{v}}\overline{\psi}_{\sigma i,r}(\mathbf{r})\Gamma_{ij}^{\sigma\sigma'}\psi_{\sigma'j,r}(\mathbf{r})\right)^{2}.\nonumber 
\end{align}
To proceed further, we must devise an expression for the $\boldsymbol{\Gamma}$-matrix
that describes all the possible couplings induced by the random potential.
Here, we assume that the random potential is scalar\textit{, i.e.},
it does not couple different bands, but it is certainly able to scatter
electrons among the different Weyl nodes. In fact, as described by
Fradkin\,\cite{Fradkin86a,Fradkin86b}, if the potential is short-ranged
correlated in real-space, then it will couple different valleys in
$\mathbf{k}$-space with a uniform strength and, therefore, we may
take

\vspace{-0.7cm}
\begin{equation}
\Gamma_{ij}^{\sigma\sigma'}\!=\!\delta_{\sigma\sigma'}\Gamma_{i,j}^{\alpha}=\!\delta_{\sigma\sigma'}\delta_{\alpha0}\delta_{ij}+\!\delta_{\sigma\sigma'}\left(1-\delta_{\alpha0}\right)T_{ij}^{\alpha},
\end{equation}
where $\mathbf{T}^{\alpha}$ are the (hermitian) generators of the
$\text{su(}N_{v}\text{)}$ algebra, with we convention to obey $\text{Tr}\left[\mathbf{T}^{\alpha}\cdot\mathbf{T}^{\beta}\right]=2N_{v}\delta_{\alpha\beta}$.
The complete statistical interaction term can then be written as,

\vspace{-0.9cm}
\begin{align}
\mathcal{L}_{I}\left[\boldsymbol{\Phi}^{\dagger},\boldsymbol{\Phi}\right] & =\frac{w^{2}}{2}\sum_{\alpha}\left(\sum_{r=1}^{n}\sum_{\sigma=\pm}\sum_{i,j=1}^{N_{v}}\overline{\psi}_{\sigma i,r}(\mathbf{r})\Gamma_{ij}^{\alpha}\psi_{\sigma j,r}(\mathbf{r})\right)^{2}\label{eq:EffeLagrangean1}\\
 & \qquad=\frac{w^{2}}{2}\sum_{r,r'=1}^{n}\sum_{\sigma,\sigma'}\sum_{ijlm}\left(\delta_{ij}\delta_{lm}+\sum_{\alpha}T_{ij}^{\alpha}T_{lm}^{\alpha}\right)\overline{\psi}_{\sigma i,r}(\mathbf{r})\psi_{\sigma j,r}(\mathbf{r})\nonumber \\
 & \qquad\qquad\qquad\qquad\qquad\qquad\qquad\qquad\qquad\qquad\times\overline{\psi}_{\sigma'l,r'}(\mathbf{r})\psi_{\sigma'm,r'}(\mathbf{r}),\nonumber 
\end{align}
which allows the use of \textit{Fierz's Identity},

\vspace{-0.7cm}

\begin{equation}
\sum_{\alpha}T_{ij}^{\alpha}T_{lm}^{\alpha}\!=\!N_{v}\delta_{im}\delta_{jl}-\delta_{ij}\delta_{lm}
\end{equation}
(see Haber\,\cite{Haber2021} for a derivation and further properties
of $\text{SU}(N)$ Lie groups), so as to reduce its form to the following:

\vspace{-0.7cm}
\begin{align}
\mathcal{L}_{I}\left[\boldsymbol{\Phi}^{\dagger},\boldsymbol{\Phi}\right] & =-\frac{w^{2}N_{v}}{2}\sum_{r,r'=1}^{n}\sum_{\sigma,\sigma'}\sum_{ij}\overline{\psi}_{\sigma i,r}(\mathbf{r})\psi_{\sigma'i,r'}(\mathbf{r})\overline{\psi}_{\sigma'j,r'}(\mathbf{r})\psi_{\sigma j,r}(\mathbf{r}).
\end{align}

All in all, we have reduced the calculation of the average SPGF of
a system with $n_{s}$ local degrees of freedom to the study of two-point
correlation functions of an interacting SFT with $n$ coupled replicas,
in the limit $n\!\to\!0$. In the next section, we will pick up this
effective Lagrangian at zero energy $E\!=\!0$, \textit{i.e.},

\vspace{-0.7cm}
\begin{align}
\mathcal{L}\left[\boldsymbol{\Phi}^{\dagger},\boldsymbol{\Phi}\right] & =\hbar v_{\text{F}}\overline{\psi}_{\sigma j,r}(\mathbf{r})\boldsymbol{\sigma}^{\sigma\sigma'}\!\cdot\!\boldsymbol{\nabla}_{\mathbf{r}}\psi_{\sigma'j,r}(\mathbf{r})\!+\!\left(\eta\!-\!iE\right)\overline{\psi}_{aj,r}(\mathbf{r})\psi_{aj,r}(\mathbf{r})\label{eq:Lagrangean}\\
 & \qquad\qquad\qquad\qquad\qquad-\frac{w^{2}N_{v}}{2}\overline{\psi}_{\sigma i,r}(\mathbf{r})\psi_{\sigma'i,r'}(\mathbf{r})\overline{\psi}_{\sigma'j,r'}(\mathbf{r})\psi_{\sigma j,r}(\mathbf{r}),\nonumber 
\end{align}
and obtain the disorder-averaged SPGF and disorder-averaged density
of states for this system at the mean-field level.

\vspace{-0.5cm}

\subsection{\label{subsec:Fradkin's-Mean-Field-Theory}Fradkin's Mean-Field Theory:
The Large-$N_{v}$ Limit}

By this point, we have all the tools to calculate the saddle-point
solutions to the field integral of Eq.\,\eqref{eq:AVERAGEDPorp}.
This was the path first followed by Fradkin\,\cite{Fradkin86a,Fradkin86b},
which eventually led to the prediction of an unconventional disorder-induced
criticality for the mean DoS of 3D nodal systems. Starting from the
interacting Lagrangian of Eq.\,\eqref{eq:Lagrangean}, we can deal
with the problem by introducing a bosonic auxiliary field $Q$ that
performs a \textit{Hubbard-Stratonovich decoupling}\,\cite{Statonovich1957,Hubbard1959}
of the 4-point contact interaction. This decoupling is based upon
the following identity for a multivariate gaussian integral:

\vspace{-0.7cm}

\begin{align}
\exp\left[\frac{\lambda^{2}}{2}\int\!\!d\mathbf{r}\!\!\int\!\!d\mathbf{r}^{'}\phi_{l}(\mathbf{r})A_{lk}(\mathbf{r},\mathbf{r}')\phi_{k}(\mathbf{r}')\right]=\text{constant \ensuremath{\times}}\hfill\hfill\\
\int\!\mathcal{D}\boldsymbol{\mu}(\mathbf{r})\exp\left[\int\!\!d\mathbf{r}\!\!\int\!\!d\mathbf{r}'\!\left(-\frac{1}{2}\mu_{l}(\mathbf{r})A_{lk}^{-1}(\mathbf{r},\mathbf{r}')\mu_{k}(\mathbf{r}')\right)\right. & \!\!\left.+\lambda\!\int\!\!d\mathbf{r}\left(\mu_{l}(\mathbf{r})\phi_{l}(\mathbf{r})\right)\right],\nonumber 
\end{align}
which can be translated to the fields appearing in Eq.\,\eqref{eq:EffeLagrangean1}
by identifying 

\vspace{-0.8cm}
\begin{align}
\lambda\to w\sqrt{N_{v}},\text{ and }\phi_{l}(\mathbf{r}) & \to\sum_{i=1}^{N_{v}}\overline{\psi}_{\sigma i,r}(\mathbf{r})\psi_{\sigma'i,r'}(\mathbf{r}),
\end{align}
where $l$ is now a \textit{super-index} that contains $(\ensuremath{\sigma,\sigma',r,r'})$,
and $A_{lk}(\mathbf{r},\mathbf{r}')\to\delta_{lk}\delta\left(\mathbf{r}-\mathbf{r}'\right)$.
For simplicity, we will write the auxiliary field $\mu_{l}(\mathbf{r})$
as a \textit{Position-Dependent Matrix Field},

\vspace{-0.9cm}
\begin{equation}
\mu_{l}(\mathbf{r})=\mu_{\ensuremath{i,\sigma,\sigma',r,r'}}(\mathbf{r})\to\frac{1}{w\sqrt{N_{v}}}Q_{(r\sigma),(r'\sigma')}(\mathbf{r}),
\end{equation}
which has $2n\times2n$ components. This way, the Hubbard-Stratonovich
transformations leads 

\vspace{-1.0cm}
\begin{align}
\!\!\!\!\!\!\!\frac{w^{2}N_{v}}{2}\overline{\psi}_{\sigma i,r}(\mathbf{r})\psi_{\sigma'i,r'}(\mathbf{r})\overline{\psi}_{\sigma'j,r'}(\mathbf{r})\psi_{\sigma j,r}(\mathbf{r}) & \to-\overline{\psi}_{\sigma i,r}(\mathbf{r})Q_{(r\sigma),(r'\sigma')}(\mathbf{r})\psi_{\sigma'i,r'}(\mathbf{r})\\
 & \qquad\qquad\qquad\quad-\frac{1}{2w^{2}N_{v}}\text{Tr}\left[\left(Q(\mathbf{r})\right)^{2}\right],\nonumber 
\end{align}
which turns the former effective Lagrangian into a new one that contains
two fields, $\boldsymbol{\Psi}(\mathbf{r})$ and $Q(\mathbf{r})$.
The new Lagrangian,

\vspace{-0.7cm}
\begin{align}
\mathcal{L}\left[\overline{\boldsymbol{\Psi}},\boldsymbol{\Psi},Q\right] & =\overline{\psi}_{\sigma j,r}(\mathbf{r})\left(\hbar v_{\text{F}}\delta_{rr'}\boldsymbol{\sigma}^{\sigma\sigma'}\!\cdot\!\boldsymbol{\nabla}_{\mathbf{r}}\!+\!\left(\eta\!-\!iE\right)\delta_{\sigma\sigma'}\delta_{rr'}\right.\label{eq:Lagrangean-1}\\
 & \qquad\qquad\left.+Q_{(r\sigma),(r'\sigma')}(\mathbf{r})\right)\psi_{\sigma'j,r'}(\mathbf{r})+\frac{1}{2w^{2}N_{v}}\text{Tr}\left[\left(Q(\mathbf{r})\right)^{2}\right],\nonumber 
\end{align}
is now simply quadratic in the $\boldsymbol{\Psi}$ fields, which
allows us to perform the integration over the fermionic fields, leaving
behind an effective action for the interacting matrix field, $Q$,
alone. More precisely, we have that

\vspace{-0.7cm}

\begin{align}
\mathcal{S}_{\text{eff}}\left[Q(\mathbf{r})\right] & =\ln\left[\iint\!\!\mathcal{D}\overline{\boldsymbol{\Psi}}(\mathbf{r})\mathcal{D}\boldsymbol{\Psi}(\mathbf{r})\exp\left[\int d\mathbf{r}\overline{\boldsymbol{\Psi}}(\mathbf{r})\cdot\boldsymbol{\mathcal{K}}\cdot\boldsymbol{\Psi}(\mathbf{r})\right]\right]\\
 & \qquad\qquad\qquad\qquad\qquad\qquad\qquad+\frac{1}{2w^{2}N_{v}}\int d\mathbf{r}\text{Tr}\left[\left(Q(\mathbf{r})\right)^{2}\right]\nonumber \\
 & =c_{1}-N_{v}\ln\det\boldsymbol{\mathcal{K}}+\frac{1}{2w^{2}N_{v}}\int d\mathbf{r}\text{Tr}\left[\left(Q(\mathbf{r})\right)^{2}\right],\nonumber 
\end{align}
which can be recast into the explicit form,

\vspace{-0.7cm}

\begin{align}
\!\!\mathcal{S}_{\text{eff}}\left[Q(\mathbf{r})\right] & \!\approx\!-N_{v}\text{Tr}\ln\left[\hbar v_{\text{F}}\delta_{rr'}\boldsymbol{\sigma}^{\sigma\sigma'}\!\!\!\cdot\!\boldsymbol{\nabla}_{\mathbf{r}}\!+\!\left(\eta\!-\!iE\right)\delta_{\sigma\sigma'}\delta_{rr'}\!+\!Q_{(r\sigma),(r'\sigma')}(\mathbf{r})\right]\\
 & \qquad\qquad\qquad\qquad\qquad\qquad\qquad\qquad+\frac{1}{2w^{2}N_{v}}\int d\mathbf{r}\text{Tr}\left[\left(Q(\mathbf{r})\right)^{2}\right].\nonumber 
\end{align}

This effective action implies that the generating functional for $n$
coupled replicas of the system, $\mathcal{Z}_{n}$, can be written
as a functional integral over a local matrix field, $Q(\mathbf{r})$,
as follows:

\vspace{-0.7cm}

\begin{equation}
\mathcal{Z}_{n}=\int\!\mathcal{D}Q(\mathbf{r})\exp\left(-N_{v}\tilde{\mathcal{S}}_{\text{eff}}\left[Q(\mathbf{r})\right]\right),
\end{equation}
where 

\vspace{-0.7cm}
\begin{align}
\tilde{\mathcal{S}}_{\text{eff}}\left[Q(\mathbf{r})\right] & =\text{Tr}\ln\left[\hbar v_{\text{F}}\delta_{rr'}\boldsymbol{\sigma}^{\sigma\sigma'}\!\cdot\!\boldsymbol{\nabla}_{\mathbf{r}}\!+\!\left(\eta\!-\!iE\right)\delta_{\sigma\sigma'}\delta_{rr'}+Q_{(r\sigma),(r'\sigma')}(\mathbf{r})\right]\\
 & \qquad\qquad\qquad\qquad\qquad\qquad\qquad\qquad-\frac{1}{2w^{2}N_{v}^{2}}\int d\mathbf{r}\text{Tr}\left[\left(Q(\mathbf{r})\right)^{2}\right].\nonumber 
\end{align}
Since the exponent comes multiplied by the number of Weyl nodes in
the system, $N_{v}$, this can be used as a \textit{``large parameter''}
that justifies the evaluation of $\mathcal{Z}_{n}$ using a \textit{Saddle-Point
Method}. Therefore, the mean-field value of $Q(\mathbf{r})$ can be
obtained from the following condition:

\vspace{-0.7cm}

\begin{equation}
\left.\frac{\delta}{\delta Q(\mathbf{r})}\tilde{\mathcal{S}}_{\text{eff}}\left[Q(\mathbf{r})\right]\right|_{Q^{{\scriptscriptstyle MF}}}\!\!\!\!=0,\label{eq:SaddlePoint}
\end{equation}
which yields the self-consistent mean-field equation,

\vspace{-0.7cm}
\begin{equation}
\!\!\!\!\!\!\!\frac{1}{8w^{2}N_{v}^{2}}Q_{(r\sigma),(r'\sigma')}^{{\scriptscriptstyle MF}}(\mathbf{r})\!=\!\left[\hbar v_{\text{F}}\mathcal{I}_{n\times n}\!\tensorproduct\!\cancel{\boldsymbol{\nabla}}_{\mathbf{r}}\!+\!\mathcal{I}_{n\times n}\!\tensorproduct\!\sigma_{0}\left(\eta\!-\!iE\right)\!+\!Q^{{\scriptscriptstyle MF}}\!(\mathbf{r})\right]_{(r\sigma),(r'\sigma')}^{-1}.\label{eq:MeanFieldEquation=0000B4}
\end{equation}
In general, Eq.\,\eqref{eq:MeanFieldEquation=0000B4} can have self-consistent
solutions that are not homogenous in space. For the sake of simplicity,
we follow Fradkin\,\,\cite{Fradkin86a,Fradkin86b} and consider
only solutions that have the very simply form, $Q_{(r\sigma),(r'\sigma')}^{{\scriptscriptstyle MF}}(\mathbf{r})\!=\!q\delta_{rr'}\delta_{\sigma\sigma'}.$
For this \textit{ansatz}, the Eq.\,\eqref{eq:MeanFieldEquation=0000B4}
reduces to

\vspace{-0.7cm}
\begin{equation}
\frac{1}{8w^{2}N_{v}^{2}}q\delta_{rr'}\delta_{\sigma\sigma'}\!=\!\delta_{rr'}\int\!\!\frac{d^{3}\mathbf{k}}{8\pi^{3}}\left[i\hbar v_{\text{F}}\boldsymbol{\sigma}\cdot\mathbf{k}+\sigma_{0}\left(\eta\!-\!iE+q\right)\right]_{\sigma\sigma'}^{-1}.\label{eq:MeanFieldEquation=0000B4-1}
\end{equation}
Fortunately, the matrix that is being integrated in $\mathbf{k}$
is now a $2\!\times\!2$, which may be explicitly inverted as 

\vspace{-0.7cm}

\begin{equation}
\frac{1}{8w^{2}N_{v}^{2}}q\delta_{\sigma\sigma'}=\int\frac{d^{3}\mathbf{k}}{8\pi^{3}}\frac{\left(\eta\!-\!iE+q\right)\delta_{\sigma\sigma'}-i\hbar v_{\text{F}}\boldsymbol{\sigma}_{\sigma\sigma'}\cdot\mathbf{k}}{\left(\eta\!-\!iE+q\right)^{2}+\hbar^{2}v_{\text{F}}^{2}\abs{\mathbf{k}}^{2}},\label{eq:MeanFieldEquation=0000B4-1-1}
\end{equation}
or, using the $\mathbf{k}\to-\mathbf{k}$ symmetry, 

\vspace{-0.7cm}

\begin{equation}
\frac{\pi^{2}\hbar^{3}v_{\text{F}}^{3}q}{w^{2}N_{v}^{2}}=\left(\eta\!-\!iE+q\right)\int_{0}^{\infty}\!\!\!\frac{x^{2}dx}{\left(\eta\!-\!iE+q\right)^{2}+x^{2}}.\label{eq:MeanFieldEquation=0000B4-1-1-1}
\end{equation}
The integral in Eq.\,\eqref{eq:MeanFieldEquation=0000B4-1-1-1} is
UV-divergent (just like it happened in our earlier diagrammatic SCBA
calculation), and one must resort to a regularization procedure. Here,
we follow the same smooth cut-off prescription used in Sect.\,\eqref{sec:Continuum-Model},
\textit{i.e.},

\vspace{-0.7cm}

\begin{equation}
\int_{0}^{\infty}\frac{x^{2}dx}{\left(\eta\!-\!iE+q\right)^{2}+x^{2}}=\frac{\pi M^{2}}{2\left(\eta\!-\!iE+q+M\right)},
\end{equation}
where $M$ is a large-momentum regularization scale. Then, the final
result will be

\vspace{-0.7cm}

\begin{equation}
\frac{\pi^{2}\hbar^{3}v_{\text{F}}^{3}q}{w^{2}N_{v}^{2}}=\frac{\eta\!-\!iE+q}{\eta\!-\!iE+q+M}.\label{eq:MeanFieldEquation=0000B4-1-1-1-1}
\end{equation}

Note that, by definition, the normalized DoS can be expressed in terms
of the mean-field $Q$-field as, $\rho(E)=\Re\left[Q^{{\scriptscriptstyle MF}}\!(\mathbf{r})\right]/4\pi w^{2}N_{v}$.
Consequently, only the mean-field solutions having a non-negative
real part are admissible. To obtain these solutions, we can manipulate
Eq.\,\eqref{eq:MeanFieldEquation=0000B4-1-1-1-1} to place it in
the form, 

\vspace{-0.7cm}

\begin{align}
q & =\left(q\!+\!\eta\!-\!iE\right)\left[g-\frac{q}{M}\right],\label{eq:MeanField}
\end{align}
where $g\!=\!w^{2}\!N_{v}^{2}M/\pi^{2}\hbar^{3}v_{\text{F}}^{3}$
is a tunable parameter, proportional to the disorder strength. Now,
if we set $\eta=E=0$ in Eq.\,\eqref{eq:MeanField}, we get the condition,

\vspace{-0.7cm}

\begin{align}
q & \left[1-g+\frac{q}{M}\right]=0,
\end{align}
which always has the admissible solution $q\!=\!0$, corresponding
to a vanishing density of states at zero-energy. However, for $g>1$,
an alternative solution becomes available, namely,

\vspace{-0.7cm}
\begin{equation}
q=M\left(g-1\right)\to\rho(0)\!=\!\frac{M}{4\pi w^{2}N_{v}}\left(g-1\right).\label{eq:CriticalDoS}
\end{equation}
Equation\,\eqref{eq:CriticalDoS} describes the appearance of a finite
DoS at nodal energy above a critical value of the disorder parameter,
$w$, just as we found in Subsect.\,\ref{subsec:SCBA}. This analysis
can then be extended to analyze the full shape of $\rho(E)$ in the
critical case of $g\!=\!1$,\textit{ i.e.},

\vspace{-0.7cm}

\begin{align}
q & =\left(q\!-\!iE\right)\left[1-\frac{q}{M}\right].
\end{align}
This can be solved explicitly, and yields

\vspace{-0.7cm}

\begin{align}
q^{2}-iEq+iME=0\to\; & q=\frac{iE\pm\sqrt{2ME}\left(i-1\right)\sqrt{1+E/2iM}}{2}\\
 & \approx\frac{iE}{2}\left(1\pm\sqrt{2M/E}\right)\mp\sqrt{ME/2}+\mathcal{O}\left[E/M\right].\nonumber 
\end{align}
By picking the admissible branch, we conclude that the density of
states, for $E\ll M$, takes the form,

\vspace{-0.7cm}

\begin{equation}
\rho(E)=\frac{1}{w^{2}N_{v}}\sqrt{\frac{ME}{32\pi^{2}}},
\end{equation}
which is the critical behavior predicted earlier from our SCBA analysis
{[}see Eq.\,\eqref{eq:DOSBranches}{]}.

\vspace{-0.5cm}

\subsection{\label{subsec:RG}Disorder Irrelevance and Renormalization Group
Results}

From the point-of-view of the effective SFT, all disorder effects
in the SPP are encapsulated in the quartic-interaction term of Eq.\,\eqref{eq:Lagrangean}.
This term is proportional to the parameter $w^{2}$, which works as
$\phi^{4}$\,-\,coupling in the fermionic effective Lagrangian.
Even tough the mean-field treatment of the functional integral gave
us identical results to the ones obtained from the diagrammatic SCBA,
this alternative formulation is much more powerful, for it allows
RG methods to be used in studying fluctuations around the saddle-point
solution. As we will see, the integration of fluctuation modes over
larger and larger length scales lead to a \textit{running behavior
of the couplings}, which quantitatively changes the universal exponents
of the critical theory that describes the unconventional SMMT point.
Without diving too deep on the subject, we will outline the way a
perturbative RG procedure can be used to improve over the mean-field
result, and recap state-of-the-art results.

Our analysis begins with the identification of the relevant couplings
associated to the effective SFT. On the one hand, we have the quadratic
curvature of $\overline{\rho(\varepsilon\!\approx\!0)}$, which defines
an effective Fermi velocity $\varkappa v_{\text{F}}$ that gets renormalized
by disorder. On the other, we have the disorder-induced $\phi^{4}$\,-\,coupling
constant, $\lambda\!=\!w^{2}$. Now that we have identified the coupling
parameters, we do a \textit{tree-level analysis} of the RG flow, which
starts by measuring every quantity in a system of natural units, \textit{i.e.},
$\hbar\!=\!v_{\text{F}}\!=\!1$, which means that $\left[E\right]=\left[\eta\right]=L^{-1}$
are inverse length scales, the Lagrangian density has a dimensions
$L^{-d}$, and the fermionic fields have dimensions,

\vspace{-0.8cm}
\begin{equation}
\left[\psi\right]=\left[\psi^{*}\right]=[L]^{\frac{1-d}{2}}.
\end{equation}
Likewise, the two coupling parameters of the theory must be rescaled
to these units which, implies that $\varkappa$ is dimensionless,
and the disorder coupling parameter has a scaling dimension of 

\vspace{-0.8cm}
\begin{equation}
[\lambda]=[L]^{-d-2+2d}=[L]^{d-2}.
\end{equation}
From the point of view of RG, what this means is that $\lambda$ is
an irrelevant deformation of the clean action for any dimension $d\!>\!2$.
This conclusion is a mere formalization the initial heuristic argument
given in the beginning of Sect.\,\ref{sec:WeaklyDisorderedNode}.
At any rate, here this takes a whole new meaning, because we pinpoint
$d\!=\!2$ as the lower critical dimension associated to this unconventional
phase transition. Therefore, we can access the properties of the non-trivial
fixed point in the RG flow of a 3D system by doing an $\varepsilon$\,-\,expansion
around $d\!=\!2$. This analysis have been the subject of many published
work\,\cite{Goswami11,Ominato14,Roy14,Syzranov15a,Syzranov15b,Syzranov16,Syzranov18},
using a variety of techniques to represent the mean DoS as an effective
SFT. For our purpose, we recap the two-loop results of Roy and Das
Sarma\,\cite{Roy14,Roy14_Err}, who have obtained the following renormalization
group equations:

\vspace{-0.8cm}

\begin{subequations}
\begin{align}
\beta_{\varkappa} & =L\frac{d\varkappa}{dL}=\varkappa\left(z-1-2\lambda\right)\\
\beta_{\lambda} & =L\frac{d\lambda}{dL}=\left(2\!-\!d\right)\lambda\!+\!2\lambda^{2}\!+\!2\lambda^{3},
\end{align}
\end{subequations}

where $L$ is a length scale that is flowing to $+\infty$. The previous
equations tell us how the two parameters of interest (the marginal
parameters in $d\!=\!2$) flow upon a coarse-graining to a double-loop
order in perturbation theory around the clean system's fixed point.
This pair of $\beta$-functions have two fixed points: \textit{(i)}
the clean fixed point, featuring $\lambda\!=\!0$ and $\chi$ arbitrary,
and\textit{ (ii) }a strong disorder fixed point where

\vspace{-0.8cm}
\begin{equation}
\beta_{\lambda^{*}}=0\Rightarrow\left(2\!-\!d\right)\!+\!2\lambda_{*}\!+\!2\lambda_{*}^{2}=0,
\end{equation}
and therefore $\lambda^{*}\!=\!\frac{\sqrt{1-2\left(2\!-\!d\right)}-1}{2}$.
Note that $\beta_{\lambda}$ is negative for $\lambda\!<\!\lambda_{*}$
but positive for $\lambda>\lambda_{*}$, which implies that the new\textit{
disordered fixed-point} is repulsive as the theory flows towards the
infra-red (\textit{i.e.}, it separates two distinct phases). The critical
exponent associated to the correlation length is given by 

\vspace{-0.7cm}
\begin{equation}
\frac{1}{\nu}=\left.\frac{d}{d\lambda}\beta_{\lambda}\right|_{\lambda=\lambda_{*}}\!\!\!=\!\left(2\!-\!d\right)+4\lambda_{*}+6\lambda_{*}^{2},
\end{equation}
which can be evaluated as an expansion in powers of $\varepsilon=d\!-\!2$.
This yields that

\vspace{-0.8cm}
\begin{align}
\lambda_{*}\! & =\!\frac{\sqrt{1+2\varepsilon}-1}{2}=\frac{\varepsilon}{2}-\frac{\varepsilon^{2}}{4}+\mathcal{O}\left[\varepsilon^{3}\right]\\
\nu & =\frac{1}{6\lambda_{*}^{2}+4\lambda_{*}-2\varepsilon}
\end{align}
and therefore, placing $\varepsilon\!=\!1$ for the three-dimensional
case in point, we arrive at 

\vspace{-0.8cm}
\begin{equation}
\lambda_{*}=\frac{1}{4}\text{ and }\nu=\frac{2}{3}.
\end{equation}
At the same time, we can obtain the resulting dynamical critical exponent,
$z$, by requiring that 

\vspace{-0.8cm}
\begin{equation}
z-1-2\lambda_{*}=0\Rightarrow z=1+\varepsilon-\frac{\varepsilon^{2}}{2}+\mathcal{O}\left[\varepsilon^{3}\right]
\end{equation}
which entails $z\!=\!3/2$ for a three-dimensional Weyl semimetal.
These results are to be compared with $z\!=\!2\text{ and }\nu\!=\!1$,
which were the overestimated values obtained within the SCBA.

\vspace{-0.5cm}

\section{\label{sec:NumericalSimulations_Anderson}Numerical Simulations of
Disordered Dirac-Weyl Systems}

\vspace{-0.2cm}

\begin{figure}[t]
\vspace{-0.4cm}
\begin{centering}
\includegraphics[scale=0.25]{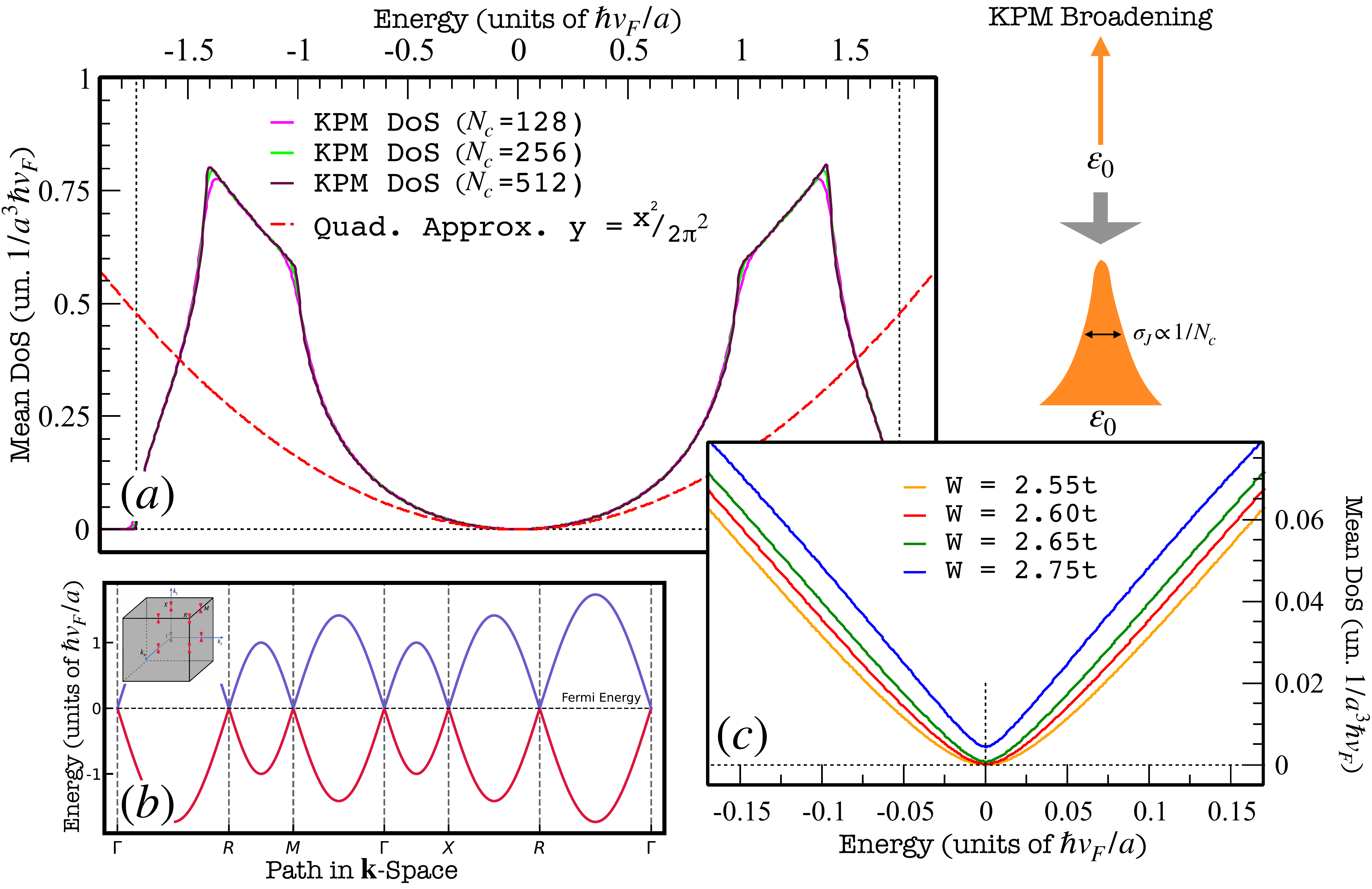}
\par\end{centering}
\vspace{-0.2cm}

\caption{\label{fig:DoS_KPMAnderson}Plots of the mean DoS calculated for the
lattice model of Eq.\,\eqref{eq:LatticeWSM_Cubic} using the KPM
approach. The spectral resolution used was of $\sigma_{J}\approx0.0003\hbar v_{\text{F}}/a$
corresponding to a total of $2^{16}$ chebyshev polynomials with a
Jackson damping kernel. (a) Overview of the clean model's density
of states. Scheme showing the energy broadening caused by the KPM
procedure. (b) Band Structure of the lattice model. (c) Lifting of
the nodal DoS as a function of disorder. The critical value lies at
about $E\!\approx\!2.65\hbar v_{\text{F}}/a$ for the box disorder
model.}

\vspace{-0.4cm}
\end{figure}

So far, the mean density of states around a disordered Dirac-Weyl
node have been evaluated by analytic methods, and defined within a
continuum low-energy model. The SCBA and SFT mean-field approaches
yielded qualitatively similar results, that served to produce a very
clear picture about what the physical effects of random fields in
DWSMs are\,\textemdash \textit{\,Disorder drives an unconventional
semimetal-to-metal phase transition (SMMT) that precedes Anderson
localization.} Despite this agreement, both approaches are not free
of limitations, chief among which is \textit{(i)} their dependence
on an artificial UV-regularization, and \textit{(ii)} the intrinsically
biased nature which assumed particularly simple forms for the mean-field
solutions. To circumvent this issue, we now present an unbiased numerical
study of the mean DoS in a tight-binding model that realizes a (eight-node)
Weyl semimetal, in the simple cubic lattice. The two-orbital Hamiltonian
reads,

\vspace{-0.7cm}

\begin{equation}
\mathcal{H}_{l}^{0}\!=\!\frac{i\hbar v_{\text{F}}}{2a}\sum_{{\scriptscriptstyle \mathbf{R}\in\mathcal{L_{\text{C}}}}}\!\left[\Psi_{\mathbf{{\scriptscriptstyle R}}}^{\dagger}\!\cdot\!\sigma^{j}\!\cdot\!\Psi_{{\scriptscriptstyle \mathbf{R}+a\mathbf{x}_{\!j}}}-\Psi_{\mathbf{{\scriptscriptstyle R}}}^{\dagger}\!\cdot\!\sigma^{j}\!\cdot\!\Psi_{{\scriptscriptstyle \mathbf{R}-a\mathbf{x}_{\!j}}}\right],\label{eq:LatticeWSM_Cubic}
\end{equation}
where $\mathcal{L}_{\text{C}}$ is a cubic lattice of parameter $a$,
$\Psi_{\mathbf{{\scriptscriptstyle R}}}^{\dagger}/\Psi_{\mathbf{{\scriptscriptstyle R}}}$
is a two-component fermionic creation/annihilation operator, and $\sigma^{j}$
are $2\!\times\!2$ Pauli matrices. This is exactly the same two-band
model that was used in the previous numerical studies reported in
Refs.\,\cite{Pixley15,Bera16,Pixley16a,Pixley16b,Pixley21,Wilson20,Pires2021}.
Formally, Eq.\,\eqref{eq:LatticeWSM_Cubic} defines a simple cubic
discretization of the single-node continuum Hamiltonian {[}Eq.\,\eqref{eq:Contin_SingleNodeAHam}{]},
which replicates the Weyl node into eight disjoint valleys that are
pinned to the \textit{Time-Reversal Invariant Momenta} (TRIM\nomenclature{TRIM}{Time-Reversal Invariant Momenta (T-Invariant Points of the First Brillouin Zone)})
of the fBz. The chiralities of the eight different valleys are not
all the same and, in fact, come in pairs that compensate each other
so as to keep an overall zero chirality. The fBz of the model is schematically
represented in Fig.\,\ref{fig:DoS_KPMAnderson}\,b, together with
the dispersion relation along a high-symmetry path. 

In the clean limit, this two-band model has the particle-hole symmetric
dispersion relation,

\vspace{-0.9cm}

\begin{equation}
E^{\text{c/v}}\!\left(\mathbf{k}\right)\!=\!\pm\sqrt{\sin^{2}\!k_{x}a\!+\!\sin^{2}\!k_{y}a\!+\!\sin^{2}\!k_{z}a},
\end{equation}
which gets reduced to $E^{\text{c/v}}\!\left(\mathbf{k}\right)\!\approx\!\pm\hbar v_{\text{F}}\abs{\mathbf{k}\!-\!\mathbf{K}_{\text{W}}}$
near each TRIM, $\mathbf{K}_{\text{W}}$. The density of states can
also be analytically calculated over the entire spectrum but, instead,
we employ the \textit{Kernel Polynomial Method} (KPM\nomenclature{KPM}{Kernel Polynomial Method})\cite{Weise2006}
which is a real-space spectral method that also allows to perform
calculations that include perturbations that break the lattice-translation
symmetry.\,\,In Fig.\,\ref{fig:DoS_KPMAnderson}a, we present the
clean DoS of the model obtained with different numbers of Chebyshev
polynomials, corresponding to an increasing spectral resolution. Further
details on the KPM, including the relationship between number of polynomials
and spectral resolution (schematically shown in Figure\,\ref{fig:DoS_KPMAnderson}b)
is provided in Appendix\,\ref{chap:Crash-Course-KPM}.

Now, we move on to the numerical observation of the SMMT in the lattice
model of Eq.\,\eqref{eq:LatticeWSM_Cubic}. This lattice model will
be used in several instances throughout this thesis but, for the present
purposes, we consider it with an\textit{ added scalar on-site random
potential}, 

\vspace{-0.8cm}

\begin{equation}
\mathcal{H}_{l}^{0}\to\mathcal{H}_{l}=\mathcal{H}_{l}^{0}+\!\sum_{{\scriptscriptstyle \mathbf{R}\in\mathcal{L_{\text{C}}}}}\mathcal{U}\left(\mathbf{R}\right)\Psi_{\mathbf{{\scriptscriptstyle R}}}^{\dagger}\!\cdot\!\Psi_{\mathbf{{\scriptscriptstyle R}}}
\end{equation}

\vspace{-0.5cm}

whose values are independently drawn out of a \textit{Uniform Box
Distribution}, \textit{i.e.}, 
\begin{figure}[t]
\vspace{-0.45cm}
\begin{centering}
\includegraphics[scale=0.3]{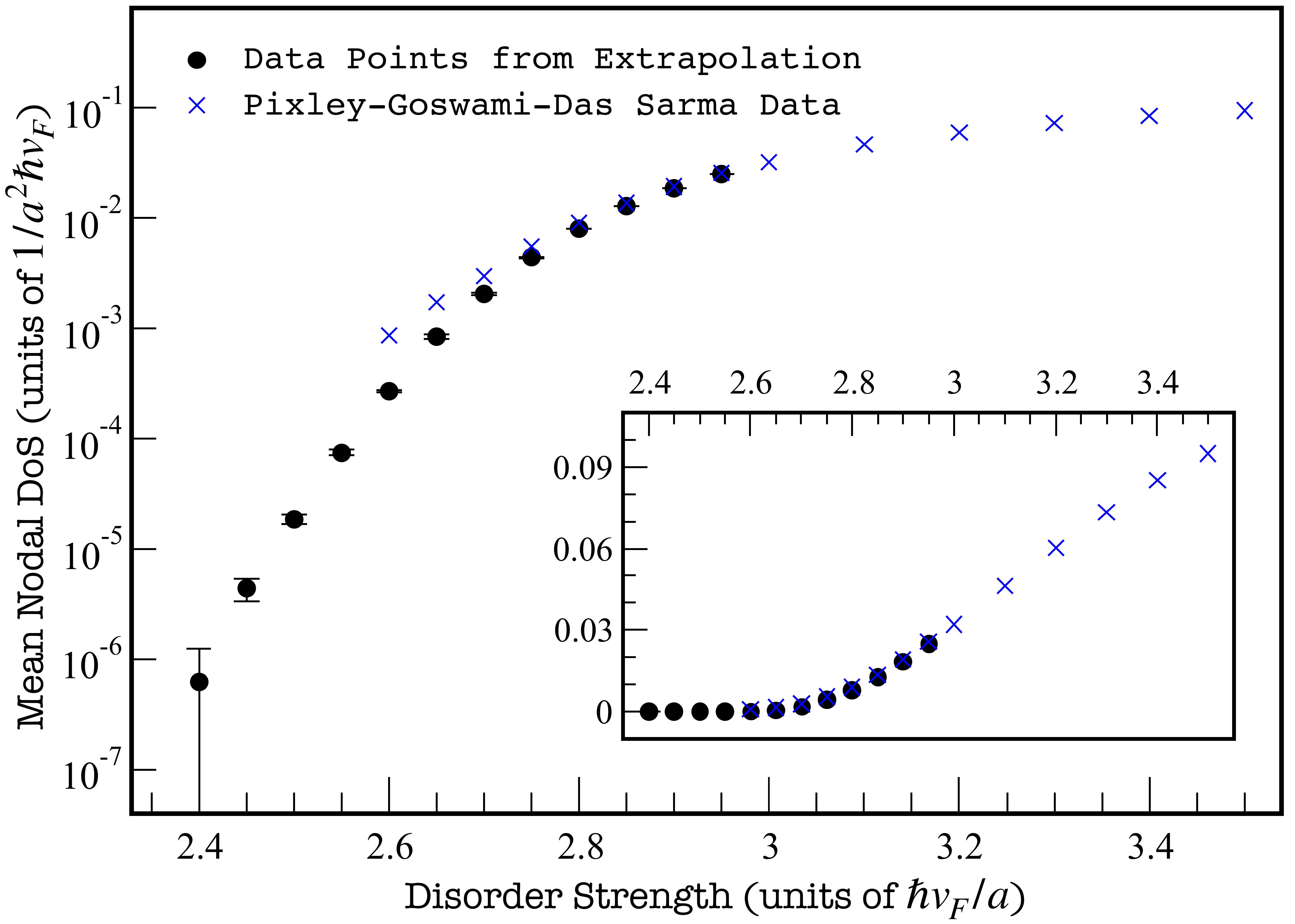}
\par\end{centering}
\vspace{-0.25cm}

\caption{\label{fig:DoS_KPMCriticalCurve}Critical curve for $\overline{\rho(E=0)}$
calculated numerically as a function of the disorder strength $W$.
Our numerical results are shown as black dots (including $2\sigma$
bilateral error bars) while the published results of Pixley \textit{et
al.}\,\cite{Pixley16b}, for the same disorder model, are represented
as blue cross markers. The inset provides the same representation
but in a double linear scale.}

\vspace{-0.45cm}
\end{figure}

\vspace{-0.8cm}
\begin{equation}
\mathcal{U}\left(\mathbf{R}\right)\leftarrow\mathcal{P}_{\mathcal{U}}=\frac{1}{W}\Theta\left(\frac{W}{2}\!-\!\abs{\mathcal{U}}\right)
\end{equation}

\vspace{-0.2cm}

where, as usual, the parameter $W$ measures the disorder strength.
In Fig.\,\ref{fig:DoS_KPMAnderson}d and Fig.\,\ref{fig:DoS_KPMCriticalCurve},
we present simulation results of the disorder-averaged DoS near $E\!=\!0$,
obtained for a system with $4$ million orbitals, using a spectral
resolution of $3\!\times\!10^{-4}\hbar v_{\text{F}}/a$ ($\text{meV}$-scale
in real samples), and the accumulated statistics of $5000$ disorder
realizations. For this calculation, we have used the efficient CPU-parallel
implementation of the KPM iteration within the open-source software
package, QuantumKITE\,\cite{Joao2020,Pires2021}, in which the smoothness
of the curves is aided by three major factors: \textit{(i)} the huge
sizes of the simulated systems, \textit{(ii)} the Jackson damping
kernel used to mitigate the \textit{Gibbs phenomenon} in a truncated
Chebyshev expansion, and \textit{(iii)} the simultaneous averaging
over disorder realizations and twisted boundary conditions. Further
details on the latter are provided in Appendix\,\ref{chap:TwistedBoundaries-1}. 

\vspace{-0.5cm}

\paragraph{Results and Discussion:}

Our main numerical results are shown in Figs.\,\ref{fig:DoS_KPMAnderson}d
and \ref{fig:DoS_KPMCriticalCurve}. In Fig.\,\ref{fig:DoS_KPMAnderson}d,
we showcase the SMMT scenario that was put forward by the early mean-field
approaches. For this specific Anderson disorder, the average DoS remains
vanishing and quadratic around the nodal energy, for disorder strengths
up to the critical value $W_{c}\!\approx\!2.5\hbar v_{\text{F}}/a$.
After that, the curve begins lifting up at the node, and a conventional
diffusive metal phase emerges. The critical curve for $\overline{\rho(E\!=\!0)}$
is shown in Fig.\,\ref{fig:DoS_KPMCriticalCurve} where the points
have been carefully obtained from an extrapolation to the limit of
infinite system size (with infinite spectral resolution). Note that
these results are in full agreement with previously published studies
(Refs.\,\cite{Kobayashi14,Pixley15,Pixley16b,Bera16,Pixley16a,Wilson20}),
and seemingly confirm that a finite-disorder unconventional critical
point exists in a lattice WSM, above which the nodal DoS acquires
a finite value. The precise value of the critical disorder is model-dependent
but, for this specific model, it has been reported to lie somewhere
in between $2.5\hbar v_{\text{F}}/a$ and $2.6\hbar v_{\text{F}}/a$\,\cite{Pixley16b}.
Our results show that this is likely over-estimated but, still, the
qualitative picture of a non-Anderson quantum critical point associated
to the SMMT remains untouched.

\global\long\def\vect#1{\overrightarrow{\mathbf{#1}}}%

\global\long\def\abs#1{\left|#1\right|}%

\global\long\def\av#1{\left\langle #1\right\rangle }%

\global\long\def\ket#1{\left|#1\right\rangle }%

\global\long\def\bra#1{\left\langle #1\right|}%

\global\long\def\tensorproduct{\otimes}%

\global\long\def\braket#1#2{\left\langle #1\mid#2\right\rangle }%

\global\long\def\omv{\overrightarrow{\Omega}}%

\global\long\def\inf{\infty}%

\lhead[\MakeUppercase{\chaptername}~\MakeUppercase{\thechapter}]{\MakeUppercase{\rightmark}}

\rhead[\MakeUppercase{Smooth Regions and Semimetal Instability}]{}

\lfoot[\thepage]{}

\cfoot[]{}

\rfoot[]{\thepage}

\chapter{\label{chap:Instability_Smooth_Regions}Instability of Weyl Semimetals
to Random Smooth Regions}

\vspace{-0.3cm}

In Chapter\,\ref{chap:Mean-Field-Quantum-Criticality}, we have presented
sound analytical and numerical evidence that random perturbations
to a DWSM give rise to an unconventional quantum criticality of the
nodal DoS, that precedes the more common Anderson localization transition.
In simple terms, $\overline{\rho(E=0)}$ is \textit{strictly zero}
in the weak disorder limit, growing finite after a critical disorder
strength in a continuous but non-differentiable way. Even though the
early unbiased numerics seemed to back up this scenario, Nandkishore
\textit{et al}.\,\cite{Nandkishore14} argued that both the SCBA
and mean-field SFT arguments may actually be missing non-perturbative
contributions that would destabilize the semi-metallic phase right
from the start. The controversy arises from the restrictive solution
to the saddle-point equation to the effective SFT, that assumes \textit{spatially
uniform fields }{[}see Subsect.\,\ref{subsec:Fradkin's-Mean-Field-Theory}{]},
on the grounds that this symmetry always gets recovered on average.
Although this seems like a sensible hypothesis, there are well-known
situations in which it does not hold at all. For example, the appearance
of in-gap bound states (\textit{Lifshitz tails}\,\cite{Lifshitz64,Halperin66,Zittartz66,Halperin67})
in weakly disordered semiconductors was shown by Cardy\,\cite{Cardy78}
to arise from non-uniform (actually spherically symmetric) solutions
to the disorder saddle-point equation. While this may seem an exotic
situation, it actually finds an important precedent in \textit{High-Energy
Physics }(HEP\nomenclature{HEP}{High-Energy Physics}) where it was
found, by Coleman\,\cite{Coleman77}, that a meta-stable classical
configuration of an interacting quantum field (\textit{a vaccum})
may actually decay into a different one that has a lower energy. This
gives rise to non-perturbative (\textit{exponential}) contributions
to the field propagators which were dubbed \textit{Instantons}. With
this perspective, the Lifshitz tails of a disordered semiconductor
are seen as instantonic effects of the disordered landscape\,\cite{Yaida16}
and, as shown in Ref.\,\cite{Nandkishore14}, a similar set of non-trivial
solutions also appears in the gapless limit, \textit{i.e.}, a Dirac-Weyl
semimetal.

Physically, the instantons of the disorder saddle-point equation correspond
to rare-events of a random potential landscape which can support nodal
bound states. By explicitly including these spherically symmetric
saddles into the ensemble-averaged SPGF, Nandkishore \textit{et al}.\,\cite{Nandkishore14}
were able to show that the net contribution to the nodal DoS is expected
to take the form,

\vspace{-0.7cm}

\begin{equation}
\overline{\rho(E\!=\!0)}\propto\exp\left(-\frac{W_{0}^{2}}{W^{2}}\right),\label{eq:NonPerturbativeDoS}
\end{equation}
provided the on-site values of the underlying disordered potential
are drawn from a gaussian white-noise potential. Unsurprisingly, the
result of Eq.\,\eqref{eq:NonPerturbativeDoS} is explicitly non-perturbative
in the disorder strength ($W$), meaning that it cannot be expanded
as Taylor series in $W$ with a finite convergence radius and, moreover,
leads to a finite nodal DoS even as $W\to0^{+}$. This mechanism leads
to an avoidance of the mean-field critical point which, now, strictly
appears as a \textit{sharp crossover behavior} in the mean nodal DoS.
While it is mathematically clear why these instantonic effects appear,
it remains unclear which kind of local environment allows these nodal
bound states to form\,\footnote{Note that, if we achieve such an interpretation, we can tailor specific
disorder models that either suppress or enhance this effect.}. In Ref.\,\cite{Nandkishore14}, these saddles are interpreted as
contributions from statistically rare \textit{smooth regions }that
may bind single-particle states within a disordered landscape. This
is inspired on the fact that the Weyl equation with a spherical potential
well (or plateau) can support nodal bound state solutions in fine-tuned
situations. If this picture holds, the smallness of the induced DoS
is explained by the rare occurrence of such smooth regions within
a white-noise disordered landscape. This mechanism of semi-metallic
destabilization by weak disorder was then dubbed a \textit{Rare-Region
Avoided Quantum-Criticality} (AQC\nomenclature{AQC}{Avoided Quantum Criticality})
and later confirmed by the accurate numerical work of Pixley \textit{et
al}.\,\cite{Pixley16a}.

In the remaining of this chapter, we directly study the effects of
smooth potential regions in the mean DoS of a DWSM. In order to segregate
the mean-field disorder effects, and the non-perturbative ones, we
analyze a tailor-made model of randomness in which a \textit{dilute
set of large spherical scalar impurities} (like spherical potential
wells/plateaux) is randomly scattered within an otherwise clean 3D
DWSM, each having a random value of the potential. The original results
presented here are based on the work published in Santos Pires \textit{et
al}.\,\cite{Pires2021}.

\vspace{-0.5cm}

\section{\label{sec:SphericalScatterers}Dirac-Weyl Fermions and Spherical
Scatterers}

\vspace{-0.2cm}

Our initial approach to the problem will be analytical and based on
a continuum model of a single-node Dirac semimetal (DSM), hosting
a single spherical well/plateau with a trivial spinor structure (a
\textit{Spherical Scatterer}).\,The choice of a DSM model, instead
of a WSM, is a matter of mathematical convenience but also an effort
to remain faithful to Ref.\,\cite{Pires2021}. Nevertheless, we antecipate
that all our conclusions carry over to the WSM case, and even to a
situation with \textit{multiple impurities} and \textit{valleys},
provided we assume a dilute impurity limit and a strong suppression
of inter-valley scattering, respectively. At any rate, our clean working
Hamiltonian is the four-band continuum model,
\begin{equation}
\mathscr{H}_{c}^{0}\!=\!-i\hbar v_{\text{F}}\int\!d\mathbf{r}\Psi_{a\mathbf{r}}^{\dagger}\left(\boldsymbol{\alpha}^{ab}\!\cdot\!\boldsymbol{\nabla}_{\mathbf{r}}\right)\Psi_{b\mathbf{r}},\label{eq:Dirac-Weyl_Hamiltonian}
\end{equation}

\vspace{-0.4cm}

which contains two copies of Eq.\,\eqref{eq:Contin_SingleNodeAHam}
with opposite chiralities. An isolated spherical scatterer is then
modeled by the addition of a spherically symmetric scalar potential,
$V(\abs{\mathbf{r}})$, to this massless Dirac Hamiltonian. Then,
the full single-particle Hamiltonian reads,

\vspace{-0.7cm}

\begin{equation}
\mathscr{H}_{c}\!=\!-i\hbar v_{\text{F}}\int\!d\mathbf{r}\Psi_{a\mathbf{r}}^{\dagger}\left(\boldsymbol{\alpha}^{ab}\!\cdot\!\boldsymbol{\nabla}_{\mathbf{r}}\right)\Psi_{b\mathbf{r}}+\int\!d\mathbf{r}V(\abs{\mathbf{r}})\Psi_{a\mathbf{r}}^{\dagger}\Psi_{a\mathbf{r}},
\end{equation}
where $\Psi_{a\mathbf{r}}^{\dagger}$/$\Psi_{a\mathbf{r}}$ are local
fermionic creation/annihilation operators, and $\boldsymbol{\alpha}=(\alpha^{x},\alpha^{y},$
$\alpha^{z})$ is a vector of Dirac matrices. The Dirac matrices can
be set up in alternative representations, as long as they obey the
anti-commutative Clifford algebra,
\begin{equation}
\left\{ \alpha^{i},\alpha^{j}\right\} =\alpha^{i}\!\cdot\!\alpha^{j}+\alpha^{j}\!\cdot\!\alpha^{i}=2\mathbb{I}\delta_{ij}.
\end{equation}

\vspace{-0.3cm}

In the following, we will use the block-diagonal (or Weyl) representation,
$\alpha^{i}\!=\!\left(-\sigma^{i}\right)\oplus\sigma^{i}$, which
evidences that the massless Dirac Hamiltonian is actually equivalent
to two uncoupled single-node Weyl Hamiltonians, in the absence of
\textit{off-diagonal perturbations} that break chiral symmetry. Mathematically,
the eigenvalue problem for the 4-component Dirac spinor $\boldsymbol{\Psi}_{E}\left(\mathbf{r}\right)$
boils down to
\begin{align}
i\hbar v_{\text{F}}\boldsymbol{\alpha}\!\cdot\!\boldsymbol{\nabla}_{\!\mathbf{r}}\boldsymbol{\Psi}_{E}\left(\mathbf{r}\right) & +\mathbb{I}_{4}V(\abs{\mathbf{r}})\boldsymbol{\Psi}_{E}\left(\mathbf{r}\right)\!=\!E\boldsymbol{\Psi}_{E}\left(\mathbf{r}\right)\label{eq:Weyl_EigenValueProblem}\\
 & \Leftrightarrow\begin{cases}
-i\hbar v_{\text{F}}\boldsymbol{\sigma}\!\cdot\!\boldsymbol{\nabla}_{\!\mathbf{r}}\psi_{E}^{{\scriptscriptstyle \text{(L)}}}\!\!\left(\mathbf{r}\right)\!+\!\mathbb{I}_{2}V(\abs{\mathbf{r}})\psi_{E}^{{\scriptscriptstyle \text{(L)}}}\!\!\left(\mathbf{r}\right) & \!\!\!\!\!=\!E\psi_{E}^{{\scriptscriptstyle \text{(L)}}}\!\!\left(\mathbf{r}\right)\\
\;\:\,i\hbar v_{\text{F}}\boldsymbol{\sigma}\!\cdot\!\boldsymbol{\nabla}_{\!\mathbf{r}}\psi_{E}^{{\scriptscriptstyle \text{(R)}}}\!\!\left(\mathbf{r}\right)\!+\!\mathbb{I}_{2}V(\abs{\mathbf{r}})\psi_{E}^{{\scriptscriptstyle \text{(R)}}}\!\!\left(\mathbf{r}\right) & \!\!\!\!\!=\!E\psi_{E}^{{\scriptscriptstyle \text{(R)}}}\!\!\left(\mathbf{r}\right)
\end{cases},
\end{align}
which is a pair of independent problems for \textit{right-handed}\,/\textit{\,left-handed}
Weyl bispinors, $\psi_{E}^{{\scriptscriptstyle \text{(R)}}}\!\!\left(\mathbf{r}\right)/\psi_{E}^{{\scriptscriptstyle \text{(L)}}}\!\!\left(\mathbf{r}\right)$.
By this point, no assumptions have been made on the scattering potential,
except for the fact that \textit{(i) }it is a \textit{scalar in spinor-space},
and \textit{(ii)} it depends only on the radial coordinate measured
from an arbitrary origin. These are enough to guarantee that Eq.\,\eqref{eq:Weyl_EigenValueProblem}
is spherically symmetric, which allows for a dramatic reduction in
the complexity of the eigenvalue problem by using angular momentum
conservation. In the following paragraphs, we trace the main steps
towards this separation of variables in spherical coordinates, leaving
further details to the discussion in Appendix\,\ref{chap:Dirac_Spherical}. 

Just like in the Schrödinger problem with a central force field, we
are required to find a complete orthonormal basis for bispinors in
the unit-sphere. This is accomplished by introducing the \textit{Spin-$\nicefrac{1}{2}$
Spherical Harmonics},

\vspace{-0.7cm}

\begin{equation}
\Theta_{j,j_{z}}^{+}\!\left(\Omega\right)=\left[\!\!\begin{array}{c}
\sqrt{\frac{j-j_{z}+1}{2j+2}}Y_{j_{z}-1/2}^{j+1/2}\left(\Omega\right)\\
-\sqrt{\frac{j+j_{z}+1}{2j+2}}Y_{j_{z}+1/2}^{j+1/2}\left(\Omega\right)
\end{array}\!\!\right]\text{ and }\Theta_{j,j_{z}}^{-}\!\left(\Omega\right)=\left[\!\!\begin{array}{c}
\sqrt{\frac{j+j_{z}}{2j}}Y_{j_{z}-1/2}^{j-1/2}\left(\Omega\right)\\
\sqrt{\frac{j-j_{z}}{2j}}Y_{j_{z}+1/2}^{j-1/2}\left(\Omega\right)
\end{array}\!\!\right],\label{eq:Normalized+-3-1-2}
\end{equation}

\vspace{-0.3cm}

where $Y_{m}^{l}\!\left(\Omega\right)$ are the usual three-dimensional
(scalar) spherical harmonics. The states defined in Eq.\,\eqref{eq:Normalized+-3-1-2}
are standard eigenfunctions\,\footnote{Meaning common eigenfunctions of $\abs{\mathbf{J}}^{2}$ and $J_{z}$.}
of the total angular momentum which, in this case, contains both an
orbital angular momentum $\mathbf{L}$, as well as a spin-$\nicefrac{1}{2}$
component ($J_{i}\!=\!L_{i}\!+\!\frac{\hbar}{2}\sigma^{i}$). Apart
from that, they are also common eigenfunctions of the \textit{Spin-Orbit
Operator,} $\mathcal{K}\!=\!\frac{\hbar}{2}\mathbf{L}\!\cdot\!\boldsymbol{\sigma}\!$,
which imposes the following useful properties:

\vspace{-0.9cm}
\begin{align}
\abs{\mathbf{J}}^{\!{\scriptscriptstyle 2}}\!\Theta_{j,j_{z}}^{{\scriptscriptstyle \pm}}\!\!\left(\Omega\right) & =\hbar^{2}j\left(j+1\right)\Theta_{j,j_{z}}^{{\scriptscriptstyle \pm}}\!\!\left(\Omega\right)\nonumber \\
J_{z}\,\Theta_{j,j_{z}}^{{\scriptscriptstyle \pm}}\!\left(\Omega\right) & =\hbar j_{z}\Theta_{j,j_{z}}^{{\scriptscriptstyle \pm}}\!\left(\Omega\right)\\
\mathcal{K}\,\Theta_{j,j_{z}}^{{\scriptscriptstyle \pm}}\!\left(\Omega\right) & =\mp\hbar\left(j+\nicefrac{1}{2}\pm1\right)\Theta_{j,j_{z}}^{{\scriptscriptstyle \pm}}\!\left(\Omega\right).\nonumber 
\end{align}

\vspace{-0.3cm}

Since one can also show that the operator $\boldsymbol{\sigma}\!\cdot\!\boldsymbol{\nabla}_{\!\mathbf{r}}$
commutes with both $\abs{\mathbf{J}}$ and $J_{z}$, but not with
$\mathcal{K}$, the general spherically symmetric solutions of Eq.\,\eqref{eq:Weyl_EigenValueProblem}
must be written as 

\vspace{-0.7cm}
\begin{subequations}
\begin{equation}
\psi_{E}^{{\scriptscriptstyle \text{(L)}}}\!\!\left(\mathbf{r}\right)\!=\!\frac{f^{{\scriptscriptstyle \text{(L)}}}\!(r)}{r}\Theta_{{\scriptscriptstyle j,j_{z}}}^{{\scriptscriptstyle +}}\!\!\left(\Omega\right)\!+\!i\frac{g^{{\scriptscriptstyle \text{(L)}}}\!(r)}{r}\Theta_{{\scriptscriptstyle j,j_{z}}}^{{\scriptscriptstyle -}}\!\!\left(\Omega\right)
\end{equation}

\vspace{-0.7cm}

\begin{equation}
\psi_{E}^{{\scriptscriptstyle \text{(R)}}}\!\!\left(\mathbf{r}\right)\!=\!\frac{f^{{\scriptscriptstyle \text{(R)}}}\!(r)}{r}\Theta_{{\scriptscriptstyle j,j_{z}}}^{{\scriptscriptstyle +}}\!\!\left(\Omega\right)\!-\!i\frac{g^{{\scriptscriptstyle \text{(R)}}}\!(r)}{r}\Theta_{{\scriptscriptstyle j,j_{z}}}^{{\scriptscriptstyle -}}\!\!\left(\Omega\right),
\end{equation}

\vspace{-0.2cm}
\end{subequations}

where $j=\nicefrac{1}{2},\nicefrac{3}{2},\cdots$, $j_{z}=-j,-j+1,\cdots,j-1,j$
and $f^{{\scriptscriptstyle \text{(L/R)}}}\!$/$g^{{\scriptscriptstyle \text{(L/R)}}}$
are complex-valued functions of the radial coordinate, $r=\abs{\mathbf{r}}$.
Finally, the radial functions can be determined by re-writing the
free Weyl operator as,
\begin{equation}
\boldsymbol{\sigma}\!\cdot\!\boldsymbol{\nabla}_{\!\mathbf{r}}=\boldsymbol{\sigma}\!\cdot\!\hat{\mathbf{r}}\left[\frac{\partial}{\partial r}-\frac{\mathbf{L}\!\cdot\!\boldsymbol{\sigma}}{\hbar r}\right]=\sigma_{r}\left[\frac{\partial}{\partial r}-\frac{\mathbf{L}\!\cdot\!\boldsymbol{\sigma}}{\hbar r}\right]
\end{equation}
and further recognize that $\sigma_{r}\Theta_{j,j_{z}}^{{\scriptscriptstyle \pm}}\!\left(\Omega\right)\!=\!\Theta_{j,j_{z}}^{{\scriptscriptstyle \mp}}\!\left(\Omega\right)$.
With these ingredients, we arrive at the following system of coupled
radial ODEs\nomenclature{ODEs}{Ordinary Differential Equations}

\vspace{-0.7cm}

\begin{equation}
\begin{cases}
\frac{d}{dr}\!\left[f^{{\scriptscriptstyle \text{(L/R)}}}\!(r)\right]\!+\!\frac{f^{{\scriptscriptstyle \text{(L/R)}}}\!\left(r\right)}{r}\!\left(j\!+\!\frac{1}{2}\right)\!=\!-\frac{1}{\hbar v_{\text{F}}}\left(E\!-\!V(r)\right)g^{{\scriptscriptstyle \text{(L/R)}}}\!(r)\\
\frac{d}{dr}\!\left[g^{{\scriptscriptstyle \text{(L/R)}}}\!(r)\right]\!-\!\frac{g^{{\scriptscriptstyle \text{(L/R)}}}\!\left(r\right)}{r}\left(j\!+\!\frac{1}{2}\right)\!=\,\,\,\,\frac{1}{\hbar v_{\text{F}}}\left(E\!-\!V(r)\right)f^{{\scriptscriptstyle \text{(L/R)}}}\!(r)
\end{cases},\label{eq:System}
\end{equation}
which can be solved for any radial profile of the central potential,
as a function of energy. In all that follows, we will assume the simplest
model for the Spherical Scatterer potential, \textit{i.e.},
\begin{equation}
V(r)\!=\!\lambda\Theta_{\text{H}}\left(b\!-\!r\right)
\end{equation}
which is a step-function that represents a spherical well (or plateaux)
of height $\lambda$ and radius $b$. In this case, the radial equations
may be independently solved for $r>b$ and $r<b$ and then glued back
continuously at the boundary\,\footnote{Since the Dirac (or Weyl) equation is a first-order differential equation
in $\mathbf{r}$, there is no need to guarantee continuity of the
derivative.}.

\vspace{-0.5cm}

\subsection{\label{subsec:Scattering-Solutions}Scattering Solutions and Phase-Shifts}

Based on Eq.\,\eqref{eq:System}, we now build the eigenstates of
the Dirac equation in the presence of the Spherical Scatterer. In
a region that has an uniform scalar potential $V_{0}$, the general
solution\,\footnote{As we will see shortly, there is a singular case \textendash{} $V_{0}\!-\!E=0$
\textemdash{} in which the system becomes decoupled. We are ignoring
this case for the time being.} takes the form

\vspace{-0.7cm}

\begin{subequations}
\begin{align}
f_{j}^{{\scriptscriptstyle \text{(L/R)}}}\!\left(r\right)\! & =\!\sqrt{r}\left[AJ_{j+1}\!\left(\!\frac{\abs{E-V_{0}}r}{\hbar v_{\text{F}}}\!\right)\!+\!BY_{j+1}\!\left(\!\frac{\abs{E-V_{0}}r}{\hbar v_{\text{F}}}\!\right)\right]\label{eq:RadialFunctions}\\
g_{j}^{{\scriptscriptstyle \text{(L/R)}}}\!\left(r\right)\! & =\!-\frac{E-V_{0}}{\abs{E-V_{0}}}\sqrt{r}\left[AJ_{j}\!\left(\!\frac{\abs{E-V_{0}}r}{\hbar v_{\text{F}}}\!\right)\!+\!BY_{j}\!\left(\!\frac{\abs{E-V_{0}}r}{\hbar v_{\text{F}}}\!\right)\right],\label{eq:RadialFunctions2}
\end{align}
\end{subequations}

which are linear superpositions of $J$ and $Y$ \textit{Bessel Functions},
with two undetermined complex-valued coefficients, $A$ and $B$.
From a physical stand-point, the acceptable solutions must decay as
$r\!\to\!+\infty$ and, also, avoid non-square-integrable divergences
as $r\to0^{{\scriptscriptstyle +}}$. This requirement imposes that
$B=0$ and therefore, the acceptable ($E\!\neq\!0$) solutions for
$r<b$ are simply,

\vspace{-0.7cm}

\begin{subequations}
\begin{align}
f_{j}^{{\scriptscriptstyle \text{(L/R)}}}\!\left(r\!<\!b\right)\! & =\!A_{\text{i}}\sqrt{r}J_{j+1}\!\left(\!\frac{\abs{E\!-\!\lambda}r}{\hbar v_{\text{F}}}\!\right)\label{eq:RadialFunctions-1-1}\\
g_{j}^{{\scriptscriptstyle \text{(L/R)}}}\!\left(r\!<\!b\right)\! & =\!-A_{\text{i}}\text{Sign}\left(E\!-\!V_{0}\right)\sqrt{r}J_{j}\!\left(\!\frac{\abs{E\!-\!\lambda}r}{\hbar v_{\text{F}}}\!\right).\label{eq:RadialFunctions2-1-1}
\end{align}
\end{subequations}

Meanwhile, outside the spherical scatterer, both Bessel components
are admissible and the general form of the solution is

\vspace{-0.7cm}

\begin{subequations}
\begin{align}
f_{j}^{{\scriptscriptstyle \text{(L/R)}}}\!\left(r\!>\!b\right)\! & =\!\sqrt{r}\left[A_{\text{o}}J_{j+1}\!\left(\!\frac{\abs Er}{\hbar v_{\text{F}}}\!\right)\!+\!B_{\text{o}}Y_{j+1}\!\left(\!\frac{\abs Er}{\hbar v_{\text{F}}}\!\right)\right]\label{eq:RadialFunctions-1}\\
g_{j}^{{\scriptscriptstyle \text{(L/R)}}}\!\left(r\!>\!b\right)\! & =\!-\text{Sign}\left(E\right)\sqrt{r}\left[A_{\text{o}}J_{j}\!\left(\!\frac{\abs Er}{\hbar v_{\text{F}}}\!\right)\!+\!B_{\text{o}}Y_{j}\!\left(\!\frac{\abs Er}{\hbar v_{\text{F}}}\!\right)\right].\label{eq:RadialFunctions2-1}
\end{align}
\end{subequations}
As mentioned before, the undetermined coefficients are mutually related
by the continuity condition of the wavefunction at $r\!=\!b$, which
yields the following set of conditions:

\vspace{-0.7cm}
\begin{align}
A_{\text{o}}J_{j+1}\!\left(\!\frac{\abs Eb}{\hbar v_{\text{F}}}\!\right)+B_{\text{o}}Y_{j+1}\!\left(\!\frac{\abs Eb}{\hbar v_{\text{F}}}\!\right) & \,\,=\,\,A_{\text{i}}J_{j+1}\!\left(\!\frac{\abs{E\!-\!\lambda}b}{\hbar v_{\text{F}}}\!\right)\label{eq:Cont_Condition}\\
A_{\text{o}}J_{j}\!\left(\!\frac{\abs Eb}{\hbar v_{\text{F}}}\!\right)+\,\,\,\,\,B_{\text{o}}Y_{j}\!\left(\!\frac{\abs Eb}{\hbar v_{\text{F}}}\!\right) & \,\,=\,\,A_{\text{i}}\text{Sign}\left(\frac{E\!-\!\lambda}{E}\right)J_{j}\!\left(\!\frac{\abs{E\!-\!\lambda}b}{\hbar v_{\text{F}}}\!\right),\label{eq:Cont_Condition2}
\end{align}
at a generic non-zero energy. As is well-known in scattering theory
(and reviewed in Appendix\,\ref{chap:Dirac_Spherical}), the net
effect of any short-ranged spherical potential is to create an \textit{energy-dependent
phase-shift}, $\delta_{j}(E)$, which appears asymptotically in the
outgoing spherical waves associated to each angular momentum channel.
By defining this phase-shift as

\vspace{-0.8cm}

\begin{subequations}
\begin{align}
f_{j}^{{\scriptscriptstyle \text{(L/R)}}}\!\left(r\right) & \!\underset{{\scriptscriptstyle r\to\infty}}{\longrightarrow}\frac{\mathcal{N}}{r}\cos\left(\frac{\abs Er}{\hbar v_{\text{F}}}-\frac{\pi}{2}(j\!+\!\frac{3}{2})\!-\!\delta_{j}\!\left(E\right)\right)\label{eq:Assymptotic1}\\
g_{j}^{{\scriptscriptstyle \text{(L/R)}}}\!\left(r\right) & \!\underset{{\scriptscriptstyle r\to\infty}}{\longrightarrow}\frac{\mathcal{N}}{r}\frac{E}{\abs E}\cos\left(\frac{\abs Er}{\hbar v_{\text{F}}}\!-\!\frac{\pi}{2}(j\!+\!\frac{1}{2})\!-\!\delta_{j}\!\left(E\right)\right),\label{eq:Assymptotic2}
\end{align}
\end{subequations}

we can employ Eqs.\,\eqref{eq:Cont_Condition}-\eqref{eq:Cont_Condition}
to determine it analytically, which gives

\vspace{-0.7cm}

\begin{align}
\!\!\!\!\!\!\!\!\!\!\!\delta_{j}(E,\lambda)\! & =\!\arctan\left(-\frac{B_{\text{o}}}{A_{\text{o}}}\right)\label{eq:PhaseShift}\\
 & =\!\arctan\!\left(\!\frac{\text{Sign}\left(E\!-\!\lambda\right)J_{j+1}\!\left(\!\frac{\abs Eb}{\hbar v_{\text{F}}}\!\right)J_{j}\!\left(\!\frac{\abs{E\!-\!\lambda}b}{\hbar v_{\text{F}}}\!\right)-\text{Sign}\left(E\right)J_{j}\!\left(\!\frac{\abs Eb}{\hbar v_{\text{F}}}\!\right)J_{j+1}\!\left(\!\frac{\abs{E\!-\!\lambda}b}{\hbar v_{\text{F}}}\!\right)}{\text{Sign}\left(\frac{E\!-\!\lambda}{E}\right)Y_{j+1}\!\left(\!\frac{\abs Eb}{\hbar v_{\text{F}}}\!\right)J_{j}\!\left(\!\frac{\abs{E\!-\!\lambda}b}{\hbar v_{\text{F}}}\!\right)-Y_{j}\!\left(\!\frac{\abs Eb}{\hbar v_{\text{F}}}\!\right)J_{j+1}\!\left(\!\frac{\abs{E\!-\!\lambda}b}{\hbar v_{\text{F}}}\!\right)}\!\right).\!\!\!\nonumber 
\end{align}
Equation.\,\eqref{eq:PhaseShift} is well known to encapsulate all
the information about the effects of the spherical scatterer in \textit{electronic
spectral structure} and \textit{scattering times}. However, it is
important to comment that the phase-shift $\delta_{j}(E,\lambda)$
is \textit{ambiguous by integer multiples of $\pi$} (see Calogero\,\cite{Calogero67}
for a general discussion on this ambiguity). For most purposes, the
precise convention is not important but, as it turns out, it will
be extremely important that we fix this phase relative to the unperturbed
(clean) Hamiltonian. This can be accomplished by fixing the \textit{UV-Assymptotic
Value} to
\begin{equation}
\delta_{j}\left(E\to\pm\infty,\lambda\right)\to-\frac{\lambda b}{\hbar v_{\text{F}}},
\end{equation}
as suggested by Ma \textit{et al}.\,\cite{Ma06,Ma85}. In this sense,
a good to define $\delta_{j}(E,\lambda)$ is by energy branches, namely{\scriptsize{}
\begin{equation}
\!\!\!\!\!\delta_{j}\left(E,\lambda\right)\!=\!-\frac{b}{\hbar v_{\text{F}}}\left[\!\lambda\!+\!\!\int_{-\infty}^{E}\!\!\!\!\!\!\!dx\frac{d}{dx}\arctan\!\left(\!\frac{\text{Sign}\left(x\right)J_{j}\!\left(\!\frac{\abs xb}{\hbar v_{\text{F}}}\!\right)J_{j\!+\!1}\!\left(\!\frac{\abs{x\!-\!\lambda}b}{\hbar v_{\text{F}}}\!\right)-\text{Sign}\left(x\!-\!\lambda\right)J_{j+1}\!\left(\!\frac{\abs xb}{\hbar v_{\text{F}}}\!\right)J_{j}\!\left(\!\frac{\abs{x\!-\!\lambda}b}{\hbar v_{\text{F}}}\!\right)}{\text{Sign}\left(\frac{x\!-\!\lambda}{x}\right)Y_{j\!+\!1}\!\left(\!\frac{\abs xb}{\hbar v_{\text{F}}}\!\right)J_{j}\!\left(\!\frac{\abs{x\!-\!\lambda}b}{\hbar v_{\text{F}}}\!\right)-Y_{j}\!\left(\!\frac{\abs xb}{\hbar v_{\text{F}}}\!\right)J_{j\!+\!1}\!\left(\!\frac{\abs{x\!-\!\lambda}b}{\hbar v_{\text{F}}}\!\right)}\!\right)\!\right],\!\label{eq:Phase_Shift_Corr_Assympt-1}
\end{equation}
}{\scriptsize\par}

for $E\!<\!0$, and{\scriptsize{}
\begin{equation}
\!\!\!\!\!\delta_{j}\left(E,\lambda\right)\!=\!-\frac{b}{\hbar v_{\text{F}}}\left[\!\lambda\!+\!\!\int_{E}^{\infty}\!\!\!\!\!\!\!dx\frac{d}{dx}\arctan\!\left(\!\frac{\text{Sign}\left(x\right)J_{j}\!\left(\!\frac{\abs xb}{\hbar v_{\text{F}}}\!\right)J_{j\!+\!1}\!\left(\!\frac{\abs{x\!-\!\lambda}b}{\hbar v_{\text{F}}}\!\right)-\text{Sign}\left(x\!-\!\lambda\right)J_{j+1}\!\left(\!\frac{\abs xb}{\hbar v_{\text{F}}}\!\right)J_{j}\!\left(\!\frac{\abs{x\!-\!\lambda}b}{\hbar v_{\text{F}}}\!\right)}{\text{Sign}\left(\frac{x\!-\!\lambda}{x}\right)Y_{j\!+\!1}\!\left(\!\frac{\abs xb}{\hbar v_{\text{F}}}\!\right)J_{j}\!\left(\!\frac{\abs{x\!-\!\lambda}b}{\hbar v_{\text{F}}}\!\right)-Y_{j}\!\left(\!\frac{\abs xb}{\hbar v_{\text{F}}}\!\right)J_{j\!+\!1}\!\left(\!\frac{\abs{x\!-\!\lambda}b}{\hbar v_{\text{F}}}\!\right)}\!\right)\!\right],\!\label{eq:Phase_Shift_Corr_Assympt-1-1}
\end{equation}
}otherwise. This was the convention used in Ref.\,\cite{Pires2021}.

\vspace{-0.5cm}

\subsection{\label{subsec:Nodal-Bound-States}Nodal Bound States of Dirac-Weyl
Fermions}

So far, our analysis have limited to scattering states of the potential,
which exist for $E\!\neq\!0$, and for which the net effect of $V(\abs{\mathbf{r}})$
is to dephase the outgoing spherical waves. In stark contrast, precisely
at the nodal energy ($E\!=\!0$), a different breed of quantum eigenstates
becomes possible by the decoupling of Eqs.\,\ref{eq:System} outside
the impurity. This way, the general radial solutions take the form,
\begin{equation}
f_{j}^{{\scriptscriptstyle \text{(L/R)}}}\!\left(r\!>\!b\right)\!=\!\frac{C}{r^{j+\frac{1}{2}}}\text{ and }g_{j}^{{\scriptscriptstyle \text{(L/R)}}}\!\left(r\!>\!b\right)\!=\!Dr^{j+\frac{1}{2}},\label{eq:RadialFunctions3}
\end{equation}
outside the scatterer, where $D\!=\!0$ so as to guarantee a (power-law)
decaying wavefunction. The continuity conditions at $r=b$ are likewise
altered and now take the form,
\begin{align}
A_{\text{i}}J_{j+1}\!\left(\!\frac{\abs{\lambda}b}{\hbar v_{\text{F}}}\!\right) & =\,\,C_{\text{o}}b^{-j-\frac{1}{2}}\text{ and }A_{\text{i}}J_{j}\!\left(\!\frac{\abs{\lambda}b}{\hbar v_{\text{F}}}\!\right)=0.\label{eq:Cont_Condition-1}
\end{align}
These equations can only be satisfied if $J_{j}\left(\abs{\lambda}b/\hbar v_{\text{F}}\right)\!=\!0$,
defining a discrete set of values for the dimensionless parameter,
$u\!=\!\abs{\lambda}b/\hbar v_{\text{F}}$, which support the presence
of zero-energy bound states in that $j$\,-\,channel. In fact, if
$u$ takes on such a \textit{``magical value''} $u_{c}$, there
will be $2j\!+\!1$ zero-energy eigenstates that correspond to squared-normalizable
wavefunctions that are bound to the spherical scatterer. Rather than
being exponentially localized wavefunctions, as happens for the states
in semiconducting Lifshitz tails, these nodal eigenstates feature
\textit{power-law tails} that fall-off as $1/r^{j+1/2}$. This is
in accordance with the absence of any intrinsic energy scale (such
as a spectral gap) in the clean Dirac Hamiltonian that could justify
the appearance of a localization length. In spite of this, we have
managed to show that these delicate nodal states appear naturally
as the \textit{gapless limit of the in-gap bound states} that are
created by isolated potential impurities in massive Dirac systems
(see Santos Pires \textit{et al}.\,\cite{Pires2021}).

\vspace{-0.5cm}

\subsection{\label{subsec:Friedel-Sum-Rule}Friedel Sum Rule and Levinson's Theorem\nomenclature{LT}{Levinson's Theorem}}

Previously, we have concluded that the effects of introducing an isolated
spherical scatterer in a DSM is to dephase the free spherical Dirac
states by a energy\,- and $j$\,-\,dependent phase-shift, $\delta_{j}(E)$.
This situation is a rather generic one in the \textit{Scattering Theory
of Spherically Symmetric Potentials}\,\cite{Calogero67}, and it
is well understood that the phase-shift induced by a scattering potential
can provide the answer to almost any question one can ask about the
corresponding single-impurity problem. For our purposes, we focus
on the spectral effects of the spherical scatterer, that is, in the
deformations it causes to the global DoS, as well as the possible
emergence of \textit{impurity bound-states}. The answers to both questions
are encapsulated in two important results that will be our subject
here: \textit{(i)} the \textit{Friedel Sum Rule} (FSR\nomenclature{FSR}{Friedel's Sum Rule}),
and \textit{(ii) }the celebrated \textit{Levinson's Theorem} (LT\nomenclature{LT}{Levinson's Theorem}).
In this Section, we will state both these results and, later, provide
a detailed proof of the FSR. This proof is an essential ingredient
to understand the results and claims of Ref.\,\cite{Pires2021} regarding
the statistical significance of zero-energy bound states for the density
of states of a DWSM.

\vspace{-0.5cm}

\paragraph{The Friedel Sum Rule (FSR):}

The Friedel Sum Rule is a general theorem of scattering theory\,\cite{Friedel52}
that relates the energy-derivative of the scattering phase-shifts,
in each angular momentum channel, to the deformation induced by the
impurity in the mean-level spacing of the whole system. In the context
of Dirac (or Weyl) particles, the FSR states that the change in the
\textit{extensive density of states} (eDoS\nomenclature{eDoS}{Density of States (per unit energy)})
is expressed as

\vspace{-0.8cm}

\begin{equation}
\delta\nu(E,\lambda,b)\!=\!\!\!\sum_{j=1/2}^{\infty}\delta\nu_{j}(E,\lambda,b)=\frac{n_{s}}{\pi}\sum_{j=1/2}^{\infty}\left(2j\!+\!1\right)\frac{\partial\delta_{j}(E,\lambda,b)}{\partial E},\label{eq:FSR}
\end{equation}
where $n_{s}\!=\!4$ ($n_{s}\!=\!2$) for Dirac (Weyl) particles,
and each $\delta\nu_{j}(E,\lambda)$ term is to be understood as the
contribution of the $j$\,-\,channel to the variation of the eDoS
at energy $E$. In Eq.\,\eqref{eq:FSR}, both $\lambda$ and $b$
work as \textit{``external parameters''} that characterize the shape
of the central impurity.

\vspace{-0.5cm}

\paragraph{Levinson's Theorem (LT):}

The Levinson's Theorem\,\cite{Levinson49} is a complementary result
to the FSR that relates the zero-momentum discontinuities in the scattering
phase-shifts, with the \textit{number of bound states} created in
the system by the impurity. In short, it is possible to prove that
any $\pi$-discontinuity in the phase-shifts at zero momentum correspond
to the creation (or destruction) of an impurity bound state. Even
though such a relationship also exists for the Schrödinger equation,
we focus exclusively on Dirac particles but take small detour to assume
that the Dirac particles are actually massive (with mass $m$). This
entails the following continuum Hamiltonian,
\begin{align}
\mathscr{H}_{c}^{m}\! & =\!-i\hbar v_{\text{F}}\int\!d\mathbf{r}\,\Psi_{a\mathbf{r}}^{\dagger}\left(\boldsymbol{\alpha}^{ab}\!\cdot\!\boldsymbol{\nabla}_{\mathbf{r}}\right)\Psi_{b\mathbf{r}}+mv_{\text{F}}^{2}\int\!d\mathbf{r}\,\Psi_{a\mathbf{r}}^{\dagger}\beta^{ab}\Psi_{b\mathbf{r}}\label{eq:MassiveDirac}\\
 & \qquad\qquad\qquad\qquad\qquad\qquad\qquad\qquad\qquad+\!\int\!d\mathbf{r}\,V(\abs{\mathbf{r}})\,\Psi_{a\mathbf{r}}^{\dagger}\Psi_{a\mathbf{r}},\nonumber 
\end{align}
where $m$ is the Dirac mass and $\beta$ is the off-diagonal $4\!\times\!4$
matrix defined as
\begin{equation}
\beta=\left[\begin{array}{cc}
\mathbb{O}_{{\scriptscriptstyle 2\times2}} & \mathbb{I}_{{\scriptscriptstyle 2\times2}}\\
\mathbb{I}_{{\scriptscriptstyle 2\times2}} & \mathbb{O}_{{\scriptscriptstyle 2\times2}}
\end{array}\right].\label{eq:DiracMassMatrix}
\end{equation}
Based on the Hamiltonian of Eq.\,\eqref{eq:MassiveDirac}, LT can
be rigorously proved for the number of bound-states lying inside the
mass-gap, by using one of the procedures described in Refs.\,\cite{Ma85,Ma06}.
In place of deriving it here, we simply state that the number of \textit{bound
states} (or \textit{quasi-bound states}\,\cite{Ma85}) of angular
momentum $j$, and a $\kappa\!=\!\pm1$ spin-orbit quantum number,
equals
\begin{align}
N_{j,\kappa}^{\text{b}}\left(m,\lambda,b\right)\! & =\!\frac{2j\!+\!1}{\pi}\left(\delta_{j}^{\kappa}\left(mv_{\text{F}}^{2},\lambda,b\right)-\delta_{j}^{\kappa}\left(-mv_{\text{F}}^{2},\lambda,b\right)\right)\label{eq:LevinsonDirac}\\
 & -\kappa\left(-1\right)^{j+\frac{1}{2}}\left(j+\frac{1}{2}\right)\left[\sin^{2}\delta_{j}^{\kappa}\left(mv_{\text{F}}^{2},\lambda,b\right)-\sin^{2}\delta_{j}^{\kappa}\left(-mv_{\text{F}}^{2},\lambda,b\right)\right],\nonumber 
\end{align}
where $\delta_{j}^{\kappa}(E,\lambda,b)$\, is the phase-shift associated
to the scattering states in that angular momentum channel. Unlike
the previous phase-shifts, we are forced to include a further spin-orbit
label, $\kappa$, because the phase-shifts of massive Dirac particles
generally depend of it. As shown in Appendix\,\ref{chap:Dirac_Spherical},
this $\kappa$\,-\,dependence disappears in the gapless limit and,
since the $m\!\to\!0^{{\scriptscriptstyle +}}$ limit of Eq.\,\eqref{eq:MassiveDirac}
leads to two uncoupled copies of a Weyl Hamiltonian, we conclude that
the LT for Dirac-Weyl fermions simply reads,
\begin{align}
n_{j}^{\text{b}}\left(\lambda,b\right)\! & =\!\frac{n_{s}}{\pi}\left(j\!+\!\frac{1}{2}\right)\left[\delta_{j}\left(0^{+}\!,\lambda,b\right)\!-\!\delta_{j}\left(0^{-}\!,\lambda,b\right)\right],\label{eq:LevinsonDirac-1}
\end{align}
where $n_{s}=2\text{ or }4$ depending on the case. This result is
in full accordance with an alternative derivation presented by Lin\,\cite{Lin06},
which does not require to take the gapless limit of a massive Dirac
Hamiltonian.

\vspace{-0.5cm}

\subsection{\label{subsec:DerivationFSR}Derivation of the Friedel Sum Rule for
Dirac-Weyl Electrons}

As referred in the beginning of the chapter, the formal derivation
of the the FSR for Dirac particles will prove to be an important step
to understand the upcoming results. In particular, such a derivation
demonstrates that the blind application of Eq.\,\eqref{eq:FSR} to
spherical scatterers, set up near a \textit{``magical value'',}
leads to somewhat confusing results which decisively contributed to
spark a recent debate on the statistical significance of these fine-tuned
nodal bound states in DWSMs (see Refs.\,\cite{Buchhold18a,Buchhold18b,Ziegler18,Wilson20,Pixley21,Pires2021}).

To derive the FSR for gapless Dirac particles, we begin by imagining
that our spherical scatterer is placed inside of a much bigger spherical
cavity of radius $R\!\gg\!b$\,\footnote{If the scattering potential has short-ranged tails, the result holds
unchanged.}. This amounts to the constrained Hamiltonian\,\cite{Mccann04}

\vspace{-0.8cm}

\begin{equation}
\mathscr{H}_{\text{c}}\!=\!-i\hbar v_{\text{F}}\int\!d\mathbf{r}\Psi_{a\mathbf{r}}^{\dagger}\left[\boldsymbol{\alpha}^{ab}\!\cdot\!\boldsymbol{\nabla}_{\!\mathbf{r}}\!-\!b^{2}\mathbb{U}^{ab}\delta^{{\scriptscriptstyle (3)}}\!\left(\abs{\mathbf{r}}\!-\!R\right)\right]\Psi_{b\mathbf{r}},\label{eq:Hamiltonian}
\end{equation}
for which one is forced to impose that the $4\!\times\!4$ matrix
$\mathbb{U}$ is unitary, hermitian and anti-commuting with $\boldsymbol{\alpha}\cdot\hat{\mathbf{r}}$.
These conditions are sufficient\cite{Berry87,alonso97} to guarantee
that Eq.\,\eqref{eq:Hamiltonian} is a hermitian Hamiltonian constrained
to the spherical cavity. While all possibilities for the boundary
matrix, $\mathbb{U}$, have been classified by McCann and Fal'ko\cite{Mccann04},
we can simply consider $\mathbb{U}=\beta$ which corresponds to an
\textit{infinite Dirac mass term} outside the cavity limits ($\abs{\mathbf{r}}>R$).
This choice, known in nuclear physics as the \textit{MIT bag model}\cite{chodos74},
avoids a \textit{Klein tunneling ``paradox}'' at the boundary but
still maintains the spherical symmetry of the problem. However, it
presents the trade-off of mixing together the two Weyl sectors of
the gapless Dirac Hamiltonian. Hence, we will work with the full 4-component
Dirac wavefunctions and argue, in the end, that this \textit{boundary-induced
valley mixing is irrelevant when $R\!\to\!+\infty$}. All in all,
the application of this boundary condition boils down to the linear
condition\cite{Mccann04}

\vspace{-0.7cm}

\begin{equation}
\left[\!\begin{array}{cc}
\mathbb{I}_{{\scriptscriptstyle 2\!\times\!2}} & -i\sigma_{r}\\
i\sigma_{r} & \mathbb{I}_{{\scriptscriptstyle 2\!\times\!2}}
\end{array}\!\right]\Psi_{E}\left(R,\theta,\phi\right)\!=\!0,\label{eq:LinearBoundaryConstraint}
\end{equation}
which translates into a linear relationship between the four, previously
independent, radial functions of Eq.\,\eqref{eq:System}, \textit{i.e.},

\vspace{-0.7cm}

\begin{equation}
\begin{cases}
f_{j}^{{\scriptscriptstyle \text{(L)}}}\!(R)\!-\!g_{j}^{{\scriptscriptstyle \text{(R)}}}\!(R)=0\\
g_{j}^{{\scriptscriptstyle \text{(L)}}}\!(R)\!-\!f_{j}^{{\scriptscriptstyle \text{(R)}}}\!(R)=0
\end{cases}\!\!\!\!.\label{eq:BoundaryConditions}
\end{equation}
Since we are assuming $E\neq0$ and $R$ very large\,\footnote{More precisely, we require $R\abs E\gg\hbar v_{\text{F}}$.},
we are entitled to impose the aforementioned boundary condition using
the asymptotic expressions for the radial functions, $f_{j}^{{\scriptscriptstyle \text{(R/L)}}}/g_{j}^{{\scriptscriptstyle \text{(R/L)}}}$,
as shown in Eqs.\,\eqref{eq:Assymptotic1}-\eqref{eq:Assymptotic2}.
This way, the system of Eq.\,\eqref{eq:BoundaryConditions} boils
down to a single condition,

\vspace{-0.7cm}

\begin{align}
\smash{\cos\left(\frac{\abs E\!R}{\hbar v_{\text{F}}}-\frac{\pi}{4}\left(2j+3\right)-\delta_{j}\!(E,\lambda,b)\right)-\text{Sign}(E)\qquad\qquad\qquad\qquad\qquad}\\
\smash{\times\cos\left(\frac{\abs E\!R}{\hbar v_{\text{F}}}-\frac{\pi}{4}\left(2j+1\right)-\delta_{j}(E,\lambda,b)\right)=0}\nonumber 
\end{align}
or, equivalently,

\vspace{-0.7cm}
\begin{align}
\smash{\sin\left(\frac{\abs E\!R}{\hbar v_{\text{F}}}-\frac{\pi}{4}\left(2j+1\right)-\delta_{j}\!(E,\lambda,b)\right)}\label{eq:Bound}\\
\!\!\!\!\!\times & \cos\left(\frac{\abs E\!R}{\hbar v_{\text{F}}}-\frac{\pi}{4}\left(2j\!+\!1\right)-\delta_{j}(E,\lambda,b)\right).\nonumber 
\end{align}
Equation\,\eqref{eq:Bound} is only obeyed if 

\vspace{-0.7cm}

\begin{subequations}
\begin{align}
E_{n} & =+\frac{hv_{\text{F}}}{4R}\left(j-1+2n+\frac{2}{\pi}\delta_{j}\!(E_{n},\lambda,b)\right)\text{ if \ensuremath{E_{n}\!>\!0}}\label{eq:General_Quantization}\\
E_{n} & =-\frac{hv_{\text{F}}}{4R}\left(j+2n+\frac{2}{\pi}\delta_{j}\!(E_{n},\lambda,b)\right)\text{ if \ensuremath{E_{n}\!<\!0}}\label{eq:General}
\end{align}
\end{subequations}

with $n\in\mathbb{Z}$. Generally, Eqs.\,\eqref{eq:General_Quantization}
and \eqref{eq:General} yield complicated implicit conditions in the
energy values that define a discrete set of levels (labelled by $n$)
corresponding to quantum states allowed by the boundary conditions
in $r\!=\!R$. Nonetheless, if $\lambda\!=\!0$ there is no spherical
scatterer in the center of the system and, consequently, the scattering
phase-shifts are zero for all energies. In this situation, the quantized
energy levels take the particularly simple form,

\vspace{-0.7cm}

\begin{align}
E_{n}^{j} & =n\frac{hv_{\text{F}}}{2R}\!\left[1-\frac{1+\left(2j\!-\!1\right)\!\!\!\!\!\!\mod\!\!4}{4\abs n}\right]\text{ with \ensuremath{n\in\left\{ \cdots,-2,-1,1,2,\cdots\right\} }}\label{eq:General_Quantization-1}
\end{align}

which gives a symmetric energy spectrum that is\textit{ uniform} with
a mean-level spacing, $hv_{\text{F}}/2R$, for all angular momentum
channel $j$. Note that this simple spectrum is only valid provided
the asymptotic forms for the radial wavefunctions {[}Eqs.\,\eqref{eq:Assymptotic1}-\eqref{eq:Assymptotic2}{]}
are valid, which implies that $\abs n\!\gg\!1$. In contrast, if a
similar calculation is done in the presence of a scatterer, the energy-dependent
phase-shift does not allow for an explicit solution of Eq.\,\eqref{eq:Bound}
at an arbitrary $R$. However, one may assume that $R$ is sufficiently
large so that $\delta_{j}(E,\lambda)$ varies slowly with energy at
the scale of $\pi\hbar v_{\text{F}}/R$. In that \textit{``pre-thermodynamic''}
limit, we would have the spectrum 

\begin{figure}[t]
\vspace{-0.6cm}
\begin{centering}
\includegraphics[scale=0.21]{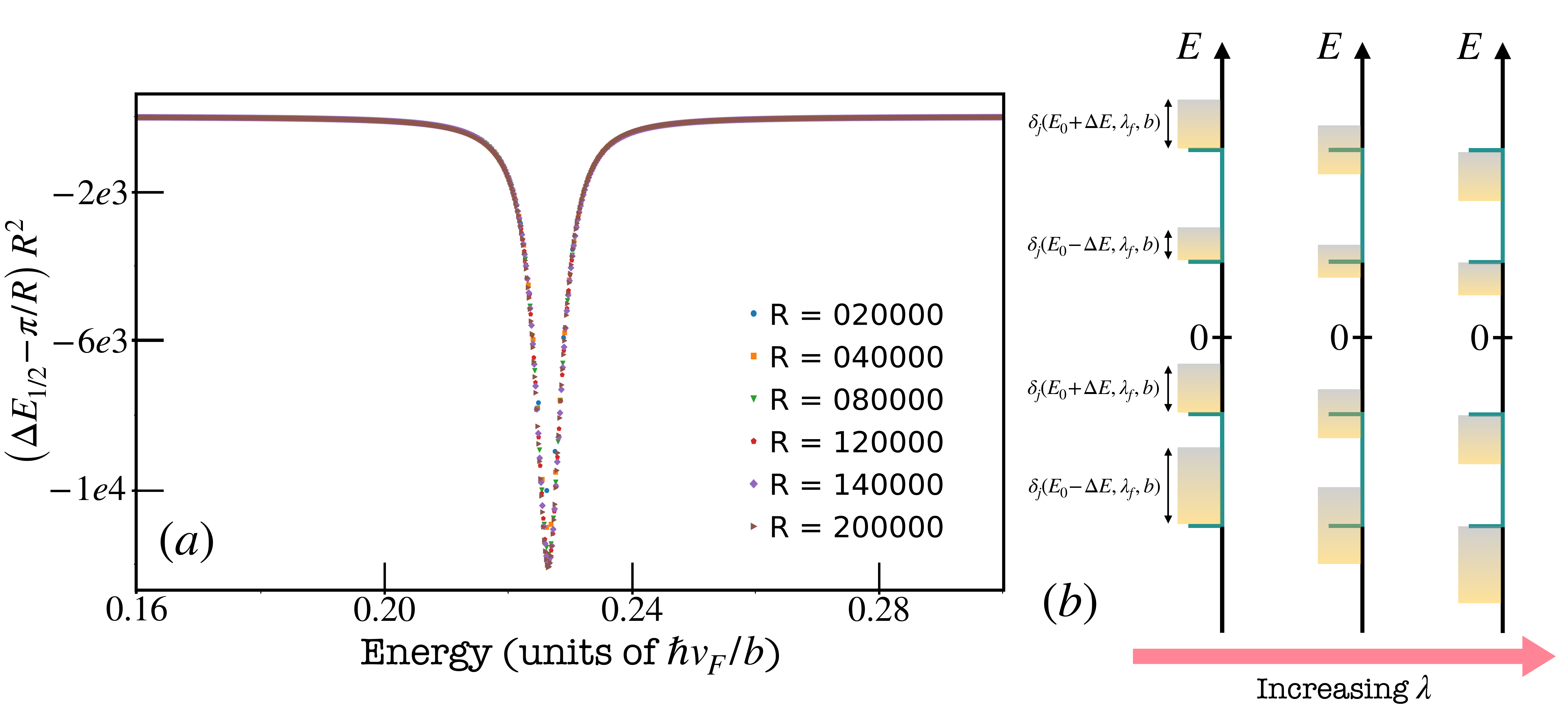}
\par\end{centering}
\vspace{-0.4cm}

\caption{\label{fig:DerivationFSR}(a) Plot of $\Delta E_{{\scriptscriptstyle 1/2}}\!=\!E_{{\scriptscriptstyle n+1}}^{{\scriptscriptstyle 1/2}}\!-\!E_{{\scriptscriptstyle n}}^{{\scriptscriptstyle 1/2}}$
the nearest-level spacings for a spherical scatterer with $\lambda b\!=\!3.1867$
calculated from the numerically found solutions of the boundary condition
{[}Eq.\,\eqref{eq:QuantizationCond}{]}. The data points are represented
as $R^{2}\!\times\!\left(\Delta E_{{\scriptscriptstyle 1/2}}\!-\!\pi R^{-1}\right)$,
such that a collapse of different values of $R$ is achieved, indicating
that $\Delta\varepsilon_{{\scriptscriptstyle 1/2}}\left(R,\varepsilon,u\right)\!\approx\!\pi R^{-1}\!-\!f\left(\varepsilon,u\right)R^{-2}$
in the presence of a central impurity. $R$ is measured in units of
$b$. (b) Pictorial representation of the motion of energy levels
triggered by the central impurity, with a value of $\lambda$ that
increases adiabatically from $0\to\lambda_{f}$.}

\vspace{-0.4cm}
\end{figure}

\vspace{-0.7cm}

\begin{equation}
E_{n}^{j}(\lambda)\!\approx n\frac{hv_{\text{F}}}{2R}\!\left[1-\frac{1+\left(2j\!-\!1\right)\!\!\!\!\!\!\mod\!\!4}{4\abs n}\right]-\frac{\hbar v_{\text{F}}}{R}\delta_{j}(E_{n}^{j},\lambda,b),\label{eq:SpacingClean-1}
\end{equation}
in which the phase-shifts enter as small displacements of the quantized
energy levels, relative to the \textit{free cavity spectrum} of Eq.\,\eqref{eq:General_Quantization-1}.
More important than the spectrum itself, Equation\,\eqref{eq:SpacingClean-1}
now entails a \textit{non-uniform spectrum} with the change in the
spacing of consecutive energy levels in the $j$-channel getting corrected
to 
\begin{equation}
E_{n+1}^{j}(\lambda)\!-\!E_{n}^{j}(\lambda)\approx\!\frac{hv_{\text{F}}}{2R}\left[1\!-\!\frac{hv_{\text{F}}}{2R}\left.\frac{\partial}{\partial E}\delta_{j}\left(E,\lambda,b\right)\right|_{E=E_{n}^{j}(\lambda)}\!\!\!+\cdots\right],
\end{equation}
due to the spherical scatterer. This result is illustrated in Fig.\,\ref{fig:DerivationFSR}\,a,
where the consecutive spacings obtained from a numerical solution
of Eq.\,\eqref{eq:QuantizationCond} are plotted for several values
of $R$ and fixed $\lambda$. Upon a re-scaling, all the spacings
fall on top of a \textit{universal curve} which is proportional to
$\partial\delta_{j}\left(E,\lambda,b\right)/\partial E$.

By this point, we have all the necessary ingredients to build up the
argument which establishes the FSR in this systems. We start by considering
a finite energy window (like the red ones in Fig.\,\ref{fig:DerivationFSR}\,b)
which is centered in an energy $E_{0}$ and has a finite width $\Delta E$.
While the number of energy levels inside the box can be hard to calculate,
the change in its value caused by the spherical scatterer is much
simpler: it is given by the inward (or outward) migration of levels
from regions of width $\simeq\hbar v_{\text{F}}\delta_{j}\left(E\pm\nicefrac{\Delta E}{2},\lambda\right)/R$
near the respective boundaries, which happens upon an adiabatic connection
of $\lambda$ from $0$ to the value that actually characterizes the
central potential. Therefore, we conclude that the change in the number
of states within that energy window is given by
\begin{align}
\Delta N\left(E_{0},\Delta E,\lambda\right)\! & =\!\sum_{j=\frac{1}{2}}^{\infty}\Delta N_{j}\left(E_{0},\Delta E,\lambda,b\right)\\
 & =\frac{n_{s}}{\pi}\sum_{j=\frac{1}{2}}^{\infty}\left(2j\!+\!1\right)\!\left[\delta_{j}\!\left(E_{0}\!+\!\frac{1}{2}\Delta E,\lambda,b\right)\!-\!\delta_{j}\left(E_{0}\!-\!\frac{1}{2}\Delta E,\lambda,b\right)\right],\nonumber 
\end{align}
where the factor $2j\!+\!1$ arises from the $j_{z}$\,-\,degeneracy
carried by each level in the $j$-channel of a gapless and spherically
symmetric Dirac system. Finally, in order to obtain the change in
the eDoS, we just take the thermodynamic ($R\!\to\!\infty$) and infinite
energy resolution limit ($\Delta E\!\to\!0$) in the appropriate order.
This yields the aforementioned FSR:

\vspace{-0.7cm}
\begin{equation}
\delta\nu\left(E_{0},\lambda\right)\!=\!\lim_{\Delta E\to0}\lim_{R\to\infty}\frac{\Delta N\left(E_{0},\Delta E,\lambda\right)}{\Delta E}=\frac{n_{s}}{\pi}\sum_{j=\frac{1}{2}}^{\infty}\left(2j\!+\!1\right)\!\frac{\partial\delta_{j}(E,\lambda)}{\partial E},\label{eq:DoS_FSR}
\end{equation}

\vspace{-0.5cm}

where $n_{s}\!=\!2,4$ depending if we are referring to a Dirac or
a Weyl system. Before proceeding, it is important to leave a note
of caution on the use of Eq.\,\eqref{eq:DoS_FSR}. The quantity calculated
here is a change in the inverse mean-level spacing (or, equivalently,
the eDoS) due to a single impurity in an otherwise infinite system.
In a real situation, there is usually a finite, albeit small, concentration
($c$) of impurities, so that one might be tempted to obtain the DoS
by simply adding up all contributions, invoking a well-defined dilute
limit. If the impurities are all identical, this procedure leads to

\vspace{-0.7cm}
\begin{equation}
\delta\rho(E,\lambda,b)=\frac{2c}{\pi}\sum_{j=\frac{1}{2}}^{\infty}\left(2j\!+\!1\right)\!\frac{\partial\delta_{j}(E,\lambda,b)}{\partial E}
\end{equation}
or, if $\lambda$ is a random variable drawn from a distribution $P(\lambda)$,
we would have 

\vspace{-0.7cm}

\begin{equation}
\overline{\delta\rho(E,b)}=\frac{2c}{\pi}\int\!\!d\lambda\,P(\lambda)\left[\sum_{j=\frac{1}{2}}^{\infty}\!\left(2j\!+\!1\right)\!\frac{\partial\delta_{j}(E,\lambda,b)}{\partial E}\right]\label{eq:MeanFSR}
\end{equation}
as a result for the mean DoS (density of states per unit volume).
As we shall see, Eq.\,\ref{eq:MeanFSR} is almost correct but, in
the thermodynamic limit, it fails badly when $E\!=\!0$ and there
is a finite probability density for an impurity to have one of the
discrete \textit{``magical values}''. As a matter of fact, the breakdown
of this FSR is one of the central point stressed in Ref.\,\cite{Pires2021},
and a major conclusion of this work.

\vspace{-0.4cm}

\section{\label{sec:Low-Energy-Resonances}Effects of Random Spherical Scatterers}

The mean-field analysis of Chapter\,\ref{chap:Mean-Field-Quantum-Criticality}
led to the clear-cut conclusion that a weak spatially-uncorrelated
scalar disorder does not destabilize a 3D semi-metallic phase (Weyl
or Dirac) to turn it into a diffusive metal with a finite DoS at the
node. Such transition was shown to occur only if the disorder becomes
sufficiently strong. This picture was first called into question by
Nandkishore \textit{et al}.\cite{Nandkishore14}, who argued that
even in a weakly disordered landscape there is a small probability
of realizing relatively smooth regions which, in a first approximation,
may be\textit{ modeled as our spherical scatterers}. If each smooth
region is taken in isolation, our earlier exact solution implies that
nodal bound states will indeed be generated for fine-tuned parameters
of that region. If such a ``magical'' combination of parameters
is found to be statistically relevant, even within a very weak disordered
potential, the implication will be that the mean nodal DoS is slightly
lifted from the start and, therefore, the later cannot not be used
as a proper order parameter for the SMMT (even though it may still
be an \textit{``approximate order parameter''} that suffers a very
sharp crossover).

\vspace{-0.4cm}

\subsection{\label{sec:Low-Energy-Resonances-1}Low-Energy Resonances of a Spherical
Scatterer}

Before considering the existence of smooth regions with random parameters,
we analyze the effects in the eDoS, caused by a single spherical scatterer
with known parameters $b$ and $\lambda$. Following Refs.\,\cite{Buchhold18a,Buchhold18b,Pires2021},
we pick up on the phase-shifts derived in Eq.\,\eqref{eq:PhaseShift}
and apply the FSR as stated in Eq.\,\eqref{eq:DoS_FSR}. For convenience,
we separate the changes in the eDoS (per Weyl node) arising from each
$j$-channel and analyze the quantity,

\begin{figure}[t]
\vspace{-0.6cm}
\begin{centering}
\includegraphics[scale=0.21]{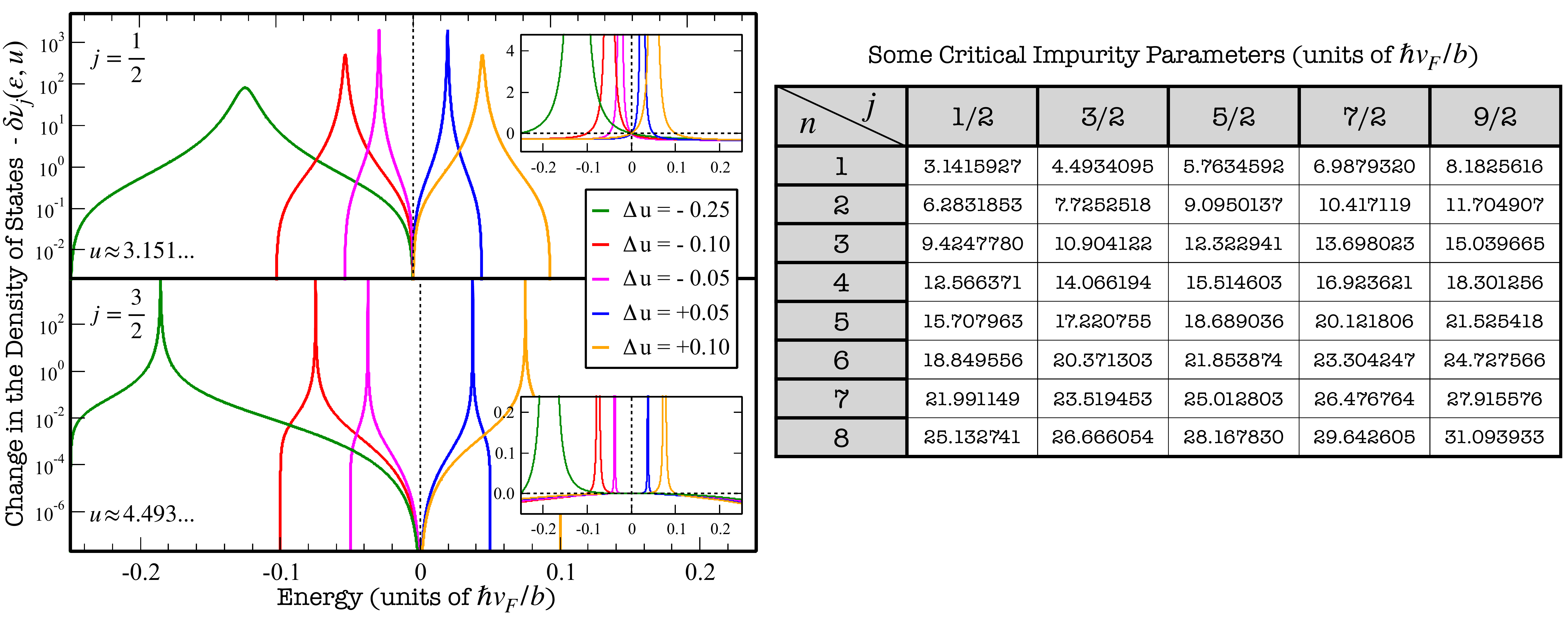}
\par\end{centering}
\vspace{-0.3cm}

\caption{\label{fig:AnalyticalWells}(a) Analytical Plots of the change in
the extensive density of states, Eq.\,\eqref{eq:DoSDeformation},
due to a near-critical impurity in the $j=1/2$ and $j=3/2$ channels
respectively. Since the resonances are extremely sharp, the main plots
represent the positive parts in a log-lin scale, while the insets
represent a zoom near the node in a linear vertical scale. (b) Table
of the lowest critical values of the impurity parameter, $u$, for
the lowest angular momentum channels.}

\vspace{-0.2cm}
\end{figure}

\vspace{-0.7cm}
\begin{equation}
\delta\nu_{j}(E,\lambda,b)=\left(\frac{2j\!+\!1}{\pi}\right)\frac{\partial\delta_{j}(E,\lambda,b)}{\partial E}.
\end{equation}
In addition, we are interested in analyzing values of $u=\lambda b/\hbar v_{\text{F}}$
that are close to one of that $j$\,-\,channel's \textit{``magical
values''}, $u_{c}^{j}$. To make these combinations clearer we reduce
the (currently redundant) number of parameters in the equations, and
define the function 

\vspace{-0.7cm}
\begin{equation}
\!\!\!f_{j}(\varepsilon,u)\!=\!\text{Sign}\left(\varepsilon\right)\frac{\text{Sign}\left(1\!-\!\frac{u}{\varepsilon}\right)J_{j+1}\!\left(\abs{\varepsilon}\right)J_{j}\!\left(\abs{\varepsilon\!-\!u}\right)-J_{j}\!\left(\abs{\varepsilon}\right)J_{j\!+\!1}\!\left(\abs{\varepsilon\!-\!u}\right)}{\text{Sign}\left(1\!-\!\frac{u}{\varepsilon}\right)Y_{j\!+\!1}\!\left(\abs{\varepsilon}\right)J_{j}\!\left(\abs{\varepsilon\!-\!u}\right)-Y_{j}\!\left(\abs{\varepsilon}\right)J_{j\!+\!1}\!\left(\abs{\varepsilon\!-\!u}\right)},\!\!\!
\end{equation}
expressed in terms of the dimensionless quantities $\varepsilon\!=\!Eb/\hbar v_{\text{F}}$
and $u\!=\!\lambda b/\hbar v_{\text{F}}$. With this, we finally arrive
at the following expression for the eDoS, 

\vspace{-0.7cm}

\begin{equation}
\delta\nu_{j}(\varepsilon,u)=\frac{2j\!+\!1}{\pi+\pi f_{j}(\varepsilon,u)^{2}}\left(\frac{\partial}{\partial\varepsilon}f_{j}(\varepsilon,u)\right),\label{eq:DoSDeformation}
\end{equation}
which is now a spectral density in $\varepsilon$ rather than in $E$.
The critical values of $u$, at which the impurities are able to create
nodal bound states are defined implicitly as $J_{j}(u_{c}^{j})\!=\!0$,
whose solutions always come in $\pm$ pairs (due to particle-hole
symmetry). Meanwhile, for each $j$ there is an\textit{ infinite number
of such solutions}, of which we list the first (positive) ones in
the table of Fig.\,\ref{fig:AnalyticalWells}b. Also in that Fig.\,\ref{fig:AnalyticalWells}a,
we plot the correction to the mean DoS, as a function of dimensionless
energy, for slightly off-tuned\,\footnote{We refer to impurities satisfying the bound state condition as \textit{``fine-tuned''},
with all the other cases being called \textit{``off-tuned}''.} spherical scatterers characterized by $u\!=\!u_{c}^{j}\!+\!\Delta u$.
In \begin{wrapfigure}[13]{o}{0.33\columnwidth}%
\vspace{-0.4cm}

\includegraphics[scale=0.145]{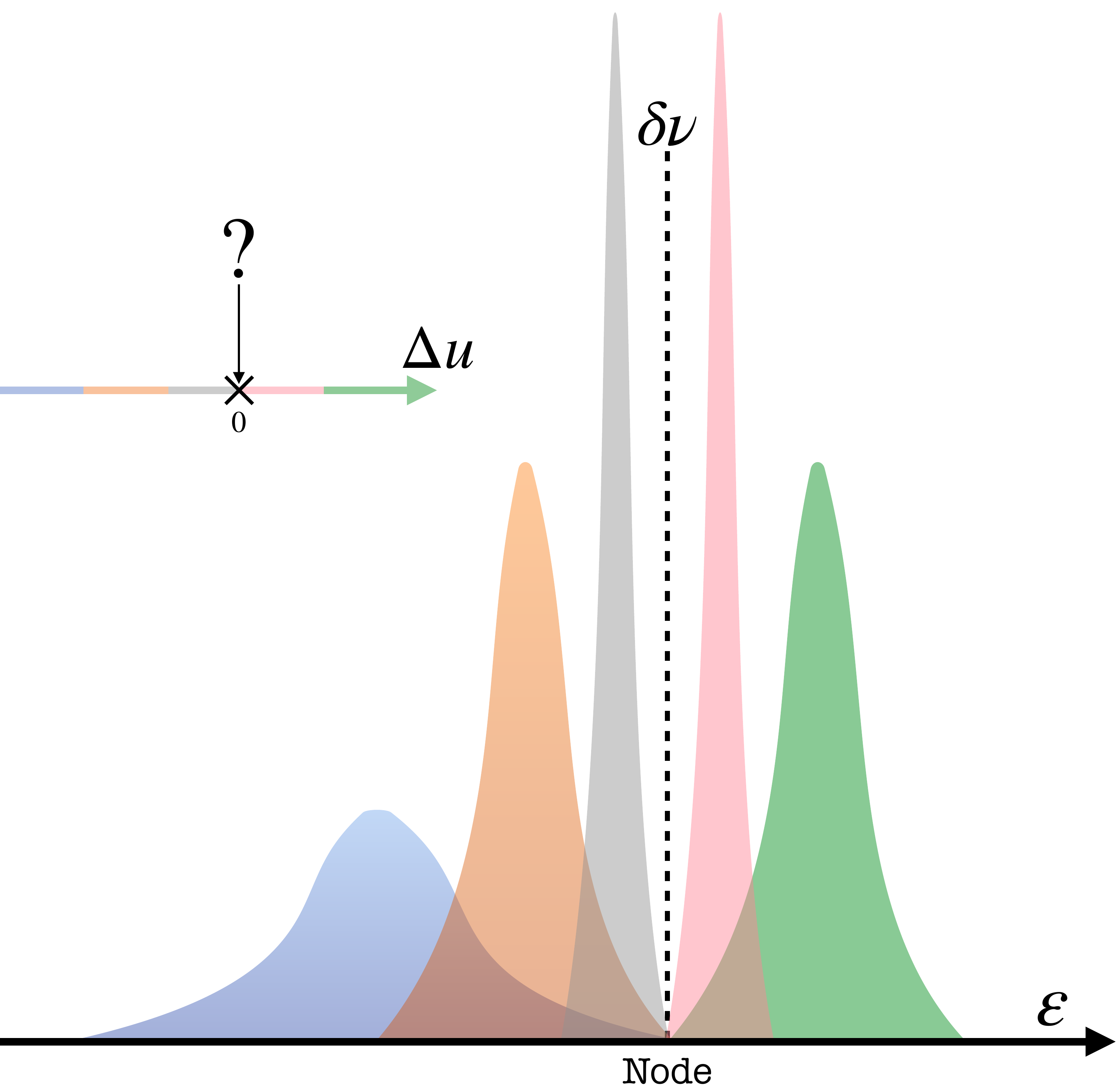}

\caption{\label{fig:Dynamics_u}Impurity resonance as $u$ crosses a magical
value (cartoon).}
\end{wrapfigure}%
 all cases, the impurity induces a \textit{very narrow resonance}
in the valence (conduction) band if $\Delta u\!<\!0$ ($\Delta u\!<\!0$)
but, remarkably, the contribution to the eDoS at the nodal energy
is \textit{precisely zero throughout}. It seems clear than no off-tuned
spherical scatterer is capable of lifting the nodal DoS, at least
in the dilute limit.

\vspace{-0.6cm}

\paragraph{Scatterer Fine-Tuning Process:}

At this point, it is useful to visualize the effect of a single impurity
in the eDoS as an adiabatic process in the parameter $u$. For that,
we consider the change in the eDoS at all energies, $\delta\nu(\varepsilon)$,
caused by a single spherical scatterer whose parameter $u$ is adiabatically
carried through one of its magical values\,\footnote{For simplicity, only the $u\!>\!0$ case will be considered. For $u\!<\!0$
the situation is analogous with a concomitant reversal of the energy
sign.}. Nearby $u=u_{c}^{j}$, the dominant contribution to $\delta\nu(\varepsilon)$
arises from the term $\delta\nu_{j}(\varepsilon)$ and the eDoS near
the nodal energy behaves as depicted in the cartoon of Fig.\,\ref{fig:Dynamics_u}.
In contrast, if $\Delta u=u\!-\!u_{c}^{j}\!\apprle\!0$ the scatterer
produces a resonance in the DoS with a sharp positive weight that
is contained within the interval $\Delta u<\!\varepsilon\!<0$. As
$\Delta u\to0^{-}$, this peak approaches the node from the valence
band and becomes sharper whilst always keeping the integral of the
positive weight region equal to $2j\!+\!1$. When $\Delta u$ crosses
$0$ into positive values, this sharp resonance \textit{suddenly shifts
and inverts relative to the node}, appearing in the conduction band
with the positive weight now contained within $0\!<\!\varepsilon\!<\!\Delta u$.
During the entirety of this process, the contribution to the nodal
DoS is exactly zero, with the exception of $\Delta u\!=\!0$ for which
we are sure that $2j\!+\!1$ exact bound states appear at the node.

Mathematically, the anomalous behavior described for the single-scatterers'
resonances as its parameter $u$ crosses a magical value, expresses
that the phase-shift $\delta_{j}(\varepsilon)$ of a $j$-channel
then becomes a \textit{non-differentiable function} at the nodal energy.
But before explaining this statement, we note that the results of
Fig.\,\ref{fig:AnalyticalWells}a seem to imply that, even though
a finite concentration of \textit{``fine-tuned''} scatterers places
a macroscopic number of bound eigenstates at the nodal energy, such
an effect \textit{does not survive statistical fluctuations} in the
values of $u$. More precisely, if one assumes a small concentration
of impurities ($c$) with the $u$ being a random variable\,\footnote{Here-forth we will refer to this as a \textit{``diversity''} in
the set of smooth regions.} with a probability density distribution, $P(u)$, then the mean DoS
reads,
\begin{equation}
\delta\rho(\varepsilon)\!=\!c\sum_{j}\int\!\!duP(u)\delta\nu_{j}(\varepsilon,u)\label{eq:Correction_DoS}
\end{equation}
which naively gives $\delta\rho(\varepsilon\!=\!0)\!=\!0$! The \textit{``fine-tuned''}
random impurities can create a nodal DoS but this happens with probability
zero. This situation was the chief motivation for the recent debate
of Refs.\,\cite{Buchhold18a,Buchhold18b,Ziegler18,Wilson20,Pires2021}
which questioned the statistical relevance of smooth-region (or instantonic)
contributions of a truly disordered landscape to the physics of a
weakly disordered Dirac or Weyl node.

\vspace{-0.5cm}

\subsection{\label{sec:Random-Impurities}Random Impurities and the Nodal DoS
Deformation}

Despite seemingly flawless, we will shown that the reasoning leading
to the previous conclusion misses one important aspect of scattering
theory: the FSR can only be applied if a standard convention\,\cite{Calogero67}
is imposed on the definition of the scattering phase-shifts. This
subtle aspect of the definition of a scattering phase shift has already
been discussed in Subsect.\,\ref{subsec:Scattering-Solutions}. As
it turns out, it was our overlook of this important fact that led
to the anomalous behavior described for the resonances induced by
slightly off-tuned impurities. 

Now, it is time to retrace our previous derivation and resolve the
ill-definition in $\delta\nu(\varepsilon\!=\!0)$ obtained from the
FSR at near a critical impurity. As referred, the essential flaw was
that we have rushed in obtaining Eq.\,\eqref{eq:DoSDeformation}
and did not verify if the phase shifts obtained in Eq.\,\eqref{eq:PhaseShift}
are defined in accordance with the appropriate convention for the
asymptotic behavior, i.e.
\begin{equation}
\delta_{j}\left(\varepsilon\to\pm\infty,u\right)\to-u.\label{eq:PhaseShifts_Assympt}
\end{equation}
In order to force this behavior, we can use the method introduced
before and define 
\begin{figure}[t]
\begin{centering}
\vspace{-0.5cm}\includegraphics[scale=0.23]{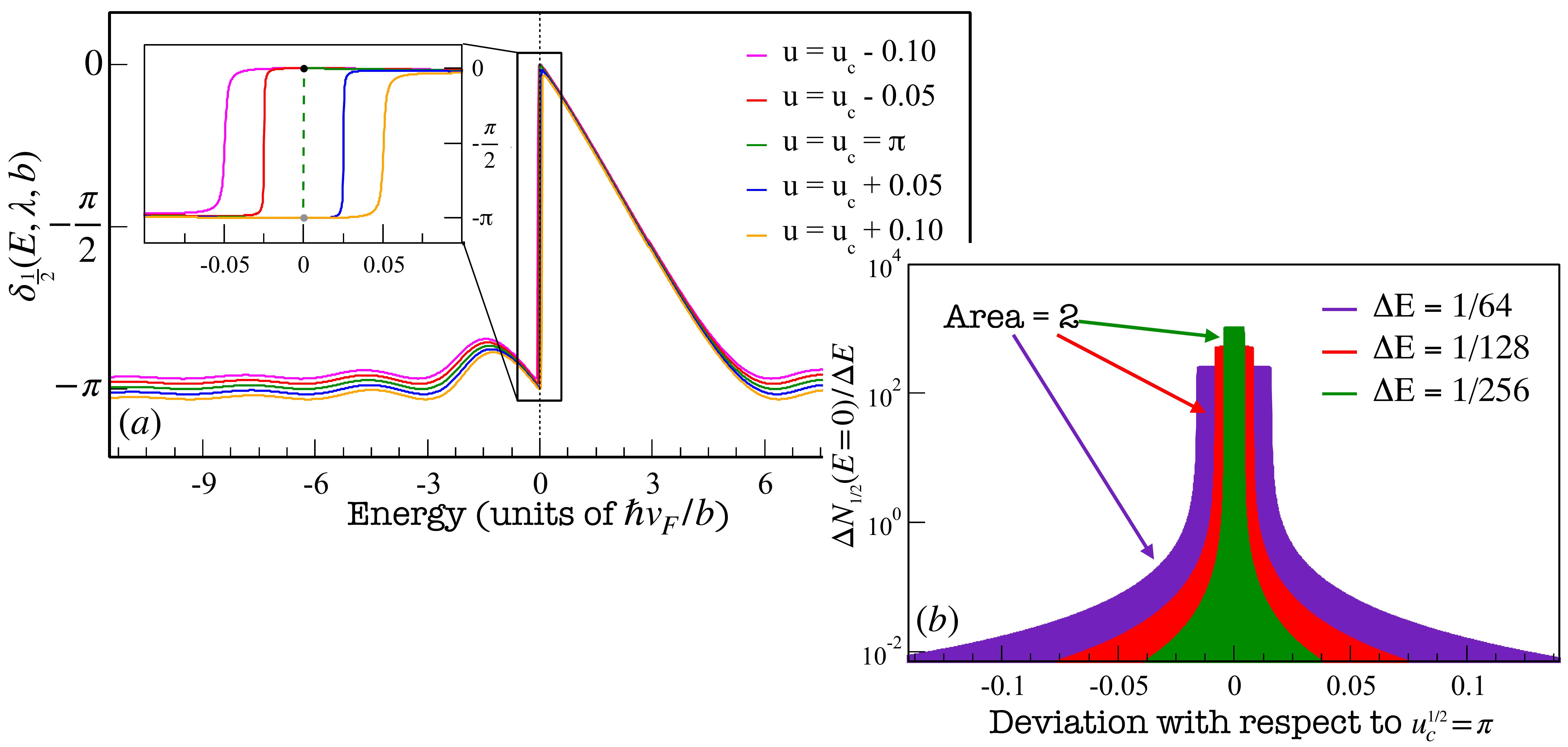}
\par\end{centering}
\begin{centering}
\vspace{-0.3cm}
\par\end{centering}
\caption{\label{fig:PhaseShifts}(a) Plots of the scattering phase shift calculated
by Eq.\,\eqref{eq:Phase_Shift_Corr_Assympt-1-2} for $j\!=\!1/2$
around the first critical value $u_{c}^{\frac{1}{2}}\!=\!\pi$. The
main panel shows that the asymptotic behavior, $\delta_{j}(\varepsilon\!\to\!\pm\infty,u)\to-u$
is indeed respected, while the inset depicts the formation of a true
$\pi$-discontinuity at $E\!=\!0$ when $u\!=\!u_{c}^{j}$ . (b) Representation
of the integrand in Eq.\,\eqref{eq:AverageDoS} as a function of
$u$ for decreasing values of the energy resolution $\Delta E$. The
integral over $u$ is conserved and equal to $2$ throughout.}

\centering{}\vspace{-0.5cm}
\end{figure}

\vspace{-1.0cm}

{\footnotesize{}
\begin{equation}
\delta_{j}\left(\varepsilon,u\right)=\begin{cases}
-u-\int_{-\infty}^{\varepsilon}dx\frac{d}{dx}\arctan\left(\frac{\text{Sign}\left(x\right)J_{j}\left(\abs x\right)J_{j\!+\!1}\left(\abs{x\!-\!u}\right)-\text{Sign}\left(x\!-\!u\right)J_{j+1}\!\left(\abs x\right)J_{j}\left(\abs{x\!-\!u}\right)}{\text{Sign}\left(1\!-\!\frac{u}{x}\right)Y_{j\!+\!1}\!\left(\abs x\right)J_{j}\!\left(\abs{x\!-\!u}\right)-Y_{j}\!\left(\abs x\right)J_{j\!+\!1}\!\left(\abs{x\!-\!u}\right)}\right) & \text{if \ensuremath{\varepsilon<0}}\\
-u-\int_{\infty}^{\varepsilon}dx\frac{d}{dx}\arctan\left(\frac{\text{Sign}\left(x\right)J_{j}\left(\abs x\right)J_{j\!+\!1}\left(\abs{x\!-\!u}\right)-\text{Sign}\left(x\!-\!u\right)J_{j+1}\left(\abs x\right)J_{j}\left(\abs{x\!-\!u}\right)}{\text{Sign}\left(1\!-\!\frac{u}{x}\right)Y_{j\!+\!1}\left(\abs x\right)J_{j}\left(\abs{x\!-\!u}\right)-Y_{j}\left(\abs x\right)J_{j\!+\!1}\left(\abs{x\!-\!u}\right)}\right) & \text{if \ensuremath{\varepsilon\geq0}}
\end{cases}.\label{eq:Phase_Shift_Corr_Assympt-1-2}
\end{equation}
}This expression for the phase-shifts can then be plotted as a function
of energy for different values of $u$. We are interested in studying
the near-critical case, $u\approx u_{c}^{j}$, for which we present
an illustrative plot in Fig.\,\ref{fig:PhaseShifts}\,a. Two essential
points can be highlighted about these results: i) If $u$ is close
to a ``magical value'' of the $j$-channel, then there is a fast
(but smooth) variation of $\pi\text{ rad}$ in the scattering phase-shift
of that same channel, and ii) if $u\!=\!u_{c}^{j}$ then there is
a proper $\pi$-discontinuity in the phase-shift at the node. Such
a large energy variation is a well-known feature of any resonant scattering
process\,\cite{Elattari99,Elattari2000} and, physically tells us
that a particle scattered at those energies by the impurity features
a particularly large scattering time. It is the fact that FSR involves
$\partial\delta_{j}/\partial\varepsilon$ which provides the connection
between large scattering times and sharp peaks in the DoS.

As the plots of Fig.\,\ref{fig:PhaseShifts}\,a already respect
the convention imposed by Eq.\,\eqref{eq:PhaseShifts_Assympt}, the
$\text{mod}\pi$ ambiguity of the phase shift no longer exists and,
therefore the $\pi$-disconuity at the node is a physical feature
of fine-tuned spherical scatterers. Such a non-analytic behavior with
energy clearly has profound implications in the use of the FSR to
extract spectral properties from the phase shifts. As it turns out,
we will show that this feature is nothing but a consequence of the
nodal bound-states of these configurations, in accordance with Levinson's
Theorem of Eq.\,\eqref{eq:LevinsonDirac-1}. In particular, we know
from Subsect.\,\ref{subsec:Nodal-Bound-States} that the number of
zero-energy bound states is either zero (for a non-critical impurity)
or $2j\!+\!1$ (if $u\!=\!u_{c}^{j}$) and, therefore, an inspection
of Eq.\,\eqref{eq:LevinsonDirac-1} leads to the conclusion that
the phase-shift of that $j$-channel must be either continuous at
$\varepsilon\!=\!0$ or have a $\pi$-discontinuity in the ``magical''
case.

Now that we have understood both the physical and mathematical origin
of the singular behavior in $\delta\nu_{j}(\varepsilon,u\!\approx\!u_{c}^{j})$,
we can now derive a consistent expression for the mean DoS of a finite
concentration of spherical scatterers with random parameter. To do
this, we must assume two hypotheses: \textit{(i)} the scatterers do
not couple different Weyl nodes, and \textit{(ii)} there is a \textit{well-defined
dilute limit} that makes meaning the contributions from different
scatterers purely additive. Under this framework, $\delta\rho(\varepsilon)$
can be evaluated by considering the change in the number of states
within an energy interval of width $\Delta\varepsilon$ around a $\varepsilon$,

\vspace{-0.7cm}
\begin{equation}
\!\!\!\!\!\!\!\!\!\!\!\delta N(\varepsilon;\Delta\varepsilon,\{u_{n}\}_{n=1,\cdots,N_{i}})\!=\!\!\sum_{j=\frac{1}{2}}^{\infty}\frac{2j\!+\!1}{\pi}\sum_{n=1}^{N_{i}}\!\left[\delta_{j}\!\left(\varepsilon\!+\!\frac{1}{2}\Delta\varepsilon,u_{n}\right)\!-\!\delta_{j}\!\left(\varepsilon\!-\!\frac{1}{2}\Delta\varepsilon,u_{n}\right)\right]\!,\!\!\!\!\label{eq:ChangeNumStates}
\end{equation}
where the summation over $n$ includes the effects of all $N_{i}$\nomenclature{$N_{i}$}{Number of Impurities in the System}
independent scatterers. Note that $\delta N$ {[}Eq.\,\ref{eq:ChangeNumStates}{]}
is still a function of the particular configuration of spherical scatterers
in the system, that is, $\{u_{n}\}_{n=1,\cdots,N_{i}}$. Assuming
that the DoS is \textit{self-averaging}, we can replace the sum over
scatterers by an integral over a single probability density for $u$,
\textit{i.e.},

\vspace{-0.7cm}
\begin{equation}
\sum_{n=1}^{N_{i}}\to N_{i}\!\!\int\!\!du\,P(u)
\end{equation}
and therefore arrive at

\vspace{-0.7cm}
\begin{equation}
\!\!\!\!\!\overline{\delta N\,(\varepsilon;\Delta\varepsilon)}\!=\!N_{i}\sum_{j=\frac{1}{2}}^{\infty}\frac{2j\!+\!1}{\pi}\int\!\!du\,P(u)\left[\delta_{j}\!\left(\varepsilon\!+\!\frac{1}{2}\Delta\varepsilon,u\right)\!-\!\delta_{j}\!\left(\varepsilon\!-\!\frac{1}{2}\Delta\varepsilon,u\right)\right].\!\!\!
\end{equation}
Then, the mean density of states (per unit volume) can be obtained
in the thermodynamic and infinite energy resolution limit,

\vspace{-0.7cm}
\begin{equation}
\overline{\delta\rho(\varepsilon)}=c\sum_{j=\frac{1}{2}}^{\infty}\frac{2j\!+\!1}{\pi}\lim_{\Delta\varepsilon\to0}\left(\int\!\!du\,P(u)\frac{\delta_{j}\!\left(\varepsilon\!+\!\frac{1}{2}\Delta\varepsilon,u\right)\!-\!\delta_{j}\!\left(\varepsilon\!-\!\frac{1}{2}\Delta\varepsilon,u\right)}{\Delta\varepsilon}\right),\label{eq:AverageDoS}
\end{equation}
where $c$ is the volume concentration of scatterers in the system.
Incidentally, Eq.\,\eqref{eq:AverageDoS} contains two situations
that require different approaches. On the one hand, if $\delta_{j}$
is a differentiable function in $E$, we may just write

\vspace{-0.7cm}
\begin{align}
\overline{\delta\rho(\varepsilon)} & =c_{i}\sum_{j=\frac{1}{2}}^{\infty}\int\!\!du\,P(u)\left(\frac{2j\!+\!1}{\pi}\frac{\partial}{\partial\varepsilon}\delta_{j}\!\left(\varepsilon,u\right)\right)=c\sum_{j=\frac{1}{2}}^{\infty}\overline{\delta\nu_{j}(\varepsilon,u)},\label{eq:Expected_AverageFSR}
\end{align}
restating that the deformation in the mean DoS is the average of $\delta\nu(E,u)$,
as obtained from the FSR. This is the expectable result, and we can
state outright that it is true for any energy $\varepsilon\neq0$
in this system. However, exactly at the nodal energy, the \textit{$\varepsilon$\,-\,differentiability
of the phase-shifts breaks down} if the integration over $u$ is allowed
to cross a \textit{``magical value}''. If that is the case, we have
shown that

\vspace{-0.7cm}
\begin{equation}
\delta_{j}\!\left(\frac{1}{2}\Delta\varepsilon,u_{c}^{j}\right)\!-\!\delta_{j}\!\left(-\!\frac{1}{2}\Delta\varepsilon,u_{c}^{j}\right)\underset{\Delta\varepsilon\to0}{\longrightarrow}\pi
\end{equation}
which entails 
\begin{figure}[t]
\vspace{-0.5cm}
\begin{centering}
\includegraphics[scale=0.23]{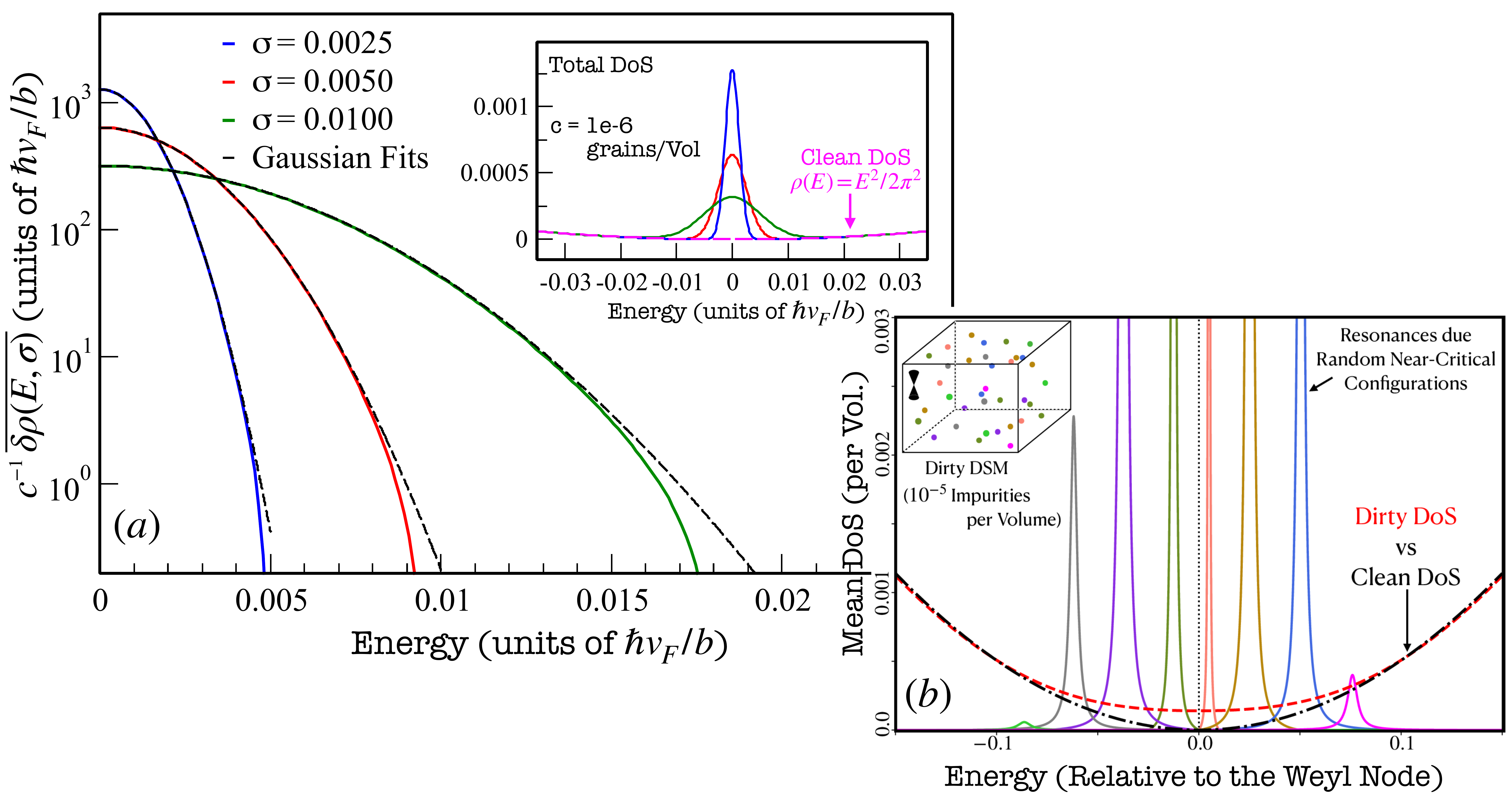}
\par\end{centering}
\vspace{-0.3cm}

\caption{\label{fig:Distribution_Emerg}(a) Prediction for $\delta\rho_{\nicefrac{1}{2}}(\varepsilon)$
due to dilute spherical impurities with a Gaussian diversity of width
$\sigma$ around $u_{\nicefrac{1}{2}}\!=\!\pi.$ The inset depicts
the total DoS for $10^{-6}b^{-1}$ impurities per volume. (b) Cartoon
of the physical justification of the lift in the DoS by a proliferation
of near-critical impurities. Figures adapted from Ref.\,\cite{Pires2021}.}

\vspace{-0.6cm}
\end{figure}

\vspace{-0.7cm}
\begin{equation}
\overline{\delta\rho(\varepsilon\!=\!0)}=c_{i}\sum_{j=\frac{1}{2}}^{\infty}\left(2j\!+\!1\right)\sum_{u_{c}^{j}}P(u_{c}^{j})\lim_{\Delta\varepsilon\to0}\left(\frac{1}{\Delta\varepsilon}\right);
\end{equation}
A clearly an \textit{ill-defined expression}! Since $\delta\rho(\varepsilon)$
is a spectral density, it is not out-of-question that it can become
infinite at a given energy, so long as it remains integrable. However,
this is not the case and to see that one has to go back to Eq.\,\eqref{eq:AverageDoS}
and calculate the $\Delta\varepsilon\to0$ limit numerically for a
set $u$s that flank a magical value. In Fig.\,\ref{fig:PhaseShifts}b,
we do this for the lowest critical value (of the $j\!=\!\nicefrac{1}{2}$
channel) and, from there, one can observe that a Dirac-$\delta$ distribution
is forming centered in $u_{c}^{j}$. Hence, the correct expression
for mean DoS should be

\vspace{-0.7cm}

\begin{align}
\!\!\!\!\!\!\!\!\!\!\!\!\overline{\delta\rho(\varepsilon)}\! & =\!\smash{N_{v}\underset{\text{Scattering States' Contribution}}{\underbrace{c_{i}\sum_{j=\frac{1}{2}}^{\infty}\int\!\!du\,P(u)\left(\frac{2j\!+\!1}{\pi}\frac{\partial}{\partial\varepsilon}\delta_{j}\!\left(\varepsilon,u\right)\right)}}\!+\!\underset{\text{Bound States' Contribution}}{N_{v}\underbrace{c_{i}\:\delta_{\varepsilon,0}\!\sum_{j=\frac{1}{2}}^{\infty}\!\sum_{u_{c}^{j}}\!P(u_{c}^{j})\left(2j\!+\!1\right)}},\!\!\!}\label{eq:CorrectedAverageDoS}
\end{align}

\vspace{0.4cm}

rather than Eq.\,\eqref{eq:Expected_AverageFSR} {[}we have included
the valley-degeneracy factor $N_{v}${]}. In other words, if $u$
is a \textit{``magical value''}, commuting the $\Delta\varepsilon\!\to\!0$
limit with the integral over $u$ is done at the expense of generating
a distribution in the integrand (containing the bound states' contribution).
This result shows that a dilute set of diverse smooth regions that
have a \textit{non-zero probability density} of \textit{``fine-tuned''}
scatterers, will lift the nodal DoS of DWSM at arbitrarily small concentrations.
In the upcoming section, we shall extract physical consequences of
Eq.\,\eqref{eq:CorrectedAverageDoS} and interpret this expression
in a brighter light.

\vspace{-0.5cm}

\section{\label{subsec:Semi-metallic-Instability}The Near-Critical Impurity
Mechanism}

\vspace{-0.2cm}

While there are no doubts that nodal bound states can be generated
by smooth potentials in a WSM, their delicate nature has sparked great
discussion the literature\,\cite{Nandkishore14,Gurarie17,Buchhold18a,Buchhold18b,Pixley16a,Ziegler18,Wilson20,Pixley21,Pires2021}.
The main point of content is the statistical relevance of such fine-tuned
states for the electronic properties within a truly disordered environment.
Focusing on the problem of random spherical scatterers, Buchhold \textit{et
al}.\,\cite{Buchhold18a,Buchhold18b} presented a calculation that
is essentially equivalent to considering only the first term in Eq.\,\eqref{eq:CorrectedAverageDoS}.
Due to a zero statistical measure of fine-tuned scatterers, this term
always yields a zero contribution for the nodal DoS. A clear way to
look at this, goes as follows: If a DWSM hosts a concentration ($c$)
of equal spherical scatterers (with the same $u$), $\rho(\varepsilon\!=\!0)\!=\!0$
will remain for almost any $u$, but will get immediately destabilized
if $u=u_{c}^{j}$. Then, one expects $\overline{\rho(\varepsilon\!=\!0)}\!\propto\!c$
in the dilute limit. However, if one allows for some \textit{continuous
statistical diversity} in the values of $u$, this effect gets completely
lost and $\rho(\varepsilon\!=\!0)\!=\!0$ in any cases.

In order to test if our corrected expression for the mean DoS changes
this picture, we consider a toy-model with a single-node Weyl hosting
dilute spherical scatterers that have random parameters around a \textit{``magical
valu}e'' $u_{c}^{j}$. Such a diversity can be modeled by values
of $u$that are drawn independently from 

\vspace{-0.7cm}
\begin{equation}
P(u)\!=\!\frac{1}{\sqrt{2\pi}\sigma}\exp\left(-\frac{\left(u-u_{c}^{j}\right)^{2}}{2\sigma^{2}}\right),\label{eq:GaussianFluctuations}
\end{equation}
where $\sigma$ measures the \textit{extent of this diversity}. Even
in this case, the first-term's contribution to Eq.\,\eqref{eq:CorrectedAverageDoS}
gives zero, because $\int P(u)\,\delta\nu_{j}(\varepsilon,u)\,du=0$
for all $j$-channels. However, if both terms are accounted for, one
finds a different result:

\vspace{-0.7cm}

\begin{equation}
\overline{\delta\rho(\varepsilon\!=\!0)}=c\frac{2j\!+\!1}{\sqrt{2\pi}\sigma},\label{eq:CorrectedAverageDoS-1}
\end{equation}
which is obviously finite for any value of $\sigma$. Therefore, the
precise criterion for a set dilute spherical scatterers to destabilize
the semi-metallic node is that the probability density in $u$ is
nonzero at some ``magical value''. To illustrate the previous reasoning,
in Fig.\,\ref{fig:Distribution_Emerg}\,a, we present results for
the mean DoS in the presence of a concentration $c=10^{-6}\text{grains/vol}$
where the parameters of each scatterer were independently drawn from
the distribution of Eq.\,\eqref{eq:GaussianFluctuations} with $u_{c}^{j}=\pi$.
This calculation was done using Eq.\,\ref{eq:CorrectedAverageDoS}
but taking only the $j\!=\!1/2$ contribution into account as, for
these values of $u$, it provides the biggest contribution by far. 

Mathematically, we have seen that the second term in Eq.\,\eqref{eq:CorrectedAverageDoS}
is explained by the emergence of a Dirac-$\delta$ distribution in
$u$, due to a consistent definition of the change in the DoS, and
it responsible for the nodal DoS to be lifted. Even so, we can also
provide a physical (and more intuitive) explanation of this result.
In fact, even though a nodal bound state can only appear in a fine-tuned
scatterer, any slight deviation from this situation still leads to
a huge spectral weight {[}$\sim\mathcal{O}(c)${]} arbitrarily close
to the nodal energy. This \textit{proliferation of needle-like resonances},
precluding the bound states in a slightly \textit{``off-tuned''}
scatterer, is what actually concentrates a finite spectral weight
arbitrarily close to the node and guarantees that the DoS gets lifted
for a dilute set of scatterers. The semi-metallic phase is then destabilized
by random smooth regions, not through a \textit{critical mechanism},
but rather a \textit{near-critical} one. Figure\,\ref{fig:Distribution_Emerg}b
depicts a cartoon of this very mechanism.

\vspace{-0.5cm}

\section{\label{sec:Numerical-Validation}Near-Critical Mechanism: A Numerical
Validation}

We have predicted that the nodal DoS is lifted by near-critical spherical
scatterers based on a continuum model of a single-node Dirac-Weyl
semimetal. As already mentioned, all of our conclusions are limited,
in principle, to a system of uncoupled nodes in which inter-impurity
interference effects are sufficiently weak so that contributions of
different smooth regions to the DoS simply add up. To see how accurate
these assumptions are, we complement the previous analysis with a
direct evaluation of the mean DoS in a lattice model of a WSM that
is decorated by a finite concentration of large spherical regions
(discretized spherical scatterers) inside of which the local potential
is given a common random strength.

\vspace{-0.5cm}

\paragraph{Lattice Model:}

The lattice model employed in this study is the same two-band model
used for the simulations presented in Sect.\,\ref{sec:NumericalSimulations_Anderson},
but now with a new tailor-made perturbation. More precisely, we have
a simple cubic lattice ($\mathcal{L}_{C}$) with the following hamiltonian,

\vspace{-0.7cm}

\begin{align}
\mathcal{H}_{l} & \!=\!\negthickspace\sum_{{\scriptscriptstyle \mathbf{R}\in\mathcal{L_{\text{C}}}}}\!\left[\frac{i\hbar v_{\text{F}}}{2a}\Psi_{\mathbf{{\scriptscriptstyle R}}}^{\dagger}\!\cdot\!\sigma^{j}\!\cdot\!\Psi_{{\scriptscriptstyle \mathbf{R}+a\hat{e}_{j}}}+\frac{\mathcal{U}\left(\mathbf{R}\right)}{2}\Psi_{\mathbf{{\scriptscriptstyle R}}}^{\dagger}\!\cdot\!\Psi_{\mathbf{{\scriptscriptstyle R}}}\right]\!+\!\text{h.c.},\label{eq:LatticeModel}
\end{align}
where $\sigma^{j}$ are the Pauli matrices and $\mathcal{U}(\mathbf{R})$
is a scalar potential. The scalar potential, in this case, is the
sum of $n$ spherical impurities, centered in positions $\mathbf{R}_{n}$,
with radius $R$ and a strength $U_{n}$, \textit{i.e.}

\begin{figure}[t]
\begin{centering}
\includegraphics[scale=0.22]{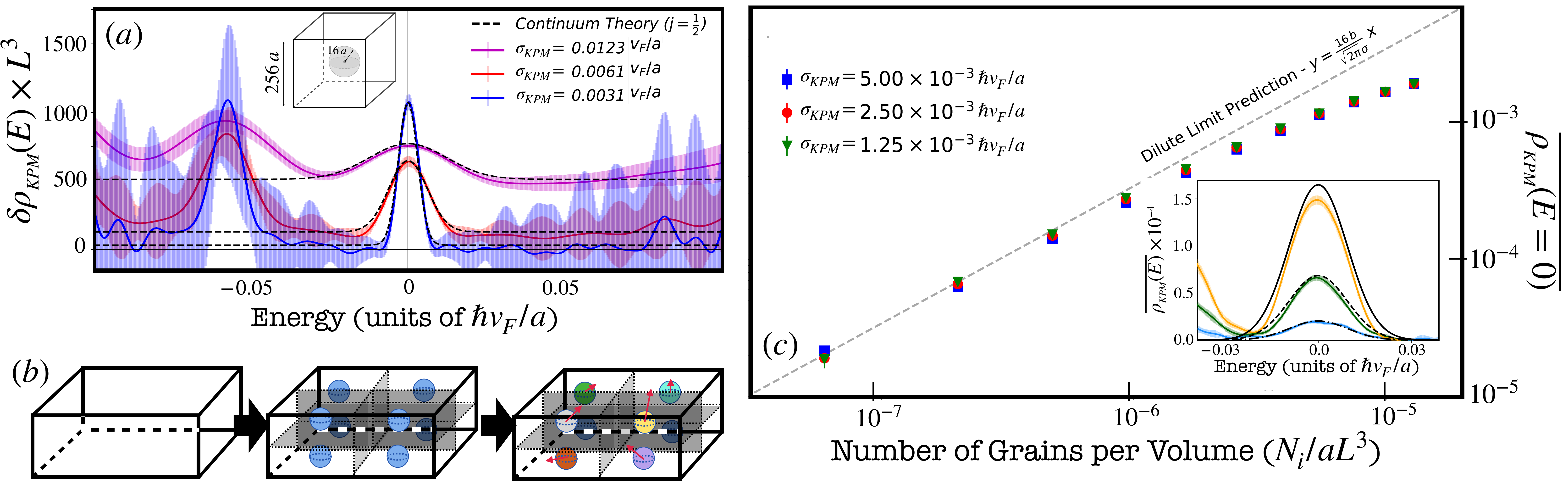}
\par\end{centering}
\caption{\label{fig:DiluteLimit}(a) DoS change due to an impurity of critical
strength $u\!=\!\pi\,v/a$ and radius $16\,a$ inside a supercell
of $256^{3}$ sites. Vertical widths are $95\%$ statistical error
bars, and dashed lines are the continuum theory predictions. (b) Scheme
of the procedure used to generate a configuration of multiple random
spheres inside the simulated supercell. (c) Plot of $\overline{\rho\!\left(E\!=\!0\right)}$
with several impurities of radius $16\,a$ inside the simulated supercell
of $512^{3}$ sites for different resolutions $\eta$. The gray line
is the dilute regime prediction. The inset shows converged $\overline{\rho\!\left(E\right)}$
for three concentrations against the predictions demonstrated in Fig.\,\ref{fig:Distribution_Emerg}
(black lines).}
\end{figure}

\vspace{-1.1cm}
\begin{equation}
\mathcal{U}(\mathbf{R})\!=\!\sum_{n=1}^{N_{i}}U_{n}\Theta_{H}\left(\abs{\mathbf{R}\!-\!\mathbf{R}_{n}}\!-\!R\right)\mathcal{I}_{2\times2}.
\end{equation}
To test the several assumptions made in previous sections, we considered
two situations: \textit{(i) }a single spherical scatterer is placed
within the simulated lattice, with a fixed position and strength,
and \textit{(ii)} a finite number of scatterers (with the same radius)
are randomly distributed in the lattice, with the values of $U_{n}$
being random variables taken from a gaussian distribution around the
corresponding magical value. In the latter case, to avoid unwanted
superposition of different spheres (highly unlikely in the dilute
limit), we have generated the impurity configurations by choosing
their centers, $\mathbf{R}_{n}$, from the positions of a randomizing
superlattice superposed to $\mathcal{L}_{C}$. This procedure is schematically
represented in Fig.\,\ref{fig:DiluteLimit}\,b. 

\vspace{-0.6cm}

\paragraph{Node Decoupling Hypotheses:}

\begin{figure}[t]
\vspace{-0.5cm}
\begin{centering}
\includegraphics[scale=0.23]{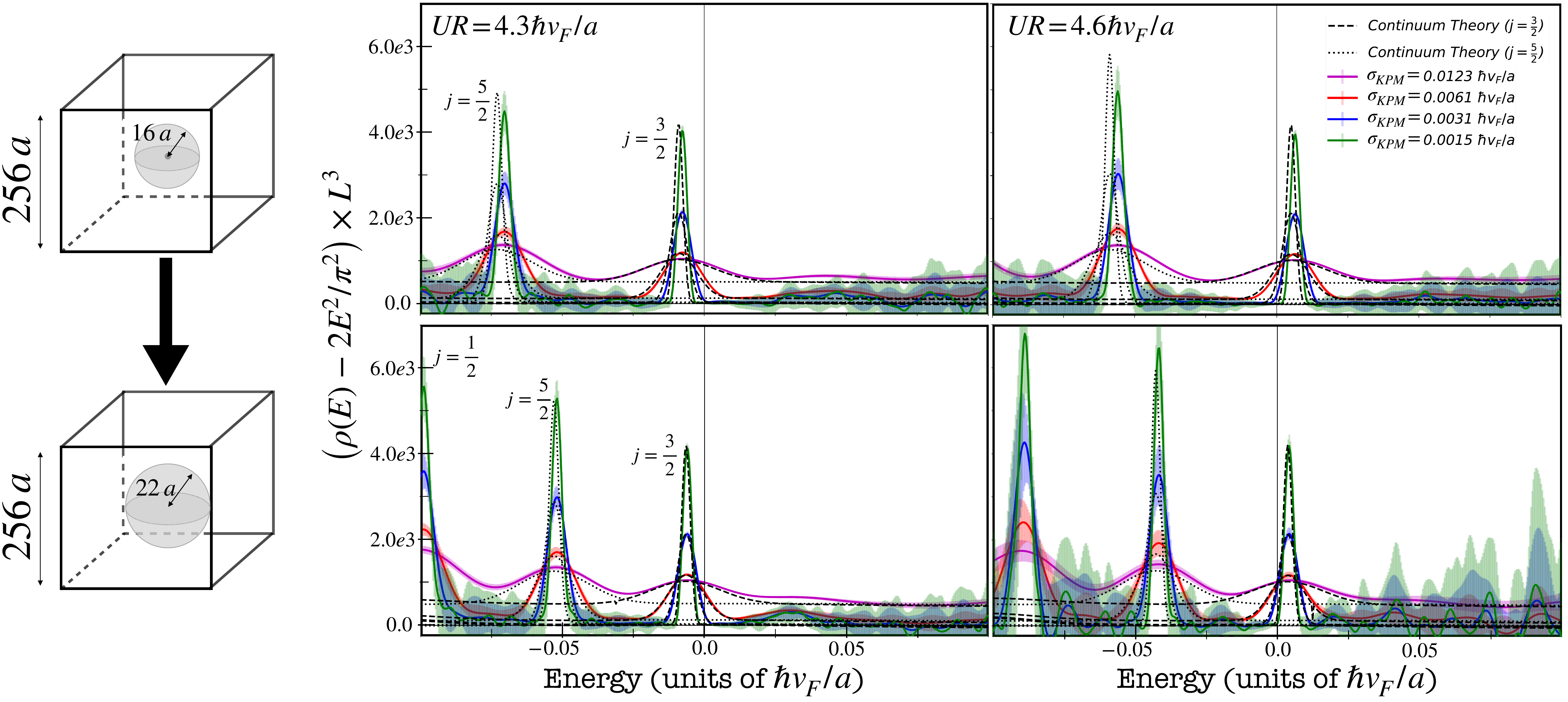}
\par\end{centering}
\vspace{-0.1cm}

\caption{\label{fig:SingleSphereNum}\textbf{Top:} Plots of the change in the
density of states due to a single spherical impurity of strength $u\!=\!\nicefrac{4.3}{R},\nicefrac{4.6}{R}\;\hbar v_{\text{F}}/a$
and radius $R\!=\!16\,a$ inside a simulated supercell of volume $256^{3}\,a^{3}$.
The vertical widths of the numerical curves are $95\%$ statistical
error bars. The two visible peaks correspond to the first resonances
associated with $j\!=\!5/2$ and $j\!=\!3/2$, from left to right.\textbf{
Bottom}: Same calculations done for an impurity of radius $R\!=\!22\,a$.
The agreement with the continuum theory is much better in this case.}

\vspace{-0.4cm}
\end{figure}
The first assumption to be tested is that internode coupling is irrelevant
for the problem of spherical scatterers. Even though this is a very
sound assumption provided the spheres are large compared to the lattice
spacing $a$, we still test it by evaluating the DoS deformation caused
by a single scatterer of radius $R$ in a lattice with side $L\!=\!256a$.
The calculations of the DoS were done using the KPM with a Jackson
kernel, as described in Appendix\,\ref{chap:Crash-Course-KPM}, within
implementation provided by the QuantumKITE\,\cite{Joao2020}, and
using averaging over twisted boundary conditions to eliminate the
mean-level spacing. As referred before, this numerical technique introduces
a finite energy resolution by effectively broadening every energy
Dirac-$\delta$ into a gaussian of variance $\sigma_{\text{KPM}}\!\propto1/N_{c}$,
where $N_{c}$ is the number of Chebyshev polynomials retained in
the expansion. Therefore, in order to compare our numerical results
for the DoS with the analytic ones (obtained from the FSR), we convolute
the latter with a gaussian that accounts for this finite spectral
resolution, \textit{i.e.},

\vspace{-0.85cm}
\begin{equation}
\rho_{\text{KPM}}(E,\sigma_{\text{KPM}})\!=\!\frac{1}{\sqrt{2\pi}\sigma_{\text{KPM}}}\int dx\rho_{\text{ex}}(x)\exp\left(-\frac{\left(E-x\right)^{2}}{2\sigma_{\text{KPM}}^{2}}\right).
\end{equation}
More precisely, we are interested in comparing the DoS deformation
obtained from Eq.\,\eqref{eq:DoSDeformation}, with $N_{v}\!=\!8$,
to the results obtained for a single spherical scatterer of radius
$R$. Thus, the right quantity to represent is not the full numerically
DoS but rather 

\vspace{-0.9cm}
\begin{equation}
\delta\rho_{\text{KPM}}\!(E,U,R,\sigma_{\text{KPM}})\!=\!\rho_{\text{KPM}}(E,U,R,\sigma_{\text{KPM}})-\frac{2E^{2}}{\pi^{2}}.
\end{equation}
This is to be compared with 

\vspace{-1.0cm}

\begin{align}
\!\!\!\!\!\delta\tilde{\rho}_{{\scriptscriptstyle \text{imp}}}\!(E,U,R,\sigma_{\text{KPM}}) & =-\frac{2E^{2}}{\pi^{2}}\!+\!\!\int_{-\infty}^{\infty}\!\!\!\!\!dx\left[\frac{4\pi R^{3}}{3L^{3}}\sum_{j}\delta\nu_{j}(x,U,R)+\frac{2x^{2}}{\pi^{2}}\right]\frac{e^{-\frac{\left(E-x\right)^{2}}{2\sigma_{\text{KPM}}^{2}}}}{\sqrt{2\pi}\sigma_{\text{KPM}}},
\end{align}

\vspace{-0.5cm}

that is the analytical FSR result for a single impurity of radius
$b\!=\!Ra$ and potential $\lambda\!=\!U\hbar v_{\text{F}}/a$, already
broadened by the finite resolution determined by $\sigma_{\text{KPM}}$.

Some illustrative results are shown in Fig.\,\ref{fig:SingleSphereNum},
where it is clear that as $R$ becomes larger (e.g. $R\!\gtrsim\!16a$)
the several resonances become well described by the continuum theory
developed in the previous sections. In fact, we can even go further
and exploit the finite resolution of the KPM calculation to analyze
the DoS deformation caused by a critical spherical scatterer. In Fig.\,\ref{fig:DiluteLimit}\,a,
we present these results for $u\!=\!\pi\,\hbar v_{\text{F}}/a$ and
a radius of $R\!=\!16a$, which are then compared to gaussians of
width $\sigma_{\text{KPM}}$ centered at the nodal energy. To the
available resolution, an exactly critical impurity introduces zero
energy bound states in the system, as predicted from our earlier.
exact solution of Weyl hamiltonian. 

In summary, we have established that, even in the more realistic lattice
WSM with several nodes in the fBz, the structure of resonances predicted
by the FSR for an isolated spherical scatterer within a single-node
continuum model are well reproduced if the latter are sufficiently
large. This is hardly surprised, as a slowly-varying potential in
real-space is not able to induce scattering processes with large momentum
transfer (see Suzuura and Ando\,\cite{Suzuura02} for a paradigmatic
example). 

\vspace{-0.4cm}

\paragraph{Dilute Limit Hypothesis:}

The numerical results clearly point to a great accuracy of the continuum
theory to describe DoS's deformations caused by isolated spherical
scatterers. The only requirement is that the corresponding lattice
perturbation is a smooth region significantly larger than the lattice
spacing. However, the validity of the single-region theory is not
enough to validate our earlier conclusions; We must also check that
there is a well-defined dilute limit associated to this model of disorder.
Even though such a result can be somewhat expected for a system with
impurities that lie very far from each other, there are known examples\,\cite{Joao2021}
of systems in which the DoS is a non-analytic function of the impurity
concentration. Therefore, this dilute limit assumption must be explicitly
verified for our system of interest.

In order to show that, we have performed KPM simulations fo large
systems ($L\!=\!512a$) in which a set of spheres of radius $R\!=\!16a$
were scattered, using the method described in Fig.\,\ref{fig:DiluteLimit}\,c.
In analogy to Sect.\,\ref{subsec:Semi-metallic-Instability}, we
considered several random configurations of scatterers and considered
random values of $U_{n}$, which have been drawn out of a gaussian
distribution centered around the ``magical value'' $U_{c}\!=\!\nicefrac{\pi}{16}\hbar v_{\text{F}}/a$,
with a standard deviation $\sigma\!=\!0.3\hbar v_{\text{F}}/a$. The
results are shown in Fig.\,\ref{fig:DiluteLimit}\,c and fully confirm
the DoS deformation predicted from the continuum theory in Fig.\,\ref{fig:Distribution_Emerg}\,a.
In fact, by analyzing the mean nodal DoS, our results further confirm
that a linear scaling with concentration, i.e. $\overline{\rho(E\!=\!0)}\propto c$,
exists in samples containing $10^{-8}$ to $10^{-6}$ random impurities
per unit volume. These results confirms that our earlier dilute limit
assumption makes sense in this context.

\global\long\def\vect#1{\overrightarrow{\mathbf{#1}}}%

\global\long\def\abs#1{\left|#1\right|}%

\global\long\def\av#1{\left\langle #1\right\rangle }%

\global\long\def\ket#1{\left|#1\right\rangle }%

\global\long\def\bra#1{\left\langle #1\right|}%

\global\long\def\tensorproduct{\otimes}%

\global\long\def\braket#1#2{\left\langle #1\mid#2\right\rangle }%

\global\long\def\omv{\overrightarrow{\Omega}}%

\global\long\def\inf{\infty}%

\lhead[\MakeUppercase{\chaptername}~\MakeUppercase{\thechapter}]{\MakeUppercase{\rightmark}}

\rhead[\MakeUppercase{Point-Like Impurities and Rare-Events}]{}

\lfoot[\thepage]{}

\cfoot[]{}

\rfoot[]{\thepage}

\chapter{\label{chap:Rare-Event-States}Rare-Event States of Point-Like Impurities
and Small Clusters}

Chapters\,\ref{chap:Mean-Field-Quantum-Criticality} and \ref{chap:Instability_Smooth_Regions}
gave us two examples of how different disorder types can result in
distinct effects on Dirac-Weyl semimetals. First, the electronic DoS
was analyzed in the presence of an \textit{Anderson random potential}.
Both the analytical and numerical calculations revealed that the semi-metallic
phase, with a vanishing nodal DoS, remains stable up to a critical
disorder strength, after which the DoS takes on a finite value. This
corresponds to a non-conventional disorder-induced SMMT\,\cite{Syzranov18,Fradkin86a,Armitage18}
that long precedes the conventional Anderson MIT. Afterwards, we also
considered the case of a DWSM that is \textit{``decorated''} by
a finite concentration of sizable scalar spherical scatterers of random
strength. This disorder model intends to mimic random smooth potential
regions that can rarely appear within a disordered landscape. These\textit{
\textquotedbl rare regions\textquotedbl} are deemed relevant in
the disordered semimetal phase because they are expected to create
a finite density of states at the node, by means of a non-perturbative
effect\,\cite{Nandkishore14,Pixley16a,Buchhold18b}. In this case,
we have concluded that, as long as there is a \textit{non-zero probability
density} for these scatterers to support bound states, the nodal DoS
gets lifted proportionally to their overall concentration in the system.
Note that our conclusions for the latter model stand in stark contrast
with our findings for the Anderson model, even when considering our
unbiased simulation results.

The two disorder models analyzed earlier are obviously probing very
different limits of what a disordered DWSM is. This partially justifies
the contrasting physical effects but, as reported by Pixley \textit{et
al}.\,\cite{Pixley16a,Pixley21}, both the mean-field and the rare-event
contributions (AQC) to the nodal DoS can be actually produced by the
same type of disorder. In effect, if we have a random scalar potential
with a local distribution that has \textit{unbounded tails}\,\footnote{Note that the numerical study of Chapter\,\ref{chap:Mean-Field-Quantum-Criticality}
dealt only with a local box-distribution potential.} (such as a gaussian or a Cauchy distribution), the DoS shows a clear
semimetal-to-metal transition at finite disorder, but its nodal value
will remain finite \textemdash{} $\overline{\rho(\varepsilon\!=\!0)}\propto\exp\left(-W_{0}^{2}/W^{2}\right)$
\textemdash{} even within the semimetal phase. From our simplified
model of random scatterers, it is not possible to understand why does
an \textit{``unbounded''} Anderson potential yield an enhanced AQC
effect, when compared to the bounded case (analyzed in Chapter\,\ref{chap:Mean-Field-Quantum-Criticality}).
Understanding this enhancement is the main subject of this chapter,
which will culminate in an alternative interpretation of what a rare-event
of a disordered landscape really is. For that, we will begin by reconsidering
the model of Chapter\,\ref{chap:Instability_Smooth_Regions} in the
limit of a vanishing scatterer radius (what we shall call a \textit{point-like
impurity}). Considering a continuous and lattice version, we will
demonstrate an essential difference relative to the model treated
in Chapter\,\ref{chap:Instability_Smooth_Regions}: bound states
\textit{do not} arise from a single point-like impurity but, instead,
require a fine-tuned configuration of (at least) a \textit{pair of
nearby impurities}. Then, we will return to the disordered lattice
WSM and use \textit{Lanczos Diagonalization} (LD\nomenclature{LD}{Lanczos Diagonalization})
to assess the appearance of rare-event nodal eigenstates in different
random on-site disorder models, pinpointing the\textit{ crucial role
played by large fluctuations} in the local potential. Finally, these
rare-event nodal states in a disordered lattice are interpreted, not
as being due to large smooth regions in the landscape (as claimed
in Ref.\,\cite{Nandkishore14}), but rather due to \textit{very small
clusters of nearby sites}, in which the conditions for a collective
bound state are sporadically met. Some results presented here are
original but still unpublished.

\vspace{-0.5cm}

\section{\label{sec:DeltaImpurities}$\delta$\,-\,Impurities in a Continuum
Weyl Semimetal}

In Chapter\,\ref{chap:Instability_Smooth_Regions}, we have shown
that a spherical scatterer can create nodal bound states of Weyl electrons,
if and only if the product of the potential ($\lambda$) with its
radius ($b$) takes on a fine-tuned discrete value. This condition
becomes problematic in the limit $b\!\to\!0^{{\scriptscriptstyle +}}$,
which drags all the \textit{``magical values''} of $\lambda$ to
$\pm\infty$. Therefore, in order to study the problem of point-like
impurities in the continuum, we follow the study of Buchhold \textit{et
al}.\,\cite{Buchhold18b} and consider the perturbed single-node
Weyl Hamiltonian,

\vspace{-0.7cm}

\begin{equation}
\mathcal{H}_{c}\!=\!-i\hbar v_{\text{F}}\int\!d\mathbf{r}\Psi_{a\mathbf{r}}^{\dagger}\left(\boldsymbol{\sigma}^{ab}\!\cdot\!\boldsymbol{\nabla}_{\mathbf{r}}\right)\Psi_{b\mathbf{r}}+\!\int\!d\mathbf{r}V(\mathbf{r})\Psi_{a\mathbf{r}}^{\dagger}\Psi_{a\mathbf{r}},\label{eq:Contin_SingleNodeAHam-1}
\end{equation}
(the same as in Sect.\,\ref{sec:Continuum-Model}), but now with
the perturbation

\vspace{-0.7cm}
\begin{equation}
V(\mathbf{r})=\sum_{n=1}^{N_{i}}U_{n}\delta^{{\scriptscriptstyle (3)}}\!\left(\mathbf{r}-\mathbf{r}_{n}\right),\label{eq:Perturbation_DeltaImp}
\end{equation}
which is a simple sum of $N_{i}$ scalar $\delta$-impurities. Unlike
the spherical scatterers, these point-like impurities certainly generate
scattering with arbitrarily large momentum transfer which would, in
principle, invalidate the assumption of independent Weyl nodes in
a lattice model. In spite of this, we can carry on with the continuum
calculation and try to compute the eDoS for a given configuration
of $\delta$-impurities\nomenclature{$\delta$-impurities}{Point-Like Impurities in a Continuum Model}.
To do this, we will take advantage of the reduced support of the perturbation
(provided $N_{i}$ is a relatively small number) and employ a projected
Green's function method.

\vspace{-0.3cm}

\subsection{Projected Green's Function Formalism}

Obtaining the single-particle properties of the Hamiltonian in Eq.\,\eqref{eq:Contin_SingleNodeAHam-1}
is a standard problem in which we have a separable Hamiltonian, $\mathcal{H}_{c}\!=\!\mathcal{H}_{c}^{0}\!+\!\mathcal{V}$,
about which we know everything in the absence of $\mathcal{V}$. To
be more precise, we know from Sect.\,\ref{sec:Continuum-Model} the
clean model's SPGF\,\footnote{All propagators here-forth will be considered as retarded, by default.},

\vspace{-0.9cm}
\begin{equation}
\mathcal{G}_{c}^{0}\left(E\right)=\left[\tilde{E}\!-\!\mathcal{H}_{c}^{0}\right]^{-1}\!\!=\frac{1}{8\pi^{3}}\int\!\!d\mathbf{k}\frac{\tilde{E}\delta_{ab}-\hbar v_{\text{F}}\boldsymbol{\sigma}^{ab}\!\cdot\!\mathbf{k}}{\tilde{E}^{2}-\hbar^{2}v_{\text{F}}^{2}\abs{\mathbf{k}}^{2}}\ket{\mathbf{k},a}\!\!\bra{\mathbf{k},b},
\end{equation}
where $\tilde{E}\!=\!E\!+\!i0^{{\scriptscriptstyle +}}$ and a summation
over $a,b$ is implicit. In the presence of any perturbation, the
system's SPGF encapsulates all single-particle properties of the perturbed
system and has a completely analogous expression:

\vspace{-0.7cm}
\begin{equation}
\mathcal{G}_{c}\left(E\right)=\left[\tilde{E}\!-\!\mathcal{H}_{c}^{0}\!-\!\mathcal{V}\right]^{-1}.\label{eq:Perturbed}
\end{equation}
However, in practice, the matrix-inversion implied by Eq.\,\eqref{eq:Perturbed}
is not feasible and one must take some type of approximation scheme,
\textit{e.g.}, perturbation theory, or use some clever trick that
allows one to bypass the full inversion in the whole Hilbert space.
Here, we will take the latter approach and, for that, we start by
writing down the exact Dyson relations, 

\vspace{-0.85cm}

\begin{subequations}
\begin{align}
\mathcal{G}_{c}\left(E\right)\! & =\!\mathcal{G}_{c}^{0}\left(E\right)\!+\mathcal{G}_{c}^{0}\left(E\right)\cdot\mathcal{V}\cdot\mathcal{G}_{c}\left(E\right)\label{eq:Dyson1}\\
\mathcal{G}_{c}\left(E\right)\! & =\!\mathcal{G}_{c}^{0}\left(E\right)\!+\mathcal{G}_{c}\left(E\right)\cdot\mathcal{V}\cdot\mathcal{G}_{c}^{0}\left(E\right),\label{eq:Dyson2}
\end{align}
\end{subequations}

\vspace{-0.4cm}

where $\cdot$ stands for a full contraction of all indices/coordinates
in the working representation. This pair of equations may be combined
and written in the different, but equivalent way,

\vspace{-0.85cm}

\begin{subequations}
\begin{align}
\mathcal{G}_{c}\left(E\right) & =\!\mathcal{G}_{c}^{0}\left(E\right)\!+\mathcal{G}_{c}^{0}\left(E\right)\cdot\mathcal{T}\left(E\right)\cdot\mathcal{G}_{c}^{0}\left(E\right)\label{eq:TMatrixEq1}\\
\mathcal{T}\left(E\right) & =\mathcal{V}+\mathcal{V}\cdot\mathcal{G}_{c}\left(E\right)\cdot\mathcal{V}\label{eq:TMatrixEq2}
\end{align}
\end{subequations}

\vspace{-0.4cm}

where $\mathcal{T}\left(E\right)$ is the (retarded) $T$-matrix,
an energy-dependent operator associated to the whole perturbation
$\mathcal{V}$. Now, we bring about the essential point of the \textit{Projected
Green's Function} (pGF\nomenclature{pGF}{Projected Green's Function})
method: Both $\mathcal{V}$ and $\mathcal{T}\left(E\right)$ only
act non-trivially in a very restricted real-space volume, the support
of the perturbation, which we generically call $\Omega$\,\footnote{Note that, since we are dealing with single-particle problems, it
is entirely equivalent to think about $\Omega$ as small subset of
real-space, or a very small subspace of the full system's Hilbert
space.} and that, in our case, is composed by the positions $\mathbf{r}_{n}$
where the $\delta$-impurities are centered. This locality property
of $\mathcal{V}$ allows us to solve Eqs.\,\eqref{eq:TMatrixEq1}-\eqref{eq:TMatrixEq2}
as a two step process:
\begin{enumerate}
\item Project the system of equations into $\Omega$, \textit{i.e.}, given
any general operator $\mathcal{O}$, and considering the projector
onto the subspace of $\Omega$ as the operator $\mathcal{P}_{\!\Omega}$,
we must transform $\mathcal{O}\to\underline{\mathcal{O}}\!=\!\mathcal{P}_{\!\Omega}\,\mathcal{O}\,\mathcal{P}_{\!\Omega}$
in all equations. For the Green's function equations, this yields

\vspace{-0.7cm}

\begin{subequations}
\begin{align}
\underline{\mathcal{G}_{c}\left(E\right)} & =\!\underline{\mathcal{G}_{c}^{0}\left(E\right)}\!+\underline{\mathcal{G}_{c}^{0}\left(E\right)}\cdot\mathcal{T}\left(E\right)\cdot\underline{\mathcal{G}_{c}^{0}\left(E\right)}\label{eq:TMatrixEq1-1}\\
\mathcal{T}\left(E\right) & =\mathcal{V}+\mathcal{V}\cdot\underline{\mathcal{G}_{c}\left(E\right)}\cdot\mathcal{V},\label{eq:TMatrixEq2-1}
\end{align}
\end{subequations}

where we have used the fact that $\mathcal{P}_{\!\Omega}\,\mathcal{V}\,\mathcal{P}_{\!\Omega}\!=\!\mathcal{V}$
and $\mathcal{P}_{\!\Omega}\,\mathcal{\mathcal{T}}\left(E\right)\,\mathcal{P}_{\!\Omega}\!=\!\mathcal{T}\left(E\right)$.
Clearly, Eqs.\,\eqref{eq:TMatrixEq1-1} and \eqref{eq:TMatrixEq2-1}
form a closed set of $N_{i}$ coupled equations that can be solved
explicitly. More precisely, we have 

\vspace{-0.8cm}
\begin{equation}
\underline{\mathcal{G}_{c}\left(E\right)}\!=\underline{\mathcal{G}_{c}^{0}\left(E\right)}\!\cdot\![\underline{\mathcal{I}}\!-\!\mathcal{V}\underline{\mathcal{G}_{c}^{0}\left(E\right)}]^{-1}
\end{equation}
and, consequently,

\vspace{-0.8cm}
\begin{equation}
\mathcal{T}\left(E\right)\!=\!\left[\mathcal{V}^{{\scriptscriptstyle -\!1}}\!\!-\!\underline{\mathcal{G}_{c}^{0}\left(E\right)}\right]^{-1}.\label{eq:T_MatProj}
\end{equation}
Note that in the previous equations, contrary to Eq.\,\eqref{eq:Perturbed},
the matrix-inversion is not a problem because the matrices involved
have small dimensions. For the specific case of the single Weyl node,
solving Eq.\,\eqref{eq:T_MatProj} would involve the inversion of
a dense $2N_{i}\!\times\!2N_{i}$ complex-valued matrix. In a common
laptop, one can numerically invert an $N\!\times\!N$ complex dense
matrix in less $2$ minutes and using about $1$\,Gb of RAM memory
for $N\!=\!8192$. Typically, these $QR$-based inversion algorithms
scale as $N^{3}$ in CPU-time and $N^{2}$ in memory usage. 
\item The perturbed SPGF for any two points outside $\Omega$ can be reconstructed
from the perturbed one by using Eq.\,\eqref{eq:TMatrixEq1}, with
the $T$-matrix calculated in the previous step.
\end{enumerate}
Note that all this process takes for granted that one knows, \textit{a
priori}, the unperturbed SPGF in real-space. If we know this quantity,
then the aforementioned $T$-\,matrix can be calculated as follows,

\vspace{-0.7cm}

{\small{}
\begin{equation}
\!\!\!\!\mathcal{T}\left(E\right)\!\to\!\!\left[\!\begin{array}{ccccc}
\frac{1}{U_{1}}-G_{11}^{\text{0r}}(E;\boldsymbol{0}) & -G_{12}^{\text{0r}}(E;\boldsymbol{0}) & -G_{11}^{\text{0r}}(E;\Delta\mathbf{r}_{21}) & -G_{12}^{\text{0r}}(E;\Delta\mathbf{r}_{21})\\
-G_{21}^{\text{0r}}(E;\boldsymbol{0}) & \frac{1}{U_{1}}-G_{22}^{\text{0r}}(E;\boldsymbol{0}) & -G_{21}^{\text{0r}}(E;\Delta\mathbf{r}_{21}) & -G_{22}^{\text{0r}}(E;\Delta\mathbf{r}_{21})\\
-G_{11}^{\text{0r}}(E;\Delta\mathbf{r}_{12}) & -G_{12}^{\text{0r}}(E;\Delta\mathbf{r}_{12}) & \frac{1}{U_{2}}-G_{11}^{\text{0r}}(E;\boldsymbol{0}) & -G_{12}^{\text{0r}}(E;\boldsymbol{0})\\
-G_{21}^{\text{0r}}(E;\Delta\mathbf{r}_{12}) & -G_{22}^{\text{0r}}(E;\Delta\mathbf{r}_{12}) & -G_{21}^{\text{0r}}(E;0) & \frac{1}{U_{2}}-G_{22}^{\text{0r}}(E;\boldsymbol{0})\\
 &  &  &  & \ddots
\end{array}\!\right]^{-1}\!\!\!\!\!,\label{eq:TMatrixProj}
\end{equation}
}where $\Delta\mathbf{r}_{ji}\!=\!\mathbf{r}_{j}\!-\!\mathbf{r}_{i}$,
and where we have assumed a real-space representation that orders
the basis of the projected subspace as

\vspace{-0.8cm}

\begin{equation}
\Omega=\text{Span}\left\{ \ket{\mathbf{r}_{1},1},\ket{\mathbf{r}_{1},2},\ket{\mathbf{r}_{2},1},\ket{\mathbf{r}_{2},2},\cdots,\ket{\mathbf{r}_{N_{i}},1},\ket{\mathbf{r}_{N_{i}},2}\right\} .\label{eq:ProjRealSpace}
\end{equation}
In summary, in the presence of a finite number of $\delta$-impurities,
the perturbed SPGF between any two points in space can be entirely
determined by calculating the matrix presented in Eq.\,\eqref{eq:TMatrixProj}.
The algorithmic complexity of this method comes from two stages: \textit{(i)}
the construction of the matrix to be inverted {[}an $\mathcal{O}(N_{i}^{2})$
process{]}, and \textit{(ii)} the inversion of this matrix. As we
shall see, this can be done analytically if the number of impurities
is sufficiently small, namely $N_{i}=1\text{ or }2$.

\vspace{-0.5cm}

\paragraph{Density of States and the Projected $T$-Matrix:}

The perturbed propagator between any two positions can be reconstructed
from the knowledge of $\mathcal{T}\left(E\right)$ in the projected
real-space representation of Eq.\,\eqref{eq:ProjRealSpace}. However,
for the purpose of calculating changes induced by $\mathcal{V}$ in
the system's global eDoS it is hardly necessary to do that. The change
in the eDoS is defined, in terms of SPGF as\,\footnote{The global sign would appear reversed for advanced SPGFs.}

\vspace{-0.7cm}

\begin{equation}
\delta\nu\left(E\right)=\nu\left(E\right)-\nu_{0}\left(E\right)=-\frac{1}{\pi}\Im\left(\text{Tr}\left[\mathcal{G}_{c}\left(E\right)\!-\!\mathcal{G}_{c}^{0}\left(E\right)\right]\right),\label{eq:DeltaRho_GF}
\end{equation}
where the trace is over the whole Hilbert space. Using the result
of Eqs.\,\eqref{eq:TMatrixEq1} and \eqref{eq:T_MatProj}, we can
cast this expression into the apparently more complicated form,

\vspace{-0.7cm}

\begin{equation}
\delta\nu\left(E\right)=-\frac{1}{\pi}\Im\left(\text{Tr}\left[\mathcal{G}_{c}^{0}\left(E\right)\cdot\left[\mathcal{V}^{{\scriptscriptstyle -\!1}}\!\!-\!\underline{\mathcal{G}_{c}^{0}\left(E\right)}\right]^{-1}\!\!\cdot\mathcal{G}_{c}^{0}\left(E\right)\right]\right),\label{eq:DeltaRho_GF-1}
\end{equation}
or, equivalently,

\vspace{-0.7cm}

\begin{align}
\delta\nu\left(E\right) & =\!-\frac{1}{\pi}\Im\left(\text{Tr}\left[\left[\underline{\mathcal{I}}\!\!-\!\underline{\mathcal{G}_{c}^{0}\left(E\right)}\cdot\mathcal{V}\right]^{-1}\!\!\cdot\left(\mathcal{G}_{c}^{0}\left(E\right)\right)^{2}\cdot\mathcal{V}\right]\right),\label{DosChange}
\end{align}
where we have used the cyclic property of the trace. Since both the
rightmost and leftmost operators inside the trace have support in
$\Omega$, one is entitled to replace the full trace by one that is
restricted to this subspace,\textit{ i.e.},

\vspace{-0.7cm}

\begin{align}
\delta\nu\left(E\right) & =\!-\frac{1}{\pi}\Im\left(\text{tr}\left[\left[\underline{\mathcal{I}}\!\!-\!\underline{\mathcal{G}_{c}^{0}\left(E\right)}\cdot\mathcal{V}\right]^{-1}\!\!\cdot\left(\underline{\mathcal{G}_{c}^{0}\left(E\right)}\right)^{2}\cdot\mathcal{V}\right]\right)\label{DosChange-1}\\
 & \qquad\qquad\qquad\qquad=\!\frac{1}{\pi}\Im\left(\text{tr}\left[\mathcal{T}\left(E\right)\cdot\left(\frac{d}{dE}\underline{\mathcal{G}_{c}^{0}\left(E\right)}\right)\right]\right),\nonumber 
\end{align}
where we have also used the fact that $d\mathcal{G}_{c}^{0}\left(E\right)/dE\!=\!-\left(\mathcal{G}_{c}^{0}\left(E\right)\right)^{2}$.
Finally, we arrive at the essential formal result of this section:
the change in the eDoS due to a cluster of $\delta$-impurities can
be obtained as a \textit{projected trace} involving the projected
$T$-matrix and the clean SPGF. Although the form of Eq.\,\eqref{DosChange-1}
is the most useful for numerical calculations, it is worth remarking
that there are alternative ways\,\cite{Buchhold18b} of writing this
result, namely,

\vspace{-0.7cm}

\begin{align}
\delta\nu\left(E\right)\! & =\!\frac{1}{\pi}\!\Im\left(\frac{d}{dE}\text{tr}\left[\ln\left[\underline{\mathcal{I}}\!\!-\!\underline{\mathcal{G}_{c}^{0}\left(E\right)}\cdot\mathcal{V}\right]\right]\right)\\
 & \qquad=\frac{1}{\pi}\!\Im\left(\frac{d}{dE}\ln\left[\det\left[\underline{\mathcal{I}}\!\!-\!\underline{\mathcal{G}_{c}^{0}\left(E\right)}\cdot\mathcal{V}\right]\right]\right),
\end{align}
where use was made of the general matrix identity, $\text{Tr}\left[\ln\left[\mathcal{M}\right]\right]=\ln\left[\det\left[\mathcal{M}\right]\right].$

\vspace{-0.4cm}

\subsection{Bound-States and the Projected Lippmann-Schwinger Equation}

Before moving on to calculate the eDoS deformation for particular
cases, it is important to take a small detour and establish a further
formal result. Here, we will show that the projected formalism can
likewise be used to identify nodal bound states in systems of a few
impurities. To see this, we start by writing the wavefunction equivalent
of Eqs.\,\eqref{eq:Dyson1}-\eqref{eq:Dyson2}; the so-called \textit{Lippmann-Schwinger
Equation} (LSE\nomenclature{LSE}{Lippmann-Schwinger Equation})\,\cite{Lippmann50,Lippmann50b}
for a scattering state $\ket{\Psi_{E}}$,

\vspace{-0.7cm}
\begin{equation}
\ket{\Psi_{E}}\!=\!\ket{\Psi_{E}^{0}}\!+\!\mathcal{G}_{c}^{0}\left(E\right)\!\cdot\!\mathcal{V}\ket{\Psi_{E}},\label{eq:LippamannSchwinger}
\end{equation}
where $\ket{\Psi_{E}^{0}}$ is an eigenstate of the unperturbed system
described by $\mathcal{H}_{c}^{0}$, being a parent extended state
of $\ket{\Psi_{E}}$. Note that Eq.\,\eqref{eq:LippamannSchwinger}
is an exact self-consistent representation of a solution to the perturbed
quantum eigenvalue problem, and has the trivial interpretation that
a perturbed solution can be seen as an unperturbed propagating wavefunction,
at that energy, to which a scattering part is added. In principle,
this equation also works to find non-propagating solutions, for which
there is no parent eigenstate of the clean Hamiltonian, \textit{i.e.},

\vspace{-1.0cm}

\begin{equation}
\left[\mathcal{I}-\mathcal{G}_{c}^{0}\left(E_{b}\right)\!\cdot\!\mathcal{V}\right]\ket{\Psi_{E_{b}}^{\text{b}}}\!=\!0,\label{eq:LippamannSchwinger-1}
\end{equation}
where $E_{b}$ is the energy of the state. Note that the previous
equation clearly indicates that such a solution must belong to the
kernel of the operator $\mathcal{I}-\mathcal{G}_{c}^{0}\left(E_{b}\right)\!\cdot\!\mathcal{V}$.
Furthermore, if the perturbation has a finite support $\Omega$, one
can project Eq.\,\eqref{eq:LippamannSchwinger-1} into this subspace
and get

\vspace{-0.7cm}

\begin{equation}
\left[\underline{\mathcal{I}}-\underline{\mathcal{G}_{c}^{0}\left(E_{b}\right)}\!\cdot\!\mathcal{V}\right]\ket{\xi_{E_{b}}^{\text{b}}}\!=\!0,\label{eq:LippamannSchwinger-1-1}
\end{equation}
where $\ket{\xi_{E_{b}}^{\text{b}}}\!=\!\mathcal{P}_{\!\Omega}\ket{\Psi_{E_{b}}^{\text{b}}}$
is the restriction of $\ket{\Psi_{\varepsilon_{b}}}$ to the support
$\Omega$. Remarkably, by analyzing the dimension of the kernel of
$\underline{\mathcal{I}}-\underline{\mathcal{G}_{c}^{0}\left(E_{b}\right)}\!\cdot\!\mathcal{V}$,
one can obtain possible bound states generated by the perturbation
$\mathcal{V}$. To prove that such a state is a bound state of the
system, it is necessary to reconstruct the corresponding wavefunction
outside $\Omega$, that is,

\vspace{-0.7cm}
\begin{equation}
\Psi_{a}^{b}\!\left(\mathbf{r}\right)\!=\!\braket{\mathbf{r},a}{\Psi_{E_{b}}^{\text{b}}}\!=\!\bra{\mathbf{r},a}\mathcal{G}_{c}^{0}\left(E_{b}\right)\!\cdot\!\mathcal{V}\ket{\xi_{\varepsilon_{b}}^{b}},
\end{equation}
where $\mathbf{r}$ is an arbitrary position. Being a bound state
or not then depends on the asymptotic properties of the clean SPGF
in real-space. For this system any solution of Eq.\,\eqref{eq:LippamannSchwinger-1-1}
with a non-zero energy will not lead to a square-normalizable wavefunction,
because $G_{ab}^{\text{0r}}(E\!\neq\!0;\abs{\boldsymbol{\Delta r}}\!\to\!\infty)\propto\abs{\boldsymbol{\Delta r}}^{-1}$.
However, if we focus only on the nodal energy ($E_{b}\!=\!0$), the
reconstructed wavefunction is guaranteed to be normalizable by the
property, $G_{ab}^{\text{0r}}(E\!=\!0;\abs{\boldsymbol{\Delta r}}\!\to\!\infty)\propto r^{-2}$\,\footnote{Despite being a fairly common situation, there is no fundamental reason
for a proper bound state to appear at energies where the clean system
has a vanishing DoS. In fact, bound states within a continuous spectrum\,\cite{Hsu16}
are known to exist since the dawn of quantum mechanics (see the seminal
paper of von Neumann and Wigner\,\cite{VonNeumann29}).}. Therefore, obtaining the null-space of the projected operator $\underline{\mathcal{I}}-\underline{\mathcal{G}_{c}^{0}\left(0\right)}\!\cdot\!\mathcal{V}$
allows us to pinpoint the existence of many-impurity nodal bound states.

\subsection{\label{subsec:PointLikeCont}The Single- and Two-Impurity Problem}

At this point, we have developed all the necessary formalism to study
the spectral effects introduced by a set of $\delta$-impurities in
a Weyl semimetal. Here, we employ it for the analytical study of two
important cases:

\vspace{-0.7cm}
\begin{equation}
\mathcal{V}_{1}\!=\!U\delta^{{\scriptscriptstyle (3)}}\!\left(\mathbf{r}\right)\text{ and }\mathcal{V}_{2}\!=\!U_{1}\delta^{{\scriptscriptstyle (3)}}\!\left(\mathbf{r}\right)+U_{2}\delta^{{\scriptscriptstyle (3)}}\!\left(\mathbf{r}-\Delta\mathbf{r}\right),
\end{equation}
which describe a system of a single or an isolated pair of $\delta$-impurities.
These problems have already been addressed in Ref.\,\cite{Buchhold18b}
but are still of great importance to understand some major qualitative
features of point-like scalar impurities in Weyl system. Namely, we
will show that nodal bound states can only arise from quantum interference
between several $\delta$-impurities, unlike what happened with the
spherical scatterers of Chapter\,\ref{chap:Instability_Smooth_Regions}.

\vspace{-0.4cm}

\paragraph{The isolated $\delta$-impurity case:}

We begin by the $\mathcal{V}_{1}$ perturbation, which is controlled
by a single real-valued parameter, $U$, measuring the strength of
the local potential. Note that $U$ has dimensions of \textit{energy
times volume}, because the Dirac-$\delta$ is a 3D distribution. Considering
the clean SPGF of Sect.\,\ref{sec:Continuum-Model}, together with
the \textit{``smooth cut-off''} UV-regularizer considered there,
we can explicitly write the $T$-matrix as 

\vspace{-0.7cm}

\begin{equation}
\mathcal{T}\left(\varepsilon\right)=\frac{4\pi\hbar v_{\text{F}}U(M-i\varepsilon)}{\varepsilon M^{2}U+4\pi\hbar v_{\text{F}}(M-i\varepsilon)}\mathcal{I}_{2\times2},
\end{equation}
where $\varepsilon\!=\!E/\hbar v_{\text{F}}$, $M$ is the smooth
cut-off scale and $\mathcal{T}\left(\varepsilon\right)$ is a matrix
defined in a support of dimension $2$. Hence, we have the following
expression for the correction to the eDoS, 
\begin{figure}[t]
\vspace{-0.5cm}
\begin{centering}
\includegraphics[scale=0.25]{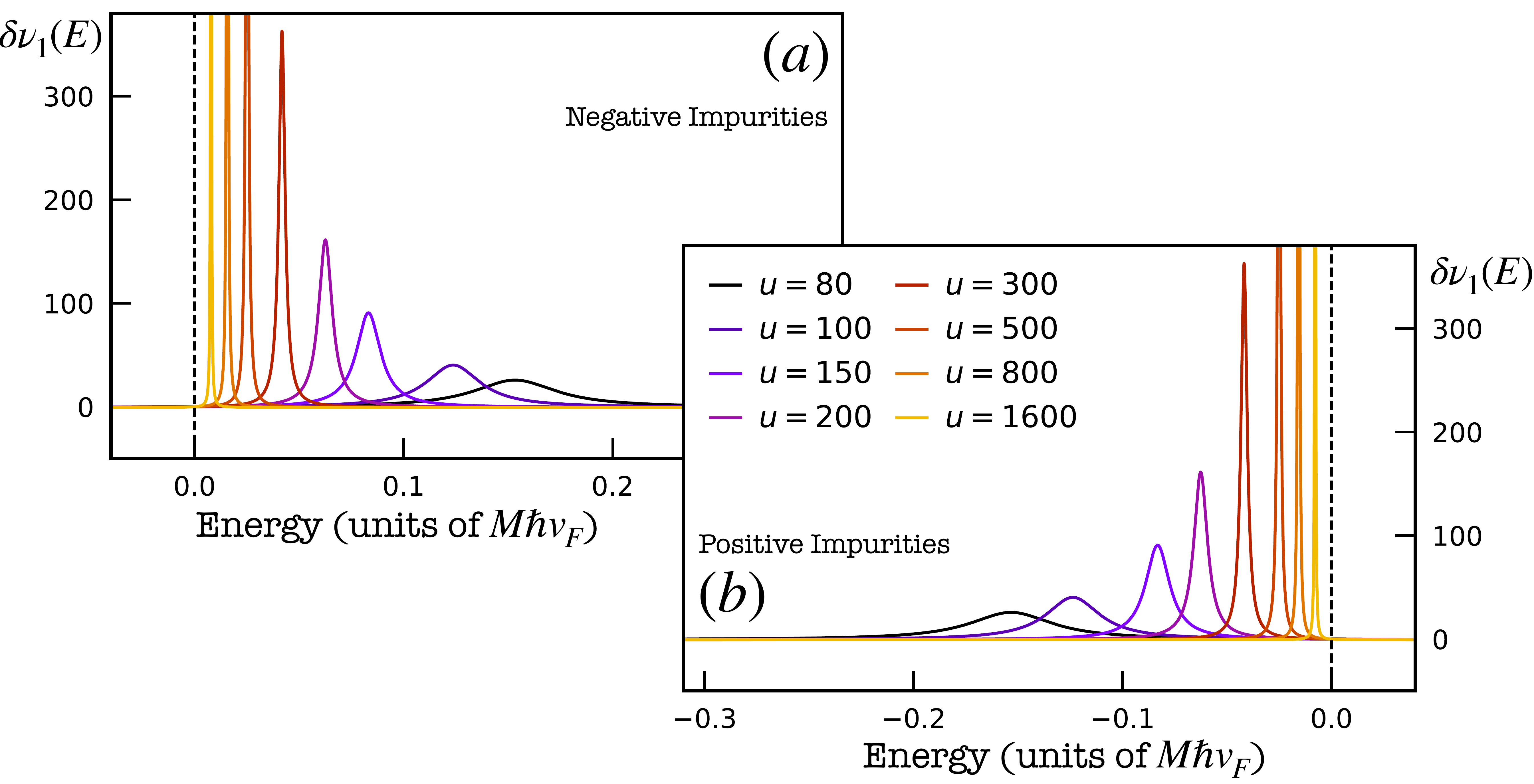}
\par\end{centering}
\vspace{-0.3cm}

\caption{\label{fig:OneDelta}Plots of the change in the eDoS induced by a
single $\delta$-impurity with different strengths. (a) Represents
the case $U\!>\!0$ while (b) stands for $U\!<\!0$.}

\vspace{-0.3cm}
\end{figure}

\vspace{-0.7cm}
\begin{align}
\delta\nu_{1}(\varkappa,u)\! & =\!-\frac{2u\varkappa\left(8\pi\!+\!u\varkappa\right)}{\pi\left(1\!+\!\varkappa^{2}\right)\left(u\varkappa\left(8\pi\!+\!u\varkappa\right)+16\pi^{2}\left(1\!+\!\varkappa^{2}\right)\right)},
\end{align}
where $\delta\nu_{1}$ is the eDoS measured in units of $M\hbar v_{\text{F}}$,
while $u\!=\!UM^{2}/\hbar v_{\text{F}}$ and $\varkappa=\varepsilon/M=E/M\hbar v_{\text{F}}$
are dimensionless parameters. In Fig.\,\ref{fig:OneDelta}, we plot
$\delta\nu_{1}$ for different values of $u$, demonstrating that
a sharp resonance appears in the valence (conduction) band as $u\!\to\!+\infty$
($u\!\to\!-\infty$). This behavior is qualitatively similar to what
was found for the spherical scatterers of Chapter\,\ref{chap:Instability_Smooth_Regions}
but with a crucial difference: the resonance \textit{never crosses
the node}. In fact, we can prove that there are no parameter $u$able
to generate a bound state of the $\delta$-impurity, because

\vspace{-0.8cm}

\begin{align}
\det\left(\underline{\mathcal{I}}-\underline{\mathcal{G}_{c}^{0}\left(E\right)}\!\cdot\!\mathcal{V}_{1}\right) & =\frac{\left(\varkappa u+4\pi(1-i\varkappa)\right)^{2}}{16\pi^{2}(1-i\varkappa)^{2}}=0
\end{align}
only has a single solution

\vspace{-0.8cm}

\begin{equation}
u\left(\varkappa\right)=-4\pi(1-i\varkappa)/\varkappa,
\end{equation}
which for zero energy ($\varkappa\!=\!0$), cannot be satisfied by
any finite $u$.

\vspace{-0.4cm}

\paragraph{Pair of $\delta$-impurities distanced by $d$:}

\begin{figure}[t]
\vspace{-0.5cm}
\begin{centering}
\includegraphics[scale=0.25]{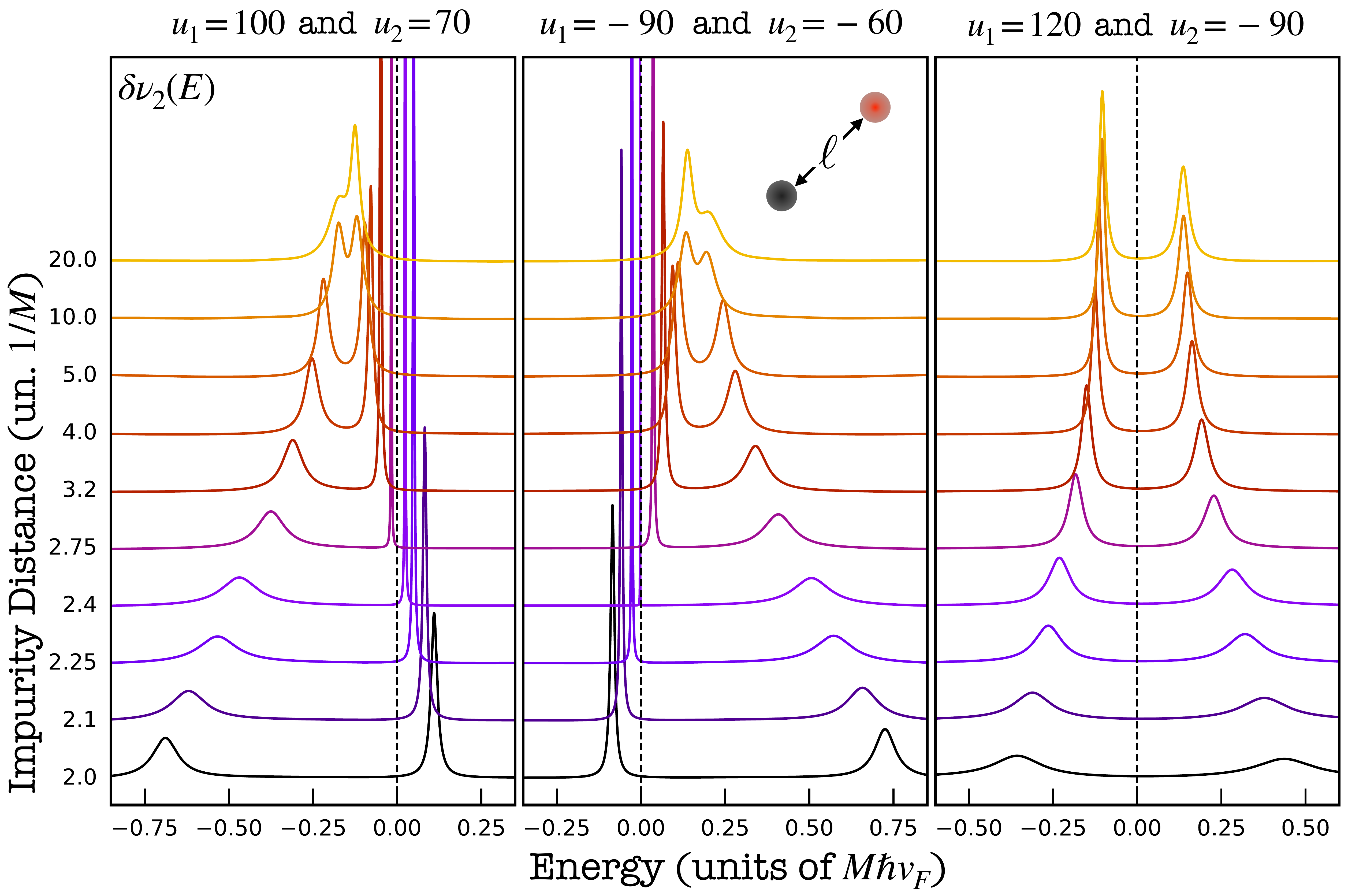}
\par\end{centering}
\vspace{-0.3cm}

\caption{\label{fig:TwoDeltas}Plots of the change in the eDoS induced by a
pair of $\delta$-impurities in three situation, with the impurity
strengths indicated above the panels. The different curves correspond
to an increasing distance between the impurities ($\ell\!=\![2,20]$),
which eventually converges towards a dilute limit result. According
to Eq.\,\eqref{eq:CriticalLenght}, every time a hybridized resonance
crosses the nodal energy ($E\!=\!0$), a nodal bound states forms
between the two impurities.}

\vspace{-0.5cm}
\end{figure}

In considering the two-impurity perturbation, $\mathcal{V}_{2}$,
it is important to first realize that the continuum model is isotropic
in space. This means that $\Delta\mathbf{r}$ can be chosen to point
in an arbitrary direction, \textit{e.g.}, $\Delta\mathbf{r}\!=\!d\,\mathbf{z}$,
with no loss of generality. With the later choice, the $T$-matrix
associated to this pair of $\delta$-impurities simply reads,

\vspace{-0.7cm}

{\small{}
\begin{equation}
\mathcal{T}\left(\varepsilon\right)\!=\!\left[\!\begin{array}{cccc}
\frac{1}{U_{1}}+\frac{\varepsilon M^{2}}{4\pi\hbar v_{\text{F}}\left(M-i\varepsilon\right)} & 0 & -\frac{ie^{id\varepsilon}}{4\pi\hbar v_{\text{F}}d^{2}} & 0\\
0 & \frac{1}{U_{1}}+\frac{\varepsilon M^{2}}{4\pi\hbar v_{\text{F}}\left(M-i\varepsilon\right)} & 0 & \frac{e^{id\varepsilon}(2d\varepsilon+i)}{4\pi\hbar v_{\text{F}}d^{2}}\\
\frac{e^{id\varepsilon}(2d\varepsilon+i)}{4\pi\hbar v_{\text{F}}d^{2}} & 0 & \frac{1}{U_{2}}+\frac{\varepsilon M^{2}}{4\pi\hbar v_{\text{F}}\left(M-i\varepsilon\right)} & 0\\
0 & -\frac{ie^{id\varepsilon}}{4\pi\hbar v_{\text{F}}d^{2}} & 0 & \frac{1}{U_{2}}+\frac{\varepsilon M^{2}}{4\pi\hbar v_{\text{F}}\left(M-i\varepsilon\right)}
\end{array}\!\right]^{-1}
\end{equation}
}which, after some lengthy algebra, yields the following result:

\vspace{-0.7cm}

{\small{}
\begin{equation}
\!\!\!\!\!\!\frac{M\,\hbar v_{\text{F}}}{\pi}\text{tr}\!\left[\mathcal{T}\left(\varepsilon\right)\!\cdot\!\left(\frac{d}{d\varepsilon}\underline{\mathcal{G}_{c}^{0}\left(\varepsilon\right)}\right)^{\!2}\!\right]\!=\!\frac{\frac{4}{\pi(\varkappa+i)^{2}}\left(\frac{iu_{1}u_{2}\varkappa}{\varkappa+i}\!+\!2\pi\left(u_{1}\!+\!u_{2}\right)\right)\!+\!\frac{8u_{1}u_{2}\varkappa}{\pi\ell^{2}}e^{2i\ell\varkappa}}{\left(\frac{iu_{1}\varkappa}{\varkappa+i}\!+\!4\pi\right)\left(\frac{iu_{2}\varkappa}{\varkappa+i}\!+\!4\pi\right)\!+\!\frac{u_{1}u_{2}}{\ell^{4}}e^{2i\ell\varkappa}\left(2i\ell\varkappa\!-\!1\right)},\!\!\label{eq:Trace_TwoImps}
\end{equation}
}where $u_{i}\!=\!U_{i}M^{2}/\hbar v_{\text{F}}$ and $\ell\!=\!d\,M$
are dimensionless quantities. From Eq.\,\eqref{eq:Trace_TwoImps},
we can obtain the correction to the eDoS due to the pair of $\delta$-impurities,
which we plot in Fig.\,\ref{fig:TwoDeltas}. Before commenting on
general results, we remark that the limit $\ell\!\to\!+\infty$ of
Eq.\,\eqref{eq:Trace_TwoImps} yields the additive result,

\vspace{-0.7cm}
\begin{equation}
\delta\nu_{2}(\varkappa,u_{1},u_{2},\ell)\underset{\ell\to\infty}{\longrightarrow}\delta\nu_{1}(\varkappa,u_{1})+\delta\nu_{1}(\varkappa,u_{2})
\end{equation}
thus confirming a well-defined dilute limit for point-like impurities.
Going back to the plots of Fig.\,\ref{fig:TwoDeltas}, we can identify
two distinct situations. If $u_{1}u_{2}\!<\!0$, the isolated impurities
would lead to two resonances in either side of the node, which hybridize
as they get closer together and are repelled away from the Weyl node.
On the contrary, if $u_{1}u_{2}\!>\!0$, the two isolated resonances
would be on the same band and hybridization tends to drive one of
them towards the node until it crosses it. When this happens, the
situation becomes analogous to the one described in Subsect.\,\ref{sec:Low-Energy-Resonances-1},
and a non-degenerate nodal bound state is formed between the impurity
pair. This argument can be proven directly by analyzing 

\vspace{-0.7cm}
\begin{equation}
\!\!\!\!\!\det\left(\underline{\mathcal{I}}-\underline{\mathcal{G}_{c}^{0}\left(E\right)}\!\cdot\!\mathcal{V}_{2}\right)\!=\!\left[\!\frac{\left(\frac{iu_{1}\varkappa}{\varkappa+i}\!+\!4\pi\right)\left(\frac{iu_{2}\varkappa}{\varkappa+i}\!+\!4\pi\right)+\frac{2i\ell\varkappa-1}{\ell^{4}}u_{1}u_{2}e^{2i\ell\varkappa}}{16\pi^{2}}\!\right]^{2}\!\!.\!\!
\end{equation}
Assuming zero energy, this determinant is null if and only if

\vspace{-0.7cm}

\begin{equation}
\ell=\ell_{c}=\left[\frac{u_{1}u_{2}}{16\pi^{2}}\right]^{\frac{1}{4}}\!\!,\label{eq:CriticalLenght}
\end{equation}
which is $3$-parameter fine-tuned situation. Note that, for the case
presented on the left (central) panel of Fig.\,\ref{fig:TwoDeltas},
we have a critical separation of $\ell_{c}\!=\!2.5803$ ($\ell_{c}\!=\!2.4182$),
which is consistent with the point at which the sharper resonance
traverses $E\!=\!0$. 

\vspace{-0.4cm}

\paragraph{The Role of Impurity Correlations:}

At this point, we can already draw some interesting conclusions about
the point-like impurity limit of the problem treated in Chapter\,\ref{chap:Instability_Smooth_Regions}.
In this case, an isolated scalar impurity is not able to bind Weyl
electrons at the nodal energy, independently of how strong the potential
is. As one takes the impurity potential to more extreme values, an
ever sharper impurity-induced resonance is observed to approach the
Weyl node without ever crossing it. In order to have such a crossing,
and the formation of a nodal bound state, it is crucial to include
interference effects between two such impurities, as previously noted
by Buchhold \textit{et al}.\,\cite{Buchhold18b}. However, even in
this case the bound states are delicate and require \textit{three
independent parameters} to be fine-tuned: the impurity strengths and
the distance between the two $\delta$-impurities.

\section{\label{sec:Atomic-Sized-Impurities}Atomic-Sized Impurities in a
Lattice Weyl Semimetal}

We have started by analyzing the case of point-like impurities in
the single-node continuum model of a Weyl semimetal. By doing so,
one is also taking for granted a set of assumptions that may not hold
in the context of a lattice system and, therefore, require confirmation
by analogous lattice calculations. In this section, we take the clean
simple cubic lattice model given by the Hamiltonian, 

\vspace{-0.7cm}

\begin{align}
\mathcal{H}_{l}^{0} & \!=\!\negthickspace\sum_{{\scriptscriptstyle \mathbf{R}\in\mathcal{L_{\text{C}}}}}\negthickspace\left[\frac{i\hbar v_{\text{F}}}{2a}\Psi_{\mathbf{{\scriptscriptstyle R}}}^{\dagger}\!\cdot\!\sigma^{j}\!\cdot\!\Psi_{{\scriptscriptstyle \mathbf{R}+a\hat{e}_{j}}}\!+\!\text{h.c.}\right]\!,\label{eq:LatticeModel-1}
\end{align}
whose properties are fully explored in the Appendix\,\ref{chap:LatticeGF}.
Since this model is discrete, the $\delta$-impurities of the continuum
model must be replaced by a set of \textit{atomic-sized perturbations},
such as,

\vspace{-0.7cm}

\begin{equation}
\mathcal{V}_{l}\!=\!\!\!\sum_{{\scriptscriptstyle \mathbf{R}\in\mathcal{L_{\text{C}}}}}\!\!V(\mathbf{R})\Psi_{a\mathbf{R}}^{\dagger}\Psi_{a\mathbf{R}},\label{eq:Contin_SingleNodeAHam-1-1}
\end{equation}
where the potential $V(\mathbf{R})$ takes the form, 

\vspace{-0.7cm}
\begin{equation}
V(\mathbf{R})=\sum_{n=1}^{N_{i}}U_{n}\delta_{\mathbf{R},\mathbf{R}_{n}}.\label{eq:Perturbation_DeltaImp-1}
\end{equation}
Unlike the continuum case, here the $U_{n}$s have dimensions of energy
and, therefore, can be naturally measured in units of $\hbar v_{\text{F}}/a$.
Fortunately, this case is not fundamentally different from the one
treated in Sect.\,\ref{sec:DeltaImpurities}, as all the pGF formalism
may be directly translated into the discrete theory, without significant
adaptations. The sole major difference will lie in the form of the
real-space SPGF (usually called a\textit{ lattice Green's Function}
(lGF\nomenclature{lGF}{Lattice Green's Function (Real-Space Propagator of a Tight-Binding Model)})
in the tight-binding context) which will differ from the continuum
one, and needs to be evaluated.

\vspace{-0.4cm}

\subsection{\label{subsec:The-Lattice-Green's}The Lattice Green's Function}

The first step to analyze the lattice problem is to calculate the
lGF associated to the model of Eq.\,\eqref{eq:LatticeModel-1}, something
that can be easily done in $\mathbf{k}$-space, where

\vspace{-0.7cm}
\begin{equation}
\mathcal{H}_{l}^{0}(\mathbf{k})=\frac{\hbar v_{\text{F}}}{a}\boldsymbol{\sigma}\cdot\boldsymbol{\sin}a\mathbf{k}
\end{equation}
is the Bloch Hamiltonian, with $\boldsymbol{\sigma}$ being a vector
of Pauli matrices and $\boldsymbol{\sin}a\mathbf{k}\!=\!\left(\sin ak_{x},\sin ak_{y},\sin ak_{z}\right)$.
Upon inversion of this $2\!\times\!2$ Hamiltonian, we can obtain
a general expression for the (retarded) lGF, in terms of an integral
over the cubic fBz:

\vspace{-0.7cm}

\begin{align}
\mathscr{G}_{ab}^{\text{0r}}(E;\boldsymbol{\Delta R})\! & =\!\bra{\boldsymbol{\Delta R},a}\left[E\!+\!i\eta\!-\!\mathcal{H}_{l}^{0}\right]^{-1}\!\ket{\mathbf{0},b}\label{eq:RealSpacePropagator-2}\\
 & \qquad=\!\frac{a^{3}}{8\pi^{3}}\int_{\text{fBz}}\!\!\!d\mathbf{k}\,\frac{E-\frac{\hbar v_{\text{F}}}{a}\boldsymbol{\sigma}\cdot\boldsymbol{\sin}a\mathbf{k}}{\left(E\!+\!i\eta\right)^{2}-\frac{\hbar^{2}v_{\text{F}}^{2}}{a^{2}}\abs{\boldsymbol{\sin}a\mathbf{k}}^{2}}e^{i\mathbf{k}\cdot\boldsymbol{\Delta R}},\nonumber 
\end{align}
which can be simplified by introducing a dimensionless energy, $\varepsilon=Ea/\hbar v_{\text{F}}$,
a dimensionless position, $\boldsymbol{\Delta L}\!=\!\boldsymbol{\Delta R}/a$,
and a dimensionless crystal momentum, $\mathbf{q}=a\mathbf{k}$. This
then yields

\begin{figure}[t]
\vspace{-0.5cm}
\begin{centering}
\includegraphics[scale=0.2]{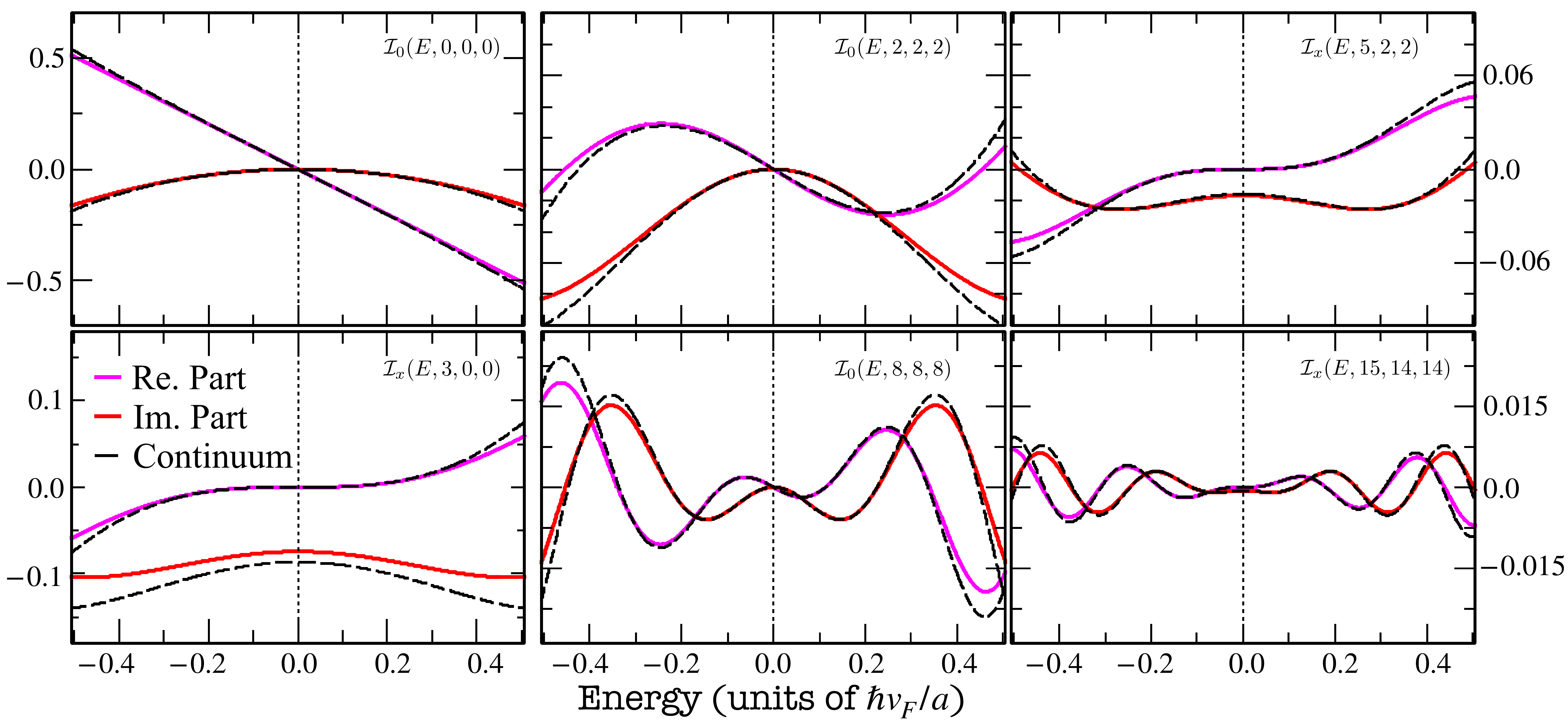}
\par\end{centering}
\vspace{-0.2cm}

\caption{\label{fig:LatticeGFvsCont}Plots of the constitutive integrals of
the lGF, as defined in Eqs.\,\eqref{eq:I0}-\eqref{eq:Ij}, calculated
as a function of energy for different $\boldsymbol{\Delta L}$ in
the lattice. The black dashed lines represent the continuum Weyl cone
approximation, corrected to account for the eight-fold degeneracy
of this lattice model. For the continuum model, the \textit{\textquotedblleft smooth
cut-off}\textquotedblright{} regularizer was chosen to be $M\!=\!1.588a^{-1}$,
in order to have achieve a perfect match with the onsite lGF (upper
left panel).}

\vspace{-0.5cm}
\end{figure}

\vspace{-0.7cm}

\begin{align}
\mathscr{G}_{ab}^{\text{0r}}(\varepsilon;\boldsymbol{\Delta L})\! & =\!\frac{a}{8\pi^{3}\hbar v_{\text{F}}}\int_{\text{C}}\!\!\!d\mathbf{q}\,\frac{\varepsilon-\boldsymbol{\sigma}\cdot\boldsymbol{\sin}\mathbf{q}}{\left(\varepsilon\!+\!i\eta\right)^{2}\!-\!\abs{\boldsymbol{\sin}\mathbf{q}}^{2}}e^{i\mathbf{q}\cdot\boldsymbol{\Delta L}},\label{eq:RealSpacePropagator-2-1}
\end{align}
where the integral is now over the 3D cube, $\text{C}\!=\![-\pi,\pi]^{3}$.
The expression of Eq.\,\eqref{eq:RealSpacePropagator-2-1} can be
suitably divided into two fundamental parts, which must be treated
independently, \textit{i.e.},

\vspace{-0.7cm}

\begin{align}
\mathscr{G}_{ab}^{\text{0r}}(\varepsilon;\boldsymbol{\Delta L})\! & =\!\frac{a\varepsilon}{8\pi^{3}\hbar v_{\text{F}}}\left(\int_{\text{C}}\!\!\!d\mathbf{q}\,\frac{\delta_{ab}e^{i\mathbf{q}\cdot\boldsymbol{\Delta L}}}{\left(\varepsilon\!+\!i\eta\right)^{2}\!-\!\abs{\boldsymbol{\sin}\mathbf{q}}^{2}}\right)\label{eq:RealSpacePropagator-2-1-1}\\
 & \qquad\qquad\qquad-\boldsymbol{\sigma}_{ab}\cdot\left(\frac{a}{8\pi^{3}\hbar v_{\text{F}}}\int_{\text{C}}\!\!\!d\mathbf{q}\,\frac{\boldsymbol{\sin}\mathbf{q}e^{i\mathbf{q}\cdot\boldsymbol{\Delta L}}}{\left(\varepsilon\!+\!i\eta\right)^{2}\!-\!\abs{\boldsymbol{\sin}\mathbf{q}}^{2}}\right),\nonumber 
\end{align}
or, equivalently,

\vspace{-0.7cm}
\begin{equation}
\mathscr{G}_{ab}^{\text{0r}}(\varepsilon;\boldsymbol{\Delta L})\!=\!\frac{a}{8\pi^{3}}\left[\varepsilon\,\delta_{ab}\,\mathcal{I}_{0}\left(\varepsilon;\boldsymbol{\Delta L}\right)-\boldsymbol{\sigma}_{ab}\cdot\boldsymbol{\mathcal{I}}\left(\varepsilon;\boldsymbol{\Delta L}\right)\right].\label{eq:DecompositionlGF}
\end{equation}
In the form of Eq.\,\eqref{eq:DecompositionlGF}, it becomes evident
that the calculation of the lGF for this model reduces to the calculation
of two constitutive (dimensionless) triple-integrals, 

\vspace{-0.7cm}

\begin{subequations}
\begin{align}
\mathcal{I}_{0}\left(\varepsilon;\boldsymbol{\Delta L}\right) & =\int_{\text{C}}\!\!\!d\mathbf{q}\,\frac{e^{i\mathbf{q}\cdot\boldsymbol{\Delta L}}}{\left(\varepsilon\!+\!i\eta\right)^{2}\!-\!\abs{\boldsymbol{\sin}\mathbf{q}}^{2}}\label{eq:I0}\\
\mathcal{I}_{j}\left(\varepsilon;\boldsymbol{\Delta L}\right) & =\int_{\text{C}}\!\!\!d\mathbf{q}\,\frac{\sin q_{j}e^{i\mathbf{q}\cdot\boldsymbol{\Delta L}}}{\left(\varepsilon\!+\!i\eta\right)^{2}\!-\!\abs{\boldsymbol{\sin}\mathbf{q}}^{2}}\label{eq:Ij}
\end{align}
\end{subequations}

which are to be taken in the limit $\eta\to0^{+}$. Note that, in
a simple cubic lattice, the vector $\boldsymbol{\Delta L}=(n_{x},n_{y},n_{z})$
is composed by a set of integer numbers and, due to cubic symmetry,
the integral of Eq.\,\eqref{eq:Ij} does not depend on the index
$j$. In Appendix\,\ref{chap:LatticeGF}, we detail a semi-analytic
procedure by which one of the $\mathbf{q}$-integrals in both Eqs.\,\eqref{eq:I0}
and \eqref{eq:Ij} can be performed analytically with the $\eta\to0^{+}$
limit taken formally. By numerically performing the two remaining
integrals, we are able to evaluate the lGF of the model to an \textit{arbitrary
energy resolution}. In Fig.\,\ref{fig:LatticeGFvsCont}, we show
these results and compare them to the continuum limit near the nodal
energy. As can be seen, the two models agree in the limit $\abs{\varepsilon}\!\ll\!\hbar v_{\text{F}}/a$
and $\boldsymbol{\Delta L}\gg1$, as they should. 

\vspace{-0.4cm}

\paragraph{Small-Scale Properties of the lGF:}

Before using the calculated lattice Green's function within the pGF
formalism to study atomic-sized impurities in this lattice model,
it is important and useful to comment on some special featured of
$\mathscr{G}^{\text{0r}}(\varepsilon;\boldsymbol{\Delta L})$ that
happen at the scale of the lattice spacing. More precisely, we can
show that given a lattice displacement $\boldsymbol{\Delta R}\!=\!(n_{x}a,n_{y}a,n_{z}a)$,
the constitutive integrals are \textit{(i) }all zero if two or three
integers are odd, \textit{(ii)} only $\mathcal{I}_{0}(E)$ is nonzero
if all integers are even and, finally \textit{(iii)} $\mathcal{I}_{j}(E)\neq0$
if and only if $n_{j}$ is odd, with the other two being even. This
weird property is clearly a feature of the discretized model\,\footnote{One can show that this actually comes from the destructive interference
of excitations from Weyl nodes at different TRIM (see Appendix\,\ref{chap:MultiValleyCont}
for further details).} and does not survive any coarse-graining procedure that leads to
a continuum limit. Nevertheless, one must be aware of this non-universal
property when making calculations in the lattice model, \textit{e.g.},
two atomic-sized impurities in the lattice model may be uncoupled
due to a vanishing lGF between their positions.

\vspace{-0.4cm}

\paragraph{The Case of Zero Energy:}

Besides the aforementioned features at the scale of the lattice spacing,
the lGF also features important symmetries referring to the reversal
of sign in the energy. Concretely, we have

\vspace{-0.95cm}

\begin{subequations}
\begin{align}
\mathcal{I}_{0}\left(-\varepsilon;\boldsymbol{\Delta L}\right) & =\left[\mathcal{I}_{0}\left(\varepsilon;\boldsymbol{\Delta L}\right)\right]^{*}\label{eq:I0-1}\\
\mathcal{I}_{j}\left(-\varepsilon;\boldsymbol{\Delta L}\right) & =-\left[\mathcal{I}_{0}\left(\varepsilon;\boldsymbol{\Delta L}\right)\right]^{*},\label{eq:Ij-1}
\end{align}
\end{subequations}

\vspace{-0.4cm}

where we employed the transformation $\mathbf{q}\to-\mathbf{q}$ on
the integrals of Eqs.\,\eqref{eq:I0}-\eqref{eq:Ij}. These symmetries
imply that lGF obeys the following relation,

\vspace{-0.85cm}

\begin{equation}
\mathscr{G}_{ab}^{\text{0r}}(-\varepsilon;\boldsymbol{\Delta L})\!=-\frac{a}{8\pi^{3}}\left[\varepsilon\,\delta_{ab}\,\left[\mathcal{I}_{0}\left(\varepsilon;\boldsymbol{\Delta L}\right)\right]^{*}-\boldsymbol{\sigma}_{ab}\cdot\left[\boldsymbol{\mathcal{I}}\left(\varepsilon;\boldsymbol{\Delta L}\right)\right]^{*}\right]=-\left[\mathscr{G}_{ba}^{\text{0r}}(\varepsilon;\boldsymbol{\Delta L})\right]^{*}.\label{eq:DecompositionlGF-1}
\end{equation}
At zero energy, $\varepsilon\!=\!0$, Eq.\,\eqref{eq:DecompositionlGF-1}
demonstrates that only the second term in the lGF survives, that is,

\vspace{-0.8cm}
\begin{equation}
\mathscr{G}_{ab}^{\text{0r}}(0;\boldsymbol{\Delta L})=-\frac{a}{8\pi^{3}}\boldsymbol{\sigma}_{ab}\cdot\Im\left[\boldsymbol{\mathcal{I}}\left(0;\boldsymbol{\Delta L}\right)\right],
\end{equation}

where the fact that $\boldsymbol{\mathcal{I}}\left(0;\boldsymbol{\Delta L}\right)$
is purely imaginary was also used.

\subsection{\label{subsec:PointsinLattice}The Single- and Two-Impurity Problem
in the Lattice}

After calculating the lGF between any two positions, we can now retrace
the whole study of Subsect.\,\ref{subsec:PointLikeCont}, made for
$\delta$-impurities in the continuum model, and now check how it
translates to a lattice model that has a \textit{built-in ultraviolet
scale} ($a^{-1}$) and several Weyl nodes are coupled by point-like
perturbations in real-space. 
\begin{figure}[t]
\vspace{-0.5cm}
\begin{centering}
\hspace{-0.5cm}\includegraphics[scale=0.23]{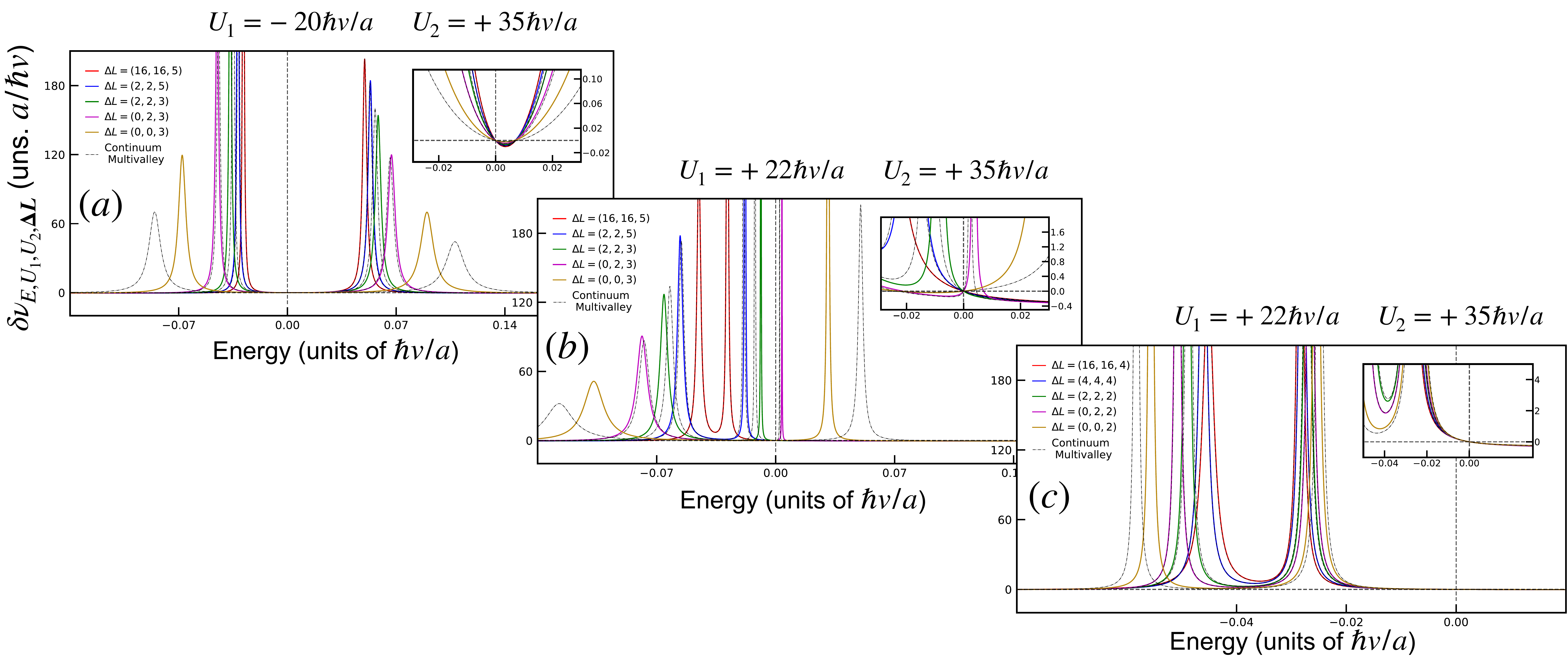}
\par\end{centering}
\vspace{-0.3cm}

\caption{\label{fig:AtomicImps}Plots of the changes induced in the eDos by
a pair of atomic-sized impurities, with onsite potentials $U_{1}$
and $U_{2}$ and distanced by a lattice vector $\boldsymbol{\Delta L}=\left(\Delta n_{x},\Delta n_{y},\Delta n_{z}\right)$.
We consider the case in which the two impurities have $U_{1}U_{2}\!<\!0$
(a) and the case in which $U_{1}U_{2}>0$ (b)-(c). Panel (c) considers
cases in which $\Delta n_{x,y,z}$ are all even integers which, in
(b) one of them is odd.}

\vspace{-0.3cm}
\end{figure}

Apart from small-scale peculiarities which are inherited from the
properties of the clean lGF, the conclusions obtained by considering
one or two atomic-sized impurities are essentially the same as what
we have seen for the continuum model. Namely, \textit{(i)} an isolated
lattice site with a scalar potential $U$ does not generate nodal
bound states, giving rise to sharp resonances that get closer to the
nodal energy as $U\!\to\!\pm\infty$, and \textit{(ii)} a pair of
impurities (distanced by $\boldsymbol{\Delta L}$ in the lattice and
having potentials $U_{1}$/$U_{2}$) can create a nodal bound state
by means of coherent scattering between them. Three representative
examples of the eDoS change caused by a pair of atomic-sized impurities
in the lattice are presented in Fig.\,\ref{fig:AtomicImps}.

In Fig.\,\ref{fig:AtomicImps}a, we show the calculation done for
two points with local potentials that have the opposite signs. Just
like the continuum $\delta$-impurities, effects of hybridization
act to further split the resonances and drive them farther from the
nodal energy. In Figs.\ref{fig:AtomicImps}b and c, an analogous calculation
is presented but now with impurities that have $U_{1}U_{2}\!>\!0$.
In this case, we find two seemingly different situations; If $\boldsymbol{\Delta L}=(\Delta n_{x},\Delta n_{y},\Delta n_{z})$
has only even coordinates, the two resonances hybridize only slightly
at short distances but there is not peak being driven towards the
nodal energy. In contrast, if one $\Delta n_{i}$ is odd, the two
resonances hybridize strongly such that one of the resonances is driven
through $\varepsilon\!=\!0$. The difference between the two cases
lies on the fact that only the latter case has a nonzero prefactor
for the term $\propto\!\boldsymbol{\sigma}$. Therefore, we conclude
that the latter term is the one ultimately responsible for creating
nodal bound states through inter-vacancy interference. Finally, it
is worth noting two further points: \textit{(i)} that, in all the
cases, the impurities \textit{do not hybridize} if $\boldsymbol{\Delta L}$
has two or three odd coordinates, and \textit{(ii)} that the $\abs{\boldsymbol{\Delta L}}\!\to\!\infty$
limit of this exact lattice calculation always coincides the analogous
calculation within the multi-valley continuum approximation derived
in Appendix \ref{chap:MultiValleyCont}. 
\begin{figure}[t]
\vspace{-0.8cm}
\begin{centering}
\hspace{-0.3cm}\includegraphics[scale=0.22]{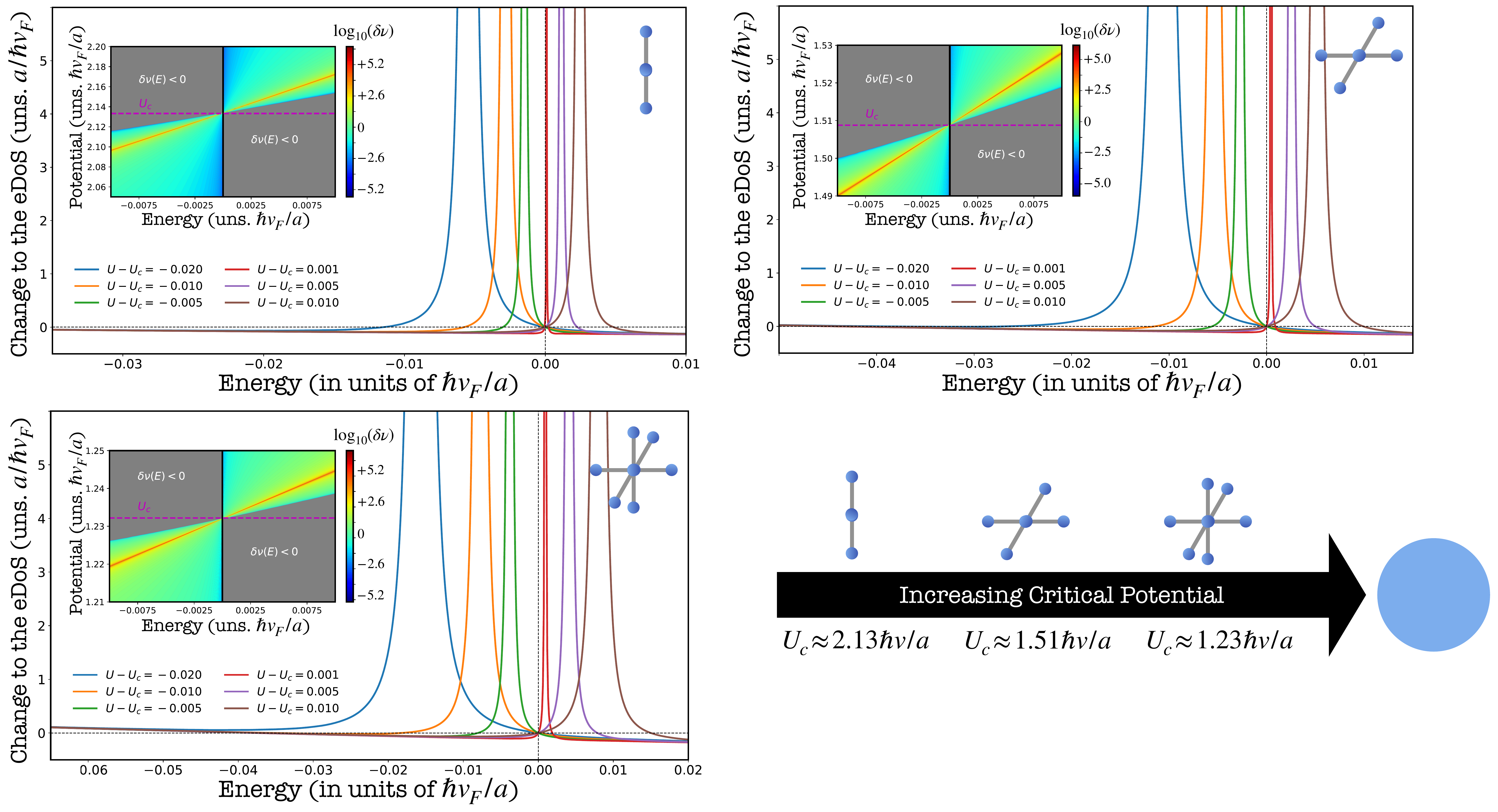}
\par\end{centering}
\vspace{-0.5cm}

\caption{\label{fig:MinimalClusters}Analysis of the near-critical mechanism
for small clusters of adjacent points in a cubic WSM lattice, with
a common potential local $U\!>\!0$. Three small clusters are presented:
a line of three sites (upper left), a cross of sites in the $xOy$-plane
(upper right), and a three-dimensional cross (lower left). In all
three cases, a bird's eye view of $\delta\nu(E,U)$ is presented as
an inset, while the behavior of the corresponding resonance as it
crosses the nodal energy is shown in the main plots. In the lower
right, we present a scheme that illustrates the evolution of the critical
value of $U$, $U_{c}$, as the size of the cluster is increased.}

\vspace{-0.4cm}
\end{figure}

\vspace{-0.4cm}

\subsection{\label{subsec:Minimal-Clusters}Minimal Clusters for Nodal Bound
States}

The same analysis that was done for two points in the lattice WSM
can also be repeated for an arbitrarily complex cluster, with each
point having its own local potential. From the discussion of Sect.\,\ref{sec:DeltaImpurities},
we are already aware that increasing the complexity of the cluster
of point-like impurities entails a very unfavorable scaling of the
method, as a larger matrix is required to be numerically built and
inverted. Therefore, we will focus on the simple case of very small
clusters of adjacent points in the lattice, as shown in the scheme
of Fig.\,\ref{fig:MinimalClusters}.

Our main point here is to push the theory of smooth regions, as presented
in Chapter\,\ref{chap:Instability_Smooth_Regions} to its small-size
limit, \textit{i.e.}, examine the possibility for a small cluster
of adjacent site with the exact same local potential, $U$, to give
rise to nodal bound states. In the plots of Fig.\,\ref{fig:MinimalClusters},
we present results for the change in the eDoS as a function of $U$
for three types of clusters: a line of three sites, a planar cross
along the $xOy$-plane, and a three-dimensional cross. In all the
cases we see there is a sharp DoS resonance that is driven through
the nodal energy as $U$ is increased\,\footnote{We have considered only positive potentials, with no loss of generality.}.
When such crossing happens, at a critical value $U_{c}$, there is
a nodal bound state that forms akin the ones that were observed for
the spherical scatterers in the previous Chapter. In fact, the latter
will be obtained as the limit of these results when the radius of
the cluster considered becomes sufficiently large. An important feature
of these critical values is that they become smaller as the cluster
includes a larger number of sites, as illustrated in the scheme of
Fig.\,\ref{fig:MinimalClusters}.

\vspace{-0.5cm}

\section{\label{sec:What-are-RareEvents}What are Rare-Regions in a Disordered
Lattice?}

Now, we are ready to establish a connection between the mesoscopic
view of nodal bound states, undertaken in previous sections, and the
rare-region-induced effects that were numerically found by Pixley
\textit{et al}.\,\cite{Pixley16a}, leading to an AQC scenario for
the mean-field transition in disordered WSMs. For that purpose, we
take a step back and carefully analyze what these non-perturbative
effects really are in the context of a disordered lattice. To do this,
we numerically study the nodal eigenstates of disordered WSM lattices
(of side dimension $L$), using the \textit{implicitly restarted Lanczos
diagonalization algorithm}\,\cite{Lanczos50,paige1980}, as implemented
in $\texttt{ARPACK}$\,\cite{Lehoucq97}. In order to separate any
putative nodal bound state, from the extended Bloch-like states, we
employ \textit{twisted boundary conditions} (with fixed twist of $\nicefrac{\pi}{3}$rad
in all directions) to open up a finite-size gap, $\Delta_{\text{f}}\!\propto\!1/L$,
inside of which the rare-region-induced states will lie. We recall
that this segregation is due to a greater degree of localization expected
for the rare-region-induced states which, as pointed out in Ref.\,\cite{Pixley16a},
makes them much \textit{more insensitive to the boundary conditions
}than the extended ones. Finally, we remark that a similar analysis
was undertaken by Pixley and Wilson in Ref.\,\cite{Pixley21}

\begin{figure}[t]
\vspace{-0.5cm}
\begin{centering}
\includegraphics[scale=0.22]{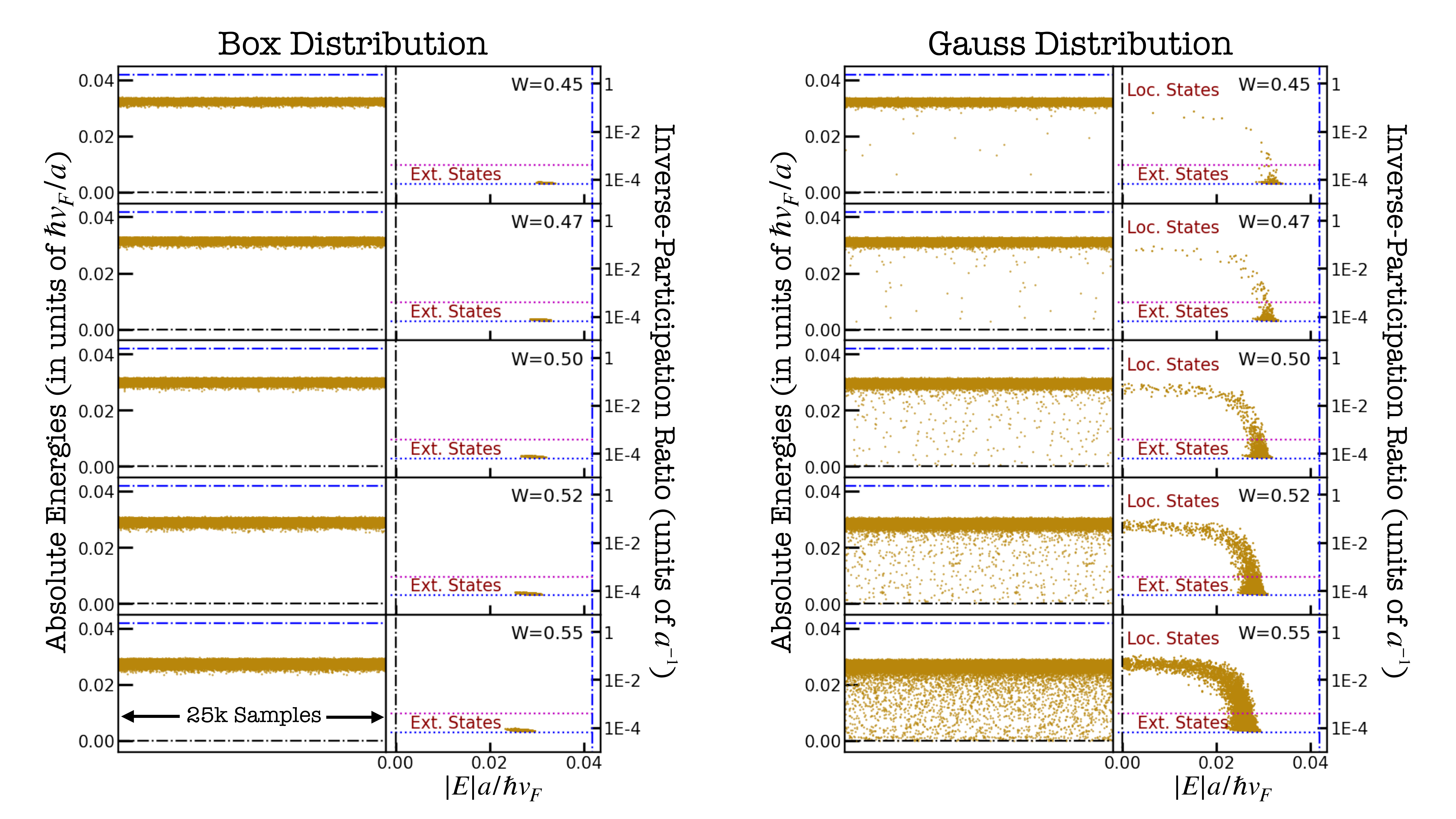}
\par\end{centering}
\vspace{-0.3cm}

\caption{\label{fig:RareEventesGauss_vs_Unif}Rare-region-induced states generated
in $25000$ random samples of a Weyl semimetal lattice of side $L\!=\!25a$,
having an uncorrelated scalar disordered potential with local energies
drawn from a BD of width $W$ (left panel) and a GD of standard deviation
$2\sqrt{3}W$ (right panel). The narrower columns on the right plot
the inverse-participation ratios for the obtained eigenstates, which
allows to distinguish between localized and extended states.}

\vspace{-0.5cm}
\end{figure}

\begin{figure}[t]
\vspace{-0.6cm}
\begin{centering}
\includegraphics[scale=0.23]{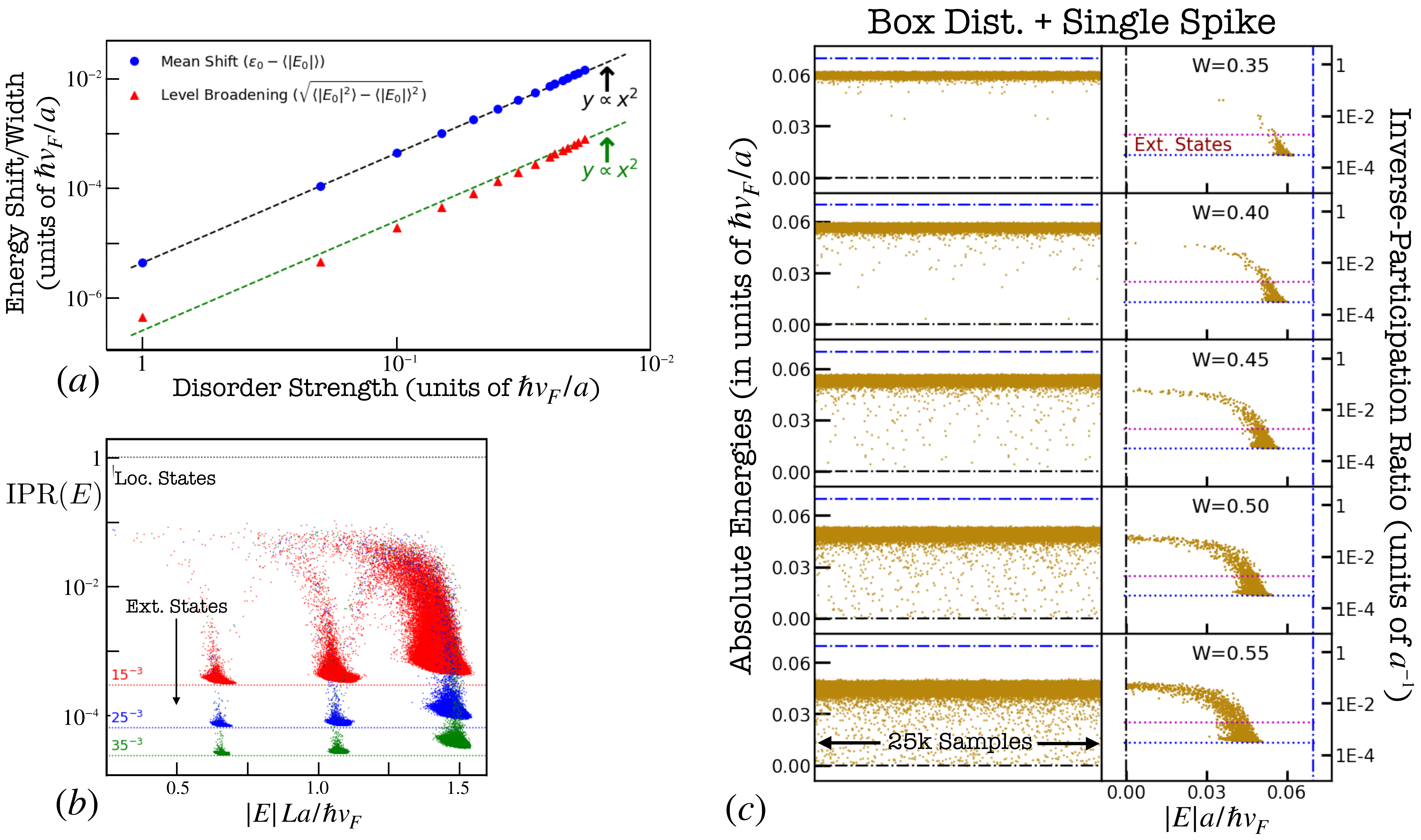}
\par\end{centering}
\vspace{-0.3cm}

\caption{\label{fig:RareEventesSingleFluct} (a) Perturbative energy broadening
of the energy level by disorder. (b) Scaling of the IPRs with $L$
{[}$L=15$ (red), $L=25$ (blue), and $L=35$(green){]}. (c) Rare-Event
nodal states generated in $25000$ random samples of a WSM lattice
of side $L\!=\!15a$, that hosts a Anderson BD disorder of strength
$W$ together with a point-like large fluctuation of $U\!=\!2\hbar v_{\text{F}}/a$
placed at the center of the simulated lattice.}

\vspace{-0.6cm}
\end{figure}

In Fig.\,\ref{fig:RareEventesGauss_vs_Unif}, we present results
from diagonalizing $25000$ independent samples of the (squared) disordered
Hamiltonian, $\mathcal{H}_{l}^{2}$. For reasons that will become
clear, we considered two different models of uncorrelated scalar disorder
that differ solely on their local potential distributions:

\vspace{-0.7cm}
\begin{equation}
P_{\!\!{\scriptscriptstyle \text{BD}}}(V)\!=\!\frac{1}{W}\Theta_{H}\left(\abs V-\frac{W}{2}\right)\text{ and }P_{\!\!{\scriptscriptstyle \text{GD}}}(V)\!=\!\frac{\exp\left[-\frac{V^{2}}{24W^{2}}\right]}{2\sqrt{6\pi}W},
\end{equation}
\textit{i.e.}, a \textit{box-distribution} (BD\nomenclature{BD}{Box-Distribution})
and a \textit{gaussian distribution} (GD\nomenclature{GD}{Gaussian Distribution}).
Note that, in both models, the parameter $W$ measures the strength
of disorder in a comparable way, that is, for a given $W$ both local
distributions have the\textit{ same standard deviation}. For the two
models, and several values of $W$\,\footnote{All chosen below the mean-field critical disorder.},
we plot in Fig.\,\ref{fig:RareEventesGauss_vs_Unif} the four lowest
(absolute) eigenvalues of all samples, in succession. Simultaneously,
we also show (in the left panels) a scatter plot of the corresponding
\textit{Inverse-Participation Ratios} (IPRs)\,\cite{Lee85,Hikami86,Kramer93}
of each determined wavefunction, \textit{i.e.},

\vspace{-0.7cm}

\begin{equation}
\text{IPR}_{\Psi}^{L}\!=\frac{\sum_{\mathbf{R}}\left(\abs{\Psi_{\mathbf{R}}^{1}}^{2}\!\!+\!\abs{\Psi_{\mathbf{R}}^{2}}^{2}\right)^{2}}{\left(\sum_{\mathbf{R}}\abs{\Psi_{\mathbf{R}}^{1}}^{2}\!\!+\!\abs{\Psi_{\mathbf{R}}^{2}}^{2}\right)^{2}},\label{eq:IPR}
\end{equation}
which characterizes the degree of locality of a quantum state in real-space
thus allowing one to correlate the eigenenergies with the localization
of the corresponding eigenstates. This quantity is perhaps the simplest
way to distinguish between localized and delocalized quantum states:
one expects an extended three-dimensional Bloch-like wavefunction
to have $\text{IPR}_{\Psi}^{L}\underset{L\to\infty}{\longrightarrow}L^{{\scriptscriptstyle -3}}$,
while $\text{IPR}_{\Psi}^{L}\underset{L\to\infty}{\longrightarrow}\text{constant}$
for a localized one. Notice that the scaling with $L$ which is presented
in Fig.\,\ref{fig:RareEventesSingleFluct}b for the data obtained
with the GD model further confirms the interpretation of the rare-region-induced
states as localized (or quasi-localized), while the ones belonging
to the energy strips are likely extended.

For the BD disorder model, we observe that the only effect of disorder
is to randomize the clean energy levels into \textit{``energy strips''}
that get wider ($\propto W^{2}/L^{2}$) and approach the nodal energy
($\propto W^{2}/L$) as $W$ increases (scalings are demonstrated
in Fig.\,\ref{fig:RareEventesSingleFluct}\,a). Actually, this is
the expectation obtained from the $2^{\text{th}}$-\,order perturbation
theory presented in Appendix \ref{chap:Perturbative-Disorder}, and
the values of the IPRs further confirm that all these states are indeed
extended. In great contrast, the GD model generates a behavior that
is qualitatively very different; Besides having the same \textit{energy
strips} of Bloch-like states that behave identically to the BD case,
there are also a different class of \textit{``rarer states''} with
energies that clearly invade the finite-size gap, $\Delta_{\text{f}}$.
These states, which have IPRs orders of magnitude higher than the
ones in the \textit{energy strips}, are the \textit{rare-region-induced
eigenstates} that create the exponentially small background DoS in
the numerical results of Ref.\,\cite{Pixley16a}. 

\vspace{-0.5cm}

\subsection{The Role of Unbounded Local Distributions}

Recapping the numerical results presented at the end of Chapter\,\ref{chap:Mean-Field-Quantum-Criticality},
we have not seen any signs of AQC in spite of having pushed the limits
of our KPM simulations in both system size and spectral resolution.
The previous analysis of the Anderson BD model makes it clear why
no rare-region-induced effects could be observed: \textit{because,
statistically speaking, there were none!} As a matter of fact, it
had already been pointed out, in Refs.\,\cite{Pixley16a,Pixley21},
that one must consider disorder models with unbounded local distributions
for allowing the rare-region-induced AQC to appear in lattice simulations.
In more direct terms, we can state that, while the mean-field criticality
is driven by the disorder strength parameter, the rare-region effects
depend on large fluctuations on the disordered potential.

One outstanding question that we can legitimately ask is the following:
How many rare fluctuations do we need to generate the rare-region
induced states? To answer this question, we repeat the analysis of
the BD disorder model, but now with an additional fixed isolated potential
spike, of strength $2.0\hbar v_{\text{F}}/a$, that is placed \textit{by
hand} in all the random samples. In Fig.\,\ref{fig:RareEventesSingleFluct}\,c,
we present these results where it becomes clear that high-IPR in-gap
states start to emerge immediately, in way that is very similar to
what was obtained for the GD model. Therefore, the short answer to
our previous question is simply: one. This surprising conclusion contradicts
the picture advocated in Nankishore \textit{et al}.\,\cite{Nandkishore14},
according to which the AQC in a disordered WSM is caused by bound
states of rare smooth regions in the potential landscape. Statistically
speaking, there are roughly as many smooth regions in the GD (or the
BD + fluctuation) disorder model, as there were in the BD Anderson
potential of Fig.\,\ref{fig:RareEventesGauss_vs_Unif}. Notwithstanding,
no rare-event eigenstate could be detected in the latter case. Thereby,
we may conclude that these rare-event states are better interpreted
as being fine-tuned bound states that emerge from the coherent scattering
of two (or maybe more) nearby lattice points, which can have very
\textit{asymmetric local potentials}. This is precisely a realization
of the mesoscopic mechanism unveiled in Subsect.\,\ref{subsec:PointsinLattice},
which singles out the isolated large potential fluctuation as a \textit{``nucleation
center''} for bound eigenstates to appear by interference with the
(much weaker) surrounding disorder.

\vspace{-0.4cm}

\section{A Consistent Picture of AQC in a Lattice Model}

From our study of the nodal eigenstates at the mesoscopic level, we
have pinpointed two main mechanisms by which these can emerge from
an uncorrelated disordered potential. On the one hand, we have a \textit{Two-Point
Asymmetric Interference }(2PAI\nomenclature{2PAI}{Two-Point Asymmetric Interference})
mechanism, by which an isolated large fluctuation of the disordered
landscape hybridizes with one (or more) of its neighboring points
generating states as the ones described in Subsect.\,\ref{subsec:PointsinLattice}.
On the other, we have a \textit{Minimal Smooth Cluster} (MSC) mechanism,
which is akin the rare-region mechanism of Chapter\,\ref{chap:Instability_Smooth_Regions},
but which we have shown to require very minute clusters of nearby
points in a lattice.

Within an arbitrary disordered landscape one expects both the 2PAI
and the MSC mechanisms to generate nodal bound states and, thus, avoid
the mean-field quantum critical point in these systems. Nevertheless,
if the local disorder distribution happens to be bounded, the system
may find itself in a situation where the 2PAI mechanism is actually
absent for weak disorder, as there may be impossible for two nearby
fluctuations to meet the criterium for the emergence of a joint nodal
bound state. In principle, the MSC mechanism is always possible for
arbitrarily weak disorder. However, as the disorder strength gets
decreased, the size of the smooth clusters must grow accordingly in
order to bind nodal electrons. Such large smooth regions are extremely
unlikely for uncorrelated disordered potentials which ultimately explains
why we could not observe any signs of AQC with the BD Anderson disorder.

\global\long\def\vect#1{\overrightarrow{\mathbf{#1}}}%

\global\long\def\abs#1{\left|#1\right|}%

\global\long\def\av#1{\left\langle #1\right\rangle }%

\global\long\def\ket#1{\left|#1\right\rangle }%

\global\long\def\bra#1{\left\langle #1\right|}%

\global\long\def\tensorproduct{\otimes}%

\global\long\def\braket#1#2{\left\langle #1\mid#2\right\rangle }%

\global\long\def\omv{\overrightarrow{\Omega}}%

\global\long\def\inf{\infty}%

\lhead[\MakeUppercase{\chaptername}~\MakeUppercase{\thechapter}]{\MakeUppercase{\rightmark}}

\rhead[\MakeUppercase{Nodal States \& Lattice Vacancies}]{}

\lfoot[\thepage]{}

\cfoot[]{}

\rfoot[]{\thepage}

\chapter{\label{chap:Vacancies}Vacancies in Weyl Semimetals}

In Chapters \ref{chap:Mean-Field-Quantum-Criticality} to \ref{chap:Rare-Event-States},
we have addressed the effects of disorder in the spectral properties
of emergent three-dimensional Weyl fermions near the Fermi level of
a quantum solid. For that, we analyzed three different models of quenched
disorder (uncorrelated random potentials, randomly placed smooth regions,
and scalar point-like impurities) in the context of continuum single-node
models, as well as multi-valley lattice models. While these simple
disorder models yield some interesting (and somewhat surprising) effects,
they \textit{do not} describe all the sources of disorder one finds
in actual crystalline samples that are grown in the laboratory. Usually,
even the most perfect crystals suffer from poor stoichiometry or flaws
in the crystallization procedure\,\cite{Crawford72,Mattox10}, that
give rise to \textit{substitutional disorder} (atomic impurities),
\textit{lattice imperfections} (\textit{e.g.}, missing atoms\,\cite{WOLLENBERGER96}
or deformed unit cells), \textit{inhomogeneous strain}, or other \textit{topological
defects}\,\cite{Mermin79,Hirth91}. Adding to all of these, real
samples will often be \textit{multi-crystalline }and always have some
\textit{vibrational dynamics} due to thermally excited phonon modes\,\cite{Paquet84,Mookerjee_1990}.
All in all, these effects break the lattice translation symmetry,
which will scatter the propagating Bloch electrons, and may significantly
change the electronic properties of the material\,\cite{Girvin80}.

By this point, we will consider another common type of lattice disorder
which is due to be present in real samples:\textit{\,dilute\,lattice\,vacancies}.\,Vacancy-disorder
in a tight-binding Hamiltonian corresponds to a set of randomly chosen
orbitals that got removed from the system\,\footnote{And, therefore, also from the single-electron Hilbert space.}.
Such a procedure attempts to model the effects caused by isolated
atoms that are missing from the crystal\,\cite{Siegel78,WOLLENBERGER96},
but without accounting for the more complicated\textit{ lattice distortions}
and\textit{ charge screening} which are also expected to appear around
the vacant site\,\cite{Massalski96}. Even so, the simplest models
of random vacancies (also called \textit{point-defects}) are able
to generate novel and remarkable physical phenomena that can serve
to explain real experimental results. Graphene is a prime example
of how point-defects can give rise to fundamentally different physics
that goes far beyond what can be reproduced by Anderson random potentials.
There, random vacancies effectively realize a chiral-symmetric (see
Sect.\,\ref{sec:Generic-Symmetries}) disorder model\,\cite{Ostrovski2014}
which originates robust nodal bound states that greatly\textit{ enhance
the density of states} at the charge neutrality point\,\cite{Gade93,Fradkin86b},
and lead to a finite \textit{universal dc conductivity}\,\cite{Ferreira2015,Joao2021}.
Regarding three-dimensional semimetals, the influence of vacancies
have been overlooked so far in theoretical literature, even though
these are common sub-products of the synthesis process\,\cite{Buckeridge2016}
and have even been experimentally observed by \textit{Transmission
Electron Microscopy} (TEM) in stoichiometric Weyl semimetals\,\cite{Besara2016}.
Additionally, the concentration of vacancies may also be externally
controlled by means of light-ion irradiation of a crystalline sample\,\cite{Zhang2019}.
While being still an unexplored territory, the study of vacancy effects
in these systems has recently received a motivational boost by the
experimental results of Xing \textit{et al}.\,\cite{Xing2020}, that
established a link between surface vacancies in the magnetic WSM,
$\text{Co}_{3}\text{Sn}_{2}\text{S}_{2}$, and the existence of exotic
\textit{localized Spin-Orbit Polarons}. In this Chapter, we will present
novel results on the effects of diluted vacancies in a cubic Weyl
semimetal focusing, not only on the spectral effects, but also in
predicting observable signatures in dc transport, magnetic response
and optical properties. This original study is published in Refs.\,\cite{Pires2022a}
and \cite{Pires2022b}.

\vspace{-0.5cm}

\section{Modeling a Vacancy in a Weyl Semimetal}

Before analyzing the case of a finite concentration of vacancies in
a macroscopic sample, we begin by considering the simpler problem
of an isolated vacancy in the tight-binding model of a cubic WSM,
which was first introduced in Eq.\,\eqref{eq:LatticeWSM_Cubic}.
Our clean Hamiltonian reads 

\vspace{-0.7cm}

\begin{align}
\mathcal{H}_{l}^{0} & \!=\!\frac{i\hbar v_{\text{F}}}{2a}\sum_{{\scriptscriptstyle \mathbf{R}\in\mathcal{L_{\text{C}}}}}\negthickspace\left[\Psi_{\mathbf{{\scriptscriptstyle R}}}^{\dagger}\!\cdot\!\sigma^{j}\!\cdot\!\Psi_{{\scriptscriptstyle \mathbf{R}+a\mathbf{x}_{\!j}}}+\Psi_{{\scriptscriptstyle \mathbf{R}+a\mathbf{x}_{\!j}}}^{\dagger}\!\cdot\!\sigma^{j}\!\cdot\!\Psi_{{\scriptscriptstyle \mathbf{R}}}\right]\!,\label{eq:LatticeModel-1-1}
\end{align}
where $\mathcal{L}_{\text{C}}$ is a simple cubic lattice of parameter
$a$, $\mathbf{x}_{j}$ are the cartesian unit vectors, $\Psi_{\mathbf{{\scriptscriptstyle R}}}^{\dagger}/\Psi_{\mathbf{{\scriptscriptstyle R}}}$
are local two-component fermionic creation/annihilation operators,
and $\sigma^{j}$ are $2\!\times\!2$ Pauli matrices. In this model,
a vacancy defect is a missing lattice site, which can happen in two
ways within the two-band Hamiltonian of Eq.\,\eqref{eq:LatticeModel-1-1}:
\textit{(i) }by removing all atoms in a unit cell (a \textit{full-vacancy}),
or \textit{(ii)} by removing only a single orbital (a \textit{half-vacancy}).
In a bipartite lattice, like a graphene monolayer, this choice is
clearly important due to the sublattice symmetry, which is broken
by a concentration imbalance between the vacancies in each sublattice\,\cite{Ostrovski2014}.
However, here the situation is less symmetric because $\mathcal{H}_{l}^{0}$
does not have a sublattice symmetry to begin with\,\footnote{A 3D Weyl semimetal is not chiral symmetric, which is obvious from
the matrix form of the corresponding Bloch Hamiltonian.}. Therefore, we shall only treat the case of a full-vacancy, keeping
in mind that no new phenomena would arise if half-vacancies were considered
instead\,\footnote{This was explicitly verified in Santos Pires \textit{et al}.\,\cite{Pires2022b},
and can be easily obtained from the same pGF method that will presented
in the remainder of this section.}. Under this framework, a single full-vacancy is implemented by removing
the hoppings connecting both orbitals within a unit cell to the remaining
lattice. This procedure actually leaves behind the \textit{disconnected
site}, something that we must take into account when looking at spectral
properties. With no loss of generality, we start by considering the
vacancy as placed at the origin ($\mathbf{R}\!=\!\boldsymbol{0}$)
and, thereby, add to $\mathcal{H}_{l}^{0}$ a term that precisely
cancels all hoppings to this site, \textit{i.e.},

\vspace{-0.8cm}
\begin{align}
\mathcal{V}_{\!\text{v}}\! & =\!\frac{i\hbar v_{\text{F}}}{2a}\left[\Psi_{\boldsymbol{0}}^{\dagger}\!\cdot\!\sigma^{i}\!\!\cdot\!\Psi_{\mathbf{x}_{\!j}}\!\!-\!\Psi_{\boldsymbol{0}}^{\dagger}\!\cdot\!\sigma^{i}\!\!\cdot\!\Psi_{-\mathbf{x}_{\!j}}\!-\text{h.c.}\right].\label{eq:Perturbation}
\end{align}
\begin{wrapfigure}[11]{o}{0.42\columnwidth}%
\includegraphics[scale=0.16]{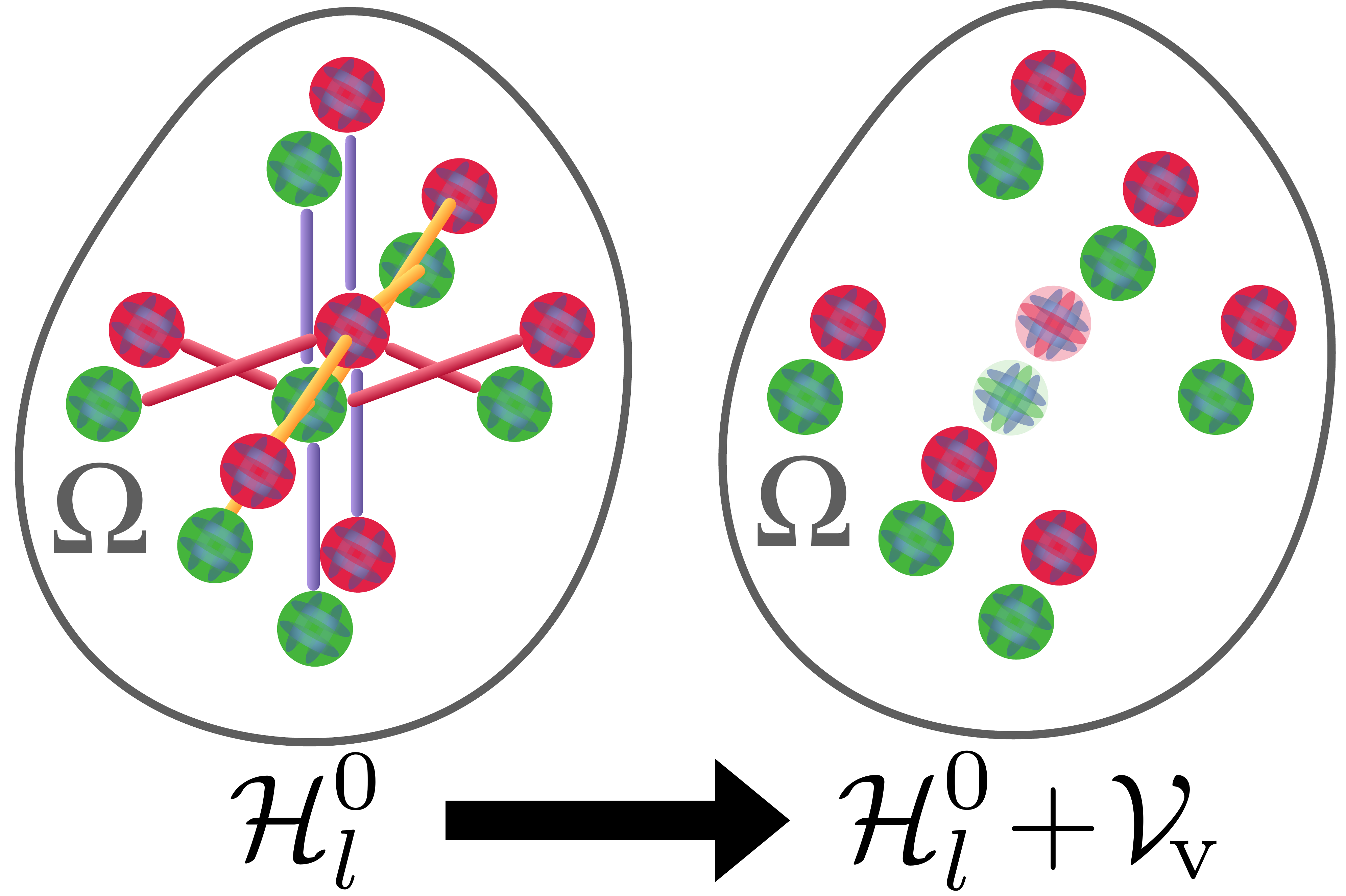}

\vspace{-0.3cm}

\caption{\label{fig:Cluster_Vacancy}Scheme of the local perturbation defined
in Eq.\,\eqref{eq:Perturbation}.}
\end{wrapfigure}%
Note that the perturbation $\mathcal{V}_{\!\text{v}}$ has the advantage
of acting only on the finite support $\Omega\!=\!\left\{ \mathbf{0},\pm a\hat{\mathbf{x}}_{1},\pm a\hat{\mathbf{x}}_{2},\pm a\hat{\mathbf{x}}_{3}\right\} $,
which forms the octahedron depicted in Fig.\,\ref{fig:Cluster_Vacancy}.
Such a local perturbation to a lattice model can be conveniently studied
by using the whole machinery of the \textit{projected Green's Function}
(pGF) method introduced in Sect.\,\ref{sec:DeltaImpurities}. To
more easily apply the pGF formalism, it is convenient to encode the
sites of $\Omega$ in the order, $\left\{ \boldsymbol{0},a\hat{x},a\hat{y},a\hat{z},-a\hat{x},-a\hat{y},-a\hat{z}\right\} $,
such that the single vacancy perturbation looks like,
\begin{equation}
\mathcal{V}_{\!\text{v}}\!=\!\frac{i\hbar v_{\text{F}}}{2a}\left[\begin{array}{ccc}
\mathbb{O}_{{\scriptscriptstyle 2\!\times\!2}} & \boldsymbol{\sigma} & \!\!\!-\boldsymbol{\sigma}\\
\!\!\!-\boldsymbol{\sigma}^{T} & \mathbb{O}_{{\scriptscriptstyle 6\!\times\!6}} & \mathbb{O}_{{\scriptscriptstyle 6\!\times\!6}}\\
\boldsymbol{\sigma}^{T} & \mathbb{O}_{{\scriptscriptstyle 6\!\times\!6}} & \mathbb{O}_{{\scriptscriptstyle 6\!\times\!6}}
\end{array}\right],\label{eq:Vacancies}
\end{equation}
\textit{i.e}, a simple $14\!\times\!14$ matrix in the projected subspace.
In Eq.\,\eqref{eq:Vacancies} (and following), we define $\mathbb{O}_{{\scriptscriptstyle n\!\times\!n}}$
as an $n\!\times\!n$ zero matrix, $\mathbb{I}_{{\scriptscriptstyle n\!\times\!n}}$
as an $n\!\times\!n$ identity matrix, and $\boldsymbol{\sigma}\!=\!(\sigma_{x},\sigma_{y},\sigma_{z})$
as a vector of Pauli matrices. At the same time, using the \textit{short-distance
properties} of the lGF and its symmetry around the nodal energy (derived
in Subsect.\,\ref{subsec:The-Lattice-Green's}), we also see that
the projected lGF reads simply as

\vspace{-0.7cm}

\begin{equation}
\underline{\mathcal{G}_{c}^{0}\left(E\right)}\!=\!\frac{a}{\hbar v_{\text{F}}}\left[\,\,\begin{array}{ccc}
f_{\varepsilon}\mathbb{I}_{{\scriptscriptstyle 2\!\times\!2}} & g_{\varepsilon}\boldsymbol{\sigma} & \!\!\!\!-g_{\varepsilon}\boldsymbol{\sigma}\\
\!\!\!\!-g_{\varepsilon}\boldsymbol{\sigma}^{T} & \,\,f_{\varepsilon}\mathbb{I}_{{\scriptscriptstyle 6\!\times\!6}} & h_{\varepsilon}\mathbb{I}_{{\scriptscriptstyle 6\!\times\!6}}\\
g_{\varepsilon}\boldsymbol{\sigma}^{T} & \,\,h_{\varepsilon}\mathbb{I}_{{\scriptscriptstyle 6\!\times\!6}} & f_{\varepsilon}\mathbb{I}_{{\scriptscriptstyle 6\!\times\!6}}
\end{array}\,\right],\label{eq:pGF_VacancyE}
\end{equation}
where $g_{\varepsilon}$, $f_{\varepsilon}$ and $h_{\varepsilon}$
are dimensionless functions of energy, that are determined from the
integrals in Eqs.\,\eqref{eq:I0}-\eqref{eq:Ij}. To be more precise,
we have

\vspace{-0.7cm}

\begin{subequations}
\begin{align}
f_{\varepsilon} & =\frac{a\varepsilon}{8\pi^{3}}\mathcal{I}_{0}\left(\varepsilon;\boldsymbol{\Delta L}\!=\!\left(0,0,0\right)\right)\!=\frac{a\varepsilon}{8\pi^{3}}\!\int_{\text{C}}\!\!\!d\mathbf{q}\,\frac{1}{\left(\varepsilon\!+\!i\eta\right)^{2}\!-\!\abs{\boldsymbol{\sin}\mathbf{q}}^{2}}\label{eq:feps}\\
g_{\varepsilon} & =\frac{a}{8\pi^{3}}\mathcal{I}_{x}\left(\varepsilon;\boldsymbol{\Delta L}\!=\!\left(1,0,0\right)\right)\!=\frac{a}{8\pi^{3}}\!\int_{\text{C}}\!\!\!d\mathbf{q}\,\frac{\sin q_{x}e^{iq_{x}}}{\left(\varepsilon\!+\!i\eta\right)^{2}\!-\!\abs{\boldsymbol{\sin}\mathbf{q}}^{2}}\\
h_{\varepsilon} & =\frac{a\varepsilon}{8\pi^{3}}\mathcal{I}_{0}\left(\varepsilon;\boldsymbol{\Delta L}\!=\!\left(2,0,0\right)\right)\!=\frac{a\varepsilon}{8\pi^{3}}\!\int_{\text{C}}\!\!\!d\mathbf{q}\,\frac{e^{2iq_{x}}}{\left(\varepsilon\!+\!i\eta\right)^{2}\!-\!\abs{\boldsymbol{\sin}\mathbf{q}}^{2}},\label{eq:heps}
\end{align}
\end{subequations}

whose integrals, over the cube $C=[-\pi,\pi]^{3}$, can be performed
semi-analytically in the limit $\eta\to0^{+}$ (as described in Appendix\,\ref{chap:LatticeGF}).
It is also relevant to emphasize that $\varepsilon\!=\!0$ is a special
point, in which the clean pGF takes on the particularly simple form,

\vspace{-0.7cm}

\begin{equation}
\underline{\mathcal{G}_{c}^{0}\left(0\right)}\!=\!\frac{ag_{0}}{\hbar v_{\text{F}}}\!\!\left[\,\,\begin{array}{ccc}
\mathbb{O}_{{\scriptscriptstyle 2\!\times\!2}} & \boldsymbol{\sigma} & \!\!\!\!-\boldsymbol{\sigma}\\
\!\!\!\!-\boldsymbol{\sigma}^{T} & \mathbb{O}_{{\scriptscriptstyle 6\!\times\!6}} & \mathbb{O}_{{\scriptscriptstyle 6\!\times\!6}}\\
\boldsymbol{\sigma}^{T} & \mathbb{O}_{{\scriptscriptstyle 6\!\times\!6}} & \mathbb{O}_{{\scriptscriptstyle 6\!\times\!6}}
\end{array}\,\right],\label{eq:pGFVacancy}
\end{equation}
where, as we shall prove, $g_{0}\!=\!\nicefrac{i}{3}$. In effect,
Eqs.\,\eqref{eq:Vacancies}-\eqref{eq:pGFVacancy} provide all the
necessary ingredients for us to make a pGF analysis of the single-vacancy
problem, on similar grounds to what was done for the atomic-sized
impurities in the lattice (see Sect.\,\ref{sec:Atomic-Sized-Impurities}).

\vspace{-0.5cm}

\section{Nodal Bound States of an Isolated Vacancy}

\begin{figure}[t]
\vspace{-0.3cm}
\begin{centering}
\includegraphics[scale=0.24]{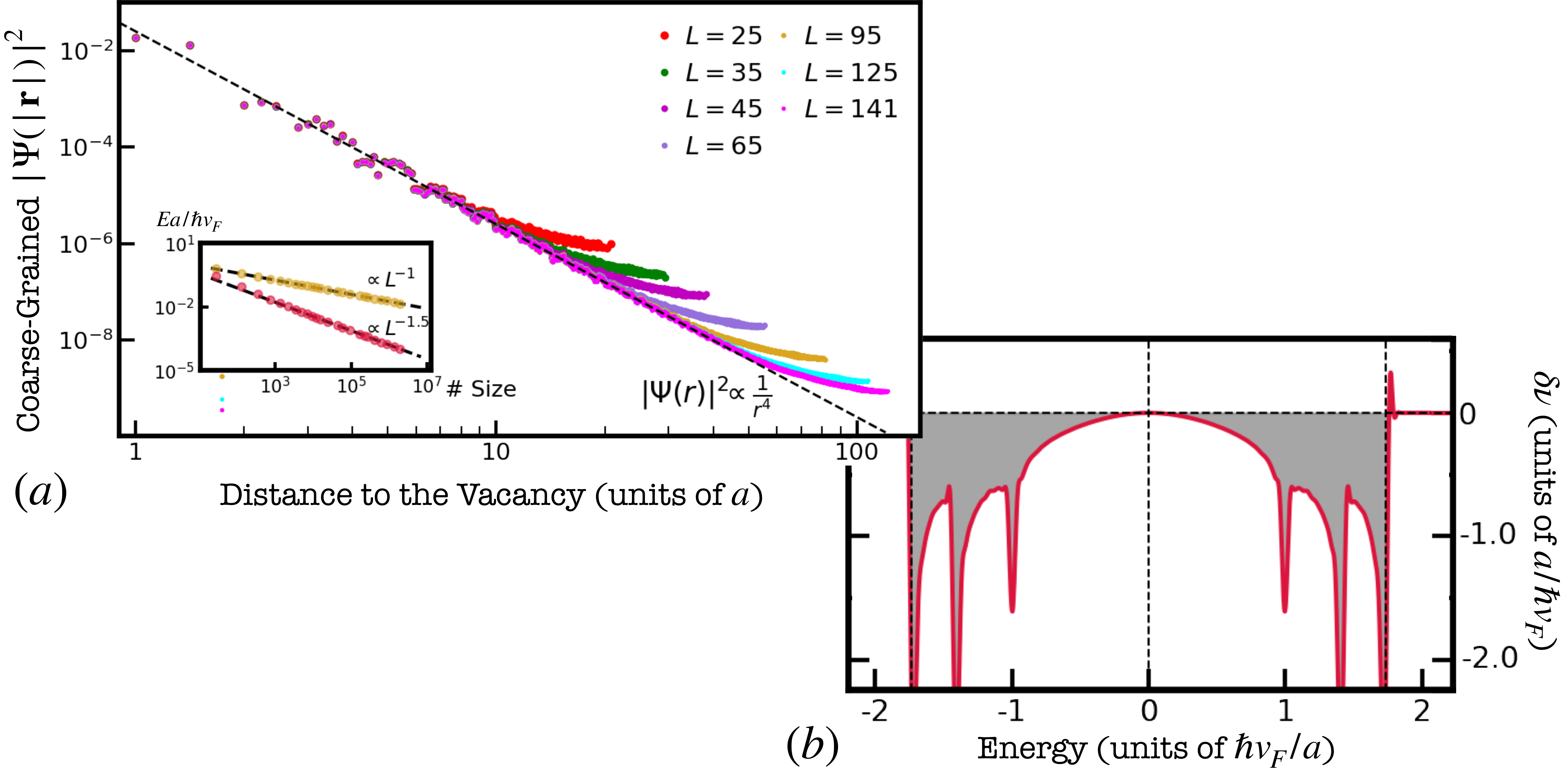}
\par\end{centering}
\vspace{-0.3cm}

\caption{\label{fig:SingleVacancy}(a) Plots of $\protect\abs{\Psi\!\left(\mathbf{R}\right)}^{2}\!\!=\!\protect\abs{\Psi_{1}\!\left(\mathbf{R}\right)}^{2}\!+\!\protect\abs{\Psi_{2}\!\left(\mathbf{R}\right)}^{2}$
for the eigenstate of a single central vacancy closest to $E\!=\!0$
and averaged over spherical shells of width $a$ centered on the vacancy.\textbf{
}In the inset, the first eigenenergy (a bound state) is shown scaling
to zero as $L^{{\scriptscriptstyle -2}}$ , while the second (an extended
state) features a $L^{{\scriptscriptstyle -1}}$ behavior. (b) Correction
to the eDoS due to a single vacancy in an infinite lattice. The whole
band is represented and the integral over the entire band is not zero
but $-2$, instead.}

\vspace{-0.3cm}
\end{figure}

The simple matrices defined in Eqs.\,\eqref{eq:Vacancies}-\eqref{eq:pGFVacancy}
can be plugged directly into the expressions derived within the general
formalism presented in Chapter \ref{chap:Rare-Event-States}, In particular,
a first question that can be asked is whether or not a vacancy is
able to create bound-states at the nodal energy. For that, we use
Eqs.\,\eqref{eq:Vacancies} and \eqref{eq:pGFVacancy} to build the
projected operator $\underline{\mathcal{I}}\!-\!\underline{\mathcal{G}_{c}^{0}\left(0\right)}\!\cdot\!\mathcal{V}_{\!\text{v}}$,
whose determinant takes on the remarkably simple form,

\vspace{-0.7cm}

\begin{equation}
\det\left(\underline{\mathcal{I}}\!-\!\underline{\mathcal{G}_{c}^{0}\left(0\right)}\!\cdot\!\mathcal{V}_{\!\text{v}}\right)\!=\!\left(i\!-\!3g_{0}\right)^{4}.\label{eq:DeterminantWeyl}
\end{equation}
The operator $\underline{\mathcal{I}}\!-\!\underline{\mathcal{G}_{c}^{0}\left(0\right)}\!\cdot\!\mathcal{V}_{\!\text{v}}$
must have, at least, a \textit{two-dimensional kernel subspace }because
of the two orbitals that were disconnected from the lattice. Hence,
with our implementation, a vacancy always produces a couple of \textit{``trivial
localized states''}, corresponding to the Wannier orbitals that were
removed. However, what Eq.\,\eqref{eq:DeterminantWeyl} is also telling
us is that the kernel is actually \textit{four-dimensional.} This
implies that an additional pair of \textit{``non-trivial bound states''}
also appears, these extending into the remaining lattice. In fact,
a full diagonalization of $\underline{\mathcal{I}}\!-\!\underline{\mathcal{G}_{c}^{0}\left(0\right)}\!\cdot\!\mathcal{V}_{\!\text{v}}$
yields the following projected wavefunctions:

\vspace{-0.7cm}

\begin{subequations}
\begin{align}
\ket{\xi_{1}^{b}} & =\frac{1}{\sqrt{6}}\left(\ket{\hat{\mathbf{x}},1}-i\ket{\hat{\mathbf{y}},1}-\ket{\hat{\mathbf{z}},2}-\ket{-\hat{\mathbf{x}},1}+i\ket{-\hat{\mathbf{y}},1}+\ket{-\hat{\mathbf{z}},2}\right)\\
\ket{\xi_{2}^{b}} & =\frac{1}{\sqrt{6}}\left(\ket{\hat{\mathbf{x}},2}+i\ket{\hat{\mathbf{y}},2}+\ket{\hat{\mathbf{z}},1}-\ket{-\hat{\mathbf{x}},2}-i\ket{-\hat{\mathbf{y}},2}-\ket{-\hat{\mathbf{z}},1}\right)
\end{align}
\end{subequations}

where $\ket{\mathbf{R},a}$ are the local Wannier states ($a$ labels
the orbital). Upon a reconstruction of the states outside of the original
support, $\Omega$, these yield the following wavefunctions:

\vspace{-0.7cm}

\begin{subequations}
\begin{align}
\Psi_{1}^{b}\!\left(\mathbf{R}\right)\! & =\!\frac{i\hbar v_{\text{F}}\sqrt{6}}{2a}\left[\begin{array}{c}
\mathscr{G}_{12}^{\text{0r}}(0;\mathbf{R})\\
\mathscr{G}_{22}^{\text{0r}}(0;\mathbf{R})
\end{array}\right]\label{eq:AVBS1}\\
\Psi_{2}^{b}\!\left(\mathbf{R}\right)\! & =\!\frac{i\hbar v_{\text{F}}\sqrt{6}}{2a}\left[\begin{array}{c}
\mathscr{G}_{11}^{\text{0r}}(0;\mathbf{R})\\
\mathscr{G}_{21}^{\text{0r}}(0;\mathbf{R})
\end{array}\right]\!,\label{eq:AVBS2}
\end{align}
\end{subequations}

where $\mathscr{G}_{ab}^{\text{0r}}(\varepsilon;\mathbf{R})$ are
spinor components of the lGF. Note that, from the continuum approximation
derived in Appendix\,\ref{chap:MultiValleyCont}, Eqs.\,\eqref{eq:AVBS1}
and \eqref{eq:AVBS2} imply that the new states generated by the vacancy
correspond to square-normalizable wavefunctions in the lattice, with
tails that decay asymptotically as $1/r^{2}$. Both the existence
of these states and their asymptotic behavior are confirmed by the
\textit{Lanczos Diagonalization} (LD) results presented in Fig.\,\ref{fig:SingleVacancy}\,a.
Further discussion is left to the next section.

\vspace{-0.5cm}

\paragraph{Deformation in the Density of States:}

Since the dimension of the system's Hilbert space was not changed,
the presence of extra nodal bound states necessarily implies that
some spectral weight was drawn out the band-continuum to the node,
by the vacancy. To characterize this process, we assess the vacancy-induced
deformation in the global extensive density of states (eDoS), through
Eq.\,\eqref{DosChange-1}, and using the matrix expression of the
clean lattice pGF at all energies, given in Eq.\,\eqref{eq:pGF_VacancyE}.
Therefore, we obtain

\vspace{-0.7cm}
\begin{equation}
\!\!\!\delta\nu(\varepsilon)\!=\!\frac{3a}{\pi\hbar v_{\text{F}}}\Im\!\left[\frac{f_{\varepsilon}\left(h_{\varepsilon}^{\prime}\!-\!2f_{\varepsilon}^{\prime}\right)\!+\!f_{\varepsilon}^{\prime}h_{\varepsilon}\!-\!4g_{\varepsilon}^{\prime}\left(i\!+\!3g_{\varepsilon}\right)}{3f_{\varepsilon}^{2}\!-\!3f_{\varepsilon}h_{\varepsilon}\!+\!2\left(i\!+\!3g_{\varepsilon}\right)^{2}}\right]\!\!,\label{eq:CorrectionDoS_Weyl}
\end{equation}
where all the functions $f_{\varepsilon}$, $g_{\varepsilon}$ and
$h_{\varepsilon}$ (and their derivatives) can be numerically calculated
from Eqs.\,\eqref{eq:feps}-\eqref{eq:heps}. A plot of $\delta\nu(\varepsilon)$
is shown in Fig.\,\ref{fig:SingleVacancy}\,b, where we see that
a negative correction to the eDoS appears across the entire band spectrum.
This is consistent with an overall transfer of spectral weight to
the emergent nodal bound states. Also, the integral of this curve
is exactly $-2$, because the number of continuum states is not conserved,
\textit{i.e.}, two states (per orbital) appear as vacancy bound-states
while two others were removed from the Hilbert space.

\vspace{-0.5cm}

\section{Density of States With Multiple Vacancies}

An isolated lattice vacancy produces a doublet of exact nodal bound
states in our model of a Weyl semimetal. Under a dilute limit argument,
a finite concentration of vacancies, $n_{\text{v}}$, would give rise
to an accumulation of $2n_{\text{v}}$ nodal eigenstates per unit
volume, which would change appreciably the electronic properties near
the Weyl node, relative to the clean WSM case. However, as happened
for the bound states analyzed in Chapter \ref{chap:Instability_Smooth_Regions},
these nodal states decay asymptotically as a power-law and, therefore,
a strong inter-vacancy interference can potentially affect the results
in the presence of multiple vacancies. To study this case, we now
employ a combination of KPM and LD calculations to tackle the many-vacancy
problem and see which features of the single-vacancy case would survive
the more realistic scenario of a macroscopic number of vacancies.
\begin{figure}[t]
\vspace{-0.5cm}
\begin{centering}
\includegraphics[scale=0.22]{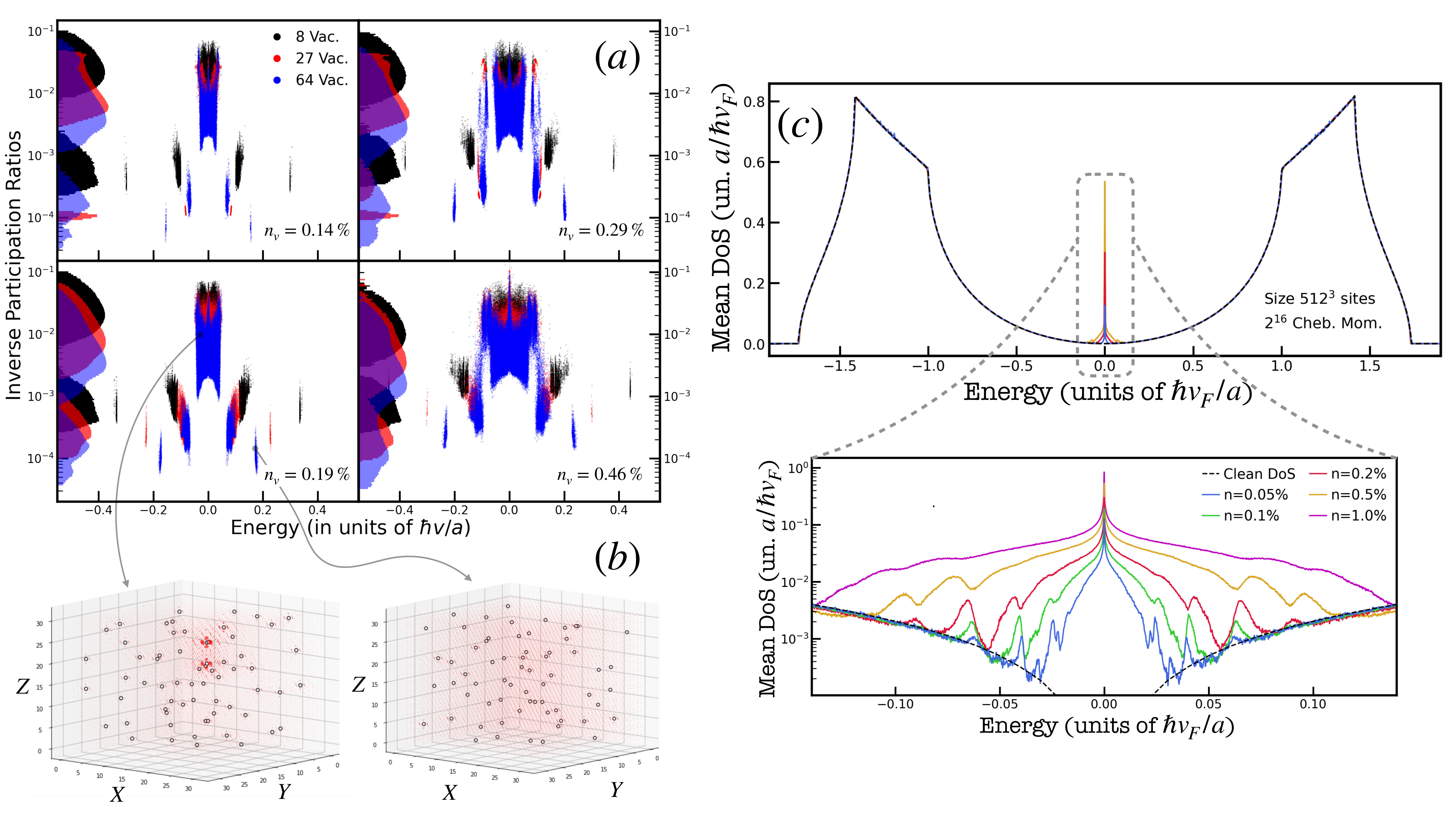}
\par\end{centering}
\vspace{-0.3cm}

\caption{\label{fig:FiniteConcentration}(a)\textbf{ }Scatter plot of $E$
vs $\text{IPR}$ for the eigenstates closest to the Weyl node in $25000$
samples, with three system sizes, and randomly placed vacancies. Histograms
of the IPRs are presented along the vertical axis. (b) Bubble chart
representation of the squared-wavefunctions for two randomly selected
eigenstates from the indicated regions. The eigenstate on left is
a heavily localized wavefunction around a few vacancies (hollow black
circles), while the right one is clearly extended across the simulated
system. (c) Mean DoS of a simulated WSM lattice of side $L\!=\!512a$
for increasing vacancy concentrations. The entire spectrum's overview
is presented in above and a close-up near the peak is shown below.}

\vspace{-0.5cm}
\end{figure}

Our starting approach is based upon a LD of lattices having $L^{3}$
unit cells, and containing a prescribed concentration ($n_{\text{v}}$)
of randomly placed vacancies. As detailed in Appendix\,\ref{chap:TwistedBoundaries-1},
the use of \textit{twisted boundary conditions} opens up a \textit{finite-size
gap} in the spectrum that serves as a separation between the nodal
bound states, and the extended ones. In principle, any state is affected
by phase-twists at the boundaries of the simulated cell, but extended
states will certainly be more sensitive. As in Sect.\,\ref{sec:What-are-RareEvents},
our finite-size gap will work as an \textit{``energy bin''} in which
the vacancy-induced states lie and, thereby, we can simply ask the
diagonalization algorithm to compute the $2n_{\text{v}}\!L^{3}\!+\!4$
eigenpairs\,\footnote{Naively, one expects that each vacancy gives rise to two bound states,
which makes this number sufficiently large to capture most states
within the finite-size gap.} that are closest to $\varepsilon\!=\!0$. In Fig.\,\ref{fig:FiniteConcentration}\,a,
we present a scatter plot of the energies and corresponding IPRs\nomenclature{IPR}{Inverse-Participation Ratio}
of every eigenpair determined for a set of $2500$ random arrangements
of vacancies, with a concentration that ranges from $0.1\%$ to $1\%$
(per unit cell). The results clearly demonstrate that, in spite of
the proximity between vacancies, the system still features a large
number of high-IPR eigenstates which are flanked by a region of extended
states. This physical interpretation is further confirmed by Fig.\,\ref{fig:FiniteConcentration}\,b,
where a 3D bubble chart of $\abs{\Psi(\mathbf{R})}^{2}$ is depicted
for two eigenstates randomly chosen from each of the regions.

The study of inter-vacancy effects by LD can provide us with a qualitative
picture on how the eigenstates near the Weyl node look like, in the
presence of many vacancies. However, this approach is severely limited
by the effective spectral resolution\,\footnote{The main issue is that the Lanczos algorithm can miss eigenstates
that are nearly-degenerate. In contrast, the KPM algorithm calculates
observables that involve the entire spectrum, albeit with an artificial
broadening of the individual energy levels.}, the finite number of eigenstates that can be considered, and the
attainable system sizes, which precludes a clear physical picture
of what to expect in the thermodynamic limit. Therefore, we now complement
the LD results with \textit{full-spectrum simulations} of the mean
DoS based on the \textit{Kernel Polynomial Method} (KPM), which is
lengthly described in Appendix\,\ref{chap:Crash-Course-KPM}.

\vspace{-0.5cm}

\paragraph{Average Density of States:}

Our KPM results for the mean DoS are summed up in Fig.\,\ref{fig:FiniteConcentration}c,
where we present an overview of the entire spectrum, for different
vacancy concentrations. The plots show a big enhancement of the number
of states in (and around) the nodal energy, which indicates that the
nodal DoS gets quickly and strongly lifted as $n_{\text{v}}$ increases\,\footnote{One can think of this as the limit $\abs{U_{n}}\to\infty$ of the
results found for atomic-sized impurities considered in Sect.\,\ref{sec:Atomic-Sized-Impurities}.}. This effect is consistent with the prevalence of the single-vacancy
nodal bound states at finite concentrations, and qualitatively agrees
with what was concluded from our previous LD analysis. Nevertheless,
as the central peak grows in height, a much wider symmetrical profile
emerges at its base. This is a telltale sign that inter-vacancy hybridization,
which becomes stronger for smaller typical distances between vacant
sites, is turning the bound states of isolated vacancies into low-energy
scattering resonances within the continuum. This, however, does not
change the total number of vacancy-induced states around the Weyl
nodes, sine the integral of the full central correction to the DoS
is seen to be proportional to $n_{\text{v}}$\,(see the inset of
Fig.\,\ref{fig:SubsidiaryResonances}c). In the bottom panel of Fig.\,\ref{fig:FiniteConcentration}c,
we present a closeup of this broadened central peak where, in addition
of showing that the vacancy-induced spectrum is diffused around $\varepsilon\!=\!0$,
we can also observe the emergence of a finer structure of subsidiary
peaks (a comb of sharp scattering resonances) around the node for
concentrations $n_{\text{v}}\lesssim1\%$.

\vspace{-0.5cm}

\subsection{\label{subsec:Inter-Vacancy-Hybridization}Inter-Vacancy Hybridization
in 3D Weyl Semimetals}

\begin{figure}[t]
\vspace{-0.5cm}
\begin{centering}
\includegraphics[scale=0.225]{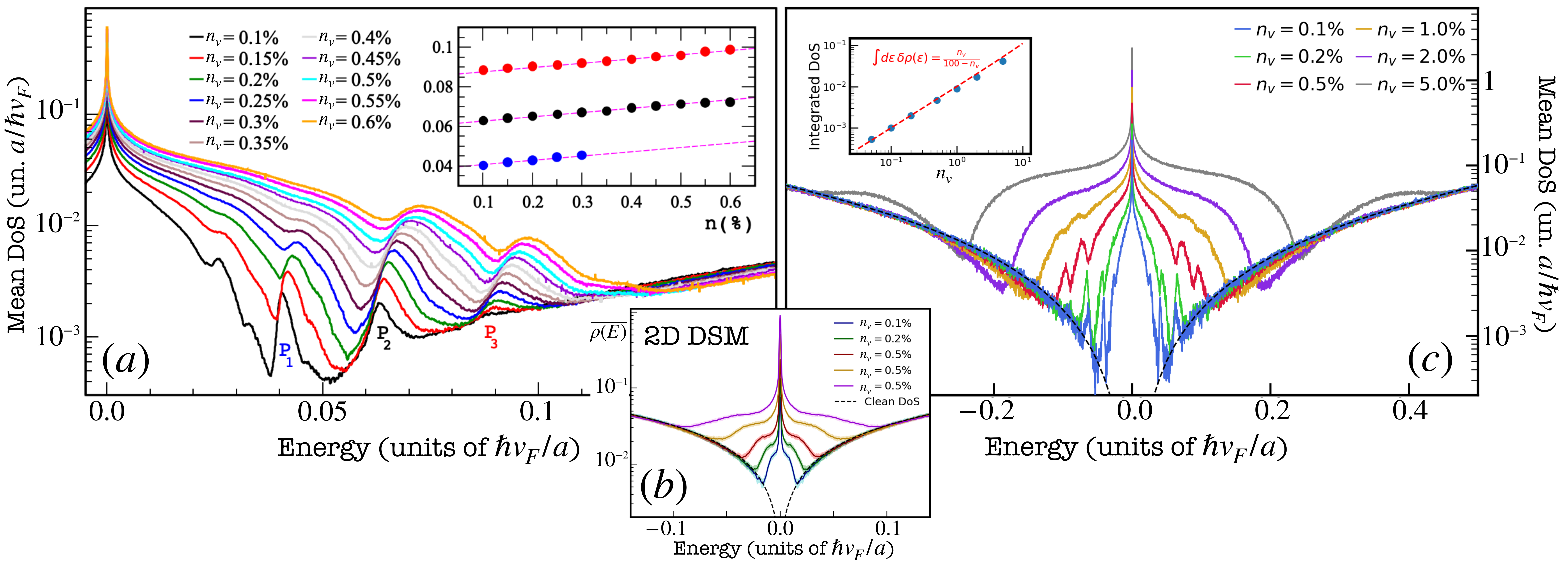}
\par\end{centering}
\vspace{-0.2cm}

\caption{\label{fig:SubsidiaryResonances}(a) Detail of mean DoS of a lattice
WSM of side $L\!=\!512a$, showing the dependence of the subsidiary
peaks in the DoS for a finer set of vacancy concentrations. Inset:
The three prominent peaks ($P_{1},P_{2}\text{ and }P_{3}$) are followed
as a function of $n$. (b) Mean DoS for a two-dimensional Dirac semimetal
in the presence of a finite vacancy concentration. (c) Correction
do the mean DoS as a function of the vacancy concentration including,
as an inset, the estimate of the integral of $\delta\nu(E)\!=\!\nu(E)\!-\!\nu_{0}(E)$
over the energy region represented plotted in the main panel.}

\vspace{-0.5cm}
\end{figure}

Our KPM results clearly demonstrate that inter-vacancy effects become
relevant in determining the induced changes in the DoS. Simply put,
isolated vacancies would place new (proper) bound states at the nodal
energy, but a finite concentration of these defects actually translates
into a broadened central peak in the DoS. This is hardly a surprising
result, as one expects quantum-interference effects to becomes relevant
in any system that hosts a sufficiently large concentration of impurities.
What is surprising here is the appearance of a superposed comb of
sharp subsidiary resonances at finite energies for $0.1\%\!<\!n_{\text{v}}\!<\!1\%$.
In Fig.\,\ref{fig:SubsidiaryResonances}a, we present a finer analysis
of this structure as a function of $n_{\text{v}}$. There it becomes
evident that, prior to being washed-out, these resonance peaks move
away from the node proportionally to the vacancy concentration. Moreover,
we also show in Fig.\,\ref{fig:SubsidiaryResonances}\,b an analogous
calculation done in a two-dimensional version of $\mathcal{H}_{l}^{0}$\,\footnote{A square lattice Hamiltonian which is the same as in Eq.\,\eqref{eq:LatticeModel-1-1},
but restricted to the $xOy$-plane.}, which realizes a (chiral-symmetric) 2D Dirac semimetal. Despite
displaying a similar broadening of the central peak due to \textit{inter-vacancy
effects}, our results show no signs of a modulated structure of subsidiary
resonances. The previous comparison between 2D and 3D analogous systems
provides a clue on the nature of these resonances. Intuitively, one
expects an increased dimensionality to decrease the quantum-interference
effects. While 2D vacancy-induced states get rapidly driven away from
the nodal energy, in 3D, the quantum-interference effects are reduced
such that inter-vacancy hybridization happens through an\textit{ intermediate
phase} with very sharp resonances around the node, that still preserve
some degree of locality around one (or a few) vacancy(ies). 

\vspace{-0.5cm}

\subsection{Magnetic Sensitivity of the Subsidiary Resonances}

The claim of locality {[}Subsect.\,\ref{subsec:Inter-Vacancy-Hybridization}{]},
done for states composing the subsidiary resonances, deserves some
further attention.\,\,In fact, from the LD results presented in
Fig.\,\ref{fig:FiniteConcentration}a, we have already some evidence
that the \textit{many-vacancy states} which accumulate around the
nodal energy are rather localized.\,\,Looking at the highest IPRs
shown in Fig.\,\ref{fig:FiniteConcentration}a, we see that these
are way too big for the states to be extended near the node.\,\,However,
it is not easy to numerically prove this statement, because the LD
algorithm can yield artificial linear combinations of
\begin{figure}[t]
\vspace{-0.3cm}
\begin{centering}
\includegraphics[scale=0.22]{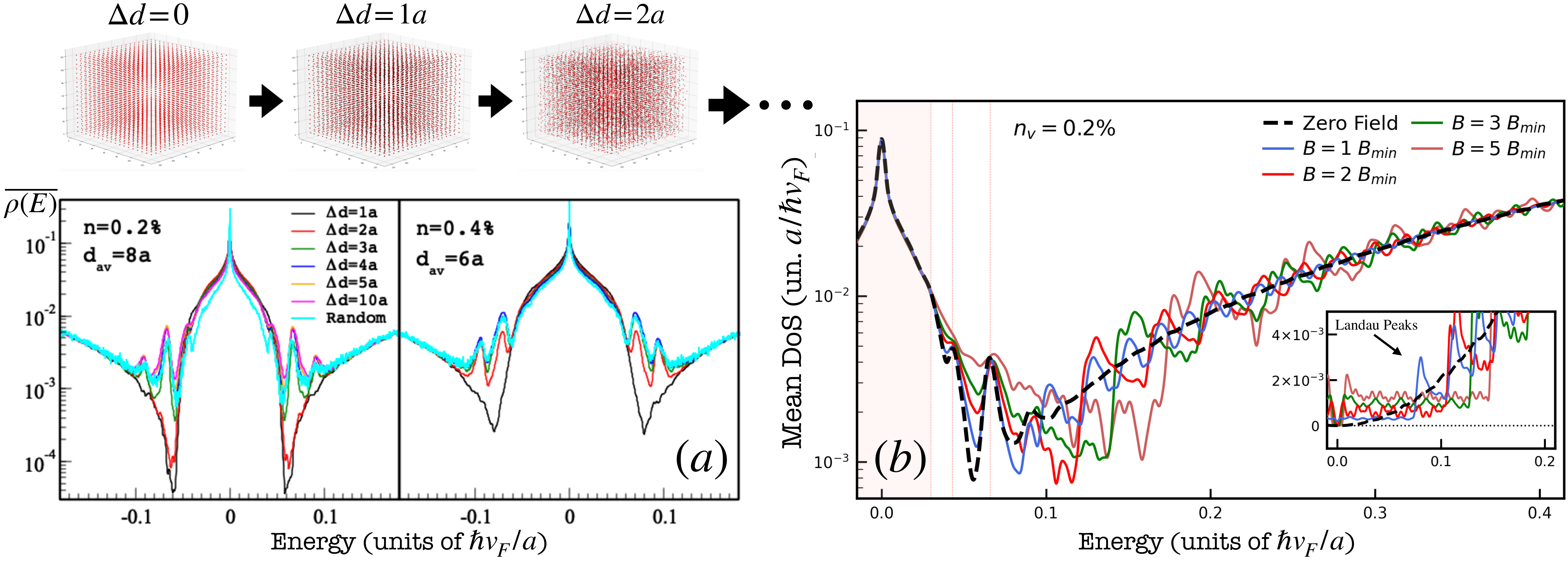}
\par\end{centering}
\vspace{-0.3cm}

\caption{\label{fig:LatticeofVacancies}(a) Mean density of states calculated
with for a system containing a superlattice of vacancies that get
progressively randomized, by increasing $\Delta d$ (see scheme on
the top). As the vacancy positions get progressively more random,
the comb of subsidiary resonance peaks emerge in the DoS. (b) DoS
of a system having a $n_{v}\!=\!0.2\%$ concentration of vacancy,
for a selection of applied magnetic fields along the $z$-axis. The
more sensitive areas of the spectrum are shaded.}

\vspace{-0.5cm}
\end{figure}
\begin{wrapfigure}[22]{o}{0.39\columnwidth}%
\vspace{-0.55cm}
\begin{centering}
\includegraphics[scale=0.26]{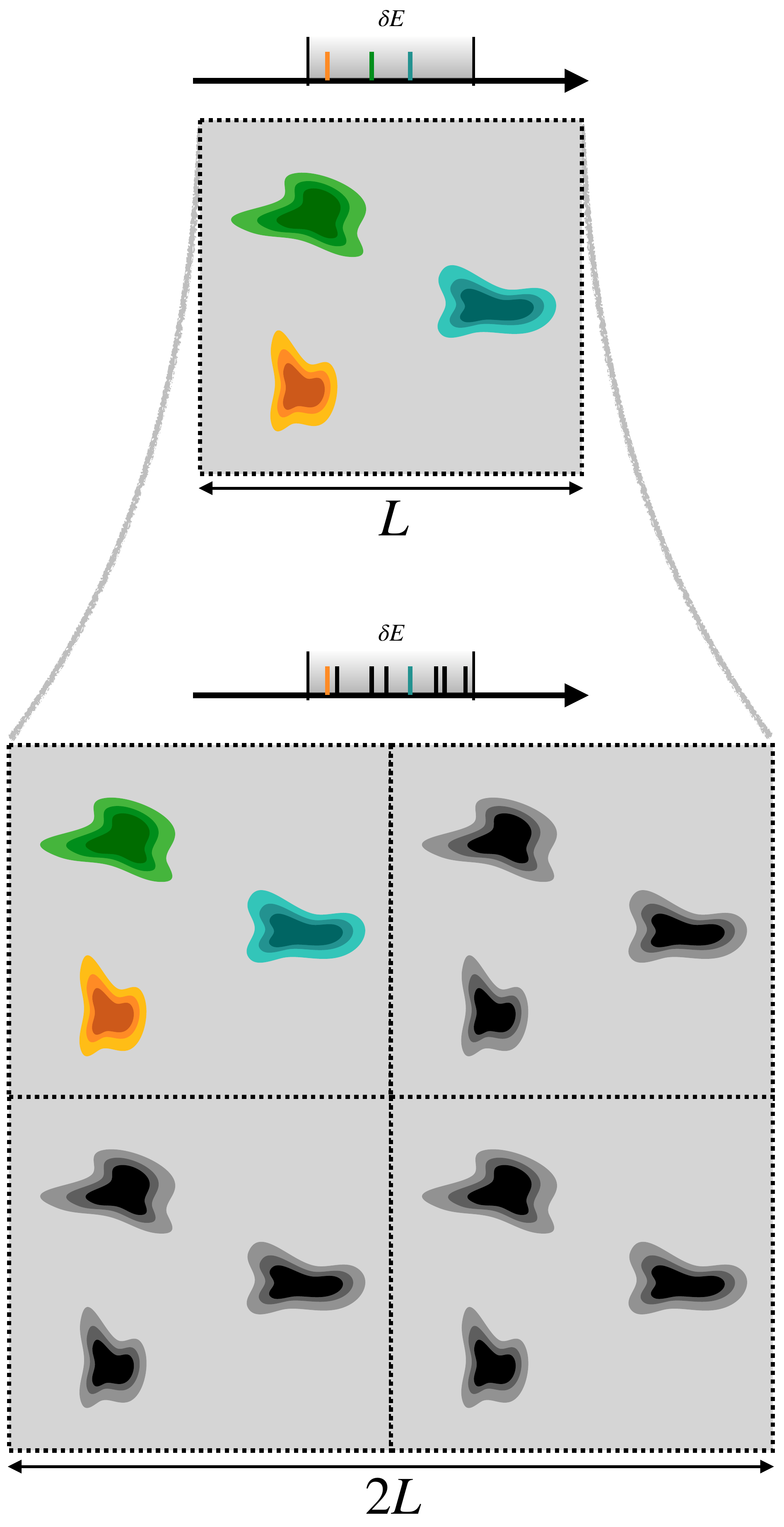}
\par\end{centering}
\vspace{-0.4cm}

\caption{\label{fig:IPRScaling}Scaling of the IPR for linear combinations
of two-dimensional localized states.}

\vspace{-0.3cm}\end{wrapfigure}%
 nearly-degenerate eigenstates, whose IPRs would scale with $L$ in
an \textit{extended-like fashion} (\textit{i.e.}, $\text{IPR}_{\Psi}^{L}\propto L^{-3}$).\,\,This
faulty behavior of the LD algorithm is illustrated by the 2D scheme
shown in Fig.\,\ref{fig:IPRScaling};\,\,There, we start with an
$L\!\times\!L$ well localized system for which an eigenstate is found
using LD with a spectral accuracy $\delta E$.\,\,Within $\delta E$,
there are $3$ exact energy levels such that the LD algorithm yields
an artificial linear combination of three localized states and, thereby,
a squared-wavefunction with three blobs in the simulated plane.\,If
we now increase the system's size from $L\!\to\!2L$ (without changing
$\delta E$), then $\sim\!9$ additional exact levels will exist within
the accuracy window and, therefore, the obtained wavefunction will
have roughly four times the number of blobs as shown in the bottom
panel of Fig.\,\ref{fig:IPRScaling}.\,\,This way, we can see that
the IPR of such a state, despite the fact that it may have small nominal
values for a fixed $L$, will scale with the system size in precisely
the same way as a typical extended state of that dimensionality.

Due to this limitation, we now change our strategy and attempt to
better understand the hybridized states by directly calculating the
DoS for a more controlled vacancy disorder model, in which a predetermined
number of vacancies are positioned in the places of an underlying
superlattice\,\footnote{The spacing parameter of this superlattice is the mean distance between
vacancies, as determined by their concentration, $n_{v}$.}. Then, the mean DoS is calculated using the KPM for a series of samples
in which the (originally periodic) vacancy positions are randomized
by a vector $\boldsymbol{\Delta}_{v}\!=\!(n_{x},n_{y},n_{z})a$, where
$-\Delta d<\!n_{x,y,z}\!<\Delta d$ are independent random integers
for each vacancy. In Fig.\,\ref{fig:LatticeofVacancies}\,a, we
showcase the mean DoS using the above procedure to position the vacancies
in the lattice. From our results, we see that an almost periodic positioning
of the vacancies, even though it still causes a broadening of the
central peak, does not create any additional structure in the density
of states. However, as soon as $\Delta d$ becomes sufficiently large
to allow two (or more) vacancies to be distanced on the order of $a$,
the structure of subsidiary resonances re-emerges. These results,
though indirect, hint that these resonances can be interpreted as
hybridized states of a few nearby vacancies (\textit{i.e.}, resonances
of small vacancy clusters).

\vspace{-0.6cm}

\paragraph{Magnetic Sensitivity:}

By this point, the reader must be rather convinced that both the central
peak and the subsidiary resonances in the mean DoS correspond to heavily
localized states. The former are likely weakly perturbed single-vacancy
states, whilst the latter probably correspond to hybridized states
between a few nearby vacancies. Nevertheless, we do a further computation
to support our earlier claim:\,\,We calculate the mean DoS with
a finite concentration of vacancies\begin{wrapfigure}[19]{o}{0.39\columnwidth}%
\vspace{-0.3cm}
\begin{centering}
\includegraphics[scale=0.26]{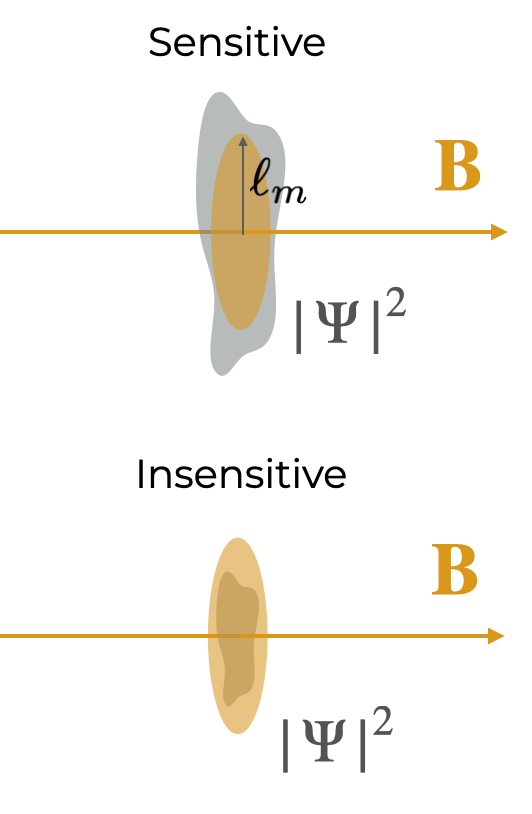}
\par\end{centering}
\vspace{-0.4cm}

\caption{\label{fig:MagneticLength}Comparison\,of\,\,the magnetic\,\,length,\,\,$\ell_{m}$,\,and\,\,the
bulk\,of\,a\,squared-wavefunction ($\protect\abs{\Psi(\mathbf{R})}^{2}$).}

\vspace{-0.3cm}\end{wrapfigure}%
 in the presence of a magnetic field $B$ along the $z$ direction. 

As depicted in the scheme of Fig.\,\ref{fig:MagneticLength}, if
a wavefunction is localized in real-space, then the effects of an
applied magnetic field will only be appreciable if the corresponding
\textit{magnetic length scale}, $\ell_{m}\!=\!\sqrt{\hbar c/eB}$,
is small when compared with the linear spatial extension of the state
itself. Therefore, a spectral region composed of more localized states
is expected to be less sensitive to applied magnetic fields. Such
a magnetic insensitivity have actually been observed near the nodal
energy of a Weyl semimetal with gaussian white-noise disorder by Lee
\textit{et al.}\,\cite{Lee2018}, being attributed to the presence
of rare-event bound states {[}like the ones shown in Fig.\,\ref{fig:RareEventesSingleFluct}{]}.
In Fig.\,\ref{fig:LatticeofVacancies}\,b, we illustrate such an
effect in the mean DoS (cubic WSM lattice of side $L\!=\!256a$\,\footnote{Under periodic boundary conditions and a typical lattice spacing of
$a\approx1\text{nm}$, this corresponds to a minimal field of $B_{\text{min}}\!\approx\!1T$.}) as a function of the magnetic field strength $B$, and with a fixed
vacancy concentration $n_{\text{v}}\!=\!0.2\%$. Clearly, the areas
identified earlier as being composed chiefly of more localized resonances
(shaded in the panel) are also the energies for which the DoS shows
\textit{lesser sensitivity }to the applied magnetic field. For completeness,
we also show corresponding calculations done in the absence of vacancies
(shown in the inset of Fig.\,\ref{fig:LatticeofVacancies}b), which
clearly place the system in a regime of well-defined Landau bands\,\cite{Shao16,Klier17}
and a finite plateau in the DoS around the node, due to the dispersion
of the lowest Landau level along the magnetic field direction (see
Subsect.\,\ref{subsec:Landau-Quantization} for a detailed discussion
of the Landau spectrum near a Weyl node).

\vspace{-0.5cm}

\section{Impact of Vacancies in DC and Optical Conductivies}

So far, we have focused on static spectral properties of Weyl electrons
under the influence of vacancy disorder. Concretely, we have characterized
the changes that are induced in the mean density of states and further
characterized the qualitative structure of the corresponding wavefunctions
in real-space. With the advent of thin-films, two-dimensional materials
and surface-state electronics, the single-electron density of states
appeared as a quantity of great experimental interest in its own right\,\cite{Li13},
because its real-space value could be measured by \textit{Scanning
Tunneling Spectroscopy}\,\cite{Park87,Zandvliet2009} (STM) or, in
$\mathbf{k}$-space, via \textit{Angle-resolved Photoemission Spectroscopy}\,\cite{Sobota21}
(ARPES). For the bulk DoS of a three-dimensional sample, direct measurements
are generally not available and one must resort to indirect measurements,
such as, low-temperature specific-heat\,\cite{Guo18,Baggioli20},
electronic transport or optical response measurements. In this section,
we present our own predictions of the effects cause by vacancy-induced
nodal bound-states (and their hybridized counterparts) on the dc conductivity
and linear optical response of a Weyl semimetal.

\vspace{-0.5cm}

\subsection{Longitudinal DC Conductivity}

The first quantity we study is the bulk linear longitudinal dc conductivity,
$\sigma_{\text{dc}}\!\left(E_{F}\right)$, which is phenomenologically
defined as 

\vspace{-0.7cm}
\begin{equation}
\mathbf{J}_{\text{dc}}=\sigma_{\text{dc}}\!\left(E_{F}\right)\mathbf{E},
\end{equation}
where $\mathbf{J}_{\text{dc}}$ is the electric current density in
the bulk, and $\mathbf{E}$ is the applied static electric field.
Note that this conductivity is a function of the Fermi energy and,
in fact, it is a property of states lying at the Fermi surface. Therefore,
provided has control over the value of $E_{F}$, this is a natural
quantity where to look for signals of both the enhancement of the
nodal DoS, as well as the subsidiary resonances arising from inter-vacancy
effects. In the diffusive transport regime, the conductivity is beautifully
related to the density of states at the Fermi level through the well-known
\textit{Einstein Relation},

\vspace{-0.7cm}

\begin{equation}
\sigma_{\text{dc}}(E_{F})=e^{2}\rho(E_{F})\mathcal{D}(E_{F}),\label{eq:Einstein}
\end{equation}
where $e$ is the elementary charge, $\mathcal{D}(E_{F})$ is the
\textit{quantum electronic diffusivity} at the Fermi level, and $\rho(E_{F})$
is the mean DoS at the Fermi level. Therefore, if the quantum diffusivity
were not affected by the vacancies, all the features found in the
mean DoS would be precisely reproduced in the conductivity as a function
of $E_{F}$. However, such an insensitivity of $\mathcal{D}(E_{F})$
to the vacancy disorder cannot be true in the light of the localized
nature of the states that are introduced in and around the nodal energy.
In fact, as pointed out by Elattari \textit{et al}.\,\cite{Elattari99,Elattari2000},
whenever sharp resonances appear in the DoS, there is a great possibility
of a strongly suppressed quantum diffusivity at those energies. Such
an effect could counteract the enhancement in the DoS and effectively
reduce the conductivity. 
\begin{figure}[t]
\vspace{-0.4cm}
\begin{centering}
\includegraphics[scale=0.23]{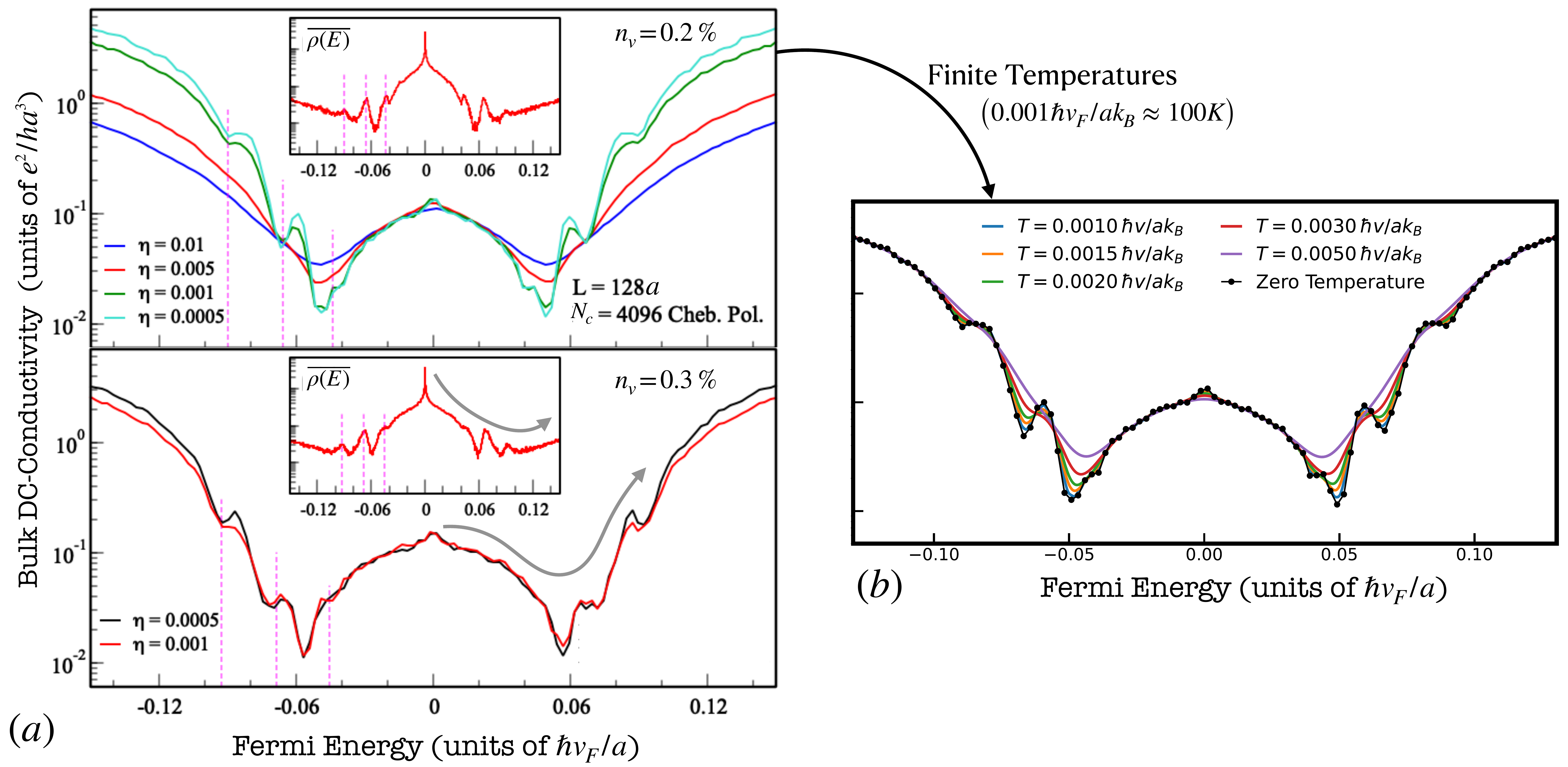}
\par\end{centering}
\vspace{-0.2cm}

\caption{\label{fig:BulkConductivity}(a) Plots of $\sigma_{\text{dc}}$ as
a function of the Fermi energy for a concentration $n_{\text{v}}\!=\!0.2\%$
(upper panel) and $n_{\text{v}}\!=\!0.3\%$ (lower panel) of randomly
placed vacancies. The different colors represent different values
of the phenomenological broadening, $\eta$, and all the data are
converged in both system size and number of Chebyshev moments. \textit{Insets:}
Mean DoS for the same concentration. The magenta lines indicate the
position of three prominent subsidiary resonances. (b) Plots of the
finite temperature dc conductivity for $n_{\text{v}}\!=\!0.2\%$.}

\vspace{-1.0cm}
\end{figure}

In order to study this quantity accurately, we performed large-scale
unbiased real-space simulations of the mean dc conductivity using
the \textit{Single-Shot Kubo-Greenwood method} implemented in the
QuantumKITE software (and described in Appendix\,\ref{chap:Crash-Course-KPM}).
Our results for $\sigma_{\text{dc}}(E_{F})$ are summarized in Fig.\,\ref{fig:BulkConductivity}a
for vacancy concentrations of $n_{\text{v}}\!=\!0.2\%$ and $0.3\%$,
in a lattice with side $L\!=\!128a$, and spectral resolutions in
the range of $\eta\!=\!10^{{\scriptscriptstyle -2}}\!-\!10^{{\scriptscriptstyle -4}}\hbar v_{\text{F}}/a$.
The results show that the dc conductivity, first decreases, but then
grows steadily as the Fermi level gets driven away from the node (see
gray lines in \ref{fig:BulkConductivity}a). However, upon a closer
inspection there is a more outstanding feature in the results: precisely
at the energies where the subsidiary resonances appear, the \textit{conductivity
displays dips} relative to its overall background value. In Fig.\,\ref{fig:BulkConductivity}a,
we complement these conclusions by showing the effect of an increasing
temperature (thermal broadening of the Fermi surface) in the conductivity.
From these results we realize that, even though the conductivity dips
cannot survive up to arbitrarily high temperatures, one is still expected
to observe them at temperatures as high as $100K$.

\vspace{-0.5cm}

\paragraph{Physical Mechanisms:}

The overall growth of the dc conductivity away from the nodal energy
is an expected result, that may be seen as the trivial effect of an
increased density of states. In contrast, the physical interpretation
of the aforementioned conductivity dips is a more subtle issue, which
cannot simply be an effect of the DoS's energy modulation. As a matter
of fact, these can only be explained by considering that the quantum
diffusivity is \textit{strongly suppressed} at the subsidiary resonances
which, in turn, is due to the localized nature of the states that
compose them. This suppression is sufficient to reduce the overall
dc conductivity, in spite of the larger number of states that exist
at that Fermi level.

\vspace{-0.5cm}

\subsection{Linear Optical Response}

We have seen that it is possible, in principle, to detect the vacancy-induced
features of the mean DoS using bulk dc transport measurements in crystalline
samples. To detect the described effects, however, an active tuning
of the system's Fermi energy is required. This is easily achieved
in two-dimensional materials, where a back gate voltage may be used
to alter the bulk charge carrier density. On the contrary, tuning
the Fermi level of a 3D material is generally a much more complex
process that employs relatively new experimental techniques, such
as a electric field-effect in thin-films\,\cite{Wang2016,Chen21},
or a precise stoichiometric control during an epitaxial growth process\,\cite{Ghosh2022}.
For this reason, it is useful to turn our attention for alternative
measurable properties that \textit{(i)} are sensitive to the existence
of vacancy-induced nodal states, and \textit{(ii)} do not depend on
modifying the Fermi energy of the system. The optical response coefficient
is such a quantity, where the frequency of the excitation field ($\omega$)
plays the role of an external control parameter that can be easily
changed.

In order to assess the effects of vacancies in the optical response
of a Weyl semimetal, we have evaluated the $\omega$-dependent linear
conductivity $\sigma_{\omega}^{ii}$ of the tight-binding model defined
in Eq.\,\eqref{eq:LatticeModel-1-1}, using the real-space approach
developed by João \textit{et al}.\cite{Joao19} and currently implemented
in the QuantumKITE package\cite{Joao2020}. This method is based upon
the spectral expansion of the linear Kubo formula,

\vspace{-0.7cm}
\begin{align}
\sigma_{\omega}^{jl} & =\frac{e^{2}}{i\hbar\omega L^{3}a^{3}}\int dEf_{\text{FD}}^{E}\text{Tr}\left[V^{j}\mathcal{G}_{\eta}^{\text{r}}\left(-\hbar^{-1}\!E-\omega\right)V^{l}\delta\left(E-\mathcal{H}_{l}\right)\right.\label{eq:DynamicalCond-2}\\
 & \qquad\qquad\quad\left.+V^{j}\delta\left(E-H_{l}\right)V^{l}\mathcal{G}_{\eta}^{\text{a}}\left(-\hbar^{-1}\!E+\omega\right)+i\hbar V^{jl}\delta\left(E-\mathcal{H}_{l}\right)\right],\nonumber 
\end{align}
where $L$ is the lateral size of the simulated system, $\mathcal{H}_{l}$
is the disordered Hamiltonian, $f_{\text{FD}}^{E}$ is the Fermi-Dirac
distribution, $V^{i}\!=\!\left[\mathbf{R}^{i},\mathcal{H}_{l}\right]$
is the velocity operator, $V^{ij}\!=\!\left[\mathbf{R}^{i},\left[\mathbf{R}^{j},\mathcal{H}_{l}\right]\right]$,
and $\mathcal{G}^{\text{a/r}}(E)=\left[E\pm i\eta-\mathcal{H}_{l}\right]$
are the advanced/retarded SPGFs of the system (in the presence of
disorder). The numerical calculation of this quantity involves an
intrinsic spectral resolution, $\eta$, relative to which one must
assure the convergence of the obtained results (numerical method detailed
in Appendix\,\ref{chap:Crash-Course-KPM}).
\begin{figure}[t]
\vspace{-0.5cm}
\begin{centering}
\includegraphics[scale=0.23]{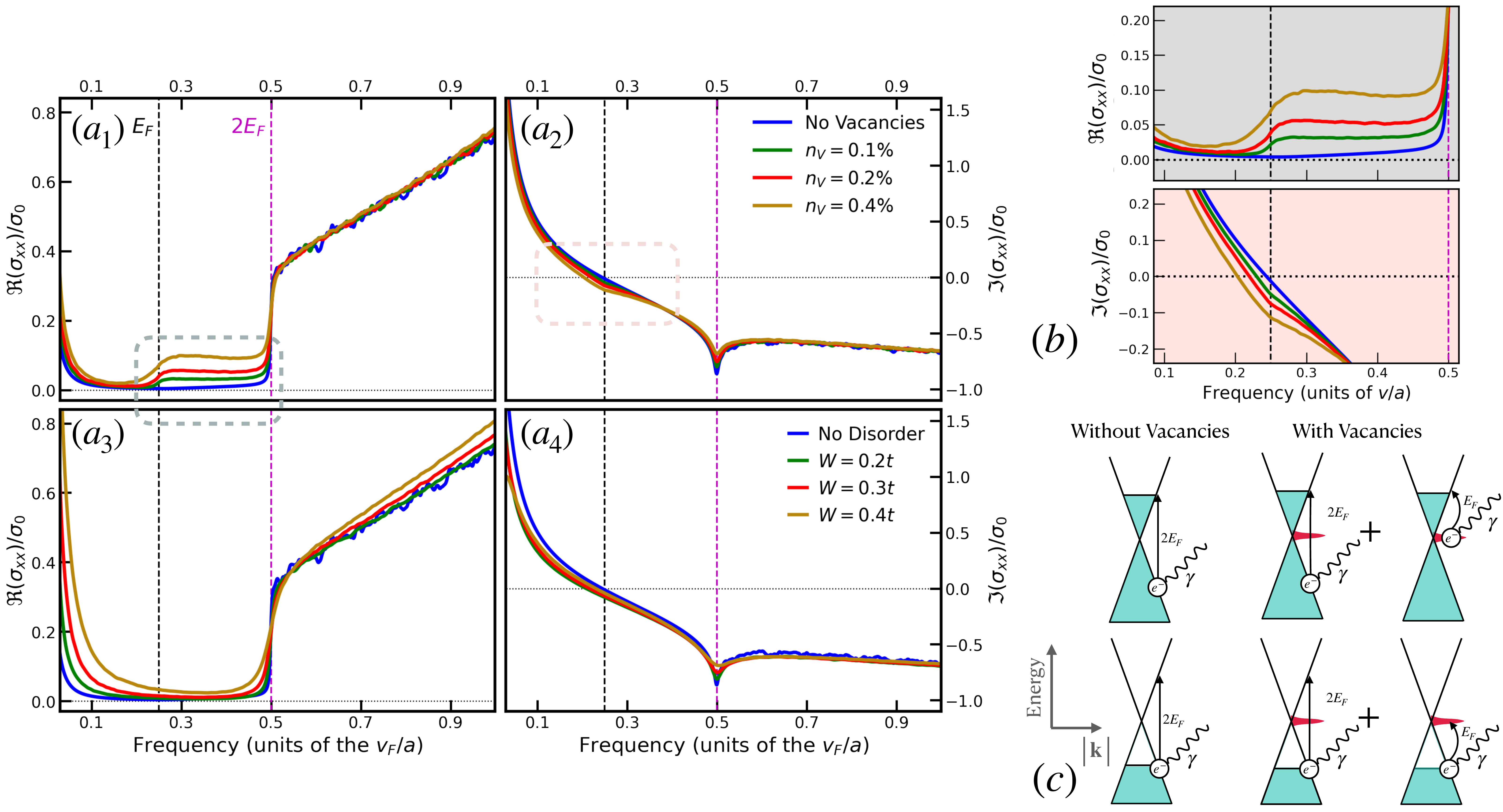}
\par\end{centering}
\vspace{-0.1cm}

\caption{\label{fig:OpticalConductivity}Plots of the longitudinal linear optical
conductivity of a doped WSM computed with an energy broadening $\eta\!=\!0.002\hbar v_{\text{F}}/a$
and a Fermi energy $E_{F}=0.25\hbar v_{\text{F}}/a$ in the presence
of a finite concentration of vacancies, in panels $\text{(a}_{\text{1-2}}\text{)}$,
and an on-site Anderson random potential of strength $W$, in panels
$\text{(a}_{\text{3-4}}\text{)}$. The clean case is shown in blue.
(b) Close-up of the vacancy-induced features in $\sigma_{xx}(\omega)$
{[}marked as dashed boxes in (a){]}. (c) Scheme of the relevant physical
mechanisms for $\omega\!>\!\protect\abs{E_{\text{F}}}$. Vertical
arrows stand for (\textit{$\mathbf{k}$-conserving}) inter-band transitions
between Bloch states, while curved arrows represent (\textit{non $\mathbf{k}$-conserving})
transitions between Bloch states and nodal quasi-bound states. Except
for the $\omega\!\to\!0^{+}$ peak, all plots are converged in the
number of Chebyshev polynomials.}

\vspace{-0.5cm}
\end{figure}

In Fig.\,\ref{fig:OpticalConductivity}\,a, we present numerical
results obtained for the real and imaginary parts of $\sigma_{\omega}^{xx}$
in the presence of either vacancies, with concentration $n_{v}$,
or a random Anderson potential of strength $W$\,\footnote{Concretely, this corresponds to an on-site random potential with a
box distribution of width $W$.} . For a pristine, but slightly doped WSM ($E_{F}\!\sim\!\pm0.1\text{eV}$),
the vertical inter-band transitions are Pauli blocked if $\omega\!<\!2\abs{E_{\text{F}}}$.
This defines the system's \textit{optical gap}, in which the imaginary
part of the optical conductivity is finite, but reverses sign at $\omega\!=\!\abs{E_{\text{F}}}$.
Our numerical results clearly show that the optical response of the
material gets altered by the presence of both types of disorder and,
more importantly, that it can be used to establish a clear distinction
between the two models (random on-site potential\textit{ versus }vacancies).
With a random potential, the optical gap remains intact at weak disorder,
with the $\omega\!=\!2\abs{E_{F}}$ transition being simply smoothened
as $W$ is increased. In addition, we also see that the typical linear
dependence of $\Re\left(\sigma_{\omega}^{xx}\right)$ is changed by
the disorder, showing a slope that increases with $W$. Remarkably,
in the presence of random vacancies, no major changes are observed
in $\sigma_{\omega}^{xx}$ except for the appearance of a \textit{new
optical gap} at $\omega\!=\!\abs{E_{F}}$ that initiates a plateau
in $\Re\left(\sigma_{\omega}^{xx}\right)$ for the interval $\abs{E_{F}}\!<\!\omega\!<\!2\abs{E_{F}}$.
This distinctive features of vacancy-induced optical response is highlighted
in Fig.\,\ref{fig:OpticalConductivity}b, where it is also clear
that the change in sign of $\Im\left(\sigma_{\omega}^{xx}\right)$
no longer happens at $\omega\!=\!\abs{E_{F}}$ but at lower frequencies. 

\vspace{-0.6cm}

\paragraph{Physical Mechanisms:}

By now, it became clear that the optical response of a bulk WSM containing
lattice vacancies, displays very distinctive features that can be
unambiguously connected to the presence of vacancy-induced nodal states.
As a matter of fact, these features can be traced back to the macroscopic
number of states that the vacancies introduce close to the nodal energy,
and the fact that these are states which strongly deviate from extended
Bloch states. As such, the vacancy-induced states efficiently participate
in momentum non-conserving (but energy conserving) transitions to/from
the extended states at the Fermi level. Since the vacancy states only
occur for $\varepsilon\!\approx\!0$, these transitions give a contribution
to the real part of the conductivity starting at $\omega\!=\!\abs{E_{\text{F}}}$.
This interpretation also explains why the width of the transition
to the vacancy-induced absorption regime indirectly probes the energy
width of the central peak in the DoS and, therefore, also the extent
of inter-vacancy hybridization. Above the clean system's optical gap,
the most relevant processes are the vertical transition between valence/conduction
Bloch states. Since vacancies, unlike Anderson disorder, do not renormalize
strongly their Fermi velocity, the slope of $\Re\left(\sigma_{\omega}^{xx}\right)$
remains unchanged.

\lhead[\chaptername~\thechapter]{\rightmark}

\rhead[\MakeUppercase{Conclusions and Outlook}]{}

\lfoot[\thepage]{}

\cfoot[]{}

\rfoot[]{\thepage}

\chapter{\label{chap:Concluding-Remarks}Concluding Remarks and Outlook}

\vspace{-0.3cm}

Topological semimetals are well known for their unique electrodynamic
properties, but also for their robustness to perturbations which are
due to topological constraints. The topology of the band-structure
guarantees, to some extent, that Weyl nodes will remain stable for
a wide range of Hamiltonian deformations. However, it does not imply
that physical properties of the emergent Weyl fermions will be unaffected
by them. In this thesis, we studied the mechanisms by which different
types of disorder can alter the basic properties of the nodal quasiparticles
in a Weyl (or Dirac) semimetal. Our main focus was placed on the destabilization
of semi-metallic phase which can be induced by disorder in these systems,
and turns an incompressible gas of Weyl fermions into a fundamentally
different diffusive phase characterized by a non-vanishing density
of states. Thereby, the electronic mean density of states emerges
here as the central physical observable that characterizes the disordered
Dirac-Weyl semimetal, and which we compute in the presence of different
types of disorder: Anderson on-site potentials, random spherical scatterers,
scalar point-like impurities, and lattice point defects. By analyzing
all these cases, we have concluded that disorder effects in these
systems can show a very \textit{non-universal character} which deems
a model-by-model analysis important for interpreting the results of
future experimental studies. Despite having focused mostly on the
density of states in the presence of disorder, we have also assessed
some more easily measurable quantities, such as the electronic specific
heat, the \textit{dc conductivity}, the \textit{electronic magnetic
properties}, and also the \textit{linear optical response}. These
studies are important since they, not only indicate where to look
for relevant signs of disorder-induced phenomena in upcomming theoretical
studies, but may guide the fundamental explanation of future experimental
results.

Before moving onto the outlook of this work, we present a short description
of what the purpose and main conclusions of each Chapter are.

\vspace{-0.6cm}

\paragraph{Chapter\,\ref{chap:Introduction}:}

\!\!\!\!\!A self-contained introduction is given to the physics of
3D semi-metallic phases, highlighting their topological features,
presenting the different variants, and reviewing the major phenomenology
that makes these systems interesting from the physical point-of-view.

\vspace{-0.6cm}

\paragraph{Chapter\,\ref{chap:Mean-Field-Quantum-Criticality}:}

\!\!\!\!\!A detailed study of the (mean-field) disorder-induced semimetal-to-metal
transition in gapless 3D systems is presented. We start by analyzing
the mean DoS of an isolated Weyl node model in the presence of a short-range
correlated on-site potential. For that, we employ a diagrammatic \textit{Self-Consistent
Born Approximation}, as well as a \textit{Statistical Field Theory}
approach which is treated at the mean-field level (with a short reference
to the effect of \textit{loop corrections}). In both cases, an \textit{unconventional
disorder-induced quantum critical point} seems to emerge in these
systems, with the mean density of states serving as the order parameter.
All these results were confirmed by accurate lattice simulations. 

\vspace{-0.6cm}

\paragraph{Chapter\,\ref{chap:Instability_Smooth_Regions}:}

\!\!\!\!\!The polemic subject of non-perturbative effects due to
smooth and statistically rare-regions of a disordered landscape is
approached. To isolate this effect, we analyze a tailor-made model
in which random smooth regions are diluted within a 3D Weyl semimetal.
Combining a continuum scattering theory approach with unbiased large-scale
lattice simulations, we demonstrate that a vanishing concentration
of smooth regions destabilizes the semi-metallic phase, leading to
a physical picture differing from mean-field theory of dirty semimetals:
The \textit{Avoided Quantum Criticality}\,(AQC) scenario. Even though
this effect is connected to fine-tuned nodal bound states generated
by critical smooth regions , we show that its statistical significance
is actually guaranteed by a \textit{near-critical mechanism}, in which
\textit{needle-like resonances proliferate} around the nodal energy
and add-up to a finite DoS on average. 

\vspace{-0.6cm}

\paragraph{Chapter\,\ref{chap:Rare-Event-States}:}

\!\!\!\!\!A model of diluted point-like impurities is studied in
both a continuum and lattice model context. No bound states, or \textit{near-critical
mechanism}, is present for isolated point-like impurities, with a
nodal bound state emerging only through a coherent multiple scattering
involving (at least) two impurities. Nevertheless, this model clarified
the conditions under which the AQC appears in a disordered lattice.
Namely, we demonstrate that very small clusters, with a few adjacent
sites of similar on-site potential, are enough to support nodal bound
states akin the ones found for smooth regions in Chapter\,\ref{chap:Instability_Smooth_Regions}.
In addition, an exact diagonalization study of the nodal eigenstates
of a disordered WSM, we have shown that the existence of rare-event
states depends on large fluctuations of the on-site energies, which
can only be provided by unbounded distributions of the local potential.
Together, these results hint that the background DoS numerically found
by Pixley \textit{et al}.\,\cite{Pixley16a} can be generated by
two different mechanisms: \textit{(i)} from smooth regions of a few
adjacent sites, and \textit{(ii)} the hybridization of a large (atypical)
fluctuation of the potential with its (typical) disordered environment.
Within a disordered landscape, one expects both mechanisms to generate
nodal bound states and, thus, avoid the mean-field quantum critical
point. The first mechanism is statistically irrelevant for a disordered
landscape whose on-site random potential is drawn from a bounded,
or unbounded, distribution. Large smooth clusters are exceedingly
rare in either case. However, the second mechanism is only active
if the on-site potential can have unbounded values. This interpretation
predicts the AQC effects to be enhanced in disordered landscapes with
unbounded (or even fat-tailed) distributions and also explains why
these were absent in the accurate large-scale lattice simulations
presented in Chapter\,\ref{chap:Mean-Field-Quantum-Criticality},
whilst Pixley \textit{et al}. \textcolor{black}{were able to pinpoint
them using gaussian distributions for the on-site Anderson potential\@.}

\vspace{-0.6cm}

\paragraph{Chapter\,\ref{chap:Vacancies}:}

\!\!\!\!\!The effects of vacancies in the electronic structure and
transport properties of a lattice Weyl semimetal are considered. In
contrast to Anderson disorder, point-like defects are found to yield
much more extreme effects in the spectrum near the Weyl node. Firstly,
an isolated vacancy is unambiguously shown to create a bound state
at the nodal energy, which survives at finite concentrations as a
strongly enhanced nodal peak in the mean DoS that gets broadened by
inter-vacancy interference. Meanwhile, these inter-vacancy effects
are fundamentally different from the analogous in 2D Dirac systems,
with the usual broadening of the peak being accompanied by a comb
of subsidiary resonances that flank the nodal energy. These resonances
are composed of scattering states that still retain a great degree
of locality in real-space, being concentrated around a few nearby
vacancies. This local character is further confirmed by a great magnetic
insensitivity of the DoS at those energies, which also translates
to a \textit{strongly suppressed quantum diffusivity} that creates
dips in the bulk dc-conductivity as a function of the Fermi level.
Finally, the optical response of a Weyl semimetal with vacancies is
also evaluated and predicted to show robust experimental signatures
of vacancy-induced states. Most notably, these are shown to generate
a shorter optical gap, ending at $\omega\!=\!\left|E_{F}\right|$,
whose threshold frequency corresponds to momentum non-conserving transitions
between scattering states and the vacancy-induced nodal eigenstates.

\vspace{-0.4cm}

\subsection*{Outlook and Further Developments}

While providing a better grasp on the disorder-induced phenomena that
have been stirring the literature on dirty Dirac-Weyl semimetals for
the past decade, this work also serves to raise a new set of questions,
which call for further research. In the following, we divide these
questions into two major avenues.

The first route is of a more fundamental nature, and concerns an improved
theoretical understanding of the unconventional phase transition found
for weakly disordered Weyl fermions. We recall that the central point
of discussion in Chapters\,\ref{chap:Mean-Field-Quantum-Criticality}
to \ref{chap:Rare-Event-States} concerned the fact that the mean
density of states at the nodal energy does not seem to be a proper
order parameter for describing this quantum phase transition\,\cite{Su17}.
Nevertheless, this does not imply the nonexistence of a phase transition,
but rather that the order parameter is probably a more complex quantity.
As it often happens in disorder-driven critical phenomena (\textit{e.g.},
see Parisi\,\cite{Parisi79}), the critical behavior only gets well-defined
on a statistical sense, being encoded in the statistical properties
of a local quantity. For instance, the statistical distribution of
the \textit{local density of states} (LDoS), which is sensitive to
the small-scale structure of wavefunctions, can be used to distinguish
a system with delocalized eigenstates from a localized phase\,\cite{Schubert10,Alvermann05,Weise2006}.
In the former, the LDoS is shown to follow a normal distribution independently
of the system size, while in the latter, it is expected to take a
log-normal distribution with a \textit{typical value} that scales
down to zero with an increasing system size\,\cite{Mirlin2000}.
Historically, such a precise statistical definition of criticality
was a crucial step to fully understand the\textit{ Anderson metal-to-insulator
transition}, culminating with the identification of the geometric
mean of the LDoS as a proper order parameter of this transition (see
Refs.\,\cite{Thouless74,Thouless75,Wegner76,Abrahams79,Wegner79,Mckane81,Lerner88,Altshuler89,VanRossum94,Janssen94,Mirlin96,Janssen2001,Schubert10}).
A similar situation may actually happen for the semimetal-to-metal
transition, which can be only perceived by thoroughly studying the
statistical properties of local observables when crossing this (mean-field)
critical point. This is a research line that is still in its infancy\,\cite{Syzranov16,Brillaux19,Goncalves20},
but certainly deserves further development.

The second route, encouraged by the results presented here, arises
from the clear demonstration that alternative models of disorder are
worth studying in the context of Dirac-Weyl semimetals. Our experience
showed that going from the simplest Anderson potential to a slightly
more realistic random arrangement of vacancies is enough to create
radical changes in the electronic structure of the emerging Weyl fermions.
These changes are deemed to have major consequences in the electrodynamic
properties of topological semimetals, which may serve to explain future
experimental results. Overall, our results hint that the theoretical
consideration of models hosting more realistic sources of disorder,
such as \textit{random alloy models}\,\cite{Fu17,Zhang2019} or more
sophisticated lattice defects\,\cite{Besara2016}, may be a fruitful
research path that is worth following.

\[
\]

\cleardoublepage{}

\appendix
\global\long\def\vect#1{\overrightarrow{\mathbf{#1}}}%

\global\long\def\abs#1{\left|#1\right|}%

\global\long\def\av#1{\left\langle #1\right\rangle }%

\global\long\def\ket#1{\left|#1\right\rangle }%

\global\long\def\bra#1{\left\langle #1\right|}%

\global\long\def\tensorproduct{\otimes}%

\global\long\def\braket#1#2{\left\langle #1\mid#2\right\rangle }%

\global\long\def\omv{\overrightarrow{\Omega}}%

\global\long\def\inf{\infty}%

\global\long\def\ketbra#1#2{\left|#1\right\rangle \!\!\left\langle #2\right|}%

\lhead[\MakeUppercase{Appendix A}]{\MakeUppercase{\rightmark}}

\rhead[\MakeUppercase{Chebyshev Methods}]{}

\lfoot[\thepage]{}

\cfoot[]{}

\rfoot[]{\thepage}

\chapter{\label{chap:Crash-Course-KPM}A Crash Course on Spectral Methods
to Disordered Lattices}

Throughout this work, we have often turned to numerical simulation
methods in order to analyze the physical behavior of a Weyl semimetal
in the presence of perturbations that break translation symmetry.
This was done mainly by studying single-particle physical observables
derived from tight-binding Hamiltonian of the general form,

\vspace{-0.7cm}

\begin{equation}
H_{\text{TB}}\!=\!\sum_{\mathbf{R}}\Psi_{\mathbf{R}}^{\dagger}\!\cdot\!\boldsymbol{V}_{\mathbf{R}}\!\cdot\!\Psi_{\mathbf{R}}-\sum_{\mathbf{R}}\sum_{\mathbf{R}\neq\mathbf{R}^{\prime}}\Psi_{\mathbf{R}^{\prime}}^{\dagger}\!\cdot\!\boldsymbol{T}_{\mathbf{R^{\prime}},\mathbf{R}}\!\cdot\!\Psi_{\mathbf{R}},\label{eq:HamiltonianTB}
\end{equation}
where $\mathbf{R}$ and $\mathbf{R}^{\prime}$ are Bravais lattice
vectors, $\Psi_{\mathbf{R}}^{\dagger}=\left[c_{1,\mathbf{R}}^{\dagger},c_{2,\mathbf{R}}^{\dagger},\cdots,c_{N_{o},\mathbf{R}}^{\dagger}\right]^{T}$is
a fermionic creation operator that acts on the space of local orbitals,
$\boldsymbol{T}_{\mathbf{R^{\prime}},\mathbf{R}}$ is the inter-cell
hopping matrix, and $\boldsymbol{V}_{\mathbf{R}}$ is the intra-cell
Hamiltonian (both in the space of local orbitals). Typically, on disordered
problems the $H_{\text{TB}}$ is not a lattice periodic operator,
for it contains some random component, \textit{e.g.}, the local energies
in the a scalar Anderson potential. In that case, one cannot use Bloch's
Theorem to block diagonalize the Hamiltonian in $\mathbf{k}$-space
and is forced to work with the full matrix, which must necessarily
refer to a finite lattice system\,\footnote{As infinite lattices have $\infty$-dimentional Hamiltonians.}.
Remarkably, in this case, it often becomes more useful to work directly
in a real-space representation. Electrons in a disordered (or even
amorphous) solid-state system still propagate across a set of Wannier
orbitals that are pinned to fixed positions in space. Since these
orbitals are well localized, only short-ranged hoppings are allowed
and, therefore, the real-space Hamiltonian is typically a very sparse
matrix.

\vspace{-0.4cm}

\section{\label{sec:Spectral-Functions}Spectral Functions}

If one knows all eigenvalue/eigenvectors of $H_{\text{TB}}$, all
single-particle properties can be suitably obtained. However, for
very large lattice (with $N$ orbitals), this route is not a viable
option as full diagonalization\,\cite{Weisse2008} have a very disadvantageous
scaling with $N$ in both CPU-time (usually cubic) and memory (usually
quadratic\,\footnote{Note that most full exact diagonalization algorithms require all eigenvectors
to be calculated and saved at once, in order to ensure numerical stability\,\cite{Fernando97}.}). Meanwhile, one is usually interested in functions of the Hamiltonian
itself\,\cite{Weise2006,Joao2020}, such as:
\begin{itemize}
\item The \textit{global density of states per unit volume} (DoS),

\vspace{-0.7cm}

\begin{equation}
\rho(E)\!=\!\frac{1}{Nv_{c}}\text{Tr}\left[\delta\left(E\!-\!H_{\text{TB}}\right)\right],\label{eq:DoS-1}
\end{equation}
where $v_{c}$ is the volume of the Bravais lattice's unit cell.
\item The \textit{longitudinal dc conductivity} given by the \textit{Kubo-Greenwood
Formula}\,\cite{Kubo57,Greenwood58},

\vspace{-0.7cm}
\begin{equation}
\sigma_{dc}^{jj}\left(E_{F}\right)\!=\!\frac{\pi\hbar e^{2}}{Nv_{c}}\text{Tr}\left[V^{j}\delta\left(E_{F}\!-\!H_{\text{TB}}\right)V^{j}\delta\left(E_{F}\!-\!H_{\text{TB}}\right)\right],\label{eq:KuboGreenwood}
\end{equation}
where $\mathbf{V}$ is the velocity operator {[}defined in Eq.\,\eqref{eq:Velocities}{]},
$e$ is the electron's charge, $E_{F}$ is the Fermi energy, and $j\!=\!x,y,z$
are cartesian indices.
\item The \textit{linear dynamical conductivity tensor} given by the \textit{Kubo
formula} in a basis-independent way (see João \textit{et al}.\,\cite{Joao19}),

\vspace{-0.7cm}
\begin{align}
\sigma^{jl}(\omega) & =\frac{e^{2}}{i\hbar\omega Nv_{c}}\int dEf_{\text{FD}}^{E}\text{Tr}\left[V^{j}\mathcal{G}_{0}^{\text{r}}\left(-\hbar^{-1}\!E-\omega\right)V^{l}\delta\left(E-H_{\text{TB}}\right)\right.\label{eq:DynamicalCond}\\
 & \qquad\quad\left.+V^{j}\delta\left(E-H_{\text{TB}}\right)V^{l}\mathcal{G}_{0}^{\text{a}}\left(-\hbar^{-1}\!E+\omega\right)+i\hbar V^{jl}\delta\left(E-H_{\text{TB}}\right)\right],\nonumber 
\end{align}
where $\mathcal{G}_{0}^{\text{r/a}}$ are the single-particle Green's
functions (SPGFs) of $H_{\text{TB}}$ {[}defined in Eq.\,\eqref{eq:SPGF}{]},
$V^{i}$ ($V^{ij}$) are cartesian components of the (generalized)
velocity operator {[}defined in Eq.\,\eqref{eq:Velocities}{]}, $f_{\text{FD}}^{E}$
is the \textit{Fermi-Dirac distribution} of the system in equilibrium,
and $\omega$ is the angular frequency of the excitation field. 
\end{itemize}
In Eqs.\,\eqref{eq:KuboGreenwood}-\eqref{eq:DynamicalCond}, almost
all essential elements are well-known by now; We have several Dirac-$\delta$
distributions of $E\!-\!H_{\text{TB}}$, as well as retarded/advanced
SPGF operators which are defined as,

\vspace{-0.7cm}
\begin{equation}
\mathcal{G}_{\eta}^{\text{r}}\left(x\right)=\left[x\!+\!i\eta\!-\!H_{\text{TB}}\right]^{-1}\text{ and }\mathcal{G}_{\eta}^{\text{a}}\left(x\right)=\left[x\!-\!i\eta\!-\!H_{\text{TB}}\right]^{-1},\label{eq:SPGF}
\end{equation}
where $\eta\to0^{+}$. But, other than these, we also have the operators
$V^{j}$ and $V^{jl}$ which are \textit{generalized velocity operators}:

\vspace{-0.7cm}
\begin{equation}
V^{j}\!=\!\frac{1}{i\hbar}\left[R^{j},H_{l}\right]\text{ and }V^{jl}\!=\!-\frac{1}{\hbar^{2}}\left[R^{j},\left[R^{l},H_{l}\right]\right],\label{eq:Velocities}
\end{equation}
where $R^{i}$ is the $i^{\text{th}}$ cartesian component of the
position operator. At this point, it is important to note two important
details:\textit{ (i)} All the large matrices involved in the strings
of operators in Eqs.\,\eqref{eq:DoS-1}-\eqref{eq:DynamicalCond}
are typically sparse real-space matrices\,\footnote{Note that, since $H_{\text{TB}}$ is sparse matrix in real-space,
so are the generalized velocities.}, and \textit{(ii)} all these quantities are expressed as traces over
the entire Hilbert space of the single-particle system. In the following
section, we will show how the aforementioned properties can be used
to efficiently calculate these spectral functions using Chebyshev
expansions and stochastic evaluation of the trace (see Weisse \textit{et
al}.\,\cite{Weise2006} for an extensive review).

\section{\label{sec:Chebyshev-Expansions}Chebyshev Expansions}

The \textit{Kernel Polynomial Method} (KPM) or, more generally, the\textit{
Chebyshev Iteration Method} (CIM\nomenclature{CIMs}{Chebyshev Iteration Methods})
is based upon the expansion of any function $h(x)$, defined for $x\in[-1,1]$
as an infinite Chebyshev series, \textit{i.e.},

\vspace{-0.7cm}
\begin{equation}
h(x)=\sum_{n=0}^{\infty}\frac{2\mu_{n}T_{n}\left(x\right)}{1+\delta_{n0}},
\end{equation}
where $T_{n}\left(x\right)\!=\!\cos\left(n\arccos x\right)$ are the
first-kind Chebyshev polynomials that obey the orthonormality condition,

\vspace{-0.7cm}
\begin{equation}
\int_{-1}^{1}\!\!\!dx\frac{T_{n}\left(x\right)T_{m}\left(x\right)}{\pi\sqrt{1+x^{2}}}=\delta_{nm}\frac{1+\delta_{n0}}{2},
\end{equation}
and the following three-term recursion relation,

\vspace{-0.7cm}

\begin{equation}
T_{n+1}\left(x\right)=2xT_{n}\left(x\right)-2T_{n-1}\left(x\right),
\end{equation}
initialized by $T_{0}\left(x\right)=1$ and $T_{1}\left(x\right)=x$.
In principle, every function defined in $x\in\left[-1,1\right]$ can
be expanded as a Chebyshev series, with the coefficients $\mu_{n}$
universally given as 

\vspace{-0.7cm}
\begin{equation}
\mu_{n}\!=\!\int_{-1}^{1}\!\!\!dx\,\frac{h(x)T_{n}(x)}{\pi\sqrt{1+x^{2}}}.
\end{equation}
For our purposes, we are interested in two particular expansions,
$h_{1}(a,x)=\delta(a-x)$ and $h_{2}(z,x)=1/(z\!-\!x)$, where $a$
is a real value and $z$ is a complex value of the form $z=a\pm i\eta$
with $\eta\to0^{+}$. In both cases, the form of the Chebyshev coefficients
is known analytically\,\cite{Huang93,Braun2014,Ferreira2015}, namely

\begin{figure}[t]
\vspace{-0.4cm}
\begin{centering}
\includegraphics[scale=0.23]{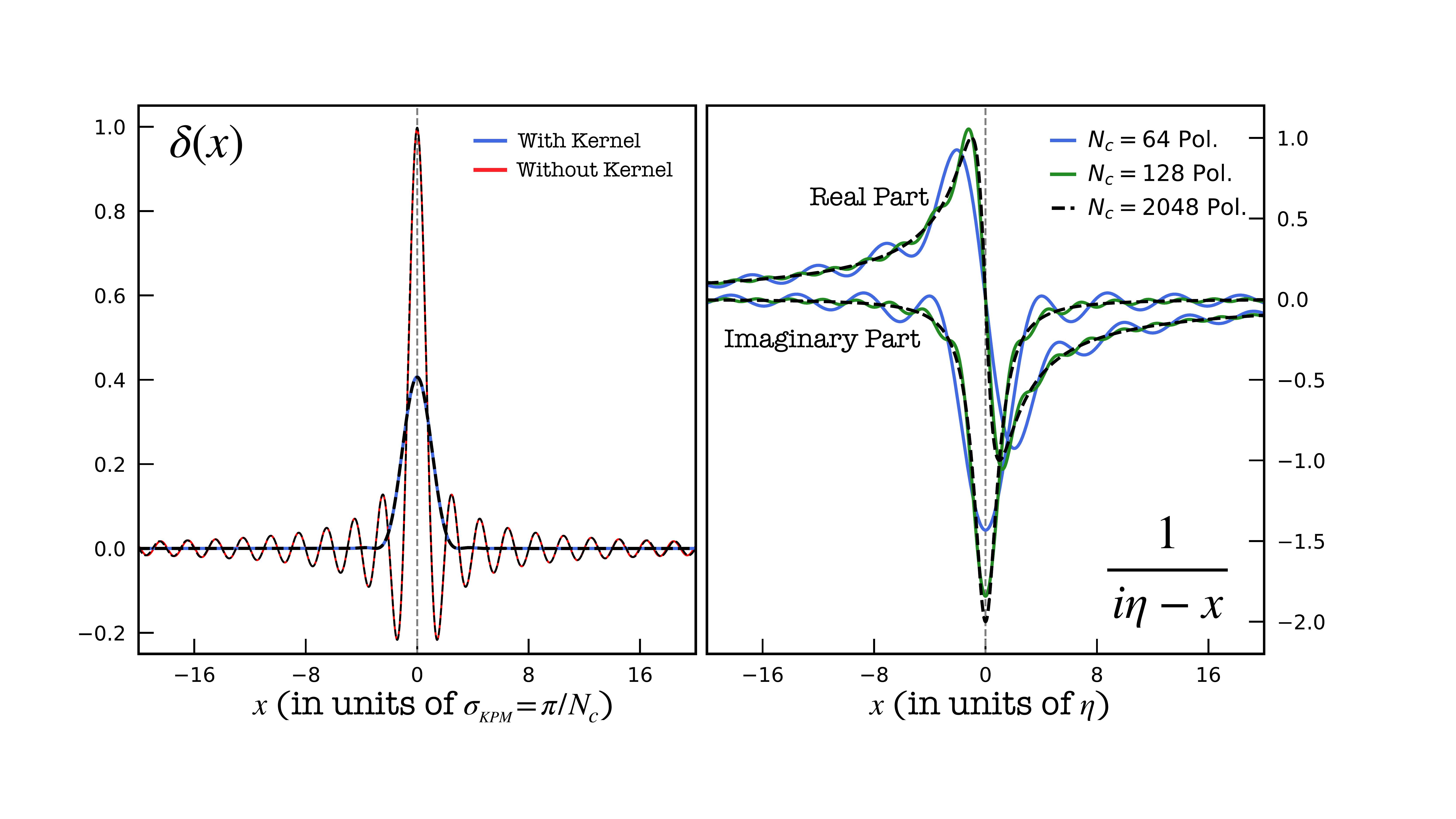}
\par\end{centering}
\caption{\label{fig:ChebyExpansion}Example of the Chebyshev approximation
of a Dirac-$\delta$ function (left) and a complex-valued green's
function, both centered in $x\!=\!0$. For $\delta(x)$ we compare
two expansions ($N_{c}\!=\!128$ and $512$ polynomials) with and
without a Jackson damping kernel. For the green's function, there
is no damping kernel.}

\vspace{-0.6cm}
\end{figure}

\vspace{-0.7cm}
\begin{align}
\delta\left(a\!-\!x\right) & =\sum_{n=0}^{\infty}\frac{2\mu_{n}^{\delta}\left(a\right)T_{n}\left(x\right)}{1+\delta_{n0}},\text{ where }\mu_{n}^{\delta}\left(a\right)\!=\!\frac{T_{n}\left(a\right)}{\pi\sqrt{1+a^{2}}}\\
\frac{1}{a\pm i\eta-x} & =\sum_{n=0}^{\infty}\frac{2\mu_{n}^{g\pm}\left(a,\eta\right)T_{n}\left(x\right)}{1+\delta_{n0}},\text{ where }\mu_{n}^{g\pm}\left(a,\eta\right)\!=\!\pm\frac{2e^{\mp in\arccos\left(a\pm i\eta\right)}}{i\sqrt{1\!-\!\left(a\pm i\eta\right)^{2}}}.\label{eq:GF_coefs}
\end{align}
In practice, such a series is only useful if it allows accurate approximations
when truncated to a finite order $N_{c}$. In Fig.\,\ref{fig:ChebyExpansion},
we show the convergence of the expansion for the function $\delta(x)$,
as a function of the number of polynomials kept in the expansion.
Since the $\delta$-function is singular, the convergence of this
series is poor due to the so-called \textit{Gibbs phenomenon}; higher-order
polynomials dominate the expansion while trying to reproduce point
discontinuities in the expanded function\,\cite{Gibbs1899,Weise2006}.
This problem, however, can be mended by introducing a damping kernel
in the expansion, that is,

\vspace{-0.7cm}
\begin{equation}
\delta\left(a-x\right)\approx\sum_{n=0}^{N_{c}}\frac{2\mu_{n}^{\delta}\left(a\right)T_{n}\left(x\right)}{1+\delta_{n0}}\to\sum_{n=0}^{N_{c}}\frac{2g_{n}\left(N_{c}\right)\mu_{n}^{\delta}\left(a\right)T_{n}\left(x\right)}{1+\delta_{n0}},
\end{equation}
where $g_{n}\left(N_{c}\right)$ depends on the truncation order but
is not unique. The optimal choice for expanding Dirac-$\delta$ functions
is to use the positive-definite \textit{Jackson's kernel}\,\cite{Jackson1912},
defined as

\vspace{-0.7cm}

\begin{equation}
g_{n}^{N_{c}}=\frac{\left(N_{c}-n+1\right)\cos\left(\frac{\pi\,n}{N_{c}+1}\right)+\sin\left(\frac{\pi\,n}{N_{c}+1}\right)\cot\left(\frac{\pi}{N_{c}+1}\right)}{N_{c}+1}.
\end{equation}
The introduction of this kernel effectively broadens the $\delta$-function
into a gaussian-like function of width $\sigma_{\text{KPM}}\approx\pi/N_{c}$\,\cite{Weise2006}.
In this case, the resolution of the expansion is therefore indexed
to the number of polynomials alone. In contrast, for the function
$1/\left(a\pm i\eta-x\right)$, Fig.\,\ref{fig:ChebyExpansion} shows
that the convergence is much more controlled so long as $\eta$ (an
effective\textit{ ``lorentzian broadening}'') is finite. Then, the
resolution is built into the expanded function. In the following,
we will employ both kinds of expansions.

\vspace{-0.4cm}

\section{Chebyshev Spectral Expansions}

The main idea of the Chebyshev spectral methods is that all the ideas
of functional expansion presented in Sect.\,\ref{sec:Chebyshev-Expansions}
can be translated to arbitrary functions of a bounded hermitian operator\,\cite{TalEzer84,Weise2006}.
The only care that must be taken is to shift and normalize the corresponding
operator, so that its spectrum lies fully inside the interval $[-1,1]$,
for which the Chebyshev series converges. More precisely, if we have
an intensive physical quantity that can be expressed as 

\vspace{-0.7cm}

\begin{equation}
Q(X)=\frac{1}{Nv_{c}}\text{Tr}\left[f(X,H_{\text{TB}})\right],\label{eq:KPM1}
\end{equation}
where $X$ is a set of parameters and $H_{l}$ is the tight-binding
Hamiltonian. We start by transforming $H_{\text{TB}}\to\tilde{H}_{\text{TB}}=\left(H_{\text{TB}}-a\right)/\lambda$,
where $a$ centers the spectrum around $0$ and $\lambda$ rescales
it. Then, we can expand the function $f$ in terms of Chebyshev polynomials
of the Hamiltonian operator, \textit{i.e.},

\vspace{-0.7cm}
\begin{equation}
Q(X)=\!\!\frac{1}{v_{c}}\sum_{n=0}^{\infty}\frac{2\mu_{n}\!(X)}{1+\delta_{n0}}\left(\frac{1}{N}\text{Tr}\left[T_{n}\left(\tilde{H}_{l}\right)\right]\right),
\end{equation}
where $\mu_{n}\left(X\right)$ are the Chebyshev-expansion coefficients
of $f$ (with respect to the second argument) and, importantly, the
term in $\left(\cdots\right)$\,\footnote{Usually called the $n^{\text{th}}$-Chebyshev Moment.}
can be efficiently calculated using the recursion relation,

\vspace{-0.7cm}

\begin{equation}
T_{n+1}\left(\tilde{H}_{l}\right)=2H_{l}T_{n}\left(\tilde{H}_{l}\right)-2T_{n-1}\left(\tilde{H}_{l}\right),
\end{equation}
with $T_{0}\left(\tilde{H}_{l}\right)=I$ and $T_{1}\left(\tilde{H}_{l}\right)=\tilde{H}_{l}$.
In addition, if $N$ is very large, the trace can be calculated using
a few stochastic vectors\,\cite{Weise2006}, namely

\vspace{-0.7cm}

\begin{equation}
\frac{1}{N}\text{Tr}\left[T_{n}\left(\tilde{H}_{l}\right)\right]\to\frac{1}{R}\sum_{r=0}^{R}\bra{\chi_{r}}T_{n}\left(\tilde{H}_{l}\right)\ket{\chi_{r}},\label{eq:KPM4}
\end{equation}
where $\ket{\chi_{r}}$ are a set of $R$ independently generated
vectors, whose elements can be chosen as random complex phases. Equations\,\eqref{eq:KPM1}-\eqref{eq:KPM4}
form the basis of any computation that uses a variant of the Kernel
Polynomial Method (KPM)\,\cite{Weise2006}. Now, we will specialize
the calculation to the three quantities indicated in Sect.\,\ref{sec:Spectral-Functions}.

\vspace{-0.4cm}

\paragraph{Global Density of States:}

The density of states is perhaps the easiest quantity of all to calculate
with KPM. The Dirac-$\delta$ of Eq.\,\eqref{eq:DoS-1} is expanded
in a truncated Chebyshev series that is complemented by a Jackson
damping kernel,

\vspace{-0.7cm}

\begin{equation}
\tilde{\rho}_{{\scriptscriptstyle \text{KPM}}}(\tilde{E},N_{c})\!=\!\frac{1}{v_{c}\sqrt{1-\tilde{E}^{2}}}\left(1+2\sum_{n=1}^{\infty}g_{n}^{N_{c}}T_{n}\left(\tilde{E}\right)\Gamma_{n}\right),\label{eq:DoS-1-1}
\end{equation}
where $N_{c}$ is the number of polynomials in the expansion, $\Gamma_{n}=\text{Tr}\left[T_{n}\left(H_{l}\right)\right]/N$
and $\tilde{E}=\left(E-a\right)/\lambda$. Since this quantity is
a spectral density, returning to the original energy variable involves
the following transformation:

\vspace{-0.7cm}

\begin{equation}
\rho_{{\scriptscriptstyle \text{KPM}}}(E,N_{c})\!=\!\lambda^{-1}\,\tilde{\rho}_{{\scriptscriptstyle \text{KPM}}}\left(\left(E-a\right)/\lambda,N_{c}\right).
\end{equation}
Before proceeding, it is important to remark that this expression
has a finite resolution $\sigma_{{\scriptscriptstyle \text{KPM}}}\!\approx\!\pi\lambda/N_{c}$
which will affect its comparison to possible analytical results that
may exists. Since, we encounter this situation in Chapter\,\ref{chap:Instability_Smooth_Regions},
here is a way to compare these results, accounting for the finite
resolution of the calculation,

\vspace{-0.7cm}
\begin{equation}
\rho_{{\scriptscriptstyle \text{analy}}}^{\sigma_{{\scriptscriptstyle \text{KPM}}}}\!(E,N_{c})=\frac{1}{\sqrt{2\pi}\sigma_{{\scriptscriptstyle \text{KPM}}}}\int_{-\infty}^{\infty}\!\!\!dx\rho_{{\scriptscriptstyle \text{analy}}}\!(x)\exp\left[-\frac{\left(E-x\right)^{2}}{2\sigma_{{\scriptscriptstyle \text{KPM}}}^{2}}\right],
\end{equation}
where the analytical expression for the DoS, $\rho_{{\scriptscriptstyle \text{analy}}}\!(x)$,
is convoluted with a gaussian that has a broadening consistent with
the spectral resolution of the KPM calculation.

\vspace{-0.4cm}

\paragraph{Longitudinal DC Conductivity:}

Like the DoS, the dc conductivity as defined in Eq.\,\eqref{eq:KuboGreenwood},
involves only velocity operators and two Dirac-$\delta$ functions
of the lattice Hamiltonian that could be expanded in a was similar
to what was done prior. However, we follow the alternative approach
of Ferreira \textit{et al}.\,\cite{Ferreira2015}, and rewrite the
dc conductivity as \vspace{-0.7cm}
\begin{equation}
\sigma_{dc}^{jj}\left(E_{F}\right)\!=\!-\frac{\hbar e^{2}}{\pi Nv_{c}}\text{Tr}\left[V^{j}\:\Im\left[\mathcal{G}_{\eta}^{\text{0r}}\!\left(E_{F}\right)\right]\:V^{j}\:\Im\left[\mathcal{G}_{\eta}^{\text{0r}}\!\left(E_{F}\right)\right]\right].\label{eq:KuboGreenwood-1}
\end{equation}
which, as we will see, allows for a more efficient computational method.
In this form, and after a proper normalization of the lattice Hamiltonian,
we can proceed with a double expansion of the SPGFs, \textit{i.e.},

\vspace{-0.7cm}

\begin{equation}
\Im\left[\mathcal{G}_{\tilde{\eta}}^{\text{KPMr}}\!\left(\tilde{E},N_{c}\right)\right]=\sum_{n=0}^{N_{c}}\mu_{n}^{g+}\left(\tilde{E},\tilde{\eta}\right)\frac{T_{n}\left(\tilde{H}_{l}\right)}{1+\delta_{n0}},
\end{equation}
where the coefficients $\mu_{n}^{g\pm}$ are defined as in Eq.\,\eqref{eq:GF_coefs}.
Using this expansion into Eq.\,\eqref{eq:KuboGreenwood-1}, one obtains

\vspace{-0.7cm}

\begin{equation}
\!\!\!\!\!\!\!\!\!\!\!\sigma_{{\scriptscriptstyle \text{KPM}}}^{jj}\!\left(\tilde{E}_{F},N_{c}\right)\!=\!-\frac{\hbar e^{2}}{\pi v_{c}}\sum_{n,m}^{N_{c}}\mu_{n}^{g+}\!\left(\tilde{E}_{F},\tilde{\eta}\right)\mu_{m}^{g+}\!\left(\tilde{E}_{F},\tilde{\eta}\right)\left(\frac{1}{N}\text{Tr}\left[V^{j}\:\frac{T_{n}\left(\tilde{H}_{l}\right)}{1+\delta_{n0}}\:V^{j}\:\frac{T_{m}\left(\tilde{H}_{l}\right)}{1+\delta_{m0}}\right]\right)\!\!\label{eq:KuboGreenwood-1-1}
\end{equation}
for which the numerically expensive part is to perform the trace over
the Hilbert space. In the notation of Ref.\,\cite{Joao19}, we can
define 

\vspace{-0.7cm}
\begin{equation}
\Gamma_{nm}^{j,l}=\frac{1}{N}\text{Tr}\left[V^{j}\:\frac{T_{n}\left(\tilde{H}_{l}\right)}{1+\delta_{n0}}\:V^{l}\:\frac{T_{m}\left(\tilde{H}_{l}\right)}{1+\delta_{m0}}\right]
\end{equation}
and therefore, the KPM dc conductivity simply reads

\vspace{-0.7cm}
\begin{equation}
\sigma_{{\scriptscriptstyle \text{KPM}}}^{jj}\!\left(E_{F},N_{c}\right)\!=\!-\frac{\hbar e^{2}}{\pi v_{c}}\sum_{n,m}^{N_{c}}\mu_{n}^{g+}\!\left(\left(E_{F}-a\right)/\lambda,\tilde{\eta}\right)\mu_{m}^{g+}\!\left(\left(E_{F}-a\right)/\lambda,\tilde{\eta}\right)\Gamma_{nm}^{j,j}.
\end{equation}
Note that the trace in $\Gamma_{nm}^{j,j}$ can be calculated using
stochastic vectors in the normal way\,\footnote{Note that here, and in all the following steps, the application of
generalized velocity operators has the exact same computational complexity
as a Hamiltonian-vector operation.} but, in the end, all elements of an $N_{c}\times N_{c}$ matrix of
double Chebyshev moments have to be build and saved in order to reconstruct
$\sigma_{{\scriptscriptstyle \text{KPM}}}^{jj}$ for all values of
the Fermi energy. However, if a single Fermi energy is of interest,
one could get away with an $\mathcal{O}(N_{c})$ number of operations
by using the so-called \textit{Single-Shot Conductivity Method} (SSCM\nomenclature{SSCM}{Single-Shot Conductivity Method})\,\cite{Ferreira2015,Joao2020}
in which one starts with a random vector $\ket{\chi_{r}}$ and simultaneously
calculate 
\begin{figure}[t]
\vspace{-0.4cm}
\begin{centering}
\includegraphics[scale=0.5]{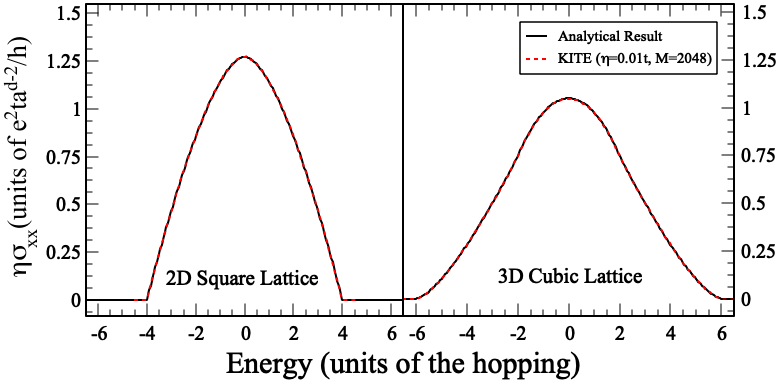}
\par\end{centering}
\vspace{-0.2cm}

\caption{\label{fig:SingleShotExample}Comparison between analytical results
and the SSCM for a range Fermi energies, $E_{F}$, in both the two-dimensional
square lattice and the three-dimensional cubic lattice. The calculation
was done using the implementation of QuantumKITE\,\cite{Joao2020}.}

\vspace{-0.5cm}
\end{figure}

\vspace{-1.0cm}

\begin{subequations}
\begin{align}
\ket{\chi_{r}^{\text{R}}}\! & =\!\frac{1}{N}\sum_{n=0}^{\infty}\mu_{n}^{g+}\!\left(\left(E_{F}-a\right)/\lambda,\tilde{\eta}\right)V^{j}\:\frac{T_{n}\left(\tilde{H}_{l}\right)}{1+\delta_{n0}}\ket{\chi_{r}}\\
\ket{\chi_{r}^{\text{L}}}\! & =\!\frac{1}{N}\sum_{n=0}^{\infty}\mu_{n}^{g+}\!\left(\left(E_{F}-a\right)/\lambda,\tilde{\eta}\right)\frac{T_{n}\left(\tilde{H}_{l}\right)}{1+\delta_{n0}}\:V^{j}\ket{\chi_{r}}
\end{align}
\end{subequations}

which can be cast in the form of a couple of independent two-point
Chebyshev recursions in the random vector. Finally, after doing this,
the conductivity can be recovered as 

\vspace{-0.7cm}
\begin{equation}
\sigma_{{\scriptscriptstyle \text{KPM}}}^{jj}\!\left(E_{F},N_{c}\right)\!=\!-\frac{\hbar e^{2}}{\pi v_{c}R}\sum_{r=1}^{R}\braket{\chi_{r}^{\text{L}}}{\chi_{r}^{\text{R}}}.
\end{equation}
In Fig.\,\ref{fig:SingleShotExample}, we showcase two calculations
done using the SSCM for different Fermi energies, in the simple models
of a 2D square lattice and 3D cubic lattice with only nearest-neighbor
hoppings. The comparison with the analytical result is perfect. 

Note that, for each value of $E_{F}$ and $\eta$, this calculation
involved $2N_{c}\times R$ operations, against the $N_{c}^{2}\times R$
operations needed for the full-spectrum dc conductivity calculation.
Physically, this simplification only arises because the longitudinal
dc conductivity only depends on the Fermi energy, which does not hold
true in general.

\paragraph{Linear Dynamical Conductivity:}

The most complex observable we compute in this work is the linear
conductivity tensor, $\sigma^{jl}\!\left(\omega\right)$, which is
a function of the electric field's frequency $\omega$. The expression
of Eq.\,\eqref{eq:DynamicalCond} involves Dirac-$\delta$ functions
of $H_{l}$ as well as SPGF. All of these can be expanded as Chebyshev
series, with analytically known coefficients and Chebyshev moments
that can be determined by recursion. This representation have already
been explored in previous work\,\cite{Garcia15,Ferreira16,Joao19,Joao2020,Sinha22}
and we limit ourselves to the final expressions. Following João et
al.\,\cite{Joao19}, we have 

\vspace{-0.7cm}
\begin{equation}
\sigma_{\!{\scriptscriptstyle \text{KPM}}}^{jl}\!\left(\tilde{E}_{F},\tilde{\omega},N_{c}\right)\!=\!\frac{e^{2}}{iv_{c}\hbar^{2}\tilde{\omega}}\left[\sum_{n=0}^{N_{c}}\Gamma_{n}^{jl}\Lambda_{n}+\!\!\sum_{n,m=0}^{N_{c}}\Gamma_{nm}^{j,l}\Lambda_{nm}^{\!\tilde{\omega}}\right]\label{eq:DynamicalCond-1}
\end{equation}
where all the energy scales are properly re-scaled by $\lambda$.
In Eq.\,\eqref{eq:DynamicalCond-1}, the symbols

\vspace{-0.7cm}

\begin{subequations}
\begin{align}
\Gamma_{n}^{jl} & =\frac{1}{N}\text{Tr}\left[V^{jl}\frac{T_{n}\left(\tilde{H}_{l}\right)}{1+\delta_{n0}}\right]\\
\Gamma_{nm}^{j,l} & =\frac{1}{N}\text{Tr}\left[V^{j}\frac{T_{n}\left(\tilde{H}_{l}\right)}{1+\delta_{n0}}V^{l}\frac{T_{m}\left(\tilde{H}_{l}\right)}{1+\delta_{m0}}\right],
\end{align}
\end{subequations}

denote traces that have to be evaluated stochastically, while

\vspace{-0.7cm}

\begin{subequations}
\hspace{-0.7cm}
\begin{align}
\!\!\!\!\!\Lambda_{n}\!= & \frac{2}{\pi}\int\!\!d\tilde{E}\,\frac{f_{\text{FD}}^{\tilde{E}}T_{n}(\tilde{E})}{\sqrt{1-\tilde{E}^{2}}}\\
\!\!\!\!\Lambda_{nm}^{\!\tilde{\omega}}\!\!\!= & \frac{2\hbar}{\pi}\!\int\!\!d\tilde{E}\,\frac{f_{\text{FD}}^{\tilde{E}}\left[\kappa_{n}^{{\scriptscriptstyle +}}\left(\hbar^{{\scriptscriptstyle -1}}\!\tilde{E}+\omega,\tilde{\eta}\right)T_{m}\left(\tilde{E}\right)\!+\!T_{n}\left(\tilde{E}\right)\kappa_{m}^{{\scriptscriptstyle -}}\left(\hbar^{{\scriptscriptstyle -1}}\!\tilde{E}-\omega,\tilde{\eta}\right)\right]}{\sqrt{1-\tilde{E}^{2}}},\!\!\!\!
\end{align}
\end{subequations}

which are energy-integrals that must be done by numerical quadrature.
All in all, the calculation of the dynamical conductivity with KPM
requires $N_{c}\!\times\!N_{c}$ operations but the scaling with system
size ($N$) remains unaltered. In addition, one can also see that,
unlike what was happened in the DoS, conductivities evaluated with
rescaled Hamiltonians retain their nominal value, with $\omega$ and
$E_{F}$ being measured in units of rescaled energies.

\vspace{-0.4cm}

\section{Comments on Implementation Efficiency}

Before closing this Appendix, it is worth remarking some important
features of the previous methods which makes them stand as go-to tools
for the study of non-interacting lattice models. In fact, the whole
power of the KPM method is that the bottleneck operation is a matrix-vector
product, $\tilde{H}_{l}\ket{\chi_{r}}$. This is a feature that the
Kernel Polynomial Method shares with other (so-called) iterative $\mathcal{O}(N)$
matrix methods, such as the \textit{Lanczos-Arnoldi Methods} (used
in Chapters\,\ref{chap:Rare-Event-States} and \ref{chap:Vacancies}
for numerical diagonalization) or the \textit{Generalized Minimal
Residual Method} (GMRES). 

\begin{figure}[t]
\vspace{-0.4cm}
\begin{centering}
\includegraphics[scale=0.23]{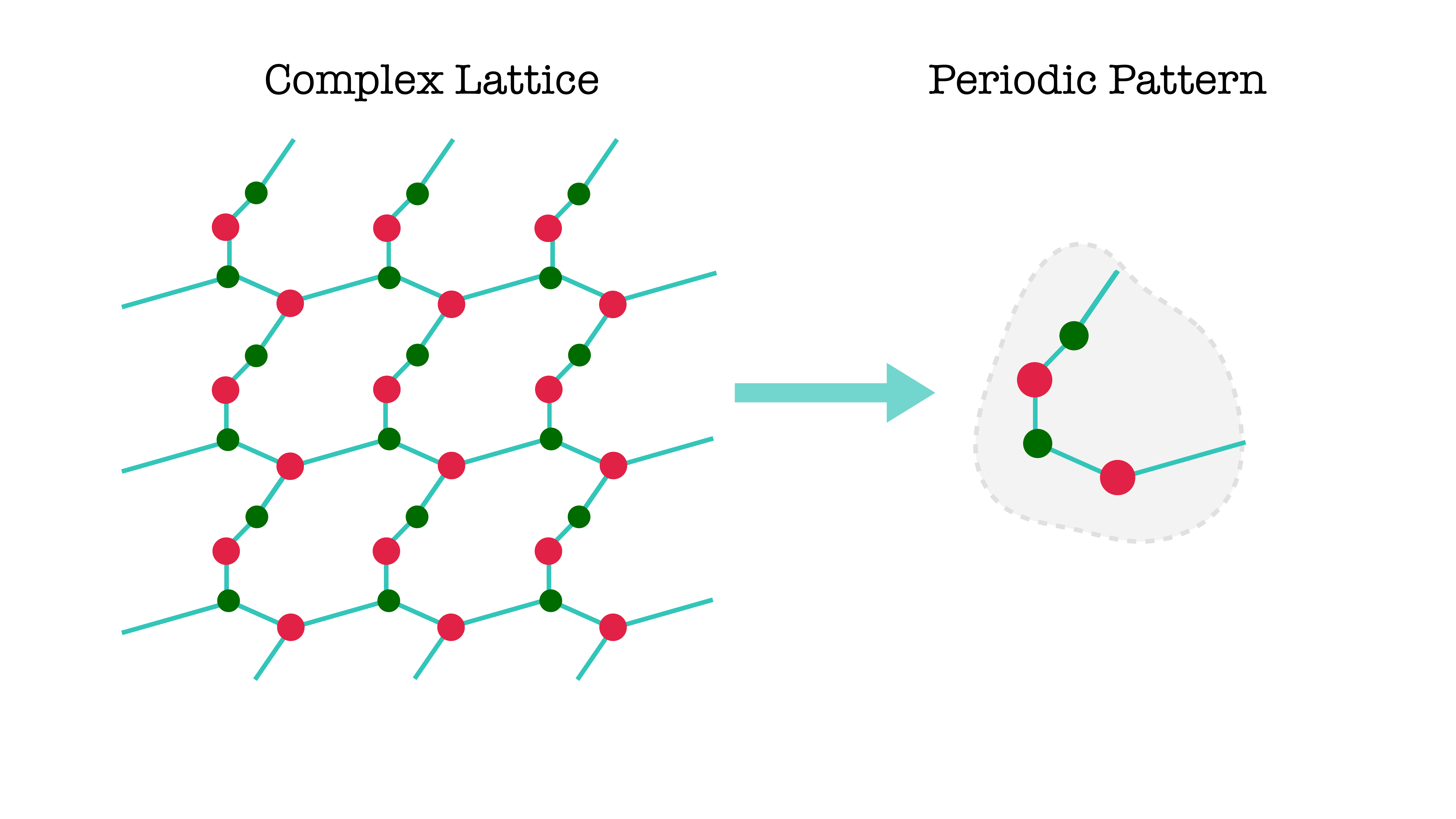}
\par\end{centering}
\caption{\label{fig:SingleShotExample-1}Scheme of the local Pattern used in
the matrix-free implementation of KPM in the QuantumKITE\,\cite{Joao2020}.}

\vspace{-0.5cm}
\end{figure}

The efficiency of these methods rely on both their iterative nature,
but crucially, on the sparse nature of the matrices. In fact, a matrix-vector
usually involves $N^{2}$ elementary operations but, if the matrix
is very sparse\,\footnote{As the Hamiltonian is in a real-space representation.}
this number is lowered to only $N$. In addition, for a disordered
or defective solid-state system, there is always an underlying periodic
lattice structure. This structure allows the matrix-vector product
to be performed without ever building the matrix explicitly, but rather
using a matrix-free approach in which a repeated local pattern acts
independently on each spacial index of $\ket{\chi_{r}}$. Any deviation
from this pattern can then be applied on top of this periodic structure,
usually involving a substantially smaller number of operations. In
simpler words, the periodic part of the Hamiltonian can be encoded
as a local graph, which is schematically represented in Fig.\,\ref{fig:SingleShotExample-1}.
These techniques, together with an efficient CPU-parallelization of
the matrix-vector product by domain-decomposition, are the main cornerstones
that make the implementation of the $\texttt{QuantumKITE}$\,\cite{Joao2020},
extensively used here, a particularly efficient and capable of reaching
enormous lattice sizes.

\lhead[\MakeUppercase{Appendix B}]{\MakeUppercase{\rightmark}}

\rhead[\MakeUppercase{Twisted Boundary Conditions}]{}

\chapter{\label{chap:TwistedBoundaries-1}Superlattices and Twisted Boundary
Conditions}

One of the main drawbacks when simulating finite Dirac-Weyl semimetal
lattices comes from the highly unfavorable scaling of the mean-level
spacing near the nodal energy with the simulated system size. For
example, if one considers our working tight-binding model, whose Bloch
Hamiltonian reads

\vspace{-0.7cm}
\begin{equation}
\mathcal{H}_{l}^{0}(\mathbf{k})=\frac{\hbar v_{\text{F}}}{a}\boldsymbol{\sigma}\cdot\boldsymbol{\sin}a\mathbf{k},
\end{equation}
the eigenstates are Bloch-waves with a crystal momentum $\mathbf{k}\in[-\pi/a,\pi/a]^{3}$
and with the dispersion relation,

\vspace{-0.7cm}
\begin{equation}
E_{\mathbf{k}}\!=\!\pm\frac{\hbar v_{\text{F}}}{a}\sqrt{\sin^{2}ak_{x}\!+\!\sin^{2}ak_{y}\!+\!\sin^{2}ak_{z}}.
\end{equation}
If the finite lattice is considered to have dimensions $L_{x}\!\times\!L_{y}\!\times\!L_{z}$
and standard \textit{Periodic Boundary Conditions} (PBC\nomenclature{PBC}{Periodic Boundary Conditions}),
the usual quantization condition in $\mathbf{k}$ applies, \textit{i.e.}\,\footnote{Note that we are assuming $L_{i}$ to be even numbers.},

\vspace{-0.7cm}
\begin{equation}
\mathbf{k}\!=\!\pi a^{-1}\left(\frac{n_{x}}{L_{x}},\frac{n_{y}}{L_{y}},\frac{n_{z}}{L_{z}}\right)\text{ with integers }n_{i}=\left\{ -\frac{L_{i}}{2}+1,\cdots,\frac{L_{i}}{2}-1,\frac{L_{i}}{2}\right\} ,\label{eq:QuantizationPBC}
\end{equation}
which translate into a corresponding quantization of the spectrum
in a discrete set of energy levels. Exactly at $E\!=\!0$, the lattice
system has $16$ degenerate energy levels corresponding two the 8
Kramers' doublets (see Subsect.\,\ref{subsec:Time-Reversal-Symmetry}
for a discussion of Kramers Theorem) placed at the TRIM of the fBz.
The nearby energy levels will be symmetrically placed around $E\!=\!0$
and will be due to $\mathbf{k}$ that are distanced by $\pm\pi/\max(aL_{x},aL_{y},aL_{z})$
which corresponds to a spacing in energy of 

\vspace{-0.7cm}
\begin{equation}
\delta E\approx\hbar v_{\text{F}}/\max(L_{x},L_{y},L_{z}).
\end{equation}
Therefore, we conclude that the linear dispersion relation of the
system around $E\!=\!0$ leads to a mean level spacing around the
node that scales as $N^{\frac{1}{3}}$, where $N$ is the total number
of lattice sites. This is obviously a problem for the use of spectral
methods in lattice simulations as, ultimately, the resolution is limited
by the mean-level spacing. Therefore, if one eight-fold increases
the simulated system, the CPU-time is also eight-folded but the mean-level
spacing gets only diminished by a factor of two.

\section{Phase-Twisted Boundary Conditions}

In $\mathbf{k}$-space, a finite lattice looks like a compact first
Brillouin zone (fBz) which is discretized in a regular lattice of
points allowed by the boundary conditions in the real-space lattice.
If one changes the boundary conditions in real-space, then so will
change the lattice of quantized $\mathbf{k}$-points. The choice to
impose PCB in a lattice simulation is mostly a matter of convenience;
These do not break any symmetries of the infinite lattice, do not
give rise to unwanted boundary phenomena and, in the limit of a very
large lattice, physical results are expected to match the ones found
in a truly infinite lattice. Nevertheless, this is not the only choice
that matches the above criteria as, for instance, one may impose the
generalized condition in the real-space wavefunctions,

\vspace{-0.7cm}
\begin{equation}
\Psi(\mathbf{R}\!+\!L_{i}\mathbf{a}_{i})=\exp\left(i\varphi_{i}\frac{\mathbf{a}_{i}\cdot\mathbf{b}_{i}}{2\pi}\right)\Psi(\mathbf{R}),\label{eq:TwistedBoundaries}
\end{equation}
where $\mathbf{a}_{i}$($\mathbf{b}_{i}$) are the primitive generators
of the real-space (dual) lattice and $\varphi_{i}\in[0,2\pi]$ is
a phase-angle. Clearly, the boundary condition defined in Eq.\,\eqref{eq:TwistedBoundaries}
includes the PBC as a particular case, for which the phase-twists
are $\boldsymbol{\varphi}\!=\!\left(\varphi_{1},\varphi_{2},\varphi_{3}\right)\!=\!\boldsymbol{0}$.
However, it further defines a much broader class of boundary conditions,
which we collectively call \textit{Twisted Boundary Conditions} (TBC\nomenclature{TBC}{Twisted Boundary Conditions}).
In fact, one can easily show that the quantization condition for $\mathbf{k}$-space,
as induced by a set of phase-twists $\boldsymbol{\varphi}$ is given
as

\begin{figure}[t]
\vspace{-0.4cm}
\begin{centering}
\includegraphics[scale=0.22]{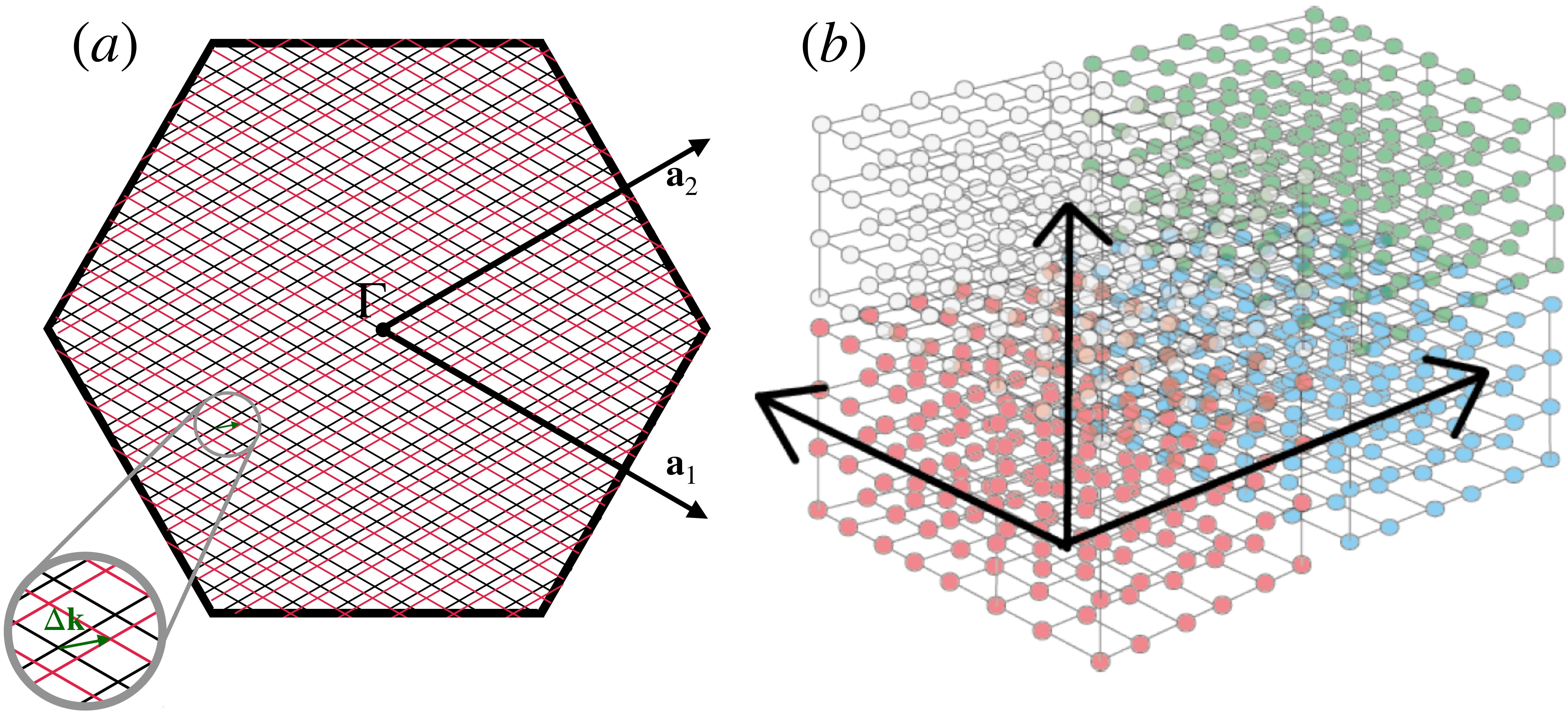}
\par\end{centering}
\caption{\label{fig:TwistedBoundaries}(a) Shift of the lattice of quantized
$\mathbf{k}$s in the presence of a boundary phase-twist. The black
(red) intersections refer to the values allowed by periodic (twisted)
boundaries (b) Superlattice used to understand the meaning of an average
over twisted boundary conditions. The arrows represent the primitive
translation vectors associated to the superlattice, while the different
colors identify four different supercells.}

\vspace{-1.0cm}
\end{figure}

\vspace{-0.7cm}

\begin{equation}
\mathbf{k}\!=\!\left(\frac{n_{1}+\nicefrac{\varphi_{1}}{2\pi}}{L_{1}}\right)\mathbf{a}_{1}+\left(\frac{n_{2}+\nicefrac{\varphi_{2}}{2\pi}}{L_{2}}\right)\mathbf{a}_{2}+\left(\frac{n_{3}+\nicefrac{\varphi_{3}}{2\pi}}{L_{3}}\right)\mathbf{a}_{3},\label{eq:k_TBC}
\end{equation}
where $n_{i}$ are the usual integers (defined in Eq.\,\ref{eq:QuantizationPBC})
and the lattice was assumed to have dimensions $L_{1}\times L_{2}\times L_{3}$
along the primitive real-space directions. Equation\,\eqref{eq:k_TBC}
offers a very transparence description of the effects of TBC in the
$\mathbf{k}$-space quantization: It leads to a global shift of the
lattice of allowed $\mathbf{k}$s by a vector $\Delta\mathbf{k}=\left(\varphi_{1}/L_{1},\varphi_{2}/L_{2},\varphi_{3}/L_{3}\right)/2\pi$.
This is depicted in Fig.\,\ref{fig:TwistedBoundaries}\,a for the
special case of a 2D hexagonal lattice.

\vspace{-0.5cm}

\section{Superlattices by Monte-Carlo Sampling}

The poor scaling of the mean-level spacing with the the system size
can be seen as caused by a ``poor sampling'' of the first Brillouin
zone near the TRIM, which is imposed by the boundary conditions. If
one fixes the boundary conditions, the only way to increase the quality
of this sampling is to increase the size of the simulated system and
that has great cost in both memory usage and CPU-time. However, the
previous analyses of the more general TBC entails a different strategy
to increase the sampling in $\mathbf{k}$: Fix the system size but
average over random boundary twists. More concretely, we define a
tight-binding Hamiltonian with a boundaries characterized by a vector
of twist phase-angles, i.e. $H_{l}(\boldsymbol{\varphi})$, and for
an arbitrary observable 

\vspace{-0.7cm}

\begin{equation}
Q(X)=\frac{1}{Nv_{c}}\av{\text{Tr}\left[f\left(X,H_{l}(\boldsymbol{\varphi})\right)\right]}_{\boldsymbol{\varphi}},\label{eq:KPM1-1}
\end{equation}
where $\av{\cdots}_{\boldsymbol{\varphi}}$ is an arithmetic average
over a uniform sampling of the vector $\boldsymbol{\varphi}\in[0,2\pi]^{3}$.
In practice, we will do a KPM evaluation of $Q(X)$, performing a
simultaneous averaging over boundary twists and random vectors to
evaluate the trace over the Hilbert space. Clearly, if $H_{l}$ is
a lattice periodic tight-binding Hamiltonian the $\text{Tr}\left[\cdots\right]$
can be replaced by a summation over the allowed $\mathbf{k}$'s in
the fBz, and therefore this boundary-averaging procedure is equivalent
to a Monte-Carlo (MC\nomenclature{MC}{Monte-Carlo}) evaluation of
a fBz integral. Hence, the method becomes exact if the system is a
periodic one.

For a non-periodic simulated system, the situation turns out to be
analogous. The easiest way to see this would be to establish an analogy
with the manipulation done to obtain Eq.\,\ref{eq:TwistedBloch}
in the introductory chapter of this thesis. There, we saw that solving
an eigenvalue problem in a compact space complemented by twisted boundary
conditions is the same as treating that same problem for a given Bloch
$\mathbf{k}$ in an infinite periodic superlattice. By summing over
the phase-twists, one is effectively summing over all $\mathbf{k}$'s
within the superlattice's fBz. In conclusion, by calculating any observable
of a lattice Hamiltonian with TBCs, and then averaging the result
over the phase-twists, one is effectively simulating that observable
in an infinite superlattice obtained by repeating the finite simulated
lattice. Although this may prove an useful resource for boosting numerical
convergence, care must taken at all instances, so as to guarantee
that there are no periodicity artifacts being introduced this way.

\lhead[\MakeUppercase{Appendix C}]{\MakeUppercase{\rightmark}}

\rhead[\MakeUppercase{Dirac Equations in Spherical Coordinates}]{}

\chapter{\label{chap:Dirac_Spherical}The Dirac Equation in Spherical Coordinates}

In Chapter\,\ref{chap:Instability_Smooth_Regions}, it was crucial
to have the analytical solution for the eigenstates of a continuum
single-node Dirac Hamiltonian in the presence of a spherical scattering
potential. The possibility of achieving a separation of variables
of a rotationally invariant Dirac (or Weyl) equation in the spherical
coordinates is a known problem in relativistic quantum mechanics'
literature (see, for example, Björken and Drell\,\cite{Bjorken65}),
whose results we have adapted for our purposes. However, due to its
technical nature, we find it useful to write this Appendix providing
the reader with clear details on how to solve this problem.

Even though in Chapter\,\ref{chap:Instability_Smooth_Regions} we
restricted ourselves to the gapless case, here we provide a more general
treatment of a gapped Dirac fermion, which is represented by a four-component
wavefunction, $\Psi\left(t,\mathbf{r}\right)$, and whose quantum
dynamics is determined by the Hamiltonian,

\vspace{-0.7cm}

\begin{equation}
\mathcal{H}_{\text{D}}=-i\hbar v_{\text{F}}\boldsymbol{\alpha}\!\cdot\!\boldsymbol{\nabla}+mv_{\text{F}}^{2}\beta+\mathcal{U}\left(\mathbf{r}\right),\label{eq:DiracHamiltonian+ScalarPotential}
\end{equation}
where $\mathcal{U}\left(\mathbf{r}\right)=eV\left(\mathbf{r}\right)\mathbb{I}_{4\times4}$
is a generic scalar potential,

\vspace{-0.7cm}
\begin{equation}
\beta=\left(\begin{array}{cc}
\mathbb{I}_{2\times2} & \mathbb{O}_{2\times2}\\
\mathbb{O}_{2\times2} & -\mathbb{I}_{2\times2}
\end{array}\right),
\end{equation}
is a Dirac-mass term and the $4\times4$ $\alpha$-matrices can be
written as

\vspace{-0.7cm}

\begin{equation}
\alpha^{i}=\left(\begin{array}{cc}
\mathbb{O}_{2\times2} & \sigma_{i}\\
\sigma_{i} & \mathbb{O}_{2\times2}
\end{array}\right),\label{eq:AlphaMatrices}
\end{equation}
in terms of the usual Pauli matrices $\sigma_{i}$. Note that, unlike
Chapter\,\ref{chap:Instability_Smooth_Regions}, here we opted to
use a Dirac representation for the spinor matrices. The two representations
are equivalent and related by a unitary transformation.

\vspace{-0.5cm}

\section{\label{sec:Angular-Momentum}Angular Momentum and Rotation Invariance}

In the absence of the scalar potential\,\footnote{Or consider that the potential is spherically symmetric, i.e., $\mathcal{U}\left(\mathbf{r}\right)=\mathcal{U}\left(r\right)$.},
the Hamiltonian of Eq.\,\eqref{eq:DiracHamiltonian+ScalarPotential}
is rotationally symmetric and thus has good quantum numbers associated
to angular momentum. Nevertheless, unlike in the Schrödinger problem,
the wavefunction $\Psi\left(t,\mathbf{r}\right)$ has itself non-trivial
rotation properties due to its spinor character. From the relativistic
Dirac theory, one has an angular momentum associated to orbital degrees
of freedom, 

\vspace{-0.7cm}

\begin{equation}
\mathbf{L}=i\hbar\left(y\partial_{z}-z\partial_{y},z\partial_{x}-x\partial_{z},x\partial_{y}-y\partial_{x}\right)\mathbb{I}_{4\times4},\label{eq:OrbitalAngularMomentum}
\end{equation}
 and an intrinsic one related to the spin of the four-component wavefunction,

\vspace{-0.7cm}

\begin{align}
\mathbf{S} & \!=\!\frac{\hbar}{2}\left(\left(\begin{array}{cc}
\sigma_{x} & \mathbb{O}_{2\times2}\\
\mathbb{O}_{2\times2} & \sigma_{x}
\end{array}\right),\left(\begin{array}{cc}
\sigma_{y} & \mathbb{O}_{2\times2}\\
\mathbb{O}_{2\times2} & \sigma_{y}
\end{array}\right),\left(\begin{array}{cc}
\sigma_{z} & \mathbb{O}_{2\times2}\\
\mathbb{O}_{2\times2} & \sigma_{z}
\end{array}\right)\right).\!\!\label{eq:SpinOperator}
\end{align}
Equation\,\eqref{eq:SpinOperator} implies that the spin (or pseudo-spin)
degree of freedom, in the Dirac representation, is a block-diagonal
operator in the four-dimensional spinor space. Joining together both
components of the angular momentum, we can establish that the cartesian
components of the total angular momentum operator read as follows,

\vspace{-0.7cm}

\begin{subequations}
\begin{align}
J_{x} & =L_{x}+S_{x}=\left(\begin{array}{cc}
L_{x}+\frac{\hbar}{2}\sigma_{x} & \mathbb{O}_{2\times2}\\
\mathbb{O}_{2\times2} & L_{x}+\frac{\hbar}{2}\sigma_{x}
\end{array}\right)\label{eq:Jx}\\
J_{y} & =L_{y}+S_{y}=\left(\begin{array}{cc}
L_{y}+\frac{\hbar}{2}\sigma_{y} & \mathbb{O}_{2\times2}\\
\mathbb{O}_{2\times2} & L_{y}+\frac{\hbar}{2}\sigma_{y}
\end{array}\right)\label{eq:Jy}\\
J_{z} & =L_{z}+S_{z}=\left(\begin{array}{cc}
L_{z}+\frac{\hbar}{2}\sigma_{z} & \mathbb{O}_{2\times2}\\
\mathbb{O}_{2\times2} & L_{z}+\frac{\hbar}{2}\sigma_{z}
\end{array}\right)\label{eq:Jz}
\end{align}
\end{subequations}

\vspace{-0.5cm}

Now, we will see that all components of the operator $\mathbf{J}=\left(J_{x},J_{y},J_{z}\right)$\,\footnote{Which are the generators of three-dimensional rotations.}
commute with $\mathcal{H}_{\text{D}}$, provided the applied potential,
$\mathcal{U}\left(\mathbf{r}\right)$, is spherically symmetric. To
prove it, we treat each term in Eq.\,\eqref{eq:DiracHamiltonian+ScalarPotential}
separately:
\begin{itemize}
\item \textbf{Massless Term: }$\mathcal{H}_{\text{D}}^{(1)}=-i\hbar v_{\text{F}}\boldsymbol{\alpha}\!\cdot\!\boldsymbol{\nabla}=v_{\text{F}}\boldsymbol{\alpha}\cdot\mathbf{p}$,
where $\mathbf{p}$ is the linear momentum operator. In this notations,
one has

\vspace{-0.7cm}

\begin{align}
J_{i}\mathcal{H}_{\text{D}}^{(1)} & =\left(\begin{array}{cc}
\mathbb{O}_{2\times2} & \sigma_{j}L_{i}p_{j}+\frac{\hbar}{2}\sigma_{i}\sigma_{j}p_{j}\\
\sigma_{j}L_{i}p_{j}+\frac{\hbar}{2}\sigma_{i}\sigma_{j}p_{j} & \mathbb{O}_{2\times2}
\end{array}\right)\\
\mathcal{H}_{\text{D}}^{(1)}J_{i} & =\left(\begin{array}{cc}
\mathbb{O}_{2\times2} & \sigma_{j}p_{j}L_{i}+\frac{\hbar}{2}\sigma_{j}\sigma_{i}p_{j}\\
\sigma_{j}p_{j}L_{i}+\frac{\hbar}{2}\sigma_{j}\sigma_{i}p_{j} & \mathbb{O}_{2\times2}
\end{array}\right),
\end{align}

\vspace{-0.5cm}

where repeated indices are summed over $1,2$ and $3$. The commutator
between the massless term and the components of the angular momentum
then read

\vspace{-0.7cm}

\begin{equation}
\left[J_{i},\mathcal{H}_{\text{D}}^{(1)}\right]=\left(\begin{array}{cc}
\mathbb{O}_{2\times2} & \sigma_{j}\left[L_{i},p_{j}\right]+\frac{\hbar}{2}\left[\sigma_{i},\sigma_{j}\right]p_{j}\\
\sigma_{j}\left[L_{i},p_{j}\right]+\frac{\hbar}{2}\left[\sigma_{i},\sigma_{j}\right]p_{j} & \mathbb{O}_{2\times2}
\end{array}\right).\label{eq:Commutator1}
\end{equation}
Now, we only need to recognize that, since $\left[x_{k},p_{l}\right]=i\hbar\delta_{k,l}$
and $L_{i}=\varepsilon_{i,j,k}x_{j}p_{k}$, we have

\vspace{-0.9cm}
\begin{equation}
\left[p_{n},L_{j}\right]=\varepsilon_{jkl}\left[p_{n},x_{k}p_{l}\right]=\varepsilon_{jkl}\left[p_{n},x_{k}\right]p_{l}=-i\hbar\varepsilon_{jkl}\delta_{nk}p_{l}=i\hbar\varepsilon_{njl}p_{l}.\label{eq:CommutatorL_P}
\end{equation}
Simultaneously, we also have $\left[\sigma_{i},\sigma_{j}\right]=2i\varepsilon_{ijk}\sigma_{k}$.
Using these $\text{su}(2)$ algebraic relations into Eq.\,\eqref{eq:Commutator1},
we arrive at

\vspace{-0.7cm}
\begin{equation}
\left[J_{i},\mathcal{H}_{\text{D}}^{(1)}\right]=i\hbar\varepsilon_{ijk}\left(\begin{array}{cc}
\mathbb{O}_{2\times2} & \sigma_{j}p_{k}+\sigma_{k}p_{j}\\
\sigma_{j}p_{k}+\sigma_{k}p_{j} & \mathbb{O}_{2\times2}
\end{array}\right)=\mathbb{O}_{4\times4},\label{eq:Commutator1-1}
\end{equation}
since $\varepsilon_{ijk}\left(\sigma_{j}p_{k}+\sigma_{k}p_{j}\right)$
is a contraction of a symmetric with an antissymetric tensor in the
$j$ and $k$ indices.
\item \textbf{Mass Term:} $\mathcal{H}_{\text{D}}^{(2)}=Mv_{\text{F}}^{2}\beta$.
For this term we have

\vspace{-0.7cm}
\begin{align}
J_{i}\mathcal{H}_{\text{D}}^{(2)} & =mv_{\text{F}}^{2}\left(\begin{array}{cc}
L_{i}+\frac{\hbar}{2}\sigma_{i} & \mathbb{O}_{2\times2}\\
\mathbb{O}_{2\times2} & -L_{i}-\frac{\hbar}{2}\sigma_{i}
\end{array}\right)\\
\mathcal{H}_{\text{D}}^{(2)}J_{i} & =mv_{\text{F}}^{2}\left(\begin{array}{cc}
L_{i}+\frac{\hbar}{2}\sigma_{i} & \mathbb{O}_{2\times2}\\
\mathbb{O}_{2\times2} & -L_{i}-\frac{\hbar}{2}\sigma_{i}
\end{array}\right),
\end{align}

which immediately yields $\left[J_{i},\mathcal{H}_{\text{D}}^{(2)}\right]=\mathbb{O}_{4\times4}$. 
\item \textbf{Potential Term:} $\mathcal{H}_{\text{D}}^{(3)}=\mathcal{U}\left(r\right)\mathbb{I}_{4\times4}$.
In this case, the result is obvious since the spinor structure is
irrelevant, as the operator is scalar and $L_{i}$, once written in
spherical coordinates, does not involve derivatives with respect to
the radial coordinate. 
\end{itemize}
Summing up all the previous results, i.e. $\mathcal{H}_{\text{D}}\!=\!\mathcal{H}_{\text{D}}^{(1)}\!+\!\mathcal{H}_{\text{D}}^{(2)}\!+\!\mathcal{H}_{\text{D}}^{(3)}$,
we have

\vspace{-0.7cm}

\begin{equation}
\left[J_{i},\mathcal{H}_{\text{D}}\right]=\mathbb{O}_{4\times4}\Longrightarrow\left[\abs{\mathbf{J}}^{2},\mathcal{H}_{\text{D}}\right]=\!\!\sum_{i=1,2,3}\left[J_{i}^{2},\mathcal{H}_{\text{D}}\right]\!=\!\mathbb{O}_{4\times4}.\label{eq:AngularMomentumConservation}
\end{equation}
Unsurprisingly, if the scalar potential term does not have angular
dependences, the massive Dirac Hamiltonian is rotationally symmetric
and, consequently, one may build eigenstates of well-defined eigenvalues
associated to both $J_{z}$ and $\abs{\mathbf{J}}^{2}$. 

\vspace{-0.5cm}

\section{Total Orbital and Spin Angular Momenta}

The spherically symmetric Dirac Hamiltonian has a conserved total
angular momentum that is composed of two components: an orbital part,
$\mathbf{L}$, and a spin-$\nicefrac{1}{2}$ part, $\mathbf{S}$.
The cartesian components of these two operators are not conserved,
by themselves, but the same does not happen for the operator,

\vspace{-0.7cm}

\begin{equation}
\abs{\mathbf{S}}^{2}=\frac{\hbar^{2}}{4}\left(\begin{array}{cc}
\sum_{i=1,2,3}\sigma_{i}^{2} & \mathbb{O}_{2\times2}\\
\mathbb{O}_{2\times2} & \sum_{i=1,2,3}\sigma_{i}^{2}
\end{array}\right)=\frac{3\hbar^{2}}{4}\mathbb{I}_{4\times4},\label{eq:S^2}
\end{equation}
which with $\mathcal{H}_{\text{D}}$, as it is proportional to the
identity operator. In contrast, although the orbital operator $\abs{\mathbf{L}}^{2}$
commutes both with $\mathcal{H}_{\text{D}}^{(2)}$ and $\mathcal{H}_{\text{D}}^{(3)}$,
we can show that it does not commute with the massless term, i.e.

\vspace{-0.7cm}

\begin{align}
\left[\abs{\mathbf{L}}^{2},\mathcal{H}_{\text{D}}^{(1)}\right] & =i\hbar\left(\!\begin{array}{cc}
\mathbb{O}_{2\times2} & \varepsilon_{ijk}\sigma_{j}L_{i}p_{k}\\
\varepsilon_{ijk}\sigma_{j}L_{i}p_{k} & \mathbb{O}_{2\times2}
\end{array}\!\right)+i\hbar\left(\!\begin{array}{cc}
\mathbb{O}_{2\times2} & \varepsilon_{ijk}\sigma_{j}p_{k}L_{i}\\
\varepsilon_{ijk}\sigma_{j}p_{k}L_{i} & \mathbb{O}_{2\times2}
\end{array}\!\right)\nonumber \\
 & =i\hbar\left(\begin{array}{cc}
\mathbb{O}_{2\times2} & \varepsilon_{ijk}\varepsilon_{ilm}\sigma_{j}x_{l}p_{m}p_{k}\\
\varepsilon_{ijk}\varepsilon_{ilm}\sigma_{j}x_{l}p_{m}p_{k} & \mathbb{O}_{2\times2}
\end{array}\right)\label{eq:CommLH}\\
 & \qquad\qquad\qquad\qquad\quad+i\hbar\left(\begin{array}{cc}
\mathbb{O}_{2\times2} & \varepsilon_{ijk}\varepsilon_{ilm}\sigma_{j}p_{k}x_{l}p_{m}\\
\varepsilon_{ijk}\varepsilon_{ilm}\sigma_{j}p_{k}x_{l}p_{m} & \mathbb{O}_{2\times2}
\end{array}\right).\nonumber 
\end{align}
Now, using the identity $\varepsilon_{ijk}\varepsilon_{ilm}=\delta_{jl}\delta_{km}-\delta_{jm}\delta_{lk}$,
and recognizing that one must have $j\neq k$, or otherwise the product
of Levi-Civita symbols would yield zero, it is clear that $p_{k}x_{j}=x_{j}p_{k}$
and $x_{k}p_{k}=p_{k}x_{k}+i\hbar$. Hence, Eq.\,\eqref{eq:CommLH}
can be written as

\vspace{-0.7cm}

\begin{align}
\left[\abs{\mathbf{L}}^{2},\mathcal{H}_{\text{D}}^{(1)}\right] & =-\hbar^{2}\left(\begin{array}{cc}
\mathbb{O}_{2\times2} & \sigma_{j}p_{j}\\
\sigma_{j}p_{j} & \mathbb{O}_{2\times2}
\end{array}\right)\neq\mathbb{O}_{4\times4},
\end{align}
which means that $\abs{\mathbf{L}}^{2}$ does not provide good quantum
numbers to label eigenfunctions of $\mathcal{H}_{\text{D}}$.

\vspace{-0.5cm}

\section{The Conserved Spin-Orbit Operator}

Till now, we have identified that $J_{z}$, $\abs{\mathbf{J}}^{2}$
and $\abs{\mathbf{S}}^{2}$ provide common good quantum numbers for
a spherically symmetric massive Dirac Hamiltonian. Notwithstanding,
there is another conserved operator built out of these two, the \textit{Spin-Orbit
Operator}, $\mathbf{S}\cdot\mathbf{L}$. Actually, it will be useful
to consider a shifted version,

\vspace{-0.7cm}

\begin{equation}
\mathcal{K}=\gamma^{0}\left(2\mathbf{S}\cdot\mathbf{L}+\hbar^{2}\mathbb{I}_{4\times4}\right)=\left(\begin{array}{cc}
\hbar\overrightarrow{\sigma}\cdot\overrightarrow{L}+\hbar^{2}\mathbb{I}_{2\times2} & \mathbb{O}_{2\times2}\\
\mathbb{O}_{2\times2} & -\hbar\overrightarrow{\sigma}\cdot\overrightarrow{L}-\hbar^{2}\mathbb{I}_{2\times2}
\end{array}\right),\label{eq:K_Operator-1-1-1}
\end{equation}
where $\gamma_{0}\!=\!\text{diag}\left(1,1,-1,-1\right)$ is a diagonal
matrix, and which can be shown directly to commute will all terms
in the Hamiltonian of Eq.\,\ref{eq:DiracHamiltonian+ScalarPotential}.
But does this quantity provide new good quantum numbers, compatible
with the previously found ones? In order to see that, we must calculate
the commutators of $\mathcal{K}$ with these operators. This is an
easy operation for each individual component of $\mathbf{J}$, i.e

\vspace{-0.7cm}

{\footnotesize{}
\begin{align}
\!\!\!\!\!\!\!\mathcal{K}J_{i} & =\left(\begin{array}{cc}
\hbar\sigma_{j}L_{j}L_{i}+\frac{\hbar^{2}}{2}L_{j}\sigma_{j}\sigma_{i}+\hbar^{2}L_{i}+\frac{\hbar}{2}\sigma_{i} & \mathbb{O}_{2\times2}\\
\mathbb{O}_{2\times2} & -\hbar\sigma_{j}L_{j}L_{i}-\frac{\hbar^{2}}{2}L_{j}\sigma_{j}\sigma_{i}-\hbar^{2}L_{i}-\frac{\hbar}{2}\sigma_{i}
\end{array}\right)\!\!\!\\
\!\!\!\!\!\!\!J_{i}\mathcal{K} & =\left(\begin{array}{cc}
\hbar\sigma_{j}L_{i}L_{j}+\frac{\hbar^{2}}{2}L_{j}\sigma_{i}\sigma_{j}+\hbar^{2}L_{i}+\frac{\hbar}{2}\sigma_{i} & \mathbb{O}_{2\times2}\\
\mathbb{O}_{2\times2} & -\hbar\sigma_{j}L_{i}L_{j}-\frac{\hbar^{2}}{2}L_{j}\sigma_{i}\sigma_{j}-\hbar^{2}L_{i}-\frac{\hbar}{2}\sigma_{i}
\end{array}\right),\!\!\!
\end{align}
}meaning that the commutator between the components of $\mathbf{J}$
and $\mathcal{K}$ take the form,

\vspace{-0.7cm}

\begin{equation}
\left[\mathcal{K},J_{i}\right]=\left(\begin{array}{cc}
\hbar\sigma_{j}\left[L_{i},L_{j}\right]+\frac{\hbar^{2}}{2}L_{j}\left[\sigma_{i},\sigma_{j}\right] & \mathbb{O}_{2\times2}\\
\mathbb{O}_{2\times2} & -\hbar\sigma_{j}\left[L_{i},L_{j}\right]-\frac{\hbar^{2}}{2}L_{j}\left[\sigma_{i},\sigma_{j}\right]
\end{array}\right).
\end{equation}
Since we know that $\left[L_{i},L_{j}\right]=i\hbar\varepsilon_{ijk}L_{k}$
and $\left[\sigma_{i},\sigma_{j}\right]=2i\varepsilon_{ijk}\sigma_{k}$,
we get

\vspace{-0.7cm}

\begin{equation}
\left[\mathcal{K},J_{i}\right]=i\hbar^{2}\left(\begin{array}{cc}
\varepsilon_{ijk}\left[L_{k},\sigma_{j}\right] & \mathbb{O}_{2\times2}\\
\mathbb{O}_{2\times2} & \varepsilon_{ijk}\left[L_{k},\sigma_{j}\right]
\end{array}\right)=\mathbb{O}_{4\times4}
\end{equation}
and then arrive at the final conclusion that $\mathcal{K}$ is also
a scalar operator. This is enough to affirm that $\mathcal{H}_{\text{D}},J_{z},\abs{\mathbf{J}}^{2},\abs{\mathbf{S}}^{2}$
and $\mathcal{K}$ are complete set of commuting observable for the
spherically symmetric Dirac equation.

\vspace{-0.5cm}

\section{Spin-$\nicefrac{1}{2}$ Spherical Harmonics}

Following Refs.\,\cite{Bjorken65,Ma85,Ma06}, we now build common
eigenspinor of $J_{z}$, $\abs{\mathbf{J}}^{2}$ and the conserved
operator $\mathcal{K}$. To do this, we being by remind the explicit
form of these operators in the question as $4\times4$ matrices in
spinor space. Namely, we have

\vspace{-0.7cm}

\begin{align}
J_{z} & =\left(\!\begin{array}{cccc}
L_{z}+\frac{\hbar}{2} & 0 & 0 & 0\\
0 & L_{z}-\frac{\hbar}{2} & 0 & 0\\
0 & 0 & L_{z}+\frac{\hbar}{2} & 0\\
0 & 0 & 0 & L_{z}-\frac{\hbar}{2}
\end{array}\!\right)\label{eq:Jz_Matrix4x4}\\
\mathcal{K} & =\left(\begin{array}{cccc}
\hbar^{2}+\hbar L_{z} & \hbar L_{-} & 0 & 0\\
\hbar L_{+} & \hbar^{2}-\hbar L_{z} & 0 & 0\\
0 & 0 & -\hbar^{2}-\hbar L_{z} & -\hbar L_{-}\\
0 & 0 & -\hbar L_{+} & -\hbar^{2}+\hbar L_{z}
\end{array}\right),\label{eq:SpinOrbit}
\end{align}

\vspace{-0.7cm}

\begin{equation}
\abs{\mathbf{J}}^{2}=\abs{\mathbf{L}}^{2}+\abs{\mathbf{S}}^{2}+2\mathbf{S}\cdot\mathbf{L}=\left(\begin{array}{cc}
\mathbb{A} & \mathbb{O}_{2\times2}\\
\mathbb{O}_{2\times2} & \mathbb{A}
\end{array}\right)\label{eq:Jz_Matrix4x4-2}
\end{equation}
with

\vspace{-0.7cm}

\begin{equation}
\mathbb{A}\!=\!\left(\!\begin{array}{cc}
\abs{\mathbf{L}}^{2}+\hbar L_{z}+\frac{3\hbar^{2}}{4} & \hbar L_{-}\\
\hbar L_{+} & \abs{\mathbf{L}}^{2}-\hbar L_{z}+\frac{3\hbar^{2}}{4}
\end{array}\!\right),
\end{equation}
and where $L_{\pm}=L_{x}\pm iL_{y}$ are the orbital angular momentum
ladder operators. The first step is to write the four-component wavefunction
in spherical coordinates, 

\vspace{-0.7cm}
\begin{equation}
\Psi(r,\theta,\varphi)=\left(\!\begin{array}{c}
\psi_{1}\left(r,\theta,\varphi\right)\\
\psi_{2}\left(r,\theta,\varphi\right)\\
\psi_{3}\left(r,\theta,\varphi\right)\\
\psi_{4}\left(r,\theta,\varphi\right)
\end{array}\!\right),
\end{equation}
and then express the $L_{z}$ operator as a differential operators
in spherical coordinates, $L_{z}=i\hbar\partial_{\varphi},$which
allows $J_{z}$ to be expressed as follows:

\vspace{-0.7cm}

\begin{equation}
J_{z}=\hbar\left(\!\!\begin{array}{cccc}
i\frac{\partial}{\partial\varphi}+\frac{1}{2} & 0 & 0 & 0\\
0 & i\frac{\partial}{\partial\varphi}-\frac{1}{2} & 0 & 0\\
0 & 0 & i\frac{\partial}{\partial\varphi}+\frac{1}{2} & 0\\
0 & 0 & 0 & i\frac{\partial}{\partial\varphi}-\frac{1}{2}
\end{array}\!\!\right).\label{eq:Jz_Matrix4x4-1}
\end{equation}
As it is, the $J_{z}$ operator is already diagonal in spinor space,
meaning that we can write the eigenvalue problem $J_{z}\Psi(r,\theta,\varphi)\!=\!\hbar\lambda\,\Psi(r,\theta,\varphi)$
as four uncoupled differential equations,

\vspace{-0.7cm}

\begin{equation}
\begin{cases}
i\frac{\partial}{\partial\varphi}\psi_{1/3}\left(r,\theta,\varphi\right)=\left(\frac{1}{2}-\hbar^{-1}\lambda\right)\psi_{1/3}\left(r,\theta,\varphi\right)\\
i\frac{\partial}{\partial\varphi}\psi_{2/4}\left(r,\theta,\varphi\right)=-\left(\frac{1}{2}-\hbar^{-1}\lambda\right)\psi_{2/4}\left(r,\theta,\varphi\right)
\end{cases}.\label{eq:DiffEqs}
\end{equation}
Separation of variables allows us to solve Eqs.\,\eqref{eq:DiffEqs},
yielding the solution

\vspace{-0.7cm}

\begin{equation}
\Psi_{\lambda}\left(r,\theta,\varphi\right)=\left(\begin{array}{c}
\phi_{1}^{\lambda}\left(r,\theta\right)e^{i\varphi\left(\hbar^{-1}\lambda-\frac{1}{2}\right)}\\
\phi_{2}^{\lambda}\left(r,\theta\right)e^{i\varphi\left(\hbar^{-1}\lambda+\frac{1}{2}\right)}\\
\phi_{3}^{\lambda}\left(r,\theta\right)e^{i\varphi\left(\hbar^{-1}\lambda-\frac{1}{2}\right)}\\
\phi_{4}^{\lambda}\left(r,\theta\right)e^{i\varphi\left(\hbar^{-1}\lambda+\frac{1}{2}\right)}
\end{array}\right),\label{eq:Eigenspinor}
\end{equation}
where we must guarantee that $\Psi_{\lambda}\left(r,\theta,\varphi+2\pi\right)=\Psi_{\lambda}\left(r,\theta,\varphi\right)$.
Note that this periodicity condition holds if and only if $\hbar^{-1}\lambda\pm\frac{1}{2}$
are integers, so that we must have 

\vspace{-0.7cm}

\begin{equation}
\Psi_{j_{z}}\left(r,\theta,\varphi\right)=\left(\begin{array}{c}
\phi_{1}^{j_{z}}\left(r,\theta\right)e^{i\varphi\left(j_{z}-\frac{1}{2}\right)}\\
\phi_{2}^{j_{z}}\left(r,\theta\right)e^{i\varphi\left(j_{z}+\frac{1}{2}\right)}\\
\phi_{3}^{j_{z}}\left(r,\theta\right)e^{i\varphi\left(j_{z}-\frac{1}{2}\right)}\\
\phi_{4}^{j_{z}}\left(r,\theta\right)e^{i\varphi\left(j_{z}+\frac{1}{2}\right)}
\end{array}\right).\label{eq:Eigenspinor-1}
\end{equation}
In this form, we have $J_{z}\Psi_{j_{z}}\left(r,\theta,\varphi\right)=\hbar j_{z}\Psi_{j_{z}}\left(r,\theta,\varphi\right)$
and $j_{z}$ is a half-integer. Equation\,\ref{eq:Eigenspinor-1}
gives the most general form of an eigenspinor of well-defined angular
momentum along $z$. Let us further restrict its form, for it have
a well-defined eigenvalue of $\abs{\mathbf{J}}^{2}$as well. The latter
operator is not originally in diagonal form, but it is block-diagonal,
being enough to solve a single $2\!\times\!2$ block. Therefore, we
have

\vspace{-0.7cm}

\begin{equation}
\begin{cases}
\left(\abs{\mathbf{L}}^{2}+\hbar L_{z}+\frac{3\hbar^{2}}{4}\right)\psi_{1}^{j_{z}}\left(r,\theta,\varphi\right)+\hbar L_{-}\psi_{2}^{j_{z}}\left(r,\theta,\varphi\right)=\mu\psi_{1}^{j_{z}}\left(r,\theta,\varphi\right)\\
\hbar L_{+}\psi_{1}^{j_{z}}\left(r,\theta,\varphi\right)+\left(\abs{\mathbf{L}}^{2}-\hbar L_{z}+\frac{3\hbar^{2}}{4}\right)\psi_{2}^{j_{z}}\left(r,\theta,\varphi\right)=\mu\psi_{2}^{j_{z}}\left(r,\theta,\varphi\right)
\end{cases}.\label{eq:Equation}
\end{equation}

All the operators in Eq.\,\eqref{eq:Equation} are purely angular,
which allows the separation of variables,

\vspace{-0.7cm}

\begin{equation}
\psi_{1}^{j_{z}}\left(r,\theta,\varphi\right)=R\left(r\right)\Theta_{j_{z}}^{1}\!\left(\theta,\varphi\right)\text{ and }\psi_{2}^{j_{z}}\left(r,\theta,\varphi\right)=R\left(r\right)\Theta_{j_{z}}^{2}\!\left(\theta,\varphi\right),
\end{equation}
where the radial function $R\left(r\right)$ is the same in both equations.
The angular functions $\Theta_{j_{z}}^{1}\!\left(\theta,\varphi\right)$
and $\Theta_{j_{z}}^{2}\!\left(\theta,\varphi\right)$ can be expanded
in the basis of normalized scalar Spherical Harmonics, $Y_{l_{z}}^{l}\!\left(\theta,\varphi\right)$,
with the restriction that $l_{z}\!=\!j_{z}\!\pm\!\nicefrac{1}{2}$,
respectively. Hence, the two first components of the Dirac spinor
take the form,

\vspace{-0.7cm}

\begin{align}
\psi_{1}^{j_{z}}\left(r,\theta,\varphi\right) & =R\left(r\right)\sum_{l=0}^{\infty}\sum_{j_{z}}a_{l,j_{z}}Y_{j_{z}-1/2}^{l}\left(\theta,\varphi\right)\\
\psi_{2}^{j_{z}}\left(r,\theta,\varphi\right) & =R\left(r\right)\sum_{l=0}^{\infty}\sum_{j_{z}}b_{l,j_{z}}Y_{j_{z}+1/2}^{l}\left(\theta,\varphi\right),
\end{align}
where the summations over $j_{z}$ are always restricted to $-l\leq j_{z}\pm1/2\leq l$.
Then, we have

\vspace{-0.7cm}

\begin{equation}
\begin{cases}
\sum_{l,l_{z}}a_{l,j_{z}}\left(\hbar^{2}l\left(l+1\right)+\hbar^{2}j_{z}+\frac{\hbar^{2}}{4}-\mu\right)Y_{j_{z}-1/2}^{l}\left(\theta,\varphi\right)\\
\qquad\qquad\qquad+b_{l,j_{z}}\hbar^{2}\sqrt{\left(l+j_{z}+\frac{1}{2}\right)\left(l-j_{z}+\frac{1}{2}\right)}Y_{j_{z}-1/2}^{l}\left(\theta,\varphi\right)=0\\
\sum_{l,l_{z}}a_{l,j_{z}}\hbar^{2}\sqrt{\left(l-j_{z}+\frac{1}{2}\right)\left(l+j_{z}+\frac{1}{2}\right)}Y_{j_{z}+1/2}^{l}\left(\theta,\varphi\right)\\
\qquad\qquad\qquad+b_{l,j_{z}}\left(\hbar^{2}l\left(l+1\right)-\hbar^{2}j_{z}+\frac{\hbar^{2}}{4}-\mu\right)Y_{j_{z}+1/2}^{l}\left(\theta,\varphi\right)=0
\end{cases},
\end{equation}
or, equivalently

\vspace{-0.7cm}

{\footnotesize{}
\begin{equation}
\begin{cases}
\sum_{l,j_{z}}\left[a_{l,j_{z}}\left(\hbar^{2}l\left(l+1\right)+\hbar^{2}j_{z}+\frac{\hbar^{2}}{4}-\mu\right)+b_{l,j_{z}}\hbar^{2}\sqrt{\left(l+j_{z}+\frac{1}{2}\right)\left(l-j_{z}+\frac{1}{2}\right)}\right]Y_{j_{z}-1/2}^{l}\left(\theta,\varphi\right)=0\\
\sum_{l,j_{z}}\left[a_{l,j_{z}}\hbar^{2}\sqrt{\left(l-j_{z}+\frac{1}{2}\right)\left(l+j_{z}+\frac{1}{2}\right)}+b_{l,j_{z}}\left(\hbar^{2}l\left(l+1\right)-\hbar^{2}j_{z}+\frac{\hbar^{2}}{4}-\mu\right)\right]Y_{j_{z}+1/2}^{l}\left(\theta,\varphi\right)=0
\end{cases}.
\end{equation}
}{\footnotesize\par}

Since the spherical Harmonics are a complete and orthogonal basis
of functions in a 3D unit spherical surface, the previous homogenous
system has a single solution; All the coefficients must be zero, which
implies that

\vspace{-0.7cm}

\begin{equation}
\!\!\!\!\begin{cases}
a_{l,j_{z}}\left(\hbar^{2}l\left(l\!+\!1\right)\!+\!\hbar^{2}j_{z}\!+\!\frac{\hbar^{2}}{4}\!-\!\mu\right)+b_{l,j_{z}}\hbar^{2}\sqrt{\left(l\!+\!j_{z}\!+\!\frac{1}{2}\right)\left(l\!-\!j_{z}\!+\!\frac{1}{2}\right)}=0\\
a_{l,j_{z}}\hbar^{2}\sqrt{\left(l\!-\!j_{z}\!+\!\frac{1}{2}\right)\left(l\!+\!j_{z}\!+\!\frac{1}{2}\right)}+b_{l,j_{z}}\left(\hbar^{2}l\left(l\!+\!1\right)\!-\!\hbar^{2}j_{z}\!+\!\frac{\hbar^{2}}{4}\!-\!\mu\right)=0
\end{cases}\!\!\!,
\end{equation}
which only has a non-trivial solution if

\vspace{-0.7cm}

\begin{equation}
\!\!\!\!\begin{cases}
l\left(l+1\right)+\frac{1}{4}-\hbar^{-2}\mu=l+\frac{1}{2}\\
l\left(l+1\right)+\frac{1}{4}-\hbar^{-2}\mu=-l-\frac{1}{2}
\end{cases}\!\!\!\!\!\!\!\!\Rightarrow\begin{cases}
\mu=\hbar^{2}\left(l^{2}-\frac{1}{4}\right)=\hbar^{2}\left(l-\frac{1}{2}\right)\left(l+\frac{1}{2}\right)\\
\mu=\hbar^{2}\left(l^{2}+2l-\frac{3}{4}\right)=\hbar^{2}\left(l+\frac{1}{2}\right)\left(l+\frac{3}{2}\right)
\end{cases}\!\!\!\!\!
\end{equation}
and therefore implies that $\mu=\hbar^{2}j\left(j+1\right)$ with
$j\!=\!l\pm1/2$. This is an expected result from the Addition Theorem
of Angular Momenta and, for each case, we have the following relation
between the $a_{l,j_{z}}$ and $b_{l,j_{z}}$ coefficients:

\vspace{-0.3cm}
\begin{itemize}
\item For $l=j+1/2$:

\vspace{-0.7cm}

\begin{equation}
b_{l,j_{z}}=-\sqrt{\frac{j+j_{z}+1}{j-j_{z}+1}}a_{l,j_{z}}
\end{equation}

\item For $l=j-1/2$:

\vspace{-0.7cm}

\begin{equation}
b_{l,j_{z}}=\sqrt{\frac{j-j_{z}}{j+j_{z}}}a_{l,j_{z}}.
\end{equation}

\end{itemize}
Hence, we have the following general form for the eigenspinors of
$J_{z}$ and $\abs{\mathbf{J}}^{2}$:

\vspace{-0.7cm}

\begin{equation}
\!\!\!\Psi_{j,j_{z}}\left(r,\theta,\varphi\right)=\left(\!\!\begin{array}{c}
R_{1}\left(r\right)Y_{j_{z}-1/2}^{j+1/2}\left(\theta,\varphi\right)+R_{2}\left(r\right)Y_{j_{z}-1/2}^{j-1/2}\left(\theta,\varphi\right)\\
-\sqrt{\frac{j+j_{z}+1}{j-j_{z}+1}}R_{1}\left(r\right)Y_{j_{z}+1/2}^{j+1/2}\left(\theta,\varphi\right)+\!\sqrt{\frac{j-j_{z}}{j+j_{z}}}R_{2}\left(r\right)Y_{j_{z}+1/2}^{j-1/2}\left(\theta,\varphi\right)\\
F_{1}\left(r\right)Y_{j_{z}-1/2}^{j+1/2}\left(\theta,\varphi\right)+F_{2}\left(r\right)Y_{j_{z}-1/2}^{j-1/2}\left(\theta,\varphi\right)\\
-\sqrt{\frac{j+j_{z}+1}{j-j_{z}+1}}F_{1}\left(r\right)Y_{j_{z}+1/2}^{j+1/2}\left(\theta,\varphi\right)+\!\sqrt{\frac{j-j_{z}}{j+j_{z}}}F_{2}\left(r\right)Y_{j_{z}+1/2}^{j-1/2}\left(\theta,\varphi\right)
\end{array}\!\!\right).\!\!\!\label{eq:Eigenspinor-2}
\end{equation}
From Eq.\,\eqref{eq:Eigenspinor-2} it is already evident that $\Psi_{j,j_{z}}\left(r,\theta,\varphi\right)$
does not have a well-defined value for $\abs{\mathbf{L}}^{2}$, because
it is a linear superposition of states with $l=j\pm1/2$, for a fixed
half-integer $j$. Therefore setting $j_{z}$ and $j$ is not enough
to fully specify the angular part of the Dirac spinor components.
However, we still have a further conserved quantity to work with,
the spin-orbit operator $\mathcal{K}$. Let us impose that

\vspace{-0.7cm}

\begin{equation}
\mathcal{K}\Psi_{j,j_{z}}\left(r,\theta,\varphi\right)=\hbar^{2}\nu\Psi_{j,j_{z}}\left(r,\theta,\varphi\right),
\end{equation}
where $\nu$ is to be determined. Despite our problem being intrinsically
$4$-dimensional, the operator $\mathcal{K}$, as written in block-diagonal
form, as shown in Eq.\,\ref{eq:SpinOrbit}. Hence, we can work with 

\vspace{-0.7cm}{\footnotesize{}
\begin{align}
\left(\begin{array}{cc}
\hbar+L_{z} & L_{-}\\
L_{+} & \hbar-L_{z}
\end{array}\right)\left(\begin{array}{c}
R_{1}\left(r\right)Y_{j_{z}-1/2}^{j+1/2}\left(\theta,\varphi\right)+R_{2}\left(r\right)Y_{j_{z}-1/2}^{j-1/2}\left(\theta,\varphi\right)\\
-\sqrt{\frac{j+j_{z}+1}{j-j_{z}+1}}R_{1}\left(r\right)Y_{j_{z}+1/2}^{j+1/2}\left(\theta,\varphi\right)+\sqrt{\frac{j-j_{z}}{j+j_{z}}}R_{2}\left(r\right)Y_{j_{z}+1/2}^{j-1/2}\left(\theta,\varphi\right)
\end{array}\right)\qquad\label{eq:K_1}\\
=\hbar\nu\left(\begin{array}{c}
R_{1}\left(r\right)Y_{j_{z}-1/2}^{j+1/2}\left(\theta,\varphi\right)+R_{2}\left(r\right)Y_{j_{z}-1/2}^{j-1/2}\left(\theta,\varphi\right)\\
-\sqrt{\frac{j+j_{z}+1}{j-j_{z}+1}}R_{1}\left(r\right)Y_{j_{z}+1/2}^{j+1/2}\left(\theta,\varphi\right)+\sqrt{\frac{j-j_{z}}{j+j_{z}}}R_{2}\left(r\right)Y_{j_{z}+1/2}^{j-1/2}\left(\theta,\varphi\right)
\end{array}\right)\nonumber 
\end{align}
}and

\vspace{-0.7cm}

{\footnotesize{}
\begin{align}
\left(\!\begin{array}{cc}
\hbar+L_{z} & L_{-}\\
L_{+} & \hbar-L_{z}
\end{array}\!\right)\left(\!\begin{array}{c}
F_{1}\left(r\right)Y_{j_{z}-1/2}^{j+1/2}\left(\theta,\varphi\right)+F_{2}\left(r\right)Y_{j_{z}-1/2}^{j-1/2}\left(\theta,\varphi\right)\\
-\sqrt{\frac{j+j_{z}+1}{j-j_{z}+1}}F_{1}\left(r\right)Y_{j_{z}+1/2}^{j+1/2}\left(\theta,\varphi\right)+\sqrt{\frac{j-j_{z}}{j+j_{z}}}F_{2}\left(r\right)Y_{j_{z}+1/2}^{j-1/2}\left(\theta,\varphi\right)
\end{array}\!\right)\qquad\label{eq:K_2}\\
=-\hbar\nu\left(\!\begin{array}{c}
F_{1}\left(r\right)Y_{j_{z}-1/2}^{j+1/2}\left(\theta,\varphi\right)+F_{2}\left(r\right)Y_{j_{z}-1/2}^{j-1/2}\left(\theta,\varphi\right)\\
-\sqrt{\frac{j+j_{z}+1}{j-j_{z}+1}}F_{1}\left(r\right)Y_{j_{z}+1/2}^{j+1/2}\left(\theta,\varphi\right)+\sqrt{\frac{j-j_{z}}{j+j_{z}}}F_{2}\left(r\right)Y_{j_{z}+1/2}^{j-1/2}\left(\theta,\varphi\right)
\end{array}\!\right),\nonumber 
\end{align}
}separately. Taking only Eq.\,\eqref{eq:K_1}, we arrive at two independent
linear systems, namely

\vspace{-0.7cm}

\begin{equation}
\begin{cases}
R_{1}\left(r\right)\left(\left(j_{z}+\frac{1}{2}-\nu\right)-\left(j+j_{z}+1\right)\right)Y_{j_{z}-1/2}^{j+1/2}\left(\theta,\varphi\right)=0\\
\sqrt{\frac{j+j_{z}+1}{j-j_{z}+1}}R_{1}\left(r\right)\left(\left(j-j_{z}+1\right)+\left(j_{z}-\frac{1}{2}+\nu\right)\right)Y_{j_{z}+1/2}^{j+1/2}\left(\theta,\varphi\right)=0
\end{cases},\label{eq:SystemR1}
\end{equation}
and

\vspace{-0.7cm}

\begin{equation}
\begin{cases}
R_{2}\left(r\right)\left(\left(j_{z}+\frac{1}{2}-\nu\right)+\left(j-j_{z}\right)\right)Y_{j_{z}-1/2}^{j-1/2}\left(\theta,\varphi\right)=0\\
\sqrt{\frac{j-j_{z}}{j+j_{z}}}R_{2}\left(r\right)\left(\left(j+j_{z}\right)+\left(\frac{1}{2}-j_{z}-\nu\right)\right)Y_{j_{z}+1/2}^{j-1/2}\left(\theta,\varphi\right)
\end{cases}.\label{eq:SystemR2}
\end{equation}
The system in Eq.\,\eqref{eq:SystemR1} admits a single non-vanishing
solution, provided

\vspace{-0.7cm}

\[
\left(j_{z}\!+\!\frac{1}{2}\!-\!\nu\!\right)-\left(j\!+\!j_{z}\!+\!1\right)=-\left(j\!-\!j_{z}\!+\!1\right)-\left(j_{z}\!-\!\frac{1}{2}\!+\!\nu\right)=j\!+\!\frac{1}{2}\!+\!\nu=0
\]
or, equivalently $\nu=-j-1/2$. On the contrary, the system of Eq.\,\eqref{eq:SystemR2}
requires that

\vspace{-0.7cm}

\[
\left(j_{z}+\frac{1}{2}-\nu\right)+\left(j-j_{z}\right)=\left(j+j_{z}\right)+\left(\frac{1}{2}-j_{z}-\nu\right)=j+\frac{1}{2}-\nu=0
\]
or $\nu=j+1/2$. In summary, $R_{1}\left(r\right)=0$ if $v=j+1/2$
and $R_{2}\left(r\right)=0$ if $\nu=-j-1/2$. Both functions will
be zero for any other value of $\nu$. 

Finally, if we turn our attention to Eq.\,\eqref{eq:K_2}, we find
an analogous situation with role of both radial functions reversed,
i.e. $F_{1}\left(r\right)=0$ if $v=-j-1/2$ and $F_{2}\left(r\right)=0$
if $\nu=j+1/2$. Hence, if we add $\nu=\pm\left(j+1/2\right)$ to
the list of angular quantum numbers $(j_{z},j)$ we uniquely classify
the angular part of the eigenspinors. Namely, we have

\vspace{-0.7cm}

\begin{equation}
\Psi_{j,j_{z},\nu=j+1/2}\left(r,\theta,\varphi\right)\!=\!\left(\begin{array}{c}
R_{2}\left(r\right)Y_{j_{z}-1/2}^{j-1/2}\left(\theta,\varphi\right)\\
\sqrt{\frac{j-j_{z}}{j+j_{z}}}R_{2}\left(r\right)Y_{j_{z}+1/2}^{j-1/2}\left(\theta,\varphi\right)\\
F_{1}\left(r\right)Y_{j_{z}-1/2}^{j+1/2}\left(\theta,\varphi\right)\\
-\sqrt{\frac{j+j_{z}+1}{j-j_{z}+1}}F_{1}\left(r\right)Y_{j_{z}+1/2}^{j+1/2}\left(\theta,\varphi\right)
\end{array}\right),
\end{equation}
and

\vspace{-0.7cm}

\begin{equation}
\Psi_{j,j_{z},\nu=-(j+1/2)}\left(r,\theta,\varphi\right)=\mathcal{N}\left(\begin{array}{c}
R_{1}\left(r\right)Y_{j_{z}-1/2}^{j+1/2}\left(\theta,\varphi\right)\\
-\sqrt{\frac{j+j_{z}+1}{j-j_{z}+1}}R_{1}\left(r\right)Y_{j_{z}+1/2}^{j+1/2}\left(\theta,\varphi\right)\\
F_{2}\left(r\right)Y_{j_{z}-1/2}^{j-1/2}\left(\theta,\varphi\right)\\
\sqrt{\frac{j-j_{z}}{j+j_{z}}}F_{2}\left(r\right)Y_{j_{z}+1/2}^{j-1/2}\left(\theta,\varphi\right)
\end{array}\right).\label{eq:Eigenspinor-2-1}
\end{equation}

Now, we can use the same notation used in the Weyl system with rotational
symmetry. Namely, we define the (normalized) spin-$1/2$ spherical
harmonics (with quantum numbers $j$ and $j_{z}$) as:

\vspace{-0.7cm}

\begin{subequations}
\begin{align}
\Theta_{j,j_{z}}^{+}\left(\theta,\varphi\right) & =\left(\begin{array}{c}
\sqrt{\frac{j-j_{z}+1}{2j+2}}Y_{j_{z}-1/2}^{j+1/2}\left(\theta,\varphi\right)\\
-\sqrt{\frac{j+j_{z}+1}{2j+2}}Y_{j_{z}+1/2}^{j+1/2}\left(\theta,\varphi\right)
\end{array}\right)\label{eq:Normalized+-3-1-1}\\
\Theta_{j,j_{z}}^{-}\left(\theta,\varphi\right) & =\left(\begin{array}{c}
\sqrt{\frac{j+j_{z}}{2j}}Y_{j_{z}-1/2}^{j-1/2}\left(\theta,\varphi\right)\\
\sqrt{\frac{j-j_{z}}{2j}}Y_{j_{z}+1/2}^{j-1/2}\left(\theta,\varphi\right)
\end{array}\right).\label{eq:Spherical2}
\end{align}
\end{subequations}

with these definitions, the previously found eigenspinors read

\vspace{-0.7cm}

\begin{align}
\Psi_{j,j_{z},\nu=+(j+1/2)}\left(r,\theta,\varphi\right) & =\Psi_{j,j_{z}}^{+}\left(r,\theta,\varphi\right)=\left(\begin{array}{c}
R^{+}\left(r\right)\Theta^{-}\left(\theta,\varphi\right)\\
F^{+}\left(r\right)\Theta^{+}\left(\theta,\varphi\right)
\end{array}\right)\\
\Psi_{j,j_{z},\nu=-(j+1/2)}\left(r,\theta,\varphi\right) & =\Psi_{j,j_{z}}^{-}\left(r,\theta,\varphi\right)=\left(\begin{array}{c}
R^{-}\left(r\right)\Theta^{+}\left(\theta,\varphi\right)\\
F^{-}\left(r\right)\Theta^{-}\left(\theta,\varphi\right)
\end{array}\right),
\end{align}
respectively. Note that $\Theta_{j,j_{z}}^{\pm}\!\left(\theta,\varphi\right)$,
as defined in Eqs.\,\eqref{eq:Normalized+-3-1-1}-\eqref{eq:Spherical2}
are orthogonal and properly normalized in the sense,

\vspace{-0.7cm}

\begin{equation}
\int_{0}^{\pi}\sin\theta d\theta\int_{0}^{2\pi}d\varphi\left[\Theta_{j,j_{z}}^{\pm}\left(\theta,\varphi\right)\right]^{\dagger}\cdot\Theta_{j,j_{z}}^{\pm}\left(\theta,\varphi\right)=1.
\end{equation}
This implies that the normalization condition for $\Psi_{j,j_{z},\nu=\pm(j+1/2)}\left(r,\theta,\varphi\right)$
read simply

\vspace{-0.7cm}

\begin{equation}
\left(\Psi_{j,j_{z}}^{\pm}\left(r,\theta,\varphi\right)\right)^{\dagger}\cdot\Psi_{j,j_{z}}^{\pm}\left(r,\theta,\varphi\right)=\!\!\int_{0}^{\infty}\!\!\!drr^{2}\left[\left(R^{\pm}\left(r\right)\right)^{*}R^{\pm}\left(r\right)+\left(F^{\pm}\left(r\right)\right)^{*}F^{\pm}\left(r\right)\right]=1,
\end{equation}
which is just a radial integral.

\vspace{-0.5cm}

\section{Spherical Separation of Variables}

Defining the spin-orbit quantum number, $\kappa\!=\!\nu/(j+\nicefrac{1}{2})=\pm1$,
we can write a general spherically symmetric Dirac eigenstate as

\vspace{-0.7cm}

\begin{equation}
\Psi_{j,j_{z}}^{\kappa}\left(r,\theta,\varphi\right)=\left(\!\!\begin{array}{c}
R^{\kappa}\left(r\right)\Theta_{jj_{z}}^{-\kappa}\left(\theta,\varphi\right)\\
F^{\kappa}\left(r\right)\Theta_{jj_{z}}^{\kappa}\left(\theta,\varphi\right)
\end{array}\!\!\right)
\end{equation}
and now we will use this general form to derive radial ODEs that allow
to determine the functions $F(r)$ and $R(r)$ in the presence of
a potential that depends only on the radial coordinate. For that purpose,
it is useful to write the Hamiltonian of Eq.\,\eqref{eq:DiracHamiltonian+ScalarPotential}
in the matrix form,

\vspace{-0.7cm}

\begin{equation}
\mathcal{H}_{\text{D}}=\left(\!\!\begin{array}{cc}
\left(\mathcal{U}\left(r\right)+mv_{\text{F}}^{2}\right)\mathbb{I}_{2\times2} & -i\hbar v_{\text{F}}\boldsymbol{\sigma}\cdot\boldsymbol{\nabla}\\
-i\hbar v_{\text{F}}\boldsymbol{\sigma}\cdot\boldsymbol{\nabla} & \left(\mathcal{U}\left(r\right)-mv_{\text{F}}^{2}\right)\mathbb{I}_{2\times2}
\end{array}\!\!\right),
\end{equation}
where we can use the fact

\vspace{-0.7cm}

\begin{equation}
\boldsymbol{\sigma}\cdot\boldsymbol{\nabla}=\boldsymbol{\sigma}\cdot\mathbf{\hat{r}}\left[\mathbf{\hat{r}}\cdot\overrightarrow{\nabla}-\frac{\overrightarrow{\sigma}\cdot\mathbf{L}}{\hbar r}\right],
\end{equation}
in order to write down

\vspace{-0.7cm}

\begin{equation}
\mathcal{H}_{\text{D}}\!=\!\left(\!\!\begin{array}{cc}
\left(\mathcal{U}\left(r\right)+mv_{\text{F}}^{2}\right)\mathbb{I}_{2\times2} & -i\hbar v_{\text{F}}\boldsymbol{\sigma}\cdot\mathbf{\hat{r}}\left[\mathbf{\hat{r}}\cdot\overrightarrow{\nabla}-\frac{\overrightarrow{\sigma}\cdot\mathbf{L}}{\hbar r}\right]\\
-i\hbar v_{\text{F}}\boldsymbol{\sigma}\cdot\mathbf{\hat{r}}\left[\mathbf{\hat{r}}\cdot\overrightarrow{\nabla}-\frac{\overrightarrow{\sigma}\cdot\mathbf{L}}{\hbar r}\right] & \left(\mathcal{U}\left(r\right)-mv_{\text{F}}^{2}\right)\mathbb{I}_{2\times2}
\end{array}\!\!\right),
\end{equation}

where $\mathbf{\hat{r}}\!=\!\mathbf{r}/r$ is a radial unit vector.
Furthermore, if we recognize that

\vspace{-0.7cm}

\begin{subequations}
\begin{align}
\boldsymbol{\sigma}\cdot\hat{\mathbf{r}}\Theta^{\pm}\left(\theta,\varphi\right) & =\left(\boldsymbol{\sigma}\cdot\hat{\mathbf{r}}\right)^{2}\Theta_{jj_{z}}^{\pm}\!\left(\theta,\varphi\right)=\Theta_{jj_{z}}^{\pm}\!\left(\theta,\varphi\right)\label{eq:MixingTheta}\\
\boldsymbol{\sigma}\cdot\mathbf{L}\Theta^{+}\left(\theta,\varphi\right) & =-\hbar\left(j+\frac{3}{2}\right)\Theta_{jj_{z}}^{+}\!\left(\theta,\varphi\right)\label{eq:J_zeigenvalues-1-1}\\
\boldsymbol{\sigma}\cdot\mathbf{L}\Theta^{-}\left(\theta,\varphi\right) & =\hbar\left(j-\frac{1}{2}\right)\Theta_{jj_{z}}^{-}\!\left(\theta,\varphi\right).\label{eq:J_zeigenvalues-1-1-1}
\end{align}
\end{subequations}

Since $\abs{\mathbf{J}}^{2}$, $J_{z}$ and $\mathcal{K}$ are all
conserved quantities of $\mathcal{H}_{\text{D}}$, we only need to
analyze how the Dirac Hamiltonian acts on the states $\Psi_{j,j_{z}}^{\kappa}\!\left(r,\theta,\varphi\right)$,
as defined in Eqs.\,\eqref{eq:Normalized+-3-1}-\eqref{eq:Normalized--3-1}.
This yields the following results:

\vspace{-0.7cm}

{\footnotesize{}
\begin{align}
\!\!\!\!\!\!\!\mathcal{H}_{\text{D}}\Psi_{j,j_{z}}^{+}\!\left(r,\theta,\varphi\right) & \!=\!\left(\!\!\!\begin{array}{c}
\left[-i\hbar v_{\text{F}}\partial_{r}F^{+}\!\left(r\right)-i\frac{\hbar v_{\text{F}}}{r}\left(j\!+\!\frac{3}{2}\right)F^{+}\!\left(r\right)+\left(\mathcal{U}\!\left(r\right)\!+\!mv_{\text{F}}^{2}\right)R^{+}\!\left(r\right)\right]\Theta_{jj_{z}}^{-}\!\left(\theta,\varphi\right)\\
\left[-i\hbar v_{\text{F}}\partial_{r}R^{+}\!\left(r\right)+i\frac{\hbar v_{\text{F}}}{r}\left(j\!-\!\frac{1}{2}\right)R^{+}\!\left(r\right)+\left(\mathcal{U}\left(r\right)\!-\!mv_{\text{F}}^{2}\right)F^{+}\!\left(r\right)\right]\Theta_{jj_{z}}^{+}\!\left(\theta,\varphi\right)
\end{array}\!\!\!\right)\!\!\!\\
\!\!\!\!\!\!\!\mathcal{H}_{\text{D}}\Psi_{j,j_{z}}^{-}\!\left(r,\theta,\varphi\right) & \!=\!\left(\!\!\!\begin{array}{c}
\left[-i\hbar v_{\text{F}}\partial_{r}F^{-}\!\left(r\right)+i\frac{\hbar v_{\text{F}}}{r}\left(j\!-\!\frac{1}{2}\right)F^{-}\!\left(r\right)+\left(\mathcal{U}\!\left(r\right)\!+\!mv_{\text{F}}^{2}\right)R^{-}\!\left(r\right)\right]\Theta_{jj_{z}}^{+}\!\left(\theta,\varphi\right)\\
\left[-i\hbar v_{\text{F}}\partial_{r}R^{-}\!\left(r\right)-i\frac{\hbar v_{\text{F}}}{r}\left(j\!+\!\frac{3}{2}\right)R^{-}\!\left(r\right)+\left(\mathcal{U}\!\left(r\right)\!-\!mv_{\text{F}}^{2}\right)F^{-}\!\left(r\right)\right]\Theta_{jj_{z}}^{-}\!\left(\theta,\varphi\right)
\end{array}\!\!\!\right).\!\!\!
\end{align}
}Finally, if we incorporate the previous results into the eigenvalue
problem for $\mathcal{H}_{D}$,

\vspace{-0.7cm}

\begin{equation}
\mathcal{H}_{\text{D}}\Psi_{j,j_{z}}^{\kappa}\left(r,\theta,\varphi\right)=E\Psi_{j,j_{z}}^{\kappa}\left(r,\theta,\varphi\right),
\end{equation}
we arrive at the radial equations,

\vspace{-0.7cm}

\begin{equation}
\!\!\!\!\!\begin{cases}
-i\hbar v_{\text{F}}\frac{d}{dr}F^{+}\left(r\right)-i\frac{\hbar v_{\text{F}}}{r}\left(j+\frac{3}{2}\right)F^{+}\left(r\right)+\left(\mathcal{U}\left(r\right)+mv_{\text{F}}^{2}\right)R^{+}\left(r\right)=ER^{+}\left(r\right)\\
-i\hbar v_{\text{F}}\frac{d}{dr}R^{+}\left(r\right)+i\frac{\hbar v_{\text{F}}}{r}\left(j-\frac{1}{2}\right)R^{+}\left(r\right)+\left(\mathcal{U}\left(r\right)-mv_{\text{F}}^{2}\right)F^{+}\left(r\right)=EF^{+}\left(r\right)
\end{cases}\!\!\!\!\!\!,\!\!
\end{equation}

for $\Psi_{j,j_{z}}^{+}\left(r,\theta,\varphi\right)$ and

\vspace{-0.7cm}

\begin{equation}
\!\!\!\!\!\!\begin{cases}
-i\hbar v_{\text{F}}\frac{d}{dr}F^{-}\left(r\right)+i\frac{\hbar v_{\text{F}}}{r}\left(j-\frac{1}{2}\right)F^{-}\left(r\right)+\left(\mathcal{U}\left(r\right)+mv_{\text{F}}^{2}\right)R^{-}\left(r\right)=ER^{-}\left(r\right)\\
-i\hbar v_{\text{F}}\frac{d}{dr}R^{-}\left(r\right)-i\frac{\hbar v_{\text{F}}}{r}\left(j+\frac{3}{2}\right)R^{-}\left(r\right)+\left(\mathcal{U}\left(r\right)-mv_{\text{F}}^{2}\right)F^{-}\left(r\right)=ER^{-}\left(r\right)
\end{cases}\!\!\!\!\!\!,\!\!
\end{equation}

for $\Psi_{j,j_{z}}^{-}\left(r,\theta,\varphi\right)$. In order to
conform with the simpler notation of Chapter\,\ref{chap:Instability_Smooth_Regions},
we define the following objects

\vspace{-0.7cm}

\begin{equation}
f^{\kappa}\!\left(r\right)\!=\!rR^{\kappa}\!\left(r\right)\quad\text{and}\quad g^{\kappa}\!\left(r\right)\!=\!-irF^{\kappa}\!\left(r\right),
\end{equation}
such that

\vspace{-0.7cm}

\begin{equation}
\frac{d}{dr}R^{\pm}\left(r\right)=\frac{1}{r}\frac{d}{dr}f^{\pm}\left(r\right)-\frac{1}{r^{2}}f^{\pm}\left(r\right)\quad\text{and}\quad\frac{d}{dr}F^{\pm}\left(r\right)=\frac{i}{r}\frac{d}{dr}g^{\pm}\left(r\right)-\frac{i}{r^{2}}g^{\pm}\left(r\right).
\end{equation}
Finally, in terms of these new functions, we arrive at the following
coupled system of radial ODEs:

\vspace{-0.7cm}

\begin{equation}
\begin{cases}
\frac{d}{dr}g_{\varepsilon}^{\kappa}\left(r\right)+\frac{\kappa}{r}\left(j+\frac{1}{2}\right)g_{\varepsilon}^{\kappa}\left(r\right)=\left(\varepsilon-u\left(r\right)-\mu\right)f_{\varepsilon}^{\kappa}\left(r\right)\\
\frac{d}{dr}f_{\varepsilon}^{\kappa}\left(r\right)-\frac{\kappa}{r}\left(j+\frac{1}{2}\right)f_{\varepsilon}^{\kappa}\left(r\right)=-\left(\varepsilon-u\left(r\right)+\mu\right)g_{\varepsilon}^{\kappa}\left(r\right)
\end{cases}.\label{eq:Sys1-1}
\end{equation}
where $u\left(r\right)=\mathcal{U}\left(r\right)/\hbar v_{\text{F}}$,
$\mu=mv_{\text{F}}/\hbar$ and $\varepsilon=E/\hbar v_{\text{F}}$.
Note that these equations are entirely equivalent to the Eq.\,\eqref{eq:System}
of Chapter\,\ref{chap:Instability_Smooth_Regions}, upon setting
$\mu\!=\!0$ and performing a unitary transformation in spinor space.
The latter difference is due to the use a (off-diagonal) Dirac representation
for the $\alpha$-matrices, rather than the (diagonal) Weyl representation. 

\lhead[\MakeUppercase{Appendix D}]{\MakeUppercase{\rightmark}}

\rhead[\MakeUppercase{Lattice Green's Function}]{}

\chapter{\label{chap:LatticeGF}Evaluation of the Lattice Green's Function}

To analyze the emergence of nodal bound states and calculate changes
in the eDoS induced by clusters of atomic-sized impurities in the
lattice WSM model, an essential ingredient was to know the clean \textit{lattice
Green's function} (lGF). Computing the lGF of an arbitrary tight-binding
Hamiltonian in an infinite lattice is, by itself, a relevant problem
and many methods have been invented to do so\,\cite{Schulman69,Horiguchi71,Morita1971a,Morita1971b,Morita1972,Joyce1994,Martinsson2002,Berciu2009,Guttmann2010,Kogan2021}.
For our particular case, the lGF can be expressed as

\vspace{-0.7cm}

\begin{align}
\mathscr{G}_{ab}^{\text{0r}}(\varepsilon;\boldsymbol{\Delta L})\! & =\!\frac{a}{8\pi^{3}\hbar v_{\text{F}}}\int_{\text{C}}\!\!\!d\mathbf{q}\,\frac{\varepsilon-\boldsymbol{\sigma}\cdot\boldsymbol{\sin}\mathbf{q}}{\left(\varepsilon\!+\!i\eta\right)^{2}\!-\!\abs{\boldsymbol{\sin}\mathbf{q}}^{2}}e^{i\mathbf{q}\cdot\boldsymbol{\Delta L}},\label{eq:RealSpacePropagator-2-1-3}
\end{align}
where $\varepsilon$ is a dimensionless energy and the triple-integral
is over $\text{C}\!=\![-\pi,\pi]^{3}$. In addition, we have further
showed that this function can be expressed in terms of two simpler
integrals, \textit{i.e.},

\vspace{-0.7cm}
\begin{equation}
\mathscr{G}_{ab}^{\text{0r}}(\varepsilon;\boldsymbol{\Delta L})\!=\!\frac{a}{8\pi^{3}}\left[\varepsilon\,\delta_{ab}\,\mathcal{I}_{0}\left(\varepsilon;\boldsymbol{\Delta L}\right)-\boldsymbol{\sigma}_{ab}\cdot\boldsymbol{\mathcal{I}}\left(\varepsilon;\boldsymbol{\Delta L}\right)\right],\label{eq:DecompositionlGF-2}
\end{equation}
with the dimensionless constitutive integrals, $\left(\mathcal{I}_{0},\mathcal{I}_{x},\mathcal{I}_{y},\mathcal{I}_{z}\right)$,
being defined as follows:

\vspace{-0.7cm}

\begin{subequations}
\begin{align}
\mathcal{I}_{0}\left(\varepsilon;\boldsymbol{\Delta L}\right) & =\int_{\text{C}}\!\!\!d\mathbf{q}\,\frac{e^{i\mathbf{q}\cdot\boldsymbol{\Delta L}}}{\left(\varepsilon\!+\!i\eta\right)^{2}\!-\!\abs{\boldsymbol{\sin}\mathbf{q}}^{2}}\label{eq:I0-2}\\
\mathcal{I}_{j}\left(\varepsilon;\boldsymbol{\Delta L}\right) & =\int_{\text{C}}\!\!\!d\mathbf{q}\,\frac{\sin q_{j}e^{i\mathbf{q}\cdot\boldsymbol{\Delta L}}}{\left(\varepsilon\!+\!i\eta\right)^{2}\!-\!\abs{\boldsymbol{\sin}\mathbf{q}}^{2}}\label{eq:Ij-2}
\end{align}
\end{subequations}

which are to be taken in the limit $\eta\to0^{+}$. In this Appendix,
we present a tailor-made semi-analytic method that we have devised
to calculate the lGF of our lattice model in the limit $\eta\to0^{+}$.
This was the method used to obtain the examples plotted in Fig.\,\ref{fig:LatticeGFvsCont}. 

\vspace{-0.5cm}

\section{Analytical Evaluation of the First Integral}

Starting directly from the integrals of Eqs.\,\eqref{eq:I0-2}-\eqref{eq:Ij-2},
it is clear that these cannot be determined analytically in any simple
way. However, some analytical progress can be achieved by rewriting
them as 

\vspace{-0.7cm}

\begin{subequations}
\begin{align}
\mathcal{I}_{0}^{{\scriptscriptstyle n,m,l}}\!\left(z\right) & =\!\!\!\int_{-\pi}^{\pi}\!\!\!\!du\!\!\int_{-\pi}^{\pi}\!\!\!\!dve^{iun}e^{ivm}\mathcal{I}_{1}^{l}\left(z^{2}\!-\sin^{2}\!u-\sin^{2}\!v\right)\label{eq:I_diag-1}\\
\mathcal{I}_{x}^{{\scriptscriptstyle n,m,l}}\!\left(z\right) & =\!\!\!\int_{-\pi}^{\pi}\!\!\!\!du\!\!\int_{-\pi}^{\pi}\!\!\!\!dv\sin\!u\,e^{iun}e^{ivm}\mathcal{I}_{1}^{l}\left(z^{2}\!-\sin^{2}\!u-\sin^{2}\!v\right),\label{eq:I_diag-1-1}
\end{align}
\end{subequations}

where $z$ is a complex variable, $\boldsymbol{\Delta L}\!=\!(n,m,l),$$(u,v)=(q_{x},q_{y})$
and $\mathcal{I}_{1}^{l}(z)$ is the complex function

\vspace{-0.7cm}
\begin{equation}
\mathcal{I}_{1}^{l}\!(z)\!=\!\!\!\int_{-\pi}^{\pi}\!\!\!\!\!\!dw\frac{e^{iwl}}{z-\!\sin^{2}\!w},
\end{equation}
that has an obvious $l\!\to\!-l$ symmetry. Due to this symmetry,
all we have to evaluate is

\vspace{-0.7cm}
\begin{equation}
\mathcal{I}_{1}^{l}\!(z)\!=\!\!\!\int_{-\pi}^{\pi}\!\!\!\!\!\!dw\frac{e^{iw\abs l}}{z-\!\sin^{2}\!w}.
\end{equation}
This latter may be cast into a contour integral along the unit circle
of the complex-plane. To accomplish this, we change the integration
variable from $\mu\!=\!e^{iw}\!\to dw\!=\!-id\mu/\mu$, which yields

\vspace{-0.7cm}
\begin{equation}
\mathcal{I}_{1}^{l}\!(z)\!=\!\ointctrclockwise_{{\scriptscriptstyle \!\!\abs{\!\mu\!}=1}}\!\!\frac{-4i\mu^{\abs l+1}d\mu}{\mu^{4}\!+\!2\left(2z\!-\!1\right)\mu^{2}\!+\!1}.\label{eq:I1_Complex}
\end{equation}
\begin{figure}[t]
\vspace{-0.5cm}
\begin{centering}
\hspace{-0.3cm}\includegraphics[scale=0.36]{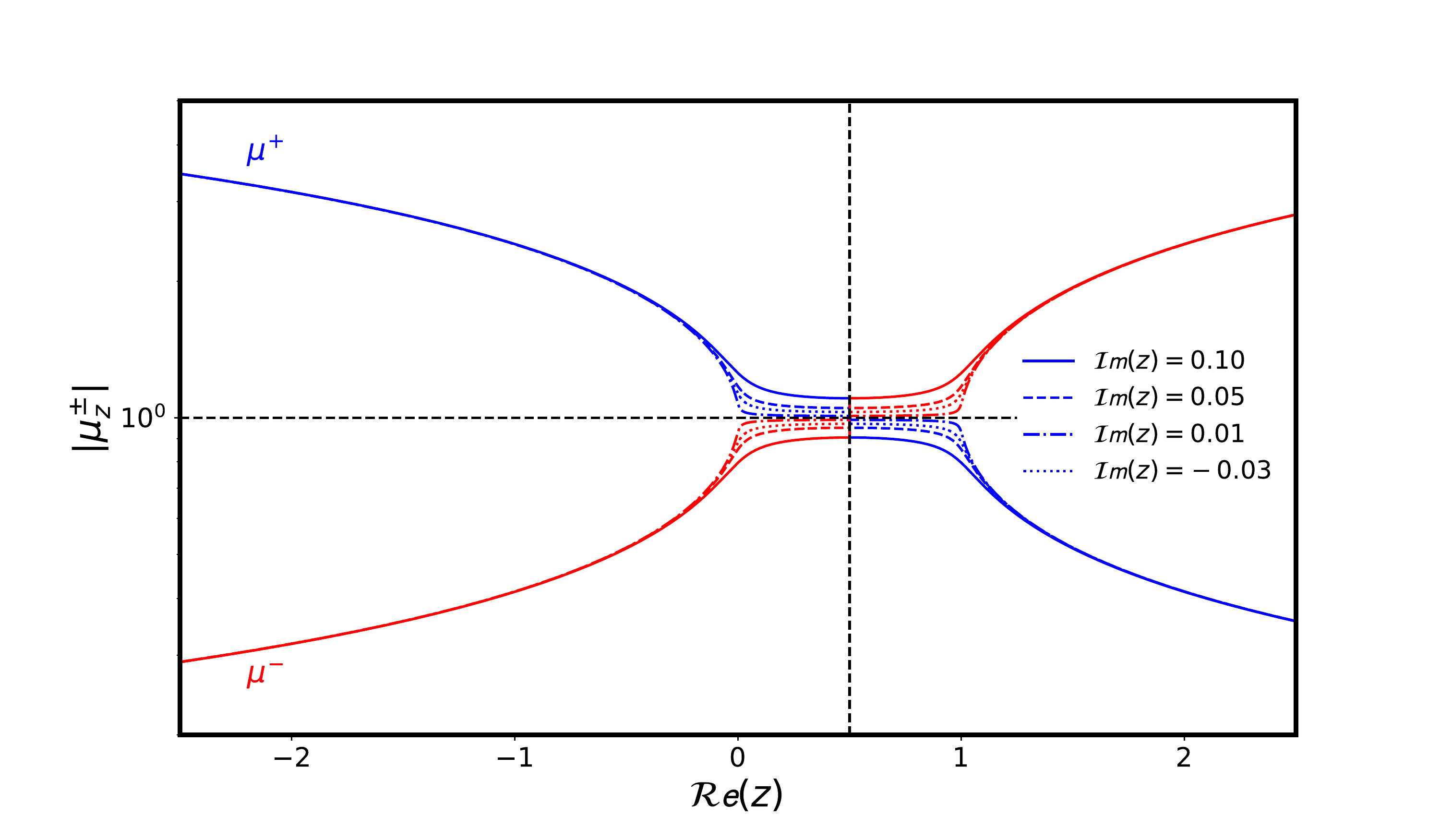}
\par\end{centering}
\vspace{-0.4cm}

\caption{\label{fig:Poles}Analysis of $\protect\abs{\rho_{z}^{\pm}}$ as a
function of a complex number $z\!=\!z'\!+\!iz''$. For $z'\!<\!0.5$
($z'\!>\!0.5$) only both $\rho^{+}$ ($\rho^{-}$) are inside the
unit circle, thus contributing to the value of the contour integral
in Eq.~\eqref{eq:I1_Complex}.}

\vspace{-0.8cm}
\end{figure}
The integral in Eq.\,\eqref{eq:I1_Complex} can be solved using the
\textit{Residue Theorem}, so that we have to pick up the poles inside
the unit circle, given any value of the complex parameter $z$. The
integrand has four poles: the complex roots of $\mu^{4}\!+\!2\left(2z\!-\!1\right)\mu^{2}\!+\!1$.
These come in two symmetric pairs, whose square is given by

\vspace{-0.7cm}
\begin{equation}
\rho_{z}^{\pm}\!=\!\left(1\!-\!2z\right)\!\pm\!\sqrt{(2z\!-\!1)^{2}\!-\!1}.
\end{equation}
By definition, the poles contributing to the integral in Eq.\,\eqref{eq:I1_Complex}
are the ones having moduli smaller than one. In Fig.\,\ref{fig:Poles},
we analyze $\abs{\rho_{z}^{\pm}}$ as a function of $z$, where one
sees that, if $\Re\left[z\right]<\nicefrac{1}{2}$( $\Re\left[z\right]>\nicefrac{1}{2}$)
only the square-roots of $\rho_{z}^{-}$ ($\rho_{z}^{+}$) lie inside
the unit circle. Furthermore, since each root corresponds to a simple
pole of the integrand, their residues are simple to calculate, and
we arrive at the final final result

\vspace{-0.7cm}
\begin{equation}
\mathcal{I}_{1}^{l}(z)\!=\!\begin{cases}
\frac{4\pi\left(\rho_{z}^{-}\right)^{\frac{\abs l}{2}}}{\rho_{z}^{-}\!-\!\rho_{z}^{+}}\left(1\!+\!(-1){}^{\abs l}\!\right) & \text{for \ensuremath{\Re[z]\!<\!\nicefrac{1}{2}}}\\
\frac{4\pi\left(\rho_{z}^{+}\right)^{\frac{\abs l}{2}}}{\rho_{z}^{+}\!-\!\rho_{z}^{-}}\left(1\!+\!(-1){}^{\abs l}\!\right) & \text{for \ensuremath{\Re[z]\!>\!\nicefrac{1}{2}}}
\end{cases}.\label{eq:I1_Calculated}
\end{equation}
From Eq.\,\eqref{eq:I1_Calculated}, it is already clear that $\mathcal{I}_{1}^{l}\!(z)\!=\!0$
whenever $l$ is an odd integer and, also using the fact that $\mathcal{I}_{1}^{-l}(z)\!=\!\mathcal{I}_{1}^{l}(z)$,
we can re-write the answer in a more condensed form:

\vspace{-0.7cm}

\begin{equation}
\mathcal{I}_{1}^{l}\!(z)\!=\!\!\begin{cases}
\frac{4\pi\left(1-2z+\text{sign}(\Re\left[2z\!-\!1\right])\sqrt{(2z\!-\!1)^{2}-1}\right)^{\frac{\abs l}{2}}}{\text{sign}(\Re\left[2z\!-\!1\right])\sqrt{(2z\!-\!1)^{2}-1}}\!\! & \!\!\!\!l\!\text{ even}\\
0 & \!\!\!\!l\text{ odd}
\end{cases}.\label{eq:I1_Calculated2}
\end{equation}
On top of all this, we can also take the formal limit of $z\!=\!x\!+\!i\eta\!\to\!x\!+\!i0^{\pm}$,
which yields

\vspace{-0.7cm}

\begin{equation}
\mathcal{I}_{1}^{l}\!(x,0^{\pm})\!=\!\!\begin{cases}
\mp\frac{2\pi i\left(1\!-\!2x\pm2i\text{sign}(2x\!-\!1)\sqrt{x(x\!-\!1)}\right)^{\frac{\abs l}{2}}}{\sqrt{x(x\!-\!1)}}\!\!\!\! & x\!\in\![0,\!1]\\
\frac{2\pi\left(1\!-\!2x+2\text{sign}(2x\!-\!1)\sqrt{x(1\!-\!x)}\right)^{\frac{\abs l}{2}}}{\text{sign}(2x\!-\!1)\sqrt{x(1\!-\!x)}}\!\!\!\! & x\!\notin\![0,\!1]
\end{cases},\label{eq:I1_Calculated3}
\end{equation}

a real-valued quantity outside the interval $[0,\!1]$, but still
complex within it. Note that, if $x\in[0,1]$, there is actually a
branch cut in the real axis, such that the $\mathcal{I}_{1}^{l}(x,0^{+})\!\neq\!\mathcal{I}_{1}^{l}(x,0^{-})$
, with two square-root singularities located at the borders of the
branch cut.

\vspace{-0.5cm}

\section{Numerical Evaluation of the Second Integral}

With the analytical expression for $\mathcal{I}_{1}^{l}(z)$, we can
now evaluate the remaining integrals by numerical quadrature. From
Eqs.\,\eqref{eq:I_diag-1}-\eqref{eq:I_diag-1-1}, the only integral
we really need is 

\vspace{-0.7cm}
\begin{equation}
\mathcal{I}_{2}^{m,l}\!(a)\!=\!\!\!\!\int_{-\pi}^{\pi}\!\!\!\!dve^{ivm}\mathcal{I}_{1}^{l}\!\left(a\!-\!\sin^{2}(v)\right)
\end{equation}
with a real parameter $a$. By the symmetry $v\!\to\!-v$, this integral
reduces to

\vspace{-0.7cm}

\begin{equation}
\mathcal{I}_{2}^{m,l}\!(a)\!=\!2\!\!\int_{0}^{\pi}\!\!\!\!dv\cos(vm)\mathcal{I}_{1}^{l}\!\left(a\!-\!\sin^{2}(v)\right)
\end{equation}
but can be further broken into,

\vspace{-0.7cm}

\begin{equation}
\mathcal{I}_{2}^{m,l}\!(a)\!=\!2\left(1\!+\!(-1)^{m}\right)\!\!\int_{0}^{\frac{\pi}{2}}\!\!\!\!dv\cos(vm)\mathcal{I}_{1}^{l}\!\left(a\!-\!\sin^{2}(v)\right).\label{eq:I2}
\end{equation}
Equation\,\ref{eq:I2} is nonzero if and only if $m$ is an even
integer, in which case it equals

\vspace{-0.7cm}

\begin{equation}
\mathcal{I}_{2}^{m,l}\!(a)\!=\!4\!\!\int_{0}^{\frac{\pi}{2}}\!\!\!\!dv\cos(vm)\mathcal{I}_{1}^{l}\!\left(a\!-\!\sin^{2}(v)\right).\label{eq:I2_Calculate}
\end{equation}
Since the integral of Eq.\,\eqref{eq:I2_Calculate} is defined in
the interval $v\!\in\![0,\nicefrac{\pi}{2}]$, we can change variables
from $\sigma\!=a\!-\!\sin^{2}(v)\to d\sigma=-2\cos v\sin vdv$, which
turns it into

\vspace{-0.7cm}

\begin{equation}
\mathcal{I}_{2}^{m,l}\!(a)\!=\!4\!\int_{a-1}^{a}\!\!\!\frac{\cos(m\arcsin\sqrt{a\!-\!\sigma})\mathcal{I}_{1}^{l}\!(\sigma)d\sigma}{\sqrt{(a\!-\!\sigma)(\sigma\!-\!a\!+\!1)}}.
\end{equation}
This expression has the advantage of clarifying that the integrand
only has integrable square-root type singularities which appear in
$\sigma\!=\!0,\!1$\,\footnote{Which may or not be inside the integration interval.}
and at the borders of the integration interval. Nevertheless, there
are three exception to this, for if $a\!=\!0,\!1,\!2$, at least two
of these singularities coincide and become (non-integrable) first-order
poles. These points appear as logarithmic divergences or discontinuities
in the values of $\mathcal{I}_{2}^{m,l}\!(a)$. In order to deal with
the integrable singularities properly in our numerical calculations,
it useful to consider three separate cases:

\vspace{-0.3cm}

\paragraph{The parameter $a$ is between $0$ and $1$:}

In this case, we have the $\sigma\!=\!0$ pole inside the integration
domain, which can be separated into four disjoint intervals of integration,
each with a single (one-sided) square-root singularity at the boundary,
\textit{i.e.},

\vspace{-0.7cm}

\begin{align}
\!\!\!\!\!\!\!\!\!\!\!\!\mathcal{I}_{2}^{m,l}\!(a)\! & =\!\!\!\int_{a-1}^{\frac{a-1}{2}}\!\!\!\!\!\!\!\!\!\!d\sigma\frac{f_{\sigma}^{m,l}}{\sqrt{\!\sigma(1\!-\!\sigma)(a\!-\!\sigma)(\sigma\!-\!a\!+\!1)}}+\int_{\frac{a-1}{2}}^{0}\!\!\!\!d\sigma\frac{f_{\sigma}^{m,l}}{\sqrt{\sigma(1\!-\!\sigma)(a\!-\!\sigma)(\sigma\!-\!a\!+\!1)}}\!\!\!\label{eq:ParcialIntegrals2}\\
 & \quad\quad+\int_{0}^{\frac{a}{2}}\!\!\!\!d\sigma\frac{f_{\sigma}^{m,l}}{\sqrt{\sigma(1\!-\!\sigma)(a\!-\!\sigma)(\sigma\!-\!a\!+\!1)}}+\int_{\frac{a}{2}}^{a}\!\!\!\!d\sigma\frac{f_{\sigma}^{m,l}}{\sqrt{\sigma(1\!-\!\sigma)(a\!-\!\sigma)(\sigma\!-\!a\!+\!1)}},\nonumber 
\end{align}
where $\!\!f_{\sigma}^{m,l}\!\!=\!\!\cos(m\arcsin\!\sqrt{\!a\!-\!\sigma})\sqrt{\!\sigma(1\!\!-\!\sigma)}\mathcal{I}_{1}^{l}\!(\sigma)$
are a regular functions of $\sigma$. In each term of Eq.\,\eqref{eq:ParcialIntegrals2},
we can change variables so as to eliminate the singularity within
that interval. Sequentially in Eq.\,\eqref{eq:ParcialIntegrals2},
we do $\tau\!=\!\sqrt{\sigma\!-\!a\!+\!1}$, $\tau\!=\!\sqrt{-\sigma}$
$\tau\!=\!\sqrt{\sigma}$ and $\tau\!=\!\sqrt{a\!-\!\sigma}$. After
these changes, we arrive at the equivalent integral

\vspace{-0.7cm}

\begin{align}
\!\!\!\!\!\!\!\!\!\!\!\!\mathcal{I}_{2}^{m,l}\!(a)\! & =\!\!\!\int_{0}^{\sqrt{\frac{1-a}{2}}}\!\!\!\!\!\!\!\!\!\!\!\!d\tau\frac{2f_{\tau^{2}\!+\!a\!-\!1}^{m,l}}{\sqrt{(1\!+\!\tau^{2})(\tau^{2}\!\!+a\!-\!1)(\tau^{2}\!\!+\!a\!+\!2)}}+\!\!\int_{0}^{\sqrt{\frac{1-a}{2}}}\!\!\!\!\!\!\!\!\!\!\!\!d\tau\frac{2f_{-\tau^{2}}^{m,l}}{\sqrt{(1\!+\!\tau^{2})(\tau^{2}\!\!+a\!-\!1)(\tau^{2}\!\!+\!a)}}\!\!\!\\
 & \quad+\int_{0}^{\sqrt{\frac{a}{2}}}\!\!\!\!\!\!\!\!\!\!d\tau\frac{2f_{\tau^{2}}^{m,l}}{\sqrt{(1\!+\!\tau^{2})(\tau^{2}\!\!+a\!-\!1)(\tau^{2}\!\!+\!a\!-\!2)}}+\int_{0}^{\sqrt{\frac{a}{2}}}\!\!\!\!\!\!\!\!d\tau\frac{2f_{a-\tau^{2}}^{m,l}}{\sqrt{(1\!-\!\tau^{2})(a\!-\!\tau^{2})(\tau^{2}\!\!-\!a\!+\!1)}}.\nonumber 
\end{align}
Note that, now, all the integrands are regular in their respective
integration domains, and can be evaluated with standard methods of
quadrature very easily. 

\vspace{-0.3cm}

\paragraph{The parameter $a$ is between $1$ and $2$:}

In this case, we have a situation analogous to the previous one, but
with the $\sigma\!=\!1$ pole inside the integration interval. To
proceed, we similarly split the integration region and perform suitable
variable changes, thus casting the integral into the form

\vspace{-0.7cm}

\begin{align}
\!\!\!\!\!\mathcal{I}_{2}^{m,l}\!(a)\! & =\!\!\!\int_{0}^{\sqrt{\frac{2-a}{2}}}\!\!\!\!\!\!\!\!\!\!\!\!\!d\tau\frac{2\left(f_{\tau^{2}\!+\!a\!-\!1}^{m,l}\!\!+\!f_{1-\tau^{2}}^{m,l}\right)}{\sqrt{(\tau^{2}\!\!-\!1)(\tau^{2}\!\!+\!a\!-\!2)(\tau^{2}\!\!+\!a\!-\!1)}}\nonumber \\
 & \qquad\qquad+\int_{0}^{\sqrt{\frac{a-1}{2}}}\!\!\!\!\!\!\!\!\!\!d\tau\frac{2f_{1\!+\!\tau^{2}}^{m,l}}{\sqrt{(1\!+\!\tau^{2})(\tau^{2}\!\!-\!a\!+\!2)(\tau^{2}\!\!-\!a\!+\!2)}}\nonumber \\
 & \qquad\qquad\qquad+\int_{0}^{\sqrt{\frac{a-1}{2}}}\!\!\!\!\!\!\!\!\!\!d\tau\frac{2f_{a-\tau^{2}}^{m,l}}{\sqrt{(\tau^{2}\!\!-\!1)(\tau^{2}\!\!-\!a)(\tau^{2}\!\!-\!a\!+\!1)}}.\label{eq:I2-1}
\end{align}
Once again, all the integrals in Eq.\,\ref{eq:I2-1} are of completely
regular functions.

\vspace{-0.3cm}

\begin{wrapfigure}[17]{o}{0.44\columnwidth}%
\includegraphics[scale=0.28]{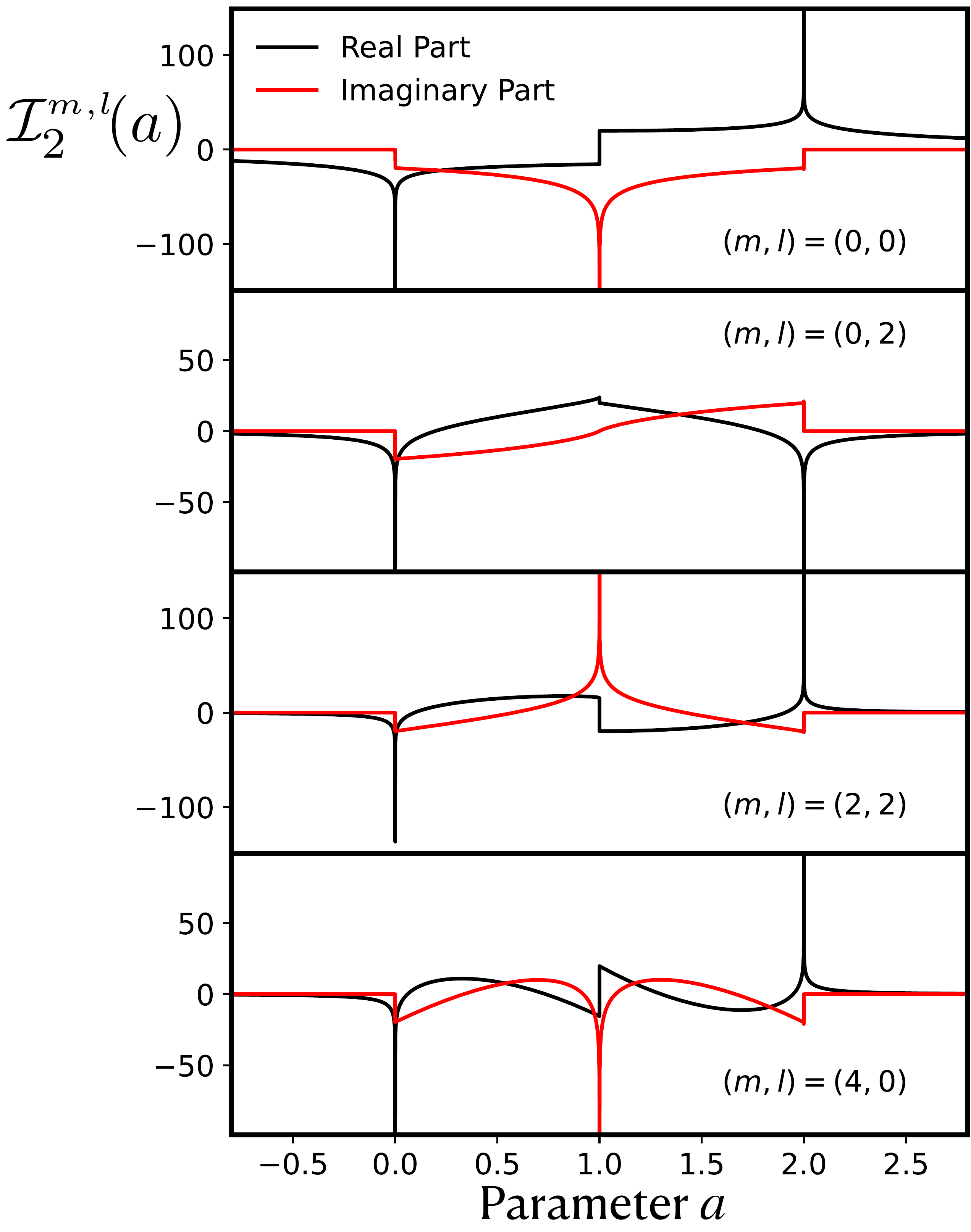}

\vspace{-0.2cm}

\caption{\label{fig:ExamplesI2}Examples of $\mathcal{I}_{2}^{m,l}\!(a)$.}

\vspace{-0.5cm}\end{wrapfigure}%

\paragraph{The parameter $a$ is not between $0$ and $2$:}

Finally, we consider the case in which there are no singularities
in the integration domain, except for the ones at the borders. In
this case, we only need to split the interval in half and change variables
in order to eliminate the integrable singularities in each case. This
procedure yields

\vspace{-0.7cm}

\begin{align}
\!\!\!\!\!\!\!\!\!\!\!\!\!\mathcal{I}_{2}^{m,l}\!(a)\! & =\!\!\!\int_{0}^{\frac{1}{2}}\!\!\!\!\!d\tau\frac{2f_{\tau^{2}\!+\!a\!-\!1}^{m,l}}{\sqrt{(\tau^{2}\!\!-\!1)(\tau^{2}\!\!+\!a\!-\!2)(\tau^{2}\!\!+\!a\!-\!1)}}+\!\!\!\nonumber \\
 & \qquad+\!\!\int_{0}^{\frac{1}{2}}\!\!\!\!\!d\tau\frac{2f_{a-\tau^{2}}^{m,l}}{\sqrt{(\tau^{2}\!\!-\!1)(\tau^{2}\!\!-\!a)(\tau^{2}\!\!-\!a\!+\!1)}}.
\end{align}

Using the previous expressions, we can easily evaluate $\mathcal{I}_{2}^{m,l}(a)$
for any real value of $a$. Some examples are shown in Fig.\,\ref{fig:ExamplesI2},
where the discontinuous/divergent behavior near the special points,
$a\!=\!0,1\text{ and }2$ are evident.

\vspace{-0.5cm}

\section{Numerical Evaluation of the Third Integral}

The two basic integrals of Eqs.\,\eqref{eq:I_diag-1}-\eqref{eq:I_diag-1-1}
may be expressed in terms of the double integral $\mathcal{I}_{2}^{m,l}(a)$
which was evaluated in the previous section. However, we provide a
detailed explanation, starting by reminding the following symmetries,

\vspace{-0.7cm}

\begin{subequations}
\begin{align}
\mathcal{I}_{0}^{{\scriptscriptstyle n,m,l}}\!(\varepsilon\!+\!i\eta)\! & =\!-\!\left[\mathcal{I}_{0}^{{\scriptscriptstyle n,m,l}}(-\varepsilon\!+\!i\eta)\right]^{*}\\
\mathcal{I}_{x}^{{\scriptscriptstyle n,m,l}}\!(\varepsilon\!+\!i\eta)\! & =\!-\!\left[\mathcal{I}_{x}^{{\scriptscriptstyle n,m,l}}(-\varepsilon\!+\!i\eta)\right]^{*}.
\end{align}
\end{subequations}

which allow us to considering only $\varepsilon\!\geq\!0$, and therefore
express both integrals, already in the $\eta\!\to\!0^{+}$ limit,
as

\vspace{-0.7cm}

\begin{subequations}
\begin{align}
\mathcal{I}_{0}^{{\scriptscriptstyle n,m,l}}\!\left(\varepsilon\right)\! & =\!\!\!\int_{-\pi}^{\pi}\!\!\!\!due^{iun}\mathcal{I}_{2}^{m,l}\!\left(\varepsilon^{2}\!\!-\!\sin^{2}\!u\!\right)\label{eq:I_diag-1-2}\\
\mathcal{I}_{x}^{{\scriptscriptstyle n,m,l}}\!\left(\varepsilon\right)\! & =\!\!\!\int_{-\pi}^{\pi}\!\!\!\!\!du\sin\!u\,e^{iun}\mathcal{I}_{2}^{m,l}\!\left(\varepsilon^{2}\!\!-\!\sin^{2}\!u\!\right).\label{eq:I_diag-1-1-1}
\end{align}
\end{subequations}

Both these integrals can be written in a reduced region, using the
same symmetries invoked in the previous section. Without surprise,
we conclude that $\mathcal{I}_{0}^{{\scriptscriptstyle n,m,l}}$ is
only nonzero iff $n,m$ and $l$ are all even, while $\mathcal{I}_{x}^{{\scriptscriptstyle n,m,l}}$
needs $n$ to be odd, with $m,l$ even. In case each integral is non-zero,
we can cast them into the forms,

\vspace{-0.7cm}

\begin{subequations}
\begin{align}
\mathcal{I}_{0}^{{\scriptscriptstyle n,m,l}}\!\left(\varepsilon\right)\! & =\!4\int_{0}^{\frac{\pi}{2}}\!\!\!\!\!du\cos(nu)\mathcal{I}_{2}^{{\scriptscriptstyle m,l}}\!(\varepsilon^{2}\!\!-\!\sin^{2}\!u\!)\label{eq:I_diag-1-2-1-1-1}\\
\mathcal{I}_{x}^{{\scriptscriptstyle n,m,l}}\!\left(\varepsilon\right)\! & =\!4i\!\int_{0}^{\frac{\pi}{2}}\!\!\!\!\!du\sin\!u\,\sin(nu)\mathcal{I}_{2}^{{\scriptscriptstyle m,l}}\!(\varepsilon^{2}\!\!-\!\sin^{2}\!u\!).
\end{align}
\end{subequations}

And, finally, we change variables from $u\to\rho\!=\!\varepsilon^{2}\!\!-\!\sin^{2}\!u$,
which yields

\vspace{-0.7cm}

\begin{subequations}
\begin{align}
\mathcal{I}_{0}^{{\scriptscriptstyle n,m,l}}\!\left(\varepsilon\right)\! & =\!4\!\int_{\varepsilon^{2}-1}^{\varepsilon^{2}}\!\!\!\!\!\frac{\cos(n\arcsin\sqrt{\varepsilon^{2}\!-\!\rho})\mathcal{I}_{2}^{{\scriptscriptstyle m,l}}\!(\rho)d\rho}{\sqrt{(\varepsilon^{2}\!-\!\rho)(\rho\!-\!\varepsilon^{2}\!+\!1)}}\label{eq:I_diag-1-2-1-1-1-1}\\
\mathcal{I}_{x}^{{\scriptscriptstyle n,m,l}}\!\left(\varepsilon\right)\! & =\!4i\!\int_{\varepsilon^{2}-1}^{\varepsilon^{2}}\!\!\!\!\!\frac{\sqrt{\varepsilon^{2}\!-\!\rho}\sin(n\arcsin\sqrt{\varepsilon^{2}\!-\!\rho})\mathcal{I}_{2}^{{\scriptscriptstyle m,l}}\!(\rho)d\rho}{\sqrt{(\varepsilon^{2}\!-\!\rho)(\rho\!-\!\varepsilon^{2}\!+\!1)}},\label{eq:I_diag-1-1-1-1-1-1-1}
\end{align}
\end{subequations}

where clearly two square-root singularities remain at the borders
of the integration interval. However, unlike our starting point, in
the integrals of Eqs.\,\eqref{eq:I_diag-1-2-1-1-1-1}-\eqref{eq:I_diag-1-2-1-1-1-1}
we are assured to only have integrable singularities, whenever the
value of $\varepsilon$ is. These singularities can be well-estimated
by numerical quadrature with a mesh of points that samples well the
values of the integrand around the said singularities.

\lhead[\MakeUppercase{Appendix E}]{\MakeUppercase{\rightmark}}

\rhead[\MakeUppercase{Multi-Valley Continuum Limit}]{}

\chapter{\label{chap:MultiValleyCont}Multi-Valley Continuum Approximation}

In this Appendix, we determine the low-energy continuum limit of the
lattice Green's function determined in Chapter\,\ref{chap:Rare-Event-States}.
Here, we properly consider the Weyl fermion excitations around all
eight inequivalent valleys fo the lattice model,

\vspace{-0.7cm}

\begin{align}
\mathcal{H}_{l}^{0} & \!=\!\negthickspace\sum_{{\scriptscriptstyle \mathbf{R}\in\mathcal{L_{\text{C}}}}}\negthickspace\left[\frac{i\hbar v_{\text{F}}}{2a}\Psi_{\mathbf{{\scriptscriptstyle R}}}^{\dagger}\!\cdot\!\sigma^{j}\!\cdot\!\Psi_{{\scriptscriptstyle \mathbf{R}+a\hat{e}_{j}}}\!\!\text{h.c.}\right]\!,\label{eq:LatticeModel-2}
\end{align}
for a Weyl semimetal in a simple cubic lattice $\mathcal{L}_{\text{C}}$.
The results obtained here are the ones used to compare with the numerically
exact results for the lGF, as shown in Fig.\,\ref{fig:LatticeGFvsCont}.
Our starting point is the original expression for the lGF as an integral
over the first Brillouin zone, \textit{i.e.},

\vspace{-0.7cm}

\begin{align}
\mathscr{G}_{ab}^{\text{0r}}(\varepsilon;\boldsymbol{\Delta L})\! & =\!\frac{a}{8\pi^{3}\hbar v_{\text{F}}}\int_{\text{C}}\!\!\!d\mathbf{q}\,\frac{\varepsilon\delta_{ab}+\boldsymbol{\sigma}_{ab}\cdot\boldsymbol{\sin}\mathbf{q}}{\left(\varepsilon\!+\!i\eta\right)^{2}\!-\!\abs{\boldsymbol{\sin}\mathbf{q}}^{2}}e^{i\mathbf{q}\cdot\boldsymbol{\Delta L}},\label{eq:RealSpacePropagator-2-1-2}
\end{align}
\begin{wrapfigure}[14]{o}{0.45\columnwidth}%
\vspace{-0.7cm}
\begin{centering}
\includegraphics[scale=0.2]{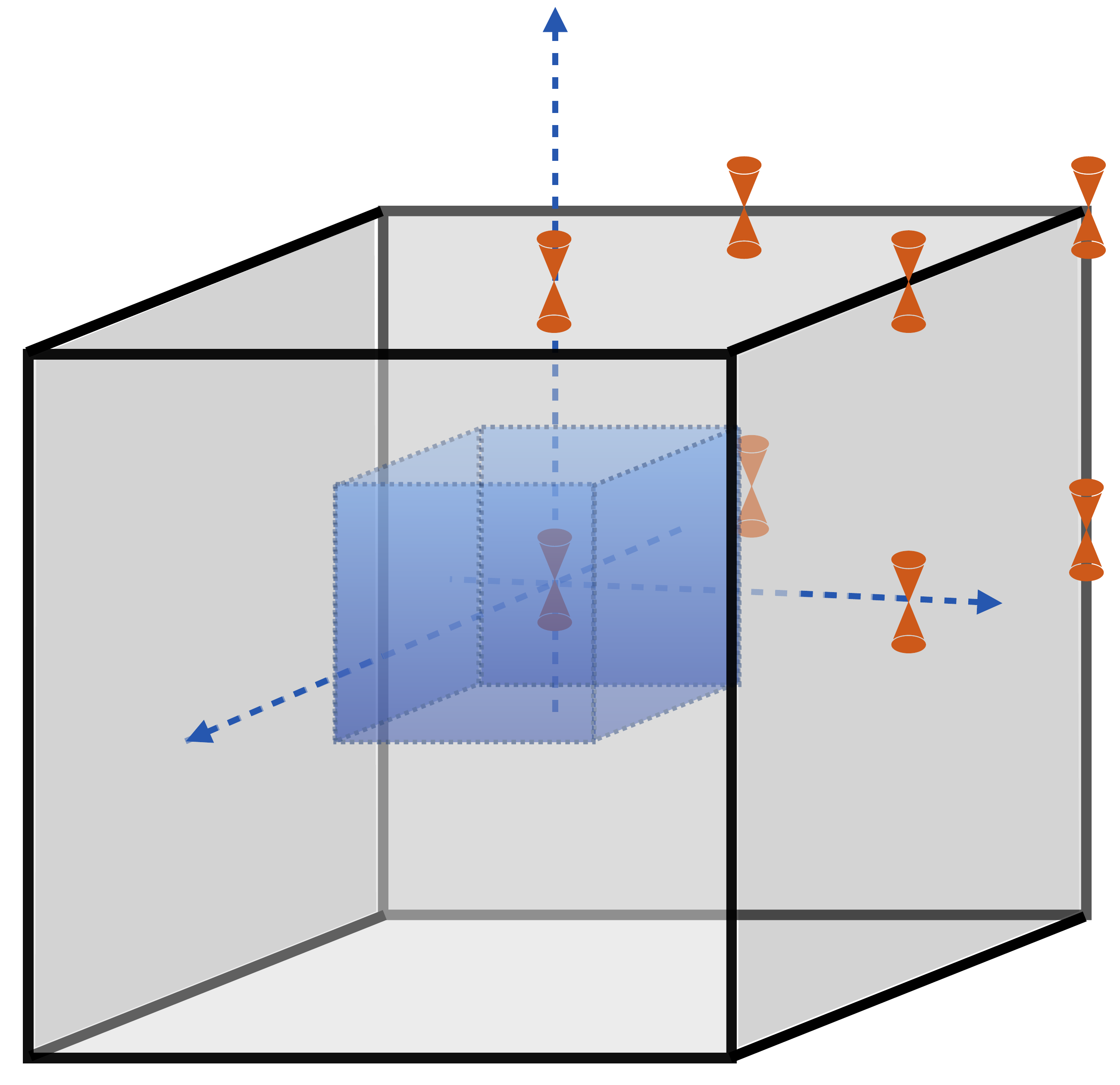}
\par\end{centering}
\caption{\label{fig:FBZ}First Brillouin zone of the lattice model. The blue
cube represents the partitioning of the fBz used to obtained the multi-valley
low-energy Green's function.}
\end{wrapfigure}%
where $\varepsilon=Ea/\hbar v_{\text{F}}$, and the crystal momenta
were normalized to the inverse lattice spacing, $\mathbf{q}\!=\!a\mathbf{k}$.
Near $\varepsilon\!=\!0$, the important contributions to the integral
will arise from states which are nearby the nodes at the TRIM. These
are represented in the scheme of Fig.\,\ref{fig:FBZ} and are simply,

\vspace{-0.7cm}
\begin{align*}
\mathbf{K}_{1}\! & =\!(0,0,0)\quad\,;\quad\mathbf{K}_{2}\!=\!(\pi,0,0)\\
\mathbf{K}_{3}\! & =\!(0,\pi,0)\quad;\quad\mathbf{K}_{4}\!=\!(0,0,\pi)\\
\mathbf{K}_{5}\! & =\!(\pi,\pi,0)\quad;\quad\mathbf{K}_{6}\!=\!(\pi,0,\pi)\\
\mathbf{K}_{7}\! & =\!(0,\pi,\pi)\quad;\quad\mathbf{K}_{8}\!=\!(\pi,\pi,\pi).
\end{align*}

In order to obtain an analytical expression for low-energies, we break
the fBz into eight equal small cubic pieces centered on each of the
$\mathbf{K}_{j}$-points. This is also depicted in Fig.\,\ref{fig:FBZ}
and turns Eq.\,\eqref{eq:RealSpacePropagator-2-1-2} into

\vspace{-0.3cm}

\begin{align}
\!\!\!\!\mathscr{G}_{ab}^{\text{0r}}(\varepsilon;\boldsymbol{\Delta L})\! & =\!\frac{a}{8\pi^{3}\hbar v_{\text{F}}}\!\sum_{\mathbf{K}_{j}}\!\int_{\text{C}^{\prime}}\!\!d\mathbf{p}\,\frac{\varepsilon\delta_{ab}+\boldsymbol{\sigma}_{ab}\cdot\boldsymbol{\sin}\left(\mathbf{K}_{j}+\mathbf{p}\right)}{\left(\varepsilon\!+\!i\eta\right)^{2}\!-\!\abs{\boldsymbol{\sin}\left(\mathbf{K}_{j}+\mathbf{p}\right)}^{2}}e^{i\left(\mathbf{K}_{j}+\mathbf{p}\right)\cdot\boldsymbol{\Delta L}},\!\!\!\label{eq:RealSpacePropagator-2-1-2-1}
\end{align}

where $\text{C}^{\prime}\!=\![-\nicefrac{\pi}{2},\nicefrac{\pi}{2}]^{3}$
and $\mathbf{p}$ are the deviations relative to the each TRIM. The
integrand can be further simplified by recognizing that 

\vspace{-0.7cm}
\begin{equation}
\sin\left(\mathbf{K}_{j}+\mathbf{p}\right)=\left(\cos K_{j}^{x}\sin p^{x},\cos K_{j}^{y}\sin p^{y},\cos K_{j}^{z}\sin p^{z}\right)=\cos\mathbf{K}_{j}\tensorproduct\sin\mathbf{p}=\pm\sin\mathbf{p},
\end{equation}

which gives

\vspace{-0.7cm}

\begin{align}
\mathscr{G}_{ab}^{\text{0r}}(\varepsilon;\boldsymbol{\Delta L})\! & =\!\frac{a}{8\pi^{3}\hbar v_{\text{F}}}\!\sum_{\mathbf{K}_{j}}e^{i\mathbf{K}_{j}\cdot\Delta\mathbf{R}}\!\!\int_{\text{C}^{\prime}}\!\!\!d\mathbf{p}\,\frac{\varepsilon\delta_{ab}+\boldsymbol{\sigma}_{ab}\cdot\left[\cos\mathbf{K}_{j}\tensorproduct\sin\mathbf{p}\right]}{\left(\varepsilon\!+\!i\eta\right)^{2}\!-\!\abs{\boldsymbol{\sin}\mathbf{p}}^{2}}e^{i\mathbf{p}\cdot\boldsymbol{\Delta L}}.\label{eq:RealSpacePropagator-2-1-2-1-1}
\end{align}

Now, in the limit $\varepsilon\!\to\!0$, we can approximate the lattice's
dispersion relation as linear, which yields 

\vspace{-0.7cm}

\begin{equation}
\mathscr{G}_{ab}^{\text{0r}}(\varepsilon;\boldsymbol{\Delta L})\approx\!\gamma_{ab}\left(\Delta\boldsymbol{L}\right)\mathcal{I}_{1c}\!\left(\varepsilon,\Delta\boldsymbol{L}\right)-\boldsymbol{\sigma}_{ab}\left(\Delta\boldsymbol{L}\right)\cdot\Delta\boldsymbol{L}\:\mathcal{I}_{2c}\!\left(\varepsilon,\Delta\boldsymbol{L}\right),
\end{equation}
where 

\vspace{-0.7cm}
\begin{equation}
\gamma\left(\Delta\boldsymbol{L}\right)\!=\!\delta_{ab}\!\sum_{\mathbf{K}_{j}}e^{i\mathbf{K}_{j}\cdot\Delta\boldsymbol{L}}\text{ and }\sigma^{l}\!\left(\Delta\mathbf{R}\right)\!=\!\sigma_{ab}^{l}\!\sum_{\mathbf{K}_{j}}e^{i\mathbf{K}_{\!j}\cdot\Delta\boldsymbol{L}}\cos K_{j}^{l}.
\end{equation}
This SPGF in real-space is fully determined by two integrals, just
like the one obtained for the continuum model of a single Weyl node.
These integrals read simply

\vspace{-0.7cm}

\begin{align}
\mathcal{I}_{1c}\left(\varepsilon,\Delta\boldsymbol{L}\right)\! & =\!\frac{1}{4\pi^{2}}\!\!\int_{0}^{\infty}\!\!dpp^{2}\!\left(\!\frac{\varepsilon}{\left(\varepsilon\!+\!i\eta\right)^{2}\!-\!p^{2}}\!\!\right)\!\int_{0}^{\pi}\!\!\!\!\!d\theta\sin\theta e^{ip\abs{\Delta\boldsymbol{L}}\cos\theta}\\
\mathcal{I}_{2c}\left(\varepsilon,\Delta\boldsymbol{L}\right)\! & =\!\frac{1}{4\pi^{2}}\!\!\int_{0}^{\infty}\!\!dpp^{2}\!\left(\!\frac{p}{\left(\varepsilon\!+\!i\eta\right)^{2}\!-\!p^{2}}\!\!\right)\!\int_{0}^{\pi}\!\!\!\!\!d\theta\cos\theta\sin\theta e^{ip\abs{\Delta\boldsymbol{L}}\cos\theta},
\end{align}
where the angular integral can be easily integrated and, therefore,

\vspace{-0.7cm}

\begin{align}
\mathcal{I}_{1c}\left(\varepsilon,\Delta\boldsymbol{L}\right)\! & =\!\frac{\varepsilon}{4\pi^{2}\abs{\Delta\boldsymbol{L}}}\!\!\int_{-\infty}^{\infty}\!\!\!\!dp\frac{p^{2}\sin\left(p\abs{\Delta\boldsymbol{L}}\right)}{\left(\varepsilon\!+\!i\eta\right)^{2}\!-\!p^{2}}\\
\mathcal{I}_{2c}\left(\varepsilon,\Delta\boldsymbol{L}\right)\! & =\!\frac{i}{4\pi^{2}\abs{\Delta\boldsymbol{L}}^{2}}\!\!\int_{-\infty}^{\infty}\!\!\!\!dp\frac{p\sin\left(p\abs{\Delta\boldsymbol{L}}\right)\!-\!p^{2}\abs{\Delta\boldsymbol{L}}\cos\left(p\abs{\Delta\boldsymbol{L}}\right)}{\left(\varepsilon\!+\!i\eta\right)^{2}\!-\!p^{2}},
\end{align}
or, equivalently,

\vspace{-0.7cm}

\begin{align}
\mathcal{I}_{1c}\left(\varepsilon,\Delta\boldsymbol{L}\right)\! & =\!\frac{\varepsilon}{8i\pi^{2}\abs{\Delta\boldsymbol{L}}}\left[\int_{-\infty}^{\infty}\!\!\frac{pe^{ip\abs{\Delta\boldsymbol{L}}}}{\left(\varepsilon\!+\!i\eta\right)^{2}\!-\!p^{2}}dp-\!\!\!\int_{-\infty}^{\infty}\!\!\frac{pe^{-ip\abs{\Delta\boldsymbol{L}}}}{\left(\varepsilon\!+\!i\eta\right)^{2}\!-\!p^{2}}dp\right],\\
\mathcal{I}_{2c}\left(\varepsilon,\Delta\boldsymbol{L}\right)\! & =\!\frac{1}{8\pi^{2}\abs{\Delta\boldsymbol{L}}^{2}}\!\left[\int_{-\infty}^{\infty}\!\!\!\!\!dp\!\left(\!\frac{p\left(1\!-\!ip\abs{\Delta\boldsymbol{L}}\right)e^{ip\abs{\Delta\boldsymbol{L}}}}{\left(\varepsilon\!+\!i\eta\right)^{2}\!-\!p^{2}}\!\right)\!\right.\nonumber \\
 & \qquad\qquad\qquad\qquad\left.-\!\int_{-\infty}^{\infty}\!\!\!\!\!dp\!\left(\!\frac{p\left(1\!+\!ip\abs{\Delta\boldsymbol{L}}\right)e^{-ip\abs{\Delta\boldsymbol{L}}}}{\left(\varepsilon\!+\!i\eta\right)^{2}\!-\!p^{2}}\!\right)\right].
\end{align}

Both these integrals can be solved by using the \textit{Residues Theorem},
and do not require any kind of regularization. In contrast, the on-site
Green's function requires us to introduce a regularization scheme,
exactly as it happened in the single node model of Chapter\,\eqref{chap:Instability_Smooth_Regions}.
The final results obtained for real-space SPGF are

\vspace{-0.7cm}

\begin{align}
\mathcal{I}_{1c}\left(\varepsilon,\Delta\boldsymbol{L}\right)\! & =\!-\frac{\varepsilon}{4\pi\abs{\Delta\boldsymbol{L}}}e^{i\left(\varepsilon\!+\!i\eta\right)\abs{\Delta\boldsymbol{L}}}\underset{\eta\to0^{+}}{\longrightarrow}-\frac{\varepsilon e^{i\varepsilon\abs{\Delta\boldsymbol{L}}}}{4\pi\abs{\Delta\boldsymbol{L}}}\\
\mathcal{I}_{2c}\left(\varepsilon,\Delta\boldsymbol{L}\right)\! & =\!-\frac{i\left[1-i\left(\varepsilon\!+\!i\eta\right)\abs{\Delta\boldsymbol{L}}\right]}{4\pi\abs{\Delta\boldsymbol{L}}^{2}}e^{i\left(\varepsilon\!+\!i\eta\right)\abs{\Delta\boldsymbol{L}}}\underset{\eta\to0^{+}}{\longrightarrow}-\frac{i\left[1-i\varepsilon\abs{\Delta\boldsymbol{L}}\right]}{4\pi\abs{\Delta\boldsymbol{L}}^{2}}e^{i\varepsilon\abs{\Delta\boldsymbol{L}}},
\end{align}
so that we can finally write down the Green's function in the continuum
approximation (and at a finite distance) as

\vspace{-0.7cm}

\begin{align}
\!\!\!\!\!\mathscr{G}_{ab}^{\text{0r}}(E;\boldsymbol{\Delta L}) & \approx\!-\frac{a^{2}E\exp\left(iE\abs{\Delta\boldsymbol{L}}/\hbar v_{\text{F}}\right)}{4\pi\hbar^{2}v_{\text{F}}^{2}\abs{\Delta\boldsymbol{L}}}\gamma_{ab}\left(\Delta\boldsymbol{L}\right)\nonumber \\
 & \quad+\left(\boldsymbol{\sigma}_{ab}\left(\Delta\boldsymbol{L}\right)\cdot\Delta\boldsymbol{L}\right)\frac{a\left[1\!-\!iE\abs{\Delta\mathbf{R}}/\hbar v_{\text{F}}\right]\exp\left(iE\abs{\Delta\boldsymbol{L}}/\hbar v_{\text{F}}\right)}{4i\pi\hbar v_{\text{F}}\abs{\Delta\boldsymbol{L}}^{2}},
\end{align}

All we have done relied on the existence of the finite distance, $\Delta\boldsymbol{L}$,
which regularizes the $p$ integrals. However, this is not the case
when dealing with $\Delta\mathbf{R}\!=\!\mathbf{0}$. Then, one has
the following expressions:

\vspace{-0.3cm}

\begin{align}
g^{d}\!\left(E,\mathbf{0}\right)\! & =\!\frac{1}{8\pi^{3}a^{3}}\!\!\int\!\!\!\!\!\int\!\!\!\!\!\int_{[-\frac{\pi}{2},\frac{\pi}{2}]^{3}}\!\!\!\!\!\!\!\!d\mathbf{p}\!\left(\!\!\frac{\left(E\!+\!i\eta\right)}{\left(E\!+\!i\eta\right)^{2}-t^{2}\abs{\mathbf{p}}^{2}}\!\!\right)\\
\boldsymbol{g}^{o}\!\!\left(E,\mathbf{0}\right) & =\frac{1}{8\pi^{3}a^{3}}\!\!\int\!\!\!\!\!\int\!\!\!\!\!\int_{[-\frac{\pi}{2},\frac{\pi}{2}]^{3}}\!\!\!\!\!\!\!\!d\mathbf{p}\!\left(\!\!\frac{t\mathbf{p}}{\left(E\!+\!i\eta\right)^{2}-t^{2}\abs{\mathbf{p}}^{2}}\!\!\right),
\end{align}

which amounts to a solution of the following integrals

\vspace{-0.5cm}

\begin{equation}
\mathcal{I}_{1c}(E,\mathbf{0})\!=\!\frac{1}{4\pi^{2}}\!\!\int_{-\infty}^{\infty}\!\!\!\!dp\!\left(\!\frac{p^{2}}{\left(E\!+\!i\eta\right)^{2}\!-\!p^{2}}\!\!\right),
\end{equation}

\vspace{-0.5cm}

\begin{equation}
\mathcal{I}_{2c}(E,\mathbf{0})\!=\!\frac{1}{4\pi^{2}}\!\!\int_{0}^{\infty}\!\!\!\!dp\left(\!\frac{p^{3}}{\left(E\!+\!i\eta\right)^{2}\!-\!p^{2}}\!\!\right)\int_{0}^{\pi}\!\!\!\!d\theta\sin\theta=0.
\end{equation}

The first integral is divergent, while the second is zero by symmetry.
The first integral then must be regularized and, for that, we make
use of a \textit{Pauli-Villars regularization} (also called \textit{``smooth
cut-off}'' in the main text), which corresponds to placing a lorentzian
filter into the integrand, \textit{i.e.},

\vspace{-0.7cm}

\begin{equation}
\mathcal{I}_{1c}(E,\mathbf{0};M)\!=\!\frac{1}{4\pi^{2}}\!\!\int_{-\infty}^{\infty}\!\!\!\!\!\!\!dp\!\left(\!\frac{p^{2}}{\left(E\!+\!i\eta\right)^{2}\!-\!p^{2}}\!\!\right)\left(\frac{M^{2}}{p^{2}+M^{2}}\right).
\end{equation}
This integral, for a finite cut-off scale $M\gg E$, can be solved
analytically by using the residues theorem, one again. This yields

\vspace{-0.7cm}
\begin{equation}
\mathcal{I}_{1c}(E,\mathbf{0};M)\!=\!\frac{-M+iE}{4\pi t^{2}}
\end{equation}
which implies that

\vspace{-0.7cm}
\begin{equation}
g^{d}\!\left(E,\mathbf{0}\right)\!=\!\frac{-ME+iE^{2}}{4\pi t^{3}a^{3}}.
\end{equation}
Furthermore, for $\Delta\mathbf{R}\!=\!\mathbf{0}$, we see that $\gamma\!=\!8\,\mathbb{I}_{4\times4}$
and the full on-site Green's function reads

\vspace{-0.7cm}
\begin{equation}
G^{0}\!\left(E,\mathbf{0}\right)\!=\!-\frac{2E}{\pi t^{3}a^{3}}\left(M+iE\right)\mathbb{I}_{4\times4}.
\end{equation}
From here, it is clear that the density of states in this system is
simply

\vspace{-0.7cm}
\begin{equation}
\rho_{c}\left(E\right)=-\frac{1}{\pi}\Im\left[G^{0}\!\left(E,\mathbf{0}\right)\right]=\frac{2E^{2}}{\pi^{2}t^{3}a^{3}},
\end{equation}
a result which is just eight times the density of states in a single-node
model, while the real part of the on-site Green's function is linearly
dependent on the cut-off scale $M$. Given a particular lattice model,
this cut-off scale can be adjusted by determining the slope of $G^{0}\left(E,\mathbf{0}\right)$
at $E=0$, \textit{i.e.},

\vspace{-0.7cm}

\begin{equation}
M=-\frac{\pi t^{3}a^{3}}{2}\frac{d}{dE}\Re\left[G^{0}\!\left(E,\mathbf{0}\right)\right].
\end{equation}
For our lattice model, this cut-off has the value $M\!\approx\!1.588t$. 

\lhead[\MakeUppercase{Appendix F}]{\MakeUppercase{\rightmark}}

\rhead[\MakeUppercase{Perturbative Statistics}]{}

\chapter{\label{chap:Perturbative-Disorder}Perturbative Dressing of a Single
Level by Disorder}

In this short Appendix, we describe the statistics of the low-lying
levels when slightly perturbed by an uncorrelated potential. This
discussion justifies some of the results presented in Sect.\,\ref{sec:What-are-RareEvents},
namely the behavior of the finite-size gap with increasing disorder
strength. Assuming a finite Weyl system with twisted boundary conditions
and odd side length ($L$), we can study the statistics of the lowest
energy levels when the lattice has an on-site scalar disordered landscape.
To be more concrete, the Hamiltonian of the system reads

\vspace{-0.7cm}

\begin{equation}
\mathcal{H}\!=\!\sum_{\mathbf{R}\in\mathcal{L}}\left(-\frac{it}{2}\sum_{i=x,y,z}\left[\Psi_{\mathbf{R}}^{\dagger}\!\cdot\!\sigma_{i}\!\cdot\!\Psi_{\mathbf{R}+a\mathbf{e}_{i}}+\Psi_{\mathbf{R}+a\mathbf{e}_{i}}^{\dagger}\!\cdot\!\sigma_{i}\!\cdot\!\Psi_{\mathbf{R}}\right]+WV_{d}(\mathbf{R})\Psi_{\mathbf{R}}^{\dagger}\!\cdot\!\Psi_{\mathbf{R}}\right),
\end{equation}
where $W\!>\!0$ is the strength of disorder and $V_{d}(\mathbf{R})$
is a random function which has uncorrelated gaussian statistics, i.e.,

\vspace{-0.7cm}

\begin{equation}
\av{V_{d}(\mathbf{R})}_{d}=0;\quad\av{V_{d}(\mathbf{R})V_{d}(\mathbf{Q})}_{d}=\delta_{\mathbf{R},\mathbf{Q}}\label{eq:V1}
\end{equation}
and all other averages of products of $V_{d}$ are obtained by Wick's
theorem. In particular, we will require the 4-point correlator

\vspace{-0.7cm}

\begin{equation}
\av{V_{d}(\mathbf{R})V_{d}(\mathbf{Q})V_{d}(\mathbf{P})V_{d}(\mathbf{L})}_{d}=\delta_{\mathbf{R},\mathbf{Q}}\delta_{\mathbf{P},\mathbf{L}}+\delta_{\mathbf{R},\mathbf{P}}\delta_{\mathbf{Q},\mathbf{L}}+\delta_{\mathbf{R},\mathbf{L}}\delta_{\mathbf{Q},\mathbf{P}}.\label{eq:V2}
\end{equation}

Since $V_{d}(\mathbf{R})$ is a random function of position, the Bloch
states are no longer a good basis to diagonalize the Hamiltonian $\mathcal{H}$.
However, one expects that for low enough $W$ a perturbative treatment
can be used to obtain the energy levels of the disordered systems.
Since we are assuming that our boundary conditions are such that the
clean energy levels are non-degenerate, we can write use a standard
non-degenerate time-independent perturbation theory to study those
in the presence of the disordered landscape. Hence, in 2nd-order perturbation
theory in $W$, we have 

\vspace{-0.7cm}
\begin{equation}
E_{0}^{{\scriptscriptstyle \pm}}[V_{d}\left(\mathbf{R}\right)]=\pm\varepsilon_{0}+W\bra{\Psi_{0}^{\pm}}\hat{V}_{d}\ket{\Psi_{0}^{\pm}}-W^{2}\!\!\!\!\sum_{\mathbf{k}\in F\!B\!Z}\sum_{s=\pm}\frac{\bra{\Psi_{0}^{\pm}}\hat{V}_{d}\ketbra{\Psi_{\mathbf{k}}^{s}}{\Psi_{\mathbf{k}}^{s}}\hat{V}_{d}\ket{\Psi_{0}^{\pm}}}{\varepsilon^{s}(\mathbf{k})\mp\varepsilon_{0}},\label{eq:PerturbationTheory}
\end{equation}
such that $\mathcal{H}_{0}\ket{\Psi_{0}^{\pm}}\!=\!\pm\varepsilon_{0}\ket{\Psi_{0}^{\pm}}.$
The previous summation over $\mathbf{k}$ excludes the lowest available
momentum, $\mathbf{k}\!=\!2\pi\boldsymbol{\varphi}/L$, where $\boldsymbol{\varphi}$
is the vector containing the boundary phase-twists applied to the
system {[}see Sect.\,\eqref{sec:What-are-RareEvents} for context{]},
and $s$ there stands for the band index.

Equation\,\eqref{eq:PerturbationTheory} basically defines the energy
of the two non-degenerate levels closest to the node as a function
of the random field $V_{d}(\mathbf{R})$. This implies that $E_{0}^{\pm}$
will also be a random variable that inherits the statistical properties
of the disordered landscape. We will now characterize their statistics
by measuring the mean and standard-deviation for a finite sample of
size $L$. Before anything, we note that, by construction

\vspace{-0.7cm}
\begin{equation}
\ket{\Psi_{0}^{\pm}}\!=\!\sum_{\mathbf{R}}\frac{e^{i2\pi\boldsymbol{\varphi}\cdot\mathbf{R}/L}}{L^{\frac{3}{2}}}\ket{\mathbf{R},\pm}\quad;\ket{\Psi_{\mathbf{k}}^{s}}\!=\sum_{\mathbf{R}}\frac{e^{i\mathbf{k}\cdot\mathbf{R}}}{L^{\frac{3}{2}}}\ket{\mathbf{R},s}
\end{equation}
and, since the disorder is assumed to act as a scalar within a unit-cell,
we have

\vspace{-0.7cm}

\begin{equation}
\bra{\Psi_{0}^{\pm}}\hat{V}_{d}\ket{\Psi_{0}^{\pm}}=\frac{1}{L^{3}}\sum_{\mathbf{R},\mathbf{Q}}e^{i2\pi\boldsymbol{\varphi}\cdot(\mathbf{R}-\mathbf{Q})/L}\bra{\mathbf{Q},\pm}\hat{V}_{d}\ket{\mathbf{R},\pm}=\frac{1}{L^{3}}\sum_{\mathbf{R}}V(\mathbf{R})
\end{equation}
and, likewise,

\vspace{-0.7cm}

\begin{align}
\bra{\Psi_{0}^{\pm}}\hat{V}_{d}\ketbra{\Psi_{\mathbf{k}}^{s}}{\Psi_{\mathbf{k}}^{s}}\hat{V}_{d}\ket{\Psi_{0}^{\pm}} & =\frac{1}{L^{6}}\sum_{\mathbf{R},\mathbf{Q}}\sum_{\mathbf{P},\mathbf{S}}\sum_{s=\pm}e^{i\mathbf{k}\cdot(\mathbf{Q}-\mathbf{P})}e^{i2\pi\boldsymbol{\varphi}\cdot(\mathbf{R}-\mathbf{S})/L}\times\\
 & \qquad\qquad\qquad\qquad\qquad\qquad\times V_{d}(\mathbf{R})V_{d}(\mathbf{Q})\delta_{\mathbf{R},\mathbf{P}}\delta_{\mathbf{Q},\mathbf{S}}\nonumber \\
 & =\frac{1}{L^{6}}\sum_{s=\pm}\sum_{\mathbf{R},\mathbf{Q}}e^{i(\mathbf{k}-2\pi\boldsymbol{\varphi}/L)\cdot(\mathbf{Q}-\mathbf{R})}V_{d}(\mathbf{R})V_{d}(\mathbf{Q}),\nonumber 
\end{align}
and finally we get to 

\vspace{-0.7cm}

\begin{align}
E_{0}^{{\scriptscriptstyle \pm}}[V_{d}\left(\mathbf{R}\right)] & =\pm\varepsilon_{0}+\frac{W}{L^{3}}\sum_{\mathbf{R}}V(\mathbf{R})-\frac{W^{2}}{L^{6}}\!\sum_{\mathbf{R},\mathbf{Q}}V_{d}(\mathbf{R})V_{d}(\mathbf{Q})\times\nonumber \\
 & \quad\times\left[\sum_{\mathbf{k}\in F\!B\!Z}\frac{e^{i(\mathbf{k}-2\pi\boldsymbol{\varphi}/L)\cdot(\mathbf{Q}-\mathbf{R})}}{\varepsilon^{+}(\mathbf{k})\mp\varepsilon_{0}}+\sum_{\mathbf{k}\in F\!B\!Z}\frac{e^{i(\mathbf{k}-2\pi\boldsymbol{\varphi}/L)\cdot(\mathbf{Q}-\mathbf{R})}}{\varepsilon^{-}(\mathbf{k})\mp\varepsilon_{0}}\right],\label{eq:PerturbationTheory-1}
\end{align}
which is the explicit formula of the functional. Considering only
the case $+$ only, we get

\vspace{-0.7cm}

\begin{align}
E_{0}^{{\scriptscriptstyle +}}[V_{d}\left(\mathbf{R}\right)] & =\varepsilon_{0}+\frac{W}{L^{3}}\sum_{\mathbf{R}}V(\mathbf{R})-\frac{W^{2}}{L^{6}}\!\sum_{\mathbf{R},\mathbf{Q}}V_{d}(\mathbf{R})V_{d}(\mathbf{Q})\times\nonumber \\
 & \quad\times\left[\sum_{\mathbf{k}\in F\!B\!Z'}\frac{e^{i(\mathbf{k}-2\pi\boldsymbol{\varphi}/L)\cdot(\mathbf{Q}-\mathbf{R})}}{\varepsilon_{\mathbf{k}}-\varepsilon_{0}}-\!\!\!\sum_{\mathbf{k}\in F\!B\!Z}\frac{e^{i(\mathbf{k}-2\pi\boldsymbol{\varphi}/L)\cdot(\mathbf{Q}-\mathbf{R})}}{\varepsilon_{\mathbf{k}}+\varepsilon_{0}}\right],\label{eq:PerturbationTheory-1-3}
\end{align}
By definition, we will be considering only samples for which the summation
over the entire lattice is zero. Hence, the functional gets an even
simpler form,

\vspace{-0.7cm}

\begin{align}
E_{0}^{{\scriptscriptstyle +}}[V_{d}\left(\mathbf{R}\right)] & =\varepsilon_{0}-\frac{W^{2}}{L^{6}}\!\sum_{\mathbf{R},\mathbf{Q}}V_{d}(\mathbf{R})V_{d}(\mathbf{Q})\times\nonumber \\
 & \qquad\times\left[\sum_{\mathbf{k}\in F\!B\!Z'}\frac{e^{i(\mathbf{k}-2\pi\boldsymbol{\varphi}/L)\cdot(\mathbf{Q}-\mathbf{R})}}{\varepsilon_{\mathbf{k}}-\varepsilon_{0}}-\!\!\!\!\sum_{\mathbf{k}\in F\!B\!Z}\frac{e^{i(\mathbf{k}-2\pi\boldsymbol{\varphi}/L)\cdot(\mathbf{Q}-\mathbf{R})}}{\varepsilon_{\mathbf{k}}+\varepsilon_{0}}\right]\nonumber \\
 & =\varepsilon_{0}-\frac{W^{2}}{L^{6}}\!\sum_{\mathbf{R},\mathbf{Q}}\!\!V_{d}(\mathbf{R})V_{d}(\mathbf{Q})\left[\!\!\sum_{\mathbf{k}\in F\!B\!Z'}\left(\!\frac{2\varepsilon_{0}e^{i(\mathbf{k}-2\pi\boldsymbol{\varphi}/L)\cdot(\mathbf{Q}-\mathbf{R})}}{\varepsilon_{\mathbf{k}}^{2}-\varepsilon_{0}^{2}}\!\right)-\frac{1}{\Delta_{\text{gap}}^{L}}\!\right]\label{eq:PerturbationTheory-1-1}
\end{align}
Now, we can evaluate the average over disorder realizations,\textit{
i.e.},

\vspace{-0.7cm}

\begin{align}
\av{E_{0}^{+}}_{d} & =\varepsilon_{0}+\frac{W^{2}}{L^{3}}\left[\sum_{\mathbf{k}\in F\!B\!Z'}\left(\frac{1}{\varepsilon_{\mathbf{k}}-\varepsilon_{0}}-\frac{1}{\varepsilon_{\mathbf{k}}+\varepsilon_{0}}\right)-\frac{1}{\Delta_{\text{gap}}^{L}}\right]\nonumber \\
 & \qquad\qquad\qquad=\varepsilon_{0}-\frac{W^{2}}{L^{3}}\left[\sum_{\mathbf{k}\in F\!B\!Z'}\left(\frac{2\varepsilon_{0}}{\varepsilon_{\mathbf{k}}^{2}-\varepsilon_{0}^{2}}\right)-\frac{1}{\Delta_{\text{gap}}^{L}}\right]
\end{align}
where we have put aside the coupling between the two lowest-lying
states. In the limit $L\to\infty$, we can see that

\vspace{-0.7cm}

\begin{equation}
\sum_{\mathbf{k}\in F\!B\!Z'}\left(\frac{2\varepsilon_{0}}{\varepsilon_{\mathbf{k}}^{2}-\varepsilon_{0}^{2}}\right)\sim L^{2}\quad;\text{ and }\frac{1}{\Delta_{\text{gap}}^{L}}\sim L,
\end{equation}
which means that the leading correction to the average value of the
first energy level will be 

\vspace{-0.7cm}

\begin{equation}
\av{E_{0}^{{\scriptscriptstyle \pm}}}_{d}=\pm\varepsilon_{0}\mp\mathcal{K}_{\boldsymbol{\varphi}}\frac{W^{2}}{L}+\mathcal{O}[\frac{1}{L^{2}}]
\end{equation}
This expression indicates what will be the average shift caused by
the disorder on the two levels closest to the Weyl node. Note that
both levels are equally drawn towards the node by the effect of disorder
and, in addition, the shift is proportional to $1/L$. Since $\varepsilon_{0}$
is also proportional to $1/L$, we conclude that the size of the finite
size gap is renormalized by disorder as follows:

\vspace{-0.7cm}

\begin{equation}
\frac{\Delta_{\text{gap}}^{d}\!(L)-\Delta_{\text{gap}}^{0,L}\!(L)}{\Delta_{\text{gap}}^{0,L}\!(L)}=-\mathcal{K}_{\boldsymbol{\theta}}\frac{W^{2}}{\varepsilon_{0}L}\propto L^{0}.
\end{equation}

From now on, we will always disregard de $1/\Delta_{\text{gap}}^{L}$,
which is always non-leading in $1/L$. Meanwhile, an analogous analysis
can be done to obtain the width of the distribution for this first
level. For this, we calculate the squared-average of $E_{0}^{\pm}$,
that is

\vspace{-0.7cm}

\begin{align}
\!\!\!\!\!\!\!\!\!\!\!\!\!\!\av{\left(E_{0}^{{\scriptscriptstyle +}}\right)^{2}}_{d} & =\varepsilon_{0}^{2}-4\varepsilon_{0}\frac{W^{2}}{L^{6}}\!\sum_{\mathbf{R},\mathbf{Q}}\av{V_{d}(\mathbf{R})V_{d}(\mathbf{Q})}_{d}\left[\sum_{\mathbf{k}\in F\!B\!Z'}\left(\frac{2\varepsilon_{0}e^{i(\mathbf{k}-2\pi\boldsymbol{\theta}/L)\cdot(\mathbf{Q}-\mathbf{R})}}{\varepsilon_{\mathbf{k}}^{2}-\varepsilon_{0}^{2}}\right)\right]\nonumber \\
 & \qquad+\frac{16W^{4}}{L^{12}}\!\sum_{\mathbf{R},\mathbf{Q}}\sum_{\mathbf{L},\mathbf{S}}\av{V_{d}(\mathbf{R})V_{d}(\mathbf{Q})V_{d}(\mathbf{L})V_{d}(\mathbf{S})}_{d}\times\nonumber \\
 & \qquad\times\left[\sum_{\mathbf{k},\mathbf{q}\in F\!B\!Z'}\left[\frac{\varepsilon_{0}^{2}}{\left(\varepsilon_{\mathbf{k}}^{2}-\varepsilon_{0}^{2}\right)\left(\varepsilon_{\mathbf{q}}^{2}-\varepsilon_{0}^{2}\right)}\right]e^{i(\mathbf{k}-2\pi\boldsymbol{\theta}/L)\cdot(\mathbf{Q}-\mathbf{R})}e^{i(\mathbf{q}-2\pi\boldsymbol{\theta}/L)\cdot(\mathbf{Q}-\mathbf{R})}\right]
\end{align}
and, now, we can make use of the statistical properties of $V(\mathbf{R})$,
given in Eqs.\,\eqref{eq:V1}-\eqref{eq:V2}, which yields a total
of five non-zero terms, \textit{i.e}.,

\vspace{-0.7cm}

\begin{align}
\!\!\!\!\!\!\!\!\!\!\!\!\!\!\!\!\!\av{\left(E_{0}^{{\scriptscriptstyle +}}\right)^{2}}_{d}\!\!\! & =\varepsilon_{0}^{2}-8\varepsilon_{0}\frac{W^{2}}{L^{3}}\!\left[\sum_{\mathbf{k}\in F\!B\!Z'}\left(\frac{\varepsilon_{0}}{\varepsilon_{\mathbf{k}}^{2}-\varepsilon_{0}^{2}}\right)\right]\!\!\!\!\!\!\!\!\!\!\nonumber \\
 & \qquad+\frac{16W^{4}}{L^{6}}\!\left[\sum_{\mathbf{k},\mathbf{q}\in F\!B\!Z'}\frac{\varepsilon_{0}^{2}}{\left(\varepsilon_{\mathbf{k}}^{2}-\varepsilon_{0}^{2}\right)\left(\varepsilon_{\mathbf{q}}^{2}-\varepsilon_{0}^{2}\right)}\right]\nonumber \\
 & \qquad\quad+\frac{16W^{4}}{L^{12}}\!\left[\sum_{\mathbf{k},\mathbf{q}\in F\!B\!Z'}\left[\frac{\varepsilon_{0}^{2}}{\left(\varepsilon_{\mathbf{k}}^{2}-\varepsilon_{0}^{2}\right)\left(\varepsilon_{\mathbf{q}}^{2}-\varepsilon_{0}^{2}\right)}\right]\sum_{\mathbf{R},\mathbf{Q}}e^{i(\mathbf{k}-\mathbf{q})\cdot(\mathbf{Q}-\mathbf{R})}\right]\nonumber \\
 & \qquad\qquad+\frac{16W^{4}}{L^{12}}\!\left[\sum_{\mathbf{k},\mathbf{q}\in F\!B\!Z'}\left[\frac{\varepsilon_{0}^{2}}{\left(\varepsilon_{\mathbf{k}}^{2}-\varepsilon_{0}^{2}\right)\left(\varepsilon_{\mathbf{q}}^{2}-\varepsilon_{0}^{2}\right)}\right]\sum_{\mathbf{R},\mathbf{Q}}e^{i(\mathbf{k}+\mathbf{q}-4\pi\boldsymbol{\varphi}/L)\cdot(\mathbf{Q}-\mathbf{R})}\right].
\end{align}
Now, the first three terms can be identified as $\av{E_{0}^{{\scriptscriptstyle \pm}}}_{d}^{2}$,
meaning that the variance of $E_{0}^{\pm}$ reads simply

\vspace{-0.7cm}

\begin{align}
\!\!\!\!\!\!\!\!\!\!\av{\left(E_{0}^{{\scriptscriptstyle \pm}}\right)^{2}}_{d} & \!\!\!-\!\av{E_{0}^{{\scriptscriptstyle \pm}}}_{d}^{2}=\frac{32W^{4}}{L^{12}}\!\left[\sum_{\mathbf{k},\mathbf{q}\in F\!B\!Z'}\left[\frac{\varepsilon_{0}^{2}}{\left(\varepsilon_{\mathbf{k}}^{2}-\varepsilon_{0}^{2}\right)\left(\varepsilon_{\mathbf{q}}^{2}-\varepsilon_{0}^{2}\right)}\right]\sum_{\mathbf{R},\mathbf{Q}}e^{i(\mathbf{k}-\mathbf{q})\cdot(\mathbf{Q}-\mathbf{R})}\right],\!\!\!\!\!\!\!\!\!\!
\end{align}
which, finally yields

\vspace{-0.7cm}

\begin{align}
\av{\left(E_{0}^{{\scriptscriptstyle \pm}}\right)^{2}}_{d}\!\!-\av{E_{0}^{{\scriptscriptstyle \pm}}}_{d}^{2} & =\frac{4W^{4}}{L^{6}}\!\!\sum_{\mathbf{k}\in F\!B\!Z'}\left[\frac{\varepsilon_{0}^{2}}{\left(\varepsilon_{\mathbf{k}}^{2}-\varepsilon_{0}^{2}\right)^{2}}\right].
\end{align}
Once again, we can show that 

\vspace{-0.7cm}

\begin{equation}
\sum_{\mathbf{k}\in F\!B\!Z'}\left[\frac{\varepsilon_{0}^{2}}{\left(\varepsilon_{\mathbf{k}}^{2}-\varepsilon_{0}^{2}\right)^{2}}\right]\sim L^{2},
\end{equation}
meaning that 

\vspace{-0.7cm}
\begin{equation}
\av{\left(E_{0}^{{\scriptscriptstyle \pm}}\right)^{2}}_{d}-\av{E_{0}^{{\scriptscriptstyle \pm}}}_{d}^{2}\propto\left(\frac{W}{L}\right)^{4},
\end{equation}
which yields a broadening that goes as $W^{2}\!/L^{2}$, in perturbation
theory.

\lhead[\MakeUppercase{Bibliography}]{}

\rhead[]{}

\lfoot[\thepage]{}

\cfoot[]{}

\rfoot[]{\thepage}

\bibliographystyle{cv}
\bibliography{thesisExample}

\cleardoublepage{}

\lhead[]{Nomenclature}

\cfoot[]{}

\rhead[Nomenclature]{}

\printnomenclature[2.5cm]

\end{document}